\documentclass[%
titlepage=false,%
toc=listof,toc=bibliography,%
chapterprefix=true,
twoside=false
]{scrbook}

\usepackage[a4paper]{geometry}
\usepackage[T1]{fontenc}

\usepackage[utf8]{inputenc}

\usepackage{subcaption} 

\usepackage{amsmath}
\numberwithin{equation}{section}

\def\nodraft

\usepackage{amssymb}

\usepackage{amsfonts}

\usepackage{amsthm}

\usepackage[bb=boondox]{mathalfa}

\usepackage[refpage,intoc]{nomencl}
\renewcommand{\nomname}{List of Symbols}
\makenomenclature

\usepackage[backend=biber,natbib]{biblatex}
\DeclareFieldFormat{postnote}{#1}
\DeclareFieldFormat{multipostnote}{#1}

\bibliography{literature}

\usepackage{tikz}

\usetikzlibrary{positioning,calc,shapes.geometric}
\usetikzlibrary{decorations.markings}
\usetikzlibrary{decorations.pathmorphing}
\usetikzlibrary{patterns}
\usepackage{pgfplots}
\pgfplotsset{compat=1.14}

\usepackage{hyperref}

\newlength\figureheight
\newlength\figurewidth
\newcommand\floor[1]{\lfloor#1\rfloor}

\newtheorem{thm}{Theorem}
\numberwithin{thm}{section}
\newtheorem{lmm}{Lemma}
\numberwithin{lmm}{section}
\newtheorem{crll}{Corollary}
\numberwithin{crll}{section}
\newtheorem{prop}{Proposition}
\numberwithin{prop}{section}

\theoremstyle{definition}
\newtheorem{defn}{Definition}
\numberwithin{defn}{section}
\newtheorem{expl}{Example}
\numberwithin{expl}{section}

\theoremstyle{remark}
\newtheorem{rmk}{Remark}
\numberwithin{rmk}{section}

\newcommand{\R}{\mathbb{R}}
\newcommand{\Q}{\mathbb{Q}}
\newcommand{\N}{\mathbb{N}}
\newcommand{\C}{\mathbb{C}}
\newcommand{\Z}{\mathbb{Z}}

\newcommand{\mulind}[1]{\mathbf{#1}}

\newcommand{\risefac}[2]{{#1}^{\overline{#2}}}

\newcommand{\bigO}{\mathcal{O}}
\newcommand{\smallO}{o}
\newcommand{\Fop}{\mathcal{F}}
\newcommand{\asyOp}{\mathcal{A}} %
\newcommand{\asyOpV}[3]{{\textbf{a}^{#1}_{#3}}}  %

\newcommand*\fring[3]{\R[[#1]]^{#2}_{#3}}
\newcommand*\cring[3]{\C[[#1]]^{#2}_{#3}}
\newcommand*\G[3]{\Gamma^{#2}_{#3}\left(#1\right)}

\newcommand{\unit}{\text{\normalfont{u}}}
\newcommand{\counit}{\epsilon}
\newcommand{\powerset}[1]{\textbf{2}^{#1}}

\newcommand{\one}{\mathbb{1}}
\renewcommand{\deg}[1]{d^{(#1)}}
\newcommand{\nvd}[1]{k^{(#1)}}

\newcommand{\Gaul}{\mathcal{G}}
\newcommand{\Gl}{\mathfrak{G}^{\text{lab}}}
\newcommand{\Gul}{\mathfrak{G}}

\newcommand{\Sact}{\mathcal{S}}

\newcommand{\allG}{\mathfrak{X}}
\newcommand{\allGc}{\mathfrak{X}^c}

\newcommand{\legs}{H^{\text{legs}}}
\newcommand{\comps}{C}

\newcommand{\insertionplaces}[2]{\mathcal{I}(#1, #2)}

\DeclareMathOperator{\Diff}{Diff}
\DeclareMathOperator{\id}{id}
\DeclareMathOperator{\im}{im}
\DeclareMathOperator{\Aut}{Aut}

\DeclareMathOperator{\res}{res}
\DeclareMathOperator{\skl}{skl}
\DeclareMathOperator{\re}{re}
\DeclareMathOperator{\sk}{sk}

\DeclareMathOperator{\orb}{Orb}
\DeclareMathOperator{\stab}{Stab}

\newcommand{\Hext}{\legs}

\newcommand{\hopffg}{\mathcal{H}^\text{\normalfont{fg}}}
\newcommand{\hopffgs}{\widetilde{\mathcal{H}}^\text{\normalfont{fg}}}
\newcommand{\hopfpos}{\mathcal{H}^\text{\normalfont{P}}}
\newcommand{\hopflat}{\mathcal{H}^\text{\normalfont{L}}}
\newcommand{\hopflats}{\widetilde{\mathcal{H}}^\text{\normalfont{L}}}

\newcommand{\subdiags}{\mathcal{P}}
\newcommand{\sdsubdiags}{\mathcal{P}^\text{s.d.}}
\newcommand{\sdsubdiagsn}{\widetilde{\mathcal{P}}^\text{s.d.}}

\newcommand{\chargroup}[2]{{\Phi^{#1}_{#2}}}
\newcommand{\fields}{F}

\newcommand{\sgset}{\mathfrak{P}}
\newcommand{\subclass}{\mathfrak{K}}
\newcommand{\residues}{\mathcal{R}}
\newcommand{\residuesstar}{\residues^*}

\newcommand{\meet}{\wedge}
\newcommand{\join}{\vee}

\newcommand{\fourvtx}{{%
\ifmmode
\usebox{\fgsimplefourvtx}
\else
\newsavebox{\fgsimplefourvtx}
\savebox{\fgsimplefourvtx}{%
\begin{tikzpicture}[x=1ex,y=1ex,baseline={([yshift=-.5ex]current bounding box.center)}] \coordinate (v) ; \def \n {4}; \def \rad {.8}; \filldraw[white] (v) circle (\rad); \foreach \s in {1,...,5} { \def \angle {45+360/\n*(\s - 1)}; \coordinate (u) at ([shift=({\angle}:\rad)]v); \draw (v) -- (u); } \filldraw (v) circle (1pt); \end{tikzpicture}%
}
\fi}}
\newcommand{\fourvtxgluon}{{%
\ifmmode
\usebox{\fgqcdfourvtx}
\else
\newsavebox{\fgqcdfourvtx}
\savebox{\fgqcdfourvtx}{%
\begin{tikzpicture}[x=1ex,y=1ex,baseline={([yshift=-.5ex]current bounding box.center)}] \coordinate (v) ; \def \n {4}; \def \rad {1.2}; \foreach \s in {1,...,5} { \def \angle {45+360/\n*(\s - 1)}; \coordinate (u) at ([shift=({\angle}:\rad)]v); \draw[gluon] (v) -- (u); } \filldraw (v) circle (1pt); \end{tikzpicture}%
}
\fi}}
\newcommand{\twothreevtxgluon}{{\ifmmode
\usebox{\fgtwothreevtxgluon}
\else
\newsavebox{\fgtwothreevtxgluon}
\savebox{\fgtwothreevtxgluon}{%
    \begin{tikzpicture}[x=1ex,y=1ex,baseline={([yshift=-.5ex]current bounding box.center)}] \coordinate (i1); \coordinate[below=1 of i1] (i2); \coordinate[right=1 of i1] (v1); \coordinate[right=1 of i2] (v2); \coordinate[right=1 of v1] (o1); \coordinate[right=1 of v2] (o2); \draw[gluon] (i1) -- (v1); \draw[gluon] (i2) -- (v2); \draw[gluon] (v1) -- (o1); \draw[gluon] (v1) -- (v2); \draw[gluon] (v2) -- (o2); \filldraw (v1) circle (1pt); \filldraw (v2) circle (1pt); \end{tikzpicture}
}
\fi}}

\pgfdeclaredecoration{complete sines}{initial}
{
    \state{initial}[
        width=+0pt,
        next state=sine,
        persistent precomputation={\pgfmathsetmacro\matchinglength{
            \pgfdecoratedinputsegmentlength / int(\pgfdecoratedinputsegmentlength/\pgfdecorationsegmentlength)}
            \setlength{\pgfdecorationsegmentlength}{\matchinglength pt}
        }] {}
    \state{sine}[width=\pgfdecorationsegmentlength]{
        \pgfpathsine{\pgfpoint{0.25\pgfdecorationsegmentlength}{0.5\pgfdecorationsegmentamplitude}}
        \pgfpathcosine{\pgfpoint{0.25\pgfdecorationsegmentlength}{-0.5\pgfdecorationsegmentamplitude}}
        \pgfpathsine{\pgfpoint{0.25\pgfdecorationsegmentlength}{-0.5\pgfdecorationsegmentamplitude}}
        \pgfpathcosine{\pgfpoint{0.25\pgfdecorationsegmentlength}{0.5\pgfdecorationsegmentamplitude}}
}
    \state{final}{}
}

\tikzset{
    photon/.style={decorate, 
        decoration={snake, amplitude=.8pt, segment length=1.5pt}
    },
    fermion/.style={postaction={decorate},
        decoration={markings, mark=at position .7 with {\arrow{>}}}
    },
    gluon/.style={decorate,  
        decoration={coil, amplitude=.8pt, segment length=1.5pt}
    },
    meson/.style={
        dash pattern=on 1pt off .5 pt
    }
}

\begin{document}

\ifmmode
\usebox{\fgtadpoletwo}
\else
\newsavebox{\fgtadpoletwo}
\savebox{\fgtadpoletwo}{%
%
}
\fi
\ifmmode
\usebox{\fgbananathree}
\else
\newsavebox{\fgbananathree}
\savebox{\fgbananathree}{%
%
%
}
\fi
\ifmmode
\usebox{\fgbananathreeandhandle}
\else
\newsavebox{\fgbananathreeandhandle}
\savebox{\fgbananathreeandhandle}{%
%
%
}
\fi
\ifmmode
\usebox{\fgcloseddunce}
\else
\newsavebox{\fgcloseddunce}
\savebox{\fgcloseddunce}{%
%
%
}
\fi
\ifmmode
\usebox{\fgdbleye}
\else
\newsavebox{\fgdbleye}
\savebox{\fgdbleye}{%
%
%
}
\fi
\ifmmode
\usebox{\fgdblhandle}
\else
\newsavebox{\fgdblhandle}
\savebox{\fgdblhandle}{%
%
%
}
\fi
\ifmmode
\usebox{\fghandle}
\else
\newsavebox{\fghandle}
\savebox{\fghandle}{%
%
%
}
\fi
\ifmmode
\usebox{\fgpropellerthree}
\else
\newsavebox{\fgpropellerthree}
\savebox{\fgpropellerthree}{%
%
%
}
\fi
\ifmmode
\usebox{\fgtwohandles}
\else
\newsavebox{\fgtwohandles}
\savebox{\fgtwohandles}{%
%
%
}
\fi
\ifmmode
\usebox{\fgtwobananasthree}
\else
\newsavebox{\fgtwobananasthree}
\savebox{\fgtwobananasthree}{%
%
%
}
\fi
\ifmmode
\usebox{\fgtadpoletwoclosed}
\else
\newsavebox{\fgtadpoletwoclosed}
\savebox{\fgtadpoletwoclosed}{%
%
%
}
\fi
\ifmmode
\usebox{\fgbananafour}
\else
\newsavebox{\fgbananafour}
\savebox{\fgbananafour}{%
%
%
}
\fi
\ifmmode
\usebox{\fgthreebubble}
\else
\newsavebox{\fgthreebubble}
\savebox{\fgthreebubble}{%
%
%
}
\fi
\ifmmode
\usebox{\fgcontractedpropellerthree}
\else
\newsavebox{\fgcontractedpropellerthree}
\savebox{\fgcontractedpropellerthree}{%
%
%
}
\fi
\ifmmode
\usebox{\fgcontracteddblhandle}
\else
\newsavebox{\fgcontracteddblhandle}
\savebox{\fgcontracteddblhandle}{%
%
%
}
\fi
\ifmmode
\usebox{\fgcontractedcloseddunce}
\else
\newsavebox{\fgcontractedcloseddunce}
\savebox{\fgcontractedcloseddunce}{%
%
%
}
\fi
\ifmmode
\usebox{\fgthreerose}
\else
\newsavebox{\fgthreerose}
\savebox{\fgthreerose}{%
%
%
}
\fi
\ifmmode
\usebox{\fgdblcontractedpropellerthree}
\else
\newsavebox{\fgdblcontractedpropellerthree}
\savebox{\fgdblcontractedpropellerthree}{%
%
%
}
\fi
\ifmmode
\usebox{\fgbananathreewithbubble}
\else
\newsavebox{\fgbananathreewithbubble}
\savebox{\fgbananathreewithbubble}{%
%
%
}
\fi
\ifmmode
\usebox{\fgtwotadpoletwos}
\else
\newsavebox{\fgtwotadpoletwos}
\savebox{\fgtwotadpoletwos}{%
%
%
}
\fi
\ifmmode
\usebox{\fgbananathreeandtadpoletwo}
\else
\newsavebox{\fgbananathreeandtadpoletwo}
\savebox{\fgbananathreeandtadpoletwo}{%
%
%
}
\fi
\ifmmode
\usebox{\fghandleandtadpoletwo}
\else
\newsavebox{\fghandleandtadpoletwo}
\savebox{\fghandleandtadpoletwo}{%
%
%
}
\fi
\ifmmode
\usebox{\fgbananathreehandle}
\else
\newsavebox{\fgbananathreehandle}
\savebox{\fgbananathreehandle}{%
%
%
}
\fi
\ifmmode
\usebox{\fgwheelthree}
\else
\newsavebox{\fgwheelthree}
\savebox{\fgwheelthree}{%
%
%
}
\fi
\ifmmode
\usebox{\fgsimpleprop}
\else
\newsavebox{\fgsimpleprop}
\savebox{\fgsimpleprop}{%
    %
%
}
\fi
\ifmmode
\usebox{\fgsimplephotonprop}
\else
\newsavebox{\fgsimplephotonprop}
\savebox{\fgsimplephotonprop}{%
    %
%
}
\fi
\ifmmode
\usebox{\fgsimplefermionprop}
\else
\newsavebox{\fgsimplefermionprop}
\savebox{\fgsimplefermionprop}{%
    %
%
}
\fi
\ifmmode
\usebox{\fgsimplemesonprop}
\else
\newsavebox{\fgsimplemesonprop}
\savebox{\fgsimplemesonprop}{%
    %
%
}
\fi
\ifmmode
\usebox{\fgsimplenullvtx}
\else
\newsavebox{\fgsimplenullvtx}
\savebox{\fgsimplenullvtx}{%
%
%
}
\fi
\ifmmode
\usebox{\fgsimpleonevtx}
\else
\newsavebox{\fgsimpleonevtx}
\savebox{\fgsimpleonevtx}{%
%
%
}
\fi
\ifmmode
\usebox{\fgsimpletwovtx}
\else
\newsavebox{\fgsimpletwovtx}
\savebox{\fgsimpletwovtx}{%
%
%
}
\fi
\ifmmode
\usebox{\fgsimplethreevtx}
\else
\newsavebox{\fgsimplethreevtx}
\savebox{\fgsimplethreevtx}{%
%
%
}
\fi
\ifmmode
\usebox{\fgsimplefourvtx}
\else
\newsavebox{\fgsimplefourvtx}
\savebox{\fgsimplefourvtx}{%
%
%
}
\fi
\ifmmode
\usebox{\fgsimplefivevtx}
\else
\newsavebox{\fgsimplefivevtx}
\savebox{\fgsimplefivevtx}{%
%
%
}
\fi
\ifmmode
\usebox{\fgsimplesixvtx}
\else
\newsavebox{\fgsimplesixvtx}
\savebox{\fgsimplesixvtx}{%
%
%
}
\fi
\ifmmode
\usebox{\fgsimpleqedvtx}
\else
\newsavebox{\fgsimpleqedvtx}
\savebox{\fgsimpleqedvtx}{%
%
%
}
\fi
\ifmmode
\usebox{\fgqcdfourvtx}
\else
\newsavebox{\fgqcdfourvtx}
\savebox{\fgqcdfourvtx}{%
%
%
}
\fi
\ifmmode
\usebox{\fgsimpleyukvtx}
\else
\newsavebox{\fgsimpleyukvtx}
\savebox{\fgsimpleyukvtx}{%
%
%
}
\fi
\ifmmode
\usebox{\fgtwothreevtxgluon}
\else
\newsavebox{\fgtwothreevtxgluon}
\savebox{\fgtwothreevtxgluon}{%
    %
}
\fi

\ifmmode
\usebox{\fgsimpletreehandle}
\else
\newsavebox{\fgsimpletreehandle}
\savebox{\fgsimpletreehandle}{%
    %
%
}
\fi
\ifmmode
\usebox{\fgjthreesimplevtx}
\else
\newsavebox{\fgjthreesimplevtx}
\savebox{\fgjthreesimplevtx}{%
%
%
}
\fi
\ifmmode
\usebox{\fgonetadpolephithree}
\else
\newsavebox{\fgonetadpolephithree}
\savebox{\fgonetadpolephithree}{%
    %
%
}
\fi
\ifmmode
\usebox{\fgconetadpolephithree}
\else
\newsavebox{\fgconetadpolephithree}
\savebox{\fgconetadpolephithree}{%
    %
%
}
\fi
\ifmmode
\usebox{\fgtwojoneloopbubblephithree}
\else
\newsavebox{\fgtwojoneloopbubblephithree}
\savebox{\fgtwojoneloopbubblephithree}{%
    %
%
}
\fi
\ifmmode
\usebox{\fgtwoconeloopbubblephithree}
\else
\newsavebox{\fgtwoconeloopbubblephithree}
\savebox{\fgtwoconeloopbubblephithree}{%
    %
%
}
\fi
\ifmmode
\usebox{\fgthreejoneltrianglephithree}
\else
\newsavebox{\fgthreejoneltrianglephithree}
\savebox{\fgthreejoneltrianglephithree}{%
%
%
}
\fi
\ifmmode
\usebox{\fgthreeconeltrianglephithree}
\else
\newsavebox{\fgthreeconeltrianglephithree}
\savebox{\fgthreeconeltrianglephithree}{%
%
%
}
\fi
\ifmmode
\usebox{\fgthreejonelpropinsphithree}
\else
\newsavebox{\fgthreejonelpropinsphithree}
\savebox{\fgthreejonelpropinsphithree}{%
    %
%
}
\fi

\title{\vspace{-1.5cm}Graphs in perturbation theory: \\Algebraic structure and asymptotics}
\author{Michael Borinsky}

\pagenumbering{roman}
\makeatletter

\thispagestyle{empty}

\date{}
\maketitle

This is an updated version of my PhD thesis which I submitted on February 5th 2018 and defended on April 27th 2018. The \href{http://dx.doi.org/10.18452/19201}{original and official version} is hosted by the Humboldt-Universität zu Berlin. 

The referees were
\begin{itemize}
\item My supervisor Dirk Kreimer (Humboldt-Universität zu Berlin),
\item David Broadhurst (Open University) and
\item Gerald Dunne (University of Connecticut).
\end{itemize}
\makeatother

\section*{Abstract}
This thesis provides an extension of the work of Dirk Kreimer and Alain Connes on the Hopf algebra structure of Feynman graphs and renormalization to general graphs. Additionally, an algebraic structure of the asymptotics of formal power series with factorial growth, which is compatible with the Hopf algebraic structure, is introduced. 

The Hopf algebraic structure on graphs permits the explicit enumeration of graphs with constraints for the allowed subgraphs. In the case of Feynman diagrams a lattice structure, which will be introduced, exposes additional unique properties for physical quantum field theories.
The differential ring of factorially divergent power series allows the extraction of asymptotic results of implicitly defined power series with vanishing radius of convergence. Together both structures provide an algebraic formulation of large graphs with constraints on the allowed subgraphs. 
These structures are motivated by and used to analyze renormalized zero-dimensional quantum field theory at high orders in perturbation theory. 

As a pure application of the Hopf algebra structure, an Hopf algebraic interpretation of the Legendre transformation in quantum field theory is given. 
The differential ring of factorially divergent power series will be used to solve two asymptotic counting problems from combinatorics: The asymptotic number of connected chord diagrams and the number of simple permutations. For both asymptotic solutions, all order asymptotic expansions are provided as generating functions in closed form. Both structures are combined in an application to zero-dimensional quantum field theory. Various quantities are explicitly given asymptotically in the zero-dimensional version of $\varphi^3$, $\varphi^4$, QED, quenched QED and Yukawa theory with their all order asymptotic expansions.

\tableofcontents
\listoffigures

\listoftables

\thispagestyle{plain}
\chapter*{Acknowledgements}

There are many people who contributed directly or indirectly to this work. Without their advice, support or both, I would not have come as far as I have. 

First and foremost, I wish to thank my supervisor Dirk Kreimer for his great support and encouragement during the process which led to this thesis. While I enjoyed much freedom to think about problems and to develop my own style of tackling them, he pushed me to write up me ideas and provided sound advice when it was necessary. I consider myself very lucky to have had his guidance during my PhD. The exceptional environment that he created with his group in Berlin gave me an inspiring starting point to dive into the kaleidoscopic world of quantum field theory. Vivid discussions between mathematicians and physicists, young students and established scientists regularly resulted in new viewpoints on tricky problems.

I am in dept to David Broadhurst for his steady encouragement. From early on, he motivated me to develop my project and encouraged me to step forward with my ideas. His wise advice and endless passion for mathematics and \textit{was die Welt im Innersten zusammen hält} has been a great source of motivation and inspiration for me. 

Equally, I wish to thank Karen Yeats for her strong support in the last years. She has been a very important source of fresh ideas and reassurance. The two research stays in her group in Vancouver and Waterloo enabled me to promote my research, to meet many inspiring people and, last but not least, experience great maple syrup based adventures. 

Special thanks also to Marko Berghoff, Erik Panzer, Inês Aniceto, Oliver Schnetz, Walter van Suijlekom, Julien Courtiel, Gerald Dunne, David Sauzin and Dominique Manchon. They helped me either by having long fruitful discussions with me, patiently explaining complicated mathematics, spotting flaws in my arguments or by having kind words in critical moments.

Without Gregor, Sylvia, David, Konrad, Claire, Iain, Ben, Lutz, Henry, Matthias, Julian, Marcel, Johannes, Markus, Lucas, David, Christian, Isabella, Susi, Dima and everyone else in the Kreimer and Yeats Gangs, my PhD would not have been half as fun as it was. I wish to thank you all.

I will not try to list them out of fear of forgetting somebody unintentionally, but all the great people, who I met during summer schools, conferences, workshops or research stays should know that I wish to thank them for the many discussions and the wonderful time we had. 

I wish to thank the Studienstiftung des deutschen Volkes, the International Max Planck Research Schools for `Mathematical and Physical Aspects of Gravitation, Cosmology and Quantum Field Theory', the Research Training Group `Mass, Spectrum, Symmetry' as well as the Erwin Schrödinger International Institute for Mathematics and Physics for generous financial support. 

Of course, I also have to thank all my other amazing friends, which I am lucky to have and who are not part of the world of propagators and imaginary dimensions. Especially, Flo, Jonas, Lorenz and Nadine took a big burden while they had to go through the ups and downs of the PhD student life with me. The whole s-crew was responsible for the best after work hours in Berlin. Thilo and Lucas seamlessly took over this role in Austria and Canada. Janna, although she had to put up with me during the last months, supported me all the time.

\thispagestyle{plain}

\chapter{Introduction}
\pagenumbering{arabic}

This thesis is about graphs and two algebraic structures which can be associated with them. The first algebraic structure appears while enumerating \textit{large graphs}. It captures the asymptotic behaviour of power series associated to graph counting problems. Second is the Hopf algebraic structure which gives an algebraic description of \textit{subgraph} structures of graphs. The Hopf algebraic structure permits the explicit enumeration of graphs with constraints for the allowed subgraphs. Together both structures give an algebraic formulation of large graphs with forbidden subgraphs. The detailed analysis of both these structures is motivated by perturbative quantum field theory. 

\section{Motivation from quantum field theory}
Perturbation theory, augmented with the powerful combinatorial method of Feynman diagrams, remains the status quo for performing quantum field theory (QFT) and therefore particle physics calculations. 
Each term in the perturbative expansion is a sum of integrals. These integrals can be depicted as Feynman diagrams and they require \textit{renormalization} to give meaningful results. 

Although the initial hurdles to use perturbation theory in quantum field theory were overcome with the invention of renormalization in the 1940s, the technique is still plagued with conceptual and practical problems which hinder progress in our understanding of matter at the fundamental level. One of these problems is the inaccessibility of information about the perturbation expansion at higher order. This inaccessibility limits the accuracy of theoretical predictions and sets the solution of intrinsically large coupling problems beyond the reach of existing theoretical tools. 
Because of the high demand for extremely accurate theoretical calculations from present day experiments, these problems are not merely unsolved academic exercises. They form a severe bottleneck for the general endeavor of understanding nature at the fundamental level.

The purpose of this thesis is to tackle these problems by studying perturbation theory with renormalization in quantum field theory at large loop orders. This will be approached by exploiting the combinatorial structure of its diagrammatic interpretation. 

\subsection{Divergent perturbation expansions}
Dyson's famous argument states that the perturbation expansions in quantum field theory are divergent \cite{dyson1952divergence}. This means that perturbative expansions of observables in those theories usually have a vanishing radius of convergence. For an observable $f$ expanded in the parameter $\hbar$,
\begin{align} \label{eqn:perturbative_expansion} f(\hbar) = \sum_{n=0}^\infty f_n \hbar^n = f_0 + f_1 \hbar + f_2 \hbar^2 + \ldots \end{align}
the sum in this expression will not converge for any value of $\hbar$ other than $0$.

This divergence can be associated with the large growth of the coefficients $f_n$ for $n \rightarrow \infty$.
In quantum field theory, this large growth of the coefficients is believed to be governed by the proliferation of Feynman diagrams, which contribute to the coefficients $f_n$, with increasing loop number. 

The analysis of this large order behavior led to many important results reaching far beyond the scope of quantum field theory \cite{bender1969anharmonic,bender1973anharmonic,le2012large}. Moreover, the divergence of the perturbation expansion in QFT is linked to non-perturbative effects \cite{alvarez2004langer,garoufalidis2012asymptotics,argyres2012semi,dunne2014generating,marino2014lectures}.

The extraction of large order results from realistic quantum field theories becomes very involved when \textit{renormalization} comes into play. For instance, the relationship between \textit{renormalons}, which are avatars of renormalization at large order, and \textit{instantons} \cite{lautrup1977high,Zichichi:1979gj}, classical field configurations, which are in close correspondence with the large order behavior of the theory \cite{lipatov1977divergence}, remains elusive \cite{suslov2005divergent}.

\subsection{The limits of explicit integration}
The most obvious way to study perturbation theory at higher order is to explicitly calculate the values of the contributing integrals. 
Although this program is impeded by the sheer difficulty of evaluating individual Feynman integrals, there has recently been significant progress in this direction. A systematic integration approach, which exploits the rich mathematical structure of Feynman integrals, has led to a breakthrough in the achievable accuracy of quantum field theory calculations. In \cite{kompaniets2017minimally}, the $\varphi^4$-theory $\beta$-function has been calculated analytically up to sixth order in perturbation theory. Additionally, the seven order calculation was recently completed \cite{schnetz2016numbers}. These new techniques make heavy use of deep mathematical insights regarding the structure of Feynman integrals. 

Considering the high amounts of intellectual energy that was and is being invested in performing these calculations at higher and higher loop orders, it seems worthwhile to look for \textit{asymptotic} alternatives for these techniques. Instead of resulting in a harder problem for each further loop order, such asymptotic methods for calculating observables should give an approximate result in the large loop order limit with successively more sophisticated corrections for lower loop orders. Such a calculation has been performed for instance in \cite{mckane1984nonperturbative,mckane1978instanton} for $\varphi^4$-theory based on a delicate combination of instanton and renormalization considerations. More elaborate results in the $\mathbb{C}\mathbb{P}^{N-1}$-model \cite{affleck1980testing,dunne2012resurgence} give further hints regarding the feasibility of this approach.

In this context, this thesis is an attempt to map the algebraic and combinatorial ground for such techniques.
This attempt is rooted in the well-explored perturbative regime of Feynman diagrams.

\section{Overview and contributions}

\subsection{Algebraic formulation}
In Chapter \ref{chp:graphs} we will start with basic definitions of graphs in an algebraic setting. The framework for perturbation theory based on the works of Kreimer and Yeats \cite{kreimer2006anatomy,Yeats2008} will be introduced in the style of the \textit{symbolic method} from combinatorics introduced by Flajolet and Sedgewick \cite{flajolet2009analytic} or its largely similar sibling, the theory of species \cite{bergeron1998combinatorial} by Bergeron, Labelle and Leroux. 
We will define expressions such as
\begin{align*} \one+ \frac18 {  \ifmmode \usebox{\fghandle} \else \newsavebox{\fghandle} \savebox{\fghandle}{ \begin{tikzpicture}[x=1ex,y=1ex,baseline={([yshift=-.5ex]current bounding box.center)}] \coordinate (v0); \coordinate [right=1.5 of v0] (v1); \coordinate [left=.7 of v0] (i0); \coordinate [right=.7 of v1] (o0); \draw (v0) -- (v1); \filldraw (v0) circle (1pt); \filldraw (v1) circle (1pt); \draw (i0) circle(.7); \draw (o0) circle(.7); \end{tikzpicture} } \fi } + \frac{1}{12} {  \ifmmode \usebox{\fgbananathree} \else \newsavebox{\fgbananathree} \savebox{\fgbananathree}{ \begin{tikzpicture}[x=1ex,y=1ex,baseline={([yshift=-.5ex]current bounding box.center)}] \coordinate (vm); \coordinate [left=1 of vm] (v0); \coordinate [right=1 of vm] (v1); \draw (v0) -- (v1); \draw (vm) circle(1); \filldraw (v0) circle (1pt); \filldraw (v1) circle (1pt); \end{tikzpicture} } \fi } + \frac{1}{8} {  \ifmmode \usebox{\fgtadpoletwo} \else \newsavebox{\fgtadpoletwo} \savebox{\fgtadpoletwo}{ \begin{tikzpicture}[x=1ex,y=1ex,baseline={([yshift=-.5ex]current bounding box.center)}] \coordinate (vm); \coordinate [left=.7 of vm] (v0); \coordinate [right=.7 of vm] (v1); \draw (v0) circle(.7); \draw (v1) circle(.7); \filldraw (vm) circle (1pt); \end{tikzpicture} } \fi } + \frac{1}{288} { ( \ifmmode \usebox{\fgbananathree} \else \newsavebox{\fgbananathree} \savebox{\fgbananathree}{ \begin{tikzpicture}[x=1ex,y=1ex,baseline={([yshift=-.5ex]current bounding box.center)}] \coordinate (vm); \coordinate [left=1 of vm] (v0); \coordinate [right=1 of vm] (v1); \draw (v0) -- (v1); \draw (vm) circle(1); \filldraw (v0) circle (1pt); \filldraw (v1) circle (1pt); \end{tikzpicture} } \fi)^2 } +\ldots \end{align*}
where we treat graphs as generators of an algebra. This algebra forms the basis of the Connes-Kreimer Hopf algebraic formulation for renormalization \cite{ConnesKreimer2000}.

Observables in quantum field theory can be expressed as \textit{algebra homomorphisms} in this context. This approach, pioneered by Connes and Kreimer \cite{ConnesKreimer2000}, gives us an algebraic formulation of perturbation theory. The perturbative expansions then arise as images of vectors of graphs such as the one above under certain homomorphisms. These specific algebra homomorphisms are called \textit{Feynman rules}.

\subsection{Zero-dimensional quantum field theory and the configuration model}
To actually obtain quantitative results to test our methods, we will use the configuration model of graph enumeration by Bender and Canfield \cite{bender1978asymptotic2} and its physical counterpart, zero-dimensional quantum field theory \cite{hurst1952enumeration,cvitanovic1978number,bender1978asymptotic1,goldberg1991tree,argyres2001zero,molinari2006enumeration}, which were both initially studied in the 1970s. Both are classic constructions which provide generating functions of multigraphs with prescribed degree distributions. These constructions will be introduced in Chapter \ref{chp:graph_enumeration}. 

Zero-dimensional quantum field theory serves as a toy-model for realistic quantum field theory calculations. Especially, the behavior of zero-dimensional quantum field theory at large order is of interest, as calculations in these regimes for realistic quantum field theories are extremely delicate if not impossible. The utility of zero-dimensional quantum field theory as a reasonable toy-model comes mainly from the interpretation of observables as \textit{combinatorial generating functions} of the number of Feynman diagrams. 

Our focus will be on the \textit{renormalization} of zero-dimensional quantum field theory and the asymptotics of the renormalization constants, which will provide the asymptotic number of \textit{skeleton} Feynman diagrams.

By asymptotics, we mean the behaviour of the coefficients of the perturbation expansion as in eq.\ \eqref{eqn:perturbative_expansion} for large $n$. In zero-dimensional quantum field theory, we will find that the asymptotics of observables are of the form 

\begin{align} \label{eqn:first_fac_asymp_expl} f_n = \alpha^{n+\beta} \Gamma(n+\beta) \left( c_0 + \frac{c_1}{\alpha(n+\beta-1)} + \frac{c_2}{\alpha^2(n+\beta-1)(n+\beta-2)} + \ldots \right), \end{align}
for large $n$ with some $\alpha \in \R_{>0}$, $\beta \in \R$ and $c_k \in \R$. 
On the combinatorial side, these quantities correspond to asymptotic expansions of multigraphs in the large \textit{excess} limit. 

The coefficients $c_k$ in these asymptotic expansions will turn out to be the perturbative expansion coefficients of an observable in \textit{another} zero-dimensional quantum field theory. This observation is due to Başar, Dunne, Ünsal \cite{basar2013resurgence}, who used techniques from Berry, Howls and Dingle \cite{berry1991hyperasymptotics,dingle1973asymptotic} to prove this.

We will use an interpretation of the zero-dimensional quantum field theory as a local expansion of a \textit{generalized hyperelliptic curve} to give an alternative proof for this asymptotic behaviour. 
This will enable us to rigorously extract complete asymptotic expansions by purely algebraic means.

\subsection{Factorially divergent power series}
Sequences with an asymptotic behaviour as in eq.\ \eqref{eqn:first_fac_asymp_expl} appear not only in graph counting, but also in many enumeration problems, which deal with coefficients of factorial growth. For instance, generating functions of some subclasses of permutations show this behaviour \cite{albert2003enumeration,bender1978asymptotic2}. 

Furthermore, there are countless examples where \textit{perturbative expansions} of physical quantities admit asymptotic expansions of this kind \cite{bender1969anharmonic,le2012large,dunne2012resurgence}.

We will study these expansions and their algebraic structure in detail in Chapter \ref{chap:facdivpow}. This analysis is independent of an interpretation as perturbation expansion in quantum field theory or some other theory, but based on the formal power series interpretation of the expansions. We will establish that the requirement to have an asymptotic expansion such as in eq.\ \eqref{eqn:first_fac_asymp_expl} exposes a well-defined subclass of power series.

The restriction to this specific class of power series is inspired by the work of Edward Bender \cite{bender1975asymptotic}. Bender's results are extended into a complete algebraic framework. This is achieved by making heavy use of generating functions in the spirit of the \textit{analytic combinatorics} or \textit{symbolic method} approach. The key step in this direction is to interpret the \textit{coefficients of the asymptotic expansion as another power series}.

These structures bear many resemblances to the theory of resurgence, which was established by {\'E}calle \cite{ecalle1981fonctions}. Resurgence assigns a special role to power series which diverge factorially, as they offer themselves to be Borel transformed. {\'E}calle's theory can be used to assign a unique function to a factorially divergent series. This function could be interpreted as the series' generating function. 
Moreover, resurgence provides a promising approach to cope with divergent perturbative expansions in physics. Its application to these problems is an active field of research \cite{alvarez2004langer,dunne2012resurgence,aniceto2011resurgence}.

The formalism can be seen as a toy model of resurgence' \textit{calcul diff\'erentiel \'etranger} \cite[Vol. 1]{ecalle1981fonctions} also called \textit{alien calculus} \cite[II.6]{mitschi2016divergent}. This toy model is unable to fully reconstruct functions from asymptotic expansions, but does not rely on analytic properties of Borel transformed functions and therefore lends itself to combinatorial applications. A detailed and illuminating account of resurgence theory is given in Sauzin's review \cite[Part II]{mitschi2016divergent}. 

We will show that power series with well-behaved asymptotic expansions, as in eq.\ \eqref{eqn:first_fac_asymp_expl}, form a subring of $\R[[x]]$, which will be denoted as $\fring{x}{\alpha}{\beta}$. This subring is also closed under composition and inversion of power series. A linear map, $\asyOp^\alpha_\beta:\fring{x}{\alpha}{\beta}\rightarrow \R[[x]]$, can be defined which \textit{maps a power series to the asymptotic expansion of its coefficients}. A natural way to define such a map is to associate the power series $\sum_{n=0}^\infty c_n x^n$ to the series $\sum_{n=0}^\infty f_n x^n$ both related as in eq.\ \eqref{eqn:first_fac_asymp_expl}. This map turns out to be a \textit{derivation}. It fulfills a \textit{product rule} and a \textit{chain rule}.
These statements will be derived from elementary properties of the $\Gamma$ function.

This new tool, the ring of factorially divergent power series, can be applied to calculate the asymptotic expansions of implicitly defined power series. This procedure is similar to the calculation of the derivative of an implicitly defined function using the implicit function theorem. 

As examples, we will discuss the asymptotic number of \textit{connected chord diagrams} and of \textit{simple permutations}, which both stem from basic combinatorial constructions. For both examples, only the first coefficients of the asymptotic expansions were known. We will deduce closed forms for the respective complete asymptotic expansions. 

\subsection{Coalgebraic structures}
The coalgebraic structure on graphs captures \textit{insertion} and \textit{contraction} operations on graphs in a natural way. In Chapter \ref{chap:coalgebra_graph} we will define a \textit{coproduct} in the Connes-Kreimer fashion \cite{ConnesKreimer2000,manchon2004hopf} that maps a graph to a formal sum of its subgraph components. The resulting \textit{Hopf algebra} structure is based on the works of Kreimer, Yeats and van Suijlekom \cite{kreimer2006anatomy,Yeats2008,van2007renormalization}.

The coproduct operation will enable us to introduces a \textit{group structure} on the set of all algebra homomorphisms from the graph algebra to some other algebra. 

Subsequently, we will define a class of \textit{Hopf ideals}, which correspond to sets of graphs which are closed under insertion and contraction of subgraphs. Some of these ideals may be used to construct algebra homomorphisms that act as projection operators on the graph algebra. They annihilate graphs which contain certain forbidden subgraphs. The group structure of algebra homomorphisms will play a central role in this construction.

In this way, we obtain generating functions for classes of graphs without certain subgraphs. 
This construction is compatible with the differential ring of factorially divergent power series. Therefore, the asymptotics of the number of graphs with certain subgraphs excluded is accessible using this method. 

The Connes-Kreimer Hopf algebra appears as a quotient Hopf algebra with respect to one of those ideals. 
In quantum field theory, the respective annihilating algebra homomorphism corresponds to the \textit{renormalized Feynman rules} of the theory. This algebra homomorphism gives us the generating functions of graphs with given edge-connectivity. 

At the end of Chapter \ref{chap:coalgebra_graph}, we will apply all these considerations to give a Hopf algebraic interpretation of the \textit{Legendre transformation} on graph generating functions. This transformation plays a central role in quantum field theory and is used to obtain the generating function of \textit{bridgeless} or \textit{1-particle-irreducible}\footnote{connected and bridgeless.} (1PI) graphs from the generating function of \textit{connected} graphs. This extends a recent study of the Legendre transformation on trees by Jackson, Kempf and Morales \cite{jackson2016robust}.

\subsection{The lattice structure of subdivergences}
Another object of interest from the perspective of quantum field theory are the \textit{counter\-terms}.
The counterterms in zero-dimensional quantum field theory have a more subtle combinatorial interpretation than the images of the renormalized Feynman rules. They `almost' count the number of primitive diagrams in the underlying theory. In the process of clarifying this statement, we will encounter the lattice structure of Feynman diagrams in Chapter \ref{chap:hopf_algebra_of_fg}, where we will also introduce the details of Kreimer's Hopf algebra of Feynman diagrams. The evaluation of the counterterms can be identified with the evaluation of the \textit{Moebius function} of the underlying subgraph poset. 

The idea to search for more properties of the subdivergence posets is inspired by the work of Berghoff \cite{berghoff2014wonderful}, who studied the posets of subdivergences in the context of Epstein-Glaser renormalization and proved that the subdivergences of diagrams with only logarithmic subdivergences form distributive lattices. Distributive lattices have already been used in \cite[Part III]{figueroa2005combinatorial} to describe subdivergences of Feynman diagrams.

We will carry the Hopf algebra structure on Feynman diagrams over to posets and lattices in the style of the incidence Hopf algebra on posets \cite{Schmitt1994}.

In distinguished renormalizable quantum field theories a \textit{join} and a \textit{meet} can be defined generally on the posets of subdivergences of Feynman diagrams, promoting the posets to \textit{algebraic lattices}. These distinguished renormalizable QFTs will be called \textit{join-meet-renormalizable}. It will be shown that a broad class of QFTs including the standard model falls into this category. $\varphi^6$-theory in $3$-dimensions will be examined as an example of a QFT, which is renormalizable, but not join-meet-renormalizable. 

The lattice structure also provides insights into the \textit{coradical filtration} which describes the hierarchy in which diagrams become important in the large-order regime. \textit{Dyson-Schwinger equations} \cite{kreimer2006etude} exploit this hierarchy to give non-perturbative results \cite{kreimer2008recursion,kruger2015filtrations}. The presentation of this structure also aims to extend the effectiveness of these methods.

Our analysis will demonstrate that in QFTs with only three-or-less-valent vertices, which are thereby join-meet-renormalizable, these lattices are \textit{semimodular}. This implies that the Hopf algebra is bigraded by the loop number of the Feynman diagram and its \textit{coradical degree}. In the language of BPHZ this means that every \textit{complete forest} of a Feynman diagram has the same length. Generally, this structure cannot be found in join-meet-renormalizable theories with also four-valent vertices as QCD or $\varphi^4$-theory. An explicit counterexample of a non-graded and non-semimodular lattice, which appears in $\varphi^4$ and Yang-Mills theories, is given. The semimodularity of the subdivergence lattices can be resurrected in these cases by dividing out the Hopf ideal generated by \textit{tadpole} diagrams. This quotient can always be formed in kinematic renormalization schemes.

\subsection{Applications to zero-dimensional quantum field theory}

In the final Chapter \ref{chap:applications_zerodim}, we will use all the aforementioned formal structures to obtain various asymptotics in zero-dimensional quantum field theory. Explicit asymptotic results of the number of disconnected, connected, 1PI and skeleton diagrams will be provided for $\varphi^3$, $\varphi^4$, QED, quenched QED and Yukawa theory.
All given results have been verified using numerical calculations.

For many of the given examples either none or only a few coefficients of the asymptotic expansions have been known. Explicit constructions of the generating functions of the asymptotic expansions will be provided in every given case.

\section{Outlook}
\subsection{Asymptotic evaluation and bounds for Feynman integrals}
Although the presented methods do not take the precise structure of the individual Feynman integrals into account, there are several ways to use the information from zero-dimensional quantum field theory to obtain estimates for the coefficients of the perturbation expansion \cite{bender1973anharmonic,simon1982large}.

An especially promising approach to give bounds and estimates for Feynman integrals is the \textit{Hepp-bound} used by Kompaniets and Panzer to estimate the $\beta$-function of $\varphi^4$-theory up to loop order thirteen \cite{kompaniets2017minimally}. This Hepp-bound, entirely combinatorial in nature, can be integrated naturally in a Hopf algebraic framework.

Ultimately, such an analysis leads to the idea of interpreting a large graph as a probabilistic object. This approach has gained much attention in the context of complex networks \cite{lovasz2012large} under the name of \textit{graphons}, but remains to be exploited in the realm of quantum field theory. An approach to quantum field theory based on a probabilistic interpretation of Feynman diagrams could lead to a new perspective on instanton and large coupling problems. 

\subsection{Series resummation}
The techniques presented in Chapter \ref{chap:facdivpow} result in various asymptotic expansions for combinatorial quantities. These asymptotic expansions lend themselves to be used in combination with resummation techniques. \textit{Hyperasymptotic} \cite{berry1990hyperasymptotics} methods can be exploited to obtain numerical results of extremely high accuracy. An augmentation of these methods with further insights from resurgence \cite{ecalle1981fonctions} should be especially beneficial, as higher order asymptotics\footnote{The asymptotic expansion of the coefficients of a first order asymptotic expansion is a second order asymptotic expansion and so on.} can be easily obtained for all given examples. This would lead to a \textit{trans-series} approach from which numerical results can be obtained via \textit{Borel-Padé} resummation.

\subsection{Lattice structure in QFT}

The lattice structure of Feynman diagrams suggests to consider groups of diagrams, which correspond to the same lattice in the lattice Hopf algebra. This gives us a reorganization of diagrams into groups which `renormalize in the same way'. The central property is the degree of the lattice, which corresponds to the coradical degree of the respective diagrams. With methods from \cite{brown2013angles} this could be used to express the log-expansion of Green functions systematically. Primitive diagrams of coradical degree one contribute to the first power in the log-expansion, 
while diagrams of coradical degree two contribute to the second and diagrams with a coradical degree equal to the loop number contribute to the leading-log coefficient \cite{kruger2015filtrations}. 

\subsection{Random graphs}
Another future line of research to pursue, which uses the results presented in this thesis, is to explore the Hopf algebra structure of \textit{random graphs} \cite{erdos1960evolution}. The set of \textit{simple graphs}, graphs without selfloops or multiple edges, can be obtained by dividing out the insertion/contraction closed set of graphs which is generated by a selfloop and a double edge. Analogously, we can construct insertion/contraction closed sets which give the generating functions of graphs with prescribed \textit{girth}, the length of a shortest cycle. In these cases, asymptotics can be obtained for the large excess limit. The advantage of the algebraic method is that all-order asymptotic expansions can be obtained. Such asymptotic expansions of random graphs have recently been studied by de Panafieu \cite{de2016analytic}. It is very plausible that the presented methods can be extended to obtain further results in this domain.

Random graphs with prescribed degree distributions form a recent and promising line of research under the heading of complex networks. Many social, economical and biological processes can be modeled as such networks \cite{albert2002statistical} and the methods presented in this thesis can be used to study them. The complete asymptotic expansions, which can be obtained, could simplify the analysis of networks of finite size significantly. 

Furthermore, the presented formalism is also capable of handling certain \textit{colorings} of graphs, as is illustrated in Chapter \ref{chap:applications_zerodim} with the examples of QED and Yukawa theory. Another example where the asymptotics of colored graphs is of interest is the \textit{Ising model} on a random graph. The Ising model has been studied on random graphs \cite{dembo2010ising}, but it would be interesting to analyze the role of the combinatorial Hopf algebra structure in this context and how it relates to the phase transition properties of this complex system.

\chapter{Graphs}
\label{chp:graphs}

\section{Definition}
The most central notion of this thesis is the graph. For reasons that will become clear later, we will not resort to the traditional definition of a graph as a set of vertices and a set of edges. Our definition includes so called multigraphs, graphs where multiple edges are allowed, in a natural way. We will consider sets of \textit{half-edges} and \textit{vertices} to be the building blocks of a graph. Based on those sets, a graph consists of a map that associates half-edges with vertices and an involution on the set of half-edges that maps a half-edge to its other half. 
Naturally, two half-edges make up an edge this way.

We will also allow some half-edges to not have an half-edge-partner, these half-edges will be called \textit{legs} of the graph. In the realm of graph cohomology, such a construction is also called a \textit{hairy} graph \cite{conant2012hairy}.

In the scope of quantum field theory, this approach based an half-edges is well known. See for instance \cite[Sec. 2.3]{Yeats2008} or \cite[Sec. 2.1]{gurau2014renormalization}. 

\begin{defn}[Graph with edges as an involution]
\label{def:graph1}
A graph is a tuple $(H,V,\nu,\iota)$ consisting of 
\begin{itemize}
\item A set of half-edges $H$.
\item A set of vertices $V$.
\item A map $\nu: H \rightarrow V$ that assigns half-edges to vertices.
\item An involution on $H$, $\iota:H\rightarrow H$ such that $\iota \circ \iota = \id$, that \textit{pairs} some half-edges.
\end{itemize}
\end{defn}
Note that we do not require the involution $\iota$ to be \textit{fixed-point free}. 
\clearpage
\begin{defn}[Graph with explicit edges]
\label{def:graph2}
A graph is a tuple $(H,V,\nu,E)$ consisting of 
\begin{itemize}
\item A set of half-edges $H$.
\item A set of vertices $V$
\item A map $\nu: H \rightarrow V$ that assigns half-edges to vertices.
\item A set of disjoint subsets of half-edges of cardinality $2$, $E \subset \powerset{H}$ such that for all 
$e_1,e_2 \in E$, $e_1 \cap e_2 =\emptyset$ and $|e| = 2$ for all $e \in E$.
\end{itemize}
\end{defn}

\begin{prop}
Definitions \ref{def:graph1} and \ref{def:graph2} are equivalent.
\end{prop}
\begin{proof}
We have to show that giving an involution $\iota$ or a set of edges $E$ is equivalent. 

The orbits of the involution $\iota:H\rightarrow H$ give a partition of $H$ into sets of cardinality $1$ and $2$. We will identify the sets of cardinality $2$ with the edges $E$. 

From a set of edges $E$, we can construct an involution $\iota$ by mapping each half-edge to its partner, if it has one, and to itself, if it has none. 
\end{proof}
Both definitions have their advantages. As Definition \ref{def:graph1} is more compact, it is slightly more useful in proofs. For the (mental) diagrammatic representation of graphs Definition \ref{def:graph2} is typically more helpful. We will switch freely between both representations. 

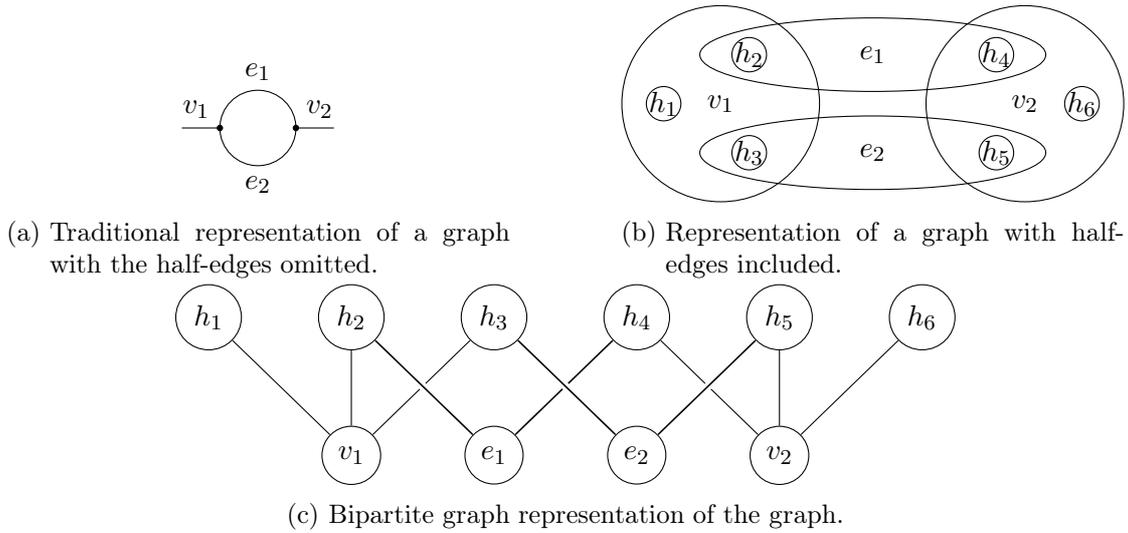
\begin{figure}
\begin{center}
    \begin{subfigure}[b]{.45\linewidth}
		\centering 
        \begin{tikzpicture}[scale=1] \coordinate (v0); \coordinate [right=1 of v0] (v1); \coordinate [left=.5 of v0] (i0); \coordinate [right=.5 of v1] (o0); \coordinate [left=.5 of v1] (vm); \draw (vm) circle(.5); \draw (i0) -- (v0); \draw (o0) -- (v1); \filldraw (v0) circle(1pt); \filldraw (v1) circle(1pt); \node at (v0) [above left] {$v_1$}; \node at (v1) [above right] {$v_2$}; \node at (vm) [above=.5] {$e_1$}; \node at (vm) [below=.5] {$e_2$}; \end{tikzpicture}%
		\caption{Traditional representation of a graph with the half-edges omitted.}
        \label{fig:graph_representation_normal}		
	\end{subfigure}%
    \hfill
    \begin{subfigure}[b]{.45\linewidth}
		\centering
        \begin{tikzpicture}[scale=.65] \node (p1) at ($({-sqrt(3)+1},0)$) {$h_1$}; \node (f1) at (1,1) {$h_2$}; \node (pb1) at (1,-1) {$h_3$}; \node (f2) at (6,1) {$h_4$}; \node (p2) at (6,-1) {$h_5$}; \node (pb2) at ($({sqrt(3)+6},0)$) {$h_6$}; \node (c1) at ($1/3*(p1)+1/3*(pb1)+1/3*(f1)$) {}; \node (c2) at ($1/3*(p2)+1/3*(pb2)+1/3*(f2)$) {}; \node at ($(c1)$) {$v_1$}; \node at ($(c2)$) {$v_2$}; \node (e1) at ($1/2*(pb1) + 1/2*(p2)$) {$e_2$}; \node (e2) at ($1/2*(f1) + 1/2*(f2)$) {$e_1$}; \draw (p1) circle (0.35); \draw (p2) circle (0.35); \draw (pb1) circle (0.35); \draw (pb2) circle (0.35); \draw (f1) circle (0.35); \draw (f2) circle (0.35); \draw (c1) circle (2.0); \draw (c2) circle (2.0); \draw (e1) ellipse [x radius=3.5, y radius=0.75]; \draw (e2) ellipse [x radius=3.5, y radius=0.75]; \end{tikzpicture}
		\caption{Representation of a graph with half-edges included.}
        \label{fig:graph_representation_half_edge}		
	\end{subfigure}
    \begin{subfigure}[b]{\linewidth}
		\centering
        \begin{tikzpicture}[main node/.style={circle,draw}] \node[main node] (h1) {$h_1$}; \node[main node] (h2) [right=of h1] {$h_2$}; \node[main node] (h3) [right=of h2] {$h_3$}; \node[main node] (h4) [right=of h3] {$h_4$}; \node[main node] (h5) [right=of h4] {$h_5$}; \node[main node] (h6) [right=of h5] {$h_6$}; \node[main node] (a1) [below=of h2] {$v_1$}; \node[main node] (a2) [below=of h3] {$e_1$}; \node[main node] (a3) [below=of h4] {$e_2$}; \node[main node] (a4) [below=of h5] {$v_2$}; \draw (h1) -- (a1); \draw (h2) -- (a1); \draw (h3) -- (a1); \draw (h4) -- (a4); \draw (h5) -- (a4); \draw (h6) -- (a4); \draw[draw=white,double=black,very thick] (h2) -- (a2); \draw[draw=white,double=black,very thick] (h4) -- (a2); \draw[draw=white,double=black,very thick] (h3) -- (a3); \draw[draw=white,double=black,very thick] (h5) -- (a3); \end{tikzpicture}
		\caption{Bipartite graph representation of the graph.}
        \label{fig:graph_representation_bipartite}
	\end{subfigure}
\end{center}
\caption{Equivalent diagrammatic representations of a graph.}
\label{fig:graph_representations}
\end{figure}

If the reference to the graph $G$ given by a tuple $(H,V,\nu,\iota)$ or equivalently $(H,V,\nu,E)$ is ambiguous, the sets $H,V,E$ and the maps $\iota, \nu$ will be referenced with the symbol for the graph in the subscript: $H_G,V_G,E_G,\iota_G,\nu_G$.

\nomenclature{$H_G$}{Half-edge set of a graph $G$}
\nomenclature{$V_G$}{Vertex set of a graph $G$}
\nomenclature{$E_G$}{Edge set of a graph $G$}
\nomenclature{$\iota_G$}{Edge involution of a graph $G$}
\nomenclature{$\nu_G$}{Half-edge to vertex mapping of a graph $G$}

\begin{defn}[Legs]
The half-edges that are not contained in any edge are called legs, as already mentioned. We will denote the set of legs of a graph as $\legs_G:= H_G \setminus \bigcup_{e\in E_G} e$.
\end{defn}
\nomenclature{$\legs_G$}{Legs of a graph $G$}

\begin{defn}[Corollas]
The preimages $\nu_G^{-1}(v)$ of the vertices $v \in V_G$ of a graph $G$ are \textit{corollas} - a subset of half-edges joined together to form a vertex. The cardinality of this set $\deg{v}_G:= | \nu_G^{-1}(v) |$ is the \textit{degree} of the vertex $v$.
Furthermore, we will denote the number of vertices with degree $d$ in $G$ as $\nvd{d}_G := \left|\{ v \in V_G : \deg{v}_G = d \}\right|$.
\end{defn}
\nomenclature{$\deg{v}_G$}{Degree of a vertex $v$ in a graph $G$}
\nomenclature{$\nvd{d}_G$}{Number of vertices of degree $d$ in a graph $G$}

\begin{defn}[Connected components]
\label{def:cntd_cmps}
In a straightforward way, we can set up an equivalence relation on the vertex set $V_G$ of a graph $G$. Two vertices $v_a,v_b \in V_G$ are in the same \textit{connected component} if we can find a path between them. A path is a sequence of half-edges $h_1, \ldots, h_{2n}$ and vertices $v_1,\ldots,v_{n-1}$, such that the odd pairs of half-edges form edges $\iota_G(h_{2k+1})=h_{2k+2}$ and the even pairs belong to the same corolla $h_{2k}, h_{2k+1} \in \nu_G^{-1}(v_k)$. If $h_1 \in \nu_G^{-1}(v_a)$ and $h_{2n} \in \nu_G^{-1}(v_b)$ then the path starts in $v_a$ and ends in $v_b$. 
The set of equivalence classes based on this relation $\comps_G := V_G / \sim$ is the set of connected components of $G$.

A graph is \textit{connected} if it has exactly one connected component.
\end{defn}
\nomenclature{$\comps_G$}{Connected components of a graph $G$}

\begin{defn}[Isomorphism]
\label{def:isomorphism}
An \textit{isomorphism} $j$, between two graphs $G_1$ and $G_2$, $j: G_1\rightarrow G_2$, is a pair of bijections $j= (j_H,j_V)$ of the respective half-edge and vertex sets which are compatible with the $\iota$ and $\nu$ maps. Formally, $j_H:H_{G_1} \rightarrow H_{G_2}$ and $j_V: V_{G_1} \rightarrow V_{G_2}$ such that $\nu_{G_2} = j_V \circ \nu_{G_1} \circ j_H^{-1}$ and $\iota_{G_2} = j_H \circ \iota_{G_1} \circ j_H^{-1}$.
\end{defn}
\begin{defn}[Automorphisms]
An isomorphism from a graph to itself is called an automorphism.
\end{defn}

Figure \ref{fig:graph_representations} illustrates different representations of a graph with two vertices which are joined by two edges and which both have one leg. The most compact representation is the traditional one in Figure \ref{fig:graph_representation_normal}. Note that legs, half-edges which are not part of an edge, are depicted as edges that are not connected to another vertex. We will use this representation throughout this thesis, but the reader should keep the individual character of the half-edges in mind. Graphs can have trivial automorphisms which come from double edges or self-loops. The graph in Figure \ref{fig:graph_representations} is an example of such a graph.

In Figure \ref{fig:graph_representation_half_edge} the automorphism of the graph that switches the half-edges $h_2$ and $h_3$ as well as $h_4$ and $h_5$ and thereby the edges $e_1$ and $e_2$ becomes more apparent, then in the traditional representation.

The representation in Figure \ref{fig:graph_representation_bipartite} is useful for computational applications, as this representation is a \textit{simple}\footnote{A simple graph is a graph without selfloops or multiple edges between the same pair of vertices.} \textit{bipartite}\footnote{A bipartite graph is a graph, whose vertex set is the union of two disjoints sets of mutually disconnected vertices.} graph. For instance, the program \textbf{nauty} \cite{mckay1981practical} only supports simple graphs.

\section{Labelled graphs}

We will consider graphs to be \textit{labelled combinatorial objects} in the context of analytic combinatorics \cite{flajolet2009analytic}. The \textit{labelled atoms} of the graph are the half-edges and the vertices. That means, we will consider the sets of half-edges and vertices to be intervals of integers. 
\begin{defn}[Labelled graph]
\label{def:labelled_graph}
A graph $G=(H,V,\nu,\iota)$ is \textit{labelled} if the sets $H$ and $V$ are intervals starting from $1$: $H = \{1,\ldots, |H|\} = [|H|]$ and $V= \{1,\ldots, |V|\} = [|V|]$. 
\end{defn}
\nomenclature{$[n]$}{The elementary interval of natural numbers with $n$ elements $[n] = \{1,\ldots,n\}\subset \N$}

An important consequence of considering labelled graphs is that there is only a finite number of labelled graphs with fixed numbers of half-edges and vertices. We will define the set of labelled graphs accordingly.
\begin{defn}[The set of labelled graphs]
Let $\Gl_{m,k}$ be the set of labelled graphs with $m$ half-edges and $k$ vertices. Explicitly, $\Gl_{m,k}$ is the set of all tuples $([m],[k],\nu,\iota)$ with some map $\nu: [m] \rightarrow [k]$ and some involution $\iota: [m] \rightarrow [m]$, where $[n]$ is the elementary interval $\{1,\ldots,n\} \subset \N$.

We will denote the set of all labelled graphs as $\Gl = \bigcup_{m,k \geq 0} \Gl_{m,k}$.
\end{defn}
\nomenclature{$\Gl$}{The set of all labelled graphs}

\subsection{Basic generating functions}

It is straightforward to find generating functions for the elements in $\Gl$, because every element is entirely determined by the numbers of half-edges, of vertices and of the two mappings $\nu$ and $\iota$. As long as we can count the number of maps $\nu$ and $\iota$, we can count the respective labelled graphs.
\begin{prop}
\label{prop:basic_counting}
The following enumeration identity holds
\begin{align} \sum_{G \in \Gl} \frac{x^{|H_G|} \lambda^{|V_G|}}{|H_G|! |V_G|!} =  \sum_{k\geq 0} e^{kx + \frac{k^2 x^2}{2}} \frac{\lambda^k}{k!}. \end{align}
\end{prop}
\begin{proof}
The number of labelled graphs on a set of half-edges $H=[m]$ and a set of vertices $V=[k]$ is equal to the number of maps $\nu: H \rightarrow V$ times the number of involutions $\iota: H \rightarrow H$. There are $k^m$ maps $\nu: [m] \rightarrow [k]$. 
The number of fixed-point-free involutions on a set of $2n$ elements is given by $(2n-1)!!$, the double factorial. Therefore, the total number of involutions\footnote{These numbers are called the telephone numbers \cite[Example II.13]{flajolet2009analytic}.} $\iota: [m] \rightarrow [m]$ is $\sum_{n \geq 0}^{\floor{\frac{m}{2}}} \binom{m}{2n} (2n-1)!!$, where each summand is the number of involutions with $m-2n$ fixed points. 

From this, we can obtain the generating function of the elements in $\Gl$, where the numbers of half-edges and vertices are marked, with a short calculation:
\begin{gather*} \sum_{G \in \Gl} \frac{x^{|H_G|} \lambda^{|V_G|}}{|H_G|! |V_G|!} = \sum_{m,k \geq 0} x^{m} \lambda^{k} \frac{k^m \sum_{n \geq 0}^{\floor{\frac{m}{2}}} \binom{m}{2n} (2n-1)!!}{m! k!} \\ = \sum_{n,k \geq 0} \sum_{m\geq 2n} x^{m} \lambda^{k} \frac{k^m (2n-1)!!}{k! (2n)! (m-2n)!} = \sum_{n,k \geq 0} \sum_{m\geq 0} x^{m+2n} \lambda^{k} \frac{k^{m+2n} (2n-1)!!}{k! (2n)! m!} \\ = \sum_{n,k \geq 0} e^{k x} x^{2n} \lambda^{k} \frac{k^{2n} (2n-1)!!}{k! (2n)!} = \sum_{n,k \geq 0} e^{k x} x^{2n} \lambda^{k} \frac{k^{2n} }{k! 2^n n!} = \sum_{k \geq 0} e^{k x+ \frac{k^2 x^2 }{2}} \frac{\lambda^{k}}{k!}, \end{gather*}
where we used $(2n-1)!! = \frac{(2n)!}{2^n n!}$ and $\sum_{n \geq 0} \frac{x^n}{n!} = e^x$.
\end{proof}

In the examples which we will discuss in the following chapters, we also want to retain some information about the degrees of the vertices in the graph. 
The following generalization of Proposition \ref{prop:basic_counting} provides a convenient way to do so. 

Instead of the total number of vertices and half-edges, we will mark the number of vertices with degree $d$, the number of legs and the number of edges: 
\begin{prop}
\label{prop:counting_with_degrees}
The following enumeration identity holds
\begin{align} \label{eqn:glcountingdegreeidentity} \sum_{G \in \Gl} \frac{\varphi_c^{|\legs_G|} a^{|E_G|} \prod_{v \in V_G} \lambda_{\deg{v}} }{|H_G|! |V_G|!} = \sum_{m\geq 0}m! [x^m y^m] e^{a \frac{y^2}{2} + \varphi_c y } e^{\sum_{d\geq 0} \lambda_d \frac{x^d}{d!} }. \end{align}
\end{prop}
\begin{proof}
The number of involutions on $m$ elements with $m - 2n$ fixed points is given by $I_{n,m}:=\binom{m}{2n} (2n-1)!!$, which we used in the proof of Proposition \ref{prop:basic_counting}. 
The exponential generating function of these involutions, which marks the total number of half-edges involved with $y$, the number of fixed-points with $\varphi_c$ and the number of pairs with $a$, is 
\begin{gather*} \sum_{n,m \geq 0} I_{n,m} \frac{y^{m}}{m!}\varphi_c^{m-2n} a^n = \sum_{n,m \geq 0} \binom{m}{2n} (2n-1)!! \frac{y^{m}}{m!} \varphi_c^{m-2n} a^{n}= \sum_{n\geq 0} \sum_{m \geq 2n } \frac{y^m \varphi_c^{m-2n} a^n}{(m-2n)!2^n n!} \\ = \sum_{n,m \geq 0} \frac{y^{m+2n} \varphi_c^m a^n}{m!2^n n!} = e^{a\frac{y^2}{2} + \varphi_c y}. \end{gather*}
The number of maps $\nu:[m] \rightarrow [k]$ with prescribed sizes of the preimages $|\nu_{-1}(v)| = \deg{v}$ such that $\sum_{v\in V} \deg{v} = m$ is given by the multinomial coefficient $\binom{m}{\deg{1}, \cdots, \deg{k}}$. 
The expression, 
\begin{gather*} M_{m,k}(\lambda_0, \lambda_1, \ldots) := \sum_{\substack{ \deg{1}, \ldots, \deg{k} \geq 0\\\deg{1}+\ldots+\deg{k} = m}} \binom{m}{\deg{1}, \ldots, \deg{k}} \prod_{i=1}^{k} \lambda_{\deg{i}}, \end{gather*}
generates the number of maps $\nu$ with marked sizes of the preimages $|\nu_{-1}(v)| = \deg{v}$. Naturally, this reduces to the expression for the numbers of maps $\nu$ from the proof of Proposition \ref{prop:basic_counting} if $\lambda_d = 1$ for all $d\in \N_0$, because 
\begin{gather*} M_{m,k}(1, 1, \ldots) = \sum_{\substack{ \deg{1}, \ldots, \deg{k} \geq 0\\\deg{1}+\ldots+\deg{k} = m}} \binom{m}{\deg{1}, \ldots, \deg{k}} = k^m, \end{gather*}
by the multinomial theorem.

Multiplying with an auxiliary parameter $x$ and summing over all possible numbers of vertices $k$ gives the exponential generating function,
\begin{gather*} \sum_{m,k\geq 0} \frac{x^{m}}{m!k!} M_{m,k}(\lambda_0, \lambda_1, \ldots) = \sum_{m,k\geq 0} \frac{x^{m}}{m!k!} \sum_{\substack{ \deg{1}, \ldots, \deg{k} \geq 0\\\deg{1}+\ldots+\deg{k} = m}} \binom{m}{\deg{1}, \ldots, \deg{k}} \prod_{i=1}^{k} \lambda_{\deg{i}} \\ = \sum_{k\geq 0} \frac{1}{k!}\sum_{\deg{1}, \ldots, \deg{k} \geq 0} x^{\sum_{i=1}^k \deg{i}} \prod_{i=1}^{k} \frac{\lambda_{\deg{i}}}{\deg{i}!} = \sum_{k\geq 0 } \frac{1}{k!}\left( \sum_{d \geq 0} \frac{\lambda_d x^d}{d!} \right)^k = e^{\sum_{d \geq 0} \lambda_d \frac{x^d}{d!} }. \end{gather*}
Therefore, we can write the generating function of graphs in $\Gl$ as a sum over the numbers of edges $n$, of vertices $k$ and of half-edges $m$,
\begin{gather*} \sum_{G \in \Gl} \frac{\varphi_c^{|\legs_G|} a^{|E_G|} \prod_{v \in V_G} \lambda_{\deg{v}} }{|H_G|! |V_G|!} = \sum_{n,m,k \geq 0} \frac{\varphi_c^{2n-m} a^{n} }{m! k!} I_{n,m} M_{m,k}(\lambda_0, \lambda_1, \ldots), \end{gather*}
which results in the statement after substitution of the respective generating functions.
\end{proof}

Especially when depicting graphs diagrammatically, it is practical to consider isomorphism classes of graphs:
It would be very cumbersome to always include the labels of vertices and half-edges into a drawing of a graph. Moreover, the properties of graphs, which we are interested in, are all invariant under an arbitrary permutation of the vertex and half-edge labels. Therefore, it is natural to consider \textit{isomorphism classes} of graphs.

\section{Graph isomorphisms and unlabelled graphs}
\label{sec:unlabelled_graphs_isos}

Two graphs can obviously only be isomorphic if they have the same number of half-edges and vertices. 
By Definition \ref{def:labelled_graph}, all elements in the subsets $\Gl_{m,k} \subset \Gl$ of graphs with $m$ half-edges and $k$ vertices have the half-edge set $[m]$ and the vertex set $[k]$. An isomorphism between two graphs $G_1,G_2 \in \Gl_{m,k}$ is a pair of bijections $j_H:[m] \rightarrow [m]$, $j_V:[k]\rightarrow [k]$ - a pair of permutations of the labels - that fulfill the condition in Definition \ref{def:isomorphism}. 
To isolate the isomorphism classes in $\Gl_{m,k}$, we will use elementary group theory. 

Let $P_{m,k} := S_{m}\times S_{k}$ be the product group of all pairs of permutations $(j_H,j_V)$ which can be used to relabel the graphs in $\Gl_{m,k}$. The group $P_{m,k}$ \textit{acts} on the sets $\Gl_{m,k}$ by permuting the half-edge and vertex labels. We define an \textit{group action} $*$ accordingly:
\begin{align*} &*& &:&&P_{m,k} \times \Gl_{m,k}& &\rightarrow& &\Gl_{m,k}, \\&&&&& ((j_V, j_H), ([m],[k],\nu,\iota))& &\mapsto& & ( [m], [k], j_V \circ \nu \circ j_H^{-1}, j_H \circ \iota \circ j_H^{-1} ). \end{align*}
Note, that a pair $j_H,j_V$ does not alter the sets $H$ and $V$. It only changes the maps $\nu$ and $\iota$ by conjugation.

Let $\orb_{P_{m,k}}(G):= \{ p * G : p \in P_{m,k} \}$ be the \textit{orbit} of the element $G \in \Gl_{m,k}$. It is the set of all graphs in $\Gl_{m,k}$ that can be obtained from $G$ by a permutation of the half-edge and vertex labels. Such a set can be interpreted as an \textit{unlabelled} graph.
\begin{defn}
We define the set of \textit{unlabelled} graphs, $\Gul$, as the set of all orbits\footnote{Equivalently, $\Gul_{m,k}$ is the \textit{quotient} of $\Gl_{m,k}$ under the group action $*$, $\Gul_{m,k} = \Gl_{m,k}/P_{m,k}$.}
\begin{align} \Gul_{m,k} &:= \left\{ \orb_{P_{m,k}}(G) : G \in \Gl_{m,k} \right\} & &\Gul:= \bigcup_{m,k \geq 0}\Gul_{m,k}, \end{align}
which is a partition of $\Gl_{m,k}$ into subsets of mutually isomorphic graphs.
\end{defn}
In an established abuse of notation we will write $\Gamma \in \Gul$ not for the orbit of specific graph under relabelling, but for a representative graph from the respective orbit. We can always find such a unique representative for each orbit in $\Gl_{m,k}$ by computing a \textit{canonical labelling} of the graph. Finding such a canonical labelling is a computationally demanding task. However, there is a powerful and established program, called \textbf{nauty}, by \citet{mckay1981practical} which solves this task sufficiently fast for most practical purposes. The explicit calculations presented in this work, which involve graph enumeration, have been cross-checked using this tool.
\nomenclature{$\Gul$}{The set of all unlabelled graphs}

Due to the \textit{orbit-stabilizer theorem}, we get the following identity:
\begin{prop}
\label{prop:graph_labelled_unlabelled}
For every $G \in \Gl_{m,k}$, 
\begin{align*} \frac{| \orb_{P_{m,k}}(G) | }{m! k!} = \frac{1}{|\Aut G|}, \end{align*}
where $\Aut G$ is the set of all automorphisms of the graph $G$. 
\end{prop}
\nomenclature{$\Aut G$}{The set of all automorphisms of a graph $G$}

\begin{proof}
The \textit{stabilizer} of $G$ under the action by $P_{m,k}$ is defined as 
$\stab_{P_{m,k}}(G):= \{ p \in P_{m,k}: p * G = G \}$. It is the set of all elements in 
$P_{m,k}$ that map the graph to itself. The elements of $\stab_{P_{m,k}}(G)$ are the automorphisms of $G$, $\stab_{P_{m,k}}(G)=\Aut G$.
From the \textit{orbit-stabilizer theorem} (see for instance \cite[Thm.\ 2.16]{holt2005handbook}), 
\begin{align} m!k!=|P_{m,k}| = \left|\orb_{P_{m,k}}(G)\right| \left|\stab_{P_{m,k}}(G)\right| = \left|\orb_{P_{m,k}}(G)\right| |\Aut G|, \end{align}
the statement follows.
\end{proof}
\begin{crll}
\label{crll:counting_unlabelled_with_degrees}
We may write the identity from Proposition \ref{prop:counting_with_degrees} as
\begin{align} \label{eqn:gulcountingdegreeidentity} \sum_{\Gamma \in \Gul} \frac{\varphi_c^{|\legs_\Gamma|} a^{|E_\Gamma|} \prod_{v \in V_\Gamma} \lambda_{\deg{v}} }{| \Aut \Gamma|} = \sum_{m\geq 0}m! [x^m y^m] e^{a \frac{y^2}{2} + \varphi_c y } e^{\sum_{d\geq 0} \lambda_d \frac{x^d}{d!} }. \end{align}
\end{crll}
\begin{proof}
\begin{gather*} \sum_{G \in \Gl} \frac{\varphi_c^{|\legs_G|} a^{|E_G|} \prod_{v \in V_G} \lambda_{\deg{v}} }{|H_G|! |V_G|!} = \sum_{m,k \geq 0} \sum_{\Gamma \in \Gul_{m,k}} \sum_{G\in \orb_{P_{m,k}}(\Gamma)}\frac{\varphi_c^{|\legs_G|} a^{|E_G|} \prod_{v \in V_G} \lambda_{\deg{v}} }{m! k!} \\ = \sum_{m,k \geq 0} \sum_{\Gamma \in \Gul_{m,k}} \frac{\varphi_c^{|\legs_\Gamma|} a^{|E_\Gamma|} \prod_{v \in V_\Gamma} \lambda_{\deg{v}} }{|\Aut \Gamma|}. \end{gather*}
\end{proof}
Note that the definitions of auto- and isomorphisms of graphs, which are based on bijections of the underlying half-edge and vertex sets, circumvent the need for `compensation factors', as introduced in \citep{janson1993birth}, when dealing with multigraph generating functions.

Comparing eqs.\ \eqref{eqn:glcountingdegreeidentity} and \eqref{eqn:gulcountingdegreeidentity}, we note that the way of writing such identities in terms of $\Gul$ has the advantage that we do not need to keep track of the explicit numbers of half-edges and vertices of the graphs in the denominators of our expressions.

We are going to write identities such as the one above in terms of unlabelled graphs in $\Gul$, although strictly speaking, we will still have exponential generating functions of labelled graphs. To translate between the unlabelled and labelled classes of graphs, we will use Proposition \ref{prop:graph_labelled_unlabelled}.

We can set up an algebraic structure both on labelled graphs and unlabelled graphs. On labelled graphs, a natural multiplication would be the \textit{labelled combinatorial product} \cite[p.\ 96]{flajolet2009analytic}. On unlabelled graphs, we will resort to the disjoint union as product. In contrast to the combinatorial product this has the disadvantage that we need to introduce $\frac{1}{|\Aut \Gamma|}$ factors in many formulas. The advantage is that the \textit{coalgebraic} structure of graphs, which will be the subject of chapter \ref{chap:coalgebra_graph}, becomes much more apparent in this case.
\section{Graph algebra}
\label{sec:graph_algebra}

Instead of the labelled combinatorial product, we are going to rely on disjoint union as a product on graphs:
\begin{defn}[Disjoint union]
For two graphs $G_1$ and $G_2$ the disjoint union $G_1 \sqcup G_2$ is the graph 
$(H_{G_1} \sqcup H_{G_2}, V_{G_1} \sqcup V_{G_2}, \eta_{G_1} \sqcup \eta_{G_2}, \iota_{G_1} \sqcup \iota_{G_2})$. 

The disjoint union of two sets $A \sqcup B$ can be explicitly constructed by multiplying the respective sets with a unique symbol: $A \sqcup B := (\heartsuit \times A) \cup (\diamondsuit \times B)$. For maps, $f: A \rightarrow C$ and $g: B \rightarrow D$, the disjoint union $f \sqcup g : A \sqcup B \rightarrow C \sqcup D$ is the unique map whose restriction on $\heartsuit \times A$ is equal to $f$ and whose restriction on $\diamondsuit \times B$ is equal to $g$.
\end{defn}
\nomenclature{$\sqcup$}{Disjoint union}

This basic notion enables us to impose our first algebraic structure on graphs:

\begin{defn}[Graph algebra]
\label{def:graph_algebra}
We define $\Gaul$ as the $\Q$-algebra generated by all elements of $\Gul$ with the following multiplication, defined on its generators:
\begin{align} &m& &:& \Gaul &\otimes \Gaul &\rightarrow& &&\Gaul, \\ && && \Gamma_1 &\otimes \Gamma_2 &\mapsto& &&\Gamma_1 \sqcup \Gamma_2, \end{align}
where $\Gamma_1 \sqcup \Gamma_2$ denotes the unlabelled graph associated to the disjoint union\footnote{Arguably, it would be clearer to use a map $\pi$ that maps an arbitrary graph to its unique isomorphic representative in $\Gul$. The product would then read, $m (\Gamma_1 \otimes \Gamma_2) = \pi( \Gamma_1 \sqcup \Gamma_2)$. We will omit this map $\pi$ to agree with the notation commonly used in the literature.} of the representatives $\Gamma_1$ and $\Gamma_2$. 
This multiplication is obviously commutative and associative.
The empty graph $\one$, $H_\one = V_\one = \emptyset$, is the neutral element of $\Gaul$.

For formal reasons it is convenient to also endow $\Gaul$ with a \textit{unit}, a linear map 
$\unit: \Q \rightarrow \Gaul, q \mapsto q \one$ that multiplies a rational number with the neutral element of $\Gaul$.

$\Gaul$ is therefore a unital commutative algebra.
\end{defn}
\nomenclature{$\Gaul$}{The (Hopf) algebra of graphs}

If we are dealing with more than one algebra at the same time, we will denote the multiplication, the neutral element and the unit of the algebra with a reference to the respective algebra in the subscript. In the case of $\Gaul$ they will be referenced as $m_\Gaul$, $\one_\Gaul$ and $\unit_\Gaul$.

Many properties of the graphs are \textit{compatible} with this multiplication. Such properties give a \textit{grading} of the algebra. 

\begin{defn}[Graded algebra]
\label{def:algebra_grading}
A grading is a decomposition of $\Gaul$ into linear subspaces
\begin{align} \Gaul = \bigoplus_{\mulind{i} \in I} \Gaul_\mulind{i} \end{align}
with an (multi-)index set $I = \N_0^n$ where $n \geq 1$, such that 
\begin{align} &m( \Gaul_\mulind{i} \otimes \Gaul_\mulind{j} )& & \subset& \Gaul_{\mulind{i}+\mulind{j}} && \text{ for all } \mulind{i},\mulind{j}\in I. \end{align}
\end{defn}
A grading with a multidimensional index set is sometimes called a multigrading. 
Such a decomposition can be constructed by fixing some graph property, for instance the number of vertices of graphs, and fixing $\Gaul_k$ to be the subspace of $\Gaul$ which is generated by all graphs with $k$ vertices. As the disjoint union of a graph with $k_1$ and a graph with $k_2$ vertices will obviously have $k_1+k_2$ vertices, this gives a grading of $\Gaul$.

The algebra $\Gaul$ is, for instance, graded by 
\begin{enumerate}
\item The number of vertices $|V_\Gamma|$.
\item The number of half-edges $|H_\Gamma|$.
\item The number of edges $|E_\Gamma|$.
\item The number of legs $|\legs_\Gamma|$.
\item The number of connected components $|\comps_\Gamma|$.
\item The number of vertices with given degree $d$, $k_\Gamma^d:= \left|\{ v \in V_\Gamma : |\nu_\Gamma^{-1}(v)| = d \}\right|$.
\item The first Betti number of the graph in the simplicial homology $h_\Gamma: = | E_\Gamma | - |V_\Gamma| + |\comps_\Gamma|$.
\end{enumerate}
as can easily be checked using Definition \ref{def:graph1}.
\nomenclature{$h_\Gamma$}{The first Betti or loop number of the graph $\Gamma$}

If there is some grading with (multi-)index set $I$, $\Gaul = \bigoplus_{\mulind{i} \in I} \Gaul_\mulind{i}$ such that the spaces $\Gaul_\mulind{i}$ are finite dimensional, we can interpret $\Gaul$ as a \textit{combinatorial class} \cite[p.\ 16]{flajolet2009analytic}.
\nomenclature{$\Gaul_i$}{Homogeneous subspace of $\Gaul$}

\subsection{A note on convergence}
We will make use of formal limits in algebras such as $\Gaul$. The grading of the algebra is a technical necessity for these considerations to make sense. Every statement involving an infinite number of generators in $\Gaul$ is required to be translatable into a statement over a finite number of generators in a suitable decomposition of $\Gaul$ that will typically be a grading. A sufficiently general grading is the bigrading by the number of half-edges $m$ and the number of vertices $k$ such that $\Gaul = \bigoplus_{m,k \in \N_0} \Gaul_{m,k}$. We will endow each $\Gaul_{m,k}$ with the discrete topology and $\Gaul$ with the product topology over all $\Gaul_{m,k}$. Statements, such as the one in the following section, that involve an infinite number of generators are required to be convergent in this topology. The convergence is usually obvious.

\subsection{The exponential formula}

The most important element of the algebra $\Gaul$ will be the sum of all graphs weighted by the cardinality of their automorphism group. We will denote this vector in $\Gaul$ by $\allG$.

\begin{align} \allG := \sum_{\substack{\Gamma \in \Gul }} \frac{\Gamma}{|\Aut \Gamma|}. \end{align}

\nomenclature{$\allG$}{Vector in $\Gaul$ consisting of all graphs}

Another important element is the respective sum of all connected graphs, that means all graphs with one connected component:
\begin{align} \allGc := \sum_{\substack{\Gamma \in \Gul\\|\comps_\Gamma|=1 }} \frac{\Gamma}{|\Aut \Gamma|}. \end{align}
Note that the empty graph has no connected component. Therefore, it does not contribute to this sum.

\nomenclature{$\allGc$}{Vector in $\Gaul$ consisting of all connected graphs}

Both infinite sums $\allG$ and $\allGc$ are clearly convergent in the product topology over the discrete topology of the subspaces $\Gaul_{m,k}$. For instance,
\begin{gather*} \allG = \sum_{m,k \geq 0} \sum_{\substack{\Gamma \in \Gul\\|H_\Gamma|=m, |V_\Gamma|=k }} \frac{\Gamma}{|\Aut \Gamma|}, \end{gather*}
where each inner sum is a finite element of $\Gaul_{m,k}$.

The following theorem is known as the `exponential formula' \cite[p. 8]{harary2014graphical}:
\begin{thm}
\label{thm:connected_disconnected}
The following algebraic identity holds in $\Gaul$:
\begin{align} \allG &= e^{ \allGc } & \allGc &= \log( \allG ). \end{align}
\end{thm}
\begin{proof}
For the sum over all graphs with $n$ connected components, we have
\begin{align*} \sum_{\substack{\Gamma \in \Gul\\|\comps_\Gamma|=n}} \frac{\Gamma}{|\Aut \Gamma|} = \frac{1}{n!}\sum_{\substack{ \gamma_1, \ldots, \gamma_n \in \Gul\\|\comps_{\gamma_i}|=1 }} \prod_{i=1}^n \frac{ \gamma_i }{|\Aut \gamma_i|}, \end{align*}
because every graph with $n$ components can be written as a disjoint union of $n$ graphs with one connected component and the factorial $\frac{1}{n!}$ accounts for overcounting symmetries between these components. 
Summing over $n \geq 0$ and using $e^{x} = \sum_{n\geq 0} \frac{x^n}{n!}$ results in the statement. 
\end{proof}

As a technical detail note that we can always reduce the statement of this theorem to a statement over a finite number of graphs with a bounded number of half-edges and vertices. For instance, 
for graphs with $m$ half-edges and $k$ vertices,
\begin{gather*} \sum_{\substack{\Gamma \in \Gul\\ |H_\Gamma|=m, |V_\Gamma|=k }} \frac{\Gamma}{|\Aut \Gamma|} = \sum_{n \geq 0} \sum_{\substack{m_1, \ldots, m_n \geq 1\\ m_1 + \ldots + m_n = m}} \sum_{\substack{k_1, \ldots, k_n \geq 1\\ k_1 + \ldots + k_n = k}} \frac{1}{n!} \sum_{\substack{ \gamma_1, \ldots, \gamma_n \in \Gul\\ |\comps_{\gamma_i}|=1\\|H_{\gamma_i}|=m_i, |V_{\gamma_i}|=k_i }} \prod_{i=1}^n \frac{ \gamma_i }{|\Aut \gamma_i|} \end{gather*}
where the sum over $n$ terminates after $k$ terms, as each connected component must at least have one vertex. %

Of course, the expression in Theorem \ref{thm:connected_disconnected} is much more convenient, but we need to keep convergence issues in mind when we think about identities involving an infinite set of graphs.

\subsection{Algebra homomorphisms}
Eventually, we will be interested in linear mappings from $\Gaul$ to some other algebra, for instance a power series algebra. Such maps, which also preserve the algebra structure, will be of special importance. They are called algebra homomorphisms. 

\begin{defn}[Algebra homomorphism]
\label{def:algebra_morphism}
A linear map $\phi: \Gaul \rightarrow \mathcal{A}$ from the algebra $\Gaul$ to another commutative algebra $\mathcal{A}$ is an algebra homomorphism if $\phi$ is compatible with the multiplication of $\Gaul$ and $\mathcal{A}$, $\phi \circ m_\Gaul = m_{\mathcal{A}}\circ (\phi \otimes \phi)$ or equivalently for all $a,b \in \Gaul$: $\phi(a) \phi(b) = \phi(a b)$. This implies that $\phi(\one_\Gaul) = \one_\mathcal{A}$.
\end{defn}
\nomenclature{$\phi$}{Homomorphism of algebras}

\begin{expl}
\label{expl:algebra_morphism_w_vertices}
The linear map defined on the generators of $\Gaul$
\begin{align} &\phi &:& &\Gaul \rightarrow \Q[[\varphi_c,a,\lambda_0, \lambda_1, \ldots]],& &\Gamma \mapsto \varphi_c^{|\legs_\Gamma|} a^{|E_\Gamma|} \prod_{v \in V_\Gamma} \lambda_{\deg{v}} \end{align}
is an algebra homomorphism from $\Gaul$ to the ring $\Q[[\varphi_c,a,\lambda_0, \lambda_1, \ldots]]$ of multivariate power series in $\varphi_c$, $a$ and the $\lambda_d$.

To verify this, observe that the algebra $\Gaul$ is graded by the number of legs $|\legs_\Gamma|$, the number of edges $|E_\Gamma|$ and for each $d \in \N_0$ by the number of vertices with degree $d$, $\nvd{d}_\Gamma$.

Therefore, $\phi(\Gamma_1 \Gamma_2) = \phi(\Gamma_1 \sqcup \Gamma_2) = \phi(\Gamma_1) \phi(\Gamma_2)$.

We can apply this map to the vector $\allG$. 
As a consequence of Corollary \ref{crll:counting_unlabelled_with_degrees}, we immediately find that 
\begin{gather*} \phi( \allG) = \phi\left( \sum_{\substack{\Gamma \in \Gul }} \frac{\Gamma}{|\Aut \Gamma|}\right) = \sum_{\substack{\Gamma \in \Gul }} \frac{\phi\left(\Gamma\right)}{|\Aut \Gamma|} \\ = \sum_{m\geq 0}m! [x^m y^m] e^{a \frac{y^2}{2} + \varphi_c y } e^{\sum_{d\geq 0} \lambda_d \frac{x^d}{d!} } . \end{gather*}

Applying Theorem \ref{thm:connected_disconnected} together with the fact that $\phi$ is an algebra homomorphism results in 
\begin{gather*} \phi( \allGc ) = \phi( \log (\allG) ) = \log \left( \sum_{m\geq 0}m! [x^m y^m] e^{a \frac{y^2}{2}+ \varphi_c y } e^{\sum_{d\geq 0} \lambda_d \frac{x^d}{d!} } \right). \end{gather*}

\end{expl}

In the next chapter, we will apply these considerations to various algebra homomorphisms that will boil down to special cases of maps such as $\phi$. Explicitly, we will use these maps to analyze \textit{zero-dimensional quantum field theories}. Calculations in these quantum field theories are essentially enumeration problems of graphs.

\chapter{Graphical enumeration}
\label{chp:graph_enumeration}
In this chapter, we will motivate our analysis of graph generating functions in detail using zero-dimensional quantum field theory. The content of this chapter is partially based on the author's article \cite{borinsky2017renormalized}.
\section{Formal integrals}
\label{sec:formalint}
Enumerating diagrams using zero-dimensional QFT is an well-established procedure with a long history \cite{hurst1952enumeration,bender1976statistical,cvitanovic1978number,bessis1980quantum,argyres2001zero} and wide-ranging applications in mathematics \cite{kontsevich1992intersection,lando2013graphs}.

The starting point for zero-dimensional QFT is the \textit{path integral}, which becomes an ordinary integral in the zero-dimensional case. For instance, in a scalar theory the \textit{partition function} is given by
\begin{align} \label{eqn:formalfuncintegral1} Z(\hbar) &:= \int_\R \frac{dx}{\sqrt{2 \pi \hbar } } e^{\frac{1}{\hbar} \left( -\frac{x^2}{2a} + V(x) \right) }, \end{align}
where $V \in x^3 \R[[x]]$, the \textit{potential}, is some power series with the first three coefficients in $x$ vanishing and $a$ is a strictly positive parameter. The whole exponent $\Sact(x) = -\frac{x^2}{2a} + V(x)$ is the \textit{action}.
\nomenclature{$\Sact$}{Action of a zero-dimensional QFT}

The integral \eqref{eqn:formalfuncintegral1} is ill-defined for general $V(x)$. If we substitute, for example, $V(x) = \frac{x^4}{4!}$, it is not integrable over $\R$. Furthermore, the power series expansion makes only limited sense as the actual function $Z(\hbar)$ will have a singularity at $\hbar=0$ - even in cases where the expression is integrable. One way to continue is to modify the integration contour, such that the integrand vanishes fast enough at the border of the integration domain. The disadvantage of this method is that the integration contour must be chosen on a case by case basis. 

Here, we are mainly concerned with the coefficients of the expansion in $\hbar$ of the integral \eqref{eqn:formalfuncintegral1}. We wish to give meaning to such an expressions in a way that highlights its \textit{power series} nature. Moreover, we would like to free ourselves from restrictions in choices of $V(x)$ as far as possible.
We therefore treat the integral \eqref{eqn:formalfuncintegral1} as a \textit{formal} expression, which is not required to yield a proper function in $\hbar$, but a formal power series in this parameter.

The procedure to obtain a power series expansion from this formal integral is well-known and widely used \cite{itzykson2005quantum}: The potential $V(x)$ is treated as a perturbation around the Gaussian kernel and the remaining integrand is expanded. 
The `integration' will be solely performed by applying the identity 
\begin{align*} \int_\R \frac{dx}{\sqrt{2 \pi \hbar}} e^{-\frac{x^2}{2 a \hbar}} x^{2n} &= \sqrt{a} (a\hbar)^n (2n-1)!! & & n \geq 0. \end{align*}
This procedure mimics the calculation of amplitudes in higher dimensions, as the above identity is the zero-dimensional version of Wick's theorem \cite{itzykson2005quantum}. This way, it directly incorporates the interpretation of the coefficients of the power series as \textit{Feynman diagrams}. 
Unfortunately, these \textit{formal integrals} seem not to have been studied in detail as isolated mathematical entities. For know, we will give a translation of the formal integral to a well-defined formal power series. This will serve as a definition of a formal integral.

We expand the exponent of $V(x)$ and exchange integration and summation and thereby \textit{define} the zero-dimensional path integral as the following expression:
\begin{defn}
\label{def:formalintegral}
Let $\Fop : x^2\R[[x]] \rightarrow \R[[\hbar]]$ be the operator that maps $\Sact(x) \in x^2 \R[[x]]$, a power series with vanishing constant and linear terms as well as a strictly negative quadratic term, $\Sact(x) = - \frac{x^2}{2a} + V(x)$, to $\mathcal{F}[\Sact(x)]\in\R[[\hbar]]$ a power series in $\hbar$, such that 
\begin{align} \label{eqn:formalintegralpwrsrs} \Fop[\Sact(x)](\hbar) = \sqrt{a} \sum_{n=0}^\infty \left(a \hbar\right)^{n} (2n-1)!! [x^{2n}] e^{\frac{1}{\hbar}V(x)}. \end{align}
\end{defn}
This gives a well-defined power series in $\hbar$, because $[x^{2n}] e^{\frac{1}{\hbar}V(x)}$ is a polynomial in $\hbar^{-1}$ of degree smaller than $n$ as $V(x) \in x^3 \R[[x]]$.
\nomenclature{$\Fop$}{Feynman transformation operator}

An advantage of applying this definition rather then using the integral itself is that Definition \ref{def:formalintegral} gives an unambiguous procedure to obtain the expansion for a given potential, whereas the integration depends heavily on the choice of the integration contour. 

The most important property of $\Fop$ and the connection to the previous chapter is that $\Fop[\Sact(x)](\hbar)$ enumerates \textit{multigraphs}. 
\subsection{Diagrammatic interpretation}

The identity from Proposition \ref{prop:counting_with_degrees}, can immediately be specialized to the identity,
\begin{align} \sum_{\substack{G \in \Gl\\|\legs_G|=0}} \frac{a^{|E_G|} \prod_{v \in V_G} \lambda_{\deg{v}} }{|H_G|! |V_G|!} = \sum_{m\geq 0}m! [x^m y^m] e^{a \frac{y^2}{2} } e^{\sum_{d\geq 0} \lambda_d \frac{x^d}{d!} } \end{align}
by restricting to graphs without legs. Applying Proposition \ref{prop:graph_labelled_unlabelled} and evaluating the coefficient extraction in $y$ gives,
\begin{align} \sum_{\substack{\Gamma \in \Gul\\|\legs_\Gamma|=0}} \frac{a^{|E_\Gamma|} \prod_{v \in V_\Gamma} \lambda_{\deg{v}} }{| \Aut \Gamma|} = \sum_{m\geq 0} a^m (2m-1)!! [x^{2m}] e^{\sum_{d\geq 0} \lambda_d \frac{x^d}{d!} } \end{align}
and scaling $a \rightarrow a \hbar$ as well as $\lambda_d \rightarrow \frac{\lambda_d}{\hbar}$ for all $d\in \N_0$ gives,
\begin{align} \sum_{\substack{\Gamma \in \Gul\\|\legs_\Gamma|=0}} \hbar^{|E_\Gamma|-|V_\Gamma|} \frac{a^{|E_\Gamma|} \prod_{v \in V_\Gamma} \lambda_{\deg{v}} }{| \Aut \Gamma|} = \sum_{m\geq 0} (\hbar a)^m (2m-1)!! [x^{2m}] e^{\frac{1}{\hbar} \sum_{d\geq 0} \lambda_d \frac{x^d}{d!} }. \end{align}
On the right hand side we recovered the expression in eq.\ \eqref{eqn:formalintegralpwrsrs} except for the $\sqrt{a}$ factor. On the left hand side, we can identify an algebra homomorphism from $\Gaul$ to $\R[[\hbar]]$:
\begin{align} &\phi_\Sact &:& &\Gaul& &\rightarrow& &&\R[[\hbar]] \\ & & & &\Gamma& &\mapsto& &&\hbar^{|E_\Gamma|-|V_\Gamma|} a^{|E_\Gamma|} \prod_{v \in V_\Gamma} \lambda_{\deg{v}}, \end{align}
which is defined for all generators $\Gamma \in \Gul$,
where $a$ and $\lambda_d$ are encoded in the action $\Sact(x) = -\frac{x^2}{2a} + \sum_{d\geq 3} \lambda_{d} \frac{x^d}{d!}$. The variables associated to the $0$-, $1$- and $2$-valent vertices are set to zero, $\lambda_0=\lambda_1=\lambda_2 = 0$, therefore all graphs with a $0$-, $1$- or $2$-valent vertex are mapped to $0$ under $\phi_\Sact$. Moreover, $\phi_\Sact$ shall map all graphs which have legs to zero: $\phi_\Sact(\Gamma) = 0$ for all $\Gamma \in \Gul$ with $\legs_\Gamma \neq \emptyset$.

We will refer to algebra homomorphisms such as $\phi_\Sact$ that emerge from an interpretation of graphs as terms in a perturbation expansion as \textit{Feynman rules}.
\nomenclature{$\phi_\Sact$}{Simple zero-dimensional Feynman rules with action $\Sact$}

Identifying this expression with the one in Definition \ref{def:formalintegral} gives the diagrammatic interpretation of $\Fop$ expressions:

\begin{prop}
\label{prop:diagraminterpretation}
If $\Sact(x) = -\frac{x^2}{2 a} + \sum_{d=3}^\infty \frac{\lambda_d}{d!} x^d$ with $a>0$, then 
\begin{align*} \Fop[\Sact(x)](\hbar) = \sqrt{a} \phi_{\Sact}( \allG) = \sqrt{a}\sum_{\substack{\Gamma \in \Gul\\|\legs_\Gamma|=0}} \hbar^{|E_\Gamma| - |V_\Gamma|} \frac{ a^{|E_\Gamma|} \prod_{v \in V_\Gamma} \lambda_{\deg{v}}}{|\Aut \Gamma|}. \end{align*}
\end{prop}
This identity can also be used as definition of $\Fop$.
It is well-known that the terms in the expansion of the integral \eqref{eqn:formalfuncintegral1} and therefore also the terms of $\Fop[\Sact(x)](\hbar)$ can be interpreted as a sum over Feynman diagrams weighted by their symmetry factor \cite{cvitanovic1978number}. To calculate the $n$-th coefficient of \eqref{eqn:formalfuncintegral1} or \eqref{eqn:formalintegralpwrsrs} with $\Sact(x) = - \frac{x^2}{2a} + V(x)$ and $V(x) = \sum_{d=3}^\infty \frac{\lambda_d}{d!} x^d$ naively
\begin{enumerate}
\item
draw all graphs with \textit{excess} $n$ and with minimal vertex degree $3$. The excess of a diagram $\Gamma$ is given by $|E_\Gamma|-|V_\Gamma|$, the number of edges minus the number of vertices. 
For connected graphs the excess is equal to the number of \textit{loops} minus 1. We say a graph has $n$ loops if it has $n$ independent cycles. The number of loops is also the first \textit{Betti number} of the graph.
\item
For each individual graph $\Gamma$ calculate the product $\prod_{v \in V_\Gamma} \lambda_{\deg{v}}$, where each vertex contributes with a factor $\lambda_{\deg{v}}$ with $\deg{v}$ the degree of the vertex. Subsequently, multiply by $a^{|E_\Gamma|}$.
\item
Calculate the cardinality of the automorphism group of the graph. Divide the result of the previous calculation by this cardinality. 
\item
Sum all monomials and multiply the obtained polynomial by a normalization factor of $\sqrt{a}$.
\end{enumerate}
We may write the power series expansion of $\Fop[\Sact(x)](\hbar)$ in a diagrammatic way as follows:
\begin{gather} \begin{gathered} \label{eqn:generalexpansion} \Fop[\Sact(x)](\hbar) = \sqrt{a} \phi_{\Sact} \Big( \one + \frac18 {  \ifmmode \usebox{\fghandle} \else \newsavebox{\fghandle} \savebox{\fghandle}{ \begin{tikzpicture}[x=1ex,y=1ex,baseline={([yshift=-.5ex]current bounding box.center)}] \coordinate (v0); \coordinate [right=1.5 of v0] (v1); \coordinate [left=.7 of v0] (i0); \coordinate [right=.7 of v1] (o0); \draw (v0) -- (v1); \filldraw (v0) circle (1pt); \filldraw (v1) circle (1pt); \draw (i0) circle(.7); \draw (o0) circle(.7); \end{tikzpicture} } \fi } + \frac{1}{12} {  \ifmmode \usebox{\fgbananathree} \else \newsavebox{\fgbananathree} \savebox{\fgbananathree}{ \begin{tikzpicture}[x=1ex,y=1ex,baseline={([yshift=-.5ex]current bounding box.center)}] \coordinate (vm); \coordinate [left=1 of vm] (v0); \coordinate [right=1 of vm] (v1); \draw (v0) -- (v1); \draw (vm) circle(1); \filldraw (v0) circle (1pt); \filldraw (v1) circle (1pt); \end{tikzpicture} } \fi } + \frac{1}{8} {  \ifmmode \usebox{\fgtadpoletwo} \else \newsavebox{\fgtadpoletwo} \savebox{\fgtadpoletwo}{ \begin{tikzpicture}[x=1ex,y=1ex,baseline={([yshift=-.5ex]current bounding box.center)}] \coordinate (vm); \coordinate [left=.7 of vm] (v0); \coordinate [right=.7 of vm] (v1); \draw (v0) circle(.7); \draw (v1) circle(.7); \filldraw (vm) circle (1pt); \end{tikzpicture} } \fi } \\ + \frac{1}{128} {  \ifmmode \usebox{\fgtwohandles} \else \newsavebox{\fgtwohandles} \savebox{\fgtwohandles}{ \begin{tikzpicture}[x=1ex,y=1ex,baseline={([yshift=-.5ex]current bounding box.center)}] \coordinate (v); \coordinate [above=1.2 of v] (v01); \coordinate [right=1.5 of v01] (v11); \coordinate [left=.7 of v01] (i01); \coordinate [right=.7 of v11] (o01); \draw (v01) -- (v11); \filldraw (v01) circle (1pt); \filldraw (v11) circle (1pt); \draw (i01) circle(.7); \draw (o01) circle(.7); \coordinate [below=1.2 of v] (v02); \coordinate [right=1.5 of v02] (v12); \coordinate [left=.7 of v02] (i02); \coordinate [right=.7 of v12] (o02); \draw (v02) -- (v12); \filldraw (v02) circle (1pt); \filldraw (v12) circle (1pt); \draw (i02) circle(.7); \draw (o02) circle(.7); \end{tikzpicture} } \fi } + \frac{1}{288} {  \ifmmode \usebox{\fgtwobananasthree} \else \newsavebox{\fgtwobananasthree} \savebox{\fgtwobananasthree}{ \begin{tikzpicture}[x=1ex,y=1ex,baseline={([yshift=-.5ex]current bounding box.center)}] \coordinate (v); \coordinate [above=1.2 of v](v01); \coordinate [right=2 of v01] (v11); \coordinate [right=1 of v01] (vm1); \draw (v01) -- (v11); \draw (vm1) circle(1); \filldraw (v01) circle (1pt); \filldraw (v11) circle (1pt); \coordinate [below=1.2 of v](v02); \coordinate [right=2 of v02] (v12); \coordinate [right=1 of v02] (vm2); \draw (v02) -- (v12); \draw (vm2) circle(1); \filldraw (v02) circle (1pt); \filldraw (v12) circle (1pt); \end{tikzpicture} } \fi } + \frac{1}{96} {  \ifmmode \usebox{\fgbananathreeandhandle} \else \newsavebox{\fgbananathreeandhandle} \savebox{\fgbananathreeandhandle}{ \begin{tikzpicture}[x=1ex,y=1ex,baseline={([yshift=-.5ex]current bounding box.center)}] \coordinate (v); \coordinate [above=1.2 of v] (vm1); \coordinate [left=1 of vm1] (v01); \coordinate [right=1 of vm1] (v11); \draw (v01) -- (v11); \draw (vm1) circle(1); \filldraw (v01) circle (1pt); \filldraw (v11) circle (1pt); \coordinate [below=1.2 of v] (vm2); \coordinate [left=.75 of vm2] (v02); \coordinate [right=.75 of vm2] (v12); \coordinate [left=.7 of v02] (i02); \coordinate [right=.7 of v12] (o02); \draw (v02) -- (v12); \filldraw (v02) circle (1pt); \filldraw (v12) circle (1pt); \draw (i02) circle(.7); \draw (o02) circle(.7); \end{tikzpicture} } \fi } + \frac{1}{48} {  \ifmmode \usebox{\fgpropellerthree} \else \newsavebox{\fgpropellerthree} \savebox{\fgpropellerthree}{ \begin{tikzpicture}[x=1ex,y=1ex,baseline={([yshift=-.5ex]current bounding box.center)}] \coordinate (v) ; \def \n {3}; \def \rad {1.2}; \def \rud {1.9}; \foreach \s in {1,...,\n} { \def \angle {360/\n*(\s - 1)}; \coordinate (s) at ([shift=({\angle}:\rad)]v); \coordinate (u) at ([shift=({\angle}:\rud)]v); \draw (v) -- (s); \filldraw (s) circle (1pt); \draw (u) circle (.7); } \filldraw (v) circle (1pt); \end{tikzpicture} } \fi } \\ + \frac{1}{16} {  \ifmmode \usebox{\fgdblhandle} \else \newsavebox{\fgdblhandle} \savebox{\fgdblhandle}{ \begin{tikzpicture}[x=1ex,y=1ex,baseline={([yshift=-.5ex]current bounding box.center)}] \coordinate (v) ; \def \n {2}; \def \rad {.7}; \def \rud {2}; \def \rid {2.7}; \foreach \s in {1,...,\n} { \def \angle {360/\n*(\s - 1)}; \coordinate (s) at ([shift=({\angle}:\rad)]v); \coordinate (u) at ([shift=({\angle}:\rud)]v); \coordinate (t) at ([shift=({\angle}:\rid)]v); \draw (s) -- (u); \filldraw (u) circle (1pt); \filldraw (s) circle (1pt); \draw (t) circle (.7); } \draw (v) circle(\rad); \end{tikzpicture} } \fi } + \frac{1}{16} {  \ifmmode \usebox{\fgdbleye} \else \newsavebox{\fgdbleye} \savebox{\fgdbleye}{ \begin{tikzpicture}[x=1ex,y=1ex,baseline={([yshift=-.5ex]current bounding box.center)}] \coordinate (v0); \coordinate[above left=1.5 of v0] (v1); \coordinate[below left=1.5 of v0] (v2); \coordinate[above right=1.5 of v0] (v3); \coordinate[below right=1.5 of v0] (v4); \draw (v1) to[bend left=80] (v2); \draw (v1) to[bend right=80] (v2); \draw (v3) to[bend right=80] (v4); \draw (v3) to[bend left=80] (v4); \draw (v1) -- (v3); \draw (v2) -- (v4); \filldraw (v1) circle(1pt); \filldraw (v2) circle(1pt); \filldraw (v3) circle(1pt); \filldraw (v4) circle(1pt); \end{tikzpicture} } \fi } + \frac{1}{8} {  \ifmmode \usebox{\fgcloseddunce} \else \newsavebox{\fgcloseddunce} \savebox{\fgcloseddunce}{ \begin{tikzpicture}[x=1ex,y=1ex,baseline={([yshift=-.5ex]current bounding box.center)}] \coordinate (v0); \coordinate[right=.5 of v0] (v3); \coordinate[right=1.5 of v3] (v4); \coordinate[above left=1.5 of v0] (v1); \coordinate[below left=1.5 of v0] (v2); \coordinate[right=.7 of v4] (o); \draw (v3) to[bend left=20] (v2); \draw (v3) to[bend right=20] (v1); \draw (v1) to[bend right=80] (v2); \draw (v1) to[bend left=80] (v2); \draw (v3) -- (v4); \filldraw (v1) circle(1pt); \filldraw (v2) circle(1pt); \filldraw (v3) circle(1pt); \filldraw (v4) circle(1pt); \draw (o) circle(.7); \end{tikzpicture} } \fi } + \frac{1}{24} {  \ifmmode \usebox{\fgwheelthree} \else \newsavebox{\fgwheelthree} \savebox{\fgwheelthree}{ \begin{tikzpicture}[x=1ex,y=1ex,baseline={([yshift=-.5ex]current bounding box.center)}] \coordinate (v) ; \def \n {3}; \def \rad {1.2}; \foreach \s in {1,...,\n} { \def \angle {360/\n*(\s - 1)}; \coordinate (s) at ([shift=({\angle}:\rad)]v); \draw (v) -- (s); \filldraw (s) circle (1pt); } \draw (v) circle (\rad); \filldraw (v) circle (1pt); \end{tikzpicture} } \fi } \\ + \frac{1}{96} {  \ifmmode \usebox{\fgbananathreeandtadpoletwo} \else \newsavebox{\fgbananathreeandtadpoletwo} \savebox{\fgbananathreeandtadpoletwo}{ \begin{tikzpicture}[x=1ex,y=1ex,baseline={([yshift=-.5ex]current bounding box.center)}] \coordinate (v); \coordinate [above=1.2 of v](vm1); \coordinate [left=1 of vm1] (v01); \coordinate [right=1 of vm1] (v11); \draw (v01) -- (v11); \draw (vm1) circle(1); \filldraw (v01) circle (1pt); \filldraw (v11) circle (1pt); \coordinate [below=1.2 of v] (vm2); \coordinate [left=.7 of vm2] (v02); \coordinate [right=.7 of vm2] (v12); \draw (v02) circle(.7); \draw (v12) circle(.7); \filldraw (vm2) circle (1pt); \end{tikzpicture} } \fi } + \frac{1}{64} {  \ifmmode \usebox{\fghandleandtadpoletwo} \else \newsavebox{\fghandleandtadpoletwo} \savebox{\fghandleandtadpoletwo}{ \begin{tikzpicture}[x=1ex,y=1ex,baseline={([yshift=-.5ex]current bounding box.center)}] \coordinate (v); \coordinate [above=1.2 of v] (vm1); \coordinate [left=.75 of vm1] (v01); \coordinate [right=.75 of vm1] (v11); \coordinate [left=.7 of v01] (i01); \coordinate [right=.7 of v11] (o01); \draw (v01) -- (v11); \filldraw (v01) circle (1pt); \filldraw (v11) circle (1pt); \draw (i01) circle(.7); \draw (o01) circle(.7); \coordinate [below=1.2 of v] (vm2); \coordinate [left=.7 of vm2] (v02); \coordinate [right=.7 of vm2] (v12); \draw (v02) circle(.7); \draw (v12) circle(.7); \filldraw (vm2) circle (1pt); \end{tikzpicture} } \fi } + \frac{1}{8} {  \ifmmode \usebox{\fgtadpoletwoclosed} \else \newsavebox{\fgtadpoletwoclosed} \savebox{\fgtadpoletwoclosed}{ \begin{tikzpicture}[x=1ex,y=1ex,baseline={([yshift=-.5ex]current bounding box.center)}] \coordinate (vm); \coordinate [left=.7 of vm] (v0); \coordinate [right=.7 of vm] (v1); \coordinate [above=.7 of v0] (v2); \coordinate [above=.7 of v1] (v3); \draw (v0) circle(.7); \draw (v1) circle(.7); \draw (v3) arc(0:180:.7) (v2); \filldraw (vm) circle (1pt); \filldraw (v2) circle (1pt); \filldraw (v3) circle (1pt); \end{tikzpicture} } \fi } + \frac{1}{16} {  \ifmmode \usebox{\fgcontractedpropellerthree} \else \newsavebox{\fgcontractedpropellerthree} \savebox{\fgcontractedpropellerthree}{ \begin{tikzpicture}[x=1ex,y=1ex,baseline={([yshift=-.5ex]current bounding box.center)}] \coordinate (v) ; \def \n {3}; \def \rad {1.2}; \def \rud {1.9}; \foreach \s in {2,...,\n} { \def \angle {360/\n*(\s - 1)}; \coordinate (s) at ([shift=({\angle}:\rad)]v); \coordinate (u) at ([shift=({\angle}:\rud)]v); \draw (v) -- (s); \filldraw (s) circle (1pt); \draw (u) circle (.7); } \filldraw (v) circle (1pt); \coordinate (s) at ([shift=({0}:.7)]v); \draw (s) circle (.7); \end{tikzpicture} } \fi } + \frac{1}{8} {  \ifmmode \usebox{\fgcontracteddblhandle} \else \newsavebox{\fgcontracteddblhandle} \savebox{\fgcontracteddblhandle}{ \begin{tikzpicture}[x=1ex,y=1ex,baseline={([yshift=-.5ex]current bounding box.center)}] \coordinate (v0); \coordinate [right=1.5 of v0] (v1); \coordinate [left=.7 of v0] (i0); \coordinate [right=.7 of v1] (o0); \coordinate [right=.7 of o0] (v2); \coordinate [right=.7 of v2] (o1); \draw (v0) -- (v1); \filldraw (v0) circle (1pt); \filldraw (v1) circle (1pt); \filldraw (v2) circle (1pt); \draw (i0) circle(.7); \draw (o0) circle(.7); \draw (o1) circle(.7); \end{tikzpicture} } \fi } + \frac{1}{12} {  \ifmmode \usebox{\fgbananathreehandle} \else \newsavebox{\fgbananathreehandle} \savebox{\fgbananathreehandle}{ \begin{tikzpicture}[x=1ex,y=1ex,baseline={([yshift=-.5ex]current bounding box.center)}] \coordinate (vm); \coordinate [left=1 of vm] (v0); \coordinate [right=1 of vm] (v1); \coordinate [right=1.5 of v1] (v2); \coordinate [right=.7 of v2] (o); \draw (v0) -- (v1); \draw (v1) -- (v2); \draw (vm) circle(1); \draw (o) circle(.7); \filldraw (v0) circle (1pt); \filldraw (v1) circle (1pt); \filldraw (v2) circle (1pt); \end{tikzpicture} } \fi } + \frac{1}{8} {  \ifmmode \usebox{\fgcontractedcloseddunce} \else \newsavebox{\fgcontractedcloseddunce} \savebox{\fgcontractedcloseddunce}{ \begin{tikzpicture}[x=1ex,y=1ex,baseline={([yshift=-.5ex]current bounding box.center)}] \coordinate (v0); \coordinate[right=.5 of v0] (v3); \coordinate[above left=1.5 of v0] (v1); \coordinate[below left=1.5 of v0] (v2); \coordinate[right=.7 of v3] (o); \draw (v3) to[bend left=20] (v2); \draw (v3) to[bend right=20] (v1); \draw (v1) to[bend right=80] (v2); \draw (v1) to[bend left=80] (v2); \filldraw (v1) circle(1pt); \filldraw (v2) circle(1pt); \filldraw (v3) circle(1pt); \draw (o) circle(.7); \end{tikzpicture} } \fi } \\ + \frac{1}{128} {  \ifmmode \usebox{\fgtwotadpoletwos} \else \newsavebox{\fgtwotadpoletwos} \savebox{\fgtwotadpoletwos}{ \begin{tikzpicture}[x=1ex,y=1ex,baseline={([yshift=-.5ex]current bounding box.center)}] \coordinate (v); \coordinate [above=1.2 of v] (vm1); \coordinate [left=.7 of vm1] (v01); \coordinate [right=.7 of vm1] (v11); \draw (v01) circle(.7); \draw (v11) circle(.7); \filldraw (vm1) circle (1pt); \coordinate [below=1.2 of v] (vm2); \coordinate [left=.7 of vm2] (v02); \coordinate [right=.7 of vm2] (v12); \draw (v02) circle(.7); \draw (v12) circle(.7); \filldraw (vm2) circle (1pt); \end{tikzpicture} } \fi } + \frac{1}{48} {  \ifmmode \usebox{\fgbananafour} \else \newsavebox{\fgbananafour} \savebox{\fgbananafour}{ \begin{tikzpicture}[x=1ex,y=1ex,baseline={([yshift=-.5ex]current bounding box.center)}] \coordinate (vm); \coordinate [left=1 of vm] (v0); \coordinate [right=1 of vm] (v1); \draw (v0) to[bend left=45] (v1); \draw (v0) to[bend right=45] (v1); \draw (vm) circle(1); \filldraw (v0) circle (1pt); \filldraw (v1) circle (1pt); \end{tikzpicture} } \fi } + \frac{1}{16} {  \ifmmode \usebox{\fgthreebubble} \else \newsavebox{\fgthreebubble} \savebox{\fgthreebubble}{ \begin{tikzpicture}[x=1ex,y=1ex,baseline={([yshift=-.5ex]current bounding box.center)}] \coordinate (vm); \coordinate [left=.7 of vm] (v0); \coordinate [right=.7 of vm] (v1); \coordinate [left=.7 of v0] (vc1); \coordinate [right=.7 of v1] (vc2); \draw (vc1) circle(.7); \draw (vc2) circle(.7); \draw (vm) circle(.7); \filldraw (v0) circle (1pt); \filldraw (v1) circle (1pt); \end{tikzpicture} } \fi } + \frac{1}{12} {  \ifmmode \usebox{\fgbananathreewithbubble} \else \newsavebox{\fgbananathreewithbubble} \savebox{\fgbananathreewithbubble}{ \begin{tikzpicture}[x=1ex,y=1ex,baseline={([yshift=-.5ex]current bounding box.center)}] \coordinate (vm); \coordinate [left=1 of vm] (v0); \coordinate [right=1 of vm] (v1); \coordinate [right=.7 of v1] (o); \draw (v0) -- (v1); \draw (vm) circle(1); \draw (o) circle(.7); \filldraw (v0) circle (1pt); \filldraw (v1) circle (1pt); \end{tikzpicture} } \fi } + \frac{1}{16} {  \ifmmode \usebox{\fgdblcontractedpropellerthree} \else \newsavebox{\fgdblcontractedpropellerthree} \savebox{\fgdblcontractedpropellerthree}{ \begin{tikzpicture}[x=1ex,y=1ex,baseline={([yshift=-.5ex]current bounding box.center)}] \coordinate (v) ; \def \rad {1.5}; \coordinate (s1) at ([shift=(0:1.2)]v); \coordinate (s2) at ([shift=(120:\rad)]v); \coordinate (s3) at ([shift=(240:\rad)]v); \coordinate [right=.7 of s1] (o); \draw (v) to[out=180,in=210] (s2) to[out=30,in=60] (v); \draw (v) to[out=300,in=330] (s3) to[out=150,in=180] (v); \draw (v) -- (s1); \filldraw (v) circle (1pt); \filldraw (s1) circle (1pt); \draw (o) circle(.7); \end{tikzpicture} } \fi } + \frac{1}{48} {  \ifmmode \usebox{\fgthreerose} \else \newsavebox{\fgthreerose} \savebox{\fgthreerose}{ \begin{tikzpicture}[x=1ex,y=1ex,baseline={([yshift=-.5ex]current bounding box.center)}] \coordinate (v) ; \def \rad {1.5}; \coordinate (s1) at ([shift=(0:\rad)]v); \coordinate (s2) at ([shift=(120:\rad)]v); \coordinate (s3) at ([shift=(240:\rad)]v); \draw (v) to[out=60,in=90] (s1) to[out=-90,in=0-60] (v); \draw (v) to[out=180,in=210] (s2) to[out=30,in=60] (v); \draw (v) to[out=300,in=330] (s3) to[out=150,in=180] (v); \filldraw (v) circle (1pt); \end{tikzpicture} } \fi } + \cdots \Big) \\ = \sqrt{a} \Big( 1 + \left( \left( \frac{1}{8} + \frac{1}{12} \right)\lambda_3^2 a^3 + \frac{1}{8} \lambda_4 a^2 \right) \hbar \\ + \left( \frac{385}{1152} \lambda_3^4 a^6 + \frac{35}{64} \lambda_3^2 \lambda_4 a^5 + \frac{35}{384} \lambda_4^2 a^4 + \frac{7}{48} \lambda_3 \lambda_5 a^4 + \frac{1}{48} \lambda_6 a^3       \right) \hbar^2 + \cdots \Big). \end{gathered} \end{gather}
Note that we already excluded graphs with $0$-, $1$- and $2$-valent vertices from the diagrammatic expansion as they are mapped to $0$ by $\phi_\Sact$.
The expression $\Fop[\Sact(x)](\hbar) = \sum_{n=0}^\infty \hbar^n P_n(\lambda_3 a^{\frac32},\lambda_4 a^{\frac42},\ldots)$ is a sequence of polynomials $P_n$ of degree $2n$.

Of course, drawing all diagrams for a specific model\footnote{A `model' in this context is a choice for $\Sact$.} and applying the zero-di\-men\-sion\-al Feynman rules $\phi_\Sact$ is not a very convenient way to calculate the power series $\Fop[\Sact(x)](\hbar)$ order by order. 
A more efficient way is to derive differential equations from the formal integral expression and solve these recursively \cite{cvitanovic1978number, argyres2001zero}. In some cases these differential equations can be solved exactly \cite{argyres2001zero} or sufficiently simple closed forms for the respective coefficients can be found. 
For example, this is possible for the zero-dimensional version of $\varphi^3$-theory, which results in the generating function of cubic multigraphs:

\begin{expl}[The partition function of $\varphi^3$-theory]
\label{expl:phi3theoryexpansion}
In $\varphi^3$-theory the potential takes the form $V(x)= \frac{x^3}{3!}$, that means $\Sact(x)= -\frac{x^2}{2}+\frac{x^3}{3!}$. From Definition \ref{def:formalintegral} it follows that,
\begin{align*} Z^{\varphi^3}(\hbar) &= \Fop\left[-\frac{x^2}{2}+\frac{x^3}{3!}\right](\hbar) = \sum_{n=0}^\infty \hbar^n (2n-1)!! [x^{2n}] e^{\frac{x^3}{3!\hbar }} = \sum_{n=0}^\infty \hbar^{n} \frac{(6n-1)!!}{(3!)^{2n} (2n)!}, \end{align*}
where we were able to expand the expression, because for all $n\in \N_0$
\begin{align*} [x^{6n}] e^{\frac{x^3}{3!\hbar }} &= \frac{1}{(3!)^{2n} \hbar^{2n} (2n)!} && \\ [x^{6n+k}] e^{\frac{x^3}{3!\hbar }} &= 0 && \forall k\in \{1,2,3,4,5\}. \end{align*}
The diagrammatic expansion starts with
\begin{gather*} Z^{\varphi^3}(\hbar) = \phi_{\Sact} \Big( \one + \frac18 {  \ifmmode \usebox{\fghandle} \else \newsavebox{\fghandle} \savebox{\fghandle}{ \begin{tikzpicture}[x=1ex,y=1ex,baseline={([yshift=-.5ex]current bounding box.center)}] \coordinate (v0); \coordinate [right=1.5 of v0] (v1); \coordinate [left=.7 of v0] (i0); \coordinate [right=.7 of v1] (o0); \draw (v0) -- (v1); \filldraw (v0) circle (1pt); \filldraw (v1) circle (1pt); \draw (i0) circle(.7); \draw (o0) circle(.7); \end{tikzpicture} } \fi } + \frac{1}{12} {  \ifmmode \usebox{\fgbananathree} \else \newsavebox{\fgbananathree} \savebox{\fgbananathree}{ \begin{tikzpicture}[x=1ex,y=1ex,baseline={([yshift=-.5ex]current bounding box.center)}] \coordinate (vm); \coordinate [left=1 of vm] (v0); \coordinate [right=1 of vm] (v1); \draw (v0) -- (v1); \draw (vm) circle(1); \filldraw (v0) circle (1pt); \filldraw (v1) circle (1pt); \end{tikzpicture} } \fi } \\ + \frac{1}{128} {  \ifmmode \usebox{\fgtwohandles} \else \newsavebox{\fgtwohandles} \savebox{\fgtwohandles}{ \begin{tikzpicture}[x=1ex,y=1ex,baseline={([yshift=-.5ex]current bounding box.center)}] \coordinate (v); \coordinate [above=1.2 of v] (v01); \coordinate [right=1.5 of v01] (v11); \coordinate [left=.7 of v01] (i01); \coordinate [right=.7 of v11] (o01); \draw (v01) -- (v11); \filldraw (v01) circle (1pt); \filldraw (v11) circle (1pt); \draw (i01) circle(.7); \draw (o01) circle(.7); \coordinate [below=1.2 of v] (v02); \coordinate [right=1.5 of v02] (v12); \coordinate [left=.7 of v02] (i02); \coordinate [right=.7 of v12] (o02); \draw (v02) -- (v12); \filldraw (v02) circle (1pt); \filldraw (v12) circle (1pt); \draw (i02) circle(.7); \draw (o02) circle(.7); \end{tikzpicture} } \fi } + \frac{1}{288} {  \ifmmode \usebox{\fgtwobananasthree} \else \newsavebox{\fgtwobananasthree} \savebox{\fgtwobananasthree}{ \begin{tikzpicture}[x=1ex,y=1ex,baseline={([yshift=-.5ex]current bounding box.center)}] \coordinate (v); \coordinate [above=1.2 of v](v01); \coordinate [right=2 of v01] (v11); \coordinate [right=1 of v01] (vm1); \draw (v01) -- (v11); \draw (vm1) circle(1); \filldraw (v01) circle (1pt); \filldraw (v11) circle (1pt); \coordinate [below=1.2 of v](v02); \coordinate [right=2 of v02] (v12); \coordinate [right=1 of v02] (vm2); \draw (v02) -- (v12); \draw (vm2) circle(1); \filldraw (v02) circle (1pt); \filldraw (v12) circle (1pt); \end{tikzpicture} } \fi } + \frac{1}{96} {  \ifmmode \usebox{\fgbananathreeandhandle} \else \newsavebox{\fgbananathreeandhandle} \savebox{\fgbananathreeandhandle}{ \begin{tikzpicture}[x=1ex,y=1ex,baseline={([yshift=-.5ex]current bounding box.center)}] \coordinate (v); \coordinate [above=1.2 of v] (vm1); \coordinate [left=1 of vm1] (v01); \coordinate [right=1 of vm1] (v11); \draw (v01) -- (v11); \draw (vm1) circle(1); \filldraw (v01) circle (1pt); \filldraw (v11) circle (1pt); \coordinate [below=1.2 of v] (vm2); \coordinate [left=.75 of vm2] (v02); \coordinate [right=.75 of vm2] (v12); \coordinate [left=.7 of v02] (i02); \coordinate [right=.7 of v12] (o02); \draw (v02) -- (v12); \filldraw (v02) circle (1pt); \filldraw (v12) circle (1pt); \draw (i02) circle(.7); \draw (o02) circle(.7); \end{tikzpicture} } \fi } + \frac{1}{48} {  \ifmmode \usebox{\fgpropellerthree} \else \newsavebox{\fgpropellerthree} \savebox{\fgpropellerthree}{ \begin{tikzpicture}[x=1ex,y=1ex,baseline={([yshift=-.5ex]current bounding box.center)}] \coordinate (v) ; \def \n {3}; \def \rad {1.2}; \def \rud {1.9}; \foreach \s in {1,...,\n} { \def \angle {360/\n*(\s - 1)}; \coordinate (s) at ([shift=({\angle}:\rad)]v); \coordinate (u) at ([shift=({\angle}:\rud)]v); \draw (v) -- (s); \filldraw (s) circle (1pt); \draw (u) circle (.7); } \filldraw (v) circle (1pt); \end{tikzpicture} } \fi } \\ + \frac{1}{16} {  \ifmmode \usebox{\fgdblhandle} \else \newsavebox{\fgdblhandle} \savebox{\fgdblhandle}{ \begin{tikzpicture}[x=1ex,y=1ex,baseline={([yshift=-.5ex]current bounding box.center)}] \coordinate (v) ; \def \n {2}; \def \rad {.7}; \def \rud {2}; \def \rid {2.7}; \foreach \s in {1,...,\n} { \def \angle {360/\n*(\s - 1)}; \coordinate (s) at ([shift=({\angle}:\rad)]v); \coordinate (u) at ([shift=({\angle}:\rud)]v); \coordinate (t) at ([shift=({\angle}:\rid)]v); \draw (s) -- (u); \filldraw (u) circle (1pt); \filldraw (s) circle (1pt); \draw (t) circle (.7); } \draw (v) circle(\rad); \end{tikzpicture} } \fi } + \frac{1}{16} {  \ifmmode \usebox{\fgdbleye} \else \newsavebox{\fgdbleye} \savebox{\fgdbleye}{ \begin{tikzpicture}[x=1ex,y=1ex,baseline={([yshift=-.5ex]current bounding box.center)}] \coordinate (v0); \coordinate[above left=1.5 of v0] (v1); \coordinate[below left=1.5 of v0] (v2); \coordinate[above right=1.5 of v0] (v3); \coordinate[below right=1.5 of v0] (v4); \draw (v1) to[bend left=80] (v2); \draw (v1) to[bend right=80] (v2); \draw (v3) to[bend right=80] (v4); \draw (v3) to[bend left=80] (v4); \draw (v1) -- (v3); \draw (v2) -- (v4); \filldraw (v1) circle(1pt); \filldraw (v2) circle(1pt); \filldraw (v3) circle(1pt); \filldraw (v4) circle(1pt); \end{tikzpicture} } \fi } + \frac{1}{8} {  \ifmmode \usebox{\fgcloseddunce} \else \newsavebox{\fgcloseddunce} \savebox{\fgcloseddunce}{ \begin{tikzpicture}[x=1ex,y=1ex,baseline={([yshift=-.5ex]current bounding box.center)}] \coordinate (v0); \coordinate[right=.5 of v0] (v3); \coordinate[right=1.5 of v3] (v4); \coordinate[above left=1.5 of v0] (v1); \coordinate[below left=1.5 of v0] (v2); \coordinate[right=.7 of v4] (o); \draw (v3) to[bend left=20] (v2); \draw (v3) to[bend right=20] (v1); \draw (v1) to[bend right=80] (v2); \draw (v1) to[bend left=80] (v2); \draw (v3) -- (v4); \filldraw (v1) circle(1pt); \filldraw (v2) circle(1pt); \filldraw (v3) circle(1pt); \filldraw (v4) circle(1pt); \draw (o) circle(.7); \end{tikzpicture} } \fi } + \frac{1}{24} {  \ifmmode \usebox{\fgwheelthree} \else \newsavebox{\fgwheelthree} \savebox{\fgwheelthree}{ \begin{tikzpicture}[x=1ex,y=1ex,baseline={([yshift=-.5ex]current bounding box.center)}] \coordinate (v) ; \def \n {3}; \def \rad {1.2}; \foreach \s in {1,...,\n} { \def \angle {360/\n*(\s - 1)}; \coordinate (s) at ([shift=({\angle}:\rad)]v); \draw (v) -- (s); \filldraw (s) circle (1pt); } \draw (v) circle (\rad); \filldraw (v) circle (1pt); \end{tikzpicture} } \fi } + \ldots \Big) \\ = 1 +\left( \frac{1}{8} + \frac{1}{12} \right) \hbar + \frac{385}{1152} \hbar^2 + \ldots                    \end{gather*}
which is the same as the expansion in \eqref{eqn:generalexpansion} with $a= \lambda_3 = 1$ and all other $\lambda_d=0$.
\end{expl}

\begin{expl}[Generating function of all multigraphs with given excess]
\label{expl:allgraphs}
The generating function of all graphs without one or two-valent vertices is given by the partition function of the `theory' with the potential $V(x)=\sum_{d=3}^\infty\frac{x^d}{d!} = e^x - 1 - x - \frac{x^2}{2}$. Therefore,
\begin{align*} Z^\text{all}(\hbar) = \Fop\left[ -x^2 - x - 1 + e^x \right](\hbar) &= \sum_{n=0}^\infty \hbar^n (2n-1)!! [x^{2n}] e^{\frac{1}{\hbar} \left( e^x - 1 - x - \frac{x^2}{2} \right)}. \end{align*}
Here, as in many cases where $V(x)$ is not merely a monomial, the extraction of coefficients is more difficult. 
Still, the power series expansion in $\hbar$ can be computed conveniently with the methods which will be established in the next section.
The diagrammatic expansion is equivalent to the one given in eq.\ \eqref{eqn:generalexpansion} with $a=1$ and the $\lambda_d=1$ for all $d \geq 3$:
\begin{align*} Z^\text{all}(\hbar) &= 1 + \frac{1}{3} \hbar + \frac{41}{36} \hbar^2 + \cdots \end{align*}
Whereas this example has no direct interpretation in QFT, except maybe for the case of gravity, where vertices with arbitrary valency appear, it shows that formal integrals are quite powerful at enumerating general graphs. Hence the techniques of zero-dimensional QFT and formal integrals can be applied to a much broader class of topics, which evolve around graph enumeration. Especially promising is the application to the theory of complex networks \cite{albert2002statistical}. 
\end{expl}

\begin{expl}[Zero-dimensional sine-Gordon model]
\label{expl:sinegordon}
For a more exotic zero-dim\-en\-sion\-al QFT take $\Sact(x)= -\frac{\sin^2(x)}{2}$ or $V(x) = \frac{x^2}{2} -\frac{\sin^2(x)}{2} = 4\frac{x^4}{4!} - 16 \frac{x^6}{6!}+ 64 \frac{x^8}{8!} +\cdots$. This can be seen as the potential of a zero-dimensional version of the sine-Gordon model \cite{cherman2014decoding}.
\begin{align*} \Fop\left[ -\frac{\sin^2(x)}{2} \right](\hbar) &= \sum_{n=0}^\infty \hbar^n (2n-1)!! [x^{2n}] e^{\frac{1}{\hbar} \left( \frac{x^2}{2} -\frac{\sin^2(x)}{2} \right)}. \end{align*}
The diagrammatic expansion starts with
\begin{gather*} Z^{\text{sine-Gordon}}(\hbar) = \phi_{\Sact} \Big( \one + \frac{1}{8} {  \ifmmode \usebox{\fgtadpoletwo} \else \newsavebox{\fgtadpoletwo} \savebox{\fgtadpoletwo}{ \begin{tikzpicture}[x=1ex,y=1ex,baseline={([yshift=-.5ex]current bounding box.center)}] \coordinate (vm); \coordinate [left=.7 of vm] (v0); \coordinate [right=.7 of vm] (v1); \draw (v0) circle(.7); \draw (v1) circle(.7); \filldraw (vm) circle (1pt); \end{tikzpicture} } \fi } + \frac{1}{128} {  \ifmmode \usebox{\fgtwotadpoletwos} \else \newsavebox{\fgtwotadpoletwos} \savebox{\fgtwotadpoletwos}{ \begin{tikzpicture}[x=1ex,y=1ex,baseline={([yshift=-.5ex]current bounding box.center)}] \coordinate (v); \coordinate [above=1.2 of v] (vm1); \coordinate [left=.7 of vm1] (v01); \coordinate [right=.7 of vm1] (v11); \draw (v01) circle(.7); \draw (v11) circle(.7); \filldraw (vm1) circle (1pt); \coordinate [below=1.2 of v] (vm2); \coordinate [left=.7 of vm2] (v02); \coordinate [right=.7 of vm2] (v12); \draw (v02) circle(.7); \draw (v12) circle(.7); \filldraw (vm2) circle (1pt); \end{tikzpicture} } \fi } + \frac{1}{48} {  \ifmmode \usebox{\fgbananafour} \else \newsavebox{\fgbananafour} \savebox{\fgbananafour}{ \begin{tikzpicture}[x=1ex,y=1ex,baseline={([yshift=-.5ex]current bounding box.center)}] \coordinate (vm); \coordinate [left=1 of vm] (v0); \coordinate [right=1 of vm] (v1); \draw (v0) to[bend left=45] (v1); \draw (v0) to[bend right=45] (v1); \draw (vm) circle(1); \filldraw (v0) circle (1pt); \filldraw (v1) circle (1pt); \end{tikzpicture} } \fi } + \frac{1}{16} {  \ifmmode \usebox{\fgthreebubble} \else \newsavebox{\fgthreebubble} \savebox{\fgthreebubble}{ \begin{tikzpicture}[x=1ex,y=1ex,baseline={([yshift=-.5ex]current bounding box.center)}] \coordinate (vm); \coordinate [left=.7 of vm] (v0); \coordinate [right=.7 of vm] (v1); \coordinate [left=.7 of v0] (vc1); \coordinate [right=.7 of v1] (vc2); \draw (vc1) circle(.7); \draw (vc2) circle(.7); \draw (vm) circle(.7); \filldraw (v0) circle (1pt); \filldraw (v1) circle (1pt); \end{tikzpicture} } \fi } + \frac{1}{48} {  \ifmmode \usebox{\fgthreerose} \else \newsavebox{\fgthreerose} \savebox{\fgthreerose}{ \begin{tikzpicture}[x=1ex,y=1ex,baseline={([yshift=-.5ex]current bounding box.center)}] \coordinate (v) ; \def \rad {1.5}; \coordinate (s1) at ([shift=(0:\rad)]v); \coordinate (s2) at ([shift=(120:\rad)]v); \coordinate (s3) at ([shift=(240:\rad)]v); \draw (v) to[out=60,in=90] (s1) to[out=-90,in=0-60] (v); \draw (v) to[out=180,in=210] (s2) to[out=30,in=60] (v); \draw (v) to[out=300,in=330] (s3) to[out=150,in=180] (v); \filldraw (v) circle (1pt); \end{tikzpicture} } \fi } + \ldots \Big) \\ = 1 + \frac{4}{8}\hbar + \left( \frac{4^2}{128} + \frac{4^2}{48} + \frac{4^2}{16} - \frac{16}{48} \right) \hbar^2 + \ldots \\ = 1 + \frac12 \hbar + \frac{9}{8} \hbar^2 + \ldots \end{gather*}
which is equal to the expansion in \eqref{eqn:generalexpansion} with $\lambda_{2d} = (-1)^d 2^{2d-2}$ and $\lambda_{2d-1}=0$ for all $d \geq 2$.

\end{expl}

\begin{expl}[Stirling's QFT]
The following example is widely used in physics. As a matrix model it is known as Penner's model \cite{penner1988perturbative}. It agrees with Stirling's asymptotic expansion of the $\Gamma$-function \cite[A. D]{kontsevich1992intersection}:

From Euler's integral for the $\Gamma$-function we can deduce with the change of variables $t \rightarrow N e^x$ ,
\begin{align*} \frac{\Gamma(N)}{\sqrt{\frac{2\pi}{N}} \left(\frac{N}{e}\right)^N } &= \frac{1}{\sqrt{\frac{2\pi}{N}} \left(\frac{N}{e}\right)^N } \int_0^\infty dt t^{N-1} e^{-t} = \frac{e^{N}}{\sqrt{\frac{2\pi}{N}}} \int_\R dx e^{-N e^x + Nx } \\ &= \int_\R \frac{dx}{\sqrt{2\pi \frac{1}{N}}} e^{N\left(1+x-e^x\right) }. \end{align*}
This is the correction term of Stirling's formula expressed as a zero-dimensional QFT. Note, that the integral is actually convergent in this case, whereas the expansion in $\frac{1}{N}$ is not.
Therefore, $\Sact(x)= 1+x-e^x$ and
\begin{align*} Z^\text{Stirling}\left(\frac{1}{N}\right)&:= \Fop[1+x-e^x]\left(\frac{1}{N}\right).     \end{align*}
We can use the well-known Stirling expansion in terms of the Bernoulli numbers $B_k$ to state the power series more explicitly \cite{whittaker1996course}:
\begin{align*} \Fop[1+x-e^x]\left(\frac{1}{N}\right) = e^{ \sum_{k=1}^\infty \frac{B_{k+1}}{k (k+1)} \frac{1}{N^{k}}}. \end{align*}
Interestingly, Proposition \ref{prop:diagraminterpretation} provides us with a combinatorial interpretation of the Stirling expansion. 
We can directly use the expansion from eq.\ \eqref{eqn:generalexpansion} by setting all $\lambda_d = -1$ for $d\geq 3$ to calculate the first terms:
\begin{gather*} Z^\text{Stirling} \left(\frac{1}{N}\right):= \phi_{\Sact} \Big( \one + \frac18 {  \ifmmode \usebox{\fghandle} \else \newsavebox{\fghandle} \savebox{\fghandle}{ \begin{tikzpicture}[x=1ex,y=1ex,baseline={([yshift=-.5ex]current bounding box.center)}] \coordinate (v0); \coordinate [right=1.5 of v0] (v1); \coordinate [left=.7 of v0] (i0); \coordinate [right=.7 of v1] (o0); \draw (v0) -- (v1); \filldraw (v0) circle (1pt); \filldraw (v1) circle (1pt); \draw (i0) circle(.7); \draw (o0) circle(.7); \end{tikzpicture} } \fi } + \frac{1}{12} {  \ifmmode \usebox{\fgbananathree} \else \newsavebox{\fgbananathree} \savebox{\fgbananathree}{ \begin{tikzpicture}[x=1ex,y=1ex,baseline={([yshift=-.5ex]current bounding box.center)}] \coordinate (vm); \coordinate [left=1 of vm] (v0); \coordinate [right=1 of vm] (v1); \draw (v0) -- (v1); \draw (vm) circle(1); \filldraw (v0) circle (1pt); \filldraw (v1) circle (1pt); \end{tikzpicture} } \fi } + \frac{1}{8} {  \ifmmode \usebox{\fgtadpoletwo} \else \newsavebox{\fgtadpoletwo} \savebox{\fgtadpoletwo}{ \begin{tikzpicture}[x=1ex,y=1ex,baseline={([yshift=-.5ex]current bounding box.center)}] \coordinate (vm); \coordinate [left=.7 of vm] (v0); \coordinate [right=.7 of vm] (v1); \draw (v0) circle(.7); \draw (v1) circle(.7); \filldraw (vm) circle (1pt); \end{tikzpicture} } \fi } + \ldots \Big) \\ = 1 + \left( \frac18 (-1)^2 + \frac{1}{12} (-1)^2 + \frac{1}{8} (-1)^1 \right) \frac{1}{N} \\ + \left( \frac{385}{1152} (-1)^4 + \frac{35}{64} (-1)^3 + \frac{35}{384} (-1)^2 + \frac{7}{48} (-1)^2 + \frac{1}{48} (-1)^1 \right) \frac{1}{N^2} + \ldots \\ =1 + \frac{1}{12} \frac{1}{N} + \frac{1}{288} \frac{1}{N^2} + \ldots \end{gather*}
which results in the well-known asymptotic expansion of the $\Gamma$ function \cite{whittaker1996course},
\begin{align*} \Gamma(N) \underset{N\rightarrow \infty}{\sim} \sqrt{\frac{2\pi}{N}} \left(\frac{N}{e}\right)^n \left( 1 +\frac{1}{12 N} + \frac{1}{288 N^2}+\ldots\right). \end{align*}
Moreover, taking the logarithm of $\Fop[1+x-e^x]\left(\frac{1}{N}\right)$, applying Theorem \ref{thm:connected_disconnected} and using the fact that the $n$-th Bernoulli number vanishes if $n$ is odd and greater than $1$, gives us the combinatorial identities for alternating sums over graphs,
\begin{align*} \frac{B_{2n}}{2 n (2n-1)} &= \sum_{\substack{ \Gamma \in \Gul\\ |\comps_\Gamma| = 1 \\ h_\Gamma = 2n}} \frac{(-1)^{|V_\Gamma|}}{|\Aut\Gamma|} & 0 &= \sum_{\substack{ \Gamma \in \Gul\\ |\comps_\Gamma| = 1 \\ h_\Gamma = 2n+1}} \frac{(-1)^{|V_\Gamma|}}{|\Aut\Gamma|}, \end{align*}
for all $n\geq 1$ and where the sum is over all \textit{connected} graphs $\Gamma$ with a fixed first Betti number, denoted by $h_\Gamma = |E_\Gamma| - |V_\Gamma| + |\comps_\Gamma|$.
\end{expl}
\section{Representation as an affine hyperelliptic curve}

Calculating the coefficients of the power series given in Definition \ref{def:formalintegral} using the expression in eq.\ \eqref{eqn:formalintegralpwrsrs} directly is inconvenient, because an intermediate bivariate quantity $e^{\frac{1}{\hbar}V(x)}$ needs to be expanded in $x$ and in $\hbar^{-1}$.

A form that is computationally more accessible can be achieved by \textit{a formal change of variables}. 
Recall that we set $\Sact(x) = -\frac{x^2}{2a} +V(x)$. Expanding the exponential in eq.\ \eqref{eqn:formalintegralpwrsrs} gives
\begin{align*} \mathcal{F}[\Sact(x)](\hbar) &= \sqrt{a} \sum_{n=0}^\infty \sum_{k=0}^\infty \hbar^{n-k} a^{n} (2n-1)!! [x^{2n}] \frac{V(x)^k}{k!}, \intertext{where the coefficients of the inner sum vanish if $n<k$, because $V(x) \in x^3 \R[[x]]$. This equation can be seen as the zero-dimensional analog of Dyson's series \cite{itzykson2005quantum}. Shifting the summation over $n$ and substituting $V(x) = \frac{x^2}{2a } + \Sact(x)$ results in,} &= \sum_{n=0}^\infty \sum_{k=0}^\infty 2^{-k} a^{n+\frac12} \hbar^{n} \frac{(2(n+k)-1)!!}{k!} [x^{2n}] \left(1 + \frac{2 a}{x^2} \Sact(x) \right)^k \intertext{ Because $2^{-k} \frac{(2(n+k)-1)!!}{(2n-1)!! k!} = { n+k-\frac12 \choose k }$ , it follows that }  &= \sum_{n=0}^\infty a^{n+\frac12} \hbar^{n} (2n-1)!! [x^{2n}] \sum_{k=0}^\infty { n+k-\frac12 \choose k } \left(1 + \frac{2a}{x^2}\Sact(x) \right)^k, \intertext{and using $\sum_{k=0}^\infty { \alpha +k-1 \choose k } x^k = \frac{1}{(1-x)^\alpha} $ gives, } &= \sum_{n=0}^\infty \hbar^{n} (2n-1)!! [x^{2n}] \left( \frac{x}{\sqrt{-2 \Sact(x)}} \right)^{2n+1}. \end{align*}
By the Lagrange inversion formula $[y^n] g(y) = \frac{1}{n} [x^{n-1}] \left(\frac{x}{f(x)}\right)^n$ \cite[A.6]{flajolet2009analytic}, where $f(g(y)) = y$, this is equivalent to
\begin{prop}
\label{prop:formalchangeofvar}
If $\Sact(x) = -\frac{x^2}{2a} + V(x)$, then
\begin{align} \mathcal{F}[\Sact(x)](\hbar)&= \sum_{n=0}^\infty \hbar^{n} (2n+1)!! [y^{2n+1}] x(y), \end{align}
where $x(y)$ is the unique power series solution of $y = \sqrt{-2 S(x(y))}$, where the positive branch of the square root is taken. 
\end{prop}

Note, that this can be seen as a formal change of variables for the formal integral from eq.\ \eqref{eqn:formalfuncintegral1}. The advantage of using the Lagrange inversion formula is that it makes clear that the formal change of variables in Proposition \ref{prop:formalchangeofvar} does not depend on the analyticity or injectiveness properties of $\Sact(x)$.

Care must be taken to ensure that $x(y)$ is interpreted as a formal power series in $\R[[y]]$, whereas $\Sact(x)$ is in $\R[[x]]$. We hope that the slight abuse of notation, where we interpret $x$ as a power series or as a variable is transparent for the reader. 

If $\Sact(x)$ is a polynomial, the equation $y = \sqrt{-2 S(x(y))}$ can be interpreted as the definition of an \textit{affine hyperelliptic curve}, 
\begin{align} \label{eqn:hyperelliptic_curve_eqn} \frac{y^2}{2} = - \Sact(x) \end{align}
with at least one singular point or \textit{ordinary double point} at the origin, because $\Sact(x) = -\frac{x^2}{2a}+\cdots$. If $\Sact(x)$ is not a polynomial, but an entire function, eq.\ \eqref{eqn:hyperelliptic_curve_eqn} describes a \textit{generalized affine hyperelliptic curve}.

This interpretation shows a surprising similarity to the theory of \textit{topological recursion} \cite{eynard2007invariants}. The affine complex curve is called the spectral curve in this realm, as it is associated to the eigenvalue distribution of a random matrix model. 
In the theory of topological recursion the \textit{branch-cut} singularities of the expansion of the curve play a vital role. They will also be important for the extraction of asymptotics from formal integrals presented in the next section. 
\begin{figure}
\centering
\begin{subfigure}[t]{0.4\textwidth}
\begin{tikzpicture} \begin{axis}[ xlabel={$x$}, ylabel={$y$}, xmin=-5, xmax=5, ymin=-5, ymax=5, axis on top, width=\figurewidth, height=\figureheight, xmajorgrids, ymajorgrids ] \addplot [black] table {%
-5 -8.16496580927726
-4.89795918367347 -7.94716036730067
-4.79591836734694 -7.73116240380083
-4.69387755102041 -7.51698638636357
-4.59183673469388 -7.30464712324638
-4.48979591836735 -7.09415977661449
-4.38775510204082 -6.88553987649383
-4.28571428571429 -6.67880333549125
-4.18367346938776 -6.47396646433575
-4.08163265306122 -6.27104598829997
-3.97959183673469 -6.07005906456591
-3.87755102040816 -5.87102330060463
-3.77551020408163 -5.67395677364627
-3.6734693877551 -5.47887805132332
-3.57142857142857 -5.28580621357825
-3.46938775510204 -5.09476087593504
-3.36734693877551 -4.90576221424379
-3.26530612244898 -4.71883099101839
-3.16326530612245 -4.53398858349917
-3.06122448979592 -4.3512570135858
-2.95918367346939 -4.17065897980091
-2.85714285714286 -3.99221789146155
-2.75510204081633 -3.81595790525484
-2.6530612244898 -3.64190396443527
-2.55102040816327 -3.47008184088574
-2.44897959183673 -3.30051818031135
-2.3469387755102 -3.1332405508664
-2.24489795918367 -2.9682774955503
-2.14285714285714 -2.80565858874847
-2.04081632653061 -2.64541449734037
-1.93877551020408 -2.48757704684978
-1.83673469387755 -2.33217929317319
-1.73469387755102 -2.17925560049212
-1.63265306122449 -2.02884172605635
-1.53061224489796 -1.88097491261865
-1.42857142857143 -1.7356939894113
-1.3265306122449 -1.59303948268164
-1.22448979591837 -1.4530537369536
-1.12244897959184 -1.31578104835766
-1.02040816326531 -1.18126781157854
-0.918367346938775 -1.04956268221574
-0.816326530612245 -0.9207167566435
-0.714285714285714 -0.794783771805981
-0.612244897959184 -0.671820327801919
-0.510204081632653 -0.551886136618102
-0.408163265306122 -0.435044300983467
-0.306122448979592 -0.321361628061933
-0.204081632653061 -0.210908983617208
-0.102040816326531 -0.103761693411383
0 -0
0 0
0.0689655172413793 0.068168201228676
0.137931034482759 0.134722897075969
0.206896551724138 0.199634747866348
0.275862068965517 0.262872955775982
0.344827586206897 0.32440514870615
0.413793103448276 0.384197251400797
0.482758620689655 0.4422133420096
0.551724137931034 0.498415491985822
0.620689655172414 0.552763586831413
0.689655172413793 0.605215124744396
0.758620689655172 0.655724989665408
0.827586206896552 0.704245194535252
0.896551724137931 0.750724589729521
0.96551724137931 0.795108530585403
1.03448275862069 0.837338496621082
1.10344827586207 0.877351653391487
1.17241379310345 0.915080345820703
1.24137931034483 0.950451509158481
1.31034482758621 0.983385980230529
1.37931034482759 1.01379768711839
1.44827586206897 1.04159268943377
1.51724137931034 1.06666803340216
1.58620689655172 1.08891037526024
1.6551724137931 1.10819431185432
1.72413793103448 1.12438033709903
1.79310344827586 1.13731231452828
1.86206896551724 1.14681431554406
1.93103448275862 1.15268661381773
2 1.15470053837925
}; \addplot [black, dashed] table {%
2 1.15470053837925
2.05263157894737 1.15347936794798
2.10526315789474 1.14972353952158
2.15789473684211 1.14328056083509
2.21052631578947 1.13397607335728
2.26315789473684 1.12160889445706
2.31578947368421 1.10594447628423
2.36842105263158 1.08670610780792
2.42105263157895 1.06356281657479
2.47368421052632 1.0361122994584
2.52631578947368 1.00385610219666
2.57894736842105 0.966162205631642
2.63157894736842 0.922206110592714
2.68421052631579 0.870872892386199
2.73684210526316 0.810582674828232
2.78947368421053 0.738949630783656
2.84210526315789 0.652023664684755
2.89473684210526 0.542233890915719
2.94736842105263 0.390388484187592
3 0
}; \addplot [black, dashed] table {%
-5 8.16496580927726
-4.89795918367347 7.94716036730067
-4.79591836734694 7.73116240380083
-4.69387755102041 7.51698638636357
-4.59183673469388 7.30464712324638
-4.48979591836735 7.09415977661449
-4.38775510204082 6.88553987649383
-4.28571428571429 6.67880333549125
-4.18367346938776 6.47396646433575
-4.08163265306122 6.27104598829997
-3.97959183673469 6.07005906456591
-3.87755102040816 5.87102330060463
-3.77551020408163 5.67395677364627
-3.6734693877551 5.47887805132332
-3.57142857142857 5.28580621357825
-3.46938775510204 5.09476087593504
-3.36734693877551 4.90576221424379
-3.26530612244898 4.71883099101839
-3.16326530612245 4.53398858349917
-3.06122448979592 4.3512570135858
-2.95918367346939 4.17065897980091
-2.85714285714286 3.99221789146155
-2.75510204081633 3.81595790525484
-2.6530612244898 3.64190396443527
-2.55102040816327 3.47008184088574
-2.44897959183673 3.30051818031135
-2.3469387755102 3.1332405508664
-2.24489795918367 2.9682774955503
-2.14285714285714 2.80565858874847
-2.04081632653061 2.64541449734037
-1.93877551020408 2.48757704684978
-1.83673469387755 2.33217929317319
-1.73469387755102 2.17925560049212
-1.63265306122449 2.02884172605635
-1.53061224489796 1.88097491261865
-1.42857142857143 1.7356939894113
-1.3265306122449 1.59303948268164
-1.22448979591837 1.4530537369536
-1.12244897959184 1.31578104835766
-1.02040816326531 1.18126781157854
-0.918367346938775 1.04956268221574
-0.816326530612245 0.9207167566435
-0.714285714285714 0.794783771805981
-0.612244897959184 0.671820327801919
-0.510204081632653 0.551886136618102
-0.408163265306122 0.435044300983467
-0.306122448979592 0.321361628061933
-0.204081632653061 0.210908983617208
-0.102040816326531 0.103761693411383
0 0
0 -0
0.0689655172413793 -0.068168201228676
0.137931034482759 -0.134722897075969
0.206896551724138 -0.199634747866348
0.275862068965517 -0.262872955775982
0.344827586206897 -0.32440514870615
0.413793103448276 -0.384197251400797
0.482758620689655 -0.4422133420096
0.551724137931034 -0.498415491985822
0.620689655172414 -0.552763586831413
0.689655172413793 -0.605215124744396
0.758620689655172 -0.655724989665408
0.827586206896552 -0.704245194535252
0.896551724137931 -0.750724589729521
0.96551724137931 -0.795108530585403
1.03448275862069 -0.837338496621082
1.10344827586207 -0.877351653391487
1.17241379310345 -0.915080345820703
1.24137931034483 -0.950451509158481
1.31034482758621 -0.983385980230529
1.37931034482759 -1.01379768711839
1.44827586206897 -1.04159268943377
1.51724137931034 -1.06666803340216
1.58620689655172 -1.08891037526024
1.6551724137931 -1.10819431185432
1.72413793103448 -1.12438033709903
1.79310344827586 -1.13731231452828
1.86206896551724 -1.14681431554406
1.93103448275862 -1.15268661381773
2 -1.15470053837925
2 -1.15470053837925
2.05263157894737 -1.15347936794798
2.10526315789474 -1.14972353952158
2.15789473684211 -1.14328056083509
2.21052631578947 -1.13397607335728
2.26315789473684 -1.12160889445706
2.31578947368421 -1.10594447628423
2.36842105263158 -1.08670610780792
2.42105263157895 -1.06356281657479
2.47368421052632 -1.0361122994584
2.52631578947368 -1.00385610219666
2.57894736842105 -0.966162205631642
2.63157894736842 -0.922206110592714
2.68421052631579 -0.870872892386199
2.73684210526316 -0.810582674828232
2.78947368421053 -0.738949630783656
2.84210526315789 -0.652023664684755
2.89473684210526 -0.542233890915719
2.94736842105263 -0.390388484187592
3 -0
}; \addplot [black, dotted] table {%
-5 1.15470053837925
5 1.15470053837925
}; \end{axis} \end{tikzpicture}
\subcaption{Plot of the elliptic curve $\frac{y^2}{2} = \frac{x^2}{2} - \frac{x^3}{3!}$, which can be associated to the perturbative expansion of zero-dimensional $\varphi^3$-theory. 
The dominant singularity can be found at $(x,y)=\left(2,\frac{2}{\sqrt{3}}\right)$.
}
\label{fig:curve_phi3}
\end{subfigure}
\quad
\begin{subfigure}[t]{0.4\textwidth}
\begin{tikzpicture} \begin{axis}[ xlabel={$x$}, ylabel={$y$}, xmin=-5, xmax=5, ymin=-5, ymax=5, axis on top, width=\figurewidth, height=\figureheight, xmajorgrids, ymajorgrids ] \addplot [black] table {%
-1.5707963267949 -1
-1.46246554563663 -0.99413795715436
-1.35413476447836 -0.976620555710087
-1.24580398332009 -0.947653171182802
-1.13747320216182 -0.907575419670957
-1.02914242100355 -0.856857176167589
-0.920811639845284 -0.796093065705644
-0.812480858687015 -0.725995491923131
-0.704150077528747 -0.647386284781828
-0.595819296370478 -0.561187065362382
-0.487488515212209 -0.46840844069979
-0.37915773405394 -0.370138155339914
-0.270826952895672 -0.267528338529221
-0.162496171737403 -0.161781996552765
-0.0541653905791344 -0.0541389085854175
0.0541653905791344 0.0541389085854175
0.162496171737403 0.161781996552765
0.270826952895672 0.267528338529221
0.379157734053941 0.370138155339915
0.487488515212209 0.46840844069979
0.595819296370478 0.561187065362382
0.704150077528747 0.647386284781828
0.812480858687016 0.725995491923131
0.920811639845284 0.796093065705644
1.02914242100355 0.856857176167589
1.13747320216182 0.907575419670957
1.24580398332009 0.947653171182803
1.35413476447836 0.976620555710087
1.46246554563663 0.99413795715436
1.5707963267949 1
}; \addplot [black, dashed] table {%
-1.5707963267949 1
-1.46246554563663 0.99413795715436
-1.35413476447836 0.976620555710087
-1.24580398332009 0.947653171182802
-1.13747320216182 0.907575419670957
-1.02914242100355 0.856857176167589
-0.920811639845284 0.796093065705644
-0.812480858687015 0.725995491923131
-0.704150077528747 0.647386284781828
-0.595819296370478 0.561187065362382
-0.487488515212209 0.46840844069979
-0.37915773405394 0.370138155339914
-0.270826952895672 0.267528338529221
-0.162496171737403 0.161781996552765
-0.0541653905791344 0.0541389085854175
0.0541653905791344 -0.0541389085854175
0.162496171737403 -0.161781996552765
0.270826952895672 -0.267528338529221
0.379157734053941 -0.370138155339915
0.487488515212209 -0.46840844069979
0.595819296370478 -0.561187065362382
0.704150077528747 -0.647386284781828
0.812480858687016 -0.725995491923131
0.920811639845284 -0.796093065705644
1.02914242100355 -0.856857176167589
1.13747320216182 -0.907575419670957
1.24580398332009 -0.947653171182803
1.35413476447836 -0.976620555710087
1.46246554563663 -0.99413795715436
1.5707963267949 -1
}; \addplot [black, dashed] table {%
-5 0.958924274663138
-4.93001625156724 0.9764125014873
-4.86003250313449 0.989120479855391
-4.79004875470173 0.996985994982286
-4.72006500626897 0.999970539457967
-4.65008125783621 0.998059501769267
-4.58009750940346 0.991262237833841
-4.5101137609707 0.979612025196152
-4.44013001253794 0.963165900109718
-4.37014626410518 0.942004378303225
-4.30016251567243 0.916231060797541
-4.23017876723967 0.885972126703468
-4.16019501880691 0.851375715483339
-4.09021127037416 0.812611201700743
-4.0202275219414 0.769868365808998
-3.95024377350864 0.723356465037984
-3.88026002507588 0.673303208927992
-3.81027627664313 0.619953644526123
-3.74029252821037 0.563568956702987
-3.67030877977761 0.504425189463042
-3.60032503134486 0.442811894508673
-3.5303412829121 0.379030713674278
-3.46035753447934 0.313393902170381
-3.39037378604658 0.246222799867555
-3.32039003761383 0.177846258104333
-3.25040628918107 0.108599029721034
-3.18042254074831 0.0388201302014438
-3.11043879231556 -0.0311488220542363
-3.0404550438828 -0.100965278199608
-2.97047129545004 -0.170287435967506
-2.90048754701728 -0.238775913040378
-2.83050379858453 -0.306095408573256
-2.76052005015177 -0.371916344734953
-2.69053630171901 -0.435916480230948
-2.62055255328625 -0.497782487908603
-2.5505688048535 -0.557211488721169
-2.48058505642074 -0.61391253454073
-2.41060130798798 -0.667608032560636
-2.34061755955523 -0.718035104313931
-2.27063381112247 -0.764946872654404
-2.20065006268971 -0.808113670399547
-2.13066631425695 -0.847324164718247
-2.0606825658242 -0.8823863917585
-1.99069881739144 -0.913128696449913
-1.92071506895868 -0.939400572879963
-1.85073132052593 -0.961073401129759
-1.78074757209317 -0.978041076961984
-1.71076382366041 -0.9902205312782
-1.64078007522765 -0.997552136802433
-1.5707963267949 -1
}; \addplot [black, dashed] table {%
-5 -0.958924274663138
-4.93001625156724 -0.9764125014873
-4.86003250313449 -0.989120479855391
-4.79004875470173 -0.996985994982286
-4.72006500626897 -0.999970539457967
-4.65008125783621 -0.998059501769267
-4.58009750940346 -0.991262237833841
-4.5101137609707 -0.979612025196152
-4.44013001253794 -0.963165900109718
-4.37014626410518 -0.942004378303225
-4.30016251567243 -0.916231060797541
-4.23017876723967 -0.885972126703468
-4.16019501880691 -0.851375715483339
-4.09021127037416 -0.812611201700743
-4.0202275219414 -0.769868365808998
-3.95024377350864 -0.723356465037984
-3.88026002507588 -0.673303208927992
-3.81027627664313 -0.619953644526123
-3.74029252821037 -0.563568956702987
-3.67030877977761 -0.504425189463042
-3.60032503134486 -0.442811894508673
-3.5303412829121 -0.379030713674278
-3.46035753447934 -0.313393902170381
-3.39037378604658 -0.246222799867555
-3.32039003761383 -0.177846258104333
-3.25040628918107 -0.108599029721034
-3.18042254074831 -0.0388201302014438
-3.11043879231556 0.0311488220542363
-3.0404550438828 0.100965278199608
-2.97047129545004 0.170287435967506
-2.90048754701728 0.238775913040378
-2.83050379858453 0.306095408573256
-2.76052005015177 0.371916344734953
-2.69053630171901 0.435916480230948
-2.62055255328625 0.497782487908603
-2.5505688048535 0.557211488721169
-2.48058505642074 0.61391253454073
-2.41060130798798 0.667608032560636
-2.34061755955523 0.718035104313931
-2.27063381112247 0.764946872654404
-2.20065006268971 0.808113670399547
-2.13066631425695 0.847324164718247
-2.0606825658242 0.8823863917585
-1.99069881739144 0.913128696449913
-1.92071506895868 0.939400572879963
-1.85073132052593 0.961073401129759
-1.78074757209317 0.978041076961984
-1.71076382366041 0.9902205312782
-1.64078007522765 0.997552136802433
-1.5707963267949 1
}; \addplot [black, dashed] table {%
1.5707963267949 1
1.64078007522765 0.997552136802433
1.71076382366041 0.9902205312782
1.78074757209317 0.978041076961984
1.85073132052593 0.961073401129759
1.92071506895868 0.939400572879963
1.99069881739144 0.913128696449913
2.0606825658242 0.8823863917585
2.13066631425695 0.847324164718246
2.20065006268971 0.808113670399547
2.27063381112247 0.764946872654404
2.34061755955523 0.718035104313931
2.41060130798798 0.667608032560636
2.48058505642074 0.613912534540729
2.5505688048535 0.557211488721168
2.62055255328626 0.497782487908603
2.69053630171901 0.435916480230948
2.76052005015177 0.371916344734952
2.83050379858453 0.306095408573256
2.90048754701728 0.238775913040378
2.97047129545004 0.170287435967506
3.0404550438828 0.100965278199608
3.11043879231556 0.0311488220542358
3.18042254074831 -0.0388201302014443
3.25040628918107 -0.108599029721034
3.32039003761383 -0.177846258104333
3.39037378604658 -0.246222799867555
3.46035753447934 -0.313393902170381
3.5303412829121 -0.379030713674278
3.60032503134486 -0.442811894508673
3.67030877977761 -0.504425189463043
3.74029252821037 -0.563568956702987
3.81027627664313 -0.619953644526123
3.88026002507588 -0.673303208927992
3.95024377350864 -0.723356465037984
4.0202275219414 -0.769868365808998
4.09021127037416 -0.812611201700743
4.16019501880691 -0.851375715483339
4.23017876723967 -0.885972126703468
4.30016251567243 -0.916231060797541
4.37014626410519 -0.942004378303225
4.44013001253794 -0.963165900109718
4.5101137609707 -0.979612025196152
4.58009750940346 -0.991262237833841
4.65008125783621 -0.998059501769267
4.72006500626897 -0.999970539457967
4.79004875470173 -0.996985994982286
4.86003250313449 -0.989120479855391
4.93001625156724 -0.9764125014873
5 -0.958924274663138
}; \addplot [black, dashed] table {%
1.5707963267949 -1
1.64078007522765 -0.997552136802433
1.71076382366041 -0.9902205312782
1.78074757209317 -0.978041076961984
1.85073132052593 -0.961073401129759
1.92071506895868 -0.939400572879963
1.99069881739144 -0.913128696449913
2.0606825658242 -0.8823863917585
2.13066631425695 -0.847324164718246
2.20065006268971 -0.808113670399547
2.27063381112247 -0.764946872654404
2.34061755955523 -0.718035104313931
2.41060130798798 -0.667608032560636
2.48058505642074 -0.613912534540729
2.5505688048535 -0.557211488721168
2.62055255328626 -0.497782487908603
2.69053630171901 -0.435916480230948
2.76052005015177 -0.371916344734952
2.83050379858453 -0.306095408573256
2.90048754701728 -0.238775913040378
2.97047129545004 -0.170287435967506
3.0404550438828 -0.100965278199608
3.11043879231556 -0.0311488220542358
3.18042254074831 0.0388201302014443
3.25040628918107 0.108599029721034
3.32039003761383 0.177846258104333
3.39037378604658 0.246222799867555
3.46035753447934 0.313393902170381
3.5303412829121 0.379030713674278
3.60032503134486 0.442811894508673
3.67030877977761 0.504425189463043
3.74029252821037 0.563568956702987
3.81027627664313 0.619953644526123
3.88026002507588 0.673303208927992
3.95024377350864 0.723356465037984
4.0202275219414 0.769868365808998
4.09021127037416 0.812611201700743
4.16019501880691 0.851375715483339
4.23017876723967 0.885972126703468
4.30016251567243 0.916231060797541
4.37014626410519 0.942004378303225
4.44013001253794 0.963165900109718
4.5101137609707 0.979612025196152
4.58009750940346 0.991262237833841
4.65008125783621 0.998059501769267
4.72006500626897 0.999970539457967
4.79004875470173 0.996985994982286
4.86003250313449 0.989120479855391
4.93001625156724 0.9764125014873
5 0.958924274663138
}; \addplot [black, dotted] table {%
-5 1
5 1
}; \addplot [black, dotted] table {%
-5 -1
5 -1
}; \end{axis} \end{tikzpicture}
\subcaption{Plot of the generalized hyperelliptic curve $\frac{y^2}{2} = \frac{\sin^2(x)}{2}$ with dom\-in\-ant sing\-ularities at 
$(x,y)=\left(\pm \frac{\pi}{2}, \pm 1\right)$.}
\label{fig:curve_sine}
\end{subfigure}
\caption{Examples of curves associated to formal integrals}
\end{figure}
\begin{expl}[$\varphi^3$-theory as the expansion of a complex curve]
\label{expl:phi3elliptic}
For $\varphi^3$-theory the complex curve takes the form
\begin{align*} \frac{y^2}{2} = \frac{x^2}{2} - \frac{x^3}{3!}. \end{align*}
This is the elliptic curve depicted in Figure \ref{fig:curve_phi3}. 
It is clearly visible that solving for $x$ will result in a multivalued function. 
With $x(y)$, we mean the power series expansion at the origin associated to the locally increasing branch. This branch is depicted as solid line.
Moreover, we see that this expansion will have a finite radius of convergence, which is dictated by the location of the branch-cut singularity which is attained at $y=\frac{2}{\sqrt{3}}$.
\end{expl}
\begin{expl}[Sine-Gordon model as expansion of a complex curve]
\label{expl:sinegordoncurve}
Consider again the action $\Sact(x)=-\frac{\sin^2(x)}{2}$ discussed in Example \ref{expl:sinegordon}. The complex curve takes the form,
\begin{align*} \frac{y^2}{2} = \frac{\sin^2(x)}{2}. \end{align*}
This curve is depicted in Figure \ref{fig:curve_sine}.
We may solve for $x(y) = \arcsin(y)$, which is the local solution around $y=0$, which is positive for $y\rightarrow 0^+$. This local solution is drawn as black line in Figure \ref{fig:curve_sine}. Obviously, $x(y)$ has singularities at $y= \pm 1$. 
From Proposition \ref{prop:formalchangeofvar} it follows that,
\begin{align*} \mathcal{F}\left[-\frac{\sin^2(x)}{2}\right](\hbar) &= \sum_{n=0}^\infty \hbar^{n} (2n+1)!! [y^{2n+1}] \arcsin(y) \\ &= \sum_{n=0}^\infty \hbar^{n} (2n-1)!! [y^{2n}] \frac{1}{\sqrt{1-y^2}}. \end{align*}
The last equality follows because $\arcsin'(y)=\frac{1}{\sqrt{1-y^2}}$. We will use this result later in Section \ref{sec:QED} to express the partition function of zero-dimensional QED using $\Fop\left[-\frac{\sin^2(x)}{2}\right]$.
\end{expl}

The representation of the coefficients of $\mathcal{F}[\Sact(x)](\hbar)$ as expansion of a generalized hyperelliptic curve can be used to calculate them efficiently. The expansion of $x(y)$ must fulfill the differential equation
\begin{align*} \frac{\partial x}{\partial y} = -\frac{y}{S'(x(y))}. \end{align*}
Using the initial condition $x(0)=0$ and $\frac{\partial x}{\partial y} > 0$, while expanding this as a power series results in the correct branch. 
\begin{expl}
For the coefficients of $Z^\text{all}(\hbar)$, where $\Sact(x) = -x^2-x-1+e^x$, we obtain the differential equation for $x(y)$:
\begin{align*} \frac{\partial x}{\partial y} = \frac{y}{1+2x-e^x} \end{align*}
The coefficients of $x(y)$ can be calculated by basic iterative methods. These coefficients can be translated into coefficients of $\Fop[-x^2-x-1+e^x](\hbar)$ using Proposition \ref{prop:formalchangeofvar}.
\end{expl}

\section{Asymptotics from singularity analysis}
One approach to calculate asymptotics of expressions such as the integral \eqref{eqn:formalfuncintegral1} is to perform the coefficient extraction with a Cauchy integral and to approximate the result using the method of steepest decent or saddle point method:
\begin{gather*} [\hbar^n] \int_\R \frac{dx}{\sqrt{2 \pi \hbar } } e^{\frac{1}{\hbar} \left( -\frac{x^2}{2a} + V(x) \right) } = \\ \oint_{|\hbar|=\epsilon} \frac{d\hbar}{\hbar^{n-1}} \int_\R \frac{dx}{\sqrt{2 \pi \hbar } } e^{\frac{1}{\hbar} \left( -\frac{x^2}{2a} + V(x) \right) } = \\ \oint_{|\hbar|=\epsilon} d\hbar \int_\R \frac{dx}{\sqrt{2 \pi } } e^{\frac{1}{\hbar} \left( -\frac{x^2}{2a} + V(x) \right) - (n-\frac32) \log \hbar } \end{gather*}
See for instance \cite{cvitanovic1978number}, where this technique was applied to $\varphi^3$-theory. This method was also applied to higher dimensional path integrals to obtain the asymptotics for realistic QFTs \cite{lipatov1977divergence}. The saddle points are solutions to the classical equations of motion and are called instantons in the realm of QFT.

The approach requires us to manipulate the integrand and to pick the right contour for the integration. A disadvantage is that this procedure will result in a complicated asymptotic expansion. 

There exists a powerful method called hyperasymptotics \cite{berry1991hyperasymptotics} to obtain large order asymptotics of integrals such as \eqref{eqn:formalfuncintegral1}. This procedure is very general, as it also provides exponentially suppressed contributions as a systematic expansion. The expansion of a specific exponential order results in an expressions involving Dingle's terminants \cite{dingle1973asymptotic}. Unfortunately, these expressions can be quite complicated \cite{berry1991hyperasymptotics}.

We will take a slightly different approach which is inspired from Ba{\c{s}}ar, Dunne and {\"U}nsal \cite{basar2013resurgence}, as we aim to obtain a complete asymptotic expansion in $n$: We will compute the large $n$ asymptotics of the coefficients $a_n$ of $\Fop[\Sact(x)](\hbar) = \sum_{n=0}a_n \hbar^n$ using \textit{singularity analysis} of the function $x(y)$. Singularity analysis was proven to be a powerful tool for asymptotics extraction even for implicitly defined power series such as $x(y)$ \cite{flajolet2009analytic}. As $x(y)$ can be interpreted as a variant of the Borel transformation of $\Fop[\Sact(x)](\hbar)$, this approach is in the spirit of resurgence \cite{ecalle1981fonctions}, where singularities of the Borel transformation are associated to the factorial divergence of expansions.

We will briefly repeat the necessary steps to compute the asymptotics of the implicitly defined power series $x(y)$. For a detailed account on singularity analysis, we refer to \cite[Ch. VI]{flajolet2009analytic}.

By Darboux's method, the asymptotics of the power series $x(y)$ are determined by the behavior of the function $x(y)$ near its \textit{dominant singularities}.
The dominant singularities of a function are the singularities which lie on the boundary of the disc of convergence of its expansion near the origin. 

Finding the actual location of the dominant singularity can be quite complicated. 
In our case we generally would need to calculate the \textit{monodromy} of the complex curve $\frac{y^2}{2}=-\Sact(x)$. 
However, in many examples the location of the dominant singularities is more or less obvious.

We will assume that the locations of the dominant singularities of $x(y)$ are known and that these singularities are of simple \textit{square root} type. Let $(\tau_i, \rho_i)$ be the coordinates of such a singularity. That means that $x(y)$ is non-analytic for $y \rightarrow \rho_i$ and that $\lim_{y\rightarrow \rho_i} x(y)= \tau_i$. The requirement that the singularity is of \textit{square root} type is equivalent to the requirement that the curve $\frac{y^2}{2}=-\Sact(x)$ is regular at $(x,y) = (\tau_i, \rho_i)$. 
\begin{expl}
\label{expl:phi3elliptic_singularity}
Consider the graph of the elliptic curve depicted in Figure \ref{fig:curve_phi3} from $\varphi^3$-theory. It is clear that $x(y)$ has a singularity at a fixed value of $y = \frac{2}{\sqrt{3}}$ indicated by the dotted line. This is in fact the unique dominant singularity in this example. 
The exponential growth of the coefficients of $x(y) = \sum_{n=0}^\infty c_n y^n$ is governed by the radius of convergence, $c_n \sim r^{-n} = \left(\frac{2}{\sqrt{3}} \right)^{-n}$. 
More precise asymptotics of the coefficients are determined by the singular expansion around the dominant singularity. 
In Figure \ref{fig:curve_phi3}, the point $(x_0,y_0) = (2, \frac{2}{\sqrt{3}})$ is the dominant singularity of $x(y)$ as well as a \textit{critical point} or \textit{saddle point} of the function $y(x)$ as expected by the implicit function theorem. This saddle point coincides with a saddle point of $\Sact(x)$. Although $x(y)$ has a singularity at this point, the curve stays regular.
\end{expl}
Having found the dominant singularity, it is surprisingly easy to obtain a complete asymptotic expansion for the large order behavior of the coefficients of $\Fop[\Sact(x)](\hbar)$. 
\begin{thm}
\label{thm:comb_int_asymp}
If $\Sact(x) \in x^2\R[[x]]$, such that the local solution $x(y)$ around the origin of $\frac{y^2}{2} = -\Sact(x)$ has only square-root type dominant singularities at the regular points $(\tau_i, \rho_i)$ with $i \in I$, then the Poincaré asymptotic expansion of the coefficients of $\Fop[\Sact(x)](\hbar)$ is given by
\begin{align} \label{eqn:asymptotic_expantion_combint} [\hbar^n] \Fop[\Sact(x)](\hbar) &= \sum_{k=0}^{R-1} \sum_{i\in I} w_{i,k} A_i^{-(n-k)} \Gamma(n-k) + \bigO\left(\sum_{i\in I} A_i^{-n} \Gamma(n-R)\right), \end{align}
for all $R \geq 0$, where $A_i = - S(\tau_i)$, the $\bigO$-notation refers to the $n\rightarrow \infty$ limit and
\begin{align} \label{eqn:asympgeneral} w_{i,n} &= \frac{1}{2\pi \mathrm{i}}[\hbar^n] \Fop[\Sact(x+\tau_i) - \Sact(\tau_i)](\hbar). \end{align}
\end{thm}

The exact shape of the asymptotic expansion can be seen as a `resurged' version of the original expansion. This was initially observed in \cite{basar2013resurgence}, where it was proven using techniques from Berry, Howls and Dingle \cite{berry1991hyperasymptotics,dingle1973asymptotic}.
Here, we will give an alternative explicit proof. It combines the Lagrange inversion formula with a Lemma by Paris concerning hypergeometric functions. 

\begin{proof}
Starting with Proposition \ref{prop:formalchangeofvar}
\begin{align} \label{eqn:appendixrefF} \mathcal{F}[\Sact(x)](\hbar)&= \sum_{n=0}^\infty \hbar^{n} (2n+1)!! [y^{2n+1}] x(y), \end{align}
we wish to compute the singular expansion of $x(y)$ at the removable singularity $(x,y)= (\tau_i, \rho_i)$ defined as the solution of $\frac{y^2}{2} = -\Sact(x)$ with positive linear coefficient, to obtain the asymptotics of $\mathcal{F}[\Sact(x)](\hbar)$.

Solving for $y$ and shifting the defining equation of the hyperelliptic curve to the point of the singularity gives,
\begin{align*} y &= \sqrt{-2\Sact(x) } \\ \rho_i - y &= \sqrt{-2\Sact(\tau_i) } - \sqrt{-2\Sact(x)} \\ 1 - \frac{y}{\rho_i} &= 1- \sqrt{\frac{\Sact(x)}{\Sact(\tau_i)}} \intertext{where $\rho_i = \sqrt{-2\Sact(\tau_i)}$. The right hand side of this equation is expected to be of the form $\approx \frac{\Sact''(\tau_i)}{2} (\tau_i-x)^2$ for $x\rightarrow \tau_i$, as we assume the singularity to be of square root type. Locally expanding around the singularity by setting $u_i=\sqrt{1-\frac{y}{\rho_i}}$ and $v_i=\tau_i-x$ gives} u_i &= \sqrt{1- \sqrt{\frac{\Sact(\tau_i-v_i)}{\Sact(\tau_i)}}}, \end{align*}
where the branch of the square root which agrees with the locally positive expansion around the origin must be taken.
We would like to solve this equation for $v_i$ to obtain the Puiseux expansion at the singular point:
\begin{align*} v_i(u_i) = \sum_{k=1}^\infty d_{i,k} u_i^k \end{align*}
The coefficients $d_{i,k}$ can be expressed using the Lagrange inversion formula,
\begin{align} \label{eqn:lagrange_singularity} d_{i,k} = [u_i^k] v_i(u_i) = \frac{1}{k} [v_i^{k-1}] \left( \frac{v_i}{u_i(v_i)} \right)^k. \end{align}
The asymptotics of $[y^n] x(y)$ are given by the singular expansions around dominant singularities,
\begin{align*} [y^n] x(y) &\underset{n\rightarrow \infty}{\sim} [y^n] \sum_{i \in I} \sum_{k=0} d_{i,k} \left(1 - \frac{y}{\rho_i}\right)^\frac{k}{2}. \intertext{Expanding using the generalized binomial theorem and noting that only odd summands in $k$ contribute asymptotically gives,} [y^n] x(y) &\underset{n\rightarrow \infty}{\sim} \sum_{i\in I} (-1)^n \rho_i^{-n} \sum_{k=0} { k+\frac12 \choose n} d_{i,2k+1}. \end{align*}
After substitution into eq.\ \eqref{eqn:appendixrefF} this results in,
\begin{align*} [\hbar^n] \mathcal{F}[S](\hbar) &= - (2n+1)!! \sum_{i \in I} \rho_i^{-2n-1} \sum_{k=0}^{R-1} { k+\frac12 \choose 2n+1 } d_{i,2k+1} \\ &+ \bigO\left(\sum_{i \in I}\left(\frac{2}{\rho_i^2}\right)^{n+\frac12}\Gamma(n-R)\right), \end{align*}
where the asymptotic behavior of the binomial ${ k+\frac12 \choose 2n+1 } \underset{n\rightarrow \infty}{\sim} \frac{C_k}{(2n+1)^{k+\frac32}}$ and the double factorial $(2n+1)!! = 2^{n+\frac32} \frac{\Gamma(n+\frac32)}{\sqrt{2 \pi}}$ were used to derive the form of the remainder term. Substituting eq.\ \eqref{eqn:lagrange_singularity} results in 
\begin{align} \begin{split} \label{eqn:Fopsingular} [\hbar^n] \mathcal{F}[S](\hbar) &= \sum_{k=0}^{R-1} \frac{(2n-1)!!}{2} { k-\frac12 \choose 2n } \sum_{i \in I} \rho_i^{-2n-1} [v_i^{2k}] \phi_i(v_i)^{2k+1} \\ &+ \bigO\left(\sum_{i \in I}(-\Sact(\tau_i))^{n}\Gamma(n-R)\right) \end{split} \end{align}
where $\phi_i(v_i):= \frac{-v_i}{\sqrt{1-\sqrt{\frac{\Sact(\tau_i-v_i)}{\Sact(\tau_i)}}}}$.

It is easily checked by the reflection and duplication formulas for the $\Gamma$-function that
\begin{align} \label{eqn:gamma_binom_ident} \frac{(2n-1)!!}{2} { k - \frac12 \choose 2n } &= \frac{(-1)^k 2^{n-k} \Gamma(k+\frac12)}{(2\pi)^{\frac32}} \frac{\Gamma(n-\frac{k}{2} +\frac14) \Gamma(n-\frac{k}{2}+\frac34) }{ \Gamma(n+1) }. \end{align}
The following identity by Paris \cite[Lemma 1]{paris1992smoothing},
\begin{align*} \frac{\Gamma(n+a)\Gamma(n+b)}{n!} &= \sum_{m=0}^{R-1} (-1)^m \frac{\risefac{(1-a)}{m} \risefac{(1-b)}{m}}{m!} \Gamma(n+a+b-1-m) && \\ &+ \bigO(\Gamma(n+a+b-1-R)) &&\forall R \geq 0, \end{align*}
can be used to expand the product of $\Gamma$ functions. The expression $\risefac{a}{n}=\frac{\Gamma(n+a)}{\Gamma(a)}$ denotes the rising factorial. 
Applying this to eq. \eqref{eqn:gamma_binom_ident} gives,
\begin{gather*} \frac{(2n-1)!!}{2} { k - \frac12 \choose 2n } = \\ \frac{(-1)^k 2^{n-k}}{(2\pi)^\frac32} \sum_{m=0}^{R-k-1} (-1)^m 2^{-2m} \frac{\Gamma(k+\frac12+2m)} { m!} \Gamma(n-k-m) + \bigO(\Gamma(n-R)) \\ = \frac{2^{n}}{2\pi} \sum_{m=0}^{R-k-1} (-1)^{m+k} 2^{-3m-2k-\frac12} (2(m+k)-1)!! \\ \times { 2m + k - \frac12 \choose m } \Gamma(n-k-m) + \bigO(\Gamma(n-R)). \end{gather*}
This can be substituted into eq.\ \eqref{eqn:Fopsingular}:
\begin{gather*} [\hbar^n] \mathcal{F}[S](\hbar) = \sum_{k=0}^{R-1} \frac{2^{n}}{2\pi} \sum_{m=0}^{R-k-1} (-1)^{m+k} 2^{-3m-2k-\frac12} (2(m+k)-1)!! \\ \times { 2m + k - \frac12 \choose m }\Gamma(n-k-m) \sum_{i \in I} \rho_i^{-2n-1} [v_i^{2k}] \phi_i(v_i)^{2k+1} \\ + \bigO\left( \sum_{i \in I}(-\Sact(\tau_i))^{n} \Gamma(n-R)\right) \\ = \sum_{m=0}^{R-1} \frac{2^{n}}{2\pi} (-1)^{m} 2^{-2m-\frac12} (2m-1)!! \Gamma(n-m) \\ \times \sum_{i \in I} \rho_i^{-2n-1} \sum_{k=0}^{m} 2^{-k} { m + k - \frac12 \choose k } [v_i^{2(m-k)}] \phi_i(v_i)^{2(m-k)+1} \\ + \bigO\left(\sum_{i \in I}(-\Sact(\tau_i))^{n}\Gamma(n-R)\right). \end{gather*}
The inner sum evaluates to,
\begin{align*} &\sum_{k=0}^{m} 2^{-k} { m + k - \frac12 \choose k } [v_i^{2(m-k)}] \phi_i(v_i)^{2(m-k)+1} \\ &= [v_i^{2m}] \phi_i(v_i)^{2m+1} \sum_{k=0}^{m} { m + k - \frac12 \choose k } 2^{-k}\left(\frac{v_i}{\phi_i(v_i)} \right)^{2k} \\ &= [v_i^{2m}] \frac{\phi_i(v_i)^{2m+1}}{\left(1-\frac12 \left(\frac{v_i}{\phi_i(v_i)} \right)^2\right)^{m+\frac12}} = [v_i^{2m}] \left( \frac{\phi_i(v_i)}{\sqrt{1-\frac12 \left(\frac{v_i}{\phi_i(v_i)} \right)^2}}\right)^{2m+1} \\ & = 2^{m+\frac12} [v_i^{2m}] \left( \frac{-v_i}{\sqrt{ 1 - \frac{\Sact(\tau_i-v_i)}{\Sact(\tau_i)} } } \right)^{2m+1} \\ & = (-2\Sact(\tau_i))^{m+\frac12} [v_i^{2m}] \left( \frac{-v_i}{\sqrt{ \Sact(\tau_i-v_i) -\Sact(\tau_i) } } \right)^{2m+1}. \end{align*}
Therefore,
\begin{align*} [\hbar^n] \mathcal{F}[S](\hbar) &= \frac{1}{2\pi} \sum_{m=0}^{R-1} (-1)^{m} (2m-1)!! \Gamma(n-m) \\ &\times \sum_{i \in I} (-\Sact(\tau_i))^{n-m} [v_i^{2m}] \left( \frac{v_i}{\sqrt{ 2\Sact(\tau_i+v_i) -2\Sact(\tau_i) } } \right)^{2m+1} \\ &+ \bigO\left(\sum_{i \in I} (-\Sact(\tau_i))^{n}\Gamma(n-R)\right) \\ &= \frac{1}{2\pi i} \sum_{m=0}^{R-1} (2m-1)!! \Gamma(n-m) \\ &\times \sum_{i \in I} (-\Sact(\tau_i))^{n-m} [v_i^{2m}] \left( \frac{v_i}{\sqrt{ - 2 \left( \Sact(\tau_i+v_i) -\Sact(\tau_i) \right) } } \right)^{2m+1} \\ &+ \bigO\left(\sum_{i \in I} (-\Sact(\tau_i))^{n}\Gamma(n-R)\right), \end{align*}
which proves the theorem after using the Lagrange inversion formula and Proposition \ref{prop:formalchangeofvar}.
\end{proof}

As was illustrated in Example \ref{expl:phi3elliptic_singularity}, a square-root type singularity of $x(y)$ coincides with a saddle point of $\Sact(x)$. This way Theorem \ref{thm:comb_int_asymp} works in a very similar way to the saddle point method.

To actually find the location of the dominant singularity in non-trivial cases powerful techniques of singularity analysis of implicitly defined functions can be applied. For instance, if $\Sact(x)$ is a polynomial, a systematic treatment given in \cite[Chap. VII]{flajolet2009analytic} can be used. With minor modifications this can also be applied to an entire function $\Sact(x)$ for the non-degenerate case \cite{banderier2015formulae}.

Note that the quadratic coefficient of $\Sact(x+\tau_i) - \Sact(\tau_i)$ in the argument for $\Fop$ in eq. \eqref{eqn:asympgeneral} is not necessarily negative. The regularity of the complex curve only guarantees that it is non-zero. We need to generalize Definition \ref{def:formalintegral} to also allow positive quadratic coefficients. 
With this generalization the choice of the branch for the square-root in eq.\ \eqref{eqn:formalintegralpwrsrs} becomes ambiguous and we have to determine the correct branch by analytic continuation. Here, we will only need a special case of Theorem \ref{thm:comb_int_asymp}, which remedies this ambiguity:
\begin{crll}
\label{crll:comb_int_asymp}
If $\Sact(x) = -\frac{x^2}{2} + \cdots$ is the power series expansion of an entire real function, which has simple critical points only on the real line, then there are not more than two dominant singularities associated with local minima of $\Sact(x)$ at $x= \tau_i$. The minima must have the same ordinate $\Sact(\tau_i) = -A$ to qualify both as dominant singularities. 
The coefficients of the asymptotic expansion are given by
\begin{align} w_{i,n} &= \frac{1}{2\pi}[\hbar^n] \Fop[\Sact(\tau_i)- \Sact(x+\tau_i)](-\hbar), \end{align}
where the argument of $\Fop$ has a strictly negative quadratic coefficient. 
\end{crll}
\begin{proof}
If $\Sact(x)$ is a real entire function, whose derivative vanishes only on isolated points of the real line, we can analytically continue $x(y)$ to a star-shaped domain excluding at most two rays on the real line. On the real line $x(y)$ can have singularities. By definition $x=0$ is a local maximum of $\Sact(x)$. It follows from Rolle's theorem that the next critical point encountered on the real line must be a local minimum. These minima are the only candidates for dominant singularities of $x(y)$. Using Theorem \ref{thm:comb_int_asymp}, we obtain $\frac{1}{2\pi i}[\hbar^n] \Fop[\Sact(x+\tau_i) - \Sact(\tau_i)](\hbar)$ as expansions around the minima. The power series $\Sact(x+\tau_i) - \Sact(\tau_i)$ starts with a positive quadratic term, resulting in a prefactor of $\sqrt{-1}$. Taking the square root in the upper half plane results in the correct expansion. Flipping the sign in the argument and in the expansion parameter absorbs the imaginary unit in eq.\ \eqref{eqn:asympgeneral}.
\end{proof}

\begin{expl}
\label{expl:phi3theoryasymptotics}
Let $\Sact(x) = -\frac{x^2}{2} + \frac{x^3}{3!}$ as in Examples \ref{expl:phi3theoryexpansion}, \ref{expl:phi3elliptic} and \ref{expl:phi3elliptic_singularity}. The location of the dominant singularity at $x= \tau = 2$ can be obtained by solving $\Sact'(\tau)=0$ (see Figure \ref{fig:curve_phi3}). There is only one non-trivial solution. Therefore, this is the only dominant singularity of $x(y)$. We have $A= -\Sact(2) = \frac23$. It follows that,
\begin{gather*} [\hbar^n] \Fop\left[-\frac{x^2}{2} + \frac{x^3}{3!}\right](\hbar) = \\ \sum_{k=0}^{R-1} w_{k} \left(\frac{2}{3} \right)^{-(n-k)} \Gamma(n-k) + \bigO\left(\left(\frac{2}{3} \right)^{-n} \Gamma(n-R)\right) \qquad \forall R\in \N_0, \end{gather*}
where
\begin{align*} w_k &= \frac{1}{2\pi}[\hbar^k] \Fop[\Sact(2) - \Sact(x+2)](-\hbar)      = \frac{1}{2\pi}[\hbar^k] \Fop\left[-\frac{x^2}{2}-\frac{x^3}{3!}\right](-\hbar). \end{align*}
Because generally $\Fop[\Sact(x)](\hbar) = \Fop[\Sact(-x)](\hbar)$, the large $n$ asymptotics of the power series
\begin{align*} \Fop\left[ -\frac{x^2}{2} + \frac{x^3}{3!} \right](\hbar) &= \sum_{n=0}z_n \hbar^n = 1 + \frac{5}{24} \hbar + \frac{385}{1152} \hbar^2 + \frac{85085}{82944} \hbar^3 \\ &+ \frac{37182145}{7962624} \hbar^4 + \frac{5391411025}{191102976} \hbar^5 + \ldots \end{align*}
are given by the same sequence with negative expansion parameter:
\begin{align*} z_n &\underset{n\rightarrow \infty}{\sim} \frac{1}{2\pi} \left( \left(\frac{2}{3} \right)^{-n} \Gamma(n) - \frac{5}{24} \left(\frac{2}{3} \right)^{-n+1} \Gamma(n-1) + \frac{385}{1152} \left(\frac{2}{3} \right)^{-n+2} \Gamma(n-2) \right. \\ & \left. - \frac{85085}{82944} \left(\frac{2}{3} \right)^{-n+3} \Gamma(n-3) + \frac{37182145}{7962624} \left(\frac{2}{3} \right)^{-n+4} \Gamma(n-4) + \ldots \right) \end{align*}
This is an occurrence of the quite general self-replicating or resurgent phenomenon of the asymptotics of power series \cite{ecalle1981fonctions}.
\end{expl}

Restricting the dominant singularities in Corollary \ref{crll:comb_int_asymp} to be regular points of the complex curve is necessary. Otherwise, it cannot be guaranteed that a critical point actually coincides with a dominant singularity of $x(y)$. We will illustrate this in
\begin{expl}
\label{expl:counterexpl}
Let $\Sact(x) = -\frac{(1-e^{-x})^2}{2}$. This action has saddle points at $\tau_k = 2\pi i k$ for all $k\in\Z$. Because $\Sact(\tau_k)=0$ using Corollary \ref{crll:comb_int_asymp} naively would imply that, $[\hbar^n] \Fop\left[ -\frac{(1-e^{-x})^2}{2} \right] \sim \left(\frac{1}{0}\right)^n$, which is clearly nonsensical. 
On the other hand, we can solve $\frac{y^2}{2} = \frac{(1-e^{-x})^2}{2}$ for $x(y)= \log\frac{1}{1-y}$. Using Proposition \ref{prop:formalchangeofvar} immediately results in
\begin{align*} \Fop\left[ -\frac{(1-e^{-x})^2}{2} \right] &= \sum_{n=0}^\infty \hbar^n (2n+1)!! [y^{2n+1}] \log\frac{1}{1-y} \\ &= \sum_{n=0}^\infty \hbar^n (2n-1)!! [y^{2n}] \frac{1}{1-y} = \sum_{n=0}^\infty \hbar^n (2n-1)!!, \end{align*}
which naturally has a sound asymptotic description. The dominant singularity of $x(y)$ is obviously at $y=1$. An association of the asymptotics with saddle points of $\frac{(1-e^{-x})^2}{2}$ is not possible in this case, due to the irregularity of the complex curve at the saddle points.

This example will be of relevance in Chapter \ref{chap:applications_zerodim}, as it gives an important generating function in zero-dimensional quenched QED and Yukawa theory.
\end{expl}

To obtain the asymptotics of less trivial counting functions of graphs or equivalently of more involved observables in zero-dimensional QFT, we will have to analyze composite power series which involve $\Fop$-expressions in non-trivial ways. 

The obvious examples are power series such as $\log \left( \Fop\left[-\frac{x^2}{2} + \frac{x^3}{3!}\right](\hbar) \right)$ which enumerates the number of connected graphs by excess. Although we know the asymptotics of the coefficients of $\Fop\left[-\frac{x^2}{2} + \frac{x^3}{3!}\right](\hbar)$ up to arbitrary high order, we do not have any information about the asymptotics in the connected case so far. 

Ultimately, we want to \textit{renormalize} quantities which essentially boils down to substituting the $\hbar$ variable for some function $f(\hbar)$. Therefore, we are also interested in the asymptotics of the coefficients of power series such as $\Fop\left[-\frac{x^2}{2} + \frac{x^3}{3!}\right](f(\hbar))$. 

The subject of the next chapter is the systematic analysis of these composition operations on power series.

\chapter{The ring of factorially divergent power series}
\label{chap:facdivpow}

The content of this chapter, which is based on the author's article \cite{borinsky2016generating},
is solely concerned with sequences $a_n$, which admit an asymptotic expansion for large $n$ of the form,
\begin{align} \label{eqn:first_example} a_n = \alpha^{n+\beta} \Gamma(n+\beta) \left( c_0 + \frac{c_1}{\alpha(n+\beta-1)} + \frac{c_2}{\alpha^2(n+\beta-1)(n+\beta-2)} + \ldots \right), \end{align}
for some $\alpha \in \R_{>0}$, $\beta \in \R$ and $c_k \in \R$ as they appeared in the statement of Theorem \ref{thm:comb_int_asymp}.
The theory of these sequences is independent of the theory of zero-dimensional QFT and graphical enumeration, but necessary to analyze the asymptotics for these problems.

We will use the following notation which is mostly standard in the context of asymptotics and combinatorics:

A (formal) power series $f \in \R[[x]]$ will be denoted in the usual `functional' notation $f(x) = \sum_{n=0}^\infty f_n x^n$. The coefficients of a power series $f$ will be expressed by the same symbol with the index attached as a subscript $f_n$ or with the coefficient extraction operator $[x^n]f(x) = f_n$. Ordinary (formal) derivatives are expressed as $f'(x) = \sum_{n=0}^\infty n f_n x^{n-1}$. The ring of power series, restricted to expansions of functions which are analytic at the origin, or equivalently power series with non-vanishing radius of convergence, will be denoted as $\R\{x\}$. The $\bigO$-notation will be used: $\bigO(a_n)$ denotes the set of all sequences $b_n$ such that $\limsup _{n\rightarrow \infty} | \frac{ b_n }{a_n} | < \infty$ and $\smallO(a_n)$ denotes all sequences $b_n$ such that $\lim _{n\rightarrow \infty} \frac{ b_n }{a_n} = 0$. Equations of the form $a_n = b_n + \bigO(c_n)$ are to be interpreted as statements $a_n-b_n \in \bigO(c_n)$ as usual. See \cite{bender1974asymptotic} for an introduction to this notation. Tuples of non-negative integers will be denoted by bold letters $\textbf{t} = (t_1,\ldots,t_L) \in \N_0^L$. The notation $|\textbf{t}|$ will be used as a short form for $\sum_{l=1}^L t_l$. We will consider the binomial coefficient ${ a \choose n}$ to be defined for all $a\in \R$ and $n\in \N_0$ such that ${ a \choose n} := [x^n] (1+x)^a$.

\nomenclature{$\R[[x]]$}{Ring of power series with real coefficients}
\nomenclature{$\bigO$}{Big $\bigO$-notation}

The only non-standard notation that will be used to improve the readability of lengthy expressions is the abbreviation $\G{n}{\alpha}{\beta} := \alpha^{n+\beta} \Gamma(n+\beta)$.

\nomenclature{$\G{n}{\alpha}{\beta}$}{Shorthand notation for $\alpha^{n+\beta} \Gamma(n+\beta)$}
\section{Prerequisites}
\label{sec:prereq}
The first step is to establish a suitable notion of a power series with a well-behaved asymptotic expansion.
\begin{defn}
\label{def:Fpowerseries}
For given $\alpha \in \R_{>0}$ and $\beta \in \R$ let $\fring{x}{{\alpha}}{\beta}$ be the subset of $\R[[x]]$, such that $f \in \fring{x}{{\alpha}}{\beta}$ if and only if there exists a sequence of real numbers $(c_{k}^f)_{k\in \N_0}$, which fulfills
\begin{align} \label{eqn:basic_asymp_C} f_n &= \sum _{k=0}^{R-1} c_{k}^f \G{n-k}{\alpha}{\beta} + \bigO\left(\G{n-R}{\alpha}{\beta}\right) && \forall R \in \N_0, \end{align}
where $\G{n}{\alpha}{\beta} = \alpha^{n+\beta} \Gamma(n+\beta)$.
\end{defn}
\nomenclature{$\fring{x}{{\alpha}}{\beta}$}{Ring of factorially divergent power series}
\begin{crll}
$\fring{x}{{\alpha}}{\beta}$ is a linear subspace of $\R[[x]]$.
\end{crll}
\begin{crll}
The sequence $(c_{k}^f)_{k\in \N_0}$ is unique for all $f \in \fring{x}{{\alpha}}{\beta}$. The coefficients can be calculated iteratively using the explicit formula 
$c^f_K = \lim \limits_{n \rightarrow \infty} \frac{f_n - \sum _{k=0}^{K-1} c_{k}^f \G{n-k}{\alpha}{\beta} }{ \G{n-K}{\alpha}{\beta}}$ for all $K\in \N_0$.
\end{crll}
\begin{rmk}
The expression in eq.\ \eqref{eqn:basic_asymp_C} represents an asymptotic expansion or Poincar\'e expansion with the asymptotic scale $\alpha^{n+\beta}\Gamma(n+\beta)$ \cite[Ch. 1.5]{bruijn1970asymptotic}.
\end{rmk}
\begin{rmk}
The subspace $\fring{x}{\alpha}{\beta}$ includes all (real) power series whose coefficients only grow exponentially: $\R\{x\} \subset \fring{x}{\alpha}{\beta}$. These with all other series with coefficients, which are in $\smallO(\G{n-R}{\alpha}{\beta})$ for all $R\in \N_0$, have an asymptotic expansion of the form in eq.\ \eqref{eqn:basic_asymp_C} with all $c_{k}^f=0$.
\end{rmk}
\begin{rmk}
Definition \ref{def:Fpowerseries} implies that if $f\in \fring{x}{\alpha}{\beta}$ then $f_n \in \bigO\left(\G{n}{\alpha}{\beta}\right)=\bigO\left(\alpha^{n} \Gamma(n+\beta)\right)$. Accordingly, the power series in $\fring{x}{\alpha}{\beta}$ are a subset of \textit{Gevrey-1} sequences \cite[Ch XI-2]{hsieh2012basic}. Being \textit{Gevrey-1} is not sufficient for a power series to be in $\fring{x}{\alpha}{\beta}$. For instance, a sequence which behaves for large $n$ as $a_n = n! (1+ \frac{1}{\sqrt{n}}+ \bigO(\frac{1}{n}))$ is \textit{Gevrey-1}, but not in $\fring{x}{\alpha}{\beta}$.
\end{rmk}
\begin{rmk}
In resurgence theory further restrictions on the allowed power series are imposed, which ensure that the Borel transformations of the sequences have proper analytic continuations or are `endless continuable' \cite[II.6]{mitschi2016divergent}. These restrictions are analogous to the requirement that, apart from $f_n$, also $c_{k}^f$ has to have a well-behaved asymptotic expansion. The coefficients of this asymptotic expansion are also required to have a well-behaved asymptotic expansion and so on. These kinds of restrictions will not be necessary for the presented algebraic considerations, which are aimed at combinatorial applications.
\end{rmk}

The central theme of this chapter is to \textit{interpret the coefficients $c_{k}^f$ of the asymptotic expansion as another power series}. 
In fact, Definition \ref{def:Fpowerseries} immediately suggests to define the following map:
\begin{defn}
\label{def:basic_asymp_definition}
Let $\asyOp^\alpha_\beta: \fring{x}{\alpha}{\beta} \rightarrow \R[[x]]$ be the map that associates a power series $\asyOp^\alpha_\beta f \in \R[[x]]$ to every power series $f \in \fring{x}{\alpha}{\beta}$ such that
\begin{align} \label{eqn:basic_asymp} (\asyOp^\alpha_\beta f)(x) = \sum_{k=0}^\infty c_{k}^f x^k, \end{align}
with the coefficients $c_{k}^f$ from Definition \ref{def:Fpowerseries}.
\end{defn}
\nomenclature{$\asyOp^\alpha_\beta$}{Asymptotic or alien derivative}
\begin{crll}
\label{crll:asyOplinear}
$\asyOp^\alpha_\beta$ is linear.
\end{crll}
\begin{rmk}
In Proposition \ref{prop:derivation} it will be proven that $\asyOp^\alpha_\beta$ is a derivation. We will adopt the usual notation for derivations and consider $\asyOp^\alpha_\beta$ to act on everything to its right.
\end{rmk}
\begin{rmk}
In the realm of resurgence such an operator is called \textit{alien derivative} or \textit{alien operator} \cite[II.6]{mitschi2016divergent}.
\end{rmk}
\begin{rmk}
$\asyOp^\alpha_\beta$ is clearly not injective. For instance, $\R\{x\} \subset \ker \asyOp^\alpha_\beta$.
\end{rmk}
\begin{expl}
The power series $f\in \R[[x]]$ associated to the sequence $f_n = n!$ clearly fulfills the requirements of Definition \ref{def:Fpowerseries} with $\alpha=1$ and $\beta=1$. Therefore, $f \in \fring{x}{1}{1}$ and $(\asyOp^1_1 f)(x) = 1$. 
\end{expl}

The asymptotic expansion in eq.\ \eqref{eqn:basic_asymp_C} is normalized such that shifts in $n$ can be absorbed by shifts in $\beta$. More specifically,
\begin{prop}
\label{prop:betashift}
For all $m \in \N_0$
\begin{align*} f\in \fring{x}{\alpha}{\beta}\text{ if and only if } f\in \fring{x}{\alpha}{\beta+m}\text{ and }\asyOp^\alpha_{\beta+m} f \in x^m \R[[x]]. \end{align*}
If either holds, then $x^m \left( \asyOp^\alpha_{\beta} f\right)(x) = \left( \asyOp^\alpha_{\beta+m} f\right)(x)$.
\end{prop}

\begin{proof}
Because $\G{n}{\alpha}{\beta} = \alpha^{n-m+\beta+m} \Gamma(n-m+\beta+m) = \G{n-m}{\alpha}{\beta+m}$, the following two relations between $f_n$ and $c^f_k$ are equivalent,
\begin{align} \label{eqn:shifteq1} f_n &= \sum _{k=0}^{R-1} c_{k}^f\G{n-k}{\alpha}{\beta} + \bigO\left(\G{n-R}{\alpha}{\beta}\right) && \forall R\in \N_0 \\ \label{eqn:shifteq2} f_n &= \sum _{k=m}^{R'-1} c_{k-m}^f\G{n-k}{\alpha}{\beta+m} + \bigO\left(\G{n-R'}{\alpha}{\beta+m}\right) && \forall R'\geq m. \end{align}
Eq.\ \eqref{eqn:shifteq1} follows from $f\in \fring{x}{\alpha}{\beta}$ by Definition \ref{def:Fpowerseries}. In that case, eq.\ \eqref{eqn:shifteq2} implies that $f\in \fring{x}{\alpha}{\beta+m}$ and that $\left( \asyOp^\alpha_{\beta+m} f\right)(x) = \sum_{k=m}^\infty c_{k-m}^f x^k= x^m\left( \asyOp^\alpha_{\beta} f\right)(x) \in x^m \R[[x]]$ by Definition \ref{def:basic_asymp_definition}.

If $f\in \fring{x}{\alpha}{\beta+m}$ and $\asyOp^\alpha_{\beta+m} f \in x^m \R[[x]]$, then we can write the asymptotic expansion of $f$ in the form of eq.\ \eqref{eqn:shifteq2}. Eq.\ \eqref{eqn:shifteq1} implies $f\in \fring{x}{\alpha}{\beta}$.
\end{proof}
\begin{prop}
\label{prop:betashiftlow}
For all $m \in \N_0$
\begin{align*} f\in x^m \fring{x}{\alpha}{\beta} \text{ if and only if } \frac{f(x)}{x^m} \in \fring{x}{\alpha}{\beta+m}. \end{align*}
If either holds, then $\left( \asyOp^\alpha_{\beta} f\right)(x) = \left( \asyOp^\alpha_{\beta+m} \frac{f(x)}{x^m} \right)(x)$.
\end{prop}
\begin{proof}
Because $\G{n+m}{\alpha}{\beta} = \alpha^{n+m+\beta} \Gamma(n+m+\beta) = \G{n}{\alpha}{\beta+m}$, the following two relations between $f_n$ and $c_k^f$ are equivalent,
\begin{align} \label{eqn:shiftloweq1} f_n &= \sum _{k=0}^{R-1} c_{k}^f\G{n-k}{\alpha}{\beta} + \bigO\left(\G{n-R}{\alpha}{\beta}\right) && \forall R\in \N_0 \\ \label{eqn:shiftloweq2} f_{n+m} &= \sum _{k=0}^{R-1} c_{k}^f\G{n-k}{\alpha}{\beta+m} + \bigO\left(\G{n-R}{\alpha}{\beta+m}\right) && \forall R\in \N_0. \end{align}
Eq.\ \eqref{eqn:shiftloweq1} follows from $f\in x^m\fring{x}{\alpha}{\beta}$. Because $f\in x^m \fring{x}{\alpha}{\beta}\subset x^m \R[[x]]$, we have $\frac{f(x)}{x^m} = \sum_{n=0}^\infty f_{n+m} x^{n}\in \R[[x]]$. Eq.\ \eqref{eqn:shiftloweq2} then implies that $\frac{f(x)}{x^m} \in \fring{x}{\alpha}{\beta+m}$ and by Definition \ref{def:basic_asymp_definition}, $\left( \asyOp^\alpha_{\beta} f\right)(x) = \left( \asyOp^\alpha_{\beta+m} \frac{f(x)}{x^m}\right)(x)$. 

If $\frac{f(x)}{x^m} \in \fring{x}{\alpha}{\beta+m}$, then $f\in x^m \R[[x]]$ and eq.\ \eqref{eqn:shiftloweq2} holds for the coefficients of $f$, which implies $f\in x^m\fring{x}{\alpha}{\beta}$ by eq.\ \eqref{eqn:shiftloweq1} and Definition \ref{def:Fpowerseries}.
\end{proof}

It follows from Proposition \ref{prop:betashift} that $\fring{x}{\alpha}{\beta} \subset \fring{x}{\alpha}{\beta+m}$ for all $m\in \N_0$. It will be convenient to only work in the spaces $\fring{x}{\alpha}{\beta}$ with $\beta > 0$ and to use Proposition \ref{prop:betashift} to verify that the subspaces $\fring{x}{\alpha}{\beta-m}$ inherit all relevant properties from $\fring{x}{\alpha}{\beta}$.
The advantage is that, with $\beta > 0$, it is easier to express uniform bounds on the remainder terms in eq.\ \eqref{eqn:basic_asymp_C}. The following definition will provide a convenient notation for these bounds.
\begin{defn}
\label{defn:basic_bigC_estimate}
For $\alpha,\beta \in \R_{>0}$ and $R \in \N_0$, let $\rho^\alpha_{\beta,R}: \fring{x}{\alpha}{\beta} \rightarrow \R_+$ be the map
\begin{align} \label{eqn:basic_bigC_estimate} \rho^\alpha_{\beta,R}(f) = \max_{0\leq K\leq R} \sup_{n\geq K} \frac{\left| f_n - \sum _{k=0}^{K-1} c_{k}^f \G{n-k}{\alpha}{\beta} \right|}{\G{n-K}{\alpha}{\beta}}, \end{align}
with the coefficients $c_k^f$ as in Definition \ref{def:Fpowerseries}.
\end{defn}
\nomenclature{$\rho^\alpha_{\beta,R}$}{Norm for $R\in \N_0$ on the ring $\fring{x}{\alpha}{\beta}$ for $\beta > 0$}
It follows directly from Definition \ref{def:Fpowerseries} that Definition \ref{defn:basic_bigC_estimate} is well-defined. 
Eq.\ \eqref{eqn:basic_bigC_estimate} can be translated into bounds for the coefficients $f_n$ and the $c_{k}^f$:
\begin{crll}
\label{crll:smallc_bigC_estimate}
If $\alpha,\beta \in \R_{>0}$ and $R\in \N_0$, then for all $f \in \fring{x}{{\alpha}}{\beta}$ and $n,K \in \N_0$ with $K \leq R$ as well as $n \geq K$, 
\begin{align} \left| f_n - \sum _{k=0}^{K-1} c_{k}^f \G{n-k}{\alpha}{\beta} \right| &\leq \rho^\alpha_{\beta,R}(f) \G{n-K}{\alpha}{\beta} &&\text{ and } &&|c^f_K| \leq \rho^\alpha_{\beta,R}(f). \end{align}
\end{crll}
\begin{rmk}
It can be verified using linearity and the triangle inequality that the maps $\rho^\alpha_{\beta,R}$ form a family of norms on $\fring{x}{\alpha}{\beta}$ with $\beta>0$. Moreover, these norms will turn out to be submultiplicative up to equivalence (see Proposition \ref{prop:submultiplicative}). However, we will not make direct use of any topological properties of the spaces $\fring{x}{{\alpha}}{\beta}$.
\end{rmk}

\section{Elementary properties of sums over \texorpdfstring{$\Gamma$}{Γ} functions}
The following lemma forms the foundation for most conclusions in this chapter. It provides an estimate for sums over $\Gamma$ functions. Moreover, it ensures that the subspace $\fring{x}{\alpha}{\beta}$ of formal power series falls into a large class of sequences studied by Bender \cite{bender1975asymptotic}.
From another perspective the lemma can be seen as an entry point to resurgence, which bypasses the necessity for analytic continuations and Borel transformations.

As with all statements in this chapter that involve estimates, we will require $\beta > 0$.
\begin{lmm}
\label{lmm:gammasum}
If $\alpha,\beta \in \R_{>0}$, then 
\begin{align} \label{eqn:centersumexpl2} \sum_{m=0}^{n} \G{m}{\alpha}{\beta} \G{n-m}{\alpha}{\beta} &\leq (2+\beta)\G{0}{\alpha}{\beta} \G{n}{\alpha}{\beta} && \forall n \in \N_0. \end{align}
\end{lmm}
\begin{proof}
Recall that $\G{n}{\alpha}{\beta} = \alpha^{n+\beta}\Gamma(n+\beta)$ and that $\Gamma:\R_{>0}\rightarrow \R_{>0}$ is a \textit{log-convex} function. If $\beta \in \R_{>0}$, then the functions $\Gamma(m+\beta)$ and $\Gamma(n-m+\beta)$ are also log-convex functions in $m$ on the interval $[0,n]$, as log-convexity is preserved under shifts and reflections. 
Furthermore, log-convexity is closed under multiplication. This implies that $\G{m}{\alpha}{\beta}\G{n-m}{\alpha}{\beta} = \alpha^{n+2\beta} \Gamma(m+\beta)\Gamma(n-m+\beta)$ is a log-convex function in $m$ on the interval $[1,n-1] \subset [0,n]$. A convex function always attains its maximum on the boundary of its domain. Accordingly,
$\G{m}{\alpha}{\beta}\G{n-m}{\alpha}{\beta} \leq \G{1}{\alpha}{\beta}\G{n-1}{\alpha}{\beta}$ for all $m \in [1,n-1]$.
This way, the sum $\sum_{m=0}^{n}\G{m}{\alpha}{\beta}\G{n-m}{\alpha}{\beta}$ can be estimated after stripping off the two boundary terms:
\begin{align} \label{eqn:sumGestimate} \sum_{m=0}^{n} \G{m}{\alpha}{\beta}\G{n-m}{\alpha}{\beta} &\leq 2 \G{0}{\alpha}{\beta}\G{n}{\alpha}{\beta} + (n-1) \G{1}{\alpha}{\beta}\G{n-1}{\alpha}{\beta} && \forall n \geq 1. \end{align}
It follows from $n \Gamma(n) = \Gamma(n+1)$ that
$ \G{1}{\alpha}{\beta}\G{n-1}{\alpha}{\beta} = \frac{\beta}{n-1+\beta} \G{0}{\alpha}{\beta}\G{n}{\alpha}{\beta} $ for all $n\geq1$.
Because $n-1+\beta \geq n-1$, substituting this into eq.\ \eqref{eqn:sumGestimate} implies the inequality in eq.\ \eqref{eqn:centersumexpl2} for all $n \geq 1$. The remaining case $n=0$ is trivially fulfilled.
\end{proof}

\begin{crll}
\label{crll:centersum}
If $\alpha,\beta \in \R_{>0}$ and $R \in \N_0$, then there exists a constant $C_R\in \R$ such that
\begin{align} \label{eqn:centersum} \sum_{m=R}^{n-R} \G{m}{\alpha}{\beta}\G{n-m}{\alpha}{\beta} &\leq C_R \G{n-R}{\alpha}{\beta} && \forall n \geq 2R \end{align}
\end{crll}
\begin{proof}
Recall that $\G{m+R}{\alpha}{\beta} = \G{m}{\alpha}{\beta+R}$.
We can shift the summation variable to rewrite the left hand side of eq.\ \eqref{eqn:centersum} as 
\begin{gather*} \sum_{m=0}^{n-2R} \G{m+R}{\alpha}{\beta}\G{n-m-R}{\alpha}{\beta} = \sum_{m=0}^{n-2R} \G{m}{\alpha}{\beta+R}\G{n-2R-m}{\alpha}{\beta+R} \\ \leq (2+\beta+R)\G{0}{\alpha}{\beta+R} \G{n-2R}{\alpha}{\beta+R}, \end{gather*}
where we applied Lemma \ref{lmm:gammasum} with the substitutions $\beta \rightarrow \beta+R$ and $n\rightarrow n-2R$. Because  $\G{n-2R}{\alpha}{\beta+R} = \G{n-R}{\alpha}{\beta}$ the statement follows.
\end{proof}

\begin{crll}
\label{crll:exponential_inO}
If $\alpha,\beta \in \R_{>0}$, $C \in \R$ and $P \in \R[m]$ is some polynomial in $m$, then 
\begin{align} \sum_{m=R}^{n} C^m P(m) \G{n-m}{\alpha}{\beta} \in \bigO(\G{n-R}{\alpha}{\beta}) && \forall R \in \N_0. \end{align}
\end{crll}
\begin{proof}
There is a $C' \in \R$ such that $|C^m P(m)|$ is bounded by $C' \G{m}{\alpha}{\beta}$ for all $m\in \N_0$. 
Therefore, Corollary \ref{crll:centersum} ensures that 
\begin{gather*} \sum_{m=R}^{n-R} C^m P(m) \G{n-m}{\alpha}{\beta} \leq C'\sum_{m=R}^{n-R} \G{m}{\alpha}{\beta} \G{n-m}{\alpha}{\beta} \in \bigO(\G{n-R}{\alpha}{\beta}). \end{gather*}
The remainder 
$\sum_{m=n-R+1}^{n} C^m P(m) \G{n-m}{\alpha}{\beta}= \sum_{m=0}^{R-1} C^{n-m} P(n-m) \G{m}{\alpha}{\beta}$ is obviously in $\bigO(\G{n-R}{\alpha}{\beta})$.
\end{proof}

\section{A derivation for asymptotics}

\begin{prop}
For all $\alpha \in \R_{>0}$ and $\beta \in \R$, the subspace $\fring{x}{\alpha}{\beta}$ is a subring of $\R[[x]]$. 

Moreover, if $f,g\in \fring{x}{\alpha}{\beta}$, then
\label{prop:derivation}
\begin{itemize}
\item
The product $f \cdot g = f(x)g(x)$ belongs to $\fring{x}{\alpha}{\beta}$.
\item
$\asyOp^\alpha_\beta$ is a derivation, that means it respects the product rule
\begin{align} \label{eqn:leipniz} ( \asyOp^\alpha_\beta ( f \cdot g ))(x) &= f(x) (\asyOp^\alpha_\beta g)(x) + g(x) (\asyOp^\alpha_\beta f)(x). \end{align}
\end{itemize}
\end{prop}

\begin{crll}
\label{crll:chain_for_product}
If $g^{1}, \ldots, g^{L} \in \fring{x}{\alpha}{\beta}$, then $\prod_{l=1}^L g^{l}(x) \in \fring{x}{\alpha}{\beta}$ and 
\begin{gather} \left( \asyOp^\alpha_\beta \left( \prod_{l=1}^L g^{l}(x) \right)\right)(x) = \sum_{l=1}^L \left(\prod_{\substack{m=1 \\ m\neq l}}^L g^{m}(x)\right) ( \asyOp^\alpha_\beta g^{l})(x).   \end{gather}
\end{crll}
\begin{proof}
Proof by induction in $L$ using the product rule.
\end{proof}
\begin{crll}
\label{crll:chain_for_product_pow}
If $g^{1}, \ldots, g^{L} \in \fring{x}{\alpha}{\beta}$ and $\textbf{t} = (t_1, \ldots, t_L) \in \N_0^L$, then $\prod_{l=1}^L (g^{l}(x))^{t_l} \in \fring{x}{\alpha}{\beta}$ and 
\begin{align} \left( \asyOp^\alpha_\beta \left( \prod_{l=1}^L (g^{l}(x))^{t_l} \right)\right)(x) = \sum_{l=1}^L t_l (g^l(x))^{t_l-1} \left(\prod_{\substack{m=1 \\ m\neq l}}^L (g^{m}(x))^{t_m}\right) ( \asyOp^\alpha_\beta g^{l})(x) . \end{align}
\end{crll}
\begin{crll}
\label{crll:polynomial_comp}
If $g^{1}, \ldots, g^{L} \in \fring{x}{\alpha}{\beta}$  and $p\in \R[y_1,\ldots,y_L]$ is polynomial in $L$ variables, then $p(g^{1}(x), \ldots, g^{L}(x))\in\fring{x}{\alpha}{\beta}$ and
\begin{align} \label{eqn:polynomial_comp} (\asyOp^\alpha_\beta (p(g^{1}, \ldots, g^{L})))(x) = \sum_{l=1}^L \frac{\partial p}{\partial g^l} (g^{1}, \ldots, g^{L}) (\asyOp^\alpha_\beta g^{l})(x). \end{align}
\end{crll}
Although the last three statements are only basic general properties of commutative derivation rings, they suggest that $\asyOp^\alpha_\beta$ fulfills a simple chain rule. In fact, Corollary \ref{crll:polynomial_comp} can still be generalized from polynomials to analytic functions, but, as already mentioned, this intuition turns out to be false in general. 

We will prove Proposition \ref{prop:derivation} alongside with another statement which will be useful to establish this general chain rule:
\begin{prop}
\label{prop:submultiplicative}
If $\alpha,\beta \in \R_{>0}$ and $R\in \N_0$, then there exists a constant $C_R \in \R$ such that 
\begin{align} \label{eqn:submultiplicative} \rho^\alpha_{\beta,R}(f \cdot g) &\leq C_R \rho^\alpha_{\beta,R}(f) \rho^\alpha_{\beta,R}(g) && \forall f,g \in \fring{x}{\alpha}{\beta}. \end{align}
\end{prop}
\begin{crll}
\label{crll:estimate_for_product_pow}
If $\alpha,\beta \in \R_{>0}$, $R\in \N_0$ and $g^{1}, \ldots, g^{L} \in \fring{x}{\alpha}{\beta}$, then there exists a constant $C_R \in \R$ such that
\begin{align} \rho^\alpha_{\beta,R} \left( \prod_{l=1}^L (g^{l}(x))^{t_l} \right) &\leq C_R^{|\textbf{t}|} && \forall \textbf{t} \in \N_0^L \text{ with } |\textbf{t}| \geq 1. \end{align}
\end{crll}
\begin{proof}
Iterating eq.\ \eqref{eqn:submultiplicative} gives a constant $C_R\in \R$ such that
\begin{align*} \rho^\alpha_{\beta,R} \left( \prod_{l=1}^L (g^{l}(x))^{t_l} \right) &\leq {C_R}^{|\textbf{t}|-1} \prod_{l=1}^L \left(\rho^\alpha_{\beta,R} (g^l) \right)^{t_l} &&\forall \textbf{t} \in \N_0^L \text{ with } |\textbf{t}| \geq 1. \end{align*} 
The right hand side is clearly bounded by ${C_R'}^{|\textbf{t}|}$ for all $|\textbf{t}| \geq 1$ with an appropriate $C_R'\in \R$ which depends on the $g^l$.
\end{proof}

We will prove Proposition \ref{prop:derivation} under the assumption that $\beta > 0$. The following Lemma shows that, as a consequence of Proposition \ref{prop:betashift}, we can do so without loss of generality. 
\begin{lmm}
\label{lmm:wlogbetaderivation}
If Proposition \ref{prop:derivation} holds for $\beta \in \R_{>0}$, then it holds for all $\beta \in \R$.
\end{lmm}
\begin{proof}
For $\beta \in \R$, choose $m\in \N_0$ such that $\beta+m >0$. If $f,g \in \fring{x}{\alpha}{\beta}$, then $f,g \in \fring{x}{\alpha}{\beta+m}$ by Proposition \ref{prop:betashift}. By the requirement $f\cdot g \in \fring{x}{\alpha}{\beta+m}$ and $( \asyOp^\alpha_{\beta+m} ( f \cdot g ))(x) = f(x) (\asyOp^\alpha_{\beta+m} g)(x) + g(x) (\asyOp^\alpha_{\beta+m} f)(x)$. Using $(\asyOp^\alpha_{\beta+m} f)(x) = x^m (\asyOp^\alpha_{\beta} f)(x)$ from Proposition \ref{prop:betashift} gives $( \asyOp^\alpha_{\beta+m} ( f \cdot g ))(x) = x^m \left( f(x) (\asyOp^\alpha_{\beta} g)(x) + g(x) (\asyOp^\alpha_{\beta} f)(x) \right)$. As $f \cdot g \in \fring{x}{\alpha}{\beta+m}$ and $\asyOp^\alpha_{\beta+m} ( f \cdot g ) \in x^m \R[[x]]$, it follows that $f \cdot g \in \fring{x}{\alpha}{\beta}$ and $\asyOp^\alpha_{\beta} ( f \cdot g ) = f(x) (\asyOp^\alpha_{\beta} g)(x) + g(x) (\asyOp^\alpha_{\beta} f)(x)$ by Proposition \ref{prop:betashift}.
\end{proof}

To prove Propositions \ref{prop:derivation} and \ref{prop:submultiplicative}, we will use some estimates for the coefficients of the product of two power series. To be able to establish these estimates, we will have to require that $\beta > 0$.
\begin{lmm}
\label{lmm:estimate_full_fg}
If $\alpha,\beta \in \R_{>0}$ and $R \in \N_0$, then there exists a constant $C_R \in \R$, such that for all $f,g \in \fring{x}{\alpha}{\beta}$ and $n,K \in \N_0$ with $K \leq R$ as well as $n \geq K$,
\begin{align} \left| \sum_{m=0}^n f_{n-m} g_m - \sum_{m=0}^{K-1} f_{n-m} g_m - \sum_{m=0}^{K-1} f_m g_{n-m} \right| &\leq C_R \rho^\alpha_{\beta,R}(f) \rho^\alpha_{\beta,R}(g) \G{n-K}{\alpha}{\beta}. \end{align}
\end{lmm}
\begin{proof} 
Corollary \ref{crll:smallc_bigC_estimate} with $K=0$ states that $\left| f_n \right| \leq \rho^\alpha_{\beta,R}(f) \G{n}{\alpha}{\beta}$ for all $f \in \fring{x}{\alpha}{\beta}$ and $n\in \N_0$.
We can use this to estimate the expression 
\begin{gather*} h_n:=\left| \sum_{m=0}^n f_{n-m} g_m - \sum_{m=0}^{K-1} f_{n-m} g_m - \sum_{m=0}^{K-1} f_m g_{n-m} \right| \end{gather*}
in different ranges for $n$,
\begin{align*} 2K> n \geq K&\Rightarrow & h_n&=\left| \sum_{m=n-K+1}^{K-1} f_{n-m} g_m\right| \\ && &\leq \rho^\alpha_{\beta,R}(f) \rho^\alpha_{\beta,R}(g) \sum_{m=n-K+1}^{K-1} \G{n-m}{\alpha}{\beta} \G{m}{\alpha}{\beta} \\ n \geq 2K&\Rightarrow & h_n&=\left| \sum_{m=K}^{n-K} f_{n-m} g_m\right| \\ && &\leq \rho^\alpha_{\beta,R}(f) \rho^\alpha_{\beta,R}(g) \sum_{m=K}^{n-K} \G{n-m}{\alpha}{\beta} \G{m}{\alpha}{\beta}. \end{align*}
It is trivial to find a constant $C_R$ such that 
\begin{gather*} \sum_{m=n-K+1}^{K-1} \G{n-m}{\alpha}{\beta} \G{m}{\alpha}{\beta} \leq C_R \G{n-K}{\alpha}{\beta}, \end{gather*}
for all $K\leq R$ and $2K> n \geq K$, because only finitely many inequalities need to be fulfilled.
Corollary \ref{crll:centersum} guarantees that we can also find a constant $C_R$ for the second case.
\end{proof}

\begin{lmm}
\label{lmm:estimate_partial_fg}
If $\alpha, \beta \in \R_{>0}$ and $R \in \N_0$, then there exists a constant $C_R \in \R$, such that for all $f,g \in \fring{x}{\alpha}{\beta}$ and $n,K \in \N_0$ with $K \leq R$ as well as $n \geq K$,
\begin{align} \left| \sum_{m=0}^{K-1} f_{n-m} g_m - \sum_{k=0}^{K-1} d^{f,g}_k \G{n-k}{\alpha}{\beta} \right| &\leq C_R \rho^\alpha_{\beta,R}(f) \rho^\alpha_{\beta,R}(g) \G{n-K}{\alpha}{\beta}, \end{align}
where $d^{f,g}_k := [x^k] g(x) (\asyOp^\alpha_\beta f)(x)$.
\end{lmm}
\begin{proof} 
Corollary \ref{crll:smallc_bigC_estimate} with the substitutions $n\rightarrow n-m$ and $K \rightarrow K-m$ implies that
\begin{align*} \left| f_{n-m} - \sum_{k=0}^{K-m-1} c^f_{k} \G{n-m-k}{\alpha}{\beta} \right| \leq \rho^\alpha_{\beta,R}(f) \G{n-K}{\alpha}{\beta}, \end{align*}
for all $f \in \fring{x}{\alpha}{\beta}$ and $n,K,m \in \N_0$ with $m \leq K \leq R$ as well as $n\geq K$ where $c_k^f = [x^k] (\asyOp^\alpha_\beta f)(x)$. It also follows from Corollary \ref{crll:smallc_bigC_estimate} that $|g_m| \leq \rho^\alpha_{\beta,R}(g) \G{m}{\alpha}{\beta}$ for all $g \in \fring{x}{\alpha}{\beta}$ and $m \in \N_0$.
Because $d^{f,g}_k = \sum_{m=0}^k c^f_{k-m} g_m$,
\begin{gather*} \left| \sum_{m=0}^{K-1} f_{n-m} g_m - \sum_{k=0}^{K-1} d^{f,g}_k \G{n-k}{\alpha}{\beta} \right| = \left| \sum_{m=0}^{K-1} f_{n-m} g_m - \sum_{k=0}^{K-1} \sum_{m=0}^k c^f_{k-m} g_m \G{n-k}{\alpha}{\beta} \right| \\ = \left| \sum_{m=0}^{K-1} \left( f_{n-m} - \sum_{k=m}^{K-1} c^f_{k-m} \G{n-k}{\alpha}{\beta} \right) g_m \right| \\ \leq \sum_{m=0}^{K-1}\left| f_{n-m} - \sum_{k=0}^{K-m-1} c^f_{k} \G{n-m-k}{\alpha}{\beta} \right| \left| g_m \right| \\ \leq \rho^\alpha_{\beta,R}(f) \rho^\alpha_{\beta,R}(g) \G{n-K}{\alpha}{\beta} \sum_{m=0}^{K-1} \G{m}{\alpha}{\beta} \qquad \forall n \geq K.   \end{gather*}
Setting $C_R = \sum_{m=0}^{R-1} \G{m}{\alpha}{\beta}$ results in the statement.
\end{proof}

\begin{lmm}
\label{lmm:estimate_final_fg}
If $\alpha, \beta \in \R_{>0}$ and $R \in \N_0$, then there exists a constant $C_R \in \R$, such that for all $f,g \in \fring{x}{\alpha}{\beta}$ and $n,K \in \N_0$ with $K \leq R$ as well as $n \geq K$,
\begin{align} \label{eqn:estimate_final_fg} \left| \sum_{m=0}^{n} f_{n-m} g_m - \sum_{k=0}^{K-1} c^{f \cdot g}_k \G{n-k}{\alpha}{\beta} \right| &\leq C_R \rho^\alpha_{\beta,R}(f) \rho^\alpha_{\beta,R}(g) \G{n-K}{\alpha}{\beta}, \end{align}
where $c^{f \cdot g}_k := [x^k] \left( f(x) (\asyOp^\alpha_\beta g)(x) + g(x) (\asyOp^\alpha_\beta f)(x) \right)$.
\end{lmm}
\begin{proof}
Note that $c^{f \cdot g}_k = d^{f,g}_k + d^{g,f}_k$ with $d^{f,g}_k$ from Lemma \ref{lmm:estimate_partial_fg} and $d^{g,f}_k$ respectively with the roles of $f$ and $g$ switched.
We can use the triangle inequality to deduce that
\begin{gather*} \left| \sum_{m=0}^n f_{n-m} g_m - \sum_{k=0}^{K-1} c^{f \cdot g}_k \G{n-k}{\alpha}{\beta} \right| \leq \left| \sum_{m=0}^n f_{n-m} g_m - \sum_{m=0}^{K-1} f_{n-m} g_m - \sum_{m=0}^{K-1} f_{m} g_{n-m} \right| \\ + \left| \sum_{m=0}^{K-1} f_{n-m} g_m - \sum_{k=0}^{K-1} d^{f,g}_k \G{n-k}{\alpha}{\beta} \right| + \left| \sum_{m=0}^{K-1} f_{m} g_{n-m} - \sum_{k=0}^{K-1} d^{g,f}_k \G{n-k}{\alpha}{\beta} \right|. \end{gather*}
Using Lemmas \ref{lmm:estimate_full_fg} and \ref{lmm:estimate_partial_fg}  on the respective terms on the right hand side of this inequality results in the statement.
\end{proof}

\begin{proof}[Proof of Proposition \ref{prop:derivation}]
By Lemma \ref{lmm:wlogbetaderivation}, it is sufficient to prove Proposition \ref{prop:derivation} for $\beta > 0$. Therefore, we can apply Lemma \ref{lmm:estimate_final_fg} for $f,g \in \fring{x}{\alpha}{\beta}$. Eq.\ \eqref{eqn:estimate_final_fg} with $K=R$ directly implies that 
\begin{align*} [x^n] f(x) g(x) = \sum_{m=0}^n f_{n-m} g_m &= \sum_{k=0}^{R-1} c^{f \cdot g}_k \G{n-k}{\alpha}{\beta} + \bigO \left(\G{n-R}{\alpha}{\beta} \right) && \forall R \in \N_0, \end{align*}
with $c^{f \cdot g}_k = [x^k] \left( f(x) (\asyOp^\alpha_\beta g)(x) + g(x) (\asyOp^\alpha_\beta f)(x) \right)$.
By the Definition \ref{def:Fpowerseries}, it follows that $f \cdot g \in \fring{x}{\alpha}{\beta}$ and from Definition \ref{def:basic_asymp_definition} follows eq.\ \eqref{eqn:leipniz}. 
\end{proof}

\begin{proof}[Proof of Proposition \ref{prop:submultiplicative}]
If $f,g \in \fring{x}{\alpha}{\beta}$, then $f \cdot g \in \fring{x}{\alpha}{\beta}$ by Proposition \ref{prop:derivation}.
Because $\beta >0$, we have by Definition \ref{defn:basic_bigC_estimate}
\begin{align*} \rho^\alpha_{\beta,R}(f \cdot g) &= \max_{0\leq K\leq R} \sup_{n\geq K} \frac{\left| \sum_{m=0}^n f_{n-m} g_m - \sum _{k=0}^{K-1} c_{k}^{f\cdot g} \G{n-k}{\alpha}{\beta} \right|}{\G{n-K}{\alpha}{\beta}} && \forall f,g \in \fring{x}{\alpha}{\beta}, \end{align*}
which is bounded by $C_R \rho^\alpha_{\beta,R}(f) \rho^\alpha_{\beta,R}(g)$ with some fixed $C_R \in \R$ as follows directly from Lemma \ref{lmm:estimate_final_fg}.
\end{proof}
\section{Composition}
\label{sec:comp}

\subsection{Composition by analytic functions}
\begin{thm}
\label{thm:chain_analytic}
If $\alpha \in \R_{>0}$, $\beta \in \R$ and $f \in \R\{y_1,\ldots,y_L\}$ is a function in $L$ variables, which is analytic at the origin, then for all $g^{1}, \ldots, g^{L} \in x\fring{x}{\alpha}{\beta}$:
\begin{itemize}
\item
The composition $f\left(g^{1}(x), \ldots, g^{L}(x)\right)$ is in $\fring{x}{\alpha}{\beta}$.
\item $\asyOp^\alpha_\beta$ fulfills a multivariate chain rule for composition with analytic functions,
\begin{gather} \label{eqn:chain_analytic} \left(\asyOp^\alpha_\beta f\left(g^{1}, \ldots, g^{L}\right) \right)(x) = \sum_{l=1}^L \frac{\partial f}{\partial g^l} \left(g^{1}, \ldots, g^{L}\right) (\asyOp^\alpha_\beta g^{l})(x). \end{gather}
\end{itemize}
\end{thm}
In \cite{bender1975asymptotic} Edward Bender established this theorem for the case $L=1$ in a less `generatingfunctionology' based notation. 
If for example $g\in \fring{x}{\alpha}{\beta}$ and $f \in \R\{x,y\}$, then his Theorem 1 allows us to calculate the asymptotics of the power series $f(g(x),x)$. 
In fact, Bender analyzed more general power series including sequences with even more rapid than factorial growth. 

The following proof of Theorem \ref{thm:chain_analytic} is a straightforward generalization of Bender's Lemma 2 and Theorem 1 in \cite{bender1975asymptotic} to the multivariate case $f \in \R\{y_1,\ldots,y_L\}$.

Again, we will start by verifying that we may assume $\beta > 0$ during the proof of Theorem \ref{eqn:chain_analytic}.
\begin{lmm}
\label{lmm:wlogbetachainanalytic}
If Theorem \ref{thm:chain_analytic} holds for $\beta\in \R_{>0}$, then it also holds for all $\beta \in \R$.
\end{lmm}
\begin{proof}
For $\beta\in \R$, choose an $m \in \N_0$ such that $\beta+m > 0$.
If $g^{1}, \ldots, g^{L} \in x\fring{x}{\alpha}{\beta}$, then by Proposition \ref{prop:betashift}, $g^{1}, \ldots, g^{L} \in x\fring{x}{\alpha}{\beta+m}$, 
$(\asyOp^\alpha_{\beta+m} g^{l})(x) = x^m (\asyOp^\alpha_{\beta} g^{l})(x)$ and by the requirement $h(x) := f(g^{1}(x), \ldots, g^{L}(x)) \in \fring{x}{\alpha}{\beta+m}$ as well as 
\begin{gather*} (\asyOp^\alpha_{\beta+m} h)(x) = \sum_{l=1}^L \frac{\partial f}{\partial g^l} (g^{1}, \ldots, g^{L}) (\asyOp^\alpha_{\beta+m} g^{l})(x) = x^m \sum_{l=1}^L \frac{\partial f}{\partial g^l} (g^{1}, \ldots, g^{L}) (\asyOp^\alpha_{\beta} g^{l})(x). \end{gather*}
From Proposition \ref{prop:betashift} it follows that $h\in \fring{x}{\alpha}{\beta}$ as well as \\
$(\asyOp^\alpha_{\beta} h)(x) = \sum_{l=1}^L \frac{\partial f}{\partial g^l} (g^{1}, \ldots, g^{L}) (\asyOp^\alpha_{\beta} g^{l})(x)$.
\end{proof}
As before, we will use our freedom to assume that $\beta>0$ to establish an estimate on the coefficients of products of power series in $x\fring{x}{\alpha}{\beta}$.
\begin{lmm}
\label{lmm:power_prod_estimate}
If $\alpha,\beta \in \R_{>0}$ and $g^{1}, \ldots, g^{L} \in x \fring{x}{\alpha}{\beta}$, then there exists a constant $C \in \R$ such that 
\begin{align} \left| [x^n] \prod_{l=1}^L \left( g^l(x) \right)^{t_l} \right| &\leq C^{|\textbf{t}|} \G{n-|\textbf{t}|+1}{\alpha}{\beta} && \forall \textbf{t} \in \N_0^L, n \in \N_0 \text{ with } n \geq |\textbf{t}| \geq 1. \end{align}
\end{lmm}

\begin{proof}

By Proposition \ref{prop:betashiftlow}, it follows from $g^l \in x \fring{x}{\alpha}{\beta}$ that $\frac{g^l(x)}{x} \in \fring{x}{\alpha}{\beta+1}$ and therefore by Corollary \ref{crll:chain_for_product_pow} that $\prod_{l=1}^L \left(\frac{g^{l}(x)}{x}\right)^{t_l} \in \fring{x}{\alpha}{\beta+1}$ for all $\textbf{t} \in \N_0^L$. 
We can apply Corollary \ref{crll:smallc_bigC_estimate} with $R=K=0$ to obtain for all $n \geq |\textbf{t}|$,
\begin{gather*} \left| [x^n] \prod_{l=1}^L \left(g^{l}(x)\right)^{t_l} \right| = \left| [x^{n-|\textbf{t}|}] \prod_{l=1}^L \left(\frac{g^l(x)}{x}\right)^{t_l} \right| \leq \rho^\alpha_{\beta+1,0} \left( \prod_{l=1}^L \left(\frac{g^l(x)}{x}\right)^{t_l} \right)\G{n-|\textbf{t}|}{\alpha}{\beta+1}. \end{gather*}
The statement follows from Corollary \ref{crll:estimate_for_product_pow} and $\G{n-|\textbf{t}|}{\alpha}{\beta+1} = \G{n-|\textbf{t}|+1}{\alpha}{\beta}$.
\end{proof}

\begin{proof}[Proof of Theorem \ref{thm:chain_analytic}]
The composition of power series $f(g^{1}(x), \ldots, g^{L}(x))$ can be expressed as the sum
$\sum_{\substack{\textbf{t} \in \N_0^{L}}} f_{t_1,\ldots,t_L} \prod_{l=1}^L \left(g^{l}(x)\right)^{t_l}$, which can be split in preparation for the extraction of asymptotics:
\begin{align*} f(g^{1}(x), \ldots, g^{L}(x))&= \sum_{\substack{\textbf{t} \in \N_0^{L} \\ |\textbf{t}| \leq R}} f_{t_1,\ldots,t_L} \prod_{l=1}^L \left(g^{l}(x)\right)^{t_l} + \sum_{\substack{\textbf{t} \in \N_0^{L} \\ |\textbf{t}| > R}} f_{t_1,\ldots,t_L} \prod_{l=1}^L \left(g^{l}(x)\right)^{t_l} && \forall R\in \N_0 . \end{align*}
The first sum is just the composition by a polynomial. 
Corollary \ref{crll:polynomial_comp} asserts that this sum is in $\fring{x}{\alpha}{\beta}$. It has the asymptotic expansion given in eq.\ \eqref{eqn:polynomial_comp} which agrees with the right hand side of eq.\ \eqref{eqn:chain_analytic} up to order $R-1$, because the partial derivative reduces the order of a polynomial by one and $g^l_0=0$.

It is left to prove that the coefficients of the power series given by the remaining sum over $|\textbf{t}|>R$ are in $\bigO(\G{n-R}{\alpha}{\beta})$. 
Because of Lemma \ref{lmm:wlogbetachainanalytic}, we may assume that $\beta > 0$ without loss of generality and apply Lemma \ref{lmm:power_prod_estimate}. Together with the fact that there is a constant $C$, such that $|f_{t_1,\ldots,t_L}|\leq {C}^{|\textbf{t}|}$ for all $\textbf{t} \in \N_0^L$, due to the analyticity of $f$, Lemma \ref{lmm:power_prod_estimate} ensures that there is a constant $C'\in \R$ such that
\begin{gather*}   \left|[x^n]\sum_{\substack{\textbf{t} \in \N_0^{L} \\ |\textbf{t}| > R}} f_{t_1,\ldots,t_L} \prod_{l=1}^L \left(g^{l}(x)\right)^{t_l}\right| \leq \sum_{\substack{\textbf{t} \in \N_0^{L} \\ n \geq |\textbf{t}| > R}} \left| f_{t_1,\ldots,t_L} \right| \left|[x^n] \prod_{l=1}^L \left(g^{l}(x)\right)^{t_l} \right|     \\ \leq \sum_{t=R+1}^n {C'}^{t} \G{n-t+1}{\alpha}{\beta} \sum_{\substack{\textbf{t} \in \N_0^{L} \\ |\textbf{t}| = t }} 1      ,      \end{gather*}
for all $n \geq R+1$. Because the last sum $| \{ t_1,\ldots, t_L \in \N_0 | t_1 + \ldots + t_L = t \}| = { t + L -1 \choose L-1 }$ is a polynomial in $t$,
Corollary \ref{crll:exponential_inO} asserts that this is in $\bigO\left(\G{n-R}{\alpha}{\beta}\right)$. 
\end{proof}
\subsection{General composition of power series in \texorpdfstring{$\fring{x}{\alpha}{\beta}$}{R[[x]]^α_β}}
Despite the fact that Bender's theorem applies to a broader range of compositions $f \circ g$, where $f$ does not need to be analytic and $g$ does not need to be an element of $\fring{x}{\alpha}{\beta}$, it cannot be used in the case $f,g \in \fring{x}{\alpha}{\beta}$, where $f \notin \ker \asyOp^\alpha_\beta$. The problem is that we cannot truncate the sum $\sum_{k=0}^\infty f_k g(x)^k$ without losing significant information. In this section, this obstacle will be confronted and the general chain rule for the asymptotic derivative will be proven. 
Let $\Diff_{\id}(\R,0):= ( \{g \in \R[[x]]: g_0=0 \text{ and } g_1=1 \}, \circ)$ denote the group of \textit{formal diffeomorphisms tangent to the identity}. It is the group of all power series with $g_0=0$ and $g_1=1$ and with composition as group operation. The restriction of this group to elements in $\fring{x}{\alpha}{\beta}$ is of special interest to us.
\begin{thm}
\label{thm:chainrule}
For all $\alpha \in \R_{>0}$ and $\beta \in \R$, 
$\Diff_{\id}(\R,0)^\alpha_\beta:= ( \{g \in \fring{x}{\alpha}{\beta}: g_0=0 \text{ and } g_1=1 \}, \circ)$ is a subgroup of $\Diff_{\id}(\R,0)$. 

Moreover, if $f,g \in \fring{x}{\alpha}{\beta}$ with $g_0=0$ and $g_1=1$, then
\begin{itemize}
\item
The composition $f \circ g$ and the inverse $g^{-1}$ belong to $\fring{x}{\alpha}{\beta}$.
\item 
$\asyOp^\alpha_\beta$ fulfills a chain rule and there is a formula for the $\asyOp^\alpha_\beta$-derivative of the compositional inverse:
\begin{align} \label{eqn:asymp_chainrule_normal} (\asyOp^\alpha_\beta (f \circ g)) (x) = f'(g(x)) (\asyOp^\alpha_\beta g)(x) + \left(\frac{x}{g(x)} \right)^{\beta} e^{\frac{g(x) - x}{\alpha x g(x)}} (\asyOp^\alpha_\beta f ) (g(x)), \\ \label{eqn:inverse_asymp} (\asyOp^\alpha_\beta g^{-1})(x) = -{g^{-1}}'(x) \left(\frac{x}{g^{-1}(x)}\right)^\beta e^{\frac{g^{-1}(x) - x}{\alpha x g^{-1}(x)}} (\asyOp^\alpha_\beta g ) (g^{-1}(x)). \end{align}
\end{itemize}
\end{thm}

\begin{crll}
\label{crll:chainimplicit}
If $f \in \R[[x]]$, $g \in \fring{x}{\alpha}{\beta}$ with $g_0=0$ and $g_1=1$ as well as $f\circ g \in \fring{x}{\alpha}{\beta}$, then $f \in \fring{x}{\alpha}{\beta}$.
\end{crll}
\begin{proof}
Theorem \ref{thm:chainrule} guarantees that $g^{-1} \in \fring{x}{\alpha}{\beta}$ and therefore also $f = (f \circ g) \circ g^{-1} \in \fring{x}{\alpha}{\beta}$. 
\end{proof}

As before, we will assume that $\beta >0$ while proving Theorem \ref{thm:chainrule}. The following lemma establishes that we can do so.

\begin{lmm}
\label{lmm:wloggeneralchainrolebeta}
If Theorem \ref{thm:chainrule} holds for $\beta \in \R_{>0}$, then it holds for all $\beta\in \R$.
\end{lmm}
\begin{proof}
For $\beta \in \R$, choose $m\in \N_0$ such that $\beta+m >0$. 
If $f,g \in \fring{x}{\alpha}{\beta}$ with $g_0=0,g_1=1$, then $f,g \in \fring{x}{\alpha}{\beta+m}$ by Proposition \ref{prop:betashift}. Because of $(\asyOp^\alpha_{\beta+m} f ) (x) = x^m(\asyOp^\alpha_{\beta} f ) (x)$ and by the requirement
\begin{gather*} (\asyOp^\alpha_{\beta+m} (f \circ g)) (x) = f'(g(x)) (\asyOp^\alpha_{\beta+m} g)(x) + \left(\frac{x}{g(x)} \right)^{\beta+m} e^{\frac{g(x) - x}{\alpha x g(x)}} (\asyOp^\alpha_{\beta+m} f ) (g(x)) \\ = x^m \left( f'(g(x)) (\asyOp^\alpha_{\beta} g)(x) + \left(\frac{x}{g(x)} \right)^{\beta} e^{\frac{g(x) - x}{\alpha x g(x)}} (\asyOp^\alpha_{\beta} f ) (g(x)) \right). \end{gather*}
Applying Proposition \ref{prop:betashift} again results in $f \circ g \in \fring{x}{\alpha}{\beta}$ and eq. \eqref{eqn:asymp_chainrule_normal}. Eq. \eqref{eqn:inverse_asymp} and $g^{-1} \in \fring{x}{\alpha}{\beta}$ follow analogously. 
Therefore, $\Diff_{\id}(\R,0)^\alpha_\beta$ is a subgroup of $\Diff_{\id}(\R,0)^\alpha_{\beta+m}$.
\end{proof}

Theorem \ref{thm:chainrule} will be proven by ensuring that if $f,g \in \fring{x}{\alpha}{\beta}$, then $f \circ g^{-1} \in \fring{x}{\alpha}{\beta}$ and by constructing the asymptotic expansion of $f \circ g^{-1}$. For this, it turns out to be convenient to work in the rings $\fring{x}{\alpha}{\beta+1}$ and $\fring{x}{\alpha}{\beta+2}$ which contain $\fring{x}{\alpha}{\beta}$ as a subring. 
We will start with some observations on intermediate quantities in $\fring{x}{\alpha}{\beta+1}$ and $\fring{x}{\alpha}{\beta+2}$. 
The following three Lemmas are basic applications of the chain rule for the composition with analytic functions and the product rule, but we will prove them in detail to get acquainted to the new notions.

\begin{lmm}
\label{lmm:chainasympterm0}
If $g \in \fring{x}{\alpha}{\beta}$ with $g_0=0,g_1=1$ and $\gamma \in \R$, then $\left(\frac{g(x)}{x}\right)^{\gamma} \in \fring{x}{\alpha}{\beta+1}$ and
\begin{align} \left( \asyOp^\alpha_{\beta+1} \left(\frac{g(x)}{x}\right)^{\gamma} \right) = \gamma \left(\frac{g(x)}{x}\right)^{\gamma-1} \left(\asyOp^\alpha_{\beta}g\right)(x). \end{align}
\end{lmm}
\begin{proof}
Observe that $F(x) := (1-x)^\gamma \in \R\{x\}$ and $F'(x) = -\gamma (1-x)^{\gamma-1}$. Proposition \ref{prop:betashiftlow} implies that $\frac{g(x)}{x} \in \fring{x}{\alpha}{\beta+1}$, because $g\in x\fring{x}{\alpha}{\beta}$. As $g_1 = 1$, we additionally have $1- \frac{g(x)}{x} \in x\fring{x}{\alpha}{\beta+1}$. Using Theorem \ref{thm:chain_analytic} results in
\begin{gather*} \left(\frac{g(x)}{x}\right)^{\gamma} = F\left( 1- \frac{g(x)}{x} \right) \in \fring{x}{\alpha}{\beta+1}, \intertext{and by the chain rule for the composition with analytic functions from eq.\ \eqref{eqn:chain_analytic},} \left( \asyOp^\alpha_{\beta+1} \left(\frac{g(x)}{x}\right)^{\gamma} \right) = F'\left( 1- \frac{g(x)}{x} \right) \left(\asyOp^\alpha_{\beta+1} \left( 1- \frac{g(x)}{x}\right) \right)(x) \\ = -\gamma \left(\frac{g(x)}{x}\right)^{\gamma-1}\left( \asyOp^\alpha_{\beta+1}\left( -\frac{g(x)}{x}\right) \right)(x) = \gamma \left(\frac{g(x)}{x}\right)^{\gamma-1} \left(\asyOp^\alpha_{\beta}g\right)(x), \end{gather*}
where we used the linearity of $\asyOp^\alpha_{\beta+1}$ and $\left( \asyOp^\alpha_{\beta+1}\frac{g(x)}{x} \right)(x) = \left( \asyOp^\alpha_{\beta} g \right)(x)$ due to Proposition \ref{prop:betashiftlow}.
\end{proof}
\begin{lmm}
\label{lmm:chainasympinterterm1}
If $g \in \fring{x}{\alpha}{\beta}$ with $g_0=0,g_1=1$, then 
\begin{gather} A(x) := \frac{\frac{x}{g(x)}-1}{x} \in \fring{x}{\alpha}{\beta+2}\text{ as well as } e^{\frac{A(x)}{\alpha}} \in \fring{x}{\alpha}{\beta+2} \text{ and } \\ \label{eqn:chainasympinterterm1} \left( \asyOp^\alpha_{\beta+2} e^{\frac{A(x)}{\alpha}} \right)(x) = -\alpha^{-1} \left( \frac{x}{g(x)}\right)^2 e^{\frac{A(x)}{\alpha}} \left( \asyOp^\alpha_{\beta} g \right)(x). \end{gather}
\end{lmm}
\begin{proof}
From Lemma \ref{lmm:chainasympterm0} with $\gamma=-1$, it follows that $\frac{x}{g(x)} \in \fring{x}{\alpha}{\beta+1}$ and 
\begin{gather*} \left( \asyOp^\alpha_{\beta+1} \frac{x}{g(x)} \right) = - \left(\frac{x}{g(x)}\right)^{2} \left(\asyOp^\alpha_{\beta}g\right)(x). \end{gather*}
Because $g_1=1$, $\frac{x}{g(x)}-1 \in x\fring{x}{\alpha}{\beta+1}$ and by Proposition \ref{prop:betashiftlow}, $A(x) = \frac{\frac{x}{g(x)}-1}{x} \in \fring{x}{\alpha}{\beta+2}$ and 
\begin{gather} \label{eqn:derivativeb2A} \left( \asyOp^\alpha_{\beta+2} A \right) (x) = \left( \asyOp^\alpha_{\beta+1} \left( \frac{x}{g(x)}-1\right) \right)(x) = - \left(\frac{x}{g(x)}\right)^{2} \left(\asyOp^\alpha_{\beta}g\right)(x). \end{gather}
Observe that $\frac{A(x)-A(0)}{\alpha} \in x\fring{x}{\alpha}{\beta+2}$. Because $e^x \in \R\{x\}$, we can apply Theorem \ref{thm:chain_analytic} to conclude that $e^{\frac{A(x)-A(0)}{\alpha}} \in \fring{x}{\alpha}{\beta+2}$ and by linearity that also $e^{\frac{A(x)}{\alpha}} \in \fring{x}{\alpha}{\beta+2}$. Finally, we can use the chain rule for the composition with analytic functions to write the left hand side of eq.\ \eqref{eqn:chainasympinterterm1} as
\begin{gather*} e^{\frac{A(0)}{\alpha}} \left( \asyOp^\alpha_{\beta+2} e^{ \frac{A(x)-A(0)}{\alpha} } \right) (x) = e^{\frac{A(0)}{\alpha}} e^{ \frac{A(x)-A(0)}{\alpha} } \left( \asyOp^\alpha_{\beta+2} \frac{A(x)-A(0)}{\alpha} \right) (x) \\ = e^{\frac{A(x)}{\alpha} } \left( \asyOp^\alpha_{\beta+2} \frac{A(x)}{\alpha} \right) (x) = \alpha^{-1} e^{\frac{A(x)}{\alpha} } \left( \asyOp^\alpha_{\beta+2} A \right) (x). \end{gather*}
The statement follows after substitution of $\left( \asyOp^\alpha_{\beta+2} A \right) (x)$ from eq.\ \eqref{eqn:derivativeb2A}.
\end{proof}

\begin{lmm}
\label{lmm:chainasympinterterm2}
If $f, g \in \fring{x}{\alpha}{\beta}$ with $g_0=0,g_1=1$ and $\gamma \in \R$, then 
\begin{gather} \label{eqn:chainasympinterterm2infring} B_\gamma(x) := f(x) g'(x) \left(\frac{g(x)}{x}\right)^{\gamma} \in \fring{x}{\alpha}{\beta+2} \qquad \text{ and } \\ \label{eqn:chainasympinterterm2} \begin{gathered} \left( \asyOp^\alpha_{\beta+2} B_\gamma \right)(x) = \\ \left(\frac{g(x)}{x}\right)^{\gamma} \left( x^2 g'(x)\left( \asyOp^\alpha_{\beta}f \right)(x) + f(x) \left( \gamma x g'(x) \frac{x}{g(x)} + \alpha^{-1} - \beta x + x^2 \frac{\partial}{\partial x} \right)(\asyOp^\alpha_{\beta} g)(x) \right). \end{gathered} \end{gather}
\end{lmm}
\begin{proof}
Due to Proposition \ref{prop:betashift}, $f\in \fring{x}{\alpha}{\beta+2}$ and $(\asyOp^\alpha_{\beta+2} f)(x) = x^2 (\asyOp^\alpha_{\beta} f)(x)$. Proposition \ref{prop:ordinary_derivative}, which will be proven in the next section, guarantees that $g' \in \fring{x}{\alpha}{\beta+2}$ and $(\asyOp^\alpha_{\beta+2} g')(x) = \left(\alpha^{-1} - x \beta + x^2 \frac{\partial}{\partial x}\right)(\asyOp^\alpha_\beta g)(x)$. Because of Lemma \ref{lmm:chainasympterm0} and Proposition \ref{prop:betashift}, $\left(\frac{g(x)}{x}\right)^{\gamma}\in \fring{x}{\alpha}{\beta+2}$ as well as $\left(\asyOp^\alpha_{\beta+2} \left(\frac{g(x)}{x}\right)^{\gamma}\right)(x) = x \gamma \left(\frac{g(x)}{x}\right)^{\gamma-1} \left(\asyOp^\alpha_{\beta}g\right)(x)$. Putting all this together we can use Corollary \ref{crll:chain_for_product} with $g^1(x) = f(x)$, $g^2(x)= g'(x)$ and $g^3(x) = \left(\frac{g(x)}{x}\right)^{\gamma}$ to obtain eqs.\ \eqref{eqn:chainasympinterterm2infring} and \eqref{eqn:chainasympinterterm2}.
\end{proof}

\begin{lmm}
\label{lmm:ABrho}
If $\alpha,\beta \in \R_{>0}$ and $A,B_\gamma$ as above, then there exists a $C_R \in \R$ such that
\begin{align} \rho^\alpha_{\beta+2,R}\left( B_\gamma(x) A(x)^m \right) &\leq C_R^{m+1} && \forall m\in \N_0 \end{align}
\end{lmm}
\begin{proof}
Apply Corollary \ref{crll:chain_for_product_pow} with $g^1(x) = B_\gamma(x), g^2(x) =A(x), t_1 = 1$ and $t_2 =m$ to verify that $B_\gamma(x) A(x)^m \in \fring{x}{\alpha}{\beta+2}$ for all $m\in \N_0$. Apply Corollary \ref{crll:estimate_for_product_pow} with the same parameters.
\end{proof}
\begin{crll}
\label{crll:ABestimate}
If $\alpha,\beta \in \R_{>0}$ and $A,B_\gamma$ as above, then there exists a $C_R \in \R$ such that 
\begin{align*} \left| [x^{n}] B_\gamma(x) A(x)^m - \sum_{k=0}^{R-1} c_{k,m} \G{n-k}{\alpha}{\beta+2} \right| &\leq C_R^{m+1} \G{n-R}{\alpha}{\beta+2} && \forall n \geq R \text{ and } m\in \N_0 \end{align*}
where $c_{k,m} = [x^k] \left( \asyOp^\alpha_{\beta+2} B_\gamma(x) A(x)^m \right)(x)$.
\end{crll}
\begin{proof}
Additionally to Lemma \ref{lmm:ABrho}, apply Corollary \ref{crll:smallc_bigC_estimate} with $K=R$.
\end{proof}

The key to the extraction of the large $n$ asymptotics of $[x^n] (f \circ g^{-1})(x)$ is a variant of the Chu-Vandermonde identity. We will prove this identity using elementary power series techniques.

\begin{lmm}
\label{lmm:identity}
For all $a\in \R$ and $m,k\in \N_0$
\begin{align} \label{eqn:chuidentity} { a \choose m } &= \sum_{l=0}^m {k+l-1 \choose l} { a -k-l \choose m-l }. \end{align}
\end{lmm}
\begin{proof}
Recall that ${ a \choose n }= [x^n] (1+x)^a$ for all $a\in \R$ and $n\in \N_0$. 
By standard generating function arguments it follows that $[x^n] \frac{1}{(1-x)^k} = { k + n - 1 \choose n }$ for all $n,k\in \N_0$.
Observe that for all $a\in \R$ and $k\in \N_0$, we have the following identities in $\R[[x]]$:
\begin{gather*} (1+x)^{a} = (1+x)^k (1+x)^{a-k} = \frac{1}{\left( 1-\frac{x}{1+x}\right)^k}(1+x)^{a-k} \\ = \sum_{l=0}^\infty { k + l - 1 \choose l }\left(\frac{x}{1+x}\right)^l (1+x)^{a-k} = \sum_{l=0}^\infty { k + l - 1 \choose l } x^l (1+x)^{a-k-l}. \end{gather*}
Extracting coefficients from the first and the last expression results in the Chu-Vander\-monde-type identity in eq.\ \eqref{eqn:chuidentity}.
\end{proof}

\begin{crll}
\label{crll:identity_rdy}
For all $\alpha,\beta \in \R_{>0}$ and $n,R,k \in \N_0$ with $n\geq R\geq k$, we have the identity in $\R[x]$
\begin{gather} \begin{gathered} \sum_{m=0}^{n-R} x^m {n+\beta+1 \choose m} \G{n-m-k}{\alpha}{\beta+2} \\ = \sum \limits_{l = 0}^{n-R} { l+k-1 \choose l } \G{n-l-k}{\alpha}{\beta+2} x^{l} \sum_{m=0}^{n-R-l} \frac{\left(\frac{x}{\alpha}\right)^{m}}{m!}. \end{gathered} \end{gather}
\end{crll}
\begin{proof}
Observe that ${ a \choose n } = \frac{1}{n!} \frac{\Gamma(a+1)}{\Gamma(a-n+1)}$ for all $a\in \R$ and $n\in \N_0$ as long as $n< a+1$.
By writing the second binomial coefficient on the right hand side of eq.\ \eqref{eqn:chuidentity} in this form and setting $a= n+\beta+1$, we get for all 
$n,m,k \in \N_0$ with 
$m+k < n+\beta+2$
\begin{align*} { n+\beta+1 \choose m } \Gamma( n - m - k+\beta+2) = \sum \limits_{l = 0}^m { k+l-1 \choose l } \frac{\Gamma(n-k-l+\beta+2)}{(m-l)!}. \end{align*}
Multiplying by $x^m \alpha^{n-m-k+\beta+2}$, summing over $m$ and using $\G{n}{\alpha}{\beta} = \alpha^{n+\beta} \Gamma(n+\beta)$ gives,
\begin{align*} \sum_{m=0}^{n-R} x^m { n+\beta+1 \choose m } \G{ n - m - k }{\alpha}{\beta+2} = \sum_{m=0}^{n-R} x^m \sum \limits_{l = 0}^m { k+l-1 \choose l } \frac{\alpha^{l-m} \G{n-k-l}{\alpha}{\beta+2}}{(m-l)!}. \end{align*}
Note that $k \leq R$ and $m \leq n-R$ imply $m+k \leq n < n+ \beta+2$.
The statement follows after changing the order of summation on the right hand side and a shift of the summation variable $m\rightarrow m+l$.
\end{proof}

We are now equipped with the necessary tools to tackle the asymptotic analysis of $f \circ g^{-1}$. The first step is to express the coefficients of $(f \circ g^{-1})(x)$ in terms of the intermediate power series $A(x)$ and $B_\gamma(x)$. We will do so using a variant of the Lagrange inversion theorem.
\begin{lmm}
\label{lmm:lagrange}
If $p,q \in \R[[x]]$ with $q_0=0$ and $q_1 = 1$, then
\begin{align} [x^n] p\left( q^{-1}(x)\right) &= [x^n] p(x) q'(x) \left(\frac{x}{q(x)}\right)^{n+1} && \forall n\in \N_0. \end{align}
\end{lmm}
\begin{proof}
Note that the identity holds for $n=0$, because $q_0=0$ and $q_1 = 1$.
It follows from the Lagrange inversion theorem \cite[A.6]{flajolet2009analytic} for $n\geq 1$,
\begin{gather*} [x^n] p\left(q^{-1}(x)\right) = \frac{1}{n}[x^{n-1}] p'(x) \left(\frac{x}{q(x)}\right)^{n} \\ = \frac{1}{n}[x^{n-1}] \frac{\partial}{\partial x} \left( p(x) \left(\frac{x}{q(x)}\right)^{n} \right) - \frac{1}{n}[x^{n-1}] p(x) \left( \frac{\partial}{\partial x} \left(\frac{x}{q(x)}\right)^{n} \right).    \end{gather*}
Using $\frac{1}{n}[x^{n-1}] \frac{\partial}{\partial x} = [x^n]$ and evaluating the derivative in the second term result in the statement.
\end{proof}

\begin{crll}
\label{crll:chainrule_lagrange}
If $f, g,A$ and $B_\gamma$ as above, then
\begin{align} \label{eqn:diffgroupsumrep1} [x^n] f(g^{-1}(x)) &= \sum_{m=0}^n {n+\beta+1 \choose m} [x^{n-m}] B_\beta(x) A(x)^m && \forall n\in \N_0. \end{align}
\end{crll}

\begin{proof}
By Lemma \ref{lmm:lagrange},
\begin{gather*} [x^n] f(g^{-1}(x)) = [x^{n}] f(x)g'(x)\left(\frac{x}{g(x)}\right)^{n+1} = [x^{n}] f(x)g'(x)\left(\frac{g(x)}{x}\right)^{\beta} \left(\frac{x}{g(x)}\right)^{n+\beta+1}. \end{gather*}
Using the definitions of $A$ and $B_\gamma$ gives $[x^n] f(g^{-1}(x)) = [x^{n}] B_\beta(x) \left(1+x A(x)\right)^{n+\beta+1}$. Expanding with the generalized binomial theorem results in eq.\ \eqref{eqn:diffgroupsumrep1}.
\end{proof}

\begin{crll}
\label{crll:diffgroupsumsubstituted_1}
If $\alpha,\beta \in \R_{>0}$ and $f,g,A,B_\gamma$ as above, then
\begin{align} \label{eqn:diffgroupsumsubstituted_1} [x^n] f(g^{-1}(x)) &= \sum_{m=0}^{n-R} {n+\beta+1 \choose m} [x^{n-m}] B_\beta(x) A(x)^m + \bigO\left(\G{n-R}{\alpha}{\beta+2}\right)&& \forall R\in \N_0. \end{align}
\end{crll}
\begin{proof}
Eq.\ \eqref{eqn:diffgroupsumsubstituted_1} follows from eq.\ \eqref{eqn:diffgroupsumrep1} and 
\begin{gather*} \left| \sum_{m=n-R+1}^{n} {n+\beta+1 \choose m} [x^{n-m}] B_\beta(x) A(x)^{m} \right| = \left|\sum_{m=0}^{R-1} {n+\beta+1 \choose n-m} [x^{m}] B_\beta(x) A(x)^{n-m} \right| \\ \leq \sum_{m=0}^{R-1} {n+\beta+1 \choose n-m} C_0^{n-m+1} \G{m}{\alpha}{\beta+2} \in \bigO\left(\G{n-R}{\alpha}{\beta+2}\right) \qquad \forall R\in \N_0, \end{gather*}
where the second step follows from Corollary \ref{crll:ABestimate} with $R=0$ and the last inclusion holds, because ${n+\beta+1 \choose n-m} = \frac{\Gamma(n+\beta+2)}{\Gamma(n-m+1) \Gamma(\beta+m+2)} \sim \frac{n^{\beta+m+1}}{\Gamma(\beta+m+2)}$ by elementary properties of the $\Gamma$ function.
\end{proof}

\begin{lmm}
\label{lmm:diffgroupsumsubstituted_3}
If $\alpha,\beta \in \R_{>0}$ and $f,g,A,B_\gamma$ as above, then for all $R\in \N_0$
\begin{align} \label{eqn:diffgroupsumsubstituted_3} [x^n] f(g^{-1}(x)) &= \sum_{k=0}^{R-1}\sum_{l = 0}^{n-R} \sum_{m=0}^{n-R-l} c_{k,l,m} { l+k-1 \choose l } \G{n-l-k}{\alpha}{\beta+2} + \bigO\left(\G{n-R}{\alpha}{\beta+2}\right), \end{align}
where $c_{k,l,m} := [x^k] \left( \asyOp^\alpha_{\beta+2} B_\beta(x) A(x)^{l} \frac{\left(\frac{A(x)}{\alpha}\right)^{m}}{m!} \right)(x)$.
\end{lmm}

\begin{proof}
For all $n,m \in \N_0$ with $n-m \geq R$ set
\begin{align*} \mathcal{R}_{n,m}:=[x^{n-m}] B_\beta(x) A(x)^m - \sum_{k=0}^{R-1} c_{k,m} \G{n-m-k}{\alpha}{\beta+2}, \end{align*}
where $c_{k,m} = [x^k] \left( \asyOp^\alpha_{\beta+2} B_\beta(x) A(x)^m \right)(x)$.
By Corollary \ref{crll:ABestimate} with $n\rightarrow n-m$, we can find a constant $C_R \in \R$ such that $\left| \mathcal{R}_{n,m} \right| \leq C_R^{m+1} \G{n-m-R}{\alpha}{\beta+2}$ for all $n-m \geq R$.
Substituting $\mathcal{R}_{n,m}$ into eq.\ \eqref{eqn:diffgroupsumsubstituted_1} gives
\begin{gather} \begin{gathered} \label{eqn:diffgroupsumsubstituted_2} [x^n] f(g^{-1}(x)) = \sum_{m=0}^{n-R} {n+\beta+1 \choose m} \sum_{k=0}^{R-1} c_{k,m} \G{n-m-k}{\alpha}{\beta+2} \\ + \sum_{m=0}^{n-R} {n+\beta+1 \choose m} \mathcal{R}_{n,m} + \bigO\left(\G{n-R}{\alpha}{\beta+2}\right) \qquad \forall R\in \N_0, \end{gathered} \end{gather}
where 
\begin{align*} |\mathcal{R}_n| :=\left| \sum_{m=0}^{n-R} {n+\beta+1 \choose m} \mathcal{R}_{n,m} \right| &\leq \sum_{m=0}^{n-R} {n+\beta+1 \choose m} C_R^{m+1} \G{n-m-R}{\alpha}{\beta+2} &&\forall n \geq R. \end{align*}
Applying Corollary \ref{crll:identity_rdy} with $x\rightarrow C_R$ on the right hand side, results in
\begin{align*} |\mathcal{R}_n| &\leq C_R \sum \limits_{l = 0}^{n-R} { l+R-1 \choose l } \G{n-l-R}{\alpha}{\beta+2} C_R^{l} \sum_{m=0}^{n-R-l} \frac{\left(\frac{C_R}{\alpha}\right)^{m}}{m!} && \forall n\geq R \end{align*}
From $\sum_{m=0}^{n-R-l} \frac{\left(\frac{C_R}{\alpha}\right)^{m}}{m!} \leq e^{\frac{C_R}{\alpha}}$ and Corollary \ref{crll:exponential_inO} follows $|\mathcal{R}_{n}| \in \bigO\left(\G{n-R}{\alpha}{\beta+2}\right)$.

Therefore, for all $R\in \N_0$:
\begin{gather*} [x^n] f(g^{-1}(x)) = \sum_{m=0}^{n-R} {n+\beta+1 \choose m} \sum_{k=0}^{R-1} c_{k,m} \G{n-m-k}{\alpha}{\beta+2} + \bigO\left(\G{n-R}{\alpha}{\beta+2}\right) \\ = \sum_{k=0}^{R-1} [x^k] \left( \asyOp^\alpha_{\beta+2} B_\beta(x) \sum_{m=0}^{n-R} {n+\beta+1 \choose m} A(x)^m \G{n-m-k}{\alpha}{\beta+2} \right) + \bigO\left(\G{n-R}{\alpha}{\beta+2}\right) , \end{gather*}
where $\asyOp^\alpha_{\beta+2}$ acts on everything on its right. Applying Corollary \ref{crll:identity_rdy} with $x \rightarrow A(x)$ to the inner sum and reordering result in the statement.
\end{proof}
\begin{lmm}
\label{lmm:diffgroupsumsubstituted_5}
If $\alpha,\beta \in \R_{>0}$ and $f,g,A,B_\gamma$ as above, then 
\begin{align} \label{eqn:diffgroupsumsubstituted_5} [x^n] f(g^{-1}(x)) &= \sum_{k=0}^{R-1}\sum_{l = 0}^{R-1-k} c_{k,l}' { l+k-1 \choose l } \G{n-l-k}{\alpha}{\beta+2} + \bigO\left(\G{n-R}{\alpha}{\beta+2}\right) \end{align}
for all $R \in \N_0$, 
where $c_{k,l}' := [x^k] \left( \asyOp^\alpha_{\beta+2} B_\beta(x) A(x)^{l} e^{ \frac{A(x)}{\alpha} } \right)(x)$.
\end{lmm}

\begin{proof}
Set $c_{k,l,m}$ as in Lemma \ref{lmm:diffgroupsumsubstituted_3}.
From Lemma \ref{lmm:ABrho} it follows that there exists a $C_R \in \R$ such that 
$\rho^\alpha_{\beta+2,R}\left( B_\beta(x) A(x)^{l+m} \right) \leq C_R^{l+m+1}$ for all $l,m\in \N_0$. It follows from Corollary \ref{crll:smallc_bigC_estimate} that
\begin{align*} |c_{k,l,m}| &= \frac{\alpha^{-m}}{m!} \left| [x^k]\left( \asyOp^\alpha_{\beta+2} B_\beta(x) A(x)^{l+m} \right)(x) \right| \leq \frac{\alpha^{-m}}{m!} C_R^{l+m+1} \end{align*}
for all $k,l,m\in \N_0$ with $k \leq R$.
Therefore, for all $k\leq R$ and $n \geq 2R - k$,
\begin{gather*} \left| \sum_{l = R-k}^{n-R} \sum_{m=0}^{n-R-l} c_{k,l,m} { l+k-1 \choose l } \G{n-l-k}{\alpha}{\beta+2} \right| \\ \leq \sum_{l = R-k}^{n-R} \sum_{m=0}^{n-R-l} \frac{\alpha^{-m}C_R^{l+m+1}}{m!} { l+k-1 \choose l } \G{n-l-k}{\alpha}{\beta+2}    \end{gather*}
which is in $\bigO\left(\G{n-R}{\alpha}{\beta+2}\right)$, because $\sum_{m=0}^{n-R-l}\frac{\alpha^{-m}C_R^{m}}{m!} \leq e^{\frac{C_R}{\alpha}}$ and by Corollary \ref{crll:exponential_inO}. Applying this to truncate the summation over $l$ in eq.\ \eqref{eqn:diffgroupsumsubstituted_3} from Lemma \ref{lmm:diffgroupsumsubstituted_3} gives for all $R \in \N_0$
\begin{gather} \begin{gathered} \label{eqn:diffgroupsumsubstituted_4} [x^n] f(g^{-1}(x)) = \sum_{k=0}^{R-1}\sum_{l = 0}^{R-k-1} \sum_{m=0}^{n-R-l} c_{k,l,m} { l+k-1 \choose l } \G{n-l-k}{\alpha}{\beta+2} \\ + \bigO\left(\G{n-R}{\alpha}{\beta+2}\right). \end{gathered} \end{gather}
Note that ${ n + m \choose n} \geq 1 \Rightarrow (n+m)! \geq n! m!$ and therefore
\begin{gather*} \sum_{m=n}^\infty \frac{C^m}{m!} = \sum_{m=0}^\infty \frac{C^{n+m}}{(n+m)!} \leq \frac{C^n}{n!} \sum_{m=0}^\infty \frac{C^{m}}{m!} = e^C \frac{C^n}{n!}. \end{gather*}
It follows that for all $n\geq R-l+1$ and $k+l\leq R$
\begin{gather*} \left| \sum_{m=n-R-l+1}^{\infty} c_{k,l,m} \G{n-l-k}{\alpha}{\beta+2}\right| \leq C_R^{l+1} \sum_{m=n-R-l+1}^{\infty} \frac{\left(\frac{C_R}{\alpha}\right)^m}{m!} \G{n-l-k}{\alpha}{\beta+2} \\ \leq e^{\frac{C_R}{\alpha}} C_R^{l+1} \left(\frac{C_R}{\alpha}\right)^{n-l-R+1} \frac{\G{n-l-k}{\alpha}{\beta+2}}{(n-R-l+1)!}, \end{gather*}
which is in $\bigO\left(\G{n-R}{\alpha}{\beta+2}\right)$ as long as $k$ and $l$ are bounded, because  $\frac{\Gamma(n-l-k+\beta+2)}{\Gamma(n-R-l+2)} \sim n^{R-k+\beta}$. Applying this to complete the summation over $m$ in eq.\ \eqref{eqn:diffgroupsumsubstituted_4} and noting that $c_{k,l}' = \sum_{m=0}^\infty c_{k,l,m}$ results in eq.\ \eqref{eqn:diffgroupsumsubstituted_5}.
\end{proof}
\begin{crll}
\label{crll:chainrule_asymp_expansion_1}
If $\alpha,\beta \in \R_{>0}$ and $f,g,A,B_\gamma$ as above, then $f \circ g^{-1} \in \fring{x}{\alpha}{\beta+2}$ and 
\begin{align} \label{eqn:chainrule_asymp_expansion_1} [x^k] \left( \asyOp^\alpha_{\beta+2} f \circ g^{-1} \right)(x) &= [x^{k}] \left( \asyOp^\alpha_{\beta+2} B_{\beta-k+1}(x) e^{\frac{A(x)}{\alpha} }\right)(x) && \forall k\in \N_0. \end{align}
\end{crll}
\begin{proof}
After the change of summation variables $k \rightarrow k+l$, eq. \eqref{eqn:diffgroupsumsubstituted_5} becomes
\begin{gather*} [x^n] f(g^{-1}(x)) = \sum_{k=0}^{R-1}\sum_{l = 0}^{k} c_{k-l,l}' { l-1 \choose l } \G{n-k}{\alpha}{\beta+2} + \bigO\left(\G{n-R}{\alpha}{\beta+2}\right) \qquad\forall R \in \N_0. \end{gather*}
By Definition \ref{def:Fpowerseries}, this equation states that $f \circ g^{-1} \in \fring{x}{\alpha}{\beta+2}$ and that 
the coefficients of the asymptotic expansion are 
\begin{gather*} c^{f \circ g^{-1}}_k = \sum_{l = 0}^{k} c_{k-l,l}' { k-1 \choose l } = \sum_{l = 0}^{k} [x^{k-l}] \left( \asyOp^\alpha_{\beta+2} B_\beta(x) A(x)^{l} { k-1 \choose l } e^{\frac{A(x)}{\alpha}} \right)(x) \\ =[x^{k}] \sum_{l = 0}^{\infty} x^l \left( \asyOp^\alpha_{\beta+2} B_\beta(x) A(x)^{l} { k-1 \choose l } e^{\frac{A(x)}{\alpha}} \right)(x) \\ = [x^{k}] \left( \asyOp^\alpha_{\beta+2} B_\beta(x) \sum_{l=0}^\infty (x A(x))^{l} { k-1 \choose l } e^{\frac{A(x)}{\alpha}} \right)(x), \end{gather*}
where $x^l \left( \asyOp^\alpha_{\beta+2} f(x)\right)(x) = \left( \asyOp^\alpha_{\beta+2} x^l f(x)\right)(x)$ for all $f\in \fring{x}{\alpha}{\beta+2}$ was used, which follows from the product rule (Proposition \ref{prop:derivation}). Because of $\sum_{l=0}^\infty { k-1 \choose l } (x A(x))^{l} = (1+xA(x))^{k-1} = \left( \frac{x}{g(x)} \right)^{k-1}$ and the definition of $B_\gamma$ in Lemma \ref{lmm:chainasympinterterm2}, the statement follows.
\end{proof}

\begin{proof}[Proof of Theorem \ref{thm:chainrule}]
Because of Lemma \ref{lmm:wloggeneralchainrolebeta}, we may assume that $\beta \in \R_{>0}$ and start with the expression from Corollary \ref{crll:chainrule_asymp_expansion_1} for $[x^k] \left( \asyOp^\alpha_{\beta+2} f \circ g^{-1} \right)(x)$. We will use Lemmas \ref{lmm:chainasympinterterm1} and \ref{lmm:chainasympinterterm2} to expand this expression. By Corollary \ref{crll:chainrule_asymp_expansion_1} and the product rule (Proposition \ref{prop:derivation}), we have for all $k\in \N_0$
\begin{align*} [x^k] \left( \asyOp^\alpha_{\beta+2} f \circ g^{-1} \right)(x) &= [x^{k}] \left( e^{\frac{A(x)}{\alpha} }\left(\asyOp^\alpha_{\beta+2} B_{\beta-k+1} \right) (x) + B_{\beta-k+1}(x) \left( \asyOp^\alpha_{\beta+2} e^{\frac{A(x)}{\alpha} }\right)(x) \right). \end{align*}
For the first term Lemma \ref{lmm:chainasympinterterm2} gives
\begin{gather*} [x^{k}] e^{\frac{A(x)}{\alpha} }\left(\asyOp^\alpha_{\beta+2} B_{\beta-k+1} \right) (x) = [x^{k}] e^{\frac{A(x)}{\alpha} }\left(\frac{g(x)}{x}\right)^{\beta-k+1} \Bigg( x^2 g'(x) \left( \asyOp^\alpha_{\beta}f \right)(x) \\ + f(x) \left( x(\beta-k+1) g'(x) \frac{x}{g(x)} + \alpha^{-1} - \beta x + x^2 \frac{\partial}{\partial x} \right)(\asyOp^\alpha_{\beta} g)(x) \Bigg) \\ = [x^{k}] e^{\frac{A(x)}{\alpha} }\left(\frac{g(x)}{x}\right)^{\beta-k+1} \Bigg( x^2 g'(x) \left( \asyOp^\alpha_{\beta}f \right)(x) \\ +\left(- x^2 f'(x) + \alpha^{-1}f(x) g'(x) \left( \frac{x}{g(x)} \right)^2 \right) (\asyOp^\alpha_{\beta} g)(x) \Bigg), \end{gather*}
where the identity $[x^k] x p'(x) q(x) = k [x^k] p(x) q(x) - [x^k] x p(x) q'(x)$ for all $p,q \in \R[[x]]$ was used in the second step to eliminate the summand which contains the $\frac{\partial}{\partial x} (\asyOp^\alpha_{\beta} g)(x)$ factor.
By Lemma \ref{lmm:chainasympinterterm1}, the second term is 
\begin{gather*} [x^k] B_{\beta-k+1}(x) \left( \asyOp^\alpha_{\beta+2} e^{\frac{A(x)}{\alpha} }\right)(x) = -[x^k] \alpha^{-1} B_{\beta-k+1}(x) \left( \frac{x}{g(x)}\right)^2 e^{\frac{A(x)}{\alpha}} \left( \asyOp^\alpha_{\beta} g \right)(x) \\ = -[x^k]\alpha^{-1} f(x) g'(x) \left(\frac{g(x)}{x}\right)^{\beta-k+1} \left( \frac{x}{g(x)}\right)^2 e^{\frac{A(x)}{\alpha}} \left( \asyOp^\alpha_{\beta} g \right)(x), \end{gather*}
where the definition of $B_{\beta-k+1}(x)$ from Lemma \ref{lmm:chainasympinterterm2} was substituted. Summing both terms and substituting the definition of $A(x)$ from Lemma \ref{lmm:chainasympinterterm1} results in 
\begin{align*}                  [x^k] \left( \asyOp^\alpha_{\beta+2} f \circ g^{-1} \right)(x)&= [x^k]x^2 e^{\frac{\frac{x}{g(x)}-1}{\alpha x} } \left(\frac{g(x)}{x}\right)^{\beta-k+1} \left( g'(x)(\asyOp^\alpha_{\beta} f)(x) - f'(x) (\asyOp^\alpha_{\beta} g)(x) \right), \end{align*}
for all $k\in \N_0$.
By Proposition \ref{prop:betashift}, the $x^2$ prefactor shows that $f \circ g^{-1}$ is actually in the subspace $\fring{x}{\alpha}{\beta} \subset \fring{x}{\alpha}{\beta+2}$ and
\begin{align*} [x^k] \left( \asyOp^\alpha_{\beta} f \circ g^{-1} \right)(x)&= [x^k]e^{\frac{\frac{x}{g(x)}-1}{\alpha x} } \left(\frac{g(x)}{x}\right)^{\beta-k-1} \left( g'(x)(\asyOp^\alpha_{\beta} f)(x) - f'(x) (\asyOp^\alpha_{\beta} g)(x) \right). \end{align*}
If we set $p(x) := e^{\frac{\frac{x}{g(x)}-1}{\alpha x} }\left(\frac{g(x)}{x}\right)^{\beta} \left( (\asyOp^\alpha_{\beta} f)(x) - \frac{f'(x)}{g'(x)} (\asyOp^\alpha_{\beta} g)(x) \right)$ and $q(x) := g(x)$, we obtain
\begin{align*} [x^k] \left( \asyOp^\alpha_{\beta} f \circ g^{-1} \right)(x) &= [x^{k}] p(x) q'(x) \left(\frac{x}{q(x)}\right)^{k+1} = [x^k]p(q^{-1}(x)) && \forall k \in \N_0, \end{align*}
by Lemma \ref{lmm:lagrange}.
After substitution, we obtain the expression
\begin{align} \label{eqn:conjugate_asymp_equation} (\asyOp^\alpha_{\beta} f \circ g^{-1} )(x) &= e^{\frac{\frac{g^{-1}(x)}{x}-1}{\alpha g^{-1}(x)} } \left(\frac{x}{g^{-1}(x)}\right)^{\beta} \left( (\asyOp^\alpha_{\beta} f)(g^{-1}(x)) -\frac{f'(g^{-1}(x))}{g'(g^{-1}(x))} (\asyOp^\alpha_{\beta} g)(g^{-1}(x)) \right). \end{align}
The special case $f(x)=x$ with application of $g'(g^{-1}(x)) = \frac{1}{{g^{-1}}'(x)}$ results in eq.\ \eqref{eqn:inverse_asymp}. Solving eq.\ \eqref{eqn:inverse_asymp} for $(\asyOp^\alpha_{\beta} g)(g^{-1}(x))$ and substituting the result into eq.\ \eqref{eqn:conjugate_asymp_equation} gives eq.\ \eqref{eqn:asymp_chainrule_normal} with the substitution $g \rightarrow g^{-1}$.

As $f \circ g^{-1} \in \fring{x}{\alpha}{\beta}$ and $x\in \fring{x}{\alpha}{\beta}$, the subset $\Diff_{\id}(\R,0)^\alpha_\beta$ is a subgroup of $\Diff_{\id}(\R,0)$.
\end{proof}

\begin{rmk}
In their article \cite{Bender1984}, Bender and Richmond established that $[x^n](1+g(x))^{\gamma n + \delta} = n \gamma e^{\frac{\gamma g_1}{\alpha}} g_n + \bigO(g_n)$ if $g_n \sim \alpha n g_{n-1}$ and $g_0 = 0$. Using Lagrange inversion, the first coefficient in the expansion of the compositional inverse in eq.\ \eqref{eqn:inverse_asymp} can be obtained from this. In this respect, Theorem \ref{thm:chainrule} is a generalization of Bender and Richmond's result.

In the same article Bender and Richmond proved a theorem similar to Theorem \ref{thm:chainrule} for the class of power series $f$ which grow more rapidly than factorial such that $n f_{n-1} \in \smallO(f_n)$. Theorem \ref{thm:chainrule} establishes a link to the excluded case $n f_{n-1} = \bigO(f_n)$.
\end{rmk}

\begin{rmk}
The chain rule in eq.\ \eqref{eqn:asymp_chainrule_normal}
exposes a peculiar algebraic structure. It would be useful to have a combinatorial interpretation of the $e^{\frac{g(x) - x}{\alpha x g(x)}}$ term. 
\end{rmk}
\section{Some remarks on differential equations}
\label{sec:dgls}

Differential equations arising from physical systems form an active field of research in the scope of resurgence \cite{garoufalidis2012asymptotics,aniceto2011resurgence}. Unfortunately, the exact calculation of an overall factor of the asymptotic expansion of a solution of an ODE, called \textit{Stokes constant}, turns out to be difficult for many problems. This fact severely limits the utility of the method for enumeration problems, as the dominant factor of the asymptotic expansion is of most interest and the detailed structure of the asymptotic expansion is secondary.

In this section it will be sketched, for the sake of completeness, how the presented combinatorial framework fits into the realm of differential equations.
The given elementary properties each have their counterpart in resurgence's alien calculus \cite[II.6]{mitschi2016divergent}.

Theorem \ref{thm:chain_analytic} serves as a good starting point to analyze differential equations with power series solutions in $\fring{x}{\alpha}{\beta}$. 
Given an analytic function $F \in \R\{x, y_0, \ldots,y_L\}$, the $\asyOp^\alpha_\beta$-derivation can be applied to the ordinary differential equation
\begin{align*} 0 = F(x, f(x), f'(x), f''(x), \ldots, f^{(L)}(x)). \end{align*}
Applying the $\asyOp$-derivative naively to both sides of this equation and using the chain rule for the composition with analytic functions results in a linear equation for the asymptotic expansions of the derivatives $f^{(l)}$. However, we do not know yet, whether $f^{(l)} \in \fring{x}{\alpha}{\beta}$ or how the asymptotic expansions of the $f^{(l)}$ relate to each other. 
\begin{prop}
\label{prop:ordinary_derivative}
If $f\in \fring{x}{\alpha}{\beta}$, then $f'(x) \in \fring{x}{\alpha}{{\beta+2}}$ and 
\begin{align} \label{eqn:derivative_commute} (\asyOp^\alpha_{\beta+2} f')(x) =\left(\alpha^{-1} - x \beta + x^2 \frac{\partial}{\partial x}\right)(\asyOp^\alpha_\beta f)(x). \end{align}
\end{prop}
\begin{proof}
Recall that $f'(x) = \sum_{n=0}^\infty n f_n x^{n-1} = \sum_{n=0}^\infty (n+1)f_{n+1} x^n$. If $f\in \fring{x}{\alpha}{\beta}$, then by Definition \ref{def:Fpowerseries}, 
\begin{align*} (n+1)f_{n+1} &= \sum _{k=0}^{R-1} c^f_k (n+1) \G{n+1-k}{\alpha}{\beta} + (n+1)\bigO\left(\G{n+1-R}{\alpha}{\beta}\right) && \forall R\in \N_0. \end{align*}
Observe that because $x \Gamma(x) = \Gamma(x+1)$ and $\G{n}{\alpha}{\beta}=\alpha^{n+\beta}\Gamma(n+\beta)$,
\begin{gather*} (n+1) \G{n+1-k}{\alpha}{\beta} \\ = \alpha^{n+1-k+\beta} \left( (n+1-k+\beta) \Gamma(n+1-k+\beta) + (k-\beta) \Gamma(n+1-k+\beta) \right)    \\ = \alpha^{-1} \G{n-k}{\alpha}{\beta+2} + (k-\beta) \G{n-k-1}{\alpha}{\beta+2}. \end{gather*}
Therefore, for all $R\in \N_0$
\begin{gather*} (n+1)f_{n+1} = \sum _{k=0}^{R-1} c^f_k \left( \alpha^{-1} \G{n-k}{\alpha}{\beta+2} + (k-\beta) \G{n-k-1}{\alpha}{\beta+2}\right) + \bigO\left(\G{n-R}{\alpha}{\beta+2}\right), \end{gather*}
and it follows from Definition \ref{def:Fpowerseries} that $f'\in \fring{x}{\alpha}{{\beta+2}}$.
Moreover, by Definition \ref{def:basic_asymp_definition}, 
\begin{gather*} (\asyOp^\alpha_{\beta+2} f')(x) = \sum_{k=0}^\infty c^{f'}_k x^k = \sum_{k=0}^\infty c^f_k \left( \alpha^{-1} x^k + (k-\beta) x^{k+1}\right) \\ = \left(\alpha^{-1} - x \beta + x^2 \frac{\partial}{\partial x}\right) (\asyOp^\alpha_\beta f)(x). \end{gather*}
\end{proof}

\begin{crll}
\label{crll:dgl_in_F}
If $F \in \R\{x, y_0, \ldots,y_L\}$ and $f \in \fring{x}{\alpha}{\beta}$ is a solution of the differential equation
\begin{align} \label{eqn:initialdgl} 0 &= F\left(x, f(x), f'(x), f''(x), \ldots, f^{(L)}(x)\right), \intertext{then $(\asyOp^\alpha_\beta f)(x)$ is a solution of the linear differential equation} \label{eqn:conddgllinasymp} 0 &= \sum_{l=0}^L x^{2L-2l} \frac{\partial F}{\partial f^{(l)}} \left(x, f^{(0)},\ldots,f^{(L)}\right) \left(\alpha^{-1} - x\beta + x^2 \frac{\partial}{\partial x}\right)^l (\asyOp^\alpha_\beta f)(x). \end{align}
\end{crll}
\begin{proof}
From Proposition \ref{prop:ordinary_derivative} and $f \in \fring{x}{\alpha}{\beta}$, it follows that $f^{(l)} \in \fring{x}{\alpha}{\beta+2l}$ and 
\begin{gather*} \left(\asyOp^\alpha_{\beta+2l} f^{(l)}\right)(x) = \left(\alpha^{-1} - x \beta + x^2 \frac{\partial}{\partial x}\right) \left(\asyOp^\alpha_{\beta+2(l-1)} f^{(l-1)}\right)(x) \\ = \left(\alpha^{-1} - x \beta + x^2 \frac{\partial}{\partial x}\right)^l (\asyOp^\alpha_\beta f)(x). \end{gather*}
By Proposition \ref{prop:betashift}, $f^{(l)} \in \fring{x}{\alpha}{\beta+2L}$ for all $L\geq l$ as well as $x^{2(L-l)}\left(\asyOp^\alpha_{\beta+2l} f^{(l)}\right)(x) = \left(\asyOp^\alpha_{\beta+2L} f^{(l)}\right)(x)$. By Theorem \ref{thm:chain_analytic}, $F\left(x, f(x), f'(x), f''(x), \ldots, f^{(L)}(x)\right) \in \fring{x}{\alpha}{\beta+2L}$. Applying $\asyOp^\alpha_{\beta+2L}$ to both sides of eq.\ \eqref{eqn:initialdgl} gives
\begin{align*} 0 &= \sum_{l=0}^L \frac{\partial F}{\partial f^{(l)}} \left(x, f^{(0)},\ldots,f^{(L)}\right) \left(\asyOp^\alpha_{\beta+2L} f^{(l)}\right)(x). \end{align*}
Substitution of $\left(\asyOp^\alpha_{\beta+2L} f^{(l)}\right)(x) =x^{2(L-l)}\left(\alpha^{-1} - x \beta + x^2 \frac{\partial}{\partial x}\right)^l (\asyOp^\alpha_\beta f)(x)$ results in eq.\ \eqref{eqn:conddgllinasymp}.
\end{proof}

\begin{rmk}
Even if it is known that the solution to a differential equation has a well-behaved asymptotic expansion, Corollary \ref{crll:dgl_in_F} provides this asymptotic expansion only up to the initial values for the linear differential equation \eqref{eqn:conddgllinasymp}. Note that the form of the asymptotic expansion can still depend non-trivially on the initial values of the solution $f$ of the nonlinear differential equation.
\end{rmk}
\begin{rmk}
The linear differential equation \eqref{eqn:conddgllinasymp} only has a non-trivial solution if $\alpha^{-1}$ is the root of a certain polynomial. More specifically, making a Frobenius ansatz for $(\asyOp^\alpha_\beta f)(x)$ in eq.\ \eqref{eqn:conddgllinasymp} gives
\begin{gather*} 0 = [x^m]\sum_{l=0}^L x^{2L-2l} \alpha^{-l} \frac{\partial F}{\partial f^{(l)}} \left(x, f^{(0)},\ldots,f^{(L)}\right), \end{gather*}
where $m$ is the smallest integer such that the equation is not trivially fulfilled.
If this root is not real or if two roots have the same modulus, the present formalism has to be generalized to complex and multiple $\alpha$ to express the asymptotic expansion of a general solution. This generalization is straightforward. We merely need to generalize Definition \ref{def:Fpowerseries} of suitable sequences to: 
\begin{defn}
For given $\beta \in \R$ and $\alpha_1, \ldots, \alpha_L \in \C$ with $|\alpha_1| = |\alpha_2| = \ldots = |\alpha_L| =: \alpha > 0$ let $\cring{x}{\alpha_1, \ldots, \alpha_L}{\beta} \subset \C[[x]]$ be the subspace of complex power series, such that $f \in \cring{x}{\alpha_1, \ldots, \alpha_L}{\beta} $  if and only if there exist sequences of complex numbers $(c_{k,l}^f)_{k\in \N_0, l \in [1,L]}$, which fulfill
\begin{align} f_n &= \sum _{k=0}^{R-1} \sum_{l=1}^L c_{k,l}^f \G{n-k}{{\alpha_l}}{\beta} + \bigO\left(\G{n-k}{{\alpha}}{\beta}\right) && \forall R \in \N_0. \end{align}
\end{defn}
\end{rmk}
\section{Applications}
\label{sec:applications}

\subsection{Connected chord diagrams}
\begin{figure}%
\begin{subfigure}[b]{0.5\textwidth}%
\centering
\begin{tikzpicture}[scale=0.6] \draw[draw=none,fill=red!20!white] (4.700000,-0.150000) rectangle (6.300000,0.750000); \draw[draw=none,fill=red!20!white] (0.600000,-0.250000) rectangle (4.400000,1.750000); \draw (0,0) arc (0:-180:-3.500000); \draw[color=red!20!white,line width=3pt] (1,0) arc (0:-180:-1.000000); \draw (1,0) arc (0:-180:-1.000000); \draw[color=red!20!white,line width=3pt] (2,0) arc (0:-180:-1.000000); \draw (2,0) arc (0:-180:-1.000000); \draw[color=red!20!white,line width=3pt] (5,0) arc (0:-180:-0.500000); \draw (5,0) arc (0:-180:-0.500000); \node at (0, -.75){$1$}; \draw[fill] (0, 0) circle (1pt); \node at (1, -.75){$2$}; \draw[fill] (1, 0) circle (1pt); \node at (2, -.75){$3$}; \draw[fill] (2, 0) circle (1pt); \node at (3, -.75){$4$}; \draw[fill] (3, 0) circle (1pt); \node at (4, -.75){$5$}; \draw[fill] (4, 0) circle (1pt); \node at (5, -.75){$6$}; \draw[fill] (5, 0) circle (1pt); \node at (6, -.75){$7$}; \draw[fill] (6, 0) circle (1pt); \node at (7, -.75){$8$}; \draw[fill] (7, 0) circle (1pt); \draw (-0.500000,0)--(7.500000,0); \end{tikzpicture}%
\subcaption{disconnected chord diagram}%
\end{subfigure}%
\begin{subfigure}[b]{0.5\textwidth}%
\centering
\begin{tikzpicture}[scale=0.6] \draw[color=white,line width=3pt] (0,0) arc (0:-180:-2.500000); \draw (0,0) arc (0:-180:-2.500000); \draw[color=white,line width=3pt] (2,0) arc (0:-180:-1.000000); \draw (2,0) arc (0:-180:-1.000000); \draw[color=white,line width=3pt] (1,0) arc (0:-180:-3.000000); \draw (1,0) arc (0:-180:-3.000000); \draw[color=white,line width=3pt] (3,0) arc (0:-180:-1.500000); \draw (3,0) arc (0:-180:-1.500000); \node at (0, -.75){$1$}; \draw[fill] (0, 0) circle (1pt); \node at (1, -.75){$2$}; \draw[fill] (1, 0) circle (1pt); \node at (2, -.75){$3$}; \draw[fill] (2, 0) circle (1pt); \node at (3, -.75){$4$}; \draw[fill] (3, 0) circle (1pt); \node at (4, -.75){$5$}; \draw[fill] (4, 0) circle (1pt); \node at (5, -.75){$6$}; \draw[fill] (5, 0) circle (1pt); \node at (6, -.75){$7$}; \draw[fill] (6, 0) circle (1pt); \node at (7, -.75){$8$}; \draw[fill] (7, 0) circle (1pt); \draw (-0.500000,0)--(7.500000,0); \end{tikzpicture}%
\subcaption{connected chord diagram}%
\end{subfigure}
\caption{Illustrations of connected and disconnected chord diagrams. The rectangles indicate the connected components of the disconnected diagram.}
\end{figure}
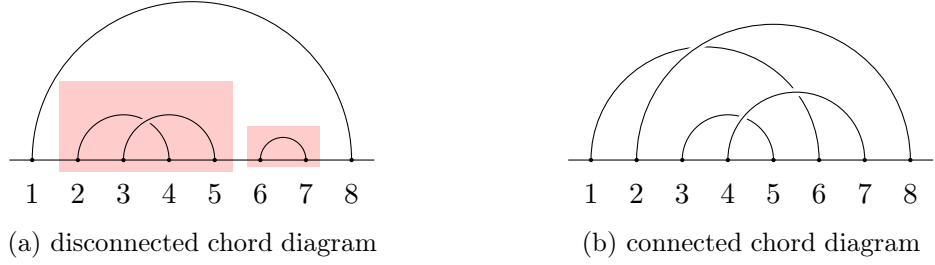

A chord diagram with $n$ chords is a circle with $2n$ points, which are labeled by integers $1,\ldots,2n$ and connected in disjoint pairs by $n$ chords. There are $(2n-1)!!$ of such diagrams. 

A chord diagram is \textit{connected} if no set of chords can be separated from the remaining chords by a line which does not cross any chords. Let $I(x) = \sum_{n=0}(2n-1)!!x^n$, the ordinary generating function of \textit{all} chord diagrams, and $C(x) = \sum_{n=0} C_n x^n$, where $C_n$ is the number of \textit{connected} chord diagrams with $n$ chords. Following \cite{Flajolet2000}, the power series $I(x)$ and $C(x)$ are related by,
\begin{align} \label{eqn:cntd_chords} I(x) &= 1+C\left(xI(x)^2\right). \end{align}
This functional equation can be solved for the coefficients of $C(x)$ by basic iterative methods. The first coefficients are 
\begin{align} C(x) = x + x^2 + 4x^3 + 27x^4 + 248 x^5 +\ldots    \end{align}
This sequence is entry \texttt{A000699} in Neil Sloane's integer sequence on-line encyclopedia \cite{oeis}.

Because $(2n-1)!!=\frac{2^{n+\frac12}}{\sqrt{2 \pi}}\Gamma(n+\frac12)=\frac{1}{\sqrt{2\pi}}\G{n}{2}{\frac12}$, the power series $I$ is in $\fring{x}{2}{{\frac12}}$ and $\left(\asyOp^2_{\frac12} I\right)(x) = \frac{1}{\sqrt{2 \pi}}$ as a direct consequence of the Definitions \ref{def:Fpowerseries} and \ref{def:basic_asymp_definition}. From eq.\ \eqref{eqn:cntd_chords}, it also follows that $C(xI(x)^2) \in \fring{x}{2}{{\frac12}}$. Because $x I(x)^2 \in \fring{x}{2}{{\frac12}}$ by the product rule (Proposition \ref{prop:derivation}), Corollary \ref{crll:chainimplicit} with $f(x) = C(x)$ and $g(x) = x I(x)^2$ implies that $C \in \fring{x}{2}{{\frac12}}$.

An application of the general chain rule from Theorem \ref{thm:chainrule} on the functional eq. \eqref{eqn:cntd_chords} results in
\begin{gather} \begin{gathered}  \left(\asyOp^2_{\frac12} I\right)(x) = \left( \asyOp^2_{\frac12} \left(1+C\left(xI(x)^2\right)\right) \right)(x) = \left( \asyOp^2_{\frac12} C\left(xI(x)^2\right) \right)(x) \\= 2 x I(x) C'\left(xI(x)^2\right) (\asyOp^2_{\frac12} I)(x) + \left(\frac{x}{xI(x)^2} \right)^\frac12 e^{\frac{xI(x)^2-x}{2 x^2 I(x)^2}} \left(\asyOp^2_{\frac12} C\right)\left(xI(x)^2\right). \end{gathered} \end{gather}
which can be solved for $\left(\asyOp^2_{\frac12} C\right)\left(xI(x)^2\right)$,
\begin{align*} \left(\asyOp^2_{\frac12} C\right)\left(xI(x)^2\right) &= \frac{I(x)-x I(x)^2 C'\left(xI(x)^2\right)}{\sqrt{2\pi}} e^{\frac{1-I(x)^2}{2 xI(x)^2}}, \end{align*}
where $\left(\asyOp^2_{\frac12} I\right)(x) = \frac{1}{\sqrt{2 \pi}}$ was used. This can be composed with the unique $y \in \R[[x]]$ which solves $y(x) I(y(x))^2 = x$,
\begin{align*} \left(\asyOp^2_{\frac12} C\right)(x) &= \frac{I(y(x))-x C'(x)}{\sqrt{2\pi}} e^{\frac{1-I(y(x))^2}{2 x}}. \end{align*}
From eq.\ \eqref{eqn:cntd_chords}, it follows that $I(y(x)) = 1+C(x)$, therefore 
\begin{align}      \left(\asyOp^2_{\frac12} C\right)(x)&= \frac{1+C(x)-2x C'(x)}{\sqrt{2\pi}}e^{-\frac{1}{2 x}( 2C(x) + C(x)^2 )}. \end{align}
Utilizing the linear differential equation  $2x^2 I'(x) + xI(x)+1 = I(x)$, from which the differential equation $C'(x) = \frac{C(x)(1+C(x))-x}{2x C(x)}$ \cite{Flajolet2000} can be deduced, results in a further simplification:
\begin{align} \label{eqn:asyC} \left(\asyOp^2_{\frac12} C\right)(x)&= \frac{1}{\sqrt{2\pi}}\frac{x}{C(x)}e^{-\frac{1}{2 x}(2C(x) + C(x)^2)}. \end{align}
This is the generating function of the full asymptotic expansion of $C_n$. 
The first coefficients are,
\begin{align} \left(\asyOp^2_{\frac12} C\right)(x) = \frac{e^{-1}}{\sqrt{2 \pi}} \left( 1 - \frac{5}{2} x - \frac{43}{8} x^2 - \frac{579}{16} x^3 - \frac{44477}{128} x^4 - \frac{5326191}{1280} x^5 + \ldots \right).   \end{align}
By Definitions \ref{def:Fpowerseries} and \ref{def:basic_asymp_definition} as well as $\frac{1}{\sqrt{2\pi}}\G{n}{2}{\frac12} = (2n-1)!!$, we get the following two equivalent expressions for the asymptotic expansion of the numbers $C_n$:
\begin{align*} C_n &=\sum_{k \geq 0}^{R-1} \G{n -k}{2}{\frac12} [x^k] \left(\asyOp^2_{\frac12} C\right)(x) + \bigO\left(\G{n -R}{2}{\frac12}\right) && \forall R \in \N_0 \\ C_n&= \sqrt{2 \pi} \sum_{k \geq 0}^{R-1} (2(n-k)-1)!! [x^k] \left(\asyOp^2_{\frac12} C\right)(x) + \bigO\left(\left(2(n-R)-1\right)!!\right)&& \forall R \in \N_0. \end{align*}
The first terms of this large $n$ expansion are
\begin{align*} C_n&= e^{-1} \left( (2n-1)!! - \frac{5}{2} (2n-3)!! - \frac{43}{8} (2n-5)!!    - \frac{579}{16} (2n-7)!!   + \ldots \right),       \end{align*}
The first term, $e^{-1}$, in this expansion has been computed by Kleitman \cite{kleitman1970proportions}, Stein and Everett \cite{stein1978class} and Bender and Richmond \cite{Bender1984} each using different methods. With the presented method an arbitrary number of coefficients can be computed.
Some additional coefficients are given in Table \ref{tab:coefficientsCM}.

\begin{table}
\centering
\tiny{
\def\arraystretch{1.5}
\begin{tabular}{|c||c||c|c|c|c|c|c|c|c|}
\hline
sequence&$0$&$1$&$2$&$3$&$4$&$5$&$6$&$7$&$8$\\
\hline\hline
$e\sqrt{2\pi}(\asyOp^2_{\frac12} C)$&$1$&$- \frac{5}{2}$&$- \frac{43}{8}$&$- \frac{579}{16}$&$- \frac{44477}{128}$&$- \frac{5326191}{1280}$&$- \frac{180306541}{3072}$&$- \frac{203331297947}{215040}$&$- \frac{58726239094693}{3440640}$\\
\hline
$e\sqrt{2\pi}(\asyOp^2_{\frac12} M)$&$1$&$-4$&$-6$&$- \frac{154}{3}$&$- \frac{1610}{3}$&$- \frac{34588}{5}$&$- \frac{4666292}{45}$&$- \frac{553625626}{315}$&$- \frac{1158735422}{35}$\\
\hline
\end{tabular}
}
\caption{First coefficients of the asymptotic expansions of $C_n$ and $M_n$.}
\label{tab:coefficientsCM}
\end{table}

The probability of a random chord diagram with $n$ chords to be connected is therefore $e^{-1}(1- \frac{5}{4n}) + \bigO(\frac{1}{n^2})$.

This result has been used by Courtiel, Yeats and Zeilberger to calculate the asymptotics of \textit{terminal chord diagrams} \cite{courtiel2016connected}. These terminal chord diagrams can be used to formulate a solution for Dyson-Schwinger equations in quantum field theory \cite{marie2013chord}.

\subsection{Monolithic chord diagrams}
A chord diagram is called monolithic if it consists only of a connected component and of isolated chords which do not `contain' each other \cite{Flajolet2000}. That means with $(a,b)$ and $(c,d)$ the labels of two chords, it is not allowed that $a < c < d< b$ and $c < a < b < d$. Let $M(x) = \sum_{n=0} M_n x^n$ be the generating function of monolithic chord diagrams. Following \cite{Flajolet2000}, $M(x)$ fulfills 
\begin{align} M(x) = C\left( \frac{x}{(1-x)^2} \right). \end{align}
Clearly, Theorem \ref{thm:chainrule} implies that $M \in \fring{x}{2}{{\frac12}}$, because $C \in \fring{x}{2}{{\frac12}}$ and $\frac{x}{(1-x)^2} \in \R\{x\} \subset \fring{x}{2}{{\frac12}}$.
Using the $\asyOp^2_{\frac12}$-derivative on both sides of this equation together with the result for $\left(\asyOp^2_{\frac12} C\right)(x)$ in eq.\ \eqref{eqn:asyC} gives
\begin{align} \label{eqn:asyM} \begin{split} \left(\asyOp^2_{\frac12} M\right)(x) &= \frac{1}{\sqrt{2\pi}}\frac{1}{(1-x)} \frac{x}{M(x)}e^{1 - \frac{x}{2}-\frac{(1-x)^2}{2x} (2M(x) + M(x)^2)} \\ &= \frac{1}{\sqrt{2\pi}}\left(1 - 4 x -6 x^2 - \frac{154}{3} x^3 - \frac{1610}{3} x^4 - \frac{34588}{5} x^5 + \ldots \right). \end{split}  \end{align}
Some additional coefficients are given in Table \ref{tab:coefficientsCM}.
The probability of a random chord diagram with $n$ chords to be non-monolithic is therefore $1-\left(1-\frac{4}{2n-1} + \bigO(\frac{1}{n^2})\right) = \frac{2}{n} + \bigO(\frac{1}{n^2})$.

\subsection{Simple permutations}

\begin{figure}%
\begin{subfigure}[b]{0.5\textwidth}%
\centering
\begin{tikzpicture}[scale=0.6] \draw[color=red!20!white,fill] (1.500000,-0.500000) rectangle (6.500000,4.500000); \draw[color=red!40!white,fill] (1.500000,-0.500000) rectangle (5.500000,3.500000); \node at (-1, 0){$1$}; \node at (0, -1){$1$}; \node at (-1, 1){$2$}; \node at (1, -1){$2$}; \node at (-1, 2){$3$}; \node at (2, -1){$3$}; \node at (-1, 3){$4$}; \node at (3, -1){$4$}; \node at (-1, 4){$5$}; \node at (4, -1){$5$}; \node at (-1, 5){$6$}; \node at (5, -1){$6$}; \node at (-1, 6){$7$}; \node at (6, -1){$7$}; \node at (-1, 7){$8$}; \node at (7, -1){$8$}; \draw[dashed] (-0.500000,-1.5)--(-0.500000,7.500000); \draw[dashed] (-1.5,-0.500000)--(7.500000,-0.500000); \draw[dashed] (0.500000,-1.5)--(0.500000,7.500000); \draw[dashed] (-1.5,0.500000)--(7.500000,0.500000); \draw[dashed] (1.500000,-1.5)--(1.500000,7.500000); \draw[dashed] (-1.5,1.500000)--(7.500000,1.500000); \draw[dashed] (2.500000,-1.5)--(2.500000,7.500000); \draw[dashed] (-1.5,2.500000)--(7.500000,2.500000); \draw[dashed] (3.500000,-1.5)--(3.500000,7.500000); \draw[dashed] (-1.5,3.500000)--(7.500000,3.500000); \draw[dashed] (4.500000,-1.5)--(4.500000,7.500000); \draw[dashed] (-1.5,4.500000)--(7.500000,4.500000); \draw[dashed] (5.500000,-1.5)--(5.500000,7.500000); \draw[dashed] (-1.5,5.500000)--(7.500000,5.500000); \draw[dashed] (6.500000,-1.5)--(6.500000,7.500000); \draw[dashed] (-1.5,6.500000)--(7.500000,6.500000); \node[circle,fill] at (0,5){}; \node[circle,fill] at (1,7){}; \node[circle,fill] at (2,2){}; \node[circle,fill] at (3,0){}; \node[circle,fill] at (4,3){}; \node[circle,fill] at (5,1){}; \node[circle,fill] at (6,4){}; \node[circle,fill] at (7,6){}; \end{tikzpicture}%
\subcaption{non-simple permutation}%
\end{subfigure}%
\begin{subfigure}[b]{0.5\textwidth}%
\centering
\begin{tikzpicture}[scale=0.6] \node at (-1, 0){$1$}; \node at (0, -1){$1$}; \node at (-1, 1){$2$}; \node at (1, -1){$2$}; \node at (-1, 2){$3$}; \node at (2, -1){$3$}; \node at (-1, 3){$4$}; \node at (3, -1){$4$}; \node at (-1, 4){$5$}; \node at (4, -1){$5$}; \node at (-1, 5){$6$}; \node at (5, -1){$6$}; \node at (-1, 6){$7$}; \node at (6, -1){$7$}; \node at (-1, 7){$8$}; \node at (7, -1){$8$}; \draw[dashed] (-0.500000,-1.5)--(-0.500000,7.500000); \draw[dashed] (-1.5,-0.500000)--(7.500000,-0.500000); \draw[dashed] (0.500000,-1.5)--(0.500000,7.500000); \draw[dashed] (-1.5,0.500000)--(7.500000,0.500000); \draw[dashed] (1.500000,-1.5)--(1.500000,7.500000); \draw[dashed] (-1.5,1.500000)--(7.500000,1.500000); \draw[dashed] (2.500000,-1.5)--(2.500000,7.500000); \draw[dashed] (-1.5,2.500000)--(7.500000,2.500000); \draw[dashed] (3.500000,-1.5)--(3.500000,7.500000); \draw[dashed] (-1.5,3.500000)--(7.500000,3.500000); \draw[dashed] (4.500000,-1.5)--(4.500000,7.500000); \draw[dashed] (-1.5,4.500000)--(7.500000,4.500000); \draw[dashed] (5.500000,-1.5)--(5.500000,7.500000); \draw[dashed] (-1.5,5.500000)--(7.500000,5.500000); \draw[dashed] (6.500000,-1.5)--(6.500000,7.500000); \draw[dashed] (-1.5,6.500000)--(7.500000,6.500000); \node[circle,fill] at (0,6){}; \node[circle,fill] at (1,3){}; \node[circle,fill] at (2,1){}; \node[circle,fill] at (3,7){}; \node[circle,fill] at (4,2){}; \node[circle,fill] at (5,5){}; \node[circle,fill] at (6,0){}; \node[circle,fill] at (7,4){}; \end{tikzpicture}%
\subcaption{simple permutation}%
\end{subfigure}%
\caption{Illustrations of simple and non-simple permutations.  The (non-trivial) intervals that map to intervals are indicated by squares.}
\end{figure}
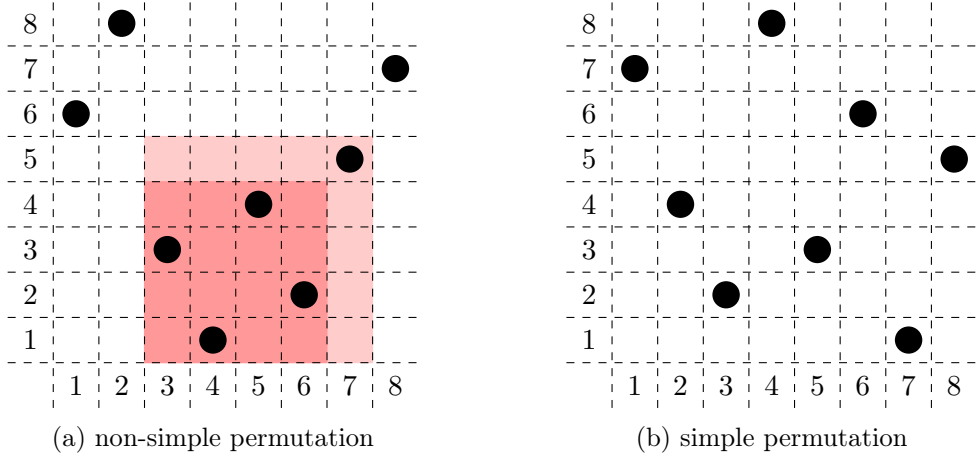

A permutation is called simple if it does not map a non-trivial interval to another interval. Expressed formally, the permutation $\pi \in S_n^\text{simple} \subset S_n$ if and only if $\pi([i,j]) \neq [k,l]$ for all $i,j,k,l \in [1,n]$ with $2\leq |[i,j]| \leq n-1$. See Albert, Atkinson and Klazar \cite{albert2003enumeration} for a detailed exposition of simple permutations.
Set $S(x) = \sum_{n=4}^\infty |S_n^\text{simple}| x^n$, the generating function of \textit{simple} permutations, and $F(x) = \sum_{n=1}^\infty n! x^n$, the generating function of \textit{all} permutations. Following \cite{albert2003enumeration}, $S(x)$ and $F(x)$ are related by the equation
\begin{align} \label{eqn:func_S} \frac{F(x) - F(x)^2}{1+F(x)} = x + S(F(x)), \end{align}
which can be solved iteratively for the coefficients of $S(x)$:
\begin{align} S(x) &= 2 x^4 + 6 x^5 + 46 x^6 + 338 x^7 + 2926 x^8 + \ldots \end{align}
This sequence is entry \texttt{A111111} of the OEIS \cite{oeis} with the slightly different convention, $\texttt{A111111}=x+2x^2+S(x)$.

As $n! = \G{n}{1}{1}$, $F(x) \in \fring{x}{1}{1}$ and $\left(\asyOp^1_1 F\right) = 1$ by the Definitions \ref{def:Fpowerseries} and \ref{def:basic_asymp_definition}. Therefore, the full asymptotic expansion of $S(x)$ can be obtained by applying the general chain rule to both sides of eq.\ \eqref{eqn:func_S}. Alternatively, eq.\ \eqref{eqn:func_S} implies $\frac{x-x^2}{1+x} = F^{-1}(x) + S(x)$ with $F^{-1}(F(x)) = x$. 
By Theorem \ref{thm:chainrule}, it follows from $F \in \fring{x}{1}{1}$ and $F_0 =0$ as well as $F_1 = 1$ that $F^{-1} \in \fring{x}{1}{1}$ and by linearity and $\frac{x-x^2}{1+x} \in \R\{x\} \subset \fring{x}{1}{1}$, we also have $S \in \fring{x}{1}{1}$.
The expression for the asymptotic expansion of $F^{-1}(x)$ in terms of $\left(\asyOp^1_1 F\right)(x)$ from eq.\ \eqref{eqn:inverse_asymp} gives
\begin{align} \left(\asyOp^1_1 S\right)(x) &= \left(\asyOp^1_1 \frac{x-x^2}{1+x}\right)(x)-\left(\asyOp^1_1 F^{-1}\right)(x) = {F^{-1}}'(x) \frac{x}{F^{-1}(x)} e^{\frac{F^{-1}(x)-x}{x F^{-1}(x)}}, \end{align}
where $\frac{x-x^2}{1+x} \in \ker \asyOp^1_1$ was used.
This can be reexpressed using the differential equation $x^2 F'(x) +(x-1)F(x)+x=0$, from which a non-linear differential equation for $F^{-1}(x)$ can be deduced, because $F'(F^{-1}(x)) {F^{-1}}'(x) = 1$:
\begin{gather*} {F^{-1}}'(x) = \frac{1}{F'(F^{-1}(x))}= \frac{F^{-1}(x)^2}{(1-F^{-1}(x))x-F^{-1}(x)}. \end{gather*}
Using this as well as $\frac{x-x^2}{1+x} = F^{-1}(x) + S(x)$ gives
\begin{align} \begin{aligned} \label{eqn:asymp_genfun_Simple} \left(\asyOp^1_1 S\right)(x) &= \frac{xF^{-1}(x)}{x-(1+x)F^{-1}(x)} e^{\frac{F^{-1}(x)-x}{x F^{-1}(x)}} \\ &= \frac{1}{1+x}\frac{1-x - (1+x)\frac{S(x)}{x} }{1 + (1+x) \frac{S(x)}{x^2} } e^{-\frac{2 + (1+x)\frac{S(x)}{x^2} }{1-x - (1+x)\frac{S(x)}{x}}}. \end{aligned} \end{align}
The coefficients of $\left(\asyOp^1_1 S\right)(x)$ can be computed iteratively. The first coefficients are
\begin{align} \left(\asyOp^1_1 S\right)(x) &= e^{-2} \left( 1 - 4 x + 2 x^2 - \frac{40}{3} x^3 - \frac{182}{3} x^4 - \frac{7624}{15} x^5  + \ldots \right). \end{align}
By the Definitions \ref{def:Fpowerseries} and \ref{def:basic_asymp_definition}, this is an expression for the asymptotics of the number of simple permutations 
\begin{align} |S_n^\text{simple}| &= \sum_{k=0}^{R-1} (n-k)! [x^k] \left(\asyOp^1_1 S\right)(x) + \bigO\left( (n-R)! \right)&& \forall R \in \N_0. \end{align}
Therefore, the asymptotic expansion starts with
\begin{align*} |S_n^\text{simple}| &= e^{-2} \left( n! - 4 (n-1)! + 2 (n-2)! - \frac{40}{3} (n-3)!   - \frac{182}{3} (n-4)!  + \ldots \right).     \end{align*}
Albert, Atkinson and Klazar \cite{albert2003enumeration} calculated the first three terms of this expansion. With the presented methods the calculation of the asymptotic expansions $\left(\asyOp^1_1 S\right)(x)$ or $\left(\asyOp^1_1 F^{-1}\right)(x)$ up to order $n$ is as easy as calculating the expansion of $S(x)$ or $F^{-1}(x)$ up to order $n+2$. 
Some additional coefficients are given in Table \ref{tab:coefficientsS}.

\begin{table}
\centering
\tiny{
\def\arraystretch{1.5}
\begin{tabular}{|c||c||c|c|c|c|c|c|c|c|c|}
\hline
sequence&$0$&$1$&$2$&$3$&$4$&$5$&$6$&$7$&$8$&$9$\\
\hline\hline
$e^{2}(\asyOp^1_{1} S)$&$1$&$-4$&$2$&$- \frac{40}{3}$&$- \frac{182}{3}$&$- \frac{7624}{15}$&$- \frac{202652}{45}$&$- \frac{14115088}{315}$&$- \frac{30800534}{63}$&$- \frac{16435427656}{2835}$\\
\hline
\end{tabular}
}
\caption{First coefficients of the asymptotic expansion of $|S_n^\text{simple}|$.}
\label{tab:coefficientsS}
\end{table}

\begin{rmk}
The examples above are chosen to demonstrate that given a (functional) equation which relates two power series in $\fring{x}{\alpha}{\beta}$, it is an easy task to calculate the full asymptotic expansion of one of the power series from the asymptotic expansion of the other power series. 
Applications include functional equations for `irreducible combinatorial objects'. The two examples fall into this category. Irreducible combinatorial objects were studied in general by Beissinger \cite{beissinger1985}. 
\end{rmk}
\begin{rmk}
Eqs. \eqref{eqn:asyC}, \eqref{eqn:asyM} and \eqref{eqn:asymp_genfun_Simple} expose another interesting algebraic property. With the expressions of the respective quantities in mind, it is obvious that Proposition \ref{prop:betashiftlow} and the chain rule imply that $(\asyOp^2_{\frac12} C)(x)\in\fring{x}{2}{\frac32}$, $(\asyOp^2_{\frac12} M)(x) \in \fring{x}{2}{\frac32}$ and $(\asyOp^1_1 S)(x) \in \fring{x}{1}{3}$. This way, the `higher-order' asymptotics of the asymptotic sequence can be calculated by iterating the application of the $\asyOp$ map. With the powerful techniques of resurgence, it might be possible to construct \textit{convergent} large-order expansions for these cases.
The fact that the asymptotics of each sequence may be expressed as a combination of polynomial and exponential expressions of the original sequence can be seen as an avatar of resurgence. 
\end{rmk}

\begin{rmk}
The power series $S(x)$ and $C(x)$ are know to be non-\text{D-finite} generating functions \cite{albert2003enumeration,Klazar2003}. Loosely speaking, this means it is considered a `hard' computational task to calculate the associated sequences up to a certain order. Using the full asymptotic expansion and resummation techniques it could be possible to obtain a `fast' algorithm to approximate these sequences.
\end{rmk}
\begin{rmk}
In quantum field theory  the \textit{coupling}, an expansion parameter, needs to be re\-para\-met\-rized in the process of \textit{renormalization} \cite{connes2001renormalization}. These reparametrizations are merely compositions of power series which are believed to be \textit{Gevrey-1}. Theorem \ref{thm:chainrule} is useful for the resummation of renormalized quantities in quantum field theory.
Dyson-Schwinger equations in quantum field theory can be stated as functional equations of a form similar to the above \cite{Broadhurst2001}. We will expand on this in the following chapters. 
\end{rmk}

\chapter{Coalgebraic graph structures}
\label{chap:coalgebra_graph}

Having discussed the algebraic structure of formal power series, we will now return to graphs and introduce a more advanced algebraic structure on them.

Ultimately, our motivation to introduce the coalgebraic structures on graphs is the formulation of renormalization of Feynman diagrams in QFT in terms of a Hopf algebra \cite{connes2001renormalization}. To also include general graphs into this framework, we will take a more general viewpoint and introduce a Hopf algebra structure on graphs that is not restricted to Feynman diagrams. This construction is very similar to the original formulation of Kreimer's Hopf algebra of Feynman diagrams. 
After we introduced the general Hopf algebra on graphs, we will present the Hopf algebra of Feynman diagrams in Chapter \ref{chap:hopf_algebra_of_fg}, which will emerge as a quotient Hopf algebra of the more general Hopf algebra of all graphs. For an in depth account on Hopf algebras consult \cite{sweedler1969hopf}. The Hopf algebra of Feynman diagrams is discussed in detail in Manchon's review article \cite{manchon2004hopf}.

An alternative motivation to introduce a Hopf algebra structure on graphs is the desire to study the \textit{subgraph structure} of graphs. Additionally to our expressions for the number of diagrams with a fixed number of vertices, edges, legs or first Betti number, we might want to get control over graphs with certain classes of \textit{subgraphs} forbidden. To achieve this the Hopf algebra structure of graphs can be used.

Therefore, we will use the notion of subgraphs as the key to the coalgebraic graph structures.

\section{Subgraphs}
We will start with a suitable definition of subgraphs, which will be \textit{edge-induced} subgraphs in graph theory terminology:
\begin{defn}[Subgraph]
A graph is a subgraph of another graph if it contains all its half-edges and vertices as well as a subset of its edges.
Equivalently, a graph $g$ is a subgraph of $G$ if $H_g = H_G$, $V_g = V_G$, $\nu_g =\nu_G$ and $E_g \subset E_G$. If $g$ is a subgraph of $G$, we will write $g \subset G$.

We will denote the set of all subgraphs of a graph $G \in \Gl$ as $\subdiags(G)$ or equivalently of a representative $\Gamma \in \Gul$ as $\subdiags(\Gamma)$. Naturally, all subgraphs of a graph are partially ordered by inclusion. In this partially ordered set, $\subdiags(\Gamma)$, the graph $\Gamma$ is the unique largest element and the subgraph without edges is the unique smallest element. This partial ordering will be of importance for the lattice structures of Feynman diagrams which will be introduced in Chapter \ref{chap:hopf_algebra_of_fg}.
\end{defn}
Note that all subgraphs of a graph have the same vertex and half-edge sets as the parent graph. 
{
For instance, the graph %
\def\scale{1ex}%
$ $ 
has, among others, three subgraphs which are isomorphic to the graph 
$  \ifmmode \usebox{\fgsimpleonevtx} \else \newsavebox{\fgsimpleonevtx} \savebox{\fgsimpleonevtx}{ %
 } \fi^2 ~ { \def\nr{2} %
 } $. Thwse are the three subgraphs which contain a single edge of 
$ %
 $. We can indicate them with thick lines in the original graph: 
$ { \def\cvsc{} %
 } $
.
}

Clearly, there is a bijection from the set of subgraphs $\subdiags(\Gamma)$ of a graph $\Gamma$ to the power set $\powerset{E_\Gamma}$. The subgraphs only differ by the edges they contain. 
\nomenclature{$\subdiags(\Gamma)$}{Set of subgraphs of a graph $\Gamma$}

\begin{figure}
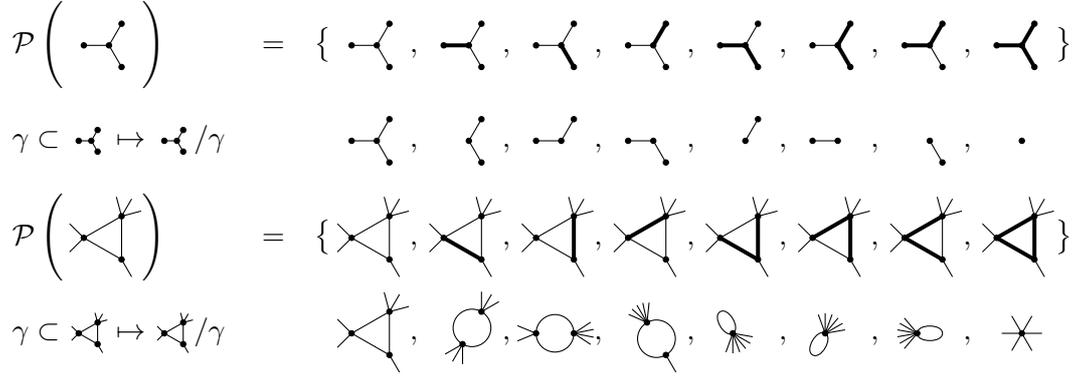

\def\scale{2ex}
\begin{align*} &\subdiags \left( %
 } \end{align*}
\caption{Examples of subgraph sets of graphs with the respective graph contractions directly underneath. The subgraphs are indicated as thick edges.}
\label{fig:subgraph_contraction}
\end{figure}

Another crucial notion is the contraction of a subgraph inside its parent graph. 
\begin{defn}[Contraction of a subgraph]
\label{def:contraction}
If $g \subset G$ with a graph $G$ and a subgraph $g$, we can contract $g$ in $G$. That means, we shrink all the edges in $G$ which also belong to $g$. 

To be more explicit, we construct a new graph $G/g$ from $g$ and $G$ by taking the set of legs of $g$ as the half-edges set of $G/g$, $H_{G/g} = \legs_g$. All edges of $G$ that are not edges of $g$ are going to be edges of $G/g$, $E_{G/g}= E_G \setminus E_g$. Observe that the edges $E_{G/g}$ only involve half-edges of $\legs_g$ as they should. As vertices we use the connected components (Definition \ref{def:cntd_cmps}) of the subgraph $V_{G/g} = \comps_g$. The map $\nu_g: H_g \rightarrow V_g$ also induces a map $\widetilde{\nu}_{G/g}: H_g \rightarrow \comps_g$, $\widetilde{\nu}_{G/g} =\pi \circ \nu_g$, where $\pi: V_g \rightarrow \comps_g$ is the projection of a vertex to its connected component.

This way we obtain a graph $G/g := \left( \legs_g, \comps_g, \pi \circ \nu_g, E_G \setminus E_g \right)$.
\end{defn}
\nomenclature{$G/g$}{Contraction of $g$ in $G$}

In Figure \ref{fig:subgraph_contraction} two examples of graphs with all their subgraphs and the respective contractions are depicted.

\begin{defn}[Residues]
An important class of graphs will be the graphs without edges: We will denote this set of unlabelled graphs without edges as $\residuesstar \subset \Gul$. The set of all graphs consisting of a single vertex\footnote{We might also call it the set of connected residues, as these are the only graphs without edges that have exactly one connected component.} will be denoted as $\residues \subset \residuesstar$:
\begin{align*} \residuesstar &:= \{ \one,  \ifmmode \usebox{\fgsimplenullvtx} \else \newsavebox{\fgsimplenullvtx} \savebox{\fgsimplenullvtx}{ \begin{tikzpicture}[x=1ex,y=1ex,baseline={([yshift=-.55ex]current bounding box.center)}] \coordinate (v) ; \filldraw[white] (v) circle (.8); \filldraw (v) circle (1pt); \end{tikzpicture} } \fi,  \ifmmode \usebox{\fgsimplenullvtx} \else \newsavebox{\fgsimplenullvtx} \savebox{\fgsimplenullvtx}{ \begin{tikzpicture}[x=1ex,y=1ex,baseline={([yshift=-.55ex]current bounding box.center)}] \coordinate (v) ; \filldraw[white] (v) circle (.8); \filldraw (v) circle (1pt); \end{tikzpicture} } \fi^2,  \ifmmode \usebox{\fgsimpleonevtx} \else \newsavebox{\fgsimpleonevtx} \savebox{\fgsimpleonevtx}{ \begin{tikzpicture}[x=1ex,y=1ex,baseline={([yshift=-.55ex]current bounding box.center)}] \coordinate (v) ; \def \n {1}; \def \rad {1}; \filldraw[white] (v) circle (\rad); \foreach \s in {1,...,\n} { \def \angle {180+360/\n*(\s - 1)}; \coordinate (u) at ([shift=({\angle}:\rad)]v); \draw (v) -- (u); } \filldraw (v) circle (1pt); \end{tikzpicture} } \fi,  \ifmmode \usebox{\fgsimplenullvtx} \else \newsavebox{\fgsimplenullvtx} \savebox{\fgsimplenullvtx}{ \begin{tikzpicture}[x=1ex,y=1ex,baseline={([yshift=-.55ex]current bounding box.center)}] \coordinate (v) ; \filldraw[white] (v) circle (.8); \filldraw (v) circle (1pt); \end{tikzpicture} } \fi ~  \ifmmode \usebox{\fgsimpleonevtx} \else \newsavebox{\fgsimpleonevtx} \savebox{\fgsimpleonevtx}{ \begin{tikzpicture}[x=1ex,y=1ex,baseline={([yshift=-.55ex]current bounding box.center)}] \coordinate (v) ; \def \n {1}; \def \rad {1}; \filldraw[white] (v) circle (\rad); \foreach \s in {1,...,\n} { \def \angle {180+360/\n*(\s - 1)}; \coordinate (u) at ([shift=({\angle}:\rad)]v); \draw (v) -- (u); } \filldraw (v) circle (1pt); \end{tikzpicture} } \fi, \ldots,  \ifmmode \usebox{\fgsimpletwovtx} \else \newsavebox{\fgsimpletwovtx} \savebox{\fgsimpletwovtx}{ \begin{tikzpicture}[x=1ex,y=1ex,baseline={([yshift=-.5ex]current bounding box.center)}] \coordinate (v) ; \def \n {2}; \def \rad {.8}; \filldraw[white] (v) circle (\rad); \foreach \s in {1,...,\n} { \def \angle {360/\n*(\s - 1)}; \coordinate (u) at ([shift=({\angle}:\rad)]v); \draw (v) -- (u); } \filldraw (v) circle (1pt); \end{tikzpicture} } \fi^2 ~  \ifmmode \usebox{\fgsimplethreevtx} \else \newsavebox{\fgsimplethreevtx} \savebox{\fgsimplethreevtx}{ \begin{tikzpicture}[x=1ex,y=1ex,baseline={([yshift=-.5ex]current bounding box.center)}] \coordinate (v) ; \def \n {3}; \def \rad {.8}; \filldraw[white] (v) circle (\rad); \foreach \s in {1,...,5} { \def \angle {180+360/\n*(\s - 1)}; \coordinate (u) at ([shift=({\angle}:\rad)]v); \draw (v) -- (u); } \filldraw (v) circle (1pt); \end{tikzpicture} } \fi, \ldots \} \\ \residues &:= \{  \ifmmode \usebox{\fgsimplenullvtx} \else \newsavebox{\fgsimplenullvtx} \savebox{\fgsimplenullvtx}{ \begin{tikzpicture}[x=1ex,y=1ex,baseline={([yshift=-.55ex]current bounding box.center)}] \coordinate (v) ; \filldraw[white] (v) circle (.8); \filldraw (v) circle (1pt); \end{tikzpicture} } \fi,  \ifmmode \usebox{\fgsimpleonevtx} \else \newsavebox{\fgsimpleonevtx} \savebox{\fgsimpleonevtx}{ \begin{tikzpicture}[x=1ex,y=1ex,baseline={([yshift=-.55ex]current bounding box.center)}] \coordinate (v) ; \def \n {1}; \def \rad {1}; \filldraw[white] (v) circle (\rad); \foreach \s in {1,...,\n} { \def \angle {180+360/\n*(\s - 1)}; \coordinate (u) at ([shift=({\angle}:\rad)]v); \draw (v) -- (u); } \filldraw (v) circle (1pt); \end{tikzpicture} } \fi,  \ifmmode \usebox{\fgsimpletwovtx} \else \newsavebox{\fgsimpletwovtx} \savebox{\fgsimpletwovtx}{ \begin{tikzpicture}[x=1ex,y=1ex,baseline={([yshift=-.5ex]current bounding box.center)}] \coordinate (v) ; \def \n {2}; \def \rad {.8}; \filldraw[white] (v) circle (\rad); \foreach \s in {1,...,\n} { \def \angle {360/\n*(\s - 1)}; \coordinate (u) at ([shift=({\angle}:\rad)]v); \draw (v) -- (u); } \filldraw (v) circle (1pt); \end{tikzpicture} } \fi,  \ifmmode \usebox{\fgsimplethreevtx} \else \newsavebox{\fgsimplethreevtx} \savebox{\fgsimplethreevtx}{ \begin{tikzpicture}[x=1ex,y=1ex,baseline={([yshift=-.5ex]current bounding box.center)}] \coordinate (v) ; \def \n {3}; \def \rad {.8}; \filldraw[white] (v) circle (\rad); \foreach \s in {1,...,5} { \def \angle {180+360/\n*(\s - 1)}; \coordinate (u) at ([shift=({\angle}:\rad)]v); \draw (v) -- (u); } \filldraw (v) circle (1pt); \end{tikzpicture} } \fi, \ldots \} \end{align*}
\end{defn}

\begin{figure}
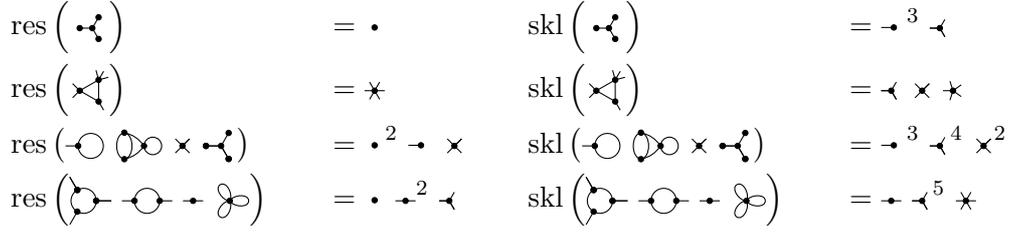

\begin{align*} &\res \left( { \def\scale{1ex} } \right)& &=  \ifmmode \usebox{\fgsimplenullvtx} \else \newsavebox{\fgsimplenullvtx} \savebox{\fgsimplenullvtx}{ %
 } \fi & &\skl \left( { \def\scale{1ex} %
 } \right)& &=  \ifmmode \usebox{\fgsimpleonevtx} \else \newsavebox{\fgsimpleonevtx} \savebox{\fgsimpleonevtx}{ %
 } \fi^3 ~  \ifmmode \usebox{\fgsimplethreevtx} \else \newsavebox{\fgsimplethreevtx} \savebox{\fgsimplethreevtx}{ %
 } \fi \\ &\res \left( { \def\scale{1ex} %
 } \right)& &=  \ifmmode \usebox{\fgsimplesixvtx} \else \newsavebox{\fgsimplesixvtx} \savebox{\fgsimplesixvtx}{ %
 } \fi & &\skl \left( { \def\scale{1ex} %
 } \right)& &=  \ifmmode \usebox{\fgsimplethreevtx} \else \newsavebox{\fgsimplethreevtx} \savebox{\fgsimplethreevtx}{ %
 } \fi ~  \ifmmode \usebox{\fgsimplefourvtx} \else \newsavebox{\fgsimplefourvtx} \savebox{\fgsimplefourvtx}{ %
 } \fi ~  \ifmmode \usebox{\fgsimplefivevtx} \else \newsavebox{\fgsimplefivevtx} \savebox{\fgsimplefivevtx}{ %
 } \fi \\ &\res \left(  \ifmmode \usebox{\fgconetadpolephithree} \else \newsavebox{\fgconetadpolephithree} \savebox{\fgconetadpolephithree}{ %
 } \fi ~  \ifmmode \usebox{\fgcontractedcloseddunce} \else \newsavebox{\fgcontractedcloseddunce} \savebox{\fgcontractedcloseddunce}{ %
 } \fi ~  \ifmmode \usebox{\fgsimplefourvtx} \else \newsavebox{\fgsimplefourvtx} \savebox{\fgsimplefourvtx}{ %
 } \fi ~  \ifmmode \usebox{\fgjthreesimplevtx} \else \newsavebox{\fgjthreesimplevtx} \savebox{\fgjthreesimplevtx}{ %
 } \fi \right)& &=  \ifmmode \usebox{\fgsimplenullvtx} \else \newsavebox{\fgsimplenullvtx} \savebox{\fgsimplenullvtx}{ %
 } \fi^2 ~  \ifmmode \usebox{\fgsimpleonevtx} \else \newsavebox{\fgsimpleonevtx} \savebox{\fgsimpleonevtx}{ %
 } \fi ~  \ifmmode \usebox{\fgsimplefourvtx} \else \newsavebox{\fgsimplefourvtx} \savebox{\fgsimplefourvtx}{ %
 } \fi & &\skl \left(  \ifmmode \usebox{\fgconetadpolephithree} \else \newsavebox{\fgconetadpolephithree} \savebox{\fgconetadpolephithree}{ %
 } \fi ~  \ifmmode \usebox{\fgcontractedcloseddunce} \else \newsavebox{\fgcontractedcloseddunce} \savebox{\fgcontractedcloseddunce}{ %
 } \fi ~  \ifmmode \usebox{\fgsimplefourvtx} \else \newsavebox{\fgsimplefourvtx} \savebox{\fgsimplefourvtx}{ %
 } \fi ~  \ifmmode \usebox{\fgjthreesimplevtx} \else \newsavebox{\fgjthreesimplevtx} \savebox{\fgjthreesimplevtx}{ %
 } \fi \right)& &=  \ifmmode \usebox{\fgsimpleonevtx} \else \newsavebox{\fgsimpleonevtx} \savebox{\fgsimpleonevtx}{ %
 } \fi^3 ~  \ifmmode \usebox{\fgsimplethreevtx} \else \newsavebox{\fgsimplethreevtx} \savebox{\fgsimplethreevtx}{ %
 } \fi^4 ~  \ifmmode \usebox{\fgsimplefourvtx} \else \newsavebox{\fgsimplefourvtx} \savebox{\fgsimplefourvtx}{ %
 } \fi^2 \\ &\res \left(  \ifmmode \usebox{\fgthreeconeltrianglephithree} \else \newsavebox{\fgthreeconeltrianglephithree} \savebox{\fgthreeconeltrianglephithree}{ %
 } \fi~  \ifmmode \usebox{\fgtwoconeloopbubblephithree} \else \newsavebox{\fgtwoconeloopbubblephithree} \savebox{\fgtwoconeloopbubblephithree}{ %
 } \fi~  \ifmmode \usebox{\fgsimpletwovtx} \else \newsavebox{\fgsimpletwovtx} \savebox{\fgsimpletwovtx}{ %
 } \fi~  \ifmmode \usebox{\fgthreerose} \else \newsavebox{\fgthreerose} \savebox{\fgthreerose}{ %
 } \fi \right)& &=  \ifmmode \usebox{\fgsimplenullvtx} \else \newsavebox{\fgsimplenullvtx} \savebox{\fgsimplenullvtx}{ %
 } \fi ~  \ifmmode \usebox{\fgsimpletwovtx} \else \newsavebox{\fgsimpletwovtx} \savebox{\fgsimpletwovtx}{ %
 } \fi^2 ~  \ifmmode \usebox{\fgsimplethreevtx} \else \newsavebox{\fgsimplethreevtx} \savebox{\fgsimplethreevtx}{ %
 } \fi & &\skl \left(  \ifmmode \usebox{\fgthreeconeltrianglephithree} \else \newsavebox{\fgthreeconeltrianglephithree} \savebox{\fgthreeconeltrianglephithree}{ %
 } \fi~  \ifmmode \usebox{\fgtwoconeloopbubblephithree} \else \newsavebox{\fgtwoconeloopbubblephithree} \savebox{\fgtwoconeloopbubblephithree}{ %
 } \fi~  \ifmmode \usebox{\fgsimpletwovtx} \else \newsavebox{\fgsimpletwovtx} \savebox{\fgsimpletwovtx}{ %
 } \fi~  \ifmmode \usebox{\fgthreerose} \else \newsavebox{\fgthreerose} \savebox{\fgthreerose}{ %
 } \fi \right)& &=  \ifmmode \usebox{\fgsimpletwovtx} \else \newsavebox{\fgsimpletwovtx} \savebox{\fgsimpletwovtx}{ %
 } \fi ~  \ifmmode \usebox{\fgsimplethreevtx} \else \newsavebox{\fgsimplethreevtx} \savebox{\fgsimplethreevtx}{ %
 } \fi^5 ~  \ifmmode \usebox{\fgsimplesixvtx} \else \newsavebox{\fgsimplesixvtx} \savebox{\fgsimplesixvtx}{ %
 } \fi \end{align*}
\caption{Examples of residues and skeletons of graphs.}
\label{fig:skl_res_examples}
\end{figure}

\begin{defn}[Residues and skeletons of graphs]
\label{def:res_skl_def}
For any graph $\Gamma$ in $\Gul$ the complete contraction $\Gamma /\Gamma$ belongs to $\residuesstar$ as all edges are contracted. We will denote this operation as the residue of $\Gamma$, $\res: \Gul \rightarrow \residuesstar$, $\Gamma \mapsto \Gamma/\Gamma$.

On the other hand there is a trivial subgraph $\gamma$ for every $\Gamma \in \Gul$ that has no edges $E_\gamma = \emptyset$. Obviously, this graph is also in $\residuesstar$. We will denote this graph as the \textit{skeleton} $\skl(\Gamma):= \gamma$ of the graph $\Gamma$.
\end{defn}
\nomenclature{$\residuesstar$}{Set of all residues}
\nomenclature{$\residues$}{Set of all residues with one connected component}
\nomenclature{$\res(\Gamma)$}{Residue of a graph $\Gamma$}
\nomenclature{$\skl(\Gamma)$}{Skeleton of a graph $\Gamma$}
In Figure \ref{fig:skl_res_examples} some examples of residues and skeletons of graphs are given.

We will use these notions to augment $\Gaul$ with a coalgebraic structure in addition to its algebraic structure. 

\section{The coalgebra of graphs}
\label{sec:coalgebra_graphs}
The center piece of the coalgebra structure is 
the coproduct. The coproduct is dual to the product $m: \Gaul\otimes \Gaul \rightarrow \Gaul$ in the sense that $\Delta: \Gaul \rightarrow \Gaul\otimes \Gaul$. Instead of combining two graphs, it decomposes the graph into its subgraphs in a natural way.
\begin{defn}[Coalgebra of graphs]
\label{def:coproduct}
\mbox{}
\begin{itemize}
\item
We define a \textit{coproduct} on $\Gaul$, a linear map defined on the generators of $\Gaul$:
\begin{align} &\Delta& &:& &\Gaul& &\rightarrow& \Gaul &\otimes \Gaul, \\ &&&& &\Gamma& &\mapsto& \sum_{\gamma \subset \Gamma} \gamma &\otimes \Gamma/\gamma, \end{align}
where we write $\gamma$ and $\Gamma/\gamma$ in the tensor product for the generators in $\Gul$ that are isomorphic to the respective graphs\footnote{As in Definition \ref{def:graph_algebra}, a more rigorous way to express this would be to set $\Delta \Gamma := \sum_{\gamma \subset \Gamma} \pi(\gamma) \otimes \pi(\Gamma/\gamma)$, where $\pi$ maps a graph to its unique unlabelled representative in $\Gul$, as both the subgraph $\gamma$ and the contraction $\Gamma/\gamma$ are labelled graphs. However, we will again adopt the established notation and omit $\pi$. }. In Figure \ref{fig:coproduct_comp_example1} this coproduct computation is illustrated on two graphs. 
\nomenclature{$\Delta$}{Coproduct in $\Gaul$}
\item
Additionally, we define a counit $\counit$ on $\Gaul$. The counit $\counit: \Gaul \rightarrow \Q$ is the projection operator, that maps all graphs \textit{without} edges in $\residuesstar$ to $1$, $\counit(r) = 1$ for all $r\in \residuesstar$ and all non-trivial graphs with edges to $0$, $\counit(\Gamma) = 0$ for all $\Gamma \notin \residuesstar$. For instance $\counit(\one) = \counit(  \ifmmode \usebox{\fgsimplethreevtx} \else \newsavebox{\fgsimplethreevtx} \savebox{\fgsimplethreevtx}{ \begin{tikzpicture}[x=1ex,y=1ex,baseline={([yshift=-.5ex]current bounding box.center)}] \coordinate (v) ; \def \n {3}; \def \rad {.8}; \filldraw[white] (v) circle (\rad); \foreach \s in {1,...,5} { \def \angle {180+360/\n*(\s - 1)}; \coordinate (u) at ([shift=({\angle}:\rad)]v); \draw (v) -- (u); } \filldraw (v) circle (1pt); \end{tikzpicture} } \fi ) = \counit(  \ifmmode \usebox{\fgsimplethreevtx} \else \newsavebox{\fgsimplethreevtx} \savebox{\fgsimplethreevtx}{ \begin{tikzpicture}[x=1ex,y=1ex,baseline={([yshift=-.5ex]current bounding box.center)}] \coordinate (v) ; \def \n {3}; \def \rad {.8}; \filldraw[white] (v) circle (\rad); \foreach \s in {1,...,5} { \def \angle {180+360/\n*(\s - 1)}; \coordinate (u) at ([shift=({\angle}:\rad)]v); \draw (v) -- (u); } \filldraw (v) circle (1pt); \end{tikzpicture} } \fi ~  \ifmmode \usebox{\fgsimpleonevtx} \else \newsavebox{\fgsimpleonevtx} \savebox{\fgsimpleonevtx}{ \begin{tikzpicture}[x=1ex,y=1ex,baseline={([yshift=-.55ex]current bounding box.center)}] \coordinate (v) ; \def \n {1}; \def \rad {1}; \filldraw[white] (v) circle (\rad); \foreach \s in {1,...,\n} { \def \angle {180+360/\n*(\s - 1)}; \coordinate (u) at ([shift=({\angle}:\rad)]v); \draw (v) -- (u); } \filldraw (v) circle (1pt); \end{tikzpicture} } \fi ) = 1$ and $\counit( \ifmmode \usebox{\fgwheelthree} \else \newsavebox{\fgwheelthree} \savebox{\fgwheelthree}{ \begin{tikzpicture}[x=1ex,y=1ex,baseline={([yshift=-.5ex]current bounding box.center)}] \coordinate (v) ; \def \n {3}; \def \rad {1.2}; \foreach \s in {1,...,\n} { \def \angle {360/\n*(\s - 1)}; \coordinate (s) at ([shift=({\angle}:\rad)]v); \draw (v) -- (s); \filldraw (s) circle (1pt); } \draw (v) circle (\rad); \filldraw (v) circle (1pt); \end{tikzpicture} } \fi)=0$. The kernel of $\counit$ is called the \textit{augmentation ideal} of $\Gaul$.
\end{itemize}
\end{defn}
\begin{prop}
Equipped with the coproduct $\Delta$ and counit $\counit$, $\Gaul$ becomes an \textit{associative coalgebra}. That means, $\Delta$ and $\counit$ fulfill
\begin{align} \label{eqn:coassociativity} (\Delta \otimes \id) \circ \Delta &= (\id \otimes \Delta) \circ \Delta \\ \label{eqn:counit_cond} (\id \otimes \counit) \circ \Delta &= (\counit \otimes \id) \circ \Delta = \id, \end{align}
where $\id: \Gaul \rightarrow \Gaul, \Gamma \mapsto \Gamma$ is the identity map.
\end{prop}

\begin{proof}
We need to prove that eqs.\ \eqref{eqn:coassociativity} and \eqref{eqn:counit_cond} follow from the definitions of $\Delta$ and $\counit$.
To prove eq.\ \eqref{eqn:coassociativity}, we just need to apply Definition \ref{def:contraction} of the contraction. Observe that for any generator $\Gamma$ of $\Gaul$,
\begin{align*} (\Delta \otimes \id) \circ \Delta (\Gamma) &= \sum_{\gamma \subset \Gamma} \sum_{\delta \subset \gamma} \delta \otimes \gamma/\delta \otimes \Gamma/\gamma \\ (\id \otimes \Delta) \circ \Delta (\Gamma) &= \sum_{\delta \subset \Gamma} \sum_{\widetilde{\gamma} \subset \Gamma/\delta} \delta \otimes \widetilde{\gamma} \otimes (\Gamma/\delta)/\widetilde{\gamma} \end{align*}
The set of subgraphs $\widetilde \gamma$ of $\Gamma/\delta$ and the set of subgraphs $\gamma$ of $\Gamma$, which contain $\delta$ as a subgraph $\delta \subset \gamma$, are in bijection. Therefore, having a subgraph $\widetilde{\gamma} \subset \Gamma/\delta$ is equivalent to having a subgraph $\gamma \subset \Gamma$ such that $\delta \subset \gamma$ where $\widetilde{\gamma} = \gamma/\delta$. Applying this to the definition of the coproduct gives
\begin{align*} (\id \otimes \Delta) \circ \Delta (\Gamma) = \sum_{\delta \subset \gamma \subset \Gamma} \delta \otimes \gamma/\delta \otimes (\Gamma/\delta)/(\gamma/\delta). \end{align*}
Definition \ref{def:contraction} of the contraction guarantees that $(\Gamma/\delta)/(\gamma/\delta) = \Gamma/\gamma$ and eq.\ \eqref{eqn:coassociativity} follows.

To make sense out of eq. \eqref{eqn:counit_cond}, note that $\Q$ is the ground field for our algebra and the tensor product. We use the usual convention to identify the naturally isomorphic spaces $\Gaul \otimes \Q \simeq \Q \otimes \Gaul \simeq \Gaul$. With that in mind, eq.\ \eqref{eqn:counit_cond} follows directly from the definitions:

For any generator $\Gamma$ of $\Gaul$, 
\begin{align*} (\id \otimes \counit) \circ \Delta(\Gamma) = \sum_{\gamma \subset \Gamma} \gamma \otimes \counit( \Gamma/\gamma). \end{align*}
The only subgraph $\gamma$ of $\Gamma$ such that $\counit(\Gamma/\gamma) \neq 0$ is $\gamma = \Gamma$. Therefore, the only term surviving on the right hand side is $\Gamma \otimes \counit( \Gamma/\Gamma ) = \Gamma \otimes 1 = \Gamma$. 
Analogously, 
\begin{align*} (\counit \otimes \id) \circ \Delta(\Gamma) = \sum_{\gamma \subset \Gamma} \counit(\gamma) \otimes \Gamma/\gamma = \Gamma, \end{align*}
as the only subgraph contributing to the sum is the one without any edges.
\end{proof}

\begin{figure}
{
\def\scale{1ex}
\begin{align*} \Delta
/\gamma \\ &=  \ifmmode \usebox{\fgsimpleonevtx} \else \newsavebox{\fgsimpleonevtx} \savebox{\fgsimpleonevtx}{ %
 } \fi^3 ~  \ifmmode \usebox{\fgsimplethreevtx} \else \newsavebox{\fgsimplethreevtx} \savebox{\fgsimplethreevtx}{ %
 } \fi \otimes { %
 } + 3~  \ifmmode \usebox{\fgsimpleonevtx} \else \newsavebox{\fgsimpleonevtx} \savebox{\fgsimpleonevtx}{ %
 } \fi^2 ~ { \def\nr{2} %
 } + 3~  \ifmmode \usebox{\fgsimpleonevtx} \else \newsavebox{\fgsimpleonevtx} \savebox{\fgsimpleonevtx}{ %
 } \fi ~ %
 \otimes {  \ifmmode \usebox{\fgsimpletreehandle} \else \newsavebox{\fgsimpletreehandle} \savebox{\fgsimpletreehandle}{ %
 } \fi } + %
 \otimes  \ifmmode \usebox{\fgsimplenullvtx} \else \newsavebox{\fgsimplenullvtx} \savebox{\fgsimplenullvtx}{ %
 } \fi \\ \Delta %
 / \gamma \\ &=  \ifmmode \usebox{\fgsimplethreevtx} \else \newsavebox{\fgsimplethreevtx} \savebox{\fgsimplethreevtx}{ %
 } \fi ~  \ifmmode \usebox{\fgsimplefourvtx} \else \newsavebox{\fgsimplefourvtx} \savebox{\fgsimplefourvtx}{ %
 } \fi ~  \ifmmode \usebox{\fgsimplefivevtx} \else \newsavebox{\fgsimplefivevtx} \savebox{\fgsimplefivevtx}{ %
 } \fi \otimes %
 +  \ifmmode \usebox{\fgsimplefivevtx} \else \newsavebox{\fgsimplefivevtx} \savebox{\fgsimplefivevtx}{ %
 } \fi ~ { \def\nl{3} \def\nr{2} %
 } +  \ifmmode \usebox{\fgsimplefourvtx} \else \newsavebox{\fgsimplefourvtx} \savebox{\fgsimplefourvtx}{ %
 } \fi ~ { \def\nl{4} \def\nr{2} %
 } +  \ifmmode \usebox{\fgsimplethreevtx} \else \newsavebox{\fgsimplethreevtx} \savebox{\fgsimplethreevtx}{ %
 } \fi ~ { \def\nl{3} \def\nr{4} %
 \otimes  \ifmmode \usebox{\fgsimplesixvtx} \else \newsavebox{\fgsimplesixvtx} \savebox{\fgsimplesixvtx}{ %
 } \fi \end{align*}
}
\caption{Examples of coproduct computations for two graphs. The subgraph and contraction operations were illustrated in Figure \ref{fig:subgraph_contraction}. Note, that we implicitly identified the subgraphs with their unlabelled counterpart and that the subgraphs retained their external leg structure.}
\label{fig:coproduct_comp_example1}
\end{figure}

Because $\Gaul$ is associative and coassociative, it makes sense to define iterations of $m$ and $\Delta$:
\begin{align} m^{0}&:= \unit, & m^1 &:= \id ,& m^{n} &:= m^{n-1} \circ ( m \otimes \id^{\otimes n-2} ) & \text{ for } n \geq 2, \\ \Delta^{0}&:= \counit, & \Delta^1 &:= \id ,& \Delta^{n} &:= ( \Delta \otimes \id^{\otimes n-2} ) \circ \Delta^{n-1} & \text{ for } n \geq 2, \end{align}
where $m^k: \Gaul^{\otimes k} \rightarrow \Gaul$ and $\Delta^{k}: \Gaul \rightarrow {\Gaul}^{\otimes k}$. 
\nomenclature{$m^{k}$}{Iterated product, $m^k: \Gaul^{\otimes k} \rightarrow \Gaul$}
\nomenclature{$\Delta^{k}$}{Iterated coproduct, $\Delta^{k}: \Gaul \rightarrow {\Gaul}^{\otimes k}$}

\begin{prop}
\label{prop:gaul_bialgebra}
$\Gaul$ is a bialgebra. That means $\Delta$ is an algebra homomorphism and $m$ is an coalgebra homomorphism:
\begin{align} \Delta \circ m = (m\otimes m) \circ \tau_{2,3} \circ ( \Delta \otimes \Delta ), \end{align}
where $\tau_{2,3}$ is the map $\Gaul \otimes \Gaul \otimes \Gaul \otimes \Gaul \rightarrow \Gaul \otimes \Gaul \otimes \Gaul \otimes \Gaul$ that switches the second and the third entry of the tensor product or equivalently, for all $a,b \in \Gaul$ we have $\Delta (ab) = (\Delta a)(\Delta b)$.
\end{prop}
\begin{proof}
As $m$ and $\Delta$ are linear maps it is sufficient to prove the statement for generators $\Gamma_1,\Gamma_2 \in \Gul$. 
By the definitions of $m$ and $\Delta$, 
\begin{gather*} \Delta \circ m(\Gamma_1 \otimes \Gamma_2) = \sum_{\gamma \subset \Gamma_1 \sqcup \Gamma_2} \gamma \otimes (\Gamma_1 \sqcup \Gamma_2)/\gamma = \sum_{\gamma_1 \subset \Gamma_1} \sum_{\gamma_2 \subset \Gamma_2} \gamma_1\sqcup \gamma_2 \otimes (\Gamma_1 \sqcup \Gamma_2)/(\gamma_1\sqcup \gamma_2). \end{gather*}
Using Definition \ref{def:contraction} it is easy to verify that $(\Gamma_1 \sqcup \Gamma_2)/(\gamma_1\sqcup \gamma_2) = (\Gamma_1/\gamma_1) \sqcup (\Gamma_2/\gamma_2)$.

Therefore, $\Gaul$ is a bialgebra \cite{sweedler1969hopf}.
\end{proof}

\section{The main identity of the graph bialgebra}

The following identity in $\Gaul$, which can be seen as the coalgebraic version of Theorem \ref{thm:connected_disconnected}, is central to our application of the coalgebra structure of graphs. It gives us an entry point to gain control over subgraph structures in graphs, whereas Theorem \ref{thm:connected_disconnected} gives us control over their connected components. 

Recall that we defined $\allG := \sum_{\Gamma \in \Gul} \frac{\Gamma}{|\Aut \Gamma|}$ in Section \ref{sec:graph_algebra}. The image of this vector in $\Gaul$ under $\Delta$ fulfills the identity:
\begin{thm}
\label{thm:coproduct_full_identity}
\begin{align} \Delta \allG = \sum_{\Gamma \in \Gul} \prod_{v \in V_{\Gamma}} (\deg{v}_\Gamma!) \allG^{(v)} \otimes \frac{\Gamma}{|\Aut \Gamma|}, \end{align}
where the product runs over all vertices of $\Gamma$ and 
\begin{align} \allG^{(v)} := \sum_{\substack{ \Gamma \in \Gul\\\res{\Gamma} = v }} \frac{\Gamma}{|\Aut \Gamma|}, \end{align}
is the sum over all graphs with the single vertex $v \in \residues$ as residue.
\end{thm}

The proof of this theorem is the main objective of this section, but first we are going to generalize the statement using another definition:

Given a subset of graphs $\sgset \subset \Gul$, we can formulate
\begin{defn}[$\sgset$-insertion/contraction closed graph set]
\label{def:insertion_contraction_closed}
We call a subset $\subclass \subset \Gul$ a $\sgset$-insertion/contraction closed graph set if it is closed under `insertion' and contraction of subgraphs from $\sgset$. That means for all $\gamma \subset \Gamma \in \Gul$, where\footnote{Note again the slight abuse of notation in the form of the silent identification of a subgraph and the respective contraction with their unlabelled representatives.} $\gamma \in \sgset$: $\Gamma \in \subclass$ iff $\Gamma/\gamma \in \subclass$.
\end{defn}
\nomenclature{$\subclass$}{Insertion/contraction closed graph set}

A set $\subclass$ is $\Gul$-insertion/contraction closed if it is closed under contraction and insertion of arbitrary subgraphs. Obviously, such a subset of graphs is completely characterized by the set of residues it includes, as contraction and insertion can not alter the residue structure of a graph.

\begin{crll}
For every $\Gul$-insertion/contraction closed class of graphs $\subclass \subset \Gul$, we have the identity
\begin{align} \Delta \allG_\subclass = \sum_{\Gamma \in \subclass} \prod_{v \in V_{\Gamma}} (\deg{v}_\Gamma!) \allG^{(v)} \otimes \frac{\Gamma}{|\Aut \Gamma|}, \end{align}
where $\allG^{(v)}$ is defined as in Theorem \ref{thm:coproduct_full_identity} and
\begin{align} \allG_\subclass := \sum_{\Gamma \in \subclass} \frac{\Gamma}{|\Aut \Gamma|}. \end{align}
\end{crll}
\begin{proof}
Let $P_\subclass:\Gaul\rightarrow \Gaul$ be the map that projects to the subspace which is generated by elements of $\subclass$. Because $\subclass$ is closed under insertion and contraction of general subgraphs, $\Delta \circ P_\subclass = (\id \otimes P_\subclass) \circ \Delta$ by Definitions \ref{def:coproduct} and \ref{def:insertion_contraction_closed}. Applying this to the result of Theorem \ref{thm:coproduct_full_identity} gives the statement.
\end{proof}

A variant for Theorem \ref{thm:coproduct_full_identity} in the context of quantum field theory was proven in \cite{van2007renormalization}. The proof presented here relies on the author's proof in \cite{borinsky2014feynman}, which is inspired from a lemma in \cite{connes2001renormalization}.
In fact, this theorem can also be seen as the Hopf algebraic version of a standard theorem in the theory of BPHZ renormalization \cite[Ch. 5.6]{collins1984renormalization}.

To prove Theorem \ref{thm:coproduct_full_identity}, we are going to use two additional notions on graphs:

As Definition \ref{def:insertion_contraction_closed} already suggests, we can perform the reverse operation of contracting a subgraph: A graph $g$ can be \textit{inserted} into a graph $G$. The natural way to do this is to replace each vertex of $G$ by a connected component of $g$ and identifying the legs of $g$ with the half-edges of $G$. Of course, there can be multiple such ways to glue a graph into another graph. One such gluing prescription will be called an \textit{insertion place}.
\begin{defn}[Insertion place]
Given two graphs $g$ and $G$, an insertion place is a set of bijections: One bijection, $\kappa:\comps_g \rightarrow V_G$ and a bijection $\xi_c$ for each connected component, $\xi_c : \legs_g \cap \widetilde{\nu}_g^{-1}(c) \rightarrow \nu_G^{-1}(\kappa(c))$ for all $c \in \comps_g$, where $\widetilde{\nu}_g:H_g \rightarrow \comps_g$ is the map $\widetilde{\nu}_g = \pi \circ \nu_g$ and $\pi$ the projection of a vertex to its connected component such that $\nu_G^{-1}(\kappa(c))$ is the set of half-edges belonging to the component $c$. The map $\kappa$ dictates which connected component of $g$ is inserted into which vertex of $G$. The bijections $\xi_c$ provide a way of gluing the respective connected component to the vertex in $G$ by identifying the legs of the connected component $c$, $\legs_g \cap \widetilde{\nu}_g^{-1}(c)$, with the half-edges associated to the target vertex given by $\nu_G^{-1}(\kappa(c))$.
We will denote the set of all insertion places of $g$ into $G$ as $\insertionplaces{g}{G}$.
\end{defn}

The number of insertion places of $g$ into $G$ is easily calculated, as there are no restrictions on the maps $\kappa$ and $\xi_c$ other that they shall be bijections. There are no insertion places if the residue of $g$ is not equal to the skeleton of $G$. That means, the external leg structures of the connected components of $g$ have to be equal to the degree structures of the vertices in $G$. In this case the number of bijections $\kappa$ is $\prod_{d\geq 0} \nvd{d}!$, as we can permute connected components with the same number of legs arbitrarily.
The number of bijections $\xi_c$ is $\deg{\kappa(c)}!$ for each $c\in \comps_\Gamma$. Because $\kappa$ is a bijection, we have for all $\xi_c$, $\prod_{v\in V_G} \deg{v}!$ choices in total. 

This gives, 
\begin{align} \label{eqn:number_of_insertion_places} |\insertionplaces{g}{G}| = \left(\prod_{d\geq 0} \nvd{d}! \right) \left( \prod_{v\in V_G} \deg{v}! \right), \end{align}
if $\res g = \skl G$ or $|\insertionplaces{g}{G}| = 0$ if $\res g \neq \skl G$.
\nomenclature{$\insertionplaces{g}{G}$}{Insertion places for $g$ into $G$}

\begin{defn}[Insertion]
Given two graphs $g$ and $G$ and an insertion place $i \in \insertionplaces{g}{G}$, we can use the insertion place to actually insert $g$ into $G$. We can construct the resulting graph $G\circ_i g := ( H_g, V_g, \nu_g, E_g \cup E')$ explicitly by adding additional edges $E'$ to the graph $g$. The edges $E'$ are constructed using the bijections $\xi_c$: The $\xi_c$ induce a bijection $\xi: \legs_g \rightarrow H_G$, $\xi = \bigsqcup_{c \in \comps_g} \xi_c$ as the connected components form a partition of the vertices of $G$. The extra edges are therefore only $E' = \xi^{-1}(E_G)$, the edges of $G$ mapped into the half-edge set of $g$.
\end{defn}

To prove Theorem \ref{thm:coproduct_full_identity} we also need the following lemma from \cite{borinsky2014feynman}, which is based on a lemma in \cite{connes2001renormalization}.

\begin{lmm}
\label{lmm:triplegraphidentity}
Given a triple of labelled graphs $g, G, \widetilde{G} \in \Gl$, the following two sets are in bijection, 
\begin{itemize}
\item
The set of all triples $(g',j_1,j_2)$ of a subgraph of $G$, $g' \subset G$, an isomorphism $j_1: g \rightarrow g'$ and an isomorphism $j_2: G/g' \rightarrow \widetilde{G}$.
\item The set of pairs $(i, j)$ of an insertion place $i \in \insertionplaces{g}{\widetilde{G}}$ and an isomorphism $j: \widetilde{G} \circ_i g \rightarrow G$.
\end{itemize}
\end{lmm}
\begin{proof}
From the triple $(g',j_1,j_2)$ we can construct an insertion place $i' \in \insertionplaces{g'}{G/g'}$ directly from the given $g' \subset G$ and Definition \ref{def:contraction} of the contraction. Using the isomorphisms $j_1,j_2$ gives us an insertion place  $i \in \insertionplaces{g}{\widetilde{G}}$. Moreover, we get an isomorphism $j: \widetilde{G} \circ_i g \rightarrow G$, because $G/g' \circ_{i'} g' = G$.

This construction is reversible. Given a pair $(i,j)$, $g$ is a subgraph of $\widetilde{G} \circ_i g$ and therefore we can identify $j(g) \subset G$ with $g'$. This also gives the isomorphism $j_1$. Contracting $g$ in $\widetilde{G} \circ_i g$ retrieves $\widetilde{G}$ and contracting $g'$ in $G$ gives $G/g'$. Therefore, we also have an isomorphism $j_2$.
\end{proof}
\begin{crll}
For given $\gamma, \widetilde{\Gamma} \in \Gul$, 
\label{crll:coproduct_insertion_identity_pre}
\begin{align} \frac{|\insertionplaces{\gamma}{\widetilde{\Gamma}}|}{|\Aut \gamma| |\Aut \widetilde{\Gamma}|} = \sum_{\Gamma \in \Gul} \frac{ |\{ \gamma' \subset \Gamma \text{ such that } \gamma' \simeq \gamma \text{ and } \Gamma/\gamma' \simeq \widetilde{\Gamma}\}|}{|\Aut \Gamma|}, \end{align}
where $\simeq$ is the equivalence relation of isomorphic graphs as defined in Section \ref{sec:unlabelled_graphs_isos}.
\end{crll}
\begin{proof}
The total number of triples $(g',j_1,j_2)$ is 
\begin{align*} |\Aut g| |\Aut \widetilde{G}| |\{ g' \subset G : g' \simeq g \text{ and } G/g' \simeq \widetilde{G}\}| \end{align*} 
and the total number of pairs $(i,j)$ is 
\begin{align*} |\Aut G | |\{ i \in \insertionplaces{g}{\widetilde{G}} : \widetilde{G} \circ_i g \simeq G \}|. \end{align*}
Both numbers are equal as guaranteed by Lemma \ref{lmm:triplegraphidentity}.

Replacing $g,\widetilde{G}$ and $G$ with the respective representative of unlabelled graphs $\gamma, \widetilde{\Gamma}$ and $\Gamma$ gives
\begin{gather*} \sum_{\Gamma \in \Gul} \frac{|\{ \gamma' \subset \Gamma \text{ such that } \gamma' \simeq \gamma \text{ and } \Gamma/\gamma' \simeq \widetilde{\Gamma}\}|}{|\Aut \Gamma|} \\ = \sum_{\Gamma \in \Gul} \frac{|\Aut \Gamma| |\{ i \in \insertionplaces{\gamma}{\widetilde{\Gamma}} : \widetilde{\Gamma} \circ_i \gamma \simeq \Gamma \}|}{|\Aut \gamma| |\Aut \widetilde{\Gamma}| |\Aut \Gamma|}, \end{gather*}
which results in the statement.
\end{proof}
This identity can be used to obtain the following identity for the coproduct of the vector $\allG$:
\begin{crll}
\label{crll:coproduct_insertion_identity}
\begin{align} \Delta \allG = \sum_{\gamma, \widetilde{\Gamma} \in \Gul} |\insertionplaces{\gamma}{\widetilde{\Gamma}}| \frac{\gamma}{|\Aut \gamma|} \otimes \frac{\widetilde{\Gamma}}{|\Aut \widetilde{\Gamma}|} \end{align}
\end{crll}
\begin{proof}
By Definition \ref{def:coproduct}, we have 
\begin{align*} \Delta \allG = \sum_{\Gamma \in \Gul} \sum_{\gamma' \subset \Gamma} \gamma' \otimes \Gamma/\gamma' = \sum_{\gamma,\widetilde{\Gamma},\Gamma \in \Gul } \frac{|\{ \gamma' \subset \Gamma: \gamma' \simeq \gamma \text{ and } \Gamma/\gamma' \simeq \widetilde{\Gamma}\}|}{| \Aut \Gamma |} \gamma \otimes \widetilde{\Gamma}, \end{align*}
which results in the statement after an application of Corollary \ref{crll:coproduct_insertion_identity_pre}.
\end{proof}

\begin{proof}[Proof of Theorem \ref{thm:coproduct_full_identity}]

The sum over all graphs with the suitable connected component structure to be inserted into a graph $\Gamma$ can be expressed as the product $\frac{\prod_{v\in V_\Gamma} \allG^{(v)}}{\prod_{d\geq 0} \nvd{d}_\Gamma!}$, where the denominator accounts for the implicit automorphisms between isomorphic graphs which have to have the same amount of legs. Applying Corollary \ref{crll:coproduct_insertion_identity} as well as the result for the number of insertion places from eq.\ \eqref{eqn:number_of_insertion_places} gives, 
\begin{align*} \Delta \allG = \sum_{\widetilde{\Gamma} \in \Gul} \frac{ \left(\prod_{d\geq 0} \nvd{d}_{\widetilde{\Gamma}}! \right) \left( \prod_{v\in V_{\widetilde{\Gamma}}} \deg{v}_{\widetilde{\Gamma}}! \right) \left(\prod_{v\in V_{\widetilde{\Gamma}}}\allG^{(v)}\right) }{\prod_{d\geq 0} \nvd{d}_{\widetilde{\Gamma}}!} \otimes \frac{\widetilde{\Gamma}}{|\Aut \widetilde{\Gamma}|}, \end{align*}
which gives the statement.
\end{proof}

\section{The Hopf algebra of graphs}

We can use the bialgebra structure to introduce a group structure on the set of algebra homomorphisms. This group enables us to manipulate algebra homomorphisms such as the Feynman rules $\phi_\Sact$ from Chapter \ref{chp:graph_enumeration} to only count graphs that do not include certain subgraphs.

We will denote the set of all algebra homomorphisms from $\Gaul$ to some commutative unital algebra $\mathcal{A}$ as $\chargroup{\Gaul}{\mathcal{A}}$.
\begin{defn}[Group of characters]
Let $\chargroup{\Gaul}{\mathcal{A}}$ be the set of all algebra homomorphisms (Definition \ref{def:algebra_morphism}) from $\Gaul$ to a unital commutative algebra $\mathcal{A}$. 

$\chargroup{\Gaul}{\mathcal{A}}$ will turn out to be a group if $\Gaul$ is a \textit{Hopf algebra}. It will be called the \textit{group of characters}.
\end{defn}
\nomenclature{$\chargroup{\Gaul}{\mathcal{A}}$}{Group of characters from $\Gaul$ to $\mathcal{A}$}

\begin{defn}[Convolution product]
Let $\star$ be the multiplication on $\chargroup{\Gaul}{\mathcal{A}}$, that maps a pair of algebra homomorphisms $\phi,\psi \in \chargroup{\Gaul}{\mathcal{A}}$ to
\begin{align} \phi \star \psi := m_{\mathcal{A}} \circ ( \phi \otimes \psi ) \circ \Delta. \end{align}
Because $\Delta$ is coassociative and $m$ is associative, the $\star$-product is associative:
\begin{align*} (\phi \star \psi) \star \zeta = \phi \star (\psi \star \zeta), \end{align*}
for all $\phi, \psi, \zeta \in \chargroup{\Gaul}{\mathcal{A}}$.
\end{defn}
\nomenclature{$\star$}{Convolution product on $\chargroup{\Gaul}{\mathcal{A}}$}

We can directly observe that there is a neutral element of $\chargroup{\Gaul}{\mathcal{A}}$ given by the algebra homomorphism $\unit_{\mathcal{A}} \circ \counit_{\Gaul}$, where $\counit_\Gaul$ is the counit of $\Gaul$ and $\unit_\mathcal{A}$ is the unit of $\mathcal{A}$. From eq.\ \eqref{eqn:counit_cond} it follows directly that 
\begin{align} (\unit_{\mathcal{A}} \circ \counit_{\Gaul}) \star \phi = \phi \star (\unit_{\mathcal{A}} \circ \counit_{\Gaul}) = \phi, \end{align}
for all algebra homomorphisms $\phi$.

\begin{expl}
\label{expl:phi_decomp}
Using the notation of the $\star$-product, we can decompose algebra homomorphisms from $\Gaul$ in a convenient way:  

Take the target algebra $\mathcal{A}=\Q[[\varphi_c, \lambda_0, \lambda_1, \ldots]]$ of power series in $\varphi_c$ and the $\lambda_d$ variables and the algebra homomorphism
\begin{align*} \phi&:& &\Gaul \rightarrow \Q[[\varphi_c, \lambda_0, \lambda_1, \ldots]], && \Gamma \mapsto \varphi_c^{|\legs_\Gamma|} \prod_{v \in V_\Gamma} \lambda_{\deg{v}}. \end{align*}

\nomenclature{$\re$}{Residual part of an algebra homomorphism decomposition}
\nomenclature{$\sk$}{Skeleton part of an algebra homomorphism decomposition}
We can define the algebra homomorphisms 
\begin{align*} \zeta&:& &\Gaul \rightarrow \Q[[\varphi_c, \lambda_0, \lambda_1, \ldots]], & &\Gamma \mapsto 1 \\ \sk&: &&\Gaul \rightarrow \Q[[\varphi_c, \lambda_0, \lambda_1, \ldots]], &&\Gamma \mapsto \begin{cases} \prod_{v \in V_\Gamma} \lambda_{\deg{v}_\Gamma}& \text{ if $\Gamma \in \residuesstar$}\\ 0& \text{ else} \end{cases} \\ \re&: &&\Gaul \rightarrow \Q[[\varphi_c, \lambda_0, \lambda_1, \ldots]], &&\Gamma \mapsto \begin{cases} \varphi_c^{|\legs_\Gamma|}& \text{ if $\Gamma \in \residuesstar$}\\ 0& \text{ else} \end{cases} \end{align*}
and observe that 
\begin{align*} \phi = \sk \star \zeta \star \re, \end{align*}
because by an application of Definition \ref{def:coproduct} of the coproduct and the $\star$-product, 
\begin{align*} \sk \star \zeta \star \re(\Gamma) &= m^{3} \circ (\sk \otimes \zeta \otimes \re ) \circ \Delta^{3} (\Gamma) = \sum_{\delta \subset \gamma \subset \Gamma} \sk(\delta) \zeta(\gamma/\delta) \re(\Gamma/\gamma) \\ &= \sk(\skl(\Gamma)) \zeta(\Gamma) \re(\res(\Gamma)) = \phi(\Gamma), \end{align*}
where $\Delta^3 = (\Delta \otimes \id) \circ \Delta = (\id \otimes \Delta) \circ \Delta$ and $m^3 = m \circ (m \otimes \id) = m \circ ( \id \otimes m)$.
\end{expl}

To actually establish that $\Gaul$ is a Hopf algebra, we need to extend the notion of the grading from the algebra setting to the bialgebra structure. 

Recall Definition \ref{def:algebra_grading} of the algebra grading. We can refine this definition to \textit{graded bialgebras}.

\begin{defn}[Graded bialgebra]
\label{def:bialgebra_grading}
A grading of $\Gaul$ as a bialgebra is a decomposition into linear subspaces
\begin{align} \Gaul = \bigoplus_{\mulind{i} \in I} \Gaul_\mulind{i} \end{align}
with an (multi-)index set $I = \N_0^n$ with some $n \geq 1$, such that 
\begin{align} &m( \Gaul_\mulind{i} \otimes \Gaul_\mulind{j} )& & \subset& \Gaul_{\mulind{i}+\mulind{j}} && \text{ for all } \mulind{i},\mulind{j}\in I \\ \text{and } & \Delta \Gaul_\mulind{k}& &\subset& \sum_{\mulind{i} + \mulind{j} = \mulind{k} } \Gaul_\mulind{i} \otimes \Gaul_\mulind{j}. && \end{align}
\end{defn}

Obviously, not every grading of the algebra is also a grading of the bialgebra. 
\begin{prop}
As a bialgebra $\Gaul$ is graded by
\begin{enumerate}
\item The number of edges $|E_\Gamma|$.
\item The first Betti number of the graph $h_\Gamma = | E_\Gamma | - |V_\Gamma| + |\comps_\Gamma|$.
\end{enumerate}
\end{prop}
\begin{proof}
It is obvious that $\Gaul$ is graded by $|E_\Gamma|$ and $h_\Gamma$ as an algebra (Definition \ref{def:algebra_grading}).

For the grading as a bialgebra by the number of edges, we just need to verify that $|E_{\Gamma}| = | E_{\gamma} | + | E_{\Gamma/\gamma}|$, for all subgraphs $\gamma \subset \Gamma$. This follows obviously from the definitions of subgraphs and contractions.

To proof that $h_\Gamma = h_\gamma + h_{\Gamma/\gamma}$ for all subgraphs $\gamma \subset \Gamma$, we substitute $h_\Gamma = | E_\Gamma | - |V_\Gamma| + |\comps_\Gamma|$,
\begin{align} h_\gamma + h_{\Gamma/\gamma} = | E_\gamma | - |V_\gamma| + |\comps_\gamma| + | E_{\Gamma/\gamma} | - |V_{\Gamma/\gamma}| + |\comps_{\Gamma/\gamma}|. \end{align}
From the definition of contractions and subgraphs, it follows that $|V_{\Gamma/\gamma}| = |\comps_\gamma|$, $|E_{\Gamma}| = | E_{\gamma} | + | E_{\Gamma/\gamma}|$, $|V_\gamma| = |V_\Gamma|$ and $|\comps_{\Gamma/\gamma}| = |\comps_\Gamma|$. The statement follows.
\end{proof}

In the light of the grading by the number of edges, the generators in $\residuesstar$, all graphs without edges, have a special role. They have degree zero as they have no edges. Moreover, the generators in $\residuesstar$ behave as \textit{group like elements} under the action of the coproduct $\Delta$:
\begin{align} \Delta r &= r \otimes r && \text{ for all } r \in \residuesstar, \end{align}
as can be checked using the definition of the coproduct.

In order to make the bialgebra $\Gaul$ into a Hopf algebra, we have to augment $\Gaul$ by formal inverses of these group like elements. In the following, we will therefore add the formal element $r^{-1}$ to $\Gaul$ for all $r\in \residuesstar$ except for the neutral element $\one \in \residuesstar$ which is its own inverse and define $r^{-1} r = r r^{-1} = \one$. 

It is easy to see that the elements in $\residuesstar$ are the only group like generators with degree $0$ in this grading. This enables us to define an \textit{antipode} on $\Gaul$ and thereby make $\Gaul$ into a Hopf algebra \cite{manchon2004hopf}. 

\begin{prop}
There exists a unique inverse $S$ of the identity map $\id:\Gamma \mapsto \Gamma$ in $\chargroup{\Gaul}{\Gaul}$, called antipode, with respect to the $\star$-product, $S \star \id = \id \star S = \unit \circ \counit$.
\end{prop}
\nomenclature{$S$}{Antipode map on the Hopf algebra $\Gaul$}
\begin{proof}
Restricted to the set of residues, it is trivial to construct such a map. Let $P_{\residuesstar}: \Gaul \rightarrow \Gaul$ and $S_{\residuesstar}: \Gaul \rightarrow \Gaul$ be the mappings $P_{\residuesstar}(r) = r$ and $S_{\residuesstar}(r) = r^{-1}$ for all $r \in \residuesstar$ as well as $P_{\residuesstar}(\Gamma) = S_{\residuesstar}(\Gamma) = 0$ for all $\Gamma \in \Gul \setminus \residuesstar$. Clearly, $P_{\residuesstar} \star S_{\residuesstar} = S_{\residuesstar} \star P_{\residuesstar} = \unit \circ \counit$.

Set $T := \unit \circ \counit - S_{\residuesstar} \star \id$ and observe that $\id = P_{\residuesstar} \star ( \unit \circ \counit - T)$. The Neumann series,
\begin{align*} \sum_{n=0}^\infty T^{\star n}, \end{align*}
where $T^{\star n} = \underbrace{T \star \ldots \star T}_{\text{$n$ times}}$ and $T^{\star 0} := \unit \circ \counit$, 
is convergent in $\Gaul$. To verify this observe that $\residuesstar \subset \ker T$ and therefore $\Gaul_k \subset \ker T^{\star k}$, where $\Gaul_k$ is the subspace of $\Gaul$ of graphs with $k$ edges. That means $\sum_{n=0}^\infty T^{\star n}$ is convergent in every subspace $\Gaul_k$ as the sum can be truncated after $k$ terms. Because $\Gaul$ is graded by the number of edges, the Neumann series is convergent in $\Gaul$.

This gives us a left inverse of $\id \in \chargroup{\Gaul}{\Gaul}$,
\begin{gather*} S := \sum_{n=0}^\infty T^{\star n} \star S_{\residuesstar} \\ S \star \id = \sum_{n=0}^\infty T^{\star n} \star S_{\residuesstar} \star P_{\residuesstar} \star ( \unit \circ \counit - T) = \sum_{n=0}^\infty T^{\star n} - \sum_{n=1}^\infty T^{\star n} = \unit \circ \counit. \end{gather*}
Analogously, we can construct a right inverse $S'$ of $\id$ by setting $T' :=\unit \circ \counit - \id \star S_{\residuesstar}$ and $S' := S_{\residuesstar} \star \sum_{n=0}^\infty (T')^{\star n}$. Both inverses must agree, because $S \star \id \star S' = (S \star \id) \star S' = S \star (\id \star S') = S = S'$.
\end{proof}
\begin{crll}
\label{crll:antipode_chargroup}
$\chargroup{\Gaul}{\mathcal{A}}$ is a group. For every algebra homomorphism
$\phi \in \chargroup{\Gaul}{\mathcal{A}}$ there exists an inverse $S^\phi = \phi \circ S$, that fulfills $S^\phi \star \phi = \unit_\mathcal{A} \circ \counit_{\Gaul}$, where $\unit_\mathcal{A}$ is the unit of the algebra $\mathcal{A}$ and $\counit_{\Gaul}$ the counit of $\Gaul$.
\end{crll}
\begin{proof}
To verify this observe that $S^\phi \star \phi = m_\mathcal{A} \circ ( S^\phi \otimes \phi ) \circ \Delta = m_\mathcal{A} \circ ( (\phi \circ S) \otimes \phi ) \circ \Delta = \phi \circ m_{\Gaul} \circ ( S \otimes \id ) \circ \Delta = \phi \circ ( S \star \id) = \phi \circ \unit_{\Gaul} \circ \counit_{\Gaul} = \unit_\mathcal{A} \circ \counit_{\Gaul}$. 
\end{proof}
\nomenclature{$S^\phi$}{Inverse of $\phi \in \chargroup{\Gaul}{\mathcal{A}}$ such that $S^\phi \star \phi = \unit_\mathcal{A} \circ \counit_{\Gaul}$}

\section{Quotient algebras and Hopf ideals}

To put this construction into action, we will presume that we are given another set of graphs $\sgset \subset \Gul$. If this subset fulfills the following conditions, it will give rise to a \textit{Hopf ideal} of $\Gaul$.
\begin{defn}[Admissible graph set]
\label{def:admissible_graphs}
We will call such a subset $\sgset \subset \Gul$ \textit{admissible} if it fulfills the conditions: 
\begin{enumerate}
\item The set $\sgset$ is a \textit{component closed graph set}. This means for all pairs of graphs $\Gamma_1, \Gamma_2 \in \Gul$, the following statements are equivalent:
\begin{align*} \Gamma_1, \Gamma_2 \in \sgset \text{ iff } \Gamma_1 \sqcup \Gamma_2 \in \sgset. \end{align*}
\item
The set $\sgset$ is $\sgset$-insertion/contraction closed (Definition \ref{def:insertion_contraction_closed}). That means for $\gamma \subset \Gamma$ with any graph $\Gamma \in \Gul$ and subgraphs $\gamma \in \sgset$, the following statements are equivalent:
\begin{align*} \Gamma \in \sgset \text{ iff } \Gamma/\gamma \in \sgset, \end{align*}
which means that $\sgset$ is closed under contraction and insertion of graphs from $\sgset$.
\item All residues $\residuesstar \subset \Gul$ are included in $\sgset$, $\residuesstar \subset \sgset$.
\end{enumerate}
\end{defn}
\nomenclature{$\sgset$}{An admissible graph subset of $\Gul$}

Each such admissible graph set gives rise to a \textit{Hopf ideal} of $\Gaul$.

\begin{defn}[Hopf ideal]
\mbox{}
\begin{enumerate}
\item
A subspace $I \subset \Gaul$ is a (two sided) ideal of $\Gaul$ if
\begin{align} \label{eqn:ideal_def} m (I \otimes \Gaul) \subset I \qquad \text{ and } \qquad m (\Gaul \otimes I) \subset I. \end{align}
\item
A subspace $I \subset \Gaul$ is a (two sided) coideal of $\Gaul$ if $\counit(I) = 0$ and
\begin{align} \label{eqn:coideal_def} \Delta I \subset I \otimes \Gaul + \Gaul \otimes I \end{align}
\item A subspace $I \subset \Gaul$ is a biideal of $\Gaul$ if it is an ideal and a coideal.
\item A subspace $I \subset \Gaul$ is a Hopf ideal if it is a biideal and $S(I) \subset I$.
\end{enumerate}
\end{defn}
The last point is implied by the third if $\Gaul$ is a commutative algebra. By the definition of the antipode $m \circ ( \id \otimes S) \circ \Delta I = \unit \circ \counit(I)$, which is equivalent to $I S(\Gaul) = S(I) \Gaul$ if $I$ is a biideal. It follows that $S(I) \subset I$.

\begin{prop}
\label{prop:restriction_ideal}
Let $I_\sgset \subset \Gaul$ be the span over generators $\Gamma \in \Gul$ that are in the complement of $\sgset$. That means $\Gamma$ is a generator of $I_\sgset$ if $\Gamma \in \Gul$ and $\Gamma \notin \sgset$. The subspace $I_\sgset$ is a \textit{Hopf ideal} of $\Gaul$.
\end{prop}
\nomenclature{$I_\sgset$}{Hopf ideal associated to an admissible graph subset of $\Gul$}
\begin{proof}
Consider the product of two generators $\Gamma_1, \Gamma_2 \in \Gul$ such that $\Gamma_1$ is also a generator of $I_\sgset$.
By Definition \ref{def:admissible_graphs}, it follows from $\Gamma_1,\Gamma_2 \in \Gul$ and $\Gamma_1 \notin \sgset$ that $\Gamma_1 \sqcup \Gamma_2 \notin \sgset$. Therefore, $I_\sgset$ is an ideal of $\Gaul$. 

As all elements of $\residuesstar$ are in $\sgset$, $\counit(I_\sgset) = 0$.

To prove eq. \eqref{eqn:coideal_def} start with Definition \ref{def:coproduct} of the coproduct $\Delta \Gamma = \sum_{\gamma \subset \Gamma} \gamma \otimes \Gamma/\gamma$ for all $\Gamma \in \Gul \setminus \sgset$, which are generators of $I_\sgset$. 

Suppose that there is a subgraph $\gamma \subset \Gamma$ such that both $\gamma$ and $\Gamma/\gamma$ were in $\sgset$. Such a subgraph would violate condition (2) from Definition \ref{def:admissible_graphs}.
Therefore, either $\gamma \otimes \Gamma / \gamma \in \Gaul \otimes I_\sgset$ or $\gamma \otimes \Gamma / \gamma \in I_\sgset \otimes \Gaul$ and $I_\sgset$ is a Hopf ideal as $\Gaul$ is a commutative algebra.
\end{proof}

\begin{defn}[Restricted graph Hopf algebras]
\label{def:restricted_graph_hopf}
Because $I_\sgset$ is a Hopf ideal the \textit{quotient} $\Gaul_\sgset:= \Gaul/I_\sgset$ will again be a Hopf algebra \cite{sweedler1969hopf}. The coproduct on this quotient $\Gaul_\sgset$ has the form
\begin{align*} &\Delta_\sgset& &:& &\Gaul_\sgset& &\rightarrow& \Gaul_\sgset &\otimes \Gaul_\sgset, \\ &&&& &\Gamma& &\mapsto& \sum_{\substack{\gamma \subset \Gamma\\\gamma \in \sgset}} \gamma &\otimes \Gamma/\gamma, \end{align*}
where the sum runs over all subgraphs that have a representative in $\sgset$.

\end{defn}
\nomenclature{$\Gaul_\sgset$}{Quotient Hopf algebra associated to an admissible graph subset of $\Gul$}
Alternatively, we may define a \textit{Hopf algebra homomorphism} $P_\sgset: \Gaul \rightarrow \Gaul_\sgset$ that projects to generators in $\Gaul_\sgset$. Clearly, $\ker P_\sgset = I_\sgset$.
\begin{defn}[Hopf algebra homomorphism]
A Hopf algebra homomorphism $\phi: \Gaul \rightarrow \mathcal{H}$ from $\Gaul$ to another Hopf algebra $\mathcal{H}$ is an algebra homomorphism that respects the algebra, $\psi \circ m_\Gaul = m_\mathcal{H} \circ (\phi \otimes \phi)$, as well as the coalgebra structure, $\Delta_\mathcal{H} \circ \phi = (\phi \otimes \phi) \circ \Delta_\Gaul$, and the antipode, $\phi \circ S_\Gaul = S_{\mathcal{H}} \circ \phi$.
\end{defn}

\begin{defn}[Comodules and coaction]
\label{def:comodules_coaction}
Moreover, we can interpret $\Gaul_\sgset$ as a \textit{left-comodule} of the algebra $\Gaul$. To do this we simply extend $\Delta_\sgset$ to the whole original Hopf algebra $\Gaul$ promoting it to a \textit{coaction}. We will use the same notation for the coproduct $\Gaul_\sgset \rightarrow \Gaul_\sgset \otimes \Gaul_\sgset$ and the coaction $\Gaul \rightarrow \Gaul_\sgset \otimes \Gaul$ as the domain should be clear from the context:
\begin{align*} &\Delta_\sgset& &:& &\Gaul& &\rightarrow& \Gaul_\sgset &\otimes \Gaul, \\ &&&& &\Gamma& &\mapsto& \sum_{\substack{\gamma \subset \Gamma\\\gamma \in \sgset}} \gamma &\otimes \Gamma/\gamma, \end{align*}
\end{defn}
\nomenclature{$\Delta_\sgset$}{Coproduct on the quotient Hopf algebra $\Gaul_\sgset$}

\begin{expl}
\label{expl:bridgeless_graphs}
One important subset of graphs is the set of \textit{bridgeless} graphs. A bridge of a graph is an edge whose removal increases the number of connected components of the graph by one. We define the subset $\sgset_{\text{bl}}$ to be the subset of graphs without bridges. Naturally, this subset is closed under disjoint union of graphs. Moreover, we can arbitrarily contract or insert bridgeless graphs into other bridgeless graphs without creating a bridge. The set $\sgset_{\text{bl}}$ therefore is an admissible graph subset, as it fulfills the requirements from Definition \ref{def:admissible_graphs}. 
The respective quotient Hopf algebra $\Gaul_{\sgset_{\text{bl}}}$ is the entry point for the Hopf algebra structure on Feynman diagrams which will be introduced in Chapter \ref{chap:hopf_algebra_of_fg}.

The Hopf algebra $\sgset_{\text{bl}}$ is also called the core Hopf algebra \cite{kreimer2010core}.
\end{expl}
\nomenclature{$\sgset_{\text{bl}}$}{Admissible graph set of bridgeless graphs}

Note that we can iterate this procedure and construct a quotient Hopf algebra of $\Gaul_\sgset$ using an admissible graph subset $\sgset' \subset \sgset$. The set of \textit{superficially divergent graphs} of a quantum field theory will be such a subset of $\sgset_{\text{bl}}$. The associated quotient will be the Hopf algebra of Feynman diagrams. 

The concept of the group of characters carries over naturally to the quotient spaces $\Gaul_\sgset$. The product will be denoted by $\star_\sgset: \chargroup{\Gaul_\sgset}{\mathcal{A}} \times \chargroup{\Gaul_\sgset}{\mathcal{A}} \rightarrow \chargroup{\Gaul_\sgset}{\mathcal{A}}, (\phi, \psi) \mapsto m_{\mathcal{A}} \circ ( \phi \otimes \psi ) \circ \Delta_\sgset$. The antipode on $\Gaul_\sgset$ will be denoted as $S_\sgset$. As before, we will write the inverse of an element $\phi \in \chargroup{\Gaul_\sgset}{\mathcal{A}}$ as the antipode with $\phi$ in the superscript, $S_\sgset^\phi := \phi \circ S_\sgset$ and $S_\sgset^\phi \star_\sgset \phi = \phi \star_\sgset S_\sgset^\phi = \unit_{\Gaul_\sgset} \circ \counit_{\Gaul_\sgset}$.
\nomenclature{$\star_\sgset$}{Convolution product on $\chargroup{\Gaul_\sgset}{\mathcal{A}}$}
\nomenclature{$S_\sgset$}{Antipode on $\Gaul_\sgset$}
\nomenclature{$S_\sgset^\phi$}{Inverse of $\phi \in \chargroup{\Gaul_\sgset}{\mathcal{A}}$ such that $S_\sgset^\phi \star_\sgset \phi = \unit_{\Gaul_\sgset} \circ \counit_{\Gaul_\sgset}$}

Using the coaction from Definition \ref{def:comodules_coaction}, we can also extend the definition of the $\star$-product to include products of the form $\star_\sgset: \chargroup{\Gaul_\sgset}{\mathcal{A}} \times \chargroup{\Gaul}{\mathcal{A}} \rightarrow \chargroup{\Gaul}{\mathcal{A}}$, where $\psi \star_\sgset \phi := m_\mathcal{A} \circ (\psi \otimes \phi) \circ \Delta_\sgset$ for $\psi \in \chargroup{\Gaul_\sgset}{\mathcal{A}}$ and $\phi \in \chargroup{\Gaul}{\mathcal{A}}$. Strictly speaking, this construction gives us a \textit{left $\chargroup{\Gaul_\sgset}{\mathcal{A}}$-module} over the group $\chargroup{\Gaul}{\mathcal{A}}$ with $\star_\sgset$ as a group action.

\section{Action on algebra homomorphisms}

With the quotient groups $\Gaul_\sgset$ and Definition \ref{def:insertion_contraction_closed} of insertion/contraction closed graph sets in hand, we can formulate an extended version of Theorem \ref{thm:coproduct_full_identity}.

\begin{thm}
\label{thm:coproduct_full_identity_refined}
For every $\sgset$-insertion/contraction closed class of graphs $\subclass \subset \Gul$, we have the identity
\begin{align} \Delta_\sgset \allG_\subclass = \sum_{\Gamma \in \subclass} \left( \prod_{v \in V_{\Gamma}} (\deg{v}_\Gamma!) \allG^{(v)}_\sgset \right) \otimes \frac{\Gamma}{|\Aut \Gamma|}, \end{align}
where
\begin{align} \allG_\subclass := \sum_{\Gamma \in \subclass} \frac{\Gamma}{|\Aut \Gamma|} \end{align}
and 
\begin{align} \allG^{(v)}_\sgset := \sum_{\substack{ \Gamma \in \sgset\\\res{\Gamma} = v }} \frac{\Gamma}{|\Aut \Gamma|}. \end{align}
\end{thm}
\begin{proof}
Observe that it follows from Definitions \ref{def:restricted_graph_hopf}, \ref{def:comodules_coaction} and \ref{def:admissible_graphs} that $(P_\sgset \otimes \id) \circ \Delta = \Delta_\sgset$, where $P_\sgset: \Gaul \rightarrow \Gaul_\sgset$ is the projection from $\Gaul$ to the subspace $\Gaul_\sgset$.
Because $\subclass$ is $\sgset$-insertion/contraction closed we additionally have $\Delta_\sgset \circ P_{\subclass} = (\id \otimes P_{\subclass}) \circ \Delta_\sgset$. Using this together with Theorem \ref{thm:coproduct_full_identity} gives the statement.
\end{proof}

Using this theorem, we may express the convolution products of characters in closed form. 
Take an algebra homomorphisms $\phi \in \chargroup{\Gaul}{R}$ from $\Gaul$ to some ring $R$, for instance some ring of power series, and $\psi \in \chargroup{\Gaul_\sgset}{R}$ some algebra homomorphism from the quotient Hopf algebra $\Gaul_\sgset$ to $R$. 

We will be interested in convolution products of the form 
\begin{align*} \psi \star_\sgset \phi \end{align*}
and specifically their evaluations of a vector such as $\allG_\subclass$: $\psi \star_\sgset \phi(\allG_\subclass)$. 

Applying Theorem \ref{thm:coproduct_full_identity_refined} and the definition of the convolution product in this case gives,
\begin{align*} (\psi \star_\sgset \phi ) ( \allG_\subclass )= \sum_{\Gamma \in \subclass} \left( \prod_{v \in V_{\Gamma}} (\deg{v}_\Gamma!) \psi \left( \allG^{(v)}_\sgset \right) \right) \frac{\phi (\Gamma)}{|\Aut \Gamma|}. \end{align*}
If we have an expression for the weighted generating function of $\phi(\allG_\subclass)$ with marked degrees of the vertices, for instance,
\begin{align*} f_{\subclass}^{\phi}(\lambda_0, \lambda_1, \ldots) := \sum_{\Gamma \in \subclass} \frac{\phi(\Gamma) \prod_{v \in V_\Gamma} \lambda_{\deg{v}_\Gamma} }{|\Aut \Gamma| }, \end{align*}
then we can express the evaluation of the convolution product $(\psi \star_\sgset \phi ) ( \allG_\subclass )$ as a multivariate composition of power series:
\begin{align*} \psi \star_\sgset \phi ( \allG_\subclass ) = f_{\subclass}^{\phi} \left( (0!) \psi \left( \allG^{(v_0)}_\sgset \right), (1!) \psi \left( \allG^{(v_1)}_\sgset \right), (2!) \psi \left( \allG^{(v_2)}_\sgset \right), \ldots \right), \end{align*}
where $v_k$ is the vertex of degree $k$.

\begin{expl}
In a couple of cases, we have such a closed form expression for $f_\subclass^{\phi}$. 
For instance, let $\zeta, \sk$ and $\re$ be the algebra homomorphisms from Example \ref{expl:phi_decomp}. 
Observe that $\zeta \star \re(\Gamma) = \varphi_c^{|\legs_\Gamma|}$ for all $\Gamma \in \Gul$. Therefore,
\begin{gather*} f_{\Gul}^{\zeta \star \re} (\lambda_0, \lambda_1, \ldots)= \sum_{\Gamma \in \Gul} \frac{\zeta \star \re(\Gamma) \prod_{v \in V_\Gamma} \lambda_{\deg{v}_\Gamma} }{|\Aut \Gamma| } \\ = \sum_{\Gamma \in \Gul} \frac{\varphi_c^{|\legs_\Gamma|} \prod_{v \in V_\Gamma} \lambda_{\deg{v}_\Gamma} }{|\Aut \Gamma| } = \sum_{m\geq 0}m! [x^m y^m] e^{\frac{y^2}{2} + \varphi_c y } e^{\sum_{d\geq 0} \frac{\lambda_d}{d!} x^d }, \end{gather*}
where the last equality follows from Corollary \ref{crll:counting_unlabelled_with_degrees}.

If we additionally have an admissible graph set $\sgset$ and an algebra homomorphism $\psi: \Gaul_\sgset \rightarrow \Q[[\varphi_c,\lambda_0, \lambda_1, \ldots]]$, then
\begin{gather*} \psi \star_\sgset \zeta \star \re ( \allG ) = f_{\Gul}^{\zeta \star \re} \left( (0!) \psi \left( \allG^{(v_0)}_\sgset \right), (1!) \psi \left( \allG^{(v_1)}_\sgset \right), (2!) \psi \left( \allG^{(v_2)}_\sgset \right), \ldots \right) \\ = \sum_{m\geq 0}m! [x^m y^m] e^{\frac{y^2}{2} + \varphi_c y } e^{\sum_{d\geq 0} \psi \left( \allG^{(v_d)}_\sgset \right) x^d }. \end{gather*}

Because the set of residues must always be included in $\sgset$, $\residuesstar \subset \sgset$, we can interpret the algebra homomorphism 
$\sk$ from Example \ref{expl:phi_decomp} as an element of $\chargroup{\Gaul_\sgset}{\Q[[\varphi_c,\lambda_0, \lambda_1,\ldots]]}$. Substituting $\psi$ with $\sk$ therefore results in 
\begin{gather*} \sk \star_\sgset \zeta \star \re ( \allG ) = \sum_{m\geq 0}m! [x^m y^m] e^{\frac{y^2}{2} + \varphi_c y } e^{\sum_{d\geq 0} \sk \left( \allG^{(v_d)}_\sgset \right) x^d } \\ = \sum_{m\geq 0}m! [x^m y^m] e^{\frac{y^2}{2} + \varphi_c y } e^{\sum_{d\geq 0} \frac{\lambda_d}{d!} x^d }, \end{gather*}
because $\sk \left( \allG^{(v_d)}_\sgset \right) = \sk( \frac{v_d}{d!} ) = \frac{\lambda_d}{d!}$, where $v_d$ is the single vertex of degree $d$. In the light of Corollary \ref{crll:counting_unlabelled_with_degrees} and Example \ref{expl:phi_decomp} this is of course obvious, as 
$\sk \star_\sgset \zeta \star \re( \Gamma) = \varphi_c^{|\legs_\Gamma|} \prod_{v\in V_\Gamma} \lambda_{\deg{v}_\Gamma}$ for all $\Gamma \in \Gul$, but it gives a first illustration of the workings of this formalism.
\end{expl}

\section{Projections to graphs without given subgraphs}
We can now use the Hopf algebra structure on $\Gaul_\sgset$ to obtain an algebra homomorphism from $\Gaul$ to some other algebra that annihilates generators in $\sgset$. For simplicity, let $\zeta: \Gaul \rightarrow \Q, \Gamma \mapsto 1$ be characteristic map\footnote{Note, that $\zeta$ does not exist on all elements of $\Gaul$, as $\Gaul$ is an infinite dimensional vector space without restriction on its elements. We will only be interested in the image of single generators in this case. Later, we will convolute $\zeta$ with other characters to make it well-defined on all elements of $\Gaul$. This operation can be seen as an instance of \textit{renormalization}.}
 from $\Gaul$ to $\Q$.

This map is an element in $\chargroup{\Gaul}{\Q}$. Obviously, we can restrict $\zeta$ to elements in $\Gaul_\sgset$, $\zeta|_\sgset: \Gaul_\sgset \rightarrow \Q$ and $\zeta|_\sgset \in \chargroup{\Gaul_\sgset}{\Q}$. As $\Gaul_\sgset$ is a Hopf algebra, $\chargroup{\Gaul_\sgset}{\Q}$ is a group and the inverse of $\zeta|_\sgset$ is given by $S^{\zeta|_\sgset}_\sgset = \zeta|_\sgset \circ S_\sgset$, where $S_\sgset$ is the antipode of the Hopf algebra $\Gaul_\sgset$. By Corollary \ref{crll:antipode_chargroup}, the inverse fulfills the convolution identity $S^{\zeta|_\sgset}_\sgset \star_\sgset \zeta |_\sgset = \unit_\Q \circ \counit_{\Gaul_\sgset}$. By Definition \ref{def:coproduct} of the counit, $\unit_\mathcal{\Q} \circ \counit_{\Gaul_\sgset}$ vanishes%
\footnote{We use the notation $\unit_{\Q}$ for the identity function $\Q \rightarrow \Q, q \mapsto q$ to agree with the previous notation.} 
on all generators of $\sgset$ except for the residues $\residuesstar$. 

As $\zeta$ is also an element of $\chargroup{\Gaul}{\Q}$, we can evaluate the product $S^{\zeta|_\sgset}_\sgset \star_\sgset \zeta$ to get a new algebra homomorphism in $\chargroup{\Gaul}{\Q}$. The domain of the map $S^{\zeta|_\sgset}_\sgset \star_\sgset \zeta$ is $\Gaul$ and it annihilates generators of $\sgset$. As 
\begin{gather*} (S^{\zeta|_\sgset}_\sgset \star_\sgset \zeta ) \big|_\sgset = S^{\zeta|_\sgset}_\sgset \star_\sgset \zeta|_\sgset=\unit_\mathcal{\Q} \circ \counit_{\Gaul_\sgset}. \end{gather*}

We consider a subgraph to be \textit{non-trivial} if it has at least one edge and therefore is not a residue. 
For $\Gamma \in \Gul$, $S^{\zeta|_\sgset}_\sgset \star_\sgset \zeta(\Gamma) = \zeta(\Gamma)$ if 
$\Gamma$ does not have any non-trivial subgraphs in $\sgset$, because 
\begin{align*} S^{\zeta|_\sgset}_\sgset \star_\sgset \zeta(\Gamma) = \sum_{\substack{\gamma \subset \Gamma\\\gamma \in \sgset}} S^{\zeta|_\sgset}_\sgset(\gamma) \zeta(\Gamma/\gamma) = S^{\zeta|_\sgset}_\sgset(\skl(\Gamma)) \zeta(\Gamma) = \zeta(\Gamma), \end{align*}
where only the empty and therefore trivial subgraph without edges was included in the sum.

We generally do not know how $S^{\zeta|_\sgset}_\sgset \star_\sgset \zeta$ acts on graphs that are not in $\sgset$, but contain a subgraph from $\sgset$. In general, the map $S^{\zeta|_\sgset}_\sgset \star_\sgset \zeta$ will not annihilate also these graphs in $\Gul$. But for certain cases of $\sgset$, we can guarantee that $S^{\zeta|_\sgset}_\sgset \star_\sgset \zeta(\Gamma)$ vanishes if $\Gamma$ has a non-trivial subgraph from $\sgset$.

\begin{defn}[Counting admissible graph set]
\label{def:counting_admissible_graphs}
We will call a subset $\sgset \subset \Gul$ \textit{counting admissible} if it, additionally to the conditions of admissibility from Definition \ref{def:admissible_graphs}, fulfills: 

For all subgraphs $\gamma_1,\gamma_2 \subset \Gamma$ of any graph $\Gamma \in \Gul$, we have
\begin{align*} \text{If }\gamma_1, \gamma_2 \in \sgset \text{ then } \gamma_1 \cup \gamma_2 \in \sgset. \end{align*}
\end{defn}
Note that this condition differs from condition (1) of Definition \ref{def:admissible_graphs} as we require the union of two subgraphs to be in $\sgset$ even if they share an edge. 

\begin{thm}
\label{thm:counting_and_projecting}
If $\sgset$ is a counting admissible graph set and $\zeta \in \chargroup{\Gaul}{\Q}$ is the characteristic function $\zeta: \Gaul \rightarrow \Q, \Gamma \mapsto 1$, then $S^{\zeta|_\sgset}_\sgset \star_\sgset \zeta \in \chargroup{\Gaul}{\Q}$ and
\begin{align} S^{\zeta|_\sgset}_\sgset \star_\sgset \zeta (\Gamma) &= \begin{cases} 1& \text{ if $\Gamma \in \residuesstar$ or $\Gamma$ does not contain any non-trivial subgraph from $\sgset$.}\\ 0& \text{ if $\Gamma \notin \residuesstar$ and $\Gamma$ has a non-trivial subgraph from $\sgset$.} \end{cases} \end{align}
where $S^{\zeta|_\sgset}_\sgset$ is the inverse of the restricted algebra homomorphism $\zeta|_\sgset$ in the group $\chargroup{\Gaul_\sgset}{\Q}$, which can be expressed using the antipode $S_\sgset$ of the Hopf algebra $\Gaul_\sgset$: $S^{\zeta|_\sgset}_\sgset = \zeta|_\sgset \circ S_\sgset$.
\end{thm}
\begin{proof}
As already stated, $S^{\zeta|_\sgset}_\sgset \star_\sgset \zeta (\Gamma) = 1$ if $\Gamma$ does not have any non-trivial subgraphs in $\sgset$. Moreover, if $\Gamma \in \sgset$ and $\Gamma \notin \residuesstar$ then 
$S^{\zeta|_\sgset}_\sgset \star_\sgset \zeta(\Gamma) = S^{\zeta|_\sgset}_\sgset \star_\sgset \zeta|_\sgset (\Gamma) = \unit_{\Gaul_\sgset} \circ \counit_{\Gaul_\sgset}(\Gamma)= 0$.
It is left to prove that $S^{\zeta|_\sgset}_\sgset \star_\sgset \zeta (\Gamma)$ vanishes if $\Gamma$ has a non-trivial subgraph from $\sgset$. 

By the definition of the $\star_\sgset$ product
\begin{align*} S^{\zeta|_\sgset}_\sgset \star_\sgset \zeta (\Gamma) = \sum_{\substack{ \gamma \subset \Gamma\\ \gamma \in \sgset }} S^{\zeta|_\sgset}_\sgset(\gamma) \zeta( \Gamma / \gamma ) = \sum_{\substack{ \gamma \subset \Gamma\\ \gamma \in \sgset }} S^{\zeta|_\sgset}_\sgset(\gamma). \end{align*}
Because $\sgset$ is counting admissible, the union of all relevant $\sgset$-subgraphs of $\Gamma$ is in $\sgset$, 
\begin{align*} \widetilde{\Gamma} := \bigcup_{\substack{ \gamma \subset \Gamma\\ \gamma \in \sgset }} \gamma \in \sgset. \end{align*}
As $\widetilde{\Gamma}$ contains all relevant subgraphs of $\Gamma$, it follows that
\begin{align*} \sum_{\substack{ \gamma \subset \Gamma\\ \gamma \in \sgset }} S^{\zeta|_\sgset}_\sgset(\gamma) = \sum_{\substack{ \gamma \subset \widetilde{\Gamma}\\ \gamma \in \sgset }} S^{\zeta|_\sgset}_\sgset(\gamma) = S^{\zeta|_\sgset}_\sgset \star_\sgset \zeta (\widetilde{\Gamma}) = 0, \end{align*}
where the last equality follows because $\widetilde{\Gamma} \in \sgset$.
\end{proof}

Combining the last theorem with the results from the last section enables us to formulate the main result of this chapter. The following theorem gives us access to the generating function of graphs without subgraphs from a counting admissible graph set. 

\begin{thm}
\label{thm:counting_relation_hopf}
If $\sgset$ is a counting admissible graph set and $\subclass$ is a $\sgset$-inser\-tion/con\-traction closed subset of graphs, then
\begin{gather} \begin{gathered} \label{eqn:counting_equation_complete} g_{\subclass} \left(\varphi_c, \lambda_0, \lambda_1, \ldots \right) = f_{\subclass} \left(\varphi_c, (0!) \psi \left( \allG^{(v_0)}_\sgset \right), (1!) \psi \left( \allG^{(v_1)}_\sgset \right), (2!) \psi \left( \allG^{(v_2)}_\sgset \right), \ldots \right) \end{gathered} \\ \label{eqn:counting_equation_complete2} f_{\subclass} \left(\varphi_c, \lambda_0, \lambda_1, \ldots \right) = g_{\subclass} \left(\varphi_c, (0!) \phi \left( \allG^{(v_0)}_\sgset \right), (1!) \phi \left( \allG^{(v_1)}_\sgset \right), (2!) \phi \left( \allG^{(v_2)}_\sgset \right), \ldots \right) \end{gather}
with the generating functions 
\begin{align} \label{eqn:counting_f_class_defn} f_{\subclass} \left(\varphi_c, \lambda_0, \lambda_1, \ldots \right) &:= \sum_{\Gamma \in \subclass} \frac{ \varphi_c^{|\legs_\Gamma|} \prod_{v\in V_\Gamma} \lambda_{\deg{v}}} {|\Aut \Gamma|} \\ \label{eqn:counting_f_class_defn2} g_{\subclass} \left(\varphi_c, \lambda_0, \lambda_1, \ldots \right) &:= \sum_{\substack{ \Gamma \in \subclass\\ \text{ such that $\Gamma$ has}\\\text{no non-trivial subgraph from $\sgset$ } } } \frac{ \varphi_c^{|\legs_\Gamma|} \prod_{v \in V_\Gamma} \lambda_{\deg{v}} }{|\Aut \Gamma|}, \intertext{and the characters} \psi &:= \sk \star_\sgset S^{\zeta|_\sgset}_\sgset, \\ \phi &:= \sk \star_\sgset \zeta|_\sgset, \end{align}
as well as $\sk$ and $\zeta$,
\begin{align*} \zeta&:& &\Gaul \rightarrow \Q[[\varphi_c, \lambda_0, \lambda_1, \ldots]], & &\Gamma \mapsto 1 \\ \sk&: &&\Gaul \rightarrow \Q[[\varphi_c, \lambda_0, \lambda_1, \ldots]], &&\Gamma \mapsto \begin{cases} \prod_{v \in V_\Gamma} \lambda_{\deg{v}_\Gamma}& \text{ if $\Gamma \in \residuesstar$}\\ 0& \text{ else} \end{cases} \end{align*}
defined as in Example \ref{expl:phi_decomp} and $S^{\zeta|_\sgset}_\sgset = \zeta|_\sgset \circ S_\sgset$ as in Theorem \ref{thm:counting_and_projecting}.
\end{thm}
\begin{proof}
Consider the convolution product
\begin{align*} \sk \star_\sgset S^{\zeta|_\sgset}_\sgset \star_\sgset \zeta \star \re, \end{align*}
where $\re$ is defined as in Example \ref{expl:phi_decomp},
\begin{align*} \re&: &&\Gaul \rightarrow \Q[[\varphi_c, \lambda_0, \lambda_1, \ldots]], &&\Gamma \mapsto \begin{cases} \varphi_c^{|\legs_\Gamma|}& \text{ if $\Gamma \in \residuesstar$}\\ 0& \text{ else} \end{cases} \end{align*}
By the same reasoning as in Example \ref{expl:phi_decomp}, we see that 
\begin{gather*} \sk \star_\sgset S^{\zeta|_\sgset}_\sgset \star_\sgset \zeta \star \re (\Gamma) = \sk( \skl(\Gamma) ) (S^{\zeta|_\sgset}_\sgset \star_\sgset \zeta)(\Gamma) \re(\res(\Gamma)). \end{gather*}
From Theorem \ref{thm:counting_and_projecting} it follows that
\begin{gather} \begin{gathered} \label{eqn:projection_identity_hopf_explicit} \sk \star_\sgset (S^{\zeta|_\sgset}_\sgset \star_\sgset \zeta) \star \re (\Gamma) \\ = \begin{cases} \varphi_c^{|\legs_\Gamma|} \prod_{v\in V_\Gamma} \lambda_{\deg{v}} & \text{ if $\Gamma$ has no non-trivial subgraph from $\sgset$.}\\ 0& \text{ else } \end{cases} \end{gathered} \end{gather}
From this and eq. \eqref{eqn:counting_f_class_defn2} it follows directly that the 
$\sk \star_\sgset S^{\zeta|_\sgset}_\sgset \star_\sgset \zeta \star \re ( \allG_\subclass )$ equals 
the left hand side of eq.\ \eqref{eqn:counting_equation_complete}.

Using the associativity of the convolution product we may apply Theorem \ref{thm:coproduct_full_identity_refined} to obtain
\begin{align*} (\sk \star_\sgset S^{\zeta|_\sgset}_\sgset) \star_\sgset (\zeta \star \re) ( \allG_\subclass ) &= \sum_{\Gamma \in \subclass} \left( \prod_{v \in V_{\Gamma}} (\deg{v}_\Gamma!) (\sk \star_\sgset S^{\zeta|_\sgset}_\sgset ) \left( \allG^{(v)}_\sgset \right) \right) \frac{(\zeta \star \re) (\Gamma)}{|\Aut \Gamma|} \\ &= \sum_{\Gamma \in \subclass} \left( \prod_{v \in V_{\Gamma}} (\deg{v}_\Gamma!) (\sk \star_\sgset S^{\zeta|_\sgset}_\sgset ) \left( \allG^{(v)}_\sgset \right) \right) \frac{\varphi_c^{|\legs_\Gamma|}}{|\Aut \Gamma|}, \end{align*}
which by comparison to eq.\ \eqref{eqn:counting_f_class_defn} shows that $\sk \star_\sgset S^{\zeta|_\sgset}_\sgset \star_\sgset \zeta \star \re ( \allG_\subclass )$ is equal to the right hand side of eq.\ \eqref{eqn:counting_equation_complete}.

Analogously, eq. \eqref{eqn:counting_equation_complete2} follows from equivalent ways of evaluating 
\begin{align*} \sk \star_\sgset \zeta|_\sgset \star_\sgset S^{\zeta|_\sgset}_\sgset \star_\sgset \zeta \star \re ( \allG_\subclass ). \end{align*}
As $\zeta|_\sgset \star_\sgset S^{\zeta|_\sgset}_\sgset = \unit_{\Q} \circ \counit_{\Gaul_\sgset}$ is the neutral element of the group $\chargroup{\Gaul_\sgset}{\Q}$, 
\begin{align*} \sk \star_\sgset \zeta|_\sgset \star_\sgset S^{\zeta|_\sgset}_\sgset \star_\sgset \zeta \star \re ( \allG_\subclass ) = \sk \star_\sgset \zeta \star \re ( \allG_\subclass ) = f_{\subclass} \left(\varphi_c, \lambda_0, \lambda_1, \ldots \right). \end{align*}
Moreover, by different bracketing, an application of Theorem \ref{thm:coproduct_full_identity_refined} and due to eq. \eqref{eqn:projection_identity_hopf_explicit}, 
\begin{gather*} (\sk \star_\sgset \zeta|_\sgset) \star_\sgset (S^{\zeta|_\sgset}_\sgset \star_\sgset \zeta \star \re) ( \allG_\subclass ) \\ = \sum_{\Gamma \in \subclass} \left( \prod_{v \in V_{\Gamma}} (\deg{v}_\Gamma!) (\sk \star_\sgset \zeta|_\sgset ) \left( \allG^{(v)}_\sgset \right) \right) \frac{(S^{\zeta|_\sgset}_\sgset \star_\sgset \zeta \star \re) (\Gamma)}{|\Aut \Gamma|} \\ = \sum_{\substack{ \Gamma \in \subclass\\ \text{ such that $\Gamma$ has}\\\text{no non-trivial subgraph from $\sgset$ } } } \left( \prod_{v \in V_{\Gamma}} (\deg{v}_\Gamma!) (\sk \star_\sgset \zeta|_\sgset ) \left( \allG^{(v)}_\sgset \right) \right) \frac{\varphi_c^{|\legs_\Gamma|}}{|\Aut \Gamma|}, \end{gather*}
which is equivalent to the right hand side of eq.\ \eqref{eqn:counting_equation_complete2}.
\end{proof}

\section{The Legendre transformation and bridgeless graphs}
\label{sec:legendre_transformation}
As an example, we will apply Theorem \ref{thm:counting_relation_hopf} to the set of bridgeless graphs and show that this application can be interpreted as a \textit{Legendre transformation}. In Example \ref{expl:bridgeless_graphs} the set of bridgeless graphs $\sgset_{\text{bl}}$ was introduced. This set of graphs is of importance as it will form the foundation for the Hopf algebra of Feynman diagrams in the next chapter. It is clear that the union of two arbitrary bridgeless subgraphs is again a bridgeless subgraph. Therefore, $\sgset_{\text{bl}}$ is counting admissible: It fulfills the conditions introduced in Definition \ref{def:counting_admissible_graphs}.

We will start in the contraction closed subset of connected graphs without external legs $\subclass_{ \ifmmode \usebox{\fgsimplenullvtx} \else \newsavebox{\fgsimplenullvtx} \savebox{\fgsimplenullvtx}{ \begin{tikzpicture}[x=1ex,y=1ex,baseline={([yshift=-.55ex]current bounding box.center)}] \coordinate (v) ; \filldraw[white] (v) circle (.8); \filldraw (v) circle (1pt); \end{tikzpicture} } \fi} := \{ \Gamma \in \Gul : \res \Gamma =  \ifmmode \usebox{\fgsimplenullvtx} \else \newsavebox{\fgsimplenullvtx} \savebox{\fgsimplenullvtx}{ \begin{tikzpicture}[x=1ex,y=1ex,baseline={([yshift=-.55ex]current bounding box.center)}] \coordinate (v) ; \filldraw[white] (v) circle (.8); \filldraw (v) circle (1pt); \end{tikzpicture} } \fi \}$.

To be specific, the corresponding generating function is 
\begin{align*} f_{\subclass_{ \ifmmode \usebox{\fgsimplenullvtx} \else \newsavebox{\fgsimplenullvtx} \savebox{\fgsimplenullvtx}{ \begin{tikzpicture}[x=1ex,y=1ex,baseline={([yshift=-.55ex]current bounding box.center)}] \coordinate (v) ; \filldraw[white] (v) circle (.8); \filldraw (v) circle (1pt); \end{tikzpicture} } \fi}}(\lambda_0, \lambda_1, \lambda_2, \ldots) = \sum_{\Gamma \in \subclass_{ \ifmmode \usebox{\fgsimplenullvtx} \else \newsavebox{\fgsimplenullvtx} \savebox{\fgsimplenullvtx}{ \begin{tikzpicture}[x=1ex,y=1ex,baseline={([yshift=-.55ex]current bounding box.center)}] \coordinate (v) ; \filldraw[white] (v) circle (.8); \filldraw (v) circle (1pt); \end{tikzpicture} } \fi}} \frac{\prod_{v \in V_\Gamma} \lambda_{\deg{v}} }{| \Aut \Gamma|} = \log \left( \sum_{m\geq 0}m! [x^m y^m] e^{\frac{y^2}{2}} e^{\sum_{d\geq 0} \lambda_d \frac{x^d}{d!} } \right), \end{align*}
which is an obvious specialization of Corollary \ref{crll:counting_unlabelled_with_degrees} and where we do not need to keep track of legs, as the graphs in $\subclass_{ \ifmmode \usebox{\fgsimplenullvtx} \else \newsavebox{\fgsimplenullvtx} \savebox{\fgsimplenullvtx}{ \begin{tikzpicture}[x=1ex,y=1ex,baseline={([yshift=-.55ex]current bounding box.center)}] \coordinate (v) ; \filldraw[white] (v) circle (.8); \filldraw (v) circle (1pt); \end{tikzpicture} } \fi}$ have no legs.

Theorem \ref{thm:counting_relation_hopf} gives us an expression for the generating function of graphs without non-trivial bridgeless subgraphs,
\begin{gather*} g_{\subclass_{ \ifmmode \usebox{\fgsimplenullvtx} \else \newsavebox{\fgsimplenullvtx} \savebox{\fgsimplenullvtx}{ \begin{tikzpicture}[x=1ex,y=1ex,baseline={([yshift=-.55ex]current bounding box.center)}] \coordinate (v) ; \filldraw[white] (v) circle (.8); \filldraw (v) circle (1pt); \end{tikzpicture} } \fi}}(\lambda_0, \lambda_1, \lambda_2, \ldots) = \sum_{\substack{ \Gamma \in \subclass_{ \ifmmode \usebox{\fgsimplenullvtx} \else \newsavebox{\fgsimplenullvtx} \savebox{\fgsimplenullvtx}{ \begin{tikzpicture}[x=1ex,y=1ex,baseline={([yshift=-.55ex]current bounding box.center)}] \coordinate (v) ; \filldraw[white] (v) circle (.8); \filldraw (v) circle (1pt); \end{tikzpicture} } \fi}\\ \text{ such that $\Gamma$ has}\\\text{no non-trivial subgraph from $\sgset_\text{bl}$ } } } \frac{\prod_{v \in V_\Gamma} \lambda_{\deg{v}} }{| \Aut \Gamma|} \\ = f_{\subclass_{ \ifmmode \usebox{\fgsimplenullvtx} \else \newsavebox{\fgsimplenullvtx} \savebox{\fgsimplenullvtx}{ \begin{tikzpicture}[x=1ex,y=1ex,baseline={([yshift=-.55ex]current bounding box.center)}] \coordinate (v) ; \filldraw[white] (v) circle (.8); \filldraw (v) circle (1pt); \end{tikzpicture} } \fi}} \left( (0!) \psi \left( \allG^{(v_0)}_{\sgset_\text{bl}} \right), (1!) \psi \left( \allG^{(v_1)}_{\sgset_\text{bl}} \right), (2!) \psi \left( \allG^{(v_2)}_{\sgset_\text{bl}} \right), \ldots \right) \\ = \log \left( \sum_{m\geq 0}m! [x^m y^m] e^{\frac{y^2}{2}} e^{\sum_{d\geq 0} \psi \left( \allG^{(v_d)}_{\sgset_\text{bl}} \right) x^d } \right), \end{gather*}
where $\psi = \sk \star_{\sgset_\text{bl}} S^{\zeta|_{\sgset_\text{bl}}}_{\sgset_\text{bl}}$.

In many cases for $\sgset$ this equation is sufficient to perform an asymptotic analysis of $g_{\subclass_{ \ifmmode \usebox{\fgsimplenullvtx} \else \newsavebox{\fgsimplenullvtx} \savebox{\fgsimplenullvtx}{ \begin{tikzpicture}[x=1ex,y=1ex,baseline={([yshift=-.55ex]current bounding box.center)}] \coordinate (v) ; \filldraw[white] (v) circle (.8); \filldraw (v) circle (1pt); \end{tikzpicture} } \fi}}(\lambda_0, \lambda_1, \lambda_2, \ldots)$ with the techniques from the last two chapters, but in the present case the generating function $g_{\subclass_{ \ifmmode \usebox{\fgsimplenullvtx} \else \newsavebox{\fgsimplenullvtx} \savebox{\fgsimplenullvtx}{ \begin{tikzpicture}[x=1ex,y=1ex,baseline={([yshift=-.55ex]current bounding box.center)}] \coordinate (v) ; \filldraw[white] (v) circle (.8); \filldraw (v) circle (1pt); \end{tikzpicture} } \fi}}(\lambda_0, \lambda_1, \lambda_2, \ldots)$ is also known explicitly. The set of connected graphs that do not contain a non-trivial bridgeless subgraph is the set of \textit{trees}: Obviously, every tree has a bridge. A connected graph which is not a tree contains at least one cycle. A cycle itself is a non-trivial bridgeless subgraph. 

\begin{lmm}
\label{lmm:trees_generating_function}
The generating function of trees $g_{\subclass_{ \ifmmode \usebox{\fgsimplenullvtx} \else \newsavebox{\fgsimplenullvtx} \savebox{\fgsimplenullvtx}{ \begin{tikzpicture}[x=1ex,y=1ex,baseline={([yshift=-.55ex]current bounding box.center)}] \coordinate (v) ; \filldraw[white] (v) circle (.8); \filldraw (v) circle (1pt); \end{tikzpicture} } \fi}}(\lambda_0, \lambda_1, \lambda_2, \ldots)$, marked by the degrees of their vertices, fulfills the identity,
\begin{align} g_{\subclass_{ \ifmmode \usebox{\fgsimplenullvtx} \else \newsavebox{\fgsimplenullvtx} \savebox{\fgsimplenullvtx}{ \begin{tikzpicture}[x=1ex,y=1ex,baseline={([yshift=-.55ex]current bounding box.center)}] \coordinate (v) ; \filldraw[white] (v) circle (.8); \filldraw (v) circle (1pt); \end{tikzpicture} } \fi}}(\lambda_0, \lambda_1, \lambda_2, \ldots) = \sum_{\substack{\Gamma \in \subclass_{ \ifmmode \usebox{\fgsimplenullvtx} \else \newsavebox{\fgsimplenullvtx} \savebox{\fgsimplenullvtx}{ \begin{tikzpicture}[x=1ex,y=1ex,baseline={([yshift=-.55ex]current bounding box.center)}] \coordinate (v) ; \filldraw[white] (v) circle (.8); \filldraw (v) circle (1pt); \end{tikzpicture} } \fi}\\\text{such that $\Gamma$ is a tree}}} \frac{\prod_{v\in V_\Gamma} \lambda_{\deg{v}}}{|\Aut \Gamma|} = -\frac{\varphi_c^2}{2} + V(\varphi_c), \end{align}
where  $V(x) = \sum_{d=0}^\infty \frac{\lambda_d}{d!} x^d$ and $\varphi_c \in \Q[[\lambda_1, \lambda_2, \lambda_3, \ldots]]$ is the unique power series solution of 
\begin{align} \varphi_c = V'(\varphi_c). \end{align}
\end{lmm}
\begin{proof}
The proof is a standard combinatorial argument for labelled tree counting \cite{flajolet2009analytic}.

The key is to observe that the power series $\varphi_c(\lambda_1, \lambda_2, \ldots)$ counts \textit{rooted trees} - trees with one leg. We can form a rooted tree by joining a set of rooted trees to a vertex while leaving one leg of the vertex free to be the new root. 
Also accounting for symmetry factors gives the equation
\begin{align*} \varphi_c = \lambda_1 + \lambda_2 \varphi_c + \lambda_3 \frac{\varphi_c^2}{2!} + \lambda_4 \frac{\varphi_c^3}{3!} + \ldots = V'(\varphi_c). \end{align*}
This is an implicit equation that can be solved for $\varphi_c(\lambda_1, \lambda_2, \ldots)$ iteratively.

In a similar way, we can obtain the generating function of trees with one vertex fixed. 
A tree with a fixed vertex can be constructed by joining a number of rooted trees together in a vertex. To get a fixed vertex of degree $d$, we have to join $d$ rooted trees together and multiply with $\lambda_d$. Summing over all possible degrees and accounting for symmetry factors gives,
\begin{align*} \lambda_0 + \lambda_1 \varphi_c + \lambda_2 \frac{\varphi_c^2}{2!} + \lambda_3 \frac{\varphi_c^3}{3!} + \ldots = V(\varphi_c). \end{align*}
The expression $V(\varphi_c)$ is therefore the generating function of trees with one fixed vertex. 

By the same reasoning, the expression $\frac{\varphi_c^2}{2}$ counts the number of trees with one edge fixed, which is just the number of pairs of rooted trees where the roots of both rooted trees are joined to an edge. 

For a tree $\Gamma$, we have the identity $|V_\Gamma| - |E_\Gamma| = 1$. Every tree has exactly one more vertex then edges. Therefore,
\begin{align*} -\frac{\varphi_c^2}{2} + V(\varphi_c) &= - \sum_{\substack{\Gamma \in \subclass_{ \ifmmode \usebox{\fgsimplenullvtx} \else \newsavebox{\fgsimplenullvtx} \savebox{\fgsimplenullvtx}{ \begin{tikzpicture}[x=1ex,y=1ex,baseline={([yshift=-.55ex]current bounding box.center)}] \coordinate (v) ; \filldraw[white] (v) circle (.8); \filldraw (v) circle (1pt); \end{tikzpicture} } \fi}\\\text{such that $\Gamma$ is a tree}}} |E_\Gamma| \frac{\lambda_{\deg{v}}}{|\Aut \Gamma|} + \sum_{\substack{\Gamma \in \subclass_{ \ifmmode \usebox{\fgsimplenullvtx} \else \newsavebox{\fgsimplenullvtx} \savebox{\fgsimplenullvtx}{ \begin{tikzpicture}[x=1ex,y=1ex,baseline={([yshift=-.55ex]current bounding box.center)}] \coordinate (v) ; \filldraw[white] (v) circle (.8); \filldraw (v) circle (1pt); \end{tikzpicture} } \fi}\\\text{such that $\Gamma$ is a tree}}} |V_\Gamma| \frac{\lambda_{\deg{v}}}{|\Aut \Gamma|}, \end{align*}
which results in the statement.
\end{proof}

Applying eq. \eqref{eqn:counting_equation_complete2} of Theorem \ref{thm:counting_relation_hopf}, gives
\begin{align*} f_{\subclass_{ \ifmmode \usebox{\fgsimplenullvtx} \else \newsavebox{\fgsimplenullvtx} \savebox{\fgsimplenullvtx}{ \begin{tikzpicture}[x=1ex,y=1ex,baseline={([yshift=-.55ex]current bounding box.center)}] \coordinate (v) ; \filldraw[white] (v) circle (.8); \filldraw (v) circle (1pt); \end{tikzpicture} } \fi}} \left(\lambda_0, \lambda_1, \ldots \right) = g_{\subclass_{ \ifmmode \usebox{\fgsimplenullvtx} \else \newsavebox{\fgsimplenullvtx} \savebox{\fgsimplenullvtx}{ \begin{tikzpicture}[x=1ex,y=1ex,baseline={([yshift=-.55ex]current bounding box.center)}] \coordinate (v) ; \filldraw[white] (v) circle (.8); \filldraw (v) circle (1pt); \end{tikzpicture} } \fi}} \left( (0!) \phi \left( \allG^{(v_0)}_{\sgset_\text{bl}} \right), (1!) \phi \left( \allG^{(v_1)}_{\sgset_\text{bl}} \right), (2!) \phi \left( \allG^{(v_2)}_{\sgset_\text{bl}} \right), \ldots \right), \end{align*}
where $\phi = \sk \star_{\sgset_\text{bl}} \zeta|_{\sgset_\text{bl}}$. An application of Lemma \ref{lmm:trees_generating_function} gives us an implicit expression for the evaluations $\phi \left( \allG^{(v_d)}_{\sgset_\text{bl}} \right)$. 
The generating functions of connected graphs without legs fulfills, 
\begin{align} \label{eqn:legendre_transform_pre} f_{\subclass_{ \ifmmode \usebox{\fgsimplenullvtx} \else \newsavebox{\fgsimplenullvtx} \savebox{\fgsimplenullvtx}{ \begin{tikzpicture}[x=1ex,y=1ex,baseline={([yshift=-.55ex]current bounding box.center)}] \coordinate (v) ; \filldraw[white] (v) circle (.8); \filldraw (v) circle (1pt); \end{tikzpicture} } \fi}} \left(\lambda_0, \lambda_1, \ldots \right) = -\frac{\varphi_c^2}{2} + \widetilde V(\varphi_c), \end{align}
where $\widetilde V(\varphi_c) = \sum_{d\geq 0} \phi \left( \allG^{(v_d)}_{\sgset_\text{bl}} \right) \varphi_c^d$ and $\varphi_c = \widetilde{V}'(\varphi_c)$.

Obviously, $\widetilde V(\varphi_c)$ can be interpreted as the generating function of connected bridgeless graphs with the number of legs marked by $\varphi_c$:
\begin{gather*} \widetilde V(\varphi_c) = \sum_{d\geq 0} \phi \left( \allG^{(v_d)}_{\sgset_\text{bl}} \right) \varphi_c^d = \sum_{d\geq 0} \sum_{\substack{ \Gamma \in \sgset_\text{bl}\\ \res\Gamma =v_d}} \frac{\varphi_c^d \phi( \Gamma ) }{|\Aut \Gamma |} \\ = \sum_{\substack{ \Gamma \in \sgset_\text{bl}\\ |\comps_\Gamma| = 1}} \frac{\varphi_c^{|\legs_\Gamma|} \sk \star_{\sgset_\text{bl}} \zeta|_{\sgset_\text{bl}}( \Gamma ) }{|\Aut \Gamma |} = \sum_{\substack{ \Gamma \in \sgset_\text{bl}\\ |\comps_\Gamma| = 1}} \frac{\varphi_c^{|\legs_\Gamma|} \prod_{v\in V_\Gamma} \lambda_{\deg{v}_\Gamma} }{|\Aut \Gamma |}. \end{gather*}

We can obtain an explicit expression for $\varphi_c$ by taking the derivative of this equation with respect to one of the formal $\lambda_1$ variable. By convention, we give $\lambda_1$ a special name $\lambda_1 =: j$. 

Taking the formal $\frac{\partial}{\partial j}$ derivative of eq. \eqref{eqn:legendre_transform_pre} results in
\begin{align*} \frac{\partial}{\partial j}f_{\subclass_{ \ifmmode \usebox{\fgsimplenullvtx} \else \newsavebox{\fgsimplenullvtx} \savebox{\fgsimplenullvtx}{ \begin{tikzpicture}[x=1ex,y=1ex,baseline={([yshift=-.55ex]current bounding box.center)}] \coordinate (v) ; \filldraw[white] (v) circle (.8); \filldraw (v) circle (1pt); \end{tikzpicture} } \fi}} \left(\lambda_0, j, \lambda_2, \ldots \right) = \frac{\partial \varphi_c}{\partial j }\frac{\partial}{\partial \varphi_c} \left( -\frac{\varphi_c^2}{2} +\widetilde V(\varphi_c) \right) + \sum_{d\geq 0} \left( \frac{\partial}{\partial j} \phi \left( \allG^{(v_d)}_{\sgset_\text{bl}} \right)\right) \varphi_c^d. \end{align*}
The first term on the right hand side vanishes as $\varphi_c = {\widetilde{V}}'(\varphi_c)$. 

The reason for the choice of $\lambda_1$ is that the only connected bridgeless graph in $\sgset_\text{bl}$, which contains a one-valent vertex, is the residue graph $ \ifmmode \usebox{\fgsimpleonevtx} \else \newsavebox{\fgsimpleonevtx} \savebox{\fgsimpleonevtx}{ \begin{tikzpicture}[x=1ex,y=1ex,baseline={([yshift=-.55ex]current bounding box.center)}] \coordinate (v) ; \def \n {1}; \def \rad {1}; \filldraw[white] (v) circle (\rad); \foreach \s in {1,...,\n} { \def \angle {180+360/\n*(\s - 1)}; \coordinate (u) at ([shift=({\angle}:\rad)]v); \draw (v) -- (u); } \filldraw (v) circle (1pt); \end{tikzpicture} } \fi$. All non-trivial connected graphs with such a vertex automatically contain a bridge which joins the one-valent vertex with the rest of the graph.
Therefore, 
\begin{align*} \sum_{d\geq 0} \left( \frac{\partial}{\partial j} \phi \left( \allG^{(v_d)}_{\sgset_\text{bl}} \right)\right) \varphi_c^d = \left( \frac{\partial}{\partial j} \phi \left(\allG^{(v_1)}_{\sgset_\text{bl}} \right)\right) \varphi_c = \left( \frac{\partial}{\partial j} \phi \left( \ifmmode \usebox{\fgsimpleonevtx} \else \newsavebox{\fgsimpleonevtx} \savebox{\fgsimpleonevtx}{ \begin{tikzpicture}[x=1ex,y=1ex,baseline={([yshift=-.55ex]current bounding box.center)}] \coordinate (v) ; \def \n {1}; \def \rad {1}; \filldraw[white] (v) circle (\rad); \foreach \s in {1,...,\n} { \def \angle {180+360/\n*(\s - 1)}; \coordinate (u) at ([shift=({\angle}:\rad)]v); \draw (v) -- (u); } \filldraw (v) circle (1pt); \end{tikzpicture} } \fi\right)\right) \varphi_c = \varphi_c, \end{align*}
because $\sk \star_{\sgset_\text{bl}} \zeta|_{\sgset_\text{bl}}(  \ifmmode \usebox{\fgsimpleonevtx} \else \newsavebox{\fgsimpleonevtx} \savebox{\fgsimpleonevtx}{ \begin{tikzpicture}[x=1ex,y=1ex,baseline={([yshift=-.55ex]current bounding box.center)}] \coordinate (v) ; \def \n {1}; \def \rad {1}; \filldraw[white] (v) circle (\rad); \foreach \s in {1,...,\n} { \def \angle {180+360/\n*(\s - 1)}; \coordinate (u) at ([shift=({\angle}:\rad)]v); \draw (v) -- (u); } \filldraw (v) circle (1pt); \end{tikzpicture} } \fi ) = \lambda_1= j$.

This way, we obtain an explicit expression for $\varphi_c$,
\begin{align*} \varphi_c = \frac{\partial}{\partial j}f_{\subclass_{ \ifmmode \usebox{\fgsimplenullvtx} \else \newsavebox{\fgsimplenullvtx} \savebox{\fgsimplenullvtx}{ \begin{tikzpicture}[x=1ex,y=1ex,baseline={([yshift=-.55ex]current bounding box.center)}] \coordinate (v) ; \filldraw[white] (v) circle (.8); \filldraw (v) circle (1pt); \end{tikzpicture} } \fi}} \left(\lambda_0, j, \lambda_2, \ldots \right). \end{align*}
and we may write eq. \eqref{eqn:legendre_transform_pre} as 
\begin{align*} W(j) = G(\varphi_c) + \varphi_c j, \end{align*} 
where $W(j):= f_{\subclass_{ \ifmmode \usebox{\fgsimplenullvtx} \else \newsavebox{\fgsimplenullvtx} \savebox{\fgsimplenullvtx}{ \begin{tikzpicture}[x=1ex,y=1ex,baseline={([yshift=-.55ex]current bounding box.center)}] \coordinate (v) ; \filldraw[white] (v) circle (.8); \filldraw (v) circle (1pt); \end{tikzpicture} } \fi}} \left(\lambda_0, j, \lambda_2, \ldots \right)$, $G(\varphi_c): = -\frac{\varphi_c^2}{2} + \widetilde V(\varphi_c) - \varphi_c j$ and $\varphi_c = \frac{\partial}{\partial j} W(j)$.

This show that $G$ and $W$ are related by a \textit{Legendre transformation} and the formal variables $j$ and $\varphi_c$ are conjugate variables. Observe that $G(\varphi_c)$ is almost the generating function of bridgeless graphs. 

Some explicit examples of the Legendre transformation in zero-dimensional quantum field theory will be given in Chapter \ref{chap:applications_zerodim}.

A more detailed analysis of the Legendre transformation on trees, which did not exploit the Hopf algebra structure of graphs but highlighted its combinatorial properties, was recently given by Jackson, Kempf and Morales \cite{jackson2016robust}.

In the following chapter we are going to analyze the maps $S^{\zeta|_\sgset}_\sgset$. We are going to specialize to the cases where $\Gaul_\sgset$ is the \textit{Hopf algebra of Feynman diagrams}. In this case the evaluations $\sk \star_\sgset S^{\zeta|_\sgset}_\sgset \left(\allG^{(v_d)}_\sgset \right)$ are called \textit{counterterms}. The evaluations $\sk \star_\sgset S^{\zeta|_\sgset}_\sgset(\Gamma)$ of individual graphs are going to turn out to be equivalent to the \textit{Moebius function} of the underlying subgraph posets.

\chapter{The Hopf algebra of Feynman diagrams}
\label{chap:hopf_algebra_of_fg}

The content of this chapter is partially based on the author's article \cite{borinsky2016lattices}.
\section{Preliminaries}
\label{sec:pre}
\subsection{Combinatorial quantum field theory}
In what follows a quantum field theory (QFT) will be characterized by its field content, its interactions, associated `weights' for these interactions and a given dimension of spacetime $D$. Let $\fields$ denote the set of fields, $\mathcal{R}_v$ the set of allowed interactions or vertex-types, represented as monomials in the fields and $\mathcal{R}_e \subset \mathcal{R}_v$ the set of propagators or edge-types, a set of distinguished interactions between two fields only. $\mathcal{R}_e$ consists of monomials of degree two and $\mathcal{R}_v$ of monomials of degree two or higher in the fields $\fields$. Additionally, a map $\omega: \mathcal{R}_e \cup \mathcal{R}_v \rightarrow \mathbb{Z}$ is given associating a weight to each interaction. 

The requirement $\mathcal{R}_e \subset \mathcal{R}_v$ ensures that there is a two-valent vertex-type for every allowed edge-type. This is not necessary for the definition of the Hopf algebra of Feynman diagrams, but it results in a simpler formula for contractions which agrees with the formalism from the previous chapters. Of course, this does not introduce a restriction to the underlying QFT: A propagator is always associated to the formal inverse of the corresponding two-valent vertex and a two-valent vertex always comes with an additional propagator in a diagram. The two valent vertex of the same type as the propagator can be canceled with the additional propagator.

In physical terms, the interactions correspond to summands in the Lagrangian of the QFT and the weights are the number of derivatives in the respective summand.

The construction above is also called a \textit{combinatorial} quantum field theory. For an in depth account on this combinatorial viewpoint on quantum field theory consult \cite{yeats2016combinatorial}.

Having clarified the important properties of a QFT for a combinatorial treatment, we can proceed to the definition of the central object of perturbative QFTs:
\subsection{Feynman diagrams}
\label{sec:feynmandiagrams}
Feynman diagrams are graphs with colored half-edges and restrictions on the allowed vertex and edge colors, which are induced by this coloring. This generalization is trivial and all previous results including the Hopf algebra structures carry over seamlessly.

\begin{defn}[Feynman diagram]
\label{def:feynmandiagram}
  A \textit{Feynman diagram} $\Gamma$ is a graph $(H,V,E,\nu)$ with a coloring of the half-edges. That is an additional map $c: H \rightarrow \fields$, which needs to be chosen such that the induced color of every vertex and edge is an allowed monomial in $\mathcal{R}_v$ or $\mathcal{R}_e$ respectively: For each vertex $v\in V$, $\prod_{h \in \nu^{-1}(v)} c(h) \in \mathcal{R}_v$ and for each edge $\{h_1,h_2\}\in E$, $c(h_1)c(h_2) \in \mathcal{R}_e$. We will call these monomials the \textit{residue} of the vertex or edge: $\res(v):= \prod_{h \in \nu^{-1}(v)} c(h)$ and $\res(\{h_1, h_2\})= c(h_1)c(h_2)$.
\end{defn}

To clarify the above definition an example is given, in which different depictions of Feynman diagrams are discussed - in analogy to the example given in Figure \ref{fig:graph_representations}.
\begin{figure}
\ifdefined\nodraft
  \subcaptionbox{Typical graph representation of a Feynman diagram.\label{fig:traditionalfg}}
  [.45\linewidth]{
    \begin{tikzpicture} \coordinate (v1); \coordinate[right=.5 of v1] (v2); \coordinate[right=of v2] (v3); \coordinate[right=.5 of v3] (v4); \draw[fermion] (v1) -- (v2); \draw[meson] (v2) to[bend left=90] (v3); \draw[fermion] (v2) -- (v3); \draw[fermion] (v3) -- (v4); \end{tikzpicture}
  }
  \subcaptionbox{Hyper graph representation of a Feynman diagram.\label{fig:hyperfeynmangraph}}
  [.45\linewidth]{
    \begin{tikzpicture}[scale=.8] \node (p1) at (0,0.75) {$\bar \psi$}; \node (f1) at (1,1.5) {$\phi$}; \node (pb1) at (1,0) {$\psi$}; \node (p2) at (5,0) {$\bar \psi$}; \node (f2) at (5,1.5) {$\phi$}; \node (pb2) at (6,0.75) {$\psi$}; \node (c1) at ($1/3*(p1)+1/3*(pb1)+1/3*(f1)$) {}; \node (c2) at ($1/3*(p2)+1/3*(pb2)+1/3*(f2)$) {}; \node at ($(c1) + (0.35,0)$) {$\bar \psi \phi \psi$}; \node at ($(c2) - (0.35,0)$) {$\bar \psi \phi \psi$}; \node (e1) at ($1/2*(pb1) + 1/2*(p2)$) {$\bar{\psi} \psi$}; \node (e2) at ($1/2*(f1) + 1/2*(f2)$) {$\phi \phi$}; \draw (p1) circle (0.25); \draw (p2) circle (0.25); \draw (pb1) circle (0.25); \draw (pb2) circle (0.25); \draw (f1) circle (0.25); \draw (f2) circle (0.25); \draw (c1) circle (1.5); \draw (c2) circle (1.5); \draw (e1) ellipse [x radius=2.5, y radius=0.5]; \draw (e2) ellipse [x radius=2.5, y radius=0.5]; \end{tikzpicture}
  }
\else

MISSING IN DRAFT MODE

\fi
  \caption{Equivalent diagrammatic representations of Feynman graphs}\label{fig:diagramaticfeynmangraphs}
\end{figure}
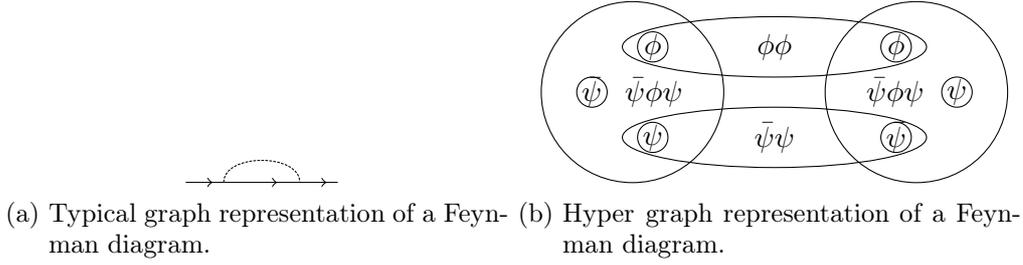
\begin{expl}[Yukawa theory]
Let $\fields = \left\{ \bar \psi, \psi, \phi \right\}$, $\mathcal R_v = \left\{ \bar \psi \psi, \phi^2, \bar \psi \phi \psi \right\}$ and $\mathcal R_e =\left\{ \bar \psi \psi, \phi^2 \right\}$, where $\bar \psi \psi$ stands for a fermion edge, $ \ifmmode \usebox{\fgsimplefermionprop} \else \newsavebox{\fgsimplefermionprop} \savebox{\fgsimplefermionprop}{ \begin{tikzpicture}[x=1ex,y=1ex,baseline={([yshift=-.5ex]current bounding box.center)}] \coordinate (v) ; \coordinate [right=1.2 of v] (u); \draw[fermion] (v) -- (u); \end{tikzpicture} } \fi$, $\phi^2$ for a meson edge, $ \ifmmode \usebox{\fgsimplemesonprop} \else \newsavebox{\fgsimplemesonprop} \savebox{\fgsimplemesonprop}{ \begin{tikzpicture}[x=1ex,y=1ex,baseline={([yshift=-.5ex]current bounding box.center)}] \coordinate (v) ; \coordinate [right=1.2 of v] (u); \draw[meson] (v) -- (u); \end{tikzpicture} } \fi$ and $\bar \psi \phi \psi$ for the fermion-fermion-meson vertex, $ \ifmmode \usebox{\fgsimpleyukvtx} \else \newsavebox{\fgsimpleyukvtx} \savebox{\fgsimpleyukvtx}{ \begin{tikzpicture}[x=1ex,y=1ex,baseline={([yshift=-.5ex]current bounding box.center)}] \coordinate (v) ; \def \rad {1}; \filldraw[white] (v) circle (\rad); \coordinate (u1) at ([shift=(180:\rad)]v); \coordinate (u2) at ([shift=(300:\rad)]v); \coordinate (u3) at ([shift=(60:\rad)]v); \draw[meson] (u1) -- (v); \draw[fermion] (u2) -- (v); \draw[fermion] (v) -- (u3); \filldraw (v) circle (1pt); \end{tikzpicture} } \fi$.
Figure \ref{fig:diagramaticfeynmangraphs} shows different graphical representations for a simple Feynman diagram in this theory. 

The usual Feynman diagram representation is given in Figure \ref{fig:traditionalfg}. The adjacency relations $E$ are represented as edges and the adjacency relations $V$ as vertices. The half-edges are omitted. 

Figure \ref{fig:hyperfeynmangraph} shows a hypergraph representation of the diagram. Its half-edges are drawn as little circles. They are colored by the corresponding field.  The adjacency relations are shown as big ellipses, enclosing the adjacent half-edges. The adjacency relations, $a\in E\cup V$ can be colored by the different allowed residues, $\res(a)$ in $\mathcal{R}_e$ and $\mathcal{R}_v$.
\end{expl}

Of course, Feynman diagrams inherit all the notions from graphs which were established in Chapters \ref{chp:graphs} and \ref{chap:coalgebra_graph}.

As in Chapter \ref{chap:coalgebra_graph}, we will be interested in the subgraphs of Feynman diagrams. For the Hopf algebra of Feynman diagrams it is convenient to start with the set of bridgeless subgraphs as defined in Example \ref{expl:bridgeless_graphs}:
\begin{align} \subdiags_{\text{bl}}(\Gamma) := \left\{\gamma \subset \Gamma \text{ such that } \gamma \text{ is bridgeless}\right\}. \end{align}
If $\Gamma$ is bridgeless, then obviously $\Gamma \in \subdiags_{\text{bl}}(\Gamma)$.
\nomenclature{$\subdiags_{\text{bl}}(\Gamma)$}{Bridgeless subgraphs of a graph $\Gamma$}

In quantum field theory language, a diagram is called \textit{one-particle-irreducible} or \textit{1PI} if it is connected and bridgeless.

\begin{expl}[Bridgeless subgraphs of a diagram in $\varphi^4$-theory]
\label{expl:subdiagrams1PI}
For the diagram 
${ \def \scale {2ex} }$
in $\varphi^4$-theory ( $\fields = \left\{ \phi \right\}$, $\mathcal R_v = \left\{ \phi^4, \phi^2 \right\}$ and $\mathcal R_e =\left\{ \phi^2 \right\}$ ).
\ifdefined\nodraft
\def \thickness {2pt}
\begin{gather*} \subdiags_{\text{bl}} \left( { \def \scale {3ex} %
 } \right\}, \end{gather*}
\else

MISSING IN DRAFT MODE

\fi
where bridgeless subgraphs are drawn with thick lines. %
\end{expl}

\paragraph{Superficial degree of divergence}
Using the map $\omega$, which is provided by the QFT, to assign a weight to every vertex and edge-type, an additional map $\omega_D$ can be defined, which assigns a weight to a Feynman diagram. This weight is called \textit{superficial degree of divergence} in the sense of \cite{Weinberg1960}:
\begin{align} \label{eqn:omega_D} \omega_D\left(\Gamma\right) &:= \sum \limits_{e\in E_\Gamma} \omega(\text{\normalfont{res}}(e)) - \sum \limits_{v\in V_\Gamma} \omega(\text{\normalfont{res}}(v)) - D h_\Gamma \end{align}
Recall that $h_\Gamma$ is the first Betti number of the diagram that fulfills $h_\Gamma = |E_\Gamma| - |V_\Gamma| + |\comps_\Gamma|$. In physics jargon $h_\Gamma$ is called the \textit{number of loops} of $\Gamma$.
Neglecting possible infrared divergences, the value of $\omega_D$ coincides with the degree of divergence of the integral associated to the diagram in the perturbation expansion of the underlying QFT in $D$-dimensions.
A 1PI diagram $\Gamma$ with $\omega_D(\Gamma) \leq 0$ is \textit{superficially divergent} (s.d.) in $D$ dimensions. 
For notational simplicity, the weight $0$ is assigned to the empty diagram, $\omega_D\left(\emptyset\right) = 0$, even though it is not divergent.
\nomenclature{$\omega_D\left(\Gamma\right)$}{Superficial degree of divergence of a graph $\Gamma$}

\begin{defn}[Renormalizable Quantum Field Theory]
\label{def:renormalizable}
A QFT is \textit{renormalizable in $D$ dimensions} if $\omega_D(\Gamma)$ depends only on the external structure of $\Gamma$ and the superficial degree of divergence of each connected diagram agrees with the weight assigned to the residue of the diagram: $\omega_D(\Gamma) = \omega(\res \Gamma)$. 
This can be expressed as the commutativity of the diagram:

\begin{center}
\begin{tikzpicture} \node (tl) {$\mathcal T$}; \node [right=of tl] (tr) {$\mathbb{Z}$}; \node [below =of tl] (bl) {$\mathbb{N}^\Phi$}; \draw[->] (tl) -- node[above] {$\omega_D$} (tr); \draw[->] (tl) to node[auto] {$\res$} (bl); \draw[->] (bl) -- node[right] {$\omega$} (tr); \end{tikzpicture}
\end{center}
\end{defn}
where $\mathcal T$ is the set of all connected Feynman diagrams of the renormalizable QFT. Specifically, $\omega_D(\Gamma)$ needs to be independent of $h_\Gamma$. 

Working with a renormalizable QFT, we need to keep track of \textit{subdivergences} or \textit{superficially divergent subdiagrams} appearing in the integrals of the perturbation expansion. The tools needed are the set of bridgeless subdiagrams and the superficial degree of divergence. The compatibility of the vertex and edge-weights and the superficial degree of divergence of the diagrams is exactly what is necessary to contract these subdivergences without leaving the space of allowed Feynman diagrams and obtain an admissible graph subset.

\paragraph{Superficially divergent subdiagrams}
The set of \textit{superficially divergent subdiagrams} or s.d.\ subdiagrams, 
\begin{align} \label{eqn:sdsubdiags} \sdsubdiags_D(\Gamma):= \left\{\gamma \in \subdiags_{\text{bl}}(\Gamma) \text{ such that } \gamma = \bigsqcup \limits_i \gamma_i \text{ and } \omega_D( \gamma_i ) \leq 0 \right\}, \end{align}
of subgraphs, whose connected components $\gamma_i$ are s.d.\ 1PI diagrams, is the object of main interest for the combinatorics of renormalization. 
The renormalizability of the QFT guarantees that for every $\gamma \in \sdsubdiags_D(\Gamma)$ the diagram resulting from the contraction $\Gamma/\gamma$ is still a valid Feynman diagram of the underlying QFT.
\nomenclature{$\sdsubdiags_D(\Gamma)$}{Superficially divergent subdiagrams of a graph $\Gamma$}

\begin{expl}[Superficially divergent subdiagrams of a diagram in $\varphi^4$-theory]
\label{expl:subdiagrams1PIsd}
Consider the same diagram as in Example \ref{expl:subdiagrams1PI} in $\varphi^4$ theory with the weights 
$\omega(\phi^2) = \omega(  \ifmmode \usebox{\fgsimpleprop} \else \newsavebox{\fgsimpleprop} \savebox{\fgsimpleprop}{ \begin{tikzpicture}[x=1ex,y=1ex,baseline={([yshift=-.5ex]current bounding box.center)}] \coordinate (v) ; \coordinate [right=1.2 of v] (u); \draw (v) -- (u); \end{tikzpicture} } \fi ) = 2$ and 
$\omega(\phi^4) = \omega(  \ifmmode \usebox{\fgsimplefourvtx} \else \newsavebox{\fgsimplefourvtx} \savebox{\fgsimplefourvtx}{ \begin{tikzpicture}[x=1ex,y=1ex,baseline={([yshift=-.5ex]current bounding box.center)}] \coordinate (v) ; \def \n {4}; \def \rad {.8}; \filldraw[white] (v) circle (\rad); \foreach \s in {1,...,5} { \def \angle {45+360/\n*(\s - 1)}; \coordinate (u) at ([shift=({\angle}:\rad)]v); \draw (v) -- (u); } \filldraw (v) circle (1pt); \end{tikzpicture} } \fi ) = 0$. The superficially divergent subdiagrams for $D=4$ are 
\ifdefined\nodraft
\def \thickness {2pt}
\begin{gather*} \sdsubdiags_4 \left( { \def \scale {3ex} \begin{tikzpicture}[x=\scale,y=\scale,baseline={([yshift=-.5ex]current bounding box.center)}] \begin{scope}[node distance=1] \coordinate (v0); \coordinate[right=.5 of v0] (v4); \coordinate[above right= of v4] (v2); \coordinate[below right= of v4] (v3); \coordinate[below right= of v2] (v5); \coordinate[right=.5 of v5] (v1); \coordinate[above right= of v2] (o1); \coordinate[below right= of v2] (o2); \coordinate[below left=.5 of v0] (i1); \coordinate[above left=.5 of v0] (i2); \coordinate[below right=.5 of v1] (o1); \coordinate[above right=.5 of v1] (o2); \draw (v0) -- (i1); \draw (v0) -- (i2); \draw (v1) -- (o1); \draw (v1) -- (o2); \draw (v0) to[bend left=20] (v2); \draw (v0) to[bend right=20] (v3); \draw (v1) to[bend left=20] (v3); \draw (v1) to[bend right=20] (v2); \draw (v2) to[bend right=60] (v3); \draw (v2) to[bend left=60] (v3); \filldraw (v0) circle(1pt); \filldraw (v1) circle(1pt); \filldraw (v2) circle(1pt); \filldraw (v3) circle(1pt); \ifdefined\cvl \draw[line width=1.5pt] (v0) to[bend left=20] (v2); \draw[line width=1.5pt] (v0) to[bend right=20] (v3); \fi \ifdefined\cvr \draw[line width=1.5pt] (v1) to[bend left=20] (v3); \draw[line width=1.5pt] (v1) to[bend right=20] (v2); \fi \ifdefined\cvml \draw[line width=1.5pt] (v2) to[bend left=60] (v3); \fi \ifdefined\cvmr \draw[line width=1.5pt] (v2) to[bend right=60] (v3); \fi \end{scope} \end{tikzpicture} } \right) = \left\{ { \def \scale {3ex} \def \cvmr {} \def \cvml {} \begin{tikzpicture}[x=\scale,y=\scale,baseline={([yshift=-.5ex]current bounding box.center)}] \begin{scope}[node distance=1] \coordinate (v0); \coordinate[right=.5 of v0] (v4); \coordinate[above right= of v4] (v2); \coordinate[below right= of v4] (v3); \coordinate[below right= of v2] (v5); \coordinate[right=.5 of v5] (v1); \coordinate[above right= of v2] (o1); \coordinate[below right= of v2] (o2); \coordinate[below left=.5 of v0] (i1); \coordinate[above left=.5 of v0] (i2); \coordinate[below right=.5 of v1] (o1); \coordinate[above right=.5 of v1] (o2); \draw (v0) -- (i1); \draw (v0) -- (i2); \draw (v1) -- (o1); \draw (v1) -- (o2); \draw (v0) to[bend left=20] (v2); \draw (v0) to[bend right=20] (v3); \draw (v1) to[bend left=20] (v3); \draw (v1) to[bend right=20] (v2); \draw (v2) to[bend right=60] (v3); \draw (v2) to[bend left=60] (v3); \filldraw (v0) circle(1pt); \filldraw (v1) circle(1pt); \filldraw (v2) circle(1pt); \filldraw (v3) circle(1pt); \ifdefined\cvl \draw[line width=1.5pt] (v0) to[bend left=20] (v2); \draw[line width=1.5pt] (v0) to[bend right=20] (v3); \fi \ifdefined\cvr \draw[line width=1.5pt] (v1) to[bend left=20] (v3); \draw[line width=1.5pt] (v1) to[bend right=20] (v2); \fi \ifdefined\cvml \draw[line width=1.5pt] (v2) to[bend left=60] (v3); \fi \ifdefined\cvmr \draw[line width=1.5pt] (v2) to[bend right=60] (v3); \fi \end{scope} \end{tikzpicture} } , { \def \scale {3ex} \def \cvmr {} \def \cvml {} \def \cvl {} \begin{tikzpicture}[x=\scale,y=\scale,baseline={([yshift=-.5ex]current bounding box.center)}] \begin{scope}[node distance=1] \coordinate (v0); \coordinate[right=.5 of v0] (v4); \coordinate[above right= of v4] (v2); \coordinate[below right= of v4] (v3); \coordinate[below right= of v2] (v5); \coordinate[right=.5 of v5] (v1); \coordinate[above right= of v2] (o1); \coordinate[below right= of v2] (o2); \coordinate[below left=.5 of v0] (i1); \coordinate[above left=.5 of v0] (i2); \coordinate[below right=.5 of v1] (o1); \coordinate[above right=.5 of v1] (o2); \draw (v0) -- (i1); \draw (v0) -- (i2); \draw (v1) -- (o1); \draw (v1) -- (o2); \draw (v0) to[bend left=20] (v2); \draw (v0) to[bend right=20] (v3); \draw (v1) to[bend left=20] (v3); \draw (v1) to[bend right=20] (v2); \draw (v2) to[bend right=60] (v3); \draw (v2) to[bend left=60] (v3); \filldraw (v0) circle(1pt); \filldraw (v1) circle(1pt); \filldraw (v2) circle(1pt); \filldraw (v3) circle(1pt); \ifdefined\cvl \draw[line width=1.5pt] (v0) to[bend left=20] (v2); \draw[line width=1.5pt] (v0) to[bend right=20] (v3); \fi \ifdefined\cvr \draw[line width=1.5pt] (v1) to[bend left=20] (v3); \draw[line width=1.5pt] (v1) to[bend right=20] (v2); \fi \ifdefined\cvml \draw[line width=1.5pt] (v2) to[bend left=60] (v3); \fi \ifdefined\cvmr \draw[line width=1.5pt] (v2) to[bend right=60] (v3); \fi \end{scope} \end{tikzpicture} } , { \def \scale {3ex} \def \cvmr {} \def \cvml {} \def \cvr {} \begin{tikzpicture}[x=\scale,y=\scale,baseline={([yshift=-.5ex]current bounding box.center)}] \begin{scope}[node distance=1] \coordinate (v0); \coordinate[right=.5 of v0] (v4); \coordinate[above right= of v4] (v2); \coordinate[below right= of v4] (v3); \coordinate[below right= of v2] (v5); \coordinate[right=.5 of v5] (v1); \coordinate[above right= of v2] (o1); \coordinate[below right= of v2] (o2); \coordinate[below left=.5 of v0] (i1); \coordinate[above left=.5 of v0] (i2); \coordinate[below right=.5 of v1] (o1); \coordinate[above right=.5 of v1] (o2); \draw (v0) -- (i1); \draw (v0) -- (i2); \draw (v1) -- (o1); \draw (v1) -- (o2); \draw (v0) to[bend left=20] (v2); \draw (v0) to[bend right=20] (v3); \draw (v1) to[bend left=20] (v3); \draw (v1) to[bend right=20] (v2); \draw (v2) to[bend right=60] (v3); \draw (v2) to[bend left=60] (v3); \filldraw (v0) circle(1pt); \filldraw (v1) circle(1pt); \filldraw (v2) circle(1pt); \filldraw (v3) circle(1pt); \ifdefined\cvl \draw[line width=1.5pt] (v0) to[bend left=20] (v2); \draw[line width=1.5pt] (v0) to[bend right=20] (v3); \fi \ifdefined\cvr \draw[line width=1.5pt] (v1) to[bend left=20] (v3); \draw[line width=1.5pt] (v1) to[bend right=20] (v2); \fi \ifdefined\cvml \draw[line width=1.5pt] (v2) to[bend left=60] (v3); \fi \ifdefined\cvmr \draw[line width=1.5pt] (v2) to[bend right=60] (v3); \fi \end{scope} \end{tikzpicture} } , { \def \scale {3ex} \def \cvmr {} \def \cvml {} \def \cvl {} \def \cvr {} \begin{tikzpicture}[x=\scale,y=\scale,baseline={([yshift=-.5ex]current bounding box.center)}] \begin{scope}[node distance=1] \coordinate (v0); \coordinate[right=.5 of v0] (v4); \coordinate[above right= of v4] (v2); \coordinate[below right= of v4] (v3); \coordinate[below right= of v2] (v5); \coordinate[right=.5 of v5] (v1); \coordinate[above right= of v2] (o1); \coordinate[below right= of v2] (o2); \coordinate[below left=.5 of v0] (i1); \coordinate[above left=.5 of v0] (i2); \coordinate[below right=.5 of v1] (o1); \coordinate[above right=.5 of v1] (o2); \draw (v0) -- (i1); \draw (v0) -- (i2); \draw (v1) -- (o1); \draw (v1) -- (o2); \draw (v0) to[bend left=20] (v2); \draw (v0) to[bend right=20] (v3); \draw (v1) to[bend left=20] (v3); \draw (v1) to[bend right=20] (v2); \draw (v2) to[bend right=60] (v3); \draw (v2) to[bend left=60] (v3); \filldraw (v0) circle(1pt); \filldraw (v1) circle(1pt); \filldraw (v2) circle(1pt); \filldraw (v3) circle(1pt); \ifdefined\cvl \draw[line width=1.5pt] (v0) to[bend left=20] (v2); \draw[line width=1.5pt] (v0) to[bend right=20] (v3); \fi \ifdefined\cvr \draw[line width=1.5pt] (v1) to[bend left=20] (v3); \draw[line width=1.5pt] (v1) to[bend right=20] (v2); \fi \ifdefined\cvml \draw[line width=1.5pt] (v2) to[bend left=60] (v3); \fi \ifdefined\cvmr \draw[line width=1.5pt] (v2) to[bend right=60] (v3); \fi \end{scope} \end{tikzpicture} } , { \def \scale {3ex} \begin{tikzpicture}[x=\scale,y=\scale,baseline={([yshift=-.5ex]current bounding box.center)}] \begin{scope}[node distance=1] \coordinate (v0); \coordinate[right=.5 of v0] (v4); \coordinate[above right= of v4] (v2); \coordinate[below right= of v4] (v3); \coordinate[below right= of v2] (v5); \coordinate[right=.5 of v5] (v1); \coordinate[above right= of v2] (o1); \coordinate[below right= of v2] (o2); \coordinate[below left=.5 of v0] (i1); \coordinate[above left=.5 of v0] (i2); \coordinate[below right=.5 of v1] (o1); \coordinate[above right=.5 of v1] (o2); \draw (v0) -- (i1); \draw (v0) -- (i2); \draw (v1) -- (o1); \draw (v1) -- (o2); \draw (v0) to[bend left=20] (v2); \draw (v0) to[bend right=20] (v3); \draw (v1) to[bend left=20] (v3); \draw (v1) to[bend right=20] (v2); \draw (v2) to[bend right=60] (v3); \draw (v2) to[bend left=60] (v3); \filldraw (v0) circle(1pt); \filldraw (v1) circle(1pt); \filldraw (v2) circle(1pt); \filldraw (v3) circle(1pt); \ifdefined\cvl \draw[line width=1.5pt] (v0) to[bend left=20] (v2); \draw[line width=1.5pt] (v0) to[bend right=20] (v3); \fi \ifdefined\cvr \draw[line width=1.5pt] (v1) to[bend left=20] (v3); \draw[line width=1.5pt] (v1) to[bend right=20] (v2); \fi \ifdefined\cvml \draw[line width=1.5pt] (v2) to[bend left=60] (v3); \fi \ifdefined\cvmr \draw[line width=1.5pt] (v2) to[bend right=60] (v3); \fi \end{scope} \end{tikzpicture} } \right\}. \end{gather*}
\else

MISSING IN DRAFT MODE

\fi
\end{expl}
\section{Hopf algebra structure of Feynman diagrams}
\label{sec:hopfalgebra}

The basis for the analysis of the lattice structure in QFTs is Kreimer's Hopf algebra of Feynman diagrams. It captures the BPHZ renormalization procedure which is necessary to obtain finite amplitudes from perturbative calculations in an algebraic framework \cite{ConnesKreimer2000}.

The Hopf algebra of Feynman diagrams will be another quotient Hopf algebra of the Hopf algebra of all graphs. 

Take $\sgset^{\text{s.d.}}_D$ to be the set of all graphs that are 
\begin{enumerate}
\item bridgeless
\item each of their non-trivial connected components is superficially divergent
\item their non-trivial connected components only contain vertices with degrees from the set $\mathcal{R}_v$.
\end{enumerate}
As illustrated in the previous section, this set is stable under insertion and contraction and fulfill the conditions of Definition \ref{def:admissible_graphs} if the underlying theory is \textit{renormalizable}. 
\nomenclature{$\sgset^{\text{s.d.}}_D$}{Set of all superficially divergent graphs of a quantum field theory in dimension $D$}

The Connes-Kreimer Hopf algebra can be identified with the quotient $\hopffg_D:=\Gaul_{\sgset^{\text{s.d.}}_D}$ from Definition \ref{def:restricted_graph_hopf}. Note that $\hopffg_D$ can also be seen as a quotient algebra of $\Gaul_{\sgset_{\text{bl}}}$ by dividing out all non-superficially divergent graphs.
\nomenclature{$\hopffg_D$}{Hopf algebra of Feynman diagrams; equivalent to the quotient $\Gaul_{\sgset^{\text{s.d.}}_D}$}

In this section, it will be illustrated how this Hopf algebra fits into the previously established framework. 

For a more detailed exposition consult \cite{manchon2004hopf} for mathematical details of Hopf algebras in general with the Connes-Kreimer Hopf algebra as a specific example. In the author's article \cite{borinsky2014feynman} computational aspects of the Connes-Kreimer Hopf algebra were discussed.

Applying Definition \ref{def:restricted_graph_hopf}, we see that the coproduct is given by
\begin{align} \label{eqn:def_cop} &\Delta_D \Gamma := \sum_{\substack{ \gamma \subset \Gamma\\ \gamma \in \sgset^{\text{s.d.}}_D }} \gamma \otimes \Gamma/\gamma& &:& &{\hopffg_D} \rightarrow {\hopffg_D} \otimes {\hopffg_D}. \end{align}
The notion of superficial degree of divergence, $\omega_D$, hidden in $\sgset^{\text{s.d.}}_D$ is the only input to the Hopf algebra structure which depends on the dimension $D$ of spacetime. We will refer to the antipode of the Hopf algebra $\hopffg_D$ as $S_D$.
\nomenclature{$\Delta_D$}{Coproduct on $\hopffg_D$}
\nomenclature{$S_D$}{Antipode on $\hopffg_D$}

\begin{expl}[Coproduct of a diagram in $\varphi^4$-theory]
\label{expl:coproducthopffg}
To illustrate the procedure of calculating the coproduct of a graph in this Hopf algebra take the same diagram from $\varphi^4$-theory as in the Examples \ref{expl:subdiagrams1PI} and \ref{expl:subdiagrams1PIsd}. 
The coproduct is calculated using the set $\sdsubdiags_4(\Gamma)$ and the definition of the contraction in Definition \ref{def:contraction}:
\ifdefined\nodraft
\begin{align*} \Delta_4 { \def \scale {2ex} } / \gamma = \\ & { \ifmmode \usebox{\fgsimplefourvtx} \else \newsavebox{\fgsimplefourvtx} \savebox{\fgsimplefourvtx}{ %
 } \fi}^4 \otimes { \def \scale {2ex} %
 } \otimes { \ifmmode \usebox{\fgsimplefourvtx} \else \newsavebox{\fgsimplefourvtx} \savebox{\fgsimplefourvtx}{ %
 } \fi} + \\ & + { \def \scale {2ex} %
 { \ifmmode \usebox{\fgsimplefourvtx} \else \newsavebox{\fgsimplefourvtx} \savebox{\fgsimplefourvtx}{ %
 } \fi} } \otimes { \def \scale {2ex} %
 } + { { \ifmmode \usebox{\fgsimplefourvtx} \else \newsavebox{\fgsimplefourvtx} \savebox{\fgsimplefourvtx}{ %
 } \fi} \def \scale {2ex} %
 } + { { \ifmmode \usebox{\fgsimplefourvtx} \else \newsavebox{\fgsimplefourvtx} \savebox{\fgsimplefourvtx}{ %
 } \fi} \def \scale {2ex} %
 { \ifmmode \usebox{\fgsimplefourvtx} \else \newsavebox{\fgsimplefourvtx} \savebox{\fgsimplefourvtx}{ %
 } \fi} } \otimes { \def \scale {2ex} %
 } = \\ & { \ifmmode \usebox{\fgsimplefourvtx} \else \newsavebox{\fgsimplefourvtx} \savebox{\fgsimplefourvtx}{ %
 } \fi}^4 \otimes { \def \scale {2ex} %
 } \otimes { \ifmmode \usebox{\fgsimplefourvtx} \else \newsavebox{\fgsimplefourvtx} \savebox{\fgsimplefourvtx}{ %
 } \fi} + 2 { { \ifmmode \usebox{\fgsimplefourvtx} \else \newsavebox{\fgsimplefourvtx} \savebox{\fgsimplefourvtx}{ %
 } \fi} ~ \def \scale {2ex} %
 } + { { \ifmmode \usebox{\fgsimplefourvtx} \else \newsavebox{\fgsimplefourvtx} \savebox{\fgsimplefourvtx}{ %
 } \fi}^2 \def \scale {2ex} %
 } \end{align*}
The last equality holds because  
${ \def \scale {1.5ex} %
 }$ are mutually isomorphic graphs.
\else

MISSING IN DRAFT MODE

\fi
\end{expl}
We could identify the residual parts of the expression with the neutral element $\one$ of ${\hopffg_D}$. That means we could set ${ \ifmmode \usebox{\fgsimplefourvtx} \else \newsavebox{\fgsimplefourvtx} \savebox{\fgsimplefourvtx}{ %
 } \fi} = \one$. As $\one - { \ifmmode \usebox{\fgsimplefourvtx} \else \newsavebox{\fgsimplefourvtx} \savebox{\fgsimplefourvtx}{ %
 } \fi}$ generates a Hopf ideal, we can work in the quotient ${\hopffg_D}/(\one - { \ifmmode \usebox{\fgsimplefourvtx} \else \newsavebox{\fgsimplefourvtx} \savebox{\fgsimplefourvtx}{ %
 } \fi})$. However, this is not necessary as mentioned in \cite{manchon2004hopf} and laid out in detail by Kock \citep{kock2015perturbative}.

As before in the general case of $\Gaul$, ${\hopffg_D}$ is graded by the loop number, $h_\Gamma$, of the diagrams, 
\begin{align} {\hopffg_D} &= \bigoplus \limits_{L \geq 0} {\hopffg_D}^{(L)}\text{ and} \label{grading_loop_1} \\ m &: {\hopffg_D}^{(L_1)} \otimes {\hopffg_D}^{(L_2)} \rightarrow {\hopffg_D}^{(L_1 + L_2)} \label{grading_loop_2} \\ \Delta_D &: {\hopffg_D}^{(L)} \rightarrow \bigoplus \limits_{\substack{L_1, L_2 \ge 0 \\ L_1 + L_2 = L} } {\hopffg_D}^{(L_1)} \otimes {\hopffg_D}^{(L_2)}, \label{grading_loop_3} \end{align}
where ${\hopffg_D}^{(L)} \subset {\hopffg_D}$ is the subspace of ${\hopffg_D}$ which is generated by diagrams $\Gamma$ with $h_\Gamma = L$.

Obviously, the result of the coproduct in the Hopf algebra is always of the form $\Delta_D \Gamma = \skl(\Gamma) \otimes \Gamma + \Gamma \otimes \res(\Gamma) + \widetilde{\Delta}_D \Gamma$ with the trivial terms $\skl(\Gamma) \otimes \Gamma + \Gamma \otimes \res(\Gamma)$ and a non-trivial part $\widetilde{\Delta}_D \Gamma$ which is called the \textit{reduced coproduct} of $\Gamma$. More formally, the reduced coproduct is defined as $\widetilde{\Delta}_D:= P^{ \otimes 2} \circ \Delta_D $, where $P$ projects into the augmentation ideal, $P: {\hopffg_D} \rightarrow \ker{\counit}$, that means it acts as the identity on all graphs that are not residues and maps residues to zero.

\begin{expl}[Reduced coproduct of a non-primitive diagram in $\varphi^4$-theory]
\ifdefined\nodraft
\begin{align} \widetilde \Delta_4 { \def \scale {2ex} \begin{tikzpicture}[x=\scale,y=\scale,baseline={([yshift=-.5ex]current bounding box.center)}] \begin{scope}[node distance=1] \coordinate (v0); \coordinate[right=.5 of v0] (v4); \coordinate[above right= of v4] (v2); \coordinate[below right= of v4] (v3); \coordinate[below right= of v2] (v5); \coordinate[right=.5 of v5] (v1); \coordinate[above right= of v2] (o1); \coordinate[below right= of v2] (o2); \coordinate[below left=.5 of v0] (i1); \coordinate[above left=.5 of v0] (i2); \coordinate[below right=.5 of v1] (o1); \coordinate[above right=.5 of v1] (o2); \draw (v0) -- (i1); \draw (v0) -- (i2); \draw (v1) -- (o1); \draw (v1) -- (o2); \draw (v0) to[bend left=20] (v2); \draw (v0) to[bend right=20] (v3); \draw (v1) to[bend left=20] (v3); \draw (v1) to[bend right=20] (v2); \draw (v2) to[bend right=60] (v3); \draw (v2) to[bend left=60] (v3); \filldraw (v0) circle(1pt); \filldraw (v1) circle(1pt); \filldraw (v2) circle(1pt); \filldraw (v3) circle(1pt); \ifdefined\cvl \draw[line width=1.5pt] (v0) to[bend left=20] (v2); \draw[line width=1.5pt] (v0) to[bend right=20] (v3); \fi \ifdefined\cvr \draw[line width=1.5pt] (v1) to[bend left=20] (v3); \draw[line width=1.5pt] (v1) to[bend right=20] (v2); \fi \ifdefined\cvml \draw[line width=1.5pt] (v2) to[bend left=60] (v3); \fi \ifdefined\cvmr \draw[line width=1.5pt] (v2) to[bend right=60] (v3); \fi \end{scope} \end{tikzpicture} } &= 2 { { \ifmmode \usebox{\fgsimplefourvtx} \else \newsavebox{\fgsimplefourvtx} \savebox{\fgsimplefourvtx}{ \begin{tikzpicture}[x=1ex,y=1ex,baseline={([yshift=-.5ex]current bounding box.center)}] \coordinate (v) ; \def \n {4}; \def \rad {.8}; \filldraw[white] (v) circle (\rad); \foreach \s in {1,...,5} { \def \angle {45+360/\n*(\s - 1)}; \coordinate (u) at ([shift=({\angle}:\rad)]v); \draw (v) -- (u); } \filldraw (v) circle (1pt); \end{tikzpicture} } \fi} ~ \def \scale {2ex} \begin{tikzpicture}[x=\scale,y=\scale,baseline={([yshift=-.5ex]current bounding box.center)}] \begin{scope}[node distance=1] \coordinate (v0); \coordinate[right=.5 of v0] (v4); \coordinate[above right= of v4] (v2); \coordinate[below right= of v4] (v3); \coordinate[above right=.5 of v2] (o1); \coordinate[below right=.5 of v3] (o2); \coordinate[below left=.5 of v0] (i1); \coordinate[above left=.5 of v0] (i2); \draw (v0) -- (i1); \draw (v0) -- (i2); \draw (v2) -- (o1); \draw (v3) -- (o2); \draw (v0) to[bend left=20] (v2); \draw (v0) to[bend right=20] (v3); \draw (v2) to[bend right=60] (v3); \draw (v2) to[bend left=60] (v3); \filldraw (v0) circle(1pt); \filldraw (v2) circle(1pt); \filldraw (v3) circle(1pt); \end{scope} \end{tikzpicture} } \otimes { \def \scale {2ex} \begin{tikzpicture}[x=\scale,y=\scale,baseline={([yshift=-.5ex]current bounding box.center)}] \begin{scope}[node distance=1] \coordinate (v0); \coordinate [right=.5 of v0] (vm); \coordinate [right=.5 of vm] (v1); \coordinate [above left=.5 of v0] (i0); \coordinate [below left=.5 of v0] (i1); \coordinate [above right=.5 of v1] (o0); \coordinate [below right=.5 of v1] (o1); \draw (vm) circle(.5); \draw (i0) -- (v0); \draw (i1) -- (v0); \draw (o0) -- (v1); \draw (o1) -- (v1); \filldraw (v0) circle(1pt); \filldraw (v1) circle(1pt); \end{scope} \end{tikzpicture} } + { { \ifmmode \usebox{\fgsimplefourvtx} \else \newsavebox{\fgsimplefourvtx} \savebox{\fgsimplefourvtx}{ \begin{tikzpicture}[x=1ex,y=1ex,baseline={([yshift=-.5ex]current bounding box.center)}] \coordinate (v) ; \def \n {4}; \def \rad {.8}; \filldraw[white] (v) circle (\rad); \foreach \s in {1,...,5} { \def \angle {45+360/\n*(\s - 1)}; \coordinate (u) at ([shift=({\angle}:\rad)]v); \draw (v) -- (u); } \filldraw (v) circle (1pt); \end{tikzpicture} } \fi}^2 \def \scale {2ex} \begin{tikzpicture}[x=\scale,y=\scale,baseline={([yshift=-.5ex]current bounding box.center)}] \begin{scope}[node distance=1] \coordinate (v0); \coordinate [below=1 of v0] (v1); \coordinate [above left=.5 of v0] (i0); \coordinate [above right=.5 of v0] (i1); \coordinate [below left=.5 of v1] (o0); \coordinate [below right=.5 of v1] (o1); \coordinate [above=.5 of v1] (vm); \draw (vm) circle(.5); \draw (i0) -- (v0); \draw (i1) -- (v0); \draw (o0) -- (v1); \draw (o1) -- (v1); \filldraw (v0) circle(1pt); \filldraw (v1) circle(1pt); \end{scope} \end{tikzpicture} } \otimes { \def \scale {2ex} \begin{tikzpicture}[x=\scale,y=\scale,baseline={([yshift=-.5ex]current bounding box.center)}] \begin{scope}[node distance=1] \coordinate (v0); \coordinate [right=.5 of v0] (vm1); \coordinate [right=.5 of vm1] (v1); \coordinate [right=.5 of v1] (vm2); \coordinate [right=.5 of vm2] (v2); \coordinate [above left=.5 of v0] (i0); \coordinate [below left=.5 of v0] (i1); \coordinate [above right=.5 of v2] (o0); \coordinate [below right=.5 of v2] (o1); \draw (vm1) circle(.5); \draw (vm2) circle(.5); \draw (i0) -- (v0); \draw (i1) -- (v0); \draw (o0) -- (v2); \draw (o1) -- (v2); \filldraw (v0) circle(1pt); \filldraw (v1) circle(1pt); \filldraw (v2) circle(1pt); \end{scope} \end{tikzpicture} } \end{align}
\else

MISSING IN DRAFT MODE

\fi
\end{expl}
Observe, that it follows immediately from the coassociativity of $\Delta_D$ that $\widetilde{\Delta}_D$ is coassociative.

The kernel of the reduced coproduct, is the space of \textit{primitive}%
\footnote{Because the coproduct is not of the form $\Delta \Gamma = \one \otimes \Gamma + \Gamma \otimes \one + \widetilde{\Delta}$, the elements in the kernel of $\widetilde{\Delta}$ are also called \textit{skew}primitive. As we can always divide out the ideal which sets all residues to $\one$, we will not treat this case differently.
} elements of the Hopf algebra, $\text{Prim} \hopffg_D := \ker\widetilde{\Delta}_D$. Primitive 1PI diagrams $\Gamma$ with $\Gamma \in \ker\widetilde{\Delta}_D$ are exactly those diagrams, which do not contain any subdivergences. They are also called \textit{skeleton diagrams} - not to be confused with the \textit{skeleton of a graph} which we defined in Definition \ref{def:res_skl_def} as the disjoint union of all vertices of a graph.

More general, we can define the iterations of the reduced coproduct $\widetilde{\Delta}_D^{n} = P^{\otimes n} \circ \Delta_D^{n}$, using the iterations of the coproduct as introduced in Section \ref{sec:coalgebra_graphs}. 

These homomorphisms give rise to an increasing filtration of $\hopffg_D$, the \textit{coradical filtration}:
\begin{gather} \begin{aligned} \label{eqn:coradfilt} ~^{(n)}\hopffg_D &:= \ker \widetilde{\Delta}_D^{n+1} &\forall& n \geq 0 \end{aligned}\\ \mathbb{Q} \simeq ~^{(0)}\hopffg_D \subset ~^{(1)}\hopffg_D \subset \ldots \subset ~^{(n)}\hopffg_D \subset \ldots \subset \hopffg_D. \end{gather}

\begin{figure}
\ifdefined\nodraft
\begin{center}
{
\def\scale {4ex}
\begin{tikzpicture}[x=\scale,y=\scale,baseline={([yshift=-.5ex]current bounding box.center)}] \begin{scope}[node distance=1] \coordinate (i); \coordinate[above right= of i] (v); \coordinate[above=.5 of v] (vm); \coordinate[above=.5 of vm] (vt); \coordinate[below right= of v] (o); \draw (vm) circle(.5); \draw (i) -- (v); \draw (o) -- (v); \filldraw (v) circle(1pt); \draw [fill=white] (vt) circle(1ex); \draw [fill=white,thick,pattern=north west lines] (vt) circle (1ex); \end{scope} \end{tikzpicture}
}
\end{center}
\else

MISSING IN DRAFT MODE

\fi
\caption{Characteristic subdiagram of every tadpole diagram}
\label{fig:snail}
\end{figure}
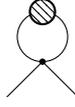

In some cases it is useful to introduce another restriction on the generator set $\sgset^{\text{s.d.}}_D$ of $\hopffg_D$. Additionally, to the already stated restrictions, we may want to restrict to Feynman diagrams without `tadpoles' (also snails or seagulls). Tadpoles are diagrams which can be split into two connected components by removing a single vertex such that one component does not contain any external leg. A tadpole diagram always has a subdiagram of a topology as depicted in Figure \ref{fig:snail}. The Hopf algebra of Feynman diagrams without tadpoles is denoted as $\hopffgs_D$.
\begin{defn}
\label{def:snailfreehopf}
We define $\hopffgs_D$ as $\hopffg_D$ with the difference that no tadpole diagrams are allowed as generators and replace $\sdsubdiags_D(\Gamma)$ in the formula for the coproduct, eq.\ \eqref{eqn:def_cop}, with
\begin{align} \sdsubdiagsn_D (\Gamma) := \left\{ \gamma \in \sdsubdiags_D(\Gamma) : \text{such that } \Gamma/\gamma \text{ is no tadpole diagram} \right\}. \end{align}
\end{defn}
Only the s.d.\ subdiagrams which do not result in a tadpole diagram upon contraction are elements of $\sdsubdiagsn_D (\Gamma)$.

A Hopf algebra homomorphism from $\hopffg_D$ to $\hopffgs_D$ is easy to set up:
\begin{align} \label{eqn:morphismpsi} &\psi& &:& &\hopffg_D & &\rightarrow& &\hopffgs_D, \\ && && &\Gamma& &\mapsto& & \begin{cases} &0 \text{ if } \Gamma \text{ is a tadpole diagram.} \\ &\Gamma \text{ else.} \end{cases} \end{align}
This map fulfills the requirements for a Hopf algebra homomorphism. The associated ideal $\ker\psi \subset \hopffg_D$ is the subspace of $\hopffg_D$ spanned by all tadpole diagrams. This ideal and the map $\psi$ are very useful, because the elements in $\ker \psi$ evaluate to zero after renormalization in kinematic subtraction schemes \cite{brown2013angles} and in minimal subtraction schemes for the massless case.
\section{Algebraic lattice structure of subdivergences}
\label{sec:posetsandlattices}
\subsection{Posets and algebraic lattices}
The set of subdivergences of a Feynman diagram is obviously partially ordered by inclusion. These partially ordered sets are quite constrained for some renormalizable QFTs: They are \textit{lattices}. In \cite[Part III]{figueroa2005combinatorial} this was studied specifically for distributive lattices. 

In this section, we will elaborate on the conditions a QFT must fulfill for these partially ordered sets to be lattices. The term \textit{join-meet-renormalizability} will be defined which characterizes QFTs in which all Feynman diagrams whose set of subdivergencies form lattices. It will be shown that this is a special property of QFTs with only four-or-less-valent vertices. 

The definitions will be illustrated with an application to the set of subdivergences of a Feynman diagram. Additionally, we will introduce the corresponding Hopf algebra for these lattices based on an incidence Hopf algebra \cite{Schmitt1994}.

First, the necessary definitions of poset and lattice theory will be introduced:

\begin{defn}[Poset]
A \textit{partially ordered set} or \textit{poset} is a finite set $P$ endowed with a partial order $\leq$. 
An interval $[x,y]$ is a subset $\left\{ z \in P : x \leq z \leq y \right\} \subset P$. If $[x,y] = \left\{x,y\right\}$, x covers y and y is covered by x.
\end{defn}
For a more detailed exposition of poset and lattice theory consult \cite{stanley1997}.

\paragraph{Hasse diagram}
A Hasse diagram of a poset $P$ is the graph with the elements of $P$ as vertices and the cover relations as edges. Larger elements are always drawn above smaller elements. 

\begin{expl}
The set of superficially divergent subdiagrams $\sdsubdiags_D(\Gamma)$ of a Feynman diagram $\Gamma$ is a poset ordered by inclusion: 
$\gamma_1 \leq \gamma_2 \Leftrightarrow \gamma_1 \subset \gamma_2$ for all $\gamma_1, \gamma_2 \in \sdsubdiags_D(\Gamma)$.

The statement that a subdiagram $\gamma_1$ covers $\gamma_2$ in $\sdsubdiags_D(\Gamma)$ is equivalent to the statement that $\gamma_1 / \gamma_2$ is primitive. The elements that are covered by the full diagram $\Gamma \in \sdsubdiags_D(\Gamma)$ are called \textit{maximal forests}; whereas, a maximal chain $\emptyset \subset \gamma_1 \subset \ldots \subset \gamma_n \subset \Gamma$, where each element is covered by the next, is a \textit{complete forest} of $\Gamma$.
\end{expl}

The Hasse diagram of a s.d.\ diagram $\Gamma$ can be constructed by following a simple procedure: Draw the diagram $\Gamma$ and find all the maximal forests $\gamma_i \in \sdsubdiags_D(\Gamma)$ such that $\Gamma/\gamma_i$ is primitive. Draw the diagrams $\gamma_i$ under $\Gamma$ and draw lines from $\Gamma$ to the $\gamma_i$. Subsequently, determine all the maximal forests $\mu_i$ of the $\gamma_i$ and draw them under the $\gamma_i$. Draw a line from $\gamma_i$ to $\mu_i$ if $\mu_i \subset \gamma_i$. Repeat this until only primitive diagrams are left. Then draw lines from the primitive subdiagrams to an additional trivial diagram without edges underneath them. Subsequently, replace diagrams with vertices.

\begin{expl}
For instance, the set of superficially divergent subdiagrams for $D=4$ of the diagram,
\ifdefined\nodraft
$ { \def\scale{2ex} \begin{tikzpicture}[x=\scale,y=\scale,baseline={([yshift=-.5ex]current bounding box.center)}] \begin{scope}[node distance=1] \coordinate (v0); \coordinate[right=.5 of v0] (v4); \coordinate[above right= of v4] (v2); \coordinate[below right= of v4] (v3); \coordinate[below right= of v2] (v5); \coordinate[right=.5 of v5] (v1); \coordinate[above right= of v2] (o1); \coordinate[below right= of v2] (o2); \coordinate[below left=.5 of v0] (i1); \coordinate[above left=.5 of v0] (i2); \coordinate[below right=.5 of v1] (o1); \coordinate[above right=.5 of v1] (o2); \draw (v0) -- (i1); \draw (v0) -- (i2); \draw (v1) -- (o1); \draw (v1) -- (o2); \draw (v0) to[bend left=20] (v2); \draw (v0) to[bend right=20] (v3); \draw (v1) to[bend left=20] (v3); \draw (v1) to[bend right=20] (v2); \draw (v2) to[bend right=60] (v3); \draw (v2) to[bend left=60] (v3); \filldraw (v0) circle(1pt); \filldraw (v1) circle(1pt); \filldraw (v2) circle(1pt); \filldraw (v3) circle(1pt); \ifdefined\cvl \draw[line width=1.5pt] (v0) to[bend left=20] (v2); \draw[line width=1.5pt] (v0) to[bend right=20] (v3); \fi \ifdefined\cvr \draw[line width=1.5pt] (v1) to[bend left=20] (v3); \draw[line width=1.5pt] (v1) to[bend right=20] (v2); \fi \ifdefined\cvml \draw[line width=1.5pt] (v2) to[bend left=60] (v3); \fi \ifdefined\cvmr \draw[line width=1.5pt] (v2) to[bend right=60] (v3); \fi \end{scope} \end{tikzpicture} }$
can be represented as the Hasse diagram
$ { \def\scale{2ex} \begin{tikzpicture}[x=\scale,y=\scale,baseline={([yshift=-.5ex]current bounding box.center)}] \begin{scope}[node distance=1] \coordinate (top) ; \coordinate [below left= of top] (v1); \coordinate [below right= of top] (v2); \coordinate [below left= of v2] (v3); \coordinate [below= of v3] (bottom); \draw (top) -- (v1); \draw (top) -- (v2); \draw (v1) -- (v3); \draw (v2) -- (v3); \draw (v3) -- (bottom); \filldraw[fill=white, draw=black] (top) circle(2pt); \filldraw[fill=white, draw=black] (v1) circle(2pt); \filldraw[fill=white, draw=black] (v2) circle(2pt); \filldraw[fill=white, draw=black] (v3) circle(2pt); \filldraw[fill=white, draw=black] (bottom) circle(2pt); \end{scope} \end{tikzpicture} } $
\else

MISSING IN DRAFT MODE

\fi
, where the vertices represent the subdiagrams in the set given in Example \ref{expl:subdiagrams1PIsd}.
\end{expl}

\begin{defn}[Lattice]
A lattice is a poset $L$ for which a unique least upper bound (\textit{join}) and a unique greatest lower bound (\textit{meet}) exists for any combination of two elements in $L$. The join of two elements $x,y \in L$ is denoted as $x \vee y$ and the meet as $x \wedge y$.
Every lattice has a unique greatest element denoted as $\hat{1}$ and a unique smallest element $\hat{0}$. Every interval of a lattice is also a lattice.
\end{defn}
In many QFTs, $\sdsubdiags_D(\Gamma)$ is a lattice for every s.d.\ diagram $\Gamma$:
\begin{defn}[Join-meet-renormalizable quantum field theory]
\label{def:joinmeetrenormalizable}
A renormalizable QFT is called join-meet-renormalizable if $\sdsubdiags_D(\Gamma)$, ordered by inclusion, is a lattice for every s.d.\ Feynman diagram $\Gamma$.
\end{defn}
\begin{thm}
A renormalizable QFT is join-meet-renormalizable if $\sdsubdiags_D(\Gamma)$ is closed under taking unions: $\gamma_1, \gamma_2 \in \sdsubdiags_D(\Gamma) \Rightarrow \gamma_1 \cup \gamma_2 \in \sdsubdiags_D(\Gamma)$ for all s.d.\ diagrams $\Gamma$.
\end{thm}
\begin{proof}
$\sdsubdiags_D(\Gamma)$ is ordered by inclusion $\gamma_1 \leq \gamma_2 \Leftrightarrow \gamma_1 \subset \gamma_2$.
The join is given by taking the union of diagrams: $\gamma_1 \join \gamma_2 := \gamma_1 \cup \gamma_2$. $\sdsubdiags_D(\Gamma)$ has a unique greatest element $\hat{1} := \Gamma$ and a unique smallest element $\hat{0} := \emptyset$. Therefore $\sdsubdiags_D(\Gamma)$ is a lattice \cite[Prop. 3.3.1]{stanley1997}. The unique meet is given by the formula, $\gamma_1 \meet \gamma_2 := \bigcup \limits_{\mu \leq \gamma_1 \text{ and } \mu \leq \gamma_2} \mu$. 
\end{proof}

A broad class of renormalizable QFTs is join-meet-renormalizable. This class includes the standard model of particle physics:
\begin{thm}
\label{thm:join_meet_std}
If all diagrams with four or more legs in a renormalizable QFT are superficially logarithmic divergent or superficially convergent, then the underlying QFT is join-meet-renormalizable.
\end{thm}
\begin{proof}
From $\gamma_1, \gamma_2 \in \sdsubdiags_D(\Gamma)$ immediately follows that $\gamma_1 \cup \gamma_2 \in \subdiags_{\text{bl}}(\Gamma)$. 
We want to verify $\gamma_1 \cup \gamma_2 \in \sdsubdiags_D(\Gamma)$.
This is obvious if all connected components of $\gamma_1$ and $\gamma_2$ are disjoint or contained in each other. 

The statement only needs to be validated if $\gamma_1$ and $\gamma_2$ are \textit{overlapping}. That means there is some connected component $\mu_1 \subset \gamma_1$ and another connected component $\mu_2 \subset \gamma_2$ such that $\mu_1 \cap \mu_2 \neq \emptyset$ and neither $\mu_1 \subset \mu_2$ nor $\mu_2 \subset \mu_1$. 

The connected subgraphs $\mu_1$ and $\mu_2$ must share at least one edge. Moreover, $\mu_1$ and $\mu_2$ are both bridgeless and connected by requirement. There must be at least two edges $e_1,e_2 \subset E_{\mu_1}$ that are not edges of $\mu_2$ that connected the subgraphs $\mu_1$ and $\mu_1 \cap \mu_2$. At least one edge is necessary as $\mu_1$ is a connected subgraph. Two edges are necessary because $\mu_1$ is bridgeless. This construction is symmetric: We can also find two edges $e_1,e_2 \in E_{\mu_2}$ which connect $\mu_2$ and $\mu_1 \cap \mu_2$. We therefore see that $\mu_1 \cap \mu_2$ must at least have four legs. 

If all diagrams with four or more legs in a renormalizable QFT are superficially logarithmic divergent or superficially convergent, then $\omega_D(\mu_1 \cap \mu_2) \geq 0$.

Observe that due to eq.\ \eqref{eqn:omega_D}, the definition of $\omega_D$, and inclusion-exclusion:
\begin{align*} \omega_D(\mu_1 \cup \mu_2) &\leq \omega_D(\mu_1) + \omega_D(\mu_2) - \omega_D(\mu_1 \cap \mu_2) \end{align*}
If $\omega_D(\mu_1)\leq0$ and $\omega_D(\mu_2)\leq 0$, then $\omega_D(\mu_1 \cup \mu_2) \leq 0$. For this reason $\mu_1 \cup \mu_2$ is superficially divergent, $\gamma_1 \cup \gamma_2 \in \sdsubdiags_D(\Gamma)$ and $\sdsubdiags_D(\Gamma)$ is closed under taking unions.
\end{proof}

In general, renormalizable QFTs are not join-meet-renormalizable. Figure \ref{fig:nolattice} shows an example of a s.d.\ diagram $\Gamma$, where $\sdsubdiags_D(\Gamma)$ is not a lattice. The diagram is depicted in Figure \ref{fig:phi6nolattice} and the corresponding poset in Figure \ref{fig:phi6nolatticeposet}.
The diagram%
\footnote{I wish to thank Erik Panzer for quickly coming up with the explicit counterexample in Figure \ref{fig:phi6nolattice}.}
appears in $\varphi^6$-theory, which is renormalizable, but not join-meet-renormalizable, in $3$-dimensions. 
{
\ifdefined\nodraft
\begin{figure}
  \subcaptionbox{Example of a diagram where $\sdsubdiags_3(\Gamma)$ is not a lattice.\label{fig:phi6nolattice}}
  [.45\linewidth]{
\def \scale{2em}
\begin{tikzpicture}[x=\scale,y=\scale,baseline={([yshift=-.5ex]current bounding box.center)}] \begin{scope}[node distance=1] \coordinate (v0); \coordinate[right=of v0] (v4); \coordinate[above right= of v4] (v2); \coordinate[below right= of v4] (v3); \coordinate[below right= of v2] (v5); \coordinate[right=of v5] (v1); \coordinate[above right= of v2] (o1); \coordinate[below right= of v2] (o2); \coordinate[below left=.5 of v0] (i1); \coordinate[above left=.5 of v0] (i2); \coordinate[below right=.5 of v1] (o1); \coordinate[above right=.5 of v1] (o2); \draw (v0) -- (i1); \draw (v0) -- (i2); \draw (v1) -- (o1); \draw (v1) -- (o2); \draw (v0) to[bend left=20] (v2); \draw (v0) to[bend left=45] (v2); \draw (v0) to[bend right=45] (v3); \draw (v0) to[bend right=20] (v3); \draw (v1) to[bend left=20] (v3); \draw (v1) to[bend left=45] (v3); \draw (v1) to[bend right=45] (v2); \draw (v1) to[bend right=20] (v2); \draw (v2) to[bend right=20] (v4); \draw (v2) to[bend left=20] (v5); \draw (v3) to[bend right=20] (v5); \draw (v3) to[bend left=20] (v4); \draw (v4) to[bend left=45] (v5); \draw (v4) to[bend left=15] (v5); \draw (v4) to[bend right=45] (v5); \draw (v4) to[bend right=15] (v5); \filldraw (v0) circle(1pt); \filldraw (v1) circle(1pt); \filldraw (v2) circle(1pt); \filldraw (v3) circle(1pt); \filldraw (v4) circle(1pt); \filldraw (v5) circle(1pt); \end{scope} \end{tikzpicture}
  }
  \subcaptionbox{The corresponding non-lattice poset. Trivial vertex multiplicities were omitted. \label{fig:phi6nolatticeposet}}
  [.45\linewidth]{
\def \scale{1em}
\begin{tikzpicture}[x=\scale,y=\scale,baseline={([yshift=-.5ex]current bounding box.center)}] \begin{scope}[node distance=1] \coordinate (top) ; \coordinate [below left=2 and 2 of top] (v1) ; \coordinate [below left=2 and 1 of top] (a2) ; \coordinate [below =2 of top] (a4) ; \coordinate [below =1 of a2] (v2) ; \coordinate [below =1 of a4] (v4) ; \coordinate [below right= 2 and 1 of top] (v5) ; \coordinate [below left = 4 and 1 of top] (w1) ; \coordinate [below= 4 of top] (w2) ; \coordinate [below left = 5 and .5 of top] (u1) ; \coordinate [below right =3 and 3 of top] (u2) ; \coordinate [below = 6 of top] (s) ; \coordinate [below = 7 of top] (t) ; \draw (top) -- (v1); \draw (top) -- (a2); \draw (top) -- (u2); \draw (top) -- (a4); \draw (top) -- (v5); \draw (a2) -- (v2); \draw (a4) -- (v4); \draw[color=white,line width=4pt] (v4) -- (w1); \draw (v4) -- (w1); \draw[color=white,line width=4pt] (v2) -- (w2); \draw (v2) -- (w1); \draw (v2) -- (w2); \draw (v4) -- (w2); \draw (v1) -- (w1); \draw (v5) -- (w2); \draw[color=white,line width=4pt] (w1) -- (u1); \draw[color=white,line width=4pt] (w2) -- (u1); \draw (w1) -- (u1); \draw (w2) -- (u1); \draw (u1) -- (s); \draw (u2) -- (s); \draw (s) -- (t); \filldraw[fill=white, draw=black] (top) circle(2pt); \filldraw[fill=white, draw=black] (v1) circle(2pt); \filldraw[fill=white, draw=black] (v2) circle(2pt); \filldraw[fill=white, draw=black] (v4) circle(2pt); \filldraw[fill=white, draw=black] (v5) circle(2pt); \filldraw[fill=white, draw=black] (a2) circle(2pt); \filldraw[fill=white, draw=black] (a4) circle(2pt); \filldraw[fill=white, draw=black] (w1) circle(2pt); \filldraw[fill=white, draw=black] (w2) circle(2pt); \filldraw[fill=white, draw=black] (u1) circle(2pt); \filldraw[fill=white, draw=black] (u2) circle(2pt); \filldraw[fill=white, draw=black] (s) circle(2pt); \filldraw[fill=white, draw=black] (t) circle(2pt); \end{scope} \end{tikzpicture}
  }
  \caption{Counterexample for a renormalizable but not join-meet-renormalizable QFT: $\varphi^6$-theory in $3$ dimensions.}
  \label{fig:nolattice}
\end{figure}
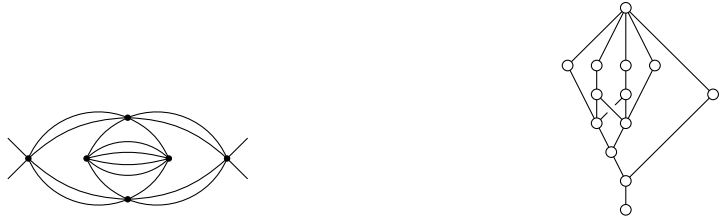
\else

MISSING IN DRAFT MODE

\fi
}

To proceed to the Hopf algebra of decorated posets some additional notation of poset and lattice theory must be introduced:

\paragraph{Order preserving maps}
A map $\sigma: P \rightarrow \N_0$ on a poset to the non-negative numbers is called strictly order preserving 
if $x < y$ implies $\sigma(x) < \sigma(y)$ for all $x,y \in P$. 
\paragraph{Cartesian product of posets}
From two posets $P_1$ and $P_2$ a new poset $P_1 \times P_2 = \left\{ (s,t) : s \in P_1 \text{ and } t \in P_2 \right\}$, the Cartesian product, with the order relation,
$(s,t) \leq (s',t')$ iff $s \leq s'$ and $t \leq t'$, can be obtained.

The Cartesian product is commutative and if $P_1$ and $P_2$ are lattices $P_1 \times P_2$ is also a lattice \cite{stanley1997}. This product is compatible with the notion of intervals:
\begin{align*} P_1 \times P_2 \supset [(s,t), (s',t')] =\left\{ (x,y) \in P_1 \times P_2 : s \leq x\leq s' \land t \leq y \leq t' \right\} = [s,s'] \times [t,t']. \end{align*}

\paragraph{Isomorphisms of posets}
An isomorphism between two posets $P_1$ and $P_2$ is a bijection $j: P_1 \rightarrow P_2$, which preserves the order relation: $j(x) \leq j(y) \Leftrightarrow x \leq y$. 

\subsection{The Hopf algebra of decorated posets}
Using the preceding notions a new Hopf algebra structure on posets, suitable for the description of the subdivergences, can be defined. This structure is essentially the one of an incidence Hopf algebra \cite{Schmitt1994} augmented by a strictly order preserving map as a decoration. This is a standard procedure as most applications of posets and lattices require an combinatorial interpretation of the elements of the posets \cite{stanley1997} - analogous to the applications of the Hopf algebras \cite{joni1979coalgebras}.
\begin{defn}[Hopf algebra of decorated posets]
Let $\mathcal{D}$ be the set of tuples $(P, \nu)$, where $P$ is a finite poset with a unique lower bound $\hat{0}$ and a unique upper bound $\hat{1}$ and a strictly order preserving map $\nu: P \rightarrow \N_0$ with $\nu(\hat{0}) = 0$. 
One can think of $\mathcal{D}$ as the set of bounded posets augmented by a strictly order preserving decoration.
An equivalence relation is set up on $\mathcal{D}$ by relating $(P_1, \nu_1) \sim (P_2, \nu_2)$ if there is an isomorphism $j: P_1 \rightarrow P_2$, which respects the decoration $\nu$: $\nu_1 = \nu_2 \circ j$. 

Let $\hopfpos$ be the $\Q$-algebra generated by all the elements in the quotient $\mathcal{D}/\sim$ with the commutative multiplication: 
\begin{align*} &m_{\hopfpos}: & \hopfpos &\otimes \hopfpos & &\rightarrow& &\hopfpos, \\ && (P_1, \nu_1) &\otimes (P_2, \nu_2) & &\mapsto& &\left( P_1 \times P_2, \nu_1 + \nu_2 \right), \end{align*}
which takes the Cartesian product of the two posets and adds the decorations $\nu$. The sum of the two functions $\nu_1$ and $\nu_2$ is to be interpreted in the sense: $(\nu_1 + \nu_2)(x,y) = \nu_1(x) + \nu_2(y)$. The singleton poset $P=\left\{\hat{0}\right\}$ with $\hat{0}=\hat{1}$ and the trivial decoration $\nu(\hat{0}) = 0$ serves as a unit: $\unit(1) = \one_{\hopfpos} := (\left\{\hat{0}\right\}, \hat{0} \mapsto 0)$.

Equipped with the coproduct,
\begin{align} \notag &\Delta_{\hopfpos}: & &\hopfpos& &\rightarrow& \hopfpos &\otimes \hopfpos, \\ \label{eqn:poset_cop} && &(P, \nu)& &\mapsto& \sum \limits_{ x \in P } ( [ \hat{0}, x ], \nu ) &\otimes \left( [x, \hat{1}], \nu - \nu(x) \right), \end{align}
where $(\nu - \nu(x))(y) = \nu(y) - \nu(x)$ 
and the counit $\counit$ which vanishes on every generator except $\one_{\hopfpos}$, the algebra $\hopfpos$ becomes a counital coalgebra. 
\end{defn}
\nomenclature{$\hopfpos$}{Hopf algebra of posets}
\begin{prop}
$\hopfpos$ is a bialgebra.
\end{prop}
\begin{proof}
As in Proposition \ref{prop:gaul_bialgebra}, the compatibility of the multiplication with the coproduct needs to be proven.
Let $(P_1, \nu_1), (P_2, \nu_2) \in \mathcal{D}$.
\begin{align*} &\Delta_{\hopfpos} \circ m_{\hopfpos} ( (P_1, \nu_1) \otimes (P_2, \nu_2) ) = \Delta_{\hopfpos} \left( P_1 \times P_2, \nu_1 + \nu_2 \right) = \\  &\sum \limits_{ x \in P_1 \times P_2 } ( [ \hat{0}, x ], \nu_1 + \nu_2 ) \otimes \left( [x, \hat{1}], \nu_1 + \nu_2 - \nu_1(x) - \nu_2(x) \right) = \\ &\sum \limits_{ y \in P_1 } \sum \limits_{ z \in P_2 }( [ \hat{0}_{P_1}, y ] \times [ \hat{0}_{P_2}, z ], \nu_1 + \nu_2 ) \otimes \left( [y, \hat{1}_{P_1}] \times [z, \hat{1}_{P_2}], \nu_1 + \nu_2 - \nu_1(x) - \nu_2(x) \right) = \\ &(m_{\hopfpos} \otimes m_{\hopfpos}) \circ \sum \limits_{ y \in P_1 } \sum \limits_{ z \in P_2 } \big[ ( [ \hat{0}_{P_1}, y ], \nu_1) \otimes ( [ \hat{0}_{P_2}, z ], \nu_2 ) \\ &\otimes ( [y, \hat{1}_{P_1}], \nu_1 - \nu_1(x) ) \otimes ( [z, \hat{1}_{P_2}], \nu_2 - \nu_2(x) ) \big] = \\ &(m_{\hopfpos} \otimes m_{\hopfpos}) \circ \tau_{2,3} \circ (\Delta_{\hopfpos} \otimes \Delta_{\hopfpos}) ( (P_1, \nu_1) \otimes (P_2, \nu_2) ), \end{align*}
where $\tau_{2,3}$ switches the second and the third factor of the tensor product.
\end{proof}
Note, that we also could have decorated the \textit{covers} of the lattices instead of the elements. We would have obtained a construction as in \cite{bergeron1999hopf} with certain restrictions on the edge-labels.
\begin{crll}
$\hopfpos$ is a connected Hopf algebra.
\end{crll}
\begin{proof}
$\hopfpos$ is graded by the value of $\nu(\hat{1})$. There is only one element of degree $0$ because $\nu$ must be strictly order preserving. It follows that $\hopfpos$ is a graded, connected bialgebra and therefore a Hopf algebra \cite{manchon2004hopf}.
\end{proof}

\subsection{A Hopf algebra homomorphism from Feynman diagrams to lattices}

\begin{thm}
Let $\nu$ map a graph to its loop number, $\nu(\gamma) = h_\gamma$.
\label{thm:hopf_alg_morph}
The map,
\begin{align*} &\chi_D:& &\hopffg_D& &\rightarrow& &\hopfpos, \\ && &\Gamma& &\mapsto& &( \sdsubdiags_D(\Gamma), \nu ), \end{align*}
which assigns to every diagram, its poset of s.d.\ subdiagrams decorated by the loop number of the subdiagram, is a Hopf algebra homomorphism.\footnote{Note that all residues $r \in \residuesstar$ map to $\one_\hopfpos$ under $\chi_D$, $\chi_D(r) = \one_\hopfpos$.}
\end{thm}
\nomenclature{$\chi_D$}{Hopf algebra homomorphism from $\hopffg_D$ to $\hopfpos$}
\begin{proof}
First, it needs to be shown that $\chi_D$ is an algebra homomorphism: $\chi_D \circ m_{\hopffg_D} = m_{\hopfpos} \circ ( \chi_D \otimes \chi_D )$. It is sufficient to prove this for the product of two generators $\Gamma_1, \Gamma_2 \in \hopffg_D$. Subdiagrams of the product, $m( \Gamma_1 \otimes \Gamma_2 ) = \Gamma_1 \sqcup \Gamma_2$, can be represented as pairs $(\gamma_1, \gamma_2)$ where $(\gamma_1, \gamma_2) \subset \Gamma_1 \sqcup \Gamma_2$ if $\gamma_1 \subset \Gamma_1$ and $\gamma_2 \subset \Gamma_2$. This corresponds to the Cartesian product regarding the poset structure of the subdivergences. The loop number of such a pair is the sum of the loop numbers of the components. Therefore,
\begin{gather*} \chi_D( \Gamma_1 \sqcup \Gamma_2 ) = ( \sdsubdiags_D( \Gamma_1 \sqcup \Gamma_2 ), \nu ) = ( \sdsubdiags_D( \Gamma_1 ) \times \sdsubdiags_D( \Gamma_2 ), \nu_1 + \nu_2 ) \\ =m_{\hopfpos} ( \chi_D(\Gamma_1) \otimes \chi_D(\Gamma_2) ). \end{gather*}
To prove that $\chi_D$ is a coalgebra homomorphism, we need to verify that, 
\begin{align} \label{eqn:comorphism} (\chi_D \otimes \chi_D) \circ \Delta_{\hopffg_D} = \Delta_{\hopfpos} \circ \chi_D. \end{align}
Choosing some generator $\Gamma$ of $\hopffg_D$ and using the definition of $\Delta_D$:
\begin{align*} (\chi_D \otimes \chi_D) \circ \Delta_{\hopffg_D} \Gamma = \sum \limits_{ \gamma \in \sdsubdiags_D(\Gamma) } \chi_D( \gamma) \otimes \chi_D( \Gamma / \gamma ), \end{align*}
the statement follows from $\chi_D(\gamma) = ( [ \hat{0}, \gamma ], \nu(\gamma) )$ and 
\begin{align*} \chi_D( \Gamma / \gamma ) = ( [ \emptyset, \Gamma / \gamma ], \nu ) \simeq ( [ \gamma, \Gamma ], \nu - \nu(\gamma) ), \end{align*}
which is a direct consequence of the definition of contractions in 
Definition \ref{def:contraction}.
\end{proof}

\begin{crll}
\label{crll:joinmeethopflat}
In a join-meet-renormalizable QFT, $\im(\chi_D) \subset \hopflat \subset \hopfpos$, where $\hopflat$ is the subspace of $\hopfpos$ which is generated by all elements $(L, \nu)$, where $L$ is a lattice. In other words: In a join-meet-renormalizable QFT, $\chi_D$ maps s.d.\ diagrams and products of them to decorated lattices. 
\end{crll}
\nomenclature{$\hopflat$}{Hopf algebra of lattices}

\begin{proof}
Follows directly from Definition \ref{def:joinmeetrenormalizable}.
\end{proof}

\begin{expl}
For any primitive diagram $\Gamma \in \text{Prim} \hopffg_D$,
\ifdefined\nodraft
\begin{align*} \chi_D( \Gamma ) = ( \sdsubdiags_D(\Gamma), \nu ) = {\def \scale {4ex} }, \end{align*}
where the vertices in the Hasse diagram are decorated by the value of $\nu$ and 
$L = h_\Gamma$ is the loop number of the primitive diagram. 

The coproduct of $\chi_D(\Gamma)$ in $\hopfpos$ can be calculated using eq.\ \eqref{eqn:poset_cop}:
\begin{align} \label{eqn:explcopposet} \Delta_{\hopfpos} {\def \scale {4ex} %
 }. \end{align}
As expected, these decorated posets are also primitive in $\hopfpos$.
\else

MISSING IN DRAFT MODE

\fi
\end{expl}
\begin{expl}
For the diagram
\ifdefined\nodraft
$ {\def \scale {1.5ex} %
 } \in \hopffg_4$,
$\chi_D$ gives the decorated poset,
\begin{align*} \chi_D\left( {\def \scale {4ex} %
 }, \end{align*}
\else

MISSING IN DRAFT MODE

\fi
\ifdefined\nodraft
of which the reduced coproduct in $\hopfpos$ can be calculated,
\begin{align} \label{eqn:explcopposet2} \widetilde{\Delta}_{\hopfpos} {\def \scale {4ex} %
 }. \end{align}
This can be compared to the coproduct calculation in Example \ref{expl:coproducthopffg}, 
\begin{align} \widetilde \Delta_4 { \def \scale {2ex} %
 } &= 2 { { \ifmmode \usebox{\fgsimplefourvtx} \else \newsavebox{\fgsimplefourvtx} \savebox{\fgsimplefourvtx}{ %
 } \fi} ~ \def \scale {2ex} %
 } + { { \ifmmode \usebox{\fgsimplefourvtx} \else \newsavebox{\fgsimplefourvtx} \savebox{\fgsimplefourvtx}{ %
 } \fi}^2 \def \scale {2ex} %
 } \end{align}
The identity from eq.\ \eqref{eqn:comorphism} is verified after computing the decorated poset of each subdiagram of 
$ {\def \scale {1.5ex} %
 }$ and comparing the previous two equations:
\begin{align*} &\chi_4 \left( { \ifmmode \usebox{\fgsimplefourvtx} \else \newsavebox{\fgsimplefourvtx} \savebox{\fgsimplefourvtx}{ %
 } \fi} ~ {\def \scale {3ex} %
 }. \end{align*}
\else

MISSING IN DRAFT MODE

\fi
\end{expl}
\section{Properties of the lattices of subdivergences}
\label{sec:properties}

Although, the Hopf algebra homomorphism $\chi_D$ can be applied in every renormalizable QFT, we shall restrict ourselves to join-meet-renormalizable QFTs, where $\chi_D$ maps to $\hopflat$, the Hopf algebra of decorated lattices, as a result of Corollary \ref{crll:joinmeethopflat}. 

The decorated lattice, which is associated to a Feynman diagram, encodes the `overlappingness' of the diagrams' subdivergences. Different join-meet-renormalizable QFTs have quite distinguished properties in this respect. Interestingly, the types of the decorated lattices appearing depend on the residues or equivalently on the superficial degree of divergence of the diagrams under consideration. For instance, it was proven by Berghoff in the context of Wonderful models that every diagram with only logarithmically divergent subdivergences (i.e. $\forall \gamma \in \sdsubdiags_D(\Gamma): \omega_D(\gamma) = 0$) is \textit{distributive}:
\begin{prop}{\cite[Prop. 3.22]{berghoff2014wonderful}}
If $\Gamma$ has only logarithmically s.d.\ subdiagrams in $D$ dimensions, (i.e. for all $\gamma \in \sdsubdiags_D(\Gamma)$ we have $\omega_D(\gamma) = 0$), then the distributivity identities,
\begin{align*} \gamma_1 \meet ( \gamma_2 \join \gamma_3) &= ( \gamma_1 \meet \gamma_2 ) \join ( \gamma_1 \meet \gamma_3 ) \\ \gamma_1 \join ( \gamma_2 \meet \gamma_3) &= ( \gamma_1 \join \gamma_2 ) \meet ( \gamma_1 \join \gamma_3 ), \end{align*}
hold for $\gamma_1, \gamma_2, \gamma_3 \in \sdsubdiags_D(\Gamma)$.
\end{prop}

Because distributive lattices are always graded \cite{stanley1997}, this implies that we have a bigrading on $\hopflat$ for these elements. One grading by the value of $\nu(\hat{1})$, corresponding to the loop number of the diagram, and one grading by the length of the maximal chains of the lattice, which coincides with the \textit{coradical degree} of the diagram in $\hopffg_D$. The coradical filtration of $\hopffg_D$, defined in eq.\ \eqref{eqn:coradfilt}, consequently becomes a grading for the subspaces generated by only logarithmically s.d.\ diagrams.

\subsection{Theories with only three-or-less-valent vertices}
From the preceding result the question arises, how much of the structure is left, if we also allow subdiagrams which are not only logarithmically divergent.
In renormalizable QFTs with only three-or-less-valent vertices, the lattices $\sdsubdiags_D(\Gamma)$ will turn out to be \textit{semimodular}. This is a weaker property than distributivity, but it still guarantees that the lattices are graded. To capture this property of $\sdsubdiags_D(\Gamma)$, some additional terms of lattice theory will be repeated following \cite{stanley1997}.

\paragraph{Join-irreducible element}
An element $x$ of a lattice $L$, $x\in L$ is called join-irreducible if $x = y \join z$ always implies $x=y$ or $x=z$.

\paragraph{Atoms and coatoms}
An element $x$ of $L$ is an atom of $L$ if it covers $\hat{0}$. It is a coatom of $L$ if $\hat{1}$ covers $x$.

\paragraph{Semimodular lattice}
A lattice $L$ is semimodular if for two elements $x,y\in L$ that cover $x \meet y$, $x$ and $y$ are covered by $x \join y$.

With these notions we can formulate
\begin{lmm}
\label{lmm:propagatorunion}
If in a renormalizable QFT with only three-or-less-valent vertices $\mu_1$ and $\mu_2$ are overlapping connected components, they must be of vertex-type and $\mu_1\cup \mu_2$ of propagator-type.
\end{lmm}
\begin{proof}
As in Theorem \ref{thm:join_meet_std} this follows from the fact that the intersection of two overlapping connected components always has four legs. In a theory with three-or-less-valent vertices, the subgraph $\mu_1 \cap \mu_2$ must therefore be superficially convergent, that means $\omega_D(\mu_1 \cap \mu_2) > 0$. From inclusion exclusion we know that 
$\omega_D(\mu_1 \cup \mu_2) \leq \omega_D(\mu_1) + \omega_D(\mu_2) - \omega_D(\mu_1 \cap \mu_2)$.
Because in a renormalizable QFT with three-or-less-valent vertices every subdivergence either has two or three external legs, we must have $\omega_D(\mu_1) = 0$, $\omega_D(\mu_2) = 0$ and $\omega_D(\mu_1 \cup \mu_2) < 0$. The statement follows.
\end{proof}

\begin{crll}
\label{crll:joinirreducible}
In a QFT with only three-or-less-valent vertices, vertex-type s.d.\ diagrams $\Gamma$ ($|\Hext(\Gamma)|=3$) are always join-irreducible elements of $\sdsubdiags_D(\Gamma)$.
\end{crll}
\begin{proof}
Suppose there were $\gamma_1, \gamma_2 \in \sdsubdiags_D(\Gamma)$ with $\gamma_1\neq \Gamma$, $\gamma_2 \neq \Gamma$ and $\gamma_1 \join \gamma_2 = \Gamma$. The subdivergences $\gamma_1$ and $\gamma_2$ are therefore overlapping. As Lemma \ref{lmm:propagatorunion} requires $\Gamma$ to be of propagator type, we have a contradiction.
\end{proof}
\begin{figure}
{
\ifdefined\nodraft
\begin{center}
${ \def\scale {10ex} \begin{tikzpicture}[x=\scale,y=\scale,baseline={([yshift=-.5ex]current bounding box.center)}] \begin{scope}[node distance=1] \coordinate (i); \coordinate[right=.5 of i] (vl); \coordinate[above right= of vl] (v1); \coordinate[below right= of vl] (v2); \coordinate[right=.5 of v1] (v3); \coordinate[right=.5 of v2] (v4); \coordinate[below right= of v3] (vr); \coordinate[right=.5 of vr] (o); \draw (i) -- (vl); \draw (o) -- (vr); \draw (vl) to[bend right=90] (vr); \draw (vl) to[bend left=90] (vr); \draw [dashed] ($(v1) - (.5,0)$) -- ($(v2) - (.5,0)$); \draw [dashed] ($(v3) + (.5,0)$) -- ($(v4) + (.5,0)$); \filldraw (vl) circle(1pt); \filldraw (vr) circle(1pt); \draw [fill=white] (v1) rectangle (v4); \draw [fill=white,thick,pattern=north west lines] (v1) rectangle (v4); \end{scope} \end{tikzpicture}}$
\end{center}
\else

MISSING IN DRAFT MODE

\fi
}
\caption{Structure of overlapping divergences in three-valent QFTs}
\label{fig:blob33}
\end{figure}
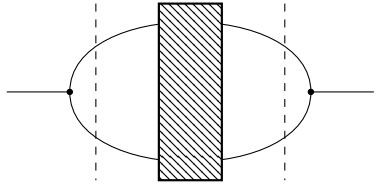

\begin{prop}
\label{prop:semimodular}
In a renormalizable QFT with only three-or-less-valent vertices, the lattice $\sdsubdiags_D(\Gamma)$ is semimodular for every Feynman diagram $\Gamma$. 
\end{prop}
\begin{proof}
Recall that 
a lattice is semimodular if for two elements $x,y\in L$ that cover $x \meet y$, $x$ and $y$ are covered by $x \join y$.

For two diagrams $\gamma_1, \gamma_2 \in \sdsubdiags_D(\Gamma)$ we can always form the contractions by $\gamma_1 \meet \gamma_2$: $\gamma_1 / (\gamma_1 \meet \gamma_2)$ and $\gamma_2 / (\gamma_1 \meet \gamma_2)$. Hence, the statement that $\gamma_1,\gamma_2$ cover $\gamma_1 \meet \gamma_2$ is equivalent to stating that $\gamma_1 / (\gamma_1 \meet \gamma_2)$ and $\gamma_2 / (\gamma_1 \meet \gamma_2)$ are primitive.

To prove that $\gamma_1 \join \gamma_2$ covers $\gamma_1$ and $\gamma_2$ if $\gamma_1$ and $\gamma_2$ cover $\gamma_1 \meet \gamma_2$, it is therefore sufficient to verify that for $\gamma_1,\gamma_2$ primitive $(\gamma_1\cup \gamma_2) / \gamma_1$ and $(\gamma_1\cup \gamma_2) / \gamma_2$ are primitive as well. This is obvious if $\gamma_1, \gamma_2$ are not overlapping. 

If $\gamma_1$ and $\gamma_2$ are overlapping and both connected, they must be of vertex-type and $\gamma_1\cup \gamma_2$ of propagator-type as proven in Lemma \ref{lmm:propagatorunion}.
Because only three-valent vertices are allowed each $\gamma_1$ and $\gamma_2$ must provide one external edge for $\gamma_1 \cup \gamma_2$. The situation is depicted in Figure \ref{fig:blob33}. For both $\gamma_1$ and $\gamma_2$ to be primitive, they must share the same four-leg kernel, depicted as a striped box. Contraction with either $\gamma_1$ or $\gamma_2$ results in a one-loop propagator, which is primitive.
\end{proof}

Semimodular lattices have a very rich structure, see for instance Stern's book \cite{stern1999semimodular}. For instance, semimodularity implies that the lattices under consideration are graded:
\begin{thm}
\label{thm:gradthree}
In a renormalizable QFT with only three-or-less-valent vertices:
\begin{itemize}
\item $\sdsubdiags_D(\Gamma)$ is a graded lattice for every propagator, vertex-type diagram or disjoint unions of both.
\item $\hopflat$ is bigraded by $\nu(\hat{1})$ and the length of the maximal chains of the lattices, which coincides with the coradical degree in $\hopflat$.
\item $\hopffg_D$ is bigraded by $h_\Gamma$ and the coradical degree of $\Gamma$. 
\item Every complete forest of $\Gamma$ has the same length.
\end{itemize}
\end{thm}
\begin{proof}
Every semimodular lattice is graded \cite[Proposition 3.3.2]{stanley1997}.
\end{proof}

\subsection{Theories with only four-or-less-valent vertices}

{
\ifdefined\nodraft
\begin{figure}
  \subcaptionbox{Diagram where $\sdsubdiags_4(\Gamma)$ forms a non-graded lattice.\label{fig:diagramnonranked}}
  [.45\linewidth]{
      $\Gamma = { \def \scale{2em} }$
  }
  \subcaptionbox{The Hasse diagram of the corresponding non-graded lattice, where the decoration was omitted.\label{fig:diagramnonrankedlattice}}
  [.45\linewidth]{
  $\chi_4(\Gamma) = { \def \scale{1em} %
 }$
  }
  \subcaptionbox{The non-trivial superficially divergent subdiagrams and the complete forests which can be formed out of them.\label{fig:diagramnonrankeddetails}}
  [\linewidth]{
\def \scale{1.5em}
$ \begin{aligned} \alpha_1 &= { \def\cvlm {} \def\cvlb {} \def\cvmt {} \def\cvmm {} \def\cvmb {} \def\cvrt {} \def\cvrm {} \def\cvrb {} %
 } & & &\end{aligned}$
      \\
with the complete forests $\emptyset \subset \delta_1 \subset \alpha_i \subset \Gamma$, 
$\emptyset \subset \delta_2 \subset \beta_i \subset \Gamma$ and
$\emptyset \subset \gamma_i \subset \Gamma$.
  }
\caption{Counterexample of a lattice, which appears in join-meet-renormalizable QFTs with four-valent vertices and is not graded.}
\label{fig:counterexplsemimod}
\end{figure}
\else

MISSING IN DRAFT MODE

\fi
}

We have shown that every lattice associated to a s.d.\ diagram in a QFT with only three-or-less-valent vertices is semimodular. For join-meet-renormalizable QFTs which also have four-valent vertices the situation is more involved as the example in Figure \ref{fig:counterexplsemimod} exposes. The depicted lattice in Figure \ref{fig:diagramnonrankedlattice} associated to the $\varphi^4$-diagram in Figure \ref{fig:diagramnonranked} is obviously not semimodular, because it is not graded. This implies that not all complete forests are of the same length in theories, where this topology can appear. This includes $\varphi^4$ and Yang-Mills theories in four dimensions. 

The s.d.\ subdiagrams of the counterexample are illustrated in Figure \ref{fig:diagramnonrankeddetails}. It can be seen that there are six complete forests of length four and three complete forests of length three.

The pleasant property of semimodularity can be recovered by working in the Hopf algebra of Feynman diagrams without tadpoles or equivalently by setting all tadpole diagrams to zero. This is quite surprising, because the independence of loops in tadpoles from external momenta and the combinatorial structure of BPHZ, encoded by the Hopf algebra of Feynman diagrams, seem independent on the first sight. 

Formally, we can transfer the restriction to tadpole-free diagrams to $\hopflat$ by the following procedure:
The Hopf algebra homomorphism $\psi: \hopffg_D \rightarrow \hopffgs_D$ defined in eq.\ \eqref{eqn:morphismpsi} gives rise to the Hopf ideal $\ker \psi \subset \hopffg_D$. Using the Hopf algebra homomorphism $\chi_D$ a Hopf ideal of $\hopflat$, $\chi_D(\ker \psi ) \subset \hopflat$, is obtained. This can be summarized in a commutative diagram:

\begin{center}
\begin{tikzpicture} \node (tl) {$\hopffg$}; \node [right=of tl] (tr) {$\hopffgs$}; \node [below =of tl] (bl) {$\hopflat$}; \node [below =of tr] (br) {$\hopflats$}; \draw[->] (tl) -- node[above] {$\psi$} (tr); \draw[->] (tl) to node[auto] {$\chi_D$} (bl); \draw[->] (bl) -- node[auto] {$\psi'$} (br); \draw[->] (tr) -- node[right] {$\chi_D'$} (br); \end{tikzpicture}
\end{center}
where $\hopflats$ is the quotient $\hopflats:= \hopflat/\chi_D(\ker\psi)$ and $\psi'$ is just the projection to $\hopflats$. 

The interesting part is the homomorphism $\chi_D': \hopffgs_D \rightarrow \hopflats$, which maps from the Hopf algebra of Feynman diagrams without tadpoles to $\hopflats$. Such a map can be constructed explicitly and for theories with only four-or-less-valent vertices, it can be ensured that $\chi_D'$ maps Feynman diagrams to decorated semimodular lattices. 

\begin{prop}
In a renormalizable QFT with only four-or-less-valent vertices, $\chi_D'$ maps elements from the Hopf algebra of Feynman diagrams without tadpoles to decorated lattices.
\end{prop}
\begin{proof}
Explicitly, $\chi_D'$ is the map,
\begin{align*} \chi_D': \Gamma \mapsto (\sdsubdiagsn_D (\Gamma), \nu), \end{align*}
where the decoration $\nu$ is the same as above.

We need to show that $\sdsubdiagsn_D(\Gamma)$ ordered by inclusion is a lattice. This is not as simple as before, because $\gamma_1, \gamma_2 \in \sdsubdiagsn_D(\Gamma)$ does not necessarily imply $\gamma_1 \cup \gamma_2 \in \sdsubdiagsn_D(\Gamma)$. From Definition \ref{def:snailfreehopf} of $\sdsubdiagsn_D(\Gamma)$, we can deduce that if $\gamma_1, \gamma_2 \in \sdsubdiagsn_D(\Gamma)$, then $\gamma_1 \cup \gamma_2 \notin \sdsubdiagsn_D(\Gamma)$ iff $\Gamma / \gamma_1 \cup \gamma_2$ is a tadpole. 

To prove that there still exists a least upper bound for every pair $\gamma_1$, $\gamma_2$ we must ensure that every element $\mu \in \sdsubdiags_D(\Gamma)$ and $\mu \notin \sdsubdiagsn_D(\Gamma)$ is only covered by only one element in $\sdsubdiagsn_D(\Gamma)$. This is equivalent to stating that if $\gamma \subset \delta \subset \Gamma$ and $\delta / \gamma$ is a primitive tadpole (i.e. a self-loop with one vertex), then there is no $\delta'$ with $\gamma \subset \delta' \subset \Gamma$ such that $\delta' / \gamma$ is a primitive tadpole. There cannot be such a second subdiagram $\delta'$. Suppose there were such $\delta$ and $\delta'$. $\delta$ and $\delta'$ are obtained from $\gamma$ by joining two of its external legs to an new edge. As only four-or-less-valent vertices are allowed, such a configuration can only be achieved if $\gamma$ is a diagram with four external legs. $\delta$ and $\delta'$ are the diagrams obtained by closing either pair of legs of $\gamma$. This would imply that $\delta_1 \cup \delta_2$ is a vacuum diagram without external legs, which is excluded.
\end{proof}

\begin{expl}
The map $\chi_D'$ can be applied to the example in Figure \ref{fig:counterexplsemimod} where the lattice obtained by $\chi_D$ is not semimodular. It can be seen from Figure \ref{fig:diagramnonrankeddetails} that the only diagrams, which do not result in tadpole diagrams upon contraction are $\delta_1$ and $\delta_2$. Accordingly, 
\begin{align*} \chi_4'\left({ \def \scale{1.5em} \begin{tikzpicture}[x=\scale,y=\scale,baseline={([yshift=-.5ex]current bounding box.center)}] \begin{scope}[node distance=1] \coordinate (i1); \coordinate[right=.5 of i1] (v0); \coordinate[right=.5 of v0] (v2); \coordinate[right=.5 of v2] (vm); \coordinate[above=of vm] (v3); \coordinate[below=of vm] (v4); \coordinate[right=.5 of vm] (v2d); \coordinate[right=.5 of v2d] (v1); \coordinate[right=.5 of v1] (o1); \ifdefined \cvlt \draw[line width=1.5pt] (v0) to[bend left=45] (v3); \else \draw (v0) to[bend left=45] (v3); \fi \ifdefined \cvlb \draw[line width=1.5pt] (v0) to[bend right=45] (v4); \else \draw (v0) to[bend right=45] (v4); \fi \ifdefined \cvlm \draw[line width=1.5pt] (v0) -- (v2); \else \draw (v0) -- (v2); \fi \ifdefined \cvrt \draw[line width=1.5pt] (v1) to[bend right=45] (v3); \else \draw (v1) to[bend right=45] (v3); \fi \ifdefined \cvrb \draw[line width=1.5pt] (v1) to[bend left=45] (v4); \else \draw (v1) to[bend left=45] (v4); \fi \ifdefined \cvmt \draw[line width=1.5pt] (v3) to[bend right=20] (v2); \else \draw (v3) to[bend right=20] (v2); \fi \ifdefined \cvmb \draw[line width=1.5pt] (v4) to[bend left=20] (v2); \else \draw (v4) to[bend left=20] (v2); \fi \ifdefined \cvmm \draw[line width=1.5pt] (v3) to[bend left=20] (v2d); \draw[line width=1.5pt] (v4) to[bend right=20] (v2d); \else \draw (v3) to[bend left=20] (v2d); \draw (v4) to[bend right=20] (v2d); \fi \filldraw[color=white] (v2d) circle(.2); \ifdefined \cvrm \draw[line width=1.5pt] (v1) -- (v2); \else \draw (v1) -- (v2); \fi \draw (v0) -- (i1); \draw (v1) -- (o1); \filldraw (v0) circle(1pt); \filldraw (v1) circle(1pt); \filldraw (v2) circle(1pt); \filldraw (v3) circle(1pt); \filldraw (v4) circle(1pt); \end{scope} \end{tikzpicture} } \right) = { \def \scale{1.5em} \begin{tikzpicture}[x=\scale,y=\scale,baseline={([yshift=-.5ex]current bounding box.center)}] \begin{scope}[node distance=1] \coordinate (top) ; \coordinate [below left= of top] (v1); \coordinate [below right= of top] (v2); \coordinate [below left= of v2] (bottom); \draw (top) -- (v1); \draw (top) -- (v2); \draw (v1) -- (bottom); \draw (v2) -- (bottom); \filldraw[fill=white, draw=black,circle] (top) node[fill,circle,draw,inner sep=1pt]{$5$}; \filldraw[fill=white, draw=black,circle] (v1) node[fill,circle,draw,inner sep=1pt]{$3$}; \filldraw[fill=white, draw=black,circle] (v2) node[fill,circle,draw,inner sep=1pt]{$3$}; \filldraw[fill=white, draw=black,circle] (bottom) node[fill,circle,draw,inner sep=1pt]{$0$}; \end{scope} \end{tikzpicture} }, \end{align*}
which is a graded lattice.
\end{expl}

\begin{prop}
In a QFT with only four-or-less-valent vertices 
$\chi_D'$ maps elements from the Hopf algebra of Feynman diagrams without tadpoles to decorated semimodular lattices.
\end{prop}
\begin{proof}
As above we only need to prove that if $\gamma_1$ and $\gamma_2$ are overlapping and primitive, then $(\gamma_1\join \gamma_2) / \gamma_1$ and $(\gamma_1\join \gamma_2) / \gamma_2$ are primitive as well.

If we have a subgraph $\gamma$ which has one connected component that connects the legs of the original graph then its contraction must be a tadpole. For this reason, we can characterize the connected components of a subgraph by the proper subset of external half-edges of the full diagram it contains. 

If $(\gamma_1 \join \gamma_2)/\gamma_1$ was not primitive, we could remove the vertex that $\gamma_1$ was contracted to and the adjacent edges. The result would be a s.d.\ subdiagram of $\gamma_2$ in contradiction with the requirement.
\end{proof}

It is interesting how important taking the quotient by the tadpole diagrams is, to obtain the property of semimodularity for the lattices of Feynman diagrams.
\begin{thm}
\label{thm:gradfour}
In a renormalizable QFT with only four-or-less-valent vertices:
\begin{itemize}
\item $\sdsubdiagsn_D(\Gamma)$ is a graded lattice for every propagator, vertex-type diagram or disjoint unions of both.
\item $\hopflats$ is bigraded by $\nu(\hat{1})$ and the length of the maximal chains of the lattices, which coincides with the coradical degree in $\hopflat$.
\item $\hopffgs_D$ is bigraded by $h_\Gamma$ and the coradical degree of $\Gamma$. 
\item Every complete forest of $\Gamma$, which does not result in a tadpole upon contraction, has the same length.
\end{itemize}
\end{thm}
\begin{proof}
Every semimodular lattice is graded \cite[Proposition 3.3.2]{stanley1997}.
\end{proof}
The overlapping diagrams in $\sdsubdiagsn_D(\Gamma)$ are characterized by the external legs of $\Gamma$ they contain. As a consequence, there is a limited number of possibilities for primitive diagrams to be overlapping. A two-leg diagram can only be the join of at most two primitive overlapping diagrams and a three-leg diagram can only be the join of at most three primitive divergent overlapping diagrams. For four-leg diagrams in theories with only four-or-less-valent vertices the restriction is even more serve: In these cases, a four-leg diagram can only by the join of at most two primitive overlapping diagrams.
\section{Applications to Zero-Dimensional QFT}
\label{sec:applications_lattices}
As an application of the lattice structure, the enumeration of some classes of primitive diagrams using techniques from zero-dimensional quantum field theories is presented. 
As in Chapter \ref{chp:graph_enumeration}, we will use the characteristic property of zero-dimensional QFT: every diagram in the perturbation expansion has the amplitude $1$.
On the Hopf algebra of Feynman diagrams such a prescription can be formulated by the character or \textit{Feynman rule}: 
\begin{align} &\phi:& &\hopffg_D& &\rightarrow& &\Q[[\hbar]],& & \\ && &\Gamma& &\mapsto& &\hbar^{h_\Gamma},& && \end{align}
which maps every Feynman diagram to $\hbar$ to the power of its number of loops in the ring of powerseries in $\hbar$. Clearly, $\phi$ is in $\chargroup{\hopffg_D}{\Q[[\hbar]]}$, the group of characters of $\hopffg_D$ to $\Q[[\hbar]]$. 
Note, that we are not setting $D=0$ even though $\phi$ are the Feynman rules for zero-dimensional QFT. Every diagram would be `convergent' and the Hopf algebra $\hopffg_0$ trivial. It might be clearer to think about $\phi$ as toy Feynman rules which assign $1$ to every Feynman diagram without any respect to kinematics. This way, we can still study the effects of renormalization on the amplitudes in an arbitrary dimension of spacetime. 

As before, we define the sum of all 1PI diagrams with a certain residue $v$ weighted by their symmetry factor as,
\begin{align} \allG^{(v)}_{\sgset^{\text{s.d.}}_D}:= \sum_{\substack{\Gamma \in \sgset^{\text{s.d.}}_D \\ \res(\Gamma) = v}} \frac{\Gamma}{|\Aut(\Gamma)|}, \end{align}
such that $\phi\left(\allG^{(v)}_{\sgset^{\text{s.d.}}_D} \right)$ is the generating function of these weighted diagrams with $\hbar$ as a counting variable. This generating function is the perturbation expansion of the Green's function for the residue $v$. 

The \textit{counterterm map} \cite{connes2001renormalization} is defined as,
\begin{align} S^R_D := R \circ \phi \circ S_D, \end{align}
in a \textit{multiplicative renormalization scheme} $R$ with the antipode $S_D$ of $\hopffg_D$.
$S^R_D$ is called the counterterm map, because it maps the sum of all 1PI diagrams with a certain residue $v$ to the corresponding counterterm, which when substituted into the Lagrangian renormalizes the QFT appropriately. The renormalized Feynman rules are given by the convolution product $\phi^R_D := S^R_D * \phi$. 

For the toy Feynman rules $\phi$, there are no kinematics to choose a multiplicative renormalization scheme from. The renormalization will be modeled as usual in the scope of zero-dimensional-QFTs by setting $R=\id$. Consequently, $S^R_D = \phi \circ S_D$. 

As was illustrated at length in Chapter \ref{chap:coalgebra_graph}, the map $\phi^R_D = S^R_D * \phi = (\phi \circ S_D) \star \phi = \unit \circ \counit$ vanishes on all generators of $\hopffg_D$ except on $\one$. This can be used to obtain differential equations for the $S^R_D\left(\allG^{(v)}_{\sgset^{\text{s.d.}}_D}\right)$ power series, which are called \textit{$z$-factors} and other interesting quantities as was done in \cite{cvitanovic1978number,argyres2001zero}. 

The antipode in the formulas above is the point where the Hopf algebra structure enters the game. The lattice structure can be used to clarify the picture even more.

We define $\phi' \in \chargroup{\hopflat}{\Q[[\hbar]]}$, a Feynman rule on the Hopf algebra of decorated lattices, analogous to $\phi \in \chargroup{{\hopffg_D}}{\Q[[\hbar]]}$:
\begin{align} &\phi':& &\hopflat& &\rightarrow& &\Q[[\hbar]],& & \\ && &(P, \nu)& &\mapsto& &\hbar^{\nu(\hat{1})},& && \end{align}
which maps a decorated lattice to the value of the decoration of the largest element. Immediately, we can see that $\phi = \phi' \circ \chi_D$.
For the counterterm map, we obtain
\begin{align} S^R_D &= \phi' \circ \chi_D \circ S_D. \end{align}
Using Theorem \ref{thm:hopf_alg_morph}, we can commute $\chi_D$ and $S_D$,
\begin{align} S^R_D &= \phi' \circ S_{\hopflat} \circ \chi_D, \end{align}
where $S_{\hopflat}$ is the antipode in the $\hopflat$.
For this reason, the evaluation of $S^R_D$ can be performed entirely in $\hopflat$. $S^R_D$ reduces to a combinatorial calculation on the lattice which is obtained by the Hopf algebra homomorphism $\chi_D$. The homomorphism $\phi' \circ S_{\hopflat}$ maps decorated lattices into the ring of powerseries in $\hbar$. Because $S_{\hopflat}$ respects the grading in $\nu(\hat{1})$, we can write
\begin{align} \phi' \circ S_{\hopflat} (L, \nu) = \hbar^{\nu(\hat{1})} \zeta \circ S_{\hopflat} (L, \nu), \end{align}
where $\zeta$ is the characteristic function $(L, \nu)\mapsto 1$. The map $\zeta \circ S_{\hopflat}$ is the \textit{Moebius function}, $\mu(\hat{0},\hat{1})$, on the lattice \cite{ehrenborg1996posets}. It is defined recursively as, 
\begin{defn}[Moebius function]
\begin{align} \label{eqn:moebius} \mu_P(x,y) &= \begin{cases} 1,&\text{if } x=y \\ - \sum_{x \leq z < y} \mu_P(x,z) & \text{if } x < y. \end{cases} \end{align}
for a poset $P$ and $x,y \in P$.
\end{defn}

We summarize these observations in 
\begin{thm}
For zero-dimensional-QFT Feynman rules as $\phi$, the counterterm map takes the form
\begin{align} {S^R_{\hopflat}}'(L,\nu) = \hbar^{\nu(\hat{1})} \mu_L( \hat{0}, \hat{1} ) \end{align}
on the Hopf algebra of lattices, where ${S^R_{\hopflat}}' = \phi' \circ S_{\hopflat}$ and with $\hat{0}$ and $\hat{1}$ the lower and upper bound of $L$.
\end{thm}
\begin{crll}
\label{crll:phimoebius}
\begin{align} \label{eqn:phimoebius} {S^R_D}(\Gamma) = \hbar^{h_\Gamma} \mu_{\sdsubdiags_D(\Gamma)}( \hat{0}, \hat{1} ) \end{align}
on the Hopf algebra of Feynman diagrams with $\hat{0}=\emptyset$ and $\hat{1}=\Gamma$, the lower and upper bound of $\sdsubdiags_D(\Gamma)$. 
\end{crll}
Note that these considerations are not limited to the Hopf algebra of Feynman diagrams. The evaluation of the character $\zeta \circ S_{\hopflat}(\Gamma)$ can be interpreted as the value of the Moebius function of the respective inclusion poset for all graph Hopf algebras from the previous chapter.

On these grounds, the counterterms in zero-dimensional QFT can be calculated only by computing the Moebius function on the lattice $\sdsubdiags_D(\Gamma)$.
The Moebius function is a well studied object in combinatorics. There are especially sophisticated techniques to calculate the Moebius functions on lattices (see \cite{stanley1997,stern1999semimodular}). 
For instance
\begin{thm}{(Rota's crosscut theorem for atoms and coatoms (special case of \cite[cor. 3.9.4]{stanley1997}))}
Let L be a finite lattice and $X$ its set of atoms and $Y$ its set of coatoms, then
\begin{align} \mu_L(\hat{0}, \hat{1}) = \sum_k (-1)^k N_k = \sum_k (-1)^k M_k, \end{align}
where $N_k$ is the number of $k$-subsets of $X$ whose join is $\hat{1}$ and 
$M_k$ is the number of $k$-subsets of $Y$ whose meet is $\hat{0}$.
\end{thm}
With this theorem the Moebius functions of all the lattices appearing in this chapter can be calculated very efficiently.

In many cases, an even simpler theorem, which is a special case of the previous one, applies:
\begin{thm}{(Hall's theorem \cite[cor. 4.1.7.]{stern1999semimodular})}
If in a lattice $\hat{1}$ is not a join of atoms or $\hat{0}$ is not a meet of coatoms, then $\mu(\hat{0}, \hat{1}) = 0$.
\end{thm}

In Corollary \ref{crll:joinirreducible}, we proved that every vertex-type subdiagram in a QFT with only three-valent vertices is join-irreducible. Hence, it is also not a join of atoms except if it is an atom itself.

\begin{thm}
\label{thm:threeQFTmoebvert}
In a renormalizable QFT with only three-or-less-valent vertices and $\Gamma$ a vertex-type s.d.\ diagram (i.e. $|\legs_\Gamma|=3$):
\begin{align} {S^R_D}(\Gamma)&= \begin{cases} - \hbar^{h_\Gamma}& \text{ if } \Gamma \text{ is primitive} \\ 0& \text{ if } \Gamma \text{ is not primitive.} \end{cases} \end{align}
\end{thm}
\begin{proof}
In both cases the element $\hat{1}=\Gamma$ in the lattice $\sdsubdiags_D(\Gamma)$ is join-irreducible (Corollary \ref{crll:joinirreducible}). If $\Gamma$ is primitive $\phi \circ S(\Gamma) = -\phi (\Gamma) = -\hbar^{h_\Gamma}$. If $\Gamma$ is not primitive, it does not cover $\hat{0}$. This implies that $\hat{1}$ is not a join of atoms. Therefore, $\mu_{\sdsubdiags_D(\Gamma)}( \hat{0}, \hat{1} )$ vanishes and so does $S^R_D(\Gamma)$ in accordance with Corollary \ref{crll:phimoebius}. 
\end{proof}

\begin{crll}
\label{crll:threevalentprimitivecounting}
In a renormalizable QFT with only three-or-less-valent vertices and $v\in \mathcal{R}_v$ a vertex-type residue:
\begin{align} {S^R_D}\left(\allG^{(v)}_{\sgset^{\text{s.d.}}_D} \right) = \frac{1}{v!} - \phi \circ P_{\text{Prim}(\hopffg_D)}\left(\allG^{(v)}_{\sgset^{\text{s.d.}}_D} \right), \end{align}
where $P_{\text{Prim}(\hopffg_D)}$ projects onto the primitive generators of $\hopffg_D$.
\end{crll}

Summarizing, we established that in a theory with only three-or-less-valent vertices the counterterm ${S^R_D}\left(\allG^{(v)}_{\sgset^{\text{s.d.}}_D} \right)$ counts the number of primitive diagrams if $v\in \mathcal{R}_v$. This fact has been used indirectly in \cite{cvitanovic1978number} to obtain the generating functions for primitive vertex diagrams in $\varphi^3$-theory. 

The conventional $z$-factor for the respective vertex is $z^{(v)} = v! {S^R_D}\left(\allG^{(v)}_{\sgset^{\text{s.d.}}_D} \right)$, where the factorial of the residue $v=\prod_{\varphi \in \fields} \varphi^{n_\varphi}$ is $v! = \prod_{\varphi \in \fields} n_{\varphi}!$. In the single colored or \textit{scalar} case the factorial reduces to $\deg{v}!$.

Further exploitation of the lattice structure leads to a statement on propagator-type diagrams in such theories:
\begin{thm}
In a renormalizable QFT with three-or-less-valent vertices the pro\-paga\-tor-type diagrams $\Gamma$, for which $S^R_D(\Gamma) \neq 0$, must have the lattice structure 
${ \def \scale{2ex} \begin{tikzpicture}[x=\scale,y=\scale,baseline={([yshift=-.5ex]current bounding box.center)}] \begin{scope}[node distance=1] \coordinate (top) ; \coordinate [below= of top] (bottom); \draw (top) -- (bottom); \filldraw[fill=white, draw=black] (top) circle(2pt); \filldraw[fill=white, draw=black] (bottom) circle(2pt); \end{scope} \end{tikzpicture} } $ or 
${ \def \scale{2ex} \begin{tikzpicture}[x=\scale,y=\scale,baseline={([yshift=-.5ex]current bounding box.center)}] \begin{scope}[node distance=1] \coordinate (top) ; \coordinate [below left= of top] (v1); \coordinate [below right= of top] (v2); \coordinate [below left= of v2] (bottom); \draw (top) -- (v1); \draw (top) -- (v2); \draw (v1) -- (bottom); \draw (v2) -- (bottom); \filldraw[fill=white, draw=black] (top) circle(2pt); \filldraw[fill=white, draw=black] (v1) circle(2pt); \filldraw[fill=white, draw=black] (v2) circle(2pt); \filldraw[fill=white, draw=black] (bottom) circle(2pt); \end{scope} \end{tikzpicture} } $.
\end{thm}
\begin{proof}
A propagator-type diagram $\Gamma$ either has a maximal forest which is the union of propagator diagrams, has at least two vertex-type subdiagrams or it is the primitive diagram of the topology $ \ifmmode \usebox{\fgtwoconeloopbubblephithree} \else \newsavebox{\fgtwoconeloopbubblephithree} \savebox{\fgtwoconeloopbubblephithree}{ \begin{tikzpicture}[x=2ex,y=2ex,baseline={([yshift=-.5ex]current bounding box.center)}] \coordinate (v0) ; \coordinate [right=1 of v0] (v1); \coordinate [left=.5 of v0] (i0); \coordinate [right=.5 of v1] (o0); \coordinate [left=.5 of v1] (vm); \draw (vm) circle(.5); \draw (i0) -- (v0); \draw (o0) -- (v1); \filldraw (v0) circle(1pt); \filldraw (v1) circle(1pt); \end{tikzpicture} } \fi$. In the first case $\Gamma$ is join-irreducible and $S^R_D(\Gamma) = 0$. In the third case the corresponding lattice is ${ \def \scale{2ex} \begin{tikzpicture}[x=\scale,y=\scale,baseline={([yshift=-.5ex]current bounding box.center)}] \begin{scope}[node distance=1] \coordinate (top) ; \coordinate [below= of top] (bottom); \draw (top) -- (bottom); \filldraw[fill=white, draw=black] (top) circle(2pt); \filldraw[fill=white, draw=black] (bottom) circle(2pt); \end{scope} \end{tikzpicture} }$. In the second case, $\Gamma$ covers at least one vertex diagram $\gamma$ which is join-irreducible. Every lattice $L$ with $\mu_L(\hat{0}, \hat{1})\neq0$ is complemented \cite[Cor. 4.1.11]{stern1999semimodular}. In a complemented lattice $L$, there is a $y \in L$ for every $x \in L$ such that $x \join y = \hat{1}$ and $y \meet x = \hat{0}$. For this reason, all the join-irreducible elements of $L$ must be atoms if $\mu_L(\hat{0}, \hat{1})\neq0$. As was shown in the proof of Proposition \ref{prop:semimodular}, a propagator cannot be the join of more than two primitive diagrams. Accordingly,
${ \def \scale{2ex} \begin{tikzpicture}[x=\scale,y=\scale,baseline={([yshift=-.5ex]current bounding box.center)}] \begin{scope}[node distance=1] \coordinate (top) ; \coordinate [below left= of top] (v1); \coordinate [below right= of top] (v2); \coordinate [below left= of v2] (bottom); \draw (top) -- (v1); \draw (top) -- (v2); \draw (v1) -- (bottom); \draw (v2) -- (bottom); \filldraw[fill=white, draw=black] (top) circle(2pt); \filldraw[fill=white, draw=black] (v1) circle(2pt); \filldraw[fill=white, draw=black] (v2) circle(2pt); \filldraw[fill=white, draw=black] (bottom) circle(2pt); \end{scope} \end{tikzpicture} } $ is the only possible lattice if $\Gamma$ is not primitive.
\end{proof}

The $z$-factors for the propagators can also be obtained using the last theorem. 
To do this, the Moebius function for each propagator diagram must be calculated using the form of the lattices and eq.\ \eqref{eqn:moebius}. The Moebius functions for the vertex-type diagrams are known from Theorem \ref{thm:threeQFTmoebvert}. 
\begin{expl}
In a renormalizable QFT with only three-or-less-valent vertices and $v\in \mathcal{R}_e$, a propagator-type residue:
\begin{align}  \begin{split} {S^R_D}\left(\allG^{(v)}_{\sgset^{\text{s.d.}}_D}\right) &= \frac{1}{v!} + \hbar \sum_{\substack{\res \Gamma_P = v \\ \left\{v_1,v_2\right\} = V(\Gamma_P)}} \frac{1}{|\Aut \Gamma_P|} \left( -1 + \frac{ v_1!}{2} \phi \circ P_{\text{Prim}(\hopffg_D)} \left(\allG^{(v_1)}_{\sgset^{\text{s.d.}}_D}\right) \right. + \\ &+ \left. \frac{v_2!}{2}\phi \circ P_{\text{Prim}(\hopffg_D)} \left(\allG^{(v_2)}_{\sgset^{\text{s.d.}}_D}\right) \right) \end{split} \end{align}
where the sum is over all primitive propagator diagrams $\Gamma_P$ with a topology as $ \ifmmode \usebox{\fgtwoconeloopbubblephithree} \else \newsavebox{\fgtwoconeloopbubblephithree} \savebox{\fgtwoconeloopbubblephithree}{ \begin{tikzpicture}[x=2ex,y=2ex,baseline={([yshift=-.5ex]current bounding box.center)}] \coordinate (v0) ; \coordinate [right=1 of v0] (v1); \coordinate [left=.5 of v0] (i0); \coordinate [right=.5 of v1] (o0); \coordinate [left=.5 of v1] (vm); \draw (vm) circle(.5); \draw (i0) -- (v0); \draw (o0) -- (v1); \filldraw (v0) circle(1pt); \filldraw (v1) circle(1pt); \end{tikzpicture} } \fi$ and exactly two vertices $v_1,v_2$. Of course, this sum is finite.

The factorials of the residues must be included to fix the external legs of the vertex-type subdiagrams. The factor of $\frac12$ is necessary, because every non-primitive diagram, which contributes to the counterterm, has exactly two maximal forests. This is an example of a simple Dyson-Schwinger equation in the style of \cite{kreimer2006anatomy}.

The conventional $z$-factor for the propagator is $z^{(v)} = {v}! {S^R_D}\left(\allG^{(v)}_{\sgset^{\text{s.d.}}_D}\right)$, where $v!$ is either $1$ or $2$.
\end{expl}

Although the counterterm map for renormalizable QFTs with only three-or-less-valent vertices enumerates primitive diagrams, we cannot assume that the situation is similar in a more general setting with also four-valent vertices. A negative result in this direction was obtained by Argyres, van Hameren, Kleiss and Papadopoulos \cite[p. 27]{argyres2001zero}. They observed that the vertex counterterm in zero-dimensional $\varphi^4$-theory does not count primitive diagrams. 

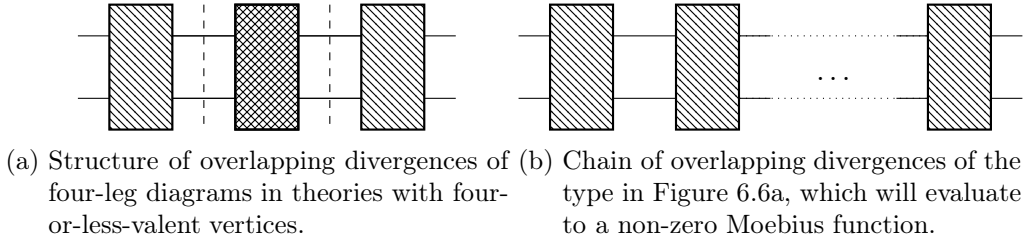
\begin{figure}
\ifdefined\nodraft
\subcaptionbox{Structure of overlapping divergences of four-leg diagrams in theories with four-or-less-valent vertices.\label{fig:fourvalblob}}[.45\linewidth]{
\def \scale {5ex}
\begin{tikzpicture}[x=\scale,y=\scale,baseline={([yshift=-.5ex]current bounding box.center)}] \begin{scope}[node distance=1] \coordinate (i0); \coordinate[below=1 of i0] (i1); \coordinate[right= of i0] (vlt); \coordinate[right= of i1] (vlb); \coordinate[right=2 of vlt] (vmt); \coordinate[right=2 of vlb] (vmb); \coordinate[right=2 of vmt] (vrt); \coordinate[right=2 of vmb] (vrb); \coordinate[right= of vrt] (o0); \coordinate[right= of vrb] (o1); \draw (i0) -- (vlt); \draw (i1) -- (vlb); \draw (vmt) -- (vlt); \draw (vmb) -- (vlb); \draw (vmt) -- (vrt); \draw (vmb) -- (vrb); \draw (o0) -- (vlt); \draw (o1) -- (vlb); \draw [dashed] ($(vlt) + (1,.5)$) -- ($(vlb) + (1,-.5)$); \draw [dashed] ($(vmt) + (1,.5)$) -- ($(vmb) + (1,-.5)$); \draw [fill=white] ($(vlt) + (.5,.5)$) rectangle ($(vlb) - (.5,.5)$) ; \draw [fill=white,thick,pattern=north west lines] ($(vlt) + (.5,.5)$) rectangle ($(vlb) - (.5,.5)$) ; \draw [fill=white] ($(vmt) + (.5,.5)$) rectangle ($(vmb) - (.5,.5)$) ; \draw [fill=white,thick,pattern=north east lines] ($(vmt) + (.5,.5)$) rectangle ($(vmb) - (.5,.5)$) ; \draw [fill=white,thick,pattern=north west lines] ($(vmt) + (.5,.5)$) rectangle ($(vmb) - (.5,.5)$) ; \draw [fill=white] ($(vrt) + (.5,.5)$) rectangle ($(vrb) - (.5,.5)$) ; \draw [fill=white,thick,pattern=north west lines] ($(vrt) + (.5,.5)$) rectangle ($(vrb) - (.5,.5)$) ; \end{scope} \end{tikzpicture}
}
\subcaptionbox{Chain of overlapping divergences of the type in Figure \ref{fig:fourvalblob}, which will evaluate to a non-zero Moebius function.\label{fig:fourvalblobchain}}[.45\linewidth]{
\def \scale {5ex}
\begin{tikzpicture}[x=\scale,y=\scale,baseline={([yshift=-.5ex]current bounding box.center)}] \begin{scope}[node distance=1] \coordinate (i0); \coordinate[below=1 of i0] (i1); \coordinate[right= of i0] (vlt); \coordinate[right= of i1] (vlb); \coordinate[right=2 of vlt] (vmt); \coordinate[right=2 of vlb] (vmb); \coordinate[right=2 of vmt] (vmmt); \node[below=.5 of vmmt] (vmm) {$\ldots$}; \coordinate[right=2 of vmb] (vmmb); \coordinate[right=2 of vmmt] (vrt); \coordinate[right=2 of vmmb] (vrb); \coordinate[right= of vrt] (o0); \coordinate[right= of vrb] (o1); \draw (i0) -- (vlt); \draw (i1) -- (vlb); \draw (vmt) -- (vlt); \draw (vmb) -- (vlb); \draw[dotted] (vmt) -- (vrt); \draw[dotted] (vmb) -- (vrb); \draw (vmt) -- ($(vmmt) - (1,0)$); \draw (vmb) -- ($(vmmb) - (1,0)$); \draw (vrt) -- ($(vmmt) + (1,0)$); \draw (vrb) -- ($(vmmb) + (1,0)$); \draw (o0) -- (vrt); \draw (o1) -- (vrb); \draw [fill=white] ($(vlt) + (.5,.5)$) rectangle ($(vlb) - (.5,.5)$) ; \draw [fill=white,thick,pattern=north west lines] ($(vlt) + (.5,.5)$) rectangle ($(vlb) - (.5,.5)$) ; \draw [fill=white] ($(vmt) + (.5,.5)$) rectangle ($(vmb) - (.5,.5)$) ; \draw [fill=white,thick,pattern=north west lines] ($(vmt) + (.5,.5)$) rectangle ($(vmb) - (.5,.5)$) ; \draw [fill=white] ($(vrt) + (.5,.5)$) rectangle ($(vrb) - (.5,.5)$) ; \draw [fill=white,thick,pattern=north west lines] ($(vrt) + (.5,.5)$) rectangle ($(vrb) - (.5,.5)$) ; \end{scope} \end{tikzpicture}
}
\else

MISSING IN DRAFT MODE

\fi
\caption{Overlapping divergences for diagrams with four legs in theories with only four-or-less-valent vertices.}
\end{figure}

From the perspective of lattice theory this result can be explained. The only way in which overlapping divergences can appear in a diagram with four legs in a QFT with only four-or-less-valent vertices is depicted in Figure \ref{fig:fourvalblob}. The dashed lines indicate the possible cuts to separate one overlapping divergence from the other. To obtain a Feynman diagram $\Gamma$ with $\mu_{\sdsubdiags_D(\Gamma)}(\hat{0},\hat{1})\neq 0$, the blob in the middle must be either of the same overlapping type as Figure \ref{fig:fourvalblob} or superficially convergent. Otherwise, a join-irreducible element would be generated, which would imply $\mu_{\sdsubdiags_D(\Gamma)}(\hat{0},\hat{1}) = 0$. 
The possible non-primitive diagrams with four legs, which give a non-vanishing Moebius function are consequently of the form depicted in Figure \ref{fig:fourvalblobchain}, where each blob must be replaced by a superficially convergent four-leg diagram such that the diagram remains 1PI. The lattice corresponding to this structure of overlapping divergences is a boolean lattice. Every superficially divergent subdiagram can be characterized by the particular set partition of `blocks' it contains. This gives a bijection from $\{0,1\}^n$ to all possible subdivergences. The Moebius function of boolean lattices evaluates to $(-1)^n$, where $n$ is the number of atoms \cite[Ex. 3.8.4.]{stanley1997}. 
Accordingly, the structure of overlapping four-leg diagrams depends on the number of superficially convergent four-leg diagrams. The situation is especially simple in $\varphi^4$-theory:
\begin{expl}[Overlapping vertex-type diagrams in $\varphi^4$-theory]
\label{expl:phi4_lattices}
The only superficially convergent four-leg diagram in $\varphi^4$-theory is the single four-leg vertex $\fourvtx$ and so the only vertex-type s.d.\ diagrams $\Gamma$ which give $\mu_{\sdsubdiags_D(\Gamma)}(\hat{0},\hat{1})\neq 0$ are 
\ifdefined\nodraft
${ \def \scale {1ex} \begin{tikzpicture}[x=2ex,y=2ex,baseline={([yshift=-.5ex]current bounding box.center)}] \coordinate (v0); \coordinate [right=.5 of v0] (vm1); \coordinate [right=.5 of vm1] (v1); \node [right=.5 of v1] (v2) {$\ldots$}; \coordinate [right=.5 of v1] (v1m); \coordinate [right=.01 of v2] (v2m); \coordinate [right=.5 of v2m] (v3); \coordinate [right=.5 of v3] (vm2); \coordinate [right=.5 of vm2] (v4); \coordinate [above left=.5 of v0] (i0); \coordinate [below left=.5 of v0] (i1); \coordinate [above right=.5 of v4] (o0); \coordinate [below right=.5 of v4] (o1); \draw (vm1) circle(.5); \draw (vm2) circle(.5); \draw (i0) -- (v0); \draw (i1) -- (v0); \draw (o0) -- (v4); \draw (o1) -- (v4); \draw ([shift=(90:.5)]v1m) arc (90:270:.5); \draw ([shift=(-90:.5)]v2m) arc (-90:90:.5); \filldraw (v0) circle(1pt); \filldraw (v1) circle(1pt); \filldraw (v3) circle(1pt); \filldraw (v4) circle(1pt); \end{tikzpicture} }$
\else

MISSING IN DRAFT MODE

\fi
, the chains of one-loop diagrams. Their generating function is $\frac18 \sum_{L\geq 0} \frac{\hbar^L }{2^{L}}$ (every diagram is weighted by its symmetry factor) and the counterterm map in $\varphi^4$ evaluates to,
\begin{align} \label{eqn:phird} {S^R_D}(\allG^{\left( \fourvtx \right)}) = \frac{1}{4!} - \phi \circ P_{\text{Prim}(\hopffg)}(\allG^{\left( \fourvtx \right)}) + \frac18 \sum_{L\geq 2} (-1)^L \left( \frac{\hbar }{2} \right)^L. \end{align}
Note again, that in this setup the legs of the diagrams are not fixed. To reobtain the numbers for the case with fixed legs, the generating function must be multiplied with the value $4!=24$. 
The formula for the usual vertex $z$-factor is $z^{\left( \fourvtx \right) } = 4! {S^R_D}(\allG^{\left( \fourvtx \right)})$.

Using this result, we can indeed use the counterterms of zero-dimensional $\varphi^4$-theory to calculate the number of primitive diagrams. We merely must include the correction term on the right-hand side of eq.\ \eqref{eqn:phird}. This calculation will be performed in the next chapter in Section \ref{subsec:varphifourtheoryapplication}.
\end{expl}

\begin{expl}[Overlapping four-leg-vertex diagrams in pure Yang-Mills theory]
In pure Yang-Mills theory there can be either the single four-valent vertex $\fourvtxgluon$ or two three-valent vertices joined by a propagator $\twothreevtxgluon$ as superficially convergent four-leg diagrams. Only chains of diagrams as in Figure \ref{fig:fourvalblobchain} or primitive diagrams give a non-zero ${S^R_D}(\Gamma)$. At two loop for instance, the non-primitive diagrams 
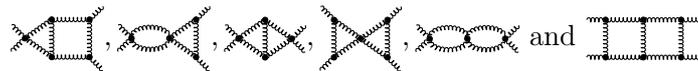
\ifdefined\nodraft
\begin{align*} { \def \scale{3ex} \begin{tikzpicture}[x=\scale,y=\scale,baseline={([yshift=-.5ex]current bounding box.center)}] \begin{scope}[node distance=1] \coordinate (v0); \coordinate [right=.707 of v0] (vm); \coordinate [above=.5 of vm] (vt1); \coordinate [below=.5 of vm] (vb1); \coordinate [right=1 of vt1] (vt2); \coordinate [right=1 of vb1] (vb2); \coordinate [above left=.5 of v0] (i0); \coordinate [below left=.5 of v0] (i1); \coordinate [above right=.5 of vt2] (o0); \coordinate [below right=.5 of vb2] (o1); \draw[gluon] (v0) -- (vt1); \draw[gluon] (v0) -- (vb1); \draw[gluon] (vt1) -- (vb1); \draw[gluon] (vt2) -- (vb2); \draw[gluon] (vt2) -- (vt1); \draw[gluon] (vb2) -- (vb1); \draw[gluon] (i0) -- (v0); \draw[gluon] (i1) -- (v0); \draw[gluon] (o0) -- (vt2); \draw[gluon] (o1) -- (vb2); \filldraw (v0) circle(1pt); \filldraw (vt1) circle(1pt); \filldraw (vb1) circle(1pt); \filldraw (vt2) circle(1pt); \filldraw (vb2) circle(1pt); \end{scope} \end{tikzpicture} }, { \def \scale{3ex} \begin{tikzpicture}[x=\scale,y=\scale,baseline={([yshift=-.5ex]current bounding box.center)}] \begin{scope}[node distance=1] \coordinate (v0); \coordinate [right=.5 of v0] (vm); \coordinate [right=.5 of vm] (v1); \coordinate [above left=.5 of v0] (i0); \coordinate [below left=.5 of v0] (i1); \coordinate [right=.707 of v1] (vm2); \coordinate [above=.5 of vm2] (vt); \coordinate [below=.5 of vm2] (vb); \coordinate [above right=.5 of vt] (o0); \coordinate [below right=.5 of vb] (o1); \draw[gluon] (v1) -- (vt); \draw[gluon] (v1) -- (vb); \draw[gluon] (vt) -- (vb); \draw[gluon] (v0) to[bend left=90] (v1); \draw[gluon] (v0) to[bend right=90] (v1); \draw[gluon] (i0) -- (v0); \draw[gluon] (i1) -- (v0); \draw[gluon] (o0) -- (vt); \draw[gluon] (o1) -- (vb); \filldraw (v0) circle(1pt); \filldraw (v1) circle(1pt); \filldraw (vt) circle(1pt); \filldraw (vb) circle(1pt); \end{scope} \end{tikzpicture} }, { \def \scale{3ex} \begin{tikzpicture}[x=\scale,y=\scale,baseline={([yshift=-.5ex]current bounding box.center)}] \begin{scope}[node distance=1] \coordinate (v0); \coordinate [right=.707 of v0] (vm); \coordinate [above=.5 of vm] (vt); \coordinate [below=.5 of vm] (vb); \coordinate [right=.707 of vm] (v1); \coordinate [above left=.5 of v0] (i0); \coordinate [below left=.5 of v0] (i1); \coordinate [above right=.5 of v1] (o0); \coordinate [below right=.5 of v1] (o1); \draw[gluon] (v0) -- (vt); \draw[gluon] (v0) -- (vb); \draw[gluon] (vt) -- (vb); \draw[gluon] (v1) -- (vt); \draw[gluon] (v1) -- (vb); \draw[gluon] (i0) -- (v0); \draw[gluon] (i1) -- (v0); \draw[gluon] (o0) -- (v1); \draw[gluon] (o1) -- (v1); \filldraw (v0) circle(1pt); \filldraw (v1) circle(1pt); \filldraw (vt) circle(1pt); \filldraw (vb) circle(1pt); \end{scope} \end{tikzpicture} }, { \def \scale{3ex} \begin{tikzpicture}[x=\scale,y=\scale,baseline={([yshift=-.5ex]current bounding box.center)}] \begin{scope}[node distance=1] \coordinate (v0); \coordinate [above=.5 of v0] (vt1); \coordinate [below=.5 of v0] (vb1); \coordinate [right=.707 of v0] (vm); \coordinate [right=.707 of vm] (v1); \coordinate [above=.5 of v1] (vt2); \coordinate [below=.5 of v1] (vb2); \coordinate [above left=.5 of vt1] (i0); \coordinate [below left=.5 of vb1] (i1); \coordinate [above right=.5 of vt2] (o0); \coordinate [below right=.5 of vb2] (o1); \draw[gluon] (vt1) -- (vm); \draw[gluon] (vb1) -- (vm); \draw[gluon] (vt2) -- (vm); \draw[gluon] (vb2) -- (vm); \draw[gluon] (vt1) -- (vb1); \draw[gluon] (vt2) -- (vb2); \draw[gluon] (i0) -- (vt1); \draw[gluon] (i1) -- (vb1); \draw[gluon] (o0) -- (vt2); \draw[gluon] (o1) -- (vb2); \filldraw (vt1) circle(1pt); \filldraw (vt2) circle(1pt); \filldraw (vm) circle(1pt); \filldraw (vb1) circle(1pt); \filldraw (vb2) circle(1pt); \end{scope} \end{tikzpicture} }, { \def \scale{3ex} \begin{tikzpicture}[x=\scale,y=\scale,baseline={([yshift=-.5ex]current bounding box.center)}] \begin{scope}[node distance=1] \coordinate (v0); \coordinate [right=.5 of v0] (vm1); \coordinate [right=.5 of vm1] (v1); \coordinate [right=.5 of v1] (vm2); \coordinate [right=.5 of vm2] (v2); \coordinate [above left=.5 of v0] (i0); \coordinate [below left=.5 of v0] (i1); \coordinate [above right=.5 of v2] (o0); \coordinate [below right=.5 of v2] (o1); \draw[gluon] (v0) to[bend left=90] (v1); \draw[gluon] (v0) to[bend right=90] (v1); \draw[gluon] (v1) to[bend left=90] (v2); \draw[gluon] (v1) to[bend right=90] (v2); \draw[gluon] (i0) -- (v0); \draw[gluon] (i1) -- (v0); \draw[gluon] (o0) -- (v2); \draw[gluon] (o1) -- (v2); \filldraw (v0) circle(1pt); \filldraw (v1) circle(1pt); \filldraw (v2) circle(1pt); \end{scope} \end{tikzpicture} } \text{ and } { \def \scale{3ex} \begin{tikzpicture}[x=\scale,y=\scale,baseline={([yshift=-.5ex]current bounding box.center)}] \begin{scope}[node distance=1] \coordinate (i0); \coordinate[below=1 of i0] (i1); \coordinate[right=.5 of i0] (vlt); \coordinate[right=.5 of i1] (vlb); \coordinate[right=1 of vlt] (vmt); \coordinate[right=1 of vlb] (vmb); \coordinate[right=1 of vmt] (vrt); \coordinate[right=1 of vmb] (vrb); \coordinate[right=.5 of vrt] (o0); \coordinate[right=.5 of vrb] (o1); \draw[gluon] (i0) -- (vlt); \draw[gluon] (i1) -- (vlb); \draw[gluon] (vmt) -- (vlt); \draw[gluon] (vmb) -- (vlb); \draw[gluon] (vmt) -- (vrt); \draw[gluon] (vmb) -- (vrb); \draw[gluon] (vlt) -- (vlb); \draw[gluon] (vmt) -- (vmb); \draw[gluon] (vrt) -- (vrb); \draw[gluon] (o0) -- (vrt); \draw[gluon] (o1) -- (vrb); \filldraw (vlt) circle(1pt); \filldraw (vlb) circle(1pt); \filldraw (vmt) circle(1pt); \filldraw (vmb) circle(1pt); \filldraw (vrt) circle(1pt); \filldraw (vrb) circle(1pt); \end{scope} \end{tikzpicture} } \end{align*}
\else

MISSING IN DRAFT MODE

\fi
contribute non-trivially to ${S^R_D}(\allG^{\left( \fourvtxgluon\right)})$. These chains of diagrams are the only four-leg diagrams which can be formed as the union of two primitive diagrams in this theory.

The generating function for $L \geq 2$ of these diagrams is 
$\frac38 \sum_{L\geq 2} \left(\frac{3 \hbar }{2}\right)^{L}$. 
Hence, the counterterm map in pure Yang-Mills theory for the four-gluon amplitude in zero-dimensional QFT evaluates to, 
\begin{align} {S^R_D}(\allG^{\left(\fourvtxgluon\right)}) = \frac{1}{4!} - \phi \circ P_{\text{Prim}(\hopffg)}(\allG^{\left( \fourvtxgluon \right) }) + \frac38 \sum_{L\geq 2} (-1)^L \left( \frac{3\hbar}{2} \right)^{L}. \end{align}
To reobtain the numbers for the case with fixed legs this generating function needs to by multiplied with the value $4!=24$ as in the example for $\varphi^4$-theory. 
The formula for the $z$-factor is $z^{\left( \fourvtxgluon \right)} = 4! {S^R_D}(\allG^{\left( \fourvtxgluon\right)} )$.
\end{expl}
The framework described in this chapter can be used to make more statements and perform explicit calculations on the weighted numbers of primitive diagrams in different QFTs and their asymptotic behavior. These aspects will be analyzed from a combinatorial perspective in the following chapter.

\chapter{Examples from zero-dimensional QFT}
The content of this chapter is partially based on the author's article \cite{borinsky2017renormalized}.
\label{chap:applications_zerodim}
\section{Overview}
\label{sec:overview}
The zero-dimensional partition function of a scalar theory with interaction given by $V(x)$ is written as a formal integral,
\begin{align*} Z(\hbar, j) := \int \limits \frac{d x}{\sqrt{2 \pi \hbar}} e^{\frac{1}{\hbar} \left( - \frac{x^2}{2} + V(x) + x j \right) }, \end{align*}
similar to eq.\ \eqref{eqn:formalfuncintegral1}.
This integral is to be understood as a formal expansion in $\hbar$ and $j$. The discussion from Section \ref{sec:formalint} does not immediately apply here, because of the additional $x j$ term, which was \textit{not} allowed in Definition \ref{def:formalintegral}. 
We can always transform the expression above into the canonical form as in Definition \ref{def:formalintegral} by formally shifting the integration variable,
\begin{align*} Z(\hbar, j) &= e^{\frac{-\frac{x_0^2}{2} + V(x_0) + x_0 j}{\hbar} } \int \limits_\mathbb{R} \frac{d x}{\sqrt{2 \pi \hbar}} e^{\frac{1}{\hbar} \left( -\frac{x^2}{2} + V(x+x_0) -V(x_0) -x V'(x_0) \right)} \\ &= e^{\frac{-\frac{x_0^2}{2} + V(x_0) + x_0 j}{\hbar} } \Fop\left[ -\frac{x^2}{2} + V(x+x_0) -V(x_0) -x V'(x_0)\right](\hbar) \end{align*}
where $x_0=x_0(j)$ is the unique power series solution of $x_0(j) = V'(x_0(j)) + j$.
Note the similarity of this shifting by a constant to the Legendre transformation described in Section \ref{sec:legendre_transformation}.

The exponential prefactor enumerates all (possibly disconnected) \textit{tree diagrams} with the prescribed vertex structure and the $\Fop$-term enumerates all diagrams with at least one cycle in each connected component. It is useful to separate the tree-level diagrams as they contribute with negative powers in $\hbar$, which spoils the simple treatment in the formalism of power series. 

Trees and diagrams with at least one cycle are isolated after restricting to connected diagrams, which are generated by the \textit{free energy} of the theory:
\begin{align*} W(\hbar, j) &:= \hbar \log Z(\hbar, j) = \\ &=-\frac{x_0^2}{2} + V(x_0) + x_0 j +\hbar \log \mathcal{F} \left[ -\frac{x^2}{2} + V(x+x_0) -V(x_0) -x V'(x_0)\right] \left(\hbar \right), \end{align*}
where we conventionally multiply by $\hbar$ to go from counting by excess to counting by loop number and $x_0=x_0(j)$. The generating function $W(\hbar,j)$ generates connected diagrams as stated in Theorem \ref{thm:connected_disconnected}.

The next step is to perform a Legendre transformation as described in detail in Section \ref{sec:legendre_transformation}, to get access to the effective action $G$, which is a generating function in $\hbar$ and $\varphi_c$, 
\begin{align*} G(\hbar,\varphi_c) &:= W - j \varphi_c \\ \varphi_c &:= \partial_j W. \end{align*}
The equation $\varphi_c = \partial_j W$ needs to be solved for $j$ to obtain $G$ as a generating function in $\hbar$ and $\varphi_c$. Explicitly, this is only necessary if the potential allows graphs with one external leg. 

The coefficients of $G$, expanded in $\varphi_c$, are called proper Green functions of the theory. More specifically, the first derivative $\partial_{\varphi_c} G |_{\varphi_c=0}$ is called the generating function of \textit{(proper) 1-point function}, the second derivative $\partial_{\varphi_c}^2 G |_{\varphi_c=0}$ is called \textit{(1PI) propagator} and higher derivatives $\partial_{\varphi_c}^k G |_{\varphi_c=0}$ are called \textit{proper $k$-point function}. 

A further step in the analysis of zero-dimensional QFT is the calculation of the \textit{renormalization constants}. The calculation is slightly artificial in zero-dimensional QFT, as there are no explicit divergences to renormalize as discussed in Section \ref{sec:applications_lattices}.
Without momentum dependence every `integral' for a graph is convergent. Thus renormalization has to be defined in analogy with higher dimensional models. 
To motivate the renormalization procedure for zero-dimensional QFT, we will use the Hopf algebra structure of Feynman diagrams in a slightly more general fashion then in Chapter \ref{chap:hopf_algebra_of_fg}.
\section{Renormalization}
\label{sec:expl_hopfalgebra}

When we speak of \textit{renormalization} in QFT, we mean the evaluation of certain products of characters on the elements 
\begin{align*} \allG_{\sgset_\text{bl}} &:= \sum_{\substack{\Gamma \in \sgset_\text{bl}}} \frac{\Gamma}{|\Aut\Gamma|}\\ \allG_{\sgset^{\text{s.d.}}_D} &:= \sum_{\substack{\Gamma \in \sgset^{\text{s.d.}}_D}} \frac{\Gamma}{|\Aut\Gamma|}\\ \allG^{(v)}_{\sgset^{\text{s.d.}}_D} &:= \sum_{\substack{\Gamma \in \sgset^{\text{s.d.}}_D\\\res\Gamma =v}} \frac{\Gamma}{|\Aut\Gamma|}, \end{align*}
of the Hopf algebra $\Gaul_{\sgset_\text{bl}}$ of bridgeless graphs from Example \ref{expl:bridgeless_graphs} and $\hopffg_D$ as defined as the quotient $\Gaul/I_{\sgset^{\text{s.d.}}_D}$ in Section \ref{sec:hopfalgebra}, which is generated by all bridgeless graphs whose connected components are superficially divergent.

The process of renormalization is in essence the inversion of a given character $\phi$ restricted to the generators in $\sgset^{\text{s.d.}}_D$ of $\hopffg_D$, $\phi|_{\sgset^{\text{s.d.}}_D} \in \chargroup{\hopffg_D}{\Q[[\hbar]]}$ and the evaluation of its convolution inverse on the $\allG^{(v)}_{\sgset^{\text{s.d.}}_D}$ vectors.

The zero-dimensional Feynman rules $\phi_\Sact:\Gaul_{\sgset_\text{bl}} \rightarrow \Q[[\varphi_c,\hbar]]$ are given by,
\begin{align} \label{eqn:simple_phi_feynman_rules} \phi_\Sact (\Gamma) = \hbar^{h_\Gamma} \varphi_c^{|\legs_\Gamma|} \prod_{v \in V_\Gamma} \lambda_{\deg{v}_\Gamma} , \end{align}
where in contrast to the preceding chapters, we interpret the $\lambda_d$ as fixed parameters that were chosen beforehand - encoded in a specific action $\Sact$. 

This way we can identify the result of the Legendre transformation from Section \ref{sec:legendre_transformation} with the evaluation,
\begin{align} \label{eqn:legendre_transformation_specific} G(\hbar, \varphi_c) = -\frac{\varphi_c^2}{2} + \phi_\Sact \left( \log \allG_{\sgset_\text{bl}} \right) - \varphi_c j. \end{align}
As described in detail in the last chapter, we will take the set of bridgeless graphs as the starting point for our Hopf algebra formulation. Here, an elegant approach is to consider the Hopf algebra of Feynman diagrams $\hopffg_D$ as comodule over the Hopf algebra $\Gaul_{\sgset_\text{bl}}$ from Chapter \ref{chap:coalgebra_graph}.

We can do this by decomposing the map $\phi_\Sact$ as in Example \ref{expl:phi_decomp} into the maps $\sk_\Sact$ and $\re$, which only act on residues, and a map $\zeta$:
\begin{align*} \sk_\Sact&: &&\hopffg_D \rightarrow \Q[[\varphi_c,\hbar]], & &\Gamma \mapsto \begin{cases} \prod_{v\in V_\Gamma} \lambda_{\deg{v}} & \text{ if $\Gamma \in \residuesstar$}\\ 0& \text{ else} \end{cases}\\ \re&: &&\Gaul_{\sgset_\text{bl}} \rightarrow \Q[[\varphi_c,\hbar]], &&\Gamma \mapsto \begin{cases} \varphi_c^{|\legs_\Gamma|} & \text{ if $\Gamma \in \residuesstar$}\\ 0& \text{ else} \end{cases}\\ \zeta&:& &\Gaul_{\sgset_\text{bl}} \rightarrow \Q[[\varphi_c,\hbar]], & &\Gamma \mapsto \hbar^{h_\Gamma}, \end{align*}
This gives us a `sandwich' decomposition of $\phi_\Sact$, that means $\phi_\Sact = \sk_\Sact \star_{\sgset^{\text{s.d.}}_D} \zeta \star_{\sgset_\text{bl}} \re$, with the respective $\star$-products, $\star_{\sgset^{\text{s.d.}}_D}: \chargroup{\hopffg_D}{\Q[[\hbar]]} \times \chargroup{\Gaul_{\sgset_\text{bl}}}{\Q[[\hbar]]} \rightarrow \chargroup{\Gaul_{\sgset_\text{bl}}}{\Q[[\hbar]]}$ and $\star_{\sgset_\text{bl}}: \chargroup{\Gaul_{\sgset_\text{bl}}}{\Q[[\hbar]]} \times \chargroup{\Gaul_{\sgset_\text{bl}}}{\Q[[\hbar]]} \rightarrow \chargroup{\Gaul_{\sgset_\text{bl}}}{\Q[[\hbar]]}$. 
To verify this, observe that for all $\Gamma \in \sgset_\text{bl}$,
\begin{align*} \sk_\Sact \star_{\sgset^{\text{s.d.}}_D} \zeta \star_{\sgset_\text{bl}} \re(\Gamma) &= \sk_\Sact(\skl(\Gamma)) \zeta(\Gamma) \re(\res(\Gamma)) = \hbar^{h_\Gamma} \varphi_c^{|\legs_\Gamma|} \prod_{v \in V_\Gamma} \lambda_{\deg{v}_\Gamma} \end{align*}
which equals $\phi_\Sact(\Gamma)$. Note that in the decomposition only $\sk_\Sact$ depends on the action $\Sact$.

The \textit{renormalized effective action} is now the evaluation,
\begin{align} \sk_\Sact \star_{\sgset^{\text{s.d.}}_D} S^{\zeta|_{\sgset^{\text{s.d.}}_D}}_D \star_{\sgset^{\text{s.d.}}_D} \zeta \star_{\sgset_\text{bl}} \re (\log \allG_{\sgset_\text{bl}}), \end{align}
where $S^{\zeta|_{\sgset^{\text{s.d.}}_D}}_D:= \zeta|_{\sgset^{\text{s.d.}}_D} \circ S_D$ with the antipode $S_D$ from $\hopffg_D$ and $\zeta|_{\sgset^{\text{s.d.}}_D}$ the restriction of $\zeta$ to the generators $\sgset^{\text{s.d.}}_D$. 

The \textit{counterterms} or $z$-factors are the evaluations 
\begin{align} z^{(v)} = v! \sk_\Sact \star_{\sgset^{\text{s.d.}}_D} S^{\zeta|_{\sgset^{\text{s.d.}}_D}}_D \left( \allG^{(v)}_{\sgset^{\text{s.d.}}_D} \right), \end{align}
where - in contrast to the last chapter - we retained the freedom to choose the allowed vertex degrees by using the $\sk_\Sact$ map.

The easiest way to calculate the $z$-factors explicitly is to use Theorem \ref{thm:coproduct_full_identity_refined} to obtain a fixed point equation in a power series ring. 
We will only consider the cases where there is a single non-zero $\lambda_d$ set to $1$, because this the only case where the inversion is possible in general. 
Furthermore, we will assume that the set $\sgset^{\text{s.d.}}_D$ only contains non-trivial diagrams with zero, two legs or $d$ legs. This is the case for the examples of $\varphi^3$- and $\varphi^4$-theory.

If the underlying QFT has multiple vertex-types, proving the existence of such a fixed-point equation can be quite involved. 
In those theories all possible different definitions of \textit{the invariant charge} must agree, as dictated by the \textit{Slavnov-Taylor-Identities} \cite{kreimer2006anatomy,van2007renormalization,pascual1984qcd,nakanishi1990covariant}.

We will use a well-known `trick' to write the generating functions of our connected bridgeless graphs as a generating function in one variable:
\begin{lmm}
\label{lmm:invariant_charge}
If $f_{\sgset_\text{bl}}^c$ is the generating function of scalar connected bridgeless (1PI) Feynman diagrams with a single allowed vertex-type of degree $\lambda_d$ besides the two-valent vertices 
$ \ifmmode \usebox{\fgsimpletwovtx} \else \newsavebox{\fgsimpletwovtx} \savebox{\fgsimpletwovtx}{ \begin{tikzpicture}[x=1ex,y=1ex,baseline={([yshift=-.5ex]current bounding box.center)}] \coordinate (v) ; \def \n {2}; \def \rad {.8}; \filldraw[white] (v) circle (\rad); \foreach \s in {1,...,\n} { \def \angle {360/\n*(\s - 1)}; \coordinate (u) at ([shift=({\angle}:\rad)]v); \draw (v) -- (u); } \filldraw (v) circle (1pt); \end{tikzpicture} } \fi$, then
\begin{gather} \begin{gathered} f_{\sgset_\text{bl}}^c( \hbar, \varphi_c, \lambda_2, \lambda_d ) =\sum_{\substack{\Gamma \in \sgset_\text{bl}\\ |\comps_\Gamma| =1 }} \frac{\hbar^{h_\Gamma} \varphi_c^{|\legs_\Gamma|}\lambda_2^{\nvd{ \ifmmode \usebox{\fgsimpletwovtx} \else \newsavebox{\fgsimpletwovtx} \savebox{\fgsimpletwovtx}{ \begin{tikzpicture}[x=1ex,y=1ex,baseline={([yshift=-.5ex]current bounding box.center)}] \coordinate (v) ; \def \n {2}; \def \rad {.8}; \filldraw[white] (v) circle (\rad); \foreach \s in {1,...,\n} { \def \angle {360/\n*(\s - 1)}; \coordinate (u) at ([shift=({\angle}:\rad)]v); \draw (v) -- (u); } \filldraw (v) circle (1pt); \end{tikzpicture} } \fi}_\Gamma} \lambda_d^{\nvd{v_d}_\Gamma}}{|\Aut \Gamma|}, \\ = \lambda_2 \frac{\varphi_c^2}{2} + q^{-2} f_{\sgset_\text{bl}}^c\left( q^2 \hbar, q \sqrt{1-\lambda_2} \varphi_c, 0, 1 \right) \end{gathered} \end{gather}
where 
\begin{align} q:= \left( \frac{\lambda_{d}}{(1-\lambda_{2})^{\frac{d}{2}}} \right)^{\frac{1}{d-2}} \end{align}
\end{lmm}
\begin{proof}
Observe that
\begin{align*} \sum_{\substack{\Gamma \in \sgset_\text{bl}\\ |\comps_\Gamma| =1 }} \frac{ \hbar^{h_\Gamma} \varphi_c^{|\legs_\Gamma|} \lambda_{2}^{\nvd{ \ifmmode \usebox{\fgsimpletwovtx} \else \newsavebox{\fgsimpletwovtx} \savebox{\fgsimpletwovtx}{ \begin{tikzpicture}[x=1ex,y=1ex,baseline={([yshift=-.5ex]current bounding box.center)}] \coordinate (v) ; \def \n {2}; \def \rad {.8}; \filldraw[white] (v) circle (\rad); \foreach \s in {1,...,\n} { \def \angle {360/\n*(\s - 1)}; \coordinate (u) at ([shift=({\angle}:\rad)]v); \draw (v) -- (u); } \filldraw (v) circle (1pt); \end{tikzpicture} } \fi}_\Gamma} \lambda_{d}^{\nvd{v_d}_\Gamma} }{|\Aut \Gamma|} &= \lambda_2 \frac{\varphi_c^2}{2} + \sum_{\substack{\Gamma \in \sgset_\text{bl}\\ |\comps_\Gamma| =1\\\nvd{ \ifmmode \usebox{\fgsimpletwovtx} \else \newsavebox{\fgsimpletwovtx} \savebox{\fgsimpletwovtx}{ \begin{tikzpicture}[x=1ex,y=1ex,baseline={([yshift=-.5ex]current bounding box.center)}] \coordinate (v) ; \def \n {2}; \def \rad {.8}; \filldraw[white] (v) circle (\rad); \foreach \s in {1,...,\n} { \def \angle {360/\n*(\s - 1)}; \coordinate (u) at ([shift=({\angle}:\rad)]v); \draw (v) -- (u); } \filldraw (v) circle (1pt); \end{tikzpicture} } \fi}_\Gamma = 0}} \frac{ \hbar^{h_\Gamma} \varphi_c^{|\legs_\Gamma|} \left(\frac{1}{1-\lambda_{2}}\right)^{|E_\Gamma|} \lambda_{d}^{\nvd{v_d}_\Gamma} }{|\Aut \Gamma|}, \end{align*}
which follows from the fact that we may distribute the two valent vertices arbitrary over every edge: Every edge serves as a `bin' for two valent vertices such that every edge contributes a $\frac{1}{1-\lambda_2}$ factor. This only works if $\Gamma$ is not a single two-valent vertex. Therefore, we have to add this exceptional case $\lambda_2 \frac{\varphi_c^2}{2}$. 

On a graph we have the identity $h_\Gamma = |E_\Gamma| - |V_\Gamma| + |\comps_\Gamma|$ by the definition of $h_\Gamma$. If the graph has only $d$-valent vertices we also have, $2|E_\Gamma| + |\legs_\Gamma| = d |V_\Gamma|$, by counting the number of half-edges of $\Gamma$. These two equations are equivalent to,
\begin{align*} |E_\Gamma| &= \frac{1}{d-2} ( d( h_\Gamma - |\comps_\Gamma| ) + |\legs_\Gamma| ) \\ |V_\Gamma| &= \frac{1}{d-2} ( 2( h_\Gamma - |\comps_\Gamma| ) + |\legs_\Gamma| ), \end{align*}
which gives us, because $|V_\Gamma| = \nvd{v_d}_\Gamma$ and $|\comps_\Gamma| = 1$,
\begin{gather*} f_{\sgset_\text{bl}}^c( \hbar, \varphi_c, \lambda_2, \lambda_d ) =\lambda_2 \frac{\varphi_c^2}{2} + \\ \left( \frac{\lambda_{d}}{(1-\lambda_{2})^{\frac{d}{2}}} \right)^{- \frac{2}{d-2}} \sum_{\substack{\Gamma \in \sgset_\text{bl}\\ |\comps_\Gamma| =1\\\nvd{ \ifmmode \usebox{\fgsimpletwovtx} \else \newsavebox{\fgsimpletwovtx} \savebox{\fgsimpletwovtx}{ \begin{tikzpicture}[x=1ex,y=1ex,baseline={([yshift=-.5ex]current bounding box.center)}] \coordinate (v) ; \def \n {2}; \def \rad {.8}; \filldraw[white] (v) circle (\rad); \foreach \s in {1,...,\n} { \def \angle {360/\n*(\s - 1)}; \coordinate (u) at ([shift=({\angle}:\rad)]v); \draw (v) -- (u); } \filldraw (v) circle (1pt); \end{tikzpicture} } \fi}_\Gamma = 0}} \frac{ \left( \hbar \left( \frac{\lambda_{d}}{(1-\lambda_{2})^{\frac{d}{2}}} \right)^{\frac{2}{d-2}} \right)^{h_\Gamma}  \left( \varphi_c \left(\frac{\lambda_{d}}{1-\lambda_{2}}\right)^{\frac{ 1}{d-2}} \right) ^{|\legs_\Gamma|}  }{|\Aut \Gamma|}, \end{gather*}
which is equivalent to the statement.
\end{proof}

Using this lemma, we can set up a fixed-point equation for the counterterms. To do this we apply the map $\sk_\Sact \star_{\sgset^{\text{s.d.}}_D} S^{\zeta|_{\sgset^{\text{s.d.}}_D}}_D \star_{\sgset^{\text{s.d.}}_D} \zeta \star_{\sgset_\text{bl}} \re$ to $\log \allG_{\sgset_\text{bl}}$ and use Theorem \ref{thm:coproduct_full_identity_refined}:
\begin{gather} \begin{gathered} \label{eqn:renormalization_condition_1} \sk_\Sact \star_{\sgset^{\text{s.d.}}_D} S^{\zeta|_{\sgset^{\text{s.d.}}_D}}_D \star_{\sgset^{\text{s.d.}}_D} \zeta \star_{\sgset_\text{bl}} \re\left( \log \allG_{\sgset_\text{bl}} \right) \\ = \sum_{\substack{\Gamma \in \sgset_\text{bl}\\ |\comps_\Gamma| =1}} \left( \prod_{v \in V_\Gamma} (\deg{v}_\Gamma!) \sk_\Sact \star_{\sgset^{\text{s.d.}}_D} S^{\zeta|_{\sgset^{\text{s.d.}}_D}}_D \left( \allG_{\sgset^{\text{s.d.}}_D}^{(v)} \right) \right) \frac{\zeta \star_{\sgset_\text{bl}} \re(\Gamma)}{|\Aut\Gamma|} \\ = \sum_{\substack{\Gamma \in \sgset_\text{bl}\\ |\comps_\Gamma| =1}} \left( (2!)\sk_\Sact \star_{\sgset^{\text{s.d.}}_D} S^{\zeta|_{\sgset^{\text{s.d.}}_D}}_D \left( \allG_{\sgset^{\text{s.d.}}_D}^{( \ifmmode \usebox{\fgsimpletwovtx} \else \newsavebox{\fgsimpletwovtx} \savebox{\fgsimpletwovtx}{ \begin{tikzpicture}[x=1ex,y=1ex,baseline={([yshift=-.5ex]current bounding box.center)}] \coordinate (v) ; \def \n {2}; \def \rad {.8}; \filldraw[white] (v) circle (\rad); \foreach \s in {1,...,\n} { \def \angle {360/\n*(\s - 1)}; \coordinate (u) at ([shift=({\angle}:\rad)]v); \draw (v) -- (u); } \filldraw (v) circle (1pt); \end{tikzpicture} } \fi)} \right) \right)^{\nvd{ \ifmmode \usebox{\fgsimpletwovtx} \else \newsavebox{\fgsimpletwovtx} \savebox{\fgsimpletwovtx}{ \begin{tikzpicture}[x=1ex,y=1ex,baseline={([yshift=-.5ex]current bounding box.center)}] \coordinate (v) ; \def \n {2}; \def \rad {.8}; \filldraw[white] (v) circle (\rad); \foreach \s in {1,...,\n} { \def \angle {360/\n*(\s - 1)}; \coordinate (u) at ([shift=({\angle}:\rad)]v); \draw (v) -- (u); } \filldraw (v) circle (1pt); \end{tikzpicture} } \fi}_\Gamma} \\ \times \left( (d!) \sk_\Sact \star_{\sgset^{\text{s.d.}}_D} S^{\zeta|_{\sgset^{\text{s.d.}}_D}}_D \left( \allG_{\sgset^{\text{s.d.}}_D}^{(v_d)} \right) \right)^{\nvd{v_d}_\Gamma} \frac{\hbar^{h_\Gamma} \varphi_c^{|\legs_\Gamma|}}{|\Aut\Gamma|} \\ = f_{\sgset_\text{bl}}^c\left( \hbar, \varphi_c, 2! \sk_\Sact \star_{\sgset^{\text{s.d.}}_D} S^{\zeta|_{\sgset^{\text{s.d.}}_D}}_D \left( \allG_{\sgset^{\text{s.d.}}_D}^{( \ifmmode \usebox{\fgsimpletwovtx} \else \newsavebox{\fgsimpletwovtx} \savebox{\fgsimpletwovtx}{ \begin{tikzpicture}[x=1ex,y=1ex,baseline={([yshift=-.5ex]current bounding box.center)}] \coordinate (v) ; \def \n {2}; \def \rad {.8}; \filldraw[white] (v) circle (\rad); \foreach \s in {1,...,\n} { \def \angle {360/\n*(\s - 1)}; \coordinate (u) at ([shift=({\angle}:\rad)]v); \draw (v) -- (u); } \filldraw (v) circle (1pt); \end{tikzpicture} } \fi)} \right), d! \sk_\Sact \star_{\sgset^{\text{s.d.}}_D} S^{\zeta|_{\sgset^{\text{s.d.}}_D}}_D \left( \allG_{\sgset^{\text{s.d.}}_D}^{(v_d)} \right) \right) \\ = z^{( \ifmmode \usebox{\fgsimpletwovtx} \else \newsavebox{\fgsimpletwovtx} \savebox{\fgsimpletwovtx}{ \begin{tikzpicture}[x=1ex,y=1ex,baseline={([yshift=-.5ex]current bounding box.center)}] \coordinate (v) ; \def \n {2}; \def \rad {.8}; \filldraw[white] (v) circle (\rad); \foreach \s in {1,...,\n} { \def \angle {360/\n*(\s - 1)}; \coordinate (u) at ([shift=({\angle}:\rad)]v); \draw (v) -- (u); } \filldraw (v) circle (1pt); \end{tikzpicture} } \fi)} \frac{\varphi_c^2}{2}+ q_\text{ren}^{-2} f_{\sgset_\text{bl}}^c\left( q_\text{ren}^2 \hbar, q_\text{ren} \sqrt{1- z^{( \ifmmode \usebox{\fgsimpletwovtx} \else \newsavebox{\fgsimpletwovtx} \savebox{\fgsimpletwovtx}{ \begin{tikzpicture}[x=1ex,y=1ex,baseline={([yshift=-.5ex]current bounding box.center)}] \coordinate (v) ; \def \n {2}; \def \rad {.8}; \filldraw[white] (v) circle (\rad); \foreach \s in {1,...,\n} { \def \angle {360/\n*(\s - 1)}; \coordinate (u) at ([shift=({\angle}:\rad)]v); \draw (v) -- (u); } \filldraw (v) circle (1pt); \end{tikzpicture} } \fi)} } \varphi_c, 0, 1 \right), \end{gathered} \end{gather}
where we used Lemma \ref{lmm:invariant_charge} as well as the assumption that only s.d.\ subgraphs with $2$ or $d$ legs appear and set
\begin{align*} q_\text{ren} &:= \left( \frac{ z^{(v_d)} }{(1-z^{( \ifmmode \usebox{\fgsimpletwovtx} \else \newsavebox{\fgsimpletwovtx} \savebox{\fgsimpletwovtx}{ \begin{tikzpicture}[x=1ex,y=1ex,baseline={([yshift=-.5ex]current bounding box.center)}] \coordinate (v) ; \def \n {2}; \def \rad {.8}; \filldraw[white] (v) circle (\rad); \foreach \s in {1,...,\n} { \def \angle {360/\n*(\s - 1)}; \coordinate (u) at ([shift=({\angle}:\rad)]v); \draw (v) -- (u); } \filldraw (v) circle (1pt); \end{tikzpicture} } \fi)})^{\frac{d}{2}}} \right)^{\frac{1}{d-2}}\\ z^{( \ifmmode \usebox{\fgsimpletwovtx} \else \newsavebox{\fgsimpletwovtx} \savebox{\fgsimpletwovtx}{ \begin{tikzpicture}[x=1ex,y=1ex,baseline={([yshift=-.5ex]current bounding box.center)}] \coordinate (v) ; \def \n {2}; \def \rad {.8}; \filldraw[white] (v) circle (\rad); \foreach \s in {1,...,\n} { \def \angle {360/\n*(\s - 1)}; \coordinate (u) at ([shift=({\angle}:\rad)]v); \draw (v) -- (u); } \filldraw (v) circle (1pt); \end{tikzpicture} } \fi)} &= 2! \sk_\Sact \star_{\sgset^{\text{s.d.}}_D} S^{\zeta|_{\sgset^{\text{s.d.}}_D}}_D \left( \allG_{\sgset^{\text{s.d.}}_D}^{( \ifmmode \usebox{\fgsimpletwovtx} \else \newsavebox{\fgsimpletwovtx} \savebox{\fgsimpletwovtx}{ \begin{tikzpicture}[x=1ex,y=1ex,baseline={([yshift=-.5ex]current bounding box.center)}] \coordinate (v) ; \def \n {2}; \def \rad {.8}; \filldraw[white] (v) circle (\rad); \foreach \s in {1,...,\n} { \def \angle {360/\n*(\s - 1)}; \coordinate (u) at ([shift=({\angle}:\rad)]v); \draw (v) -- (u); } \filldraw (v) circle (1pt); \end{tikzpicture} } \fi)} \right)\\ z^{(v_d)} &= d! \sk_\Sact \star_{\sgset^{\text{s.d.}}_D} S^{\zeta|_{\sgset^{\text{s.d.}}_D}}_D \left( \allG_{\sgset^{\text{s.d.}}_D}^{(v_d)} \right). \end{align*}

The evaluation of $\sk_\Sact \star_{\sgset^{\text{s.d.}}_D} S^{\zeta|_{\sgset^{\text{s.d.}}_D}}_D \star_{\sgset^{\text{s.d.}}_D} \zeta \star_{\sgset_\text{bl}} \re$ restricted on the s.d.\ graphs in $\sgset^{\text{s.d.}}_D$ is trivial as
\begin{gather*} \sk_\Sact \star_{\sgset^{\text{s.d.}}_D} S^{\zeta|_{\sgset^{\text{s.d.}}_D}}_D \star_{\sgset^{\text{s.d.}}_D} \zeta \star_{\sgset_\text{bl}} \re \big|_{\sgset^{\text{s.d.}}_D} = \sk_\Sact \star_{\sgset^{\text{s.d.}}_D} S^{\zeta|_{\sgset^{\text{s.d.}}_D}}_D \star_{\sgset^{\text{s.d.}}_D} \zeta|_{\sgset^{\text{s.d.}}_D} \star_{\sgset^{\text{s.d.}}_D} \re \intertext{and therefore} \sk_\Sact \star_{\sgset^{\text{s.d.}}_D} S^{\zeta|_{\sgset^{\text{s.d.}}_D}}_D \star_{\sgset^{\text{s.d.}}_D} \zeta|_{\sgset^{\text{s.d.}}_D} \star_{\sgset^{\text{s.d.}}_D} \re\left( \log \allG_{\sgset^{\text{s.d.}}_D} \right) \\ = \sk_\Sact \star_{\sgset^{\text{s.d.}}_D} \left( S^{\zeta|_{\sgset^{\text{s.d.}}_D}}_D \star_{\sgset^{\text{s.d.}}_D} \zeta|_{\sgset^{\text{s.d.}}_D} \right) \star_{\sgset^{\text{s.d.}}_D} \re\left( \log \allG_{\sgset^{\text{s.d.}}_D} \right) \\ = \sk_\Sact \star_{\sgset^{\text{s.d.}}_D} (\unit_{\hopffg_D} \circ \counit_{\hopffg_D}) \star_{\sgset^{\text{s.d.}}_D} \re\left( \log \allG_{\sgset^{\text{s.d.}}_D} \right) \\ = \frac{1}{d!} \varphi_c^d, \end{gather*}
because the vertex of degree $d$ is the only residue graph that is not mapped to zero by $\sk_\Sact$.

This identity is also called \textit{renormalization condition} in the physics literature.

The set of connected bridgeless graphs with two or $d$ legs in $\sgset_\text{bl}$ with only $d$-valent vertices agrees with the set of
superficially divergent graphs in $\sgset^{\text{s.d.}}_D$, because our theories are required to be renormalizable and only s.d.\ subgraphs with $2$ or $d$ legs may appear. Therefore,
\begin{gather} \begin{gathered} \label{eqn:renormalization_condition_2} \sk_\Sact \star_{\sgset^{\text{s.d.}}_D} S^{\zeta|_{\sgset^{\text{s.d.}}_D}}_D \star_{\sgset^{\text{s.d.}}_D} \zeta \star_{\sgset_\text{bl}} \re\left( \log \allG_{\sgset_\text{bl}} \right) \\ = \sk_\Sact \star_{\sgset^{\text{s.d.}}_D} S^{\zeta|_{\sgset^{\text{s.d.}}_D}}_D \star_{\sgset^{\text{s.d.}}_D} \zeta|_{\sgset^{\text{s.d.}}_D} \star_{\sgset^{\text{s.d.}}_D} \re\left( \log \allG_{\sgset^{\text{s.d.}}_D} \right) + \bigO(\varphi_c^{d+1}) = \frac{1}{d!} \varphi_c^d + \bigO(\varphi_c^{d+1}). \end{gathered} \end{gather}
We can also express this with our generating function $f_{\sgset_\text{bl}}^c$ by eq.\ \eqref{eqn:renormalization_condition_1},
\begin{align} \label{eqn:renormalization_condition_3} z^{( \ifmmode \usebox{\fgsimpletwovtx} \else \newsavebox{\fgsimpletwovtx} \savebox{\fgsimpletwovtx}{ \begin{tikzpicture}[x=1ex,y=1ex,baseline={([yshift=-.5ex]current bounding box.center)}] \coordinate (v) ; \def \n {2}; \def \rad {.8}; \filldraw[white] (v) circle (\rad); \foreach \s in {1,...,\n} { \def \angle {360/\n*(\s - 1)}; \coordinate (u) at ([shift=({\angle}:\rad)]v); \draw (v) -- (u); } \filldraw (v) circle (1pt); \end{tikzpicture} } \fi)} \frac{\varphi_c^2}{2}+ q_\text{ren}^{-2} f_{\sgset_\text{bl}}^c\left( q_\text{ren}^2 \hbar, q_\text{ren} \sqrt{1- z^{( \ifmmode \usebox{\fgsimpletwovtx} \else \newsavebox{\fgsimpletwovtx} \savebox{\fgsimpletwovtx}{ \begin{tikzpicture}[x=1ex,y=1ex,baseline={([yshift=-.5ex]current bounding box.center)}] \coordinate (v) ; \def \n {2}; \def \rad {.8}; \filldraw[white] (v) circle (\rad); \foreach \s in {1,...,\n} { \def \angle {360/\n*(\s - 1)}; \coordinate (u) at ([shift=({\angle}:\rad)]v); \draw (v) -- (u); } \filldraw (v) circle (1pt); \end{tikzpicture} } \fi)} } \varphi_c, 0, 1 \right) = \frac{1}{d!} \varphi_c^d + \bigO(\varphi_c^{d+1}). \end{align}
Taking the second and the $d$-th $\varphi_c$-derivative on both sides of eq. \eqref{eqn:renormalization_condition_3} and setting $\varphi_c$ to zero subsequently gives,
\begin{gather} \begin{gathered} \label{eqn:counterterm_relation1} 0 = \left. \frac{\partial^2}{\partial \varphi_c^2} \left( z^{( \ifmmode \usebox{\fgsimpletwovtx} \else \newsavebox{\fgsimpletwovtx} \savebox{\fgsimpletwovtx}{ \begin{tikzpicture}[x=1ex,y=1ex,baseline={([yshift=-.5ex]current bounding box.center)}] \coordinate (v) ; \def \n {2}; \def \rad {.8}; \filldraw[white] (v) circle (\rad); \foreach \s in {1,...,\n} { \def \angle {360/\n*(\s - 1)}; \coordinate (u) at ([shift=({\angle}:\rad)]v); \draw (v) -- (u); } \filldraw (v) circle (1pt); \end{tikzpicture} } \fi)} \frac{\varphi_c^2}{2} + q_\text{ren}^{-2} f_{\sgset_\text{bl}}^c\left( q_\text{ren}^2 \hbar, q_\text{ren} \sqrt{1- z^{( \ifmmode \usebox{\fgsimpletwovtx} \else \newsavebox{\fgsimpletwovtx} \savebox{\fgsimpletwovtx}{ \begin{tikzpicture}[x=1ex,y=1ex,baseline={([yshift=-.5ex]current bounding box.center)}] \coordinate (v) ; \def \n {2}; \def \rad {.8}; \filldraw[white] (v) circle (\rad); \foreach \s in {1,...,\n} { \def \angle {360/\n*(\s - 1)}; \coordinate (u) at ([shift=({\angle}:\rad)]v); \draw (v) -- (u); } \filldraw (v) circle (1pt); \end{tikzpicture} } \fi)} } \varphi_c, 0, 1 \right) \right) \right|_{\varphi_c=0} \\ = z^{( \ifmmode \usebox{\fgsimpletwovtx} \else \newsavebox{\fgsimpletwovtx} \savebox{\fgsimpletwovtx}{ \begin{tikzpicture}[x=1ex,y=1ex,baseline={([yshift=-.5ex]current bounding box.center)}] \coordinate (v) ; \def \n {2}; \def \rad {.8}; \filldraw[white] (v) circle (\rad); \foreach \s in {1,...,\n} { \def \angle {360/\n*(\s - 1)}; \coordinate (u) at ([shift=({\angle}:\rad)]v); \draw (v) -- (u); } \filldraw (v) circle (1pt); \end{tikzpicture} } \fi)} + (1- z^{( \ifmmode \usebox{\fgsimpletwovtx} \else \newsavebox{\fgsimpletwovtx} \savebox{\fgsimpletwovtx}{ \begin{tikzpicture}[x=1ex,y=1ex,baseline={([yshift=-.5ex]current bounding box.center)}] \coordinate (v) ; \def \n {2}; \def \rad {.8}; \filldraw[white] (v) circle (\rad); \foreach \s in {1,...,\n} { \def \angle {360/\n*(\s - 1)}; \coordinate (u) at ([shift=({\angle}:\rad)]v); \draw (v) -- (u); } \filldraw (v) circle (1pt); \end{tikzpicture} } \fi)}) \left( 1 + \left. \frac{\partial^2}{\partial \varphi_c^2} G(q_\text{ren}^2 \hbar,\varphi_c) \right|_{\varphi_c=0} \right) \\ = 1 + (1- z^{( \ifmmode \usebox{\fgsimpletwovtx} \else \newsavebox{\fgsimpletwovtx} \savebox{\fgsimpletwovtx}{ \begin{tikzpicture}[x=1ex,y=1ex,baseline={([yshift=-.5ex]current bounding box.center)}] \coordinate (v) ; \def \n {2}; \def \rad {.8}; \filldraw[white] (v) circle (\rad); \foreach \s in {1,...,\n} { \def \angle {360/\n*(\s - 1)}; \coordinate (u) at ([shift=({\angle}:\rad)]v); \draw (v) -- (u); } \filldraw (v) circle (1pt); \end{tikzpicture} } \fi)}) \left. \frac{\partial^2}{\partial \varphi_c^2} G(q_\text{ren}^2 \hbar,\varphi_c) \right|_{\varphi_c=0}  \end{gathered} \\  \begin{gathered} \label{eqn:counterterm_relation2} 1= \left. \frac{\partial^d}{\partial \varphi_c^d} \left( z^{( \ifmmode \usebox{\fgsimpletwovtx} \else \newsavebox{\fgsimpletwovtx} \savebox{\fgsimpletwovtx}{ \begin{tikzpicture}[x=1ex,y=1ex,baseline={([yshift=-.5ex]current bounding box.center)}] \coordinate (v) ; \def \n {2}; \def \rad {.8}; \filldraw[white] (v) circle (\rad); \foreach \s in {1,...,\n} { \def \angle {360/\n*(\s - 1)}; \coordinate (u) at ([shift=({\angle}:\rad)]v); \draw (v) -- (u); } \filldraw (v) circle (1pt); \end{tikzpicture} } \fi)} \frac{\varphi_c^2}{2} + q_\text{ren}^{-2} f_{\sgset_\text{bl}}^c\left( q_\text{ren}^2 \hbar, q_\text{ren} \sqrt{1- z^{( \ifmmode \usebox{\fgsimpletwovtx} \else \newsavebox{\fgsimpletwovtx} \savebox{\fgsimpletwovtx}{ \begin{tikzpicture}[x=1ex,y=1ex,baseline={([yshift=-.5ex]current bounding box.center)}] \coordinate (v) ; \def \n {2}; \def \rad {.8}; \filldraw[white] (v) circle (\rad); \foreach \s in {1,...,\n} { \def \angle {360/\n*(\s - 1)}; \coordinate (u) at ([shift=({\angle}:\rad)]v); \draw (v) -- (u); } \filldraw (v) circle (1pt); \end{tikzpicture} } \fi)} } \varphi_c, 0, 1 \right) \right) \right|_{\varphi_c=0} \\ = q_\text{ren} (1- z^{( \ifmmode \usebox{\fgsimpletwovtx} \else \newsavebox{\fgsimpletwovtx} \savebox{\fgsimpletwovtx}{ \begin{tikzpicture}[x=1ex,y=1ex,baseline={([yshift=-.5ex]current bounding box.center)}] \coordinate (v) ; \def \n {2}; \def \rad {.8}; \filldraw[white] (v) circle (\rad); \foreach \s in {1,...,\n} { \def \angle {360/\n*(\s - 1)}; \coordinate (u) at ([shift=({\angle}:\rad)]v); \draw (v) -- (u); } \filldraw (v) circle (1pt); \end{tikzpicture} } \fi)})^{\frac{d}{2}} \left. \frac{\partial^d}{\partial \varphi_c^d} G(q_\text{ren}^2 \hbar,\varphi_c) \right|_{\varphi_c=0} = z^{(v_d)} \left. \frac{\partial^d}{\partial \varphi_c^d} G(q_\text{ren}^2 \hbar,\varphi_c) \right|_{\varphi_c=0}, \end{gathered} \end{gather}
where we used the information from the Legendre transformation in eq.\ \eqref{eqn:legendre_transformation_specific},
\begin{align*} \left. \frac{\partial^2}{\partial \varphi_c^2} f_{\sgset_\text{bl}}^c\left( \hbar, \varphi_c, 0, 1 \right) \right|_{\varphi_c=0} &= 1 + \left. \frac{\partial^2 G}{\partial \varphi_c^2} (\hbar,\varphi_c) \right|_{\varphi_c=0} \\ \left. \frac{\partial^d}{\partial \varphi_c^d} f_{\sgset_\text{bl}}^c\left( \hbar, \varphi_c, 0, 1 \right) \right|_{\varphi_c=0} &= \left. \frac{\partial^d G}{\partial \varphi_c^d} (\hbar,\varphi_c) \right|_{\varphi_c=0}. \end{align*}
Combining both identities from eq.\ \eqref{eqn:counterterm_relation1} and \eqref{eqn:counterterm_relation2} results in,
\begin{align*} q_\text{ren} = \left( \frac{ z^{(v_d)} }{(1-z^{( \ifmmode \usebox{\fgsimpletwovtx} \else \newsavebox{\fgsimpletwovtx} \savebox{\fgsimpletwovtx}{ \begin{tikzpicture}[x=1ex,y=1ex,baseline={([yshift=-.5ex]current bounding box.center)}] \coordinate (v) ; \def \n {2}; \def \rad {.8}; \filldraw[white] (v) circle (\rad); \foreach \s in {1,...,\n} { \def \angle {360/\n*(\s - 1)}; \coordinate (u) at ([shift=({\angle}:\rad)]v); \draw (v) -- (u); } \filldraw (v) circle (1pt); \end{tikzpicture} } \fi)})^{\frac{d}{2}}} \right)^{\frac{1}{d-2}} = \left( \frac{ \left. \frac{\partial^d}{\partial \varphi_c^d} G(q_\text{ren}^2 \hbar,\varphi_c) \right|_{\varphi_c=0} }{ \left(-\left. \frac{\partial^2}{\partial \varphi_c^2} G(q_\text{ren}^2 \hbar,\varphi_c) \right|_{\varphi_c=0}\right)^{\frac{d}{2}}} \right)^{-\frac{1}{d-2}}, \end{align*}
from which follows $q_\text{ren}^2 \alpha(q_\text{ren}^2 \hbar) = 1$, where 
\begin{align} \label{eqn:invariantchargedef} \alpha(\hbar) := \left( \frac{ \left. \frac{\partial^d}{\partial \varphi_c^d} G(\hbar,\varphi_c) \right|_{\varphi_c=0} }{ \left(-\left. \frac{\partial^2}{\partial \varphi_c^2} G(\hbar,\varphi_c) \right|_{\varphi_c=0}\right)^{\frac{d}{2}}} \right)^{\frac{2}{d-2}}. \end{align}
The quantity $\alpha(\hbar)$ is called the \textit{invariant charge} of our theory. 

We will use $q_\text{ren}^2 \alpha(q_\text{ren}^2 \hbar) = 1$ as a fixed-point equation by interpreting $\hbar$ as a formal power series in another 
variable, $\hbar_\text{ren}$. The quantities $\hbar$ and $\hbar_\text{ren}$ are related by $\hbar(\hbar_\text{ren}) \alpha( \hbar(\hbar_\text{ren})) = \hbar_\text{ren}$.

Note that this is the classic and critical insight to renormalization theory: The expansion parameter $\hbar$ is interpreted as a function of an \textit{renormalized} expansion parameter \cite{gell1954quantum}. 

Using the equations for the counterterms in eqs.\ \eqref{eqn:counterterm_relation1} and \eqref{eqn:counterterm_relation2}, we may express the $z$-factors as
\begin{align*} z^{( \ifmmode \usebox{\fgsimpletwovtx} \else \newsavebox{\fgsimpletwovtx} \savebox{\fgsimpletwovtx}{ \begin{tikzpicture}[x=1ex,y=1ex,baseline={([yshift=-.5ex]current bounding box.center)}] \coordinate (v) ; \def \n {2}; \def \rad {.8}; \filldraw[white] (v) circle (\rad); \foreach \s in {1,...,\n} { \def \angle {360/\n*(\s - 1)}; \coordinate (u) at ([shift=({\angle}:\rad)]v); \draw (v) -- (u); } \filldraw (v) circle (1pt); \end{tikzpicture} } \fi)}(\hbar_\text{ren}) &= \frac{1}{-\left. \frac{\partial^2}{\partial \varphi_c^2} G(\hbar(\hbar_\text{ren})) \right|_{\varphi_c=0} } \\ z^{(v_d)}(\hbar_\text{ren}) &= \frac{1}{\left. \frac{\partial^d}{\partial \varphi_c^d} G(\hbar(\hbar_\text{ren})) \right|_{\varphi_c=0} } \end{align*}

Therefore, we can obtain the $z$-factors in zero-dimensional QFT from the proper Green functions and from the solution of the equation for the renormalized expansion parameter $\hbar_\text{ren}$. This computation can be performed in $\R[[\hbar]]$ and $\R[[\hbar_\text{ren}]]$. The asymptotics of these quantities can be obtained explicitly using of the $\asyOp{}{}{}$-derivative.
\section{Factorially divergent power series in zero-dimensional QFT}
\label{sec:ring2}
In this section, we will briefly recapitulate the notions from Chapter \ref{chap:facdivpow} and introduce additional notation tailored for our application to zero-dimensional QFT.
The algebraic formulation of the ring of factorially divergent power series not only will give us access to the asymptotic expansions of composite quantities, but also will provide us with a compact notation for lengthy asymptotic expressions.

We repeat the central Definition \ref{def:Fpowerseries} with $A = \frac{1}{\alpha}$ as this change of variables simplifies the notation:
\begin{defn}
Define $\fring{x}{A}{\beta}$ with $A \in \R_{>0}$ to be the subset of the ring of power series $f \in \R[[x]]$, whose coefficients have a Poincaré asymptotic expansion of the form,
\begin{align} f_n = \sum_{k=0}^{R-1} c_{k} A^{-n-\beta+k} \Gamma(n+\beta-k) + \bigO(A^{-n} \Gamma(n+\beta-R)), \end{align}
with coefficients $c_{k} \in \R$ and $\beta \in \R$. This subset forms a subring of $\R[[x]]$ as was shown in Chapter \ref{chap:facdivpow}. 
\end{defn}
Note that $A$ corresponds to $\alpha^{-1}$ from Chapter \ref{chap:facdivpow} (Definition \ref{def:Fpowerseries}). This different notation was also chosen to comply with the standard notation in the resurgence literature. 

We will introduce an additional operator similar to the operator $\asyOp^\alpha_\beta$ defined in Chapter \ref{chap:facdivpow} to simplify the notation:
\begin{defn}
\label{def:asymp}
Let $\asyOpV{A}{}{x}: \fring{x}{A}{\beta} \rightarrow x^{-\beta}\R[[x]]$ be the operator which maps a power series $f(x) = \sum_{n=0}^\infty f_n x^n$ to the generalized power series $\left(\asyOpV{A}{}{x} f\right)(x) = x^{-\beta}\sum_{k=0}^\infty c_k x^{k}$ such that, 
\begin{align} f_n &= \sum_{k=0}^{R-1} c_{k} A^{-n-\beta+k} \Gamma(n+\beta-k) + \bigO(A^{-n} \Gamma(n+\beta-R)) &&\forall R\geq 0. \end{align}
\end{defn}
\nomenclature{$\asyOpV{A}{}{x}$}{Modified asymptotic derivative}

The monomial $x^{-\beta}$ is included into the definition of the $\asyOpV{}{}{}$-operator, which maps to power series with a fixed monomial prefactor or equivalently generalized Laurent series. 
Moreover, we explicitly include the formal parameter $x$ into the notation. 
The former change simplifies the notation of the chain rule for compositions of power series heavily. 
The later change enables us to use the formalism on multivariate power series.

Both operators are related as, $\asyOp^\alpha_\beta (f(x)) = x^{\beta} \asyOpV{A}{}{x} f(x)$, where $A= \alpha^{-1}$.

\begin{expl}
Let $f(x) = \sum_{n=m+1}^\infty \Gamma(n+m) x^n$. It follows that $f \in \fring{x}{1}{m}$ and $\left(\asyOpV{1}{}{x} f\right)(x) = \frac{1}{x^m}$.
\end{expl}
\begin{expl}
For certain QED-type theories, we will need sequences which do not behave as an integer shift of the $\Gamma$-function. If for instance, 
$f(x) = \sum_{n=0}^\infty (2n-1)!! x^n = \frac{1}{\sqrt{2 \pi}} \sum_{n=0}^\infty 2^{n+\frac12} \Gamma\left(n+\frac12 \right) x^n$, then  
$\left(\asyOpV{\frac12}{}{x} f\right)(x) = \frac{1}{\sqrt{2 \pi x}}$ in agreement with Definition \ref{def:asymp}.
\end{expl}

As $\asyOp^\alpha_\beta$, the $\asyOpV{}{}{}$-operator is a derivative, which obeys the following identities for $f,g \in \fring{x}{A}{\beta}$. These identities follow directly from the properties of the $\asyOp^\alpha_\beta$-operator, which were established in Chapter \ref{chap:facdivpow} (Corollary \ref{crll:asyOplinear}, Proposition \ref{prop:derivation}, Theorem \ref{thm:chain_analytic} and Theorem \ref{thm:chainrule}):
\begin{align*} &\asyOpV{A}{\beta}{x} (f(x) + g(x))& &=& &\asyOpV{A}{\beta}{x} f(x) + \asyOpV{A}{\beta}{x} g(x) && \textit{Linearity} \\ &\asyOpV{A}{\beta}{x} (f(x) g(x)) & &=& &g(x) \asyOpV{A}{\beta}{x} f(x) + f(x) \asyOpV{A}{\beta}{x} g(x) && \textit{Product rule} \\ &\asyOpV{A}{\beta}{x} f(g(x)) & &=& &f'(g(x)) \asyOpV{A}{\beta}{x} g(x) + e^{A \left( \frac{1}{x} - \frac{1}{\xi}\right)} (\asyOpV{A}{\beta}{\xi} f(\xi)) \big|_{\xi=g(x)} && \textit{Chain rule} \\ &\asyOpV{A}{\beta}{x} g^{-1}(x) & &=& -& {g^{-1}}'(x) e^{A \left( \frac{1}{x} - \frac{1}{\xi}\right)} (\asyOpV{A}{\beta}{\xi} g(\xi)) \big|_{\xi=g^{-1}(x)} && \textit{Inverse} \\ & & &=& -& e^{A \left( \frac{1}{x} - \frac{1}{\xi}\right)} \left. \frac{\asyOpV{A}{\beta}{\xi} g(\xi)} { \partial_\xi g(\xi) } \right|_{\xi=g^{-1}(x)} &&    \end{align*}
where $f'(x)$ denotes the usual derivative of $f(x)$. We require $g_0=0$ and $g_1=1$ for the chain rule and the formula for the inverse. 

With this notation at hand, the asymptotics of a formal integral, which fulfills the restrictions of Corollary \ref{crll:comb_int_asymp}, may be written in compact form as,
\begin{align*} \asyOpV{A}{}{\hbar} \Fop\left[\Sact(x)\right](\hbar) = \frac{1}{2\pi} \sum_{i \in I} \Fop\left[\Sact(\tau_i) - \Sact(x+\tau_i)\right](-\hbar), \end{align*}
where $\tau_i$ are the locations of the dominant saddle points, $A = -\Sact(\tau_i)$ and 
$\Fop\left[\Sact(x)\right](\hbar) \in \fring{\hbar}{A}{0}$.
The important property is that $\Fop$-expressions are stable under application of an $\asyOpV{}{}{}$-derivative. This makes the calculation of the asymptotics as easy as calculating the expansion at low-order.
\begin{expl}
\label{expl:phi3theorycompactasymptotics}
The asymptotics deduced in Example \ref{expl:phi3theoryasymptotics} can be written in compact form as,
\begin{align*} \asyOpV{\frac23}{}{\hbar} \Fop\left[-\frac{x^2}{2} + \frac{x^3}{3!}\right](\hbar) &= \frac{1}{2\pi} \Fop\left[-\frac{x^2}{2}+\frac{x^3}{3!}\right](-\hbar), \end{align*}
where $\Fop\left[-\frac{x^2}{2}+\frac{x^3}{3!}\right](\hbar) \in \fring{\hbar}{\frac23}{0}$.
\end{expl}
\section{Notation and verification}
The coefficients of asymptotic expansions in the following section are given in the notation of Section \ref{sec:ring2}. That means, a row in a table such as, 

\begin{minipage}{\linewidth}%
\vspace{1ex}%
\centering
\def\arraystretch{1.5}%
\begin{tabular}{|c||c||c|c|c|c|c|c|}
\hline
&prefactor&$\hbar^{0}$&$\hbar^{1}$&$\hbar^{2}$&$\hbar^{3}$&$\hbar^{4}$&$\hbar^{5}$\\
\hline\hline
$\asyOpV{A}{0}{\hbar} f$&$ C \hbar^{-\beta}$&$c_0$&$c_1$&$c_2$&$c_3$&$c_4$&$c_5$\\
\hline
\end{tabular}%
\vspace{1ex}%
\end{minipage}
corresponds to an asymptotic expansion of the coefficients of the power series $f(\hbar)$:
\begin{align*} [\hbar^n] f(\hbar) = C \sum_{k=0}^{R-1} c_k A^{-n-\beta+k} \Gamma(n+\beta-k) + \bigO(A^{-n} \Gamma(n+\beta-R)). \end{align*}
The redundant prefactor $C$ was included to highlight the overall transcendental number that will be the same for every expansion in a single theory.

The given low-order expansions were checked by explicitly counting diagrams with the program \texttt{feyngen} \cite{borinsky2014feynman}. 
All given expansions were computed up to at least $100$ coefficients using basic computer algebra.
Although the asymptotics were completely obtained by analytic means, numerical computations were used to verify the analytic results. 
All given asymptotic expansions were checked by computing the asymptotics from the original expansions using the Richardson-extrapolation of the first $100$ coefficients.

\section{Examples from scalar theories}
\subsection{\texorpdfstring{$\varphi^3$}{ϕ³}-theory}

\paragraph{Disconnected diagrams}
We start with an analysis of the asymptotics of zero-dimensional $\varphi^3$-theory, which has been analyzed in \cite{cvitanovic1978number} using differential equations. For the sake of completeness, we will repeat the calculation with different methods and obtain all-order asymptotics in terms of $\Fop$ expressions.

The partition function with sources is given by the formal integral,
\begin{align} Z^{\varphi^3}(\hbar, j) &:= \int \frac{dx}{\sqrt{2 \pi \hbar}} e^{\frac{1}{\hbar} \left( -\frac{x^2}{2} + \frac{x^3}{3!} + x j \right) } = 1 + \frac{j^2}{2\hbar} + \frac{j^3}{3! \hbar} + \frac{1}{2} j + \frac{5}{24}\hbar + \ldots \end{align}
This expansion may be depicted as,
\begin{align*} Z^{\varphi^3}(\hbar,j) &= \phi_\Sact' \Big( \one +  \frac12 {  \ifmmode \usebox{\fgsimpletreehandle} \else \newsavebox{\fgsimpletreehandle} \savebox{\fgsimpletreehandle}{ \begin{tikzpicture}[x=1ex,y=1ex,baseline={([yshift=-.5ex]current bounding box.center)}] \coordinate (v) ; \coordinate [right=1.2 of v] (u); \draw (v) -- (u); \filldraw (v) circle (1pt); \filldraw (u) circle (1pt); \end{tikzpicture} } \fi } + \frac16 {  \ifmmode \usebox{\fgjthreesimplevtx} \else \newsavebox{\fgjthreesimplevtx} \savebox{\fgjthreesimplevtx}{ \begin{tikzpicture}[x=1ex,y=1ex,baseline={([yshift=-.5ex]current bounding box.center)}] \coordinate (v) ; \def \n {3}; \def \rad {1.2}; \filldraw[white] (v) circle (\rad); \foreach \s in {1,...,5} { \def \angle {180+360/\n*(\s - 1)}; \coordinate (u) at ([shift=({\angle}:\rad)]v); \draw (v) -- (u); \filldraw (u) circle (1pt); } \filldraw (v) circle (1pt); \end{tikzpicture} } \fi } + \frac18 {  \begin{tikzpicture}[x=1ex,y=1ex,baseline={([yshift=-.5ex]current bounding box.center)}] \coordinate (v0) ; \coordinate [right=1.2 of v0] (u0); \coordinate [below=1.2 of v0] (v1) ; \coordinate [right=1.2 of v1] (u1); \draw (v0) -- (u0); \filldraw (v0) circle (1pt); \filldraw (u0) circle (1pt); \draw (v1) -- (u1); \filldraw (v1) circle (1pt); \filldraw (u1) circle (1pt); \end{tikzpicture} } + \frac12 {  \ifmmode \usebox{\fgonetadpolephithree} \else \newsavebox{\fgonetadpolephithree} \savebox{\fgonetadpolephithree}{ \begin{tikzpicture}[x=2ex,y=2ex,baseline={([yshift=-.5ex]current bounding box.center)}] \coordinate (v0) ; \coordinate [right=1 of v0] (v1); \coordinate [left=.7 of v0] (i0); \coordinate [left=.5 of v1] (vm); \draw (vm) circle(.5); \draw (i0) -- (v0); \filldraw (v0) circle(1pt); \filldraw (i0) circle (1pt); \end{tikzpicture} } \fi } + \frac18 {  \begin{tikzpicture}[x=2ex,y=2ex,baseline={([yshift=-.5ex]current bounding box.center)}] \coordinate (v00); \coordinate [below=1.2 of v00] (v01); \coordinate [right=1 of v00] (v10); \coordinate [right=1 of v01] (v11); \coordinate [left=.7 of v00] (i00); \coordinate [left=.7 of v01] (i01); \coordinate [left=.5 of v10] (vm0); \coordinate [left=.5 of v11] (vm1); \draw (vm0) circle(.45); \draw (vm1) circle(.45); \draw (i00) -- (v00); \draw (i01) -- (v01); \filldraw (v00) circle(1pt); \filldraw (v01) circle(1pt); \filldraw (i00) circle (1pt); \filldraw (i01) circle(1pt); \end{tikzpicture} } + \frac14 {  \ifmmode \usebox{\fgtwojoneloopbubblephithree} \else \newsavebox{\fgtwojoneloopbubblephithree} \savebox{\fgtwojoneloopbubblephithree}{ \begin{tikzpicture}[x=2ex,y=2ex,baseline={([yshift=-.5ex]current bounding box.center)}] \coordinate (v0) ; \coordinate [right=1 of v0] (v1); \coordinate [left=.7 of v0] (i0); \coordinate [right=.7 of v1] (o0); \coordinate [left=.5 of v1] (vm); \draw (vm) circle(.5); \draw (i0) -- (v0); \draw (o0) -- (v1); \filldraw (v0) circle(1pt); \filldraw (v1) circle(1pt); \filldraw (i0) circle (1pt); \filldraw (o0) circle (1pt); \end{tikzpicture} } \fi } \\ &+ \frac16 {  \ifmmode \usebox{\fgthreejoneltrianglephithree} \else \newsavebox{\fgthreejoneltrianglephithree} \savebox{\fgthreejoneltrianglephithree}{ \begin{tikzpicture}[x=1ex,y=1ex,baseline={([yshift=-.5ex]current bounding box.center)}] \coordinate (v) ; \def \n {3}; \def \rad {1}; \def \rud {2.2}; \foreach \s in {1,...,5} { \def \angle {360/\n*(\s - 1)}; \def \ungle {360/\n*\s}; \coordinate (s) at ([shift=({\angle}:\rad)]v); \coordinate (t) at ([shift=({\ungle}:\rad)]v); \coordinate (u) at ([shift=({\angle}:\rud)]v); \draw (s) -- (u); \filldraw (u) circle (1pt); \filldraw (s) circle (1pt); } \draw (v) circle(\rad); \end{tikzpicture} } \fi } + \frac14 {  \ifmmode \usebox{\fgthreejonelpropinsphithree} \else \newsavebox{\fgthreejonelpropinsphithree} \savebox{\fgthreejonelpropinsphithree}{ \begin{tikzpicture}[x=2ex,y=2ex,baseline={([yshift=-.5ex]current bounding box.center)}] \coordinate (v0) ; \coordinate [right=1 of v0] (v1); \coordinate [right=.7 of v1] (v2); \coordinate [left=.7 of v0] (i0); \coordinate [above right=.7 of v2] (o0); \coordinate [below right=.7 of v2] (o1); \coordinate [left=.5 of v1] (vm); \draw (vm) circle(.5); \draw (i0) -- (v0); \draw (v1) -- (v2); \draw (o0) -- (v2); \draw (o1) -- (v2); \filldraw (v0) circle(1pt); \filldraw (v1) circle(1pt); \filldraw (v2) circle(1pt); \filldraw (i0) circle (1pt); \filldraw (o0) circle (1pt); \filldraw (o1) circle (1pt); \end{tikzpicture} } \fi } + \frac18 {  \ifmmode \usebox{\fghandle} \else \newsavebox{\fghandle} \savebox{\fghandle}{ \begin{tikzpicture}[x=1ex,y=1ex,baseline={([yshift=-.5ex]current bounding box.center)}] \coordinate (v0); \coordinate [right=1.5 of v0] (v1); \coordinate [left=.7 of v0] (i0); \coordinate [right=.7 of v1] (o0); \draw (v0) -- (v1); \filldraw (v0) circle (1pt); \filldraw (v1) circle (1pt); \draw (i0) circle(.7); \draw (o0) circle(.7); \end{tikzpicture} } \fi } + \frac{1}{12} {  \ifmmode \usebox{\fgbananathree} \else \newsavebox{\fgbananathree} \savebox{\fgbananathree}{ \begin{tikzpicture}[x=1ex,y=1ex,baseline={([yshift=-.5ex]current bounding box.center)}] \coordinate (vm); \coordinate [left=1 of vm] (v0); \coordinate [right=1 of vm] (v1); \draw (v0) -- (v1); \draw (vm) circle(1); \filldraw (v0) circle (1pt); \filldraw (v1) circle (1pt); \end{tikzpicture} } \fi } + \ldots \Big) \end{align*}
with the Feynman rule $\phi_\Sact':\Gamma \mapsto \hbar^{|E_\Gamma|-|V_\Gamma|} j^{\nvd{ \ifmmode \usebox{\fgsimpleonevtx} \else \newsavebox{\fgsimpleonevtx} \savebox{\fgsimpleonevtx}{ \begin{tikzpicture}[x=1ex,y=1ex,baseline={([yshift=-.55ex]current bounding box.center)}] \coordinate (v) ; \def \n {1}; \def \rad {1}; \filldraw[white] (v) circle (\rad); \foreach \s in {1,...,\n} { \def \angle {180+360/\n*(\s - 1)}; \coordinate (u) at ([shift=({\angle}:\rad)]v); \draw (v) -- (u); } \filldraw (v) circle (1pt); \end{tikzpicture} } \fi}_\Gamma}$, which also assigns a power of $j$ to a graph for every $1$-valent vertex it has.
After a shift and rescaling of the integration variable $Z^{\varphi^3}(\hbar, j)$ takes the form,
\begin{align} \begin{split} \label{eqn:Zphi3_as_Zphi0} Z^{\varphi^3}(\hbar, j) &=  e^{\frac{-\frac{x_0^2}{2} + \frac{x_0^3}{3!}+ x_0 j}{\hbar} } \frac{1}{(1-2j)^{\frac14} } \int \frac{dx}{\sqrt{2 \pi \frac{\hbar}{(1-2j)^{\frac32}}}} e^{\frac{(1-2j)^{\frac32}}{\hbar} \left( - \frac{x^2}{2 } + \frac{x^3}{3!} \right) } \\  &= e^{\frac{(1-2 j)^{\frac32}-1+3j}{3\hbar} } \frac{1}{(1-2j)^{\frac14} } Z^{\varphi^3}_0 \left(\frac{\hbar}{(1-2j)^{\frac32}} \right) \end{split} \end{align}
where $x_0 := 1- \sqrt{1-2j}$ and $Z^{\varphi^3}_0(\hbar) := Z^{\varphi^3}(\hbar, 0)$. The last equality gives a significant simplification, because we are effectively left with a univariate generating function. The combinatorial explanation for this is that we can always `dress' a graph without external legs, a vacuum graph, by attaching an arbitrary number of rooted trees to the edges of the original graph, similar to the argument for Lemma \ref{lmm:invariant_charge}. Note that $-\frac{x_0^2}{2} + \frac{x_0^3}{3!}+ x_0 j = \frac13 ((1-2 j)^{\frac32}-1+3j)$, sequence \texttt{A001147} in the OEIS \cite{oeis}, is the generating function of all connected trees build out of three valent vertices. 

The generating function of $\varphi^3$-graphs without legs is given by
\begin{align*} Z^{\varphi^3}_0 \left(\hbar \right) &= \Fop\left[-\frac{x^2}{2}+\frac{x^3}{3!}\right]\left(\hbar\right), \end{align*}
which has been discussed in Examples \ref{expl:phi3theoryexpansion}, \ref{expl:phi3elliptic}, \ref{expl:phi3elliptic_singularity}, \ref{expl:phi3theoryasymptotics} and \ref{expl:phi3theorycompactasymptotics}. The first coefficients of $Z^{\varphi^3}(\hbar, j)$ are given in Table \ref{tab:Zphi3}.
\begin{table}
\begin{subtable}[c]{\textwidth}
\centering
\tiny
\def\arraystretch{1.5}
\begin{tabular}{|c||c||c|c|c|c|c|c|}
\hline
&prefactor&$\hbar^{0}$&$\hbar^{1}$&$\hbar^{2}$&$\hbar^{3}$&$\hbar^{4}$&$\hbar^{5}$\\
\hline\hline
$\partial_j^{0} Z^{\varphi^3} \big|_{j=0}$&$\hbar^{0}$&$1$&$ \frac{5}{24}$&$ \frac{385}{1152}$&$ \frac{85085}{82944}$&$ \frac{37182145}{7962624}$&$ \frac{5391411025}{191102976}$\\
\hline
$\partial_j^{1} Z^{\varphi^3} \big|_{j=0}$&$\hbar^{0}$&$ \frac{1}{2}$&$ \frac{35}{48}$&$ \frac{5005}{2304}$&$ \frac{1616615}{165888}$&$ \frac{929553625}{15925248}$&$ \frac{167133741775}{382205952}$\\
\hline
$\partial_j^{2} Z^{\varphi^3} \big|_{j=0}$&$\hbar^{-1}$&$1$&$ \frac{35}{24}$&$ \frac{5005}{1152}$&$ \frac{1616615}{82944}$&$ \frac{929553625}{7962624}$&$ \frac{167133741775}{191102976}$\\
\hline
$\partial_j^{3} Z^{\varphi^3} \big|_{j=0}$&$\hbar^{-1}$&$ \frac{5}{2}$&$ \frac{385}{48}$&$ \frac{85085}{2304}$&$ \frac{37182145}{165888}$&$ \frac{26957055125}{15925248}$&$ \frac{5849680962125}{382205952}$\\
\hline
\end{tabular}
\subcaption{The first coefficients of the bivariate generating function $Z^{\varphi^3}(\hbar, j)$.}
\label{tab:Zphi3}
\end{subtable}
\begin{subtable}[c]{\textwidth}
\centering
\tiny
\def\arraystretch{1.5}
\begin{tabular}{|c||c||c|c|c|c|c|c|}
\hline
&prefactor&$\hbar^{0}$&$\hbar^{1}$&$\hbar^{2}$&$\hbar^{3}$&$\hbar^{4}$&$\hbar^{5}$\\
\hline\hline
$\asyOpV{\frac23}{0}{\hbar} \partial_j^{0} Z^{\varphi^3} \big|_{j=0}$&$\frac{\hbar^{0}}{2\pi}$&$1$&$- \frac{5}{24}$&$ \frac{385}{1152}$&$- \frac{85085}{82944}$&$ \frac{37182145}{7962624}$&$- \frac{5391411025}{191102976}$\\
\hline
$\asyOpV{\frac23}{0}{\hbar} \partial_j^{1} Z^{\varphi^3} \big|_{j=0}$&$\frac{\hbar^{-1}}{2\pi}$&$2$&$ \frac{1}{12}$&$- \frac{35}{576}$&$ \frac{5005}{41472}$&$- \frac{1616615}{3981312}$&$ \frac{185910725}{95551488}$\\
\hline
$\asyOpV{\frac23}{0}{\hbar} \partial_j^{2} Z^{\varphi^3} \big|_{j=0}$&$\frac{\hbar^{-2}}{2\pi}$&$4$&$ \frac{1}{6}$&$- \frac{35}{288}$&$ \frac{5005}{20736}$&$- \frac{1616615}{1990656}$&$ \frac{185910725}{47775744}$\\
\hline
$\asyOpV{\frac23}{0}{\hbar} \partial_j^{3} Z^{\varphi^3} \big|_{j=0}$&$\frac{\hbar^{-3}}{2\pi}$&$8$&$- \frac{5}{3}$&$ \frac{25}{144}$&$- \frac{1925}{10368}$&$ \frac{425425}{995328}$&$- \frac{37182145}{23887872}$\\
\hline
\end{tabular}
\subcaption{The first coefficients of the bivariate generating function $\asyOpV{\frac23}{0}{\hbar} Z^{\varphi^3}(\hbar, j)$.}
\label{tab:Zphi3asymp}
\end{subtable}
\caption{Partition function in $\varphi^3$-theory.}
\end{table}
Using Theorem \ref{thm:comb_int_asymp} the generating function of the asymptotics of $Z^{\varphi^3}_0 $ were calculated in Example \ref{expl:phi3theoryasymptotics}. Written in the notation of Section \ref{sec:ring2}. We have $Z^{\varphi^3}_0 \in \fring{\hbar}{\frac{2}{3}}{0}$ and
\begin{align*} \asyOpV{\frac23}{0}{\hbar} Z^{\varphi^3}_0 (\hbar) = \frac{1}{2\pi} \Fop\left[-\frac{x^2}{2}+\frac{x^3}{3!}\right]\left(-\hbar\right) = \frac{1}{2\pi} Z^{\varphi^3}_0 (-\hbar). \end{align*}
This very simple form for this generating function can of course be traced back to the simple structure of $\varphi^3$-theory, which is almost invariant under the $\asyOpV{}{}{}$-derivative. 

The bivariate generating function of the asymptotics is obtained by using the $\asyOpV{}{}{}$-derivative on eq.\ \eqref{eqn:Zphi3_as_Zphi0} and applying the chain rule from Section \ref{sec:ring2}:
\begin{align} \begin{split} \label{eqn:Zphi3asymp} \asyOpV{\frac23}{0}{\hbar} Z^{\varphi^3}(\hbar, j) &= e^{\frac{(1-2 j)^{\frac32}-1+3j}{3\hbar} } \frac{1}{(1-2j)^{\frac14} } e^{\frac23 \frac{1 - (1-2 j)^{\frac32}}{\hbar}} \asyOpV{\frac23}{0}{\widetilde \hbar} Z^{\varphi^3}_0 \left(\widetilde \hbar \right) \big|_{\widetilde \hbar = \frac{\hbar}{(1-2j)^{\frac32}}} \\ &= \frac{1}{2\pi} e^{\frac{1-(1-2 j)^{\frac32}+3j}{3\hbar} } \frac{1}{(1-2j)^{\frac14} } Z^{\varphi^3}_0 \left(-\frac{\hbar}{(1-2j)^{\frac32}} \right). \end{split} \end{align}
Note that the $\asyOpV{}{}{}$-derivative commutes with expansions in $j$, as we leave the number of external legs fixed while taking the limit to large loop order.
The first coefficients of the asymptotics of $Z^{\varphi^3}(\hbar, j)$ are listed in Table \ref{tab:Zphi3asymp}.

We may also expand the expression for the asymptotics in eq.\ \eqref{eqn:Zphi3asymp} in $\hbar$ to obtain a generating function for the first coefficient of the asymptotic expansions of the derivatives by $j$:
\begin{align*} \asyOpV{\frac23}{0}{\hbar} Z^{\varphi^3}(\hbar, j) &=\frac{1}{2\pi} e^{\frac{2j}{\hbar}} \left( 1 + \left(-\frac{5}{24} + \frac14 \frac{2j}{\hbar} - \frac18 \frac{(2j)^2}{\hbar^2} \right) \hbar +\ldots\right) \\ \asyOpV{\frac23}{0}{\hbar} \partial_j^m Z^{\varphi^3}(\hbar, j) \big|_{j=0} &= \frac{1}{2\pi} \left(\frac{2}{\hbar}\right)^{m}\left(1 + \left( -\frac{5}{24} + \frac{3m}{8} - \frac{m^2}{8} \right) \hbar + \ldots \right) \end{align*}
By Definition \ref{def:asymp} this can be translated into an asymptotic expression for large order coefficients. With $\partial_j^m Z^{\varphi^3}(\hbar, j)\big|_{j=0} = \sum_{n=0}^\infty z_{m,n} \hbar^n$:
\begin{align*} z_{m,n} &= \sum_{k=0}^{R-1} c_{m,k} \left(\frac{2}{3} \right)^{-m-n+k} \Gamma(n+m-k) \\ &+\bigO\left( \left(\frac{2}{3} \right)^{-m-n+R} \Gamma(n+m-R)\right) , \intertext{for all $R \geq 0$, where $c_{m,k}= [\hbar^k] \hbar^m \asyOpV{\frac23}{0}{\hbar}\partial_j^m Z^{\varphi^3}(\hbar, j)\big|_{j=0}$ or more explicitly, } z_{m,n} & \underset{n\rightarrow \infty}{\sim} \frac{2^m}{2\pi} \left(\frac{2}{3} \right)^{-m-n} \Gamma(n+m) \\ & \times \left( 1 + \frac23 \left( -\frac{5}{24} + \frac{3m}{8} - \frac{m^2}{8} \right) \frac{1}{n+m-1} +\ldots \right), \end{align*}
which agrees with the coefficients, which were given in \cite{cvitanovic1978number} in a different notation.
\paragraph{Connected diagrams}
\begin{table}
\begin{subtable}[c]{\textwidth}
\centering
\tiny
\def\arraystretch{1.5}
\begin{tabular}{|c||c|c|c|c|c|c|}
\hline
&$\hbar^{0}$&$\hbar^{1}$&$\hbar^{2}$&$\hbar^{3}$&$\hbar^{4}$&$\hbar^{5}$\\
\hline\hline
$\partial_j^{0} W^{\varphi^3} \big|_{j=0}$&$0$&$0$&$ \frac{5}{24}$&$ \frac{5}{16}$&$ \frac{1105}{1152}$&$ \frac{565}{128}$\\
\hline
$\partial_j^{1} W^{\varphi^3} \big|_{j=0}$&$0$&$ \frac{1}{2}$&$ \frac{5}{8}$&$ \frac{15}{8}$&$ \frac{1105}{128}$&$ \frac{1695}{32}$\\
\hline
$\partial_j^{2} W^{\varphi^3} \big|_{j=0}$&$1$&$1$&$ \frac{25}{8}$&$15$&$ \frac{12155}{128}$&$ \frac{11865}{16}$\\
\hline
$\partial_j^{3} W^{\varphi^3} \big|_{j=0}$&$1$&$4$&$ \frac{175}{8}$&$150$&$ \frac{158015}{128}$&$11865$\\
\hline
\end{tabular}
\subcaption{Table of the first coefficients of the bivariate generating function $W^{\varphi^3}(\hbar, j)$.}
\label{tab:Wphi3}
\end{subtable}
\begin{subtable}[c]{\textwidth}
\centering
\tiny
\def\arraystretch{1.5}
\begin{tabular}{|c||c||c|c|c|c|c|c|}
\hline
&prefactor&$\hbar^{0}$&$\hbar^{1}$&$\hbar^{2}$&$\hbar^{3}$&$\hbar^{4}$&$\hbar^{5}$\\
\hline\hline
$\asyOpV{\frac23}{0}{\hbar} \partial_j^{0} W^{\varphi^3}\big|_{j=0}$&$\frac{\hbar^{1}}{2 \pi}$&$1$&$- \frac{5}{12}$&$ \frac{25}{288}$&$- \frac{20015}{10368}$&$ \frac{398425}{497664}$&$- \frac{323018725}{5971968}$\\
\hline
$\asyOpV{\frac23}{0}{\hbar} \partial_j^{1} W^{\varphi^3}\big|_{j=0}$&$\frac{\hbar^{0}}{2 \pi}$&$2$&$- \frac{5}{6}$&$- \frac{155}{144}$&$- \frac{17315}{5184}$&$- \frac{3924815}{248832}$&$- \frac{294332125}{2985984}$\\
\hline
$\asyOpV{\frac23}{0}{\hbar} \partial_j^{2} W^{\varphi^3}\big|_{j=0}$&$\frac{\hbar^{-1}}{2 \pi}$&$4$&$- \frac{11}{3}$&$- \frac{275}{72}$&$- \frac{31265}{2592}$&$- \frac{7249295}{124416}$&$- \frac{553369915}{1492992}$\\
\hline
$\asyOpV{\frac23}{0}{\hbar} \partial_j^{3} W^{\varphi^3}\big|_{j=0}$&$\frac{\hbar^{-2}}{2 \pi}$&$8$&$- \frac{46}{3}$&$- \frac{407}{36}$&$- \frac{51065}{1296}$&$- \frac{12501815}{62208}$&$- \frac{988327615}{746496}$\\
\hline
\end{tabular}
\subcaption{Table of the first coefficients of the bivariate generating function $\asyOpV{\frac23}{0}{\hbar} W^{\varphi^3}(\hbar, j)$.}
\label{tab:Wphi3asymp}
\end{subtable}
\caption{Free energy in $\varphi^3$-theory.}
\end{table}

The generating function of the connected graphs can be obtained by taking the logarithm of $Z^{\varphi^3}$:
\begin{align} \begin{split} \label{eqn:Wphi3_explicit} W^{\varphi^3} ( \hbar, j ) &:= \hbar \log Z^{\varphi^3}(\hbar, j) \\ &= \frac13 ((1-2 j)^{\frac32}-1+3j) + \frac14 \hbar \log \frac{1}{1-2j } + \hbar \log Z^{\varphi^3}_0 \left(\frac{\hbar}{(1-2j)^{\frac32}} \right) \end{split} \\ \notag &= \frac{5}{24}\hbar^{2} + \frac{1}{2} j \hbar + \frac{5}{8} j \hbar^{2} + \frac{1}{2} j^{2} + \frac{1}{2} j^{2} \hbar + \frac{25}{16} j^{2} \hbar^{2} + \ldots \end{align}
This can be written as the diagrammatic expansion,
\begin{align*} W^{\varphi^3}(\hbar,j) &=  \phi_\Sact \Big( \frac12 {  \ifmmode \usebox{\fgsimpletreehandle} \else \newsavebox{\fgsimpletreehandle} \savebox{\fgsimpletreehandle}{ \begin{tikzpicture}[x=1ex,y=1ex,baseline={([yshift=-.5ex]current bounding box.center)}] \coordinate (v) ; \coordinate [right=1.2 of v] (u); \draw (v) -- (u); \filldraw (v) circle (1pt); \filldraw (u) circle (1pt); \end{tikzpicture} } \fi } + \frac16 {  \ifmmode \usebox{\fgjthreesimplevtx} \else \newsavebox{\fgjthreesimplevtx} \savebox{\fgjthreesimplevtx}{ \begin{tikzpicture}[x=1ex,y=1ex,baseline={([yshift=-.5ex]current bounding box.center)}] \coordinate (v) ; \def \n {3}; \def \rad {1.2}; \filldraw[white] (v) circle (\rad); \foreach \s in {1,...,5} { \def \angle {180+360/\n*(\s - 1)}; \coordinate (u) at ([shift=({\angle}:\rad)]v); \draw (v) -- (u); \filldraw (u) circle (1pt); } \filldraw (v) circle (1pt); \end{tikzpicture} } \fi } + \frac12 {  \ifmmode \usebox{\fgonetadpolephithree} \else \newsavebox{\fgonetadpolephithree} \savebox{\fgonetadpolephithree}{ \begin{tikzpicture}[x=2ex,y=2ex,baseline={([yshift=-.5ex]current bounding box.center)}] \coordinate (v0) ; \coordinate [right=1 of v0] (v1); \coordinate [left=.7 of v0] (i0); \coordinate [left=.5 of v1] (vm); \draw (vm) circle(.5); \draw (i0) -- (v0); \filldraw (v0) circle(1pt); \filldraw (i0) circle (1pt); \end{tikzpicture} } \fi } + \frac14 {  \ifmmode \usebox{\fgtwojoneloopbubblephithree} \else \newsavebox{\fgtwojoneloopbubblephithree} \savebox{\fgtwojoneloopbubblephithree}{ \begin{tikzpicture}[x=2ex,y=2ex,baseline={([yshift=-.5ex]current bounding box.center)}] \coordinate (v0) ; \coordinate [right=1 of v0] (v1); \coordinate [left=.7 of v0] (i0); \coordinate [right=.7 of v1] (o0); \coordinate [left=.5 of v1] (vm); \draw (vm) circle(.5); \draw (i0) -- (v0); \draw (o0) -- (v1); \filldraw (v0) circle(1pt); \filldraw (v1) circle(1pt); \filldraw (i0) circle (1pt); \filldraw (o0) circle (1pt); \end{tikzpicture} } \fi } \\ &+ \frac16 {  \ifmmode \usebox{\fgthreejoneltrianglephithree} \else \newsavebox{\fgthreejoneltrianglephithree} \savebox{\fgthreejoneltrianglephithree}{ \begin{tikzpicture}[x=1ex,y=1ex,baseline={([yshift=-.5ex]current bounding box.center)}] \coordinate (v) ; \def \n {3}; \def \rad {1}; \def \rud {2.2}; \foreach \s in {1,...,5} { \def \angle {360/\n*(\s - 1)}; \def \ungle {360/\n*\s}; \coordinate (s) at ([shift=({\angle}:\rad)]v); \coordinate (t) at ([shift=({\ungle}:\rad)]v); \coordinate (u) at ([shift=({\angle}:\rud)]v); \draw (s) -- (u); \filldraw (u) circle (1pt); \filldraw (s) circle (1pt); } \draw (v) circle(\rad); \end{tikzpicture} } \fi } + \frac14 {  \ifmmode \usebox{\fgthreejonelpropinsphithree} \else \newsavebox{\fgthreejonelpropinsphithree} \savebox{\fgthreejonelpropinsphithree}{ \begin{tikzpicture}[x=2ex,y=2ex,baseline={([yshift=-.5ex]current bounding box.center)}] \coordinate (v0) ; \coordinate [right=1 of v0] (v1); \coordinate [right=.7 of v1] (v2); \coordinate [left=.7 of v0] (i0); \coordinate [above right=.7 of v2] (o0); \coordinate [below right=.7 of v2] (o1); \coordinate [left=.5 of v1] (vm); \draw (vm) circle(.5); \draw (i0) -- (v0); \draw (v1) -- (v2); \draw (o0) -- (v2); \draw (o1) -- (v2); \filldraw (v0) circle(1pt); \filldraw (v1) circle(1pt); \filldraw (v2) circle(1pt); \filldraw (i0) circle (1pt); \filldraw (o0) circle (1pt); \filldraw (o1) circle (1pt); \end{tikzpicture} } \fi } + \frac18 {  \ifmmode \usebox{\fghandle} \else \newsavebox{\fghandle} \savebox{\fghandle}{ \begin{tikzpicture}[x=1ex,y=1ex,baseline={([yshift=-.5ex]current bounding box.center)}] \coordinate (v0); \coordinate [right=1.5 of v0] (v1); \coordinate [left=.7 of v0] (i0); \coordinate [right=.7 of v1] (o0); \draw (v0) -- (v1); \filldraw (v0) circle (1pt); \filldraw (v1) circle (1pt); \draw (i0) circle(.7); \draw (o0) circle(.7); \end{tikzpicture} } \fi } + \frac{1}{12} {  \ifmmode \usebox{\fgbananathree} \else \newsavebox{\fgbananathree} \savebox{\fgbananathree}{ \begin{tikzpicture}[x=1ex,y=1ex,baseline={([yshift=-.5ex]current bounding box.center)}] \coordinate (vm); \coordinate [left=1 of vm] (v0); \coordinate [right=1 of vm] (v1); \draw (v0) -- (v1); \draw (vm) circle(1); \filldraw (v0) circle (1pt); \filldraw (v1) circle (1pt); \end{tikzpicture} } \fi } + \ldots \Big) \end{align*}
where we now assign the slightly modified Feynman rules $\phi_\Sact:\Gamma \mapsto \hbar^{h_\Gamma} j^{\nvd{ \ifmmode \usebox{\fgsimpleonevtx} \else \newsavebox{\fgsimpleonevtx} \savebox{\fgsimpleonevtx}{ \begin{tikzpicture}[x=1ex,y=1ex,baseline={([yshift=-.55ex]current bounding box.center)}] \coordinate (v) ; \def \n {1}; \def \rad {1}; \filldraw[white] (v) circle (\rad); \foreach \s in {1,...,\n} { \def \angle {180+360/\n*(\s - 1)}; \coordinate (u) at ([shift=({\angle}:\rad)]v); \draw (v) -- (u); } \filldraw (v) circle (1pt); \end{tikzpicture} } \fi}_\Gamma}$ to every $\varphi^3$-graph.
The large-$n$ asymptotics of the coefficients $w_n(j) = [\hbar^n] W^{\varphi^3} ( \hbar, j )$ can be obtained by using the chain rule for $\asyOpV{}{}{}$:
\begin{align} \asyOpV{\frac23}{0}{\hbar} W^{\varphi^3} (\hbar, j) &= \hbar \left[e^{\frac{2}{3} \left(\frac{1}{\hbar} - \frac{1}{\widetilde \hbar} \right) } \asyOpV{\frac23}{0}{\widetilde \hbar} \log Z^{\varphi^3}_0 \left( \widetilde \hbar \right)\right]_{\widetilde \hbar =\frac{\hbar}{(1-2j)^{\frac32}}}. \end{align}
Some coefficients of the bivariate generating functions $W^{\varphi^3} (\hbar, j)$ and $\asyOpV{\frac23}{0}{\hbar} W^{\varphi^3} (\hbar, j)$ are given in Tables \ref{tab:Wphi3} and \ref{tab:Wphi3asymp}. Comparing Tables \ref{tab:Zphi3asymp} and \ref{tab:Wphi3asymp}, we can observe the classic result, proven by Wright \cite{wright1970asymptotic}, that the asymptotics of connected and disconnected graphs differ only by a subdominant contribution. 

With the expressions above, we have explicit generating functions for the connected \textit{$n$-point functions} and their all-order asymptotics. 
For instance,
\begin{align*} \left. W^{\varphi^3} \right|_{j=0} &= \hbar \log Z^{\varphi^3}_0(\hbar) & \asyOpV{\frac23}{0}{\hbar} \left. W^{\varphi^3} \right|_{j=0} &= \hbar \asyOpV{\frac23}{0}{\hbar} \log Z^{\varphi^3}_0(\hbar) \\ \left. \frac{\partial W^{\varphi^3} }{\partial j} \right|_{j=0} &= \frac12 \hbar + 3 \hbar^2 \partial_\hbar \log Z^{\varphi^3}_0(\hbar) & \asyOpV{\frac23}{0}{\hbar} \left. \frac{\partial W^{\varphi^3} }{\partial j} \right|_{j=0} &= \left(2 + 3 \hbar^2 \partial_\hbar \right) \asyOpV{\frac23}{0}{\hbar} \log Z^{\varphi^3}_0(\hbar) . \end{align*}
Every $n$-point function is a linear combination of $\log Z^{\varphi^3}(\hbar)$ and its derivatives and the asymptotics are linear combinations of $\asyOpV{\frac23}{0}{\hbar} \log Z^{\varphi^3}(\hbar) = \frac{1}{2\pi}\frac{Z^{\varphi^3}_0(-\hbar)}{Z^{\varphi^3}_0(\hbar)}$ and its derivatives. 

We could derive differential equations, which are fulfilled by $Z^{\varphi^3}_0(\hbar)$, $\log Z^{\varphi^3}(\hbar)$ and $\asyOpV{\frac23}{0}{\hbar} \log Z^{\varphi^3}(\hbar)$ to simplify the expressions above. This would have to be done in a very model specific manner. We will not pursue this path in the scope of this thesis, as we aim for providing machinery which can be used for general models. 

\paragraph{1PI diagrams}

The next object of interest is the effective action,
\begin{align} G^{\varphi^3}(\hbar, \varphi_c) &= W^{\varphi^3}(\hbar, j(\hbar, \varphi_c)) - j(\hbar, \varphi_c) \varphi_c, \end{align}
which is the Legendre transform of $W$ as described in Section \ref{sec:legendre_transformation}, where $j(\hbar, \varphi_c)$ is the solution of  
$\varphi_c = \partial_j W^{\varphi^3} \left( \hbar, j \right)$.
A small calculation reveals what for the special case of $\varphi^3$-theory this can be written explicitly in terms of 
$\varphi_c$. It is convenient to define $\gamma^{\varphi^3}_0(\hbar) := \frac{G^{\varphi^3}(\hbar,0)}{\hbar} = \frac{W^{\varphi^3}(\hbar, j(\hbar, 0))}{\hbar}$. Eq.\ \eqref{eqn:Wphi3_explicit} gives us the more explicit form,
\begin{align*} \gamma^{\varphi^3}_0(\hbar)&= \frac{(1-2 j_0(\hbar))^{\frac32}-1+3j_0(\hbar)}{3\hbar} \\ &+ \frac14 \log \frac{1}{1-2j_0(\hbar) } + \log Z^{\varphi^3}_0 \left(\frac{\hbar}{(1-2j_0(\hbar))^{\frac32}}\right). \end{align*}
where $j_0(\hbar)=j(\hbar,0)$ is the unique power series solution of the equation 
\begin{align*} 0= \frac{\partial W^{\varphi^3}}{\partial j}\left(\hbar, j_0(\hbar)\right). \end{align*}
The bivariate generating function $G^{\varphi^3}(\hbar, \varphi_c)$ is then,
\begin{align} G^{\varphi^3}(\hbar, \varphi_c) &= -\frac{\varphi_c^2}{2} + \frac{\varphi_c^3}{3!} + \frac12 \hbar \log \frac{1}{1-\varphi_c} + \hbar \gamma^{\varphi^3}_0\left(\frac{\hbar}{(1-\varphi_c)^3}\right). \end{align}
The combinatorial interpretation of the identity is the following: A 1PI diagram either has no or only one loop, or it can be reduced to a vacuum diagram by removing all external legs and the attached vertices. 
This bivariate generating function can be depicted diagrammatically as,
\begin{align*} G^{\varphi^3}(\hbar,\varphi_c) &= -\frac{\varphi_c^2}{2} + \phi_\Sact \Big(  \frac16 {  \ifmmode \usebox{\fgsimplethreevtx} \else \newsavebox{\fgsimplethreevtx} \savebox{\fgsimplethreevtx}{ \begin{tikzpicture}[x=1ex,y=1ex,baseline={([yshift=-.5ex]current bounding box.center)}] \coordinate (v) ; \def \n {3}; \def \rad {.8}; \filldraw[white] (v) circle (\rad); \foreach \s in {1,...,5} { \def \angle {180+360/\n*(\s - 1)}; \coordinate (u) at ([shift=({\angle}:\rad)]v); \draw (v) -- (u); } \filldraw (v) circle (1pt); \end{tikzpicture} } \fi } + \frac12 {  \ifmmode \usebox{\fgconetadpolephithree} \else \newsavebox{\fgconetadpolephithree} \savebox{\fgconetadpolephithree}{ \begin{tikzpicture}[x=2ex,y=2ex,baseline={([yshift=-.5ex]current bounding box.center)}] \coordinate (v0) ; \coordinate [right=1 of v0] (v1); \coordinate [left=.5 of v0] (i0); \coordinate [left=.5 of v1] (vm); \draw (vm) circle(.5); \draw (i0) -- (v0); \filldraw (v0) circle(1pt); \end{tikzpicture} } \fi } + \frac14 {  \ifmmode \usebox{\fgtwoconeloopbubblephithree} \else \newsavebox{\fgtwoconeloopbubblephithree} \savebox{\fgtwoconeloopbubblephithree}{ \begin{tikzpicture}[x=2ex,y=2ex,baseline={([yshift=-.5ex]current bounding box.center)}] \coordinate (v0) ; \coordinate [right=1 of v0] (v1); \coordinate [left=.5 of v0] (i0); \coordinate [right=.5 of v1] (o0); \coordinate [left=.5 of v1] (vm); \draw (vm) circle(.5); \draw (i0) -- (v0); \draw (o0) -- (v1); \filldraw (v0) circle(1pt); \filldraw (v1) circle(1pt); \end{tikzpicture} } \fi } + \frac16 {  \ifmmode \usebox{\fgthreeconeltrianglephithree} \else \newsavebox{\fgthreeconeltrianglephithree} \savebox{\fgthreeconeltrianglephithree}{ \begin{tikzpicture}[x=1ex,y=1ex,baseline={([yshift=-.5ex]current bounding box.center)}] \coordinate (v) ; \def \n {3}; \def \rad {1}; \def \rud {2.2}; \foreach \s in {1,...,5} { \def \angle {360/\n*(\s - 1)}; \def \ungle {360/\n*\s}; \coordinate (s) at ([shift=({\angle}:\rad)]v); \coordinate (t) at ([shift=({\ungle}:\rad)]v); \coordinate (u) at ([shift=({\angle}:\rud)]v); \draw (s) -- (u); \filldraw (s) circle (1pt); } \draw (v) circle(\rad); \end{tikzpicture} } \fi } + \frac{1}{12} {  \ifmmode \usebox{\fgbananathree} \else \newsavebox{\fgbananathree} \savebox{\fgbananathree}{ \begin{tikzpicture}[x=1ex,y=1ex,baseline={([yshift=-.5ex]current bounding box.center)}] \coordinate (vm); \coordinate [left=1 of vm] (v0); \coordinate [right=1 of vm] (v1); \draw (v0) -- (v1); \draw (vm) circle(1); \filldraw (v0) circle (1pt); \filldraw (v1) circle (1pt); \end{tikzpicture} } \fi } + \ldots \Big), \end{align*}
where we finally arrived at the required form of $\phi_\Sact:\Gamma \mapsto \hbar^{h_\Gamma} \varphi_c^{|\legs_\Gamma|}$, which additionally assigns a $\varphi_c$ to every leg of an 1PI graph.

Acting with the $\asyOpV{}{}{}$-derivative on $\gamma^{\varphi^3}_0$ gives,
\begin{align*} \asyOpV{\frac23}{0}{\hbar} \gamma^{\varphi^3}_0(\hbar) &= \asyOpV{\frac23}{0}{\hbar} \frac{W^{\varphi^3}(\hbar, j_0(\hbar))}{\hbar} \\ &= \left(\asyOpV{\frac23}{0}{\hbar} \frac{W^{\varphi^3}(\hbar, j)}{\hbar} \right)_{j=j_0(\hbar)} + \frac{\frac{\partial W^{\varphi^3}}{\partial j}(\hbar, j)\big|_{j=j_0(\hbar)}}{\hbar} \asyOpV{\frac23}{0}{\hbar} j_0(\hbar), \intertext{where the second term vanishes by the definition of $j_0$. Therefore,} \asyOpV{\frac23}{0}{\hbar} \gamma^{\varphi^3}_0(\hbar) &= \left[e^{\frac{2}{3} \left(\frac{1}{\hbar} - \frac{1}{\widetilde \hbar} \right) } \asyOpV{\frac23}{0}{\widetilde \hbar} \log Z^{\varphi^3}_0 \left( \widetilde \hbar \right)\right]_{\widetilde \hbar =\frac{\hbar}{(1-2j_0(\hbar))^{\frac32}}}, \end{align*}
and
\begin{align} \begin{split} \asyOpV{\frac23}{0}{\hbar} G^{\varphi^3}(\hbar, \varphi_c) &= \hbar \left[e^{\frac{2}{3} \left(\frac{1}{\hbar} - \frac{1}{\widetilde \hbar} \right) } \asyOpV{\frac23}{0}{\widetilde \hbar} \gamma^{\varphi^3}_0(\widetilde \hbar) \right]_{\widetilde \hbar = \frac{\hbar}{(1-\varphi_c)^3}} \\ &= \hbar \left[e^{\frac{2}{3} \left(\frac{1}{\hbar} - \frac{1}{\widetilde \hbar} \right) } \asyOpV{\frac23}{0}{\widetilde \hbar} \log Z^{\varphi^3}_0 \left( \widetilde \hbar \right)\right]_{\widetilde \hbar =\frac{\hbar}{(1-\varphi_c)^3\left(1-2j_0\left(\frac{\hbar}{(1-\varphi_c)^3}\right)\right)^{\frac32}}}. \end{split} \end{align}

\begin{table}
\begin{subtable}[c]{\textwidth}
\centering
\tiny
\def\arraystretch{1.5}
\begin{tabular}{|c||c|c|c|c|c|c|}
\hline
&$\hbar^{0}$&$\hbar^{1}$&$\hbar^{2}$&$\hbar^{3}$&$\hbar^{4}$&$\hbar^{5}$\\
\hline\hline
$\partial_{\varphi_c}^{0} G^{\varphi^3} \big|_{\varphi_c=0}$&$0$&$0$&$ \frac{1}{12}$&$ \frac{5}{48}$&$ \frac{11}{36}$&$ \frac{539}{384}$\\
\hline
$\partial_{\varphi_c}^{1} G^{\varphi^3} \big|_{\varphi_c=0}$&$0$&$ \frac{1}{2}$&$ \frac{1}{4}$&$ \frac{5}{8}$&$ \frac{11}{4}$&$ \frac{539}{32}$\\
\hline
$\partial_{\varphi_c}^{2} G^{\varphi^3} \big|_{\varphi_c=0}$&$-1$&$ \frac{1}{2}$&$1$&$ \frac{35}{8}$&$ \frac{55}{2}$&$ \frac{7007}{32}$\\
\hline
$\partial_{\varphi_c}^{3} G^{\varphi^3} \big|_{\varphi_c=0}$&$1$&$1$&$5$&$35$&$ \frac{605}{2}$&$ \frac{49049}{16}$\\
\hline
\end{tabular}
\subcaption{Table of the first coefficients of the bivariate generating function $G^{\varphi^3}(\hbar, \varphi_c)$.}
\label{tab:Gphi3}
\end{subtable}
\begin{subtable}[c]{\textwidth}
\centering
\tiny
\def\arraystretch{1.5}
\begin{tabular}{|c||c||c|c|c|c|c|c|}
\hline
&prefactor&$\hbar^{0}$&$\hbar^{1}$&$\hbar^{2}$&$\hbar^{3}$&$\hbar^{4}$&$\hbar^{5}$\\
\hline\hline
$\asyOpV{\frac23}{0}{\hbar} \partial_{\varphi_c}^{0} G^{\varphi^3} \big|_{\varphi_c=0}$&$e^{-1} \frac{\hbar^{1}}{2 \pi}$&$1$&$- \frac{7}{6}$&$- \frac{11}{72}$&$- \frac{10135}{1296}$&$- \frac{536087}{31104}$&$- \frac{296214127}{933120}$\\
\hline
$\asyOpV{\frac23}{0}{\hbar} \partial_{\varphi_c}^{1} G^{\varphi^3} \big|_{\varphi_c=0}$&$e^{-1} \frac{\hbar^{0}}{2 \pi}$&$2$&$- \frac{7}{3}$&$- \frac{137}{36}$&$- \frac{10729}{648}$&$- \frac{1630667}{15552}$&$- \frac{392709787}{466560}$\\
\hline
$\asyOpV{\frac23}{0}{\hbar} \partial_{\varphi_c}^{2} G^{\varphi^3} \big|_{\varphi_c=0}$&$e^{-1} \frac{\hbar^{-1}}{2 \pi}$&$4$&$- \frac{26}{3}$&$- \frac{179}{18}$&$- \frac{15661}{324}$&$- \frac{2531903}{7776}$&$- \frac{637309837}{233280}$\\
\hline
$\asyOpV{\frac23}{0}{\hbar} \partial_{\varphi_c}^{3} G^{\varphi^3} \big|_{\varphi_c=0}$&$e^{-1} \frac{\hbar^{-2}}{2 \pi}$&$8$&$- \frac{100}{3}$&$- \frac{101}{9}$&$- \frac{18883}{162}$&$- \frac{3471563}{3888}$&$- \frac{940175917}{116640}$\\
\hline
\end{tabular}
\subcaption{Table of the first coefficients of the bivariate generating function $\asyOpV{\frac23}{0}{\hbar} G^{\varphi^3}(\hbar, \varphi_c)$.}
\label{tab:Gphi3asymp}
\end{subtable}
\caption{Effective action in $\varphi^3$-theory.}
\end{table}

This can be expanded in $\varphi_c$ to obtain the asymptotics of the 1PI or `proper' $n$-point functions. Some coefficients of the bivariate generating function $G^{\varphi^3}$ and its asymptotics are listed in Tables \ref{tab:Gphi3} and \ref{tab:Gphi3asymp}.

As for the disconnected diagrams, we can also expand $\asyOpV{\frac23}{0}{\hbar} G^{\varphi^3}(\hbar, \varphi_c)$ in $\hbar$ to obtain an asymptotic expansion for general $m$ with $\partial_{\varphi_c}^m G^{\varphi^3}(\hbar, \varphi_c)\big|_{\varphi_c=0} = \sum_{n=0}^\infty g_{m,n} \hbar^n$.
Expanding gives,
\begin{gather*} \asyOpV{\frac23}{0}{\hbar} G^{\varphi^3}(\hbar, \varphi_c) = \hbar \frac{e^{-1+\frac{2 \varphi_c}{\hbar}}}{2\pi} \left( 1 - \frac16 \left( 7+ 3 (2\varphi_c)^2 \right) \hbar \right. \\ \left. -\frac{1}{72} \left( 11 + 126 (2\varphi_c) - 42 (2\varphi_c)^2 - 8 (2\varphi_c)^3 - 9 (2\varphi_c)^4 \right) \hbar^2 + \ldots \right). \intertext{Translated into an asymptotic expansion this becomes, } g_{m,n} \underset{n\rightarrow \infty}{\sim} \frac{e^{-1}}{2\pi} 2^m \left(\frac{2}{3} \right)^{-m-n} \Gamma(n+m-1) \\ \times\left( 1 - \frac19 \frac{ 7 - 3 m + 3 m^2}{n+m-2} -\frac{1}{162} \frac{ 11 + 210 m - 123 m^2 + 48 m^3 - 9 m^4 }{(n+m-3)(n+m-2)} + \ldots \right). \end{gather*}

\paragraph{Renormalization constants and skeleton diagrams}
To perform the \textit{renormalization} as explained in detail in Section \ref{sec:expl_hopfalgebra}, the invariant charge in $\varphi^3$-theory needs to be defined in accordance to eq.\ \eqref{eqn:invariantchargedef},
\begin{align} \alpha(\hbar) := \left( \frac{ \partial_{\varphi_c}^3 G^{\varphi^3} |_{\varphi_c = 0} (\hbar) }{ \left(- \partial_{\varphi_c}^2 G^{\varphi^3} |_{\varphi_c = 0}(\hbar) \right)^{\frac32}} \right)^2. \end{align}
The exponents in the expression above are a consequence of the combinatorial fact, that a 1PI $\varphi^3$-graph has two additional vertices and three additional propagators for each additional loop.
We need to solve 
\begin{align*} \hbar_{\text{ren}} = \hbar(\hbar_\text{ren}) \alpha(\hbar(\hbar_\text{ren})) \end{align*}
for $\hbar(\hbar_{\text{ren}})$. The asymptotics in $\hbar_{\text{ren}}$ can be obtained by using the formula for the compositional inverse of the $\asyOpV{}{}{}$-derivative given in Section \ref{sec:ring2} on this expression:
\begin{align*} \begin{split} (\asyOpV{\frac23}{0}{\hbar_{\text{ren}}} \hbar(\hbar_{\text{ren}}))&=       - \left[ e^{\frac23 \left(\frac{1}{\hbar_{\text{ren}}} - \frac{1}{\hbar}\right)} \frac{ \asyOpV{\frac23}{0}{\hbar} \left(\hbar \alpha(\hbar) \right) } { \partial_\hbar \left( \hbar \alpha(\hbar)\right) }\right]_{\hbar = \hbar(\hbar_\text{ren})}.           \end{split} \end{align*}
The $z$-factors are then obtained as explained in Section \ref{sec:expl_hopfalgebra}. They fulfill the identities,
\begin{align*} -1 &= z^{( \ifmmode \usebox{\fgsimpletwovtx} \else \newsavebox{\fgsimpletwovtx} \savebox{\fgsimpletwovtx}{ \begin{tikzpicture}[x=1ex,y=1ex,baseline={([yshift=-.5ex]current bounding box.center)}] \coordinate (v) ; \def \n {2}; \def \rad {.8}; \filldraw[white] (v) circle (\rad); \foreach \s in {1,...,\n} { \def \angle {360/\n*(\s - 1)}; \coordinate (u) at ([shift=({\angle}:\rad)]v); \draw (v) -- (u); } \filldraw (v) circle (1pt); \end{tikzpicture} } \fi)}(\hbar_\text{ren}) \partial_{\varphi_c}^2 G^{\varphi^3} \big|_{\varphi_c = 0} \left(\hbar(\hbar_\text{ren})\right) \\ 1 &= z^{( \ifmmode \usebox{\fgsimplethreevtx} \else \newsavebox{\fgsimplethreevtx} \savebox{\fgsimplethreevtx}{ \begin{tikzpicture}[x=1ex,y=1ex,baseline={([yshift=-.5ex]current bounding box.center)}] \coordinate (v) ; \def \n {3}; \def \rad {.8}; \filldraw[white] (v) circle (\rad); \foreach \s in {1,...,5} { \def \angle {180+360/\n*(\s - 1)}; \coordinate (u) at ([shift=({\angle}:\rad)]v); \draw (v) -- (u); } \filldraw (v) circle (1pt); \end{tikzpicture} } \fi)}(\hbar_\text{ren}) \partial_{\varphi_c}^3 G^{\varphi^3} \big|_{\varphi_c = 0} \left(\hbar(\hbar_\text{ren})\right) \end{align*}
\begin{table}
\begin{subtable}[c]{\textwidth}
\centering
\tiny
\def\arraystretch{1.5}
\begin{tabular}{|c||c|c|c|c|c|c|}
\hline
&$\hbar_{\text{ren}}^{0}$&$\hbar_{\text{ren}}^{1}$&$\hbar_{\text{ren}}^{2}$&$\hbar_{\text{ren}}^{3}$&$\hbar_{\text{ren}}^{4}$&$\hbar_{\text{ren}}^{5}$\\
\hline\hline
$\hbar(\hbar_{\text{ren}})$&$0$&$1$&$- \frac{7}{2}$&$6$&$- \frac{33}{2}$&$- \frac{345}{16}$\\
\hline
$z^{( \ifmmode \usebox{\fgsimpletwovtx} \else \newsavebox{\fgsimpletwovtx} \savebox{\fgsimpletwovtx}{ \begin{tikzpicture}[x=1ex,y=1ex,baseline={([yshift=-.5ex]current bounding box.center)}] \coordinate (v) ; \def \n {2}; \def \rad {.8}; \filldraw[white] (v) circle (\rad); \foreach \s in {1,...,\n} { \def \angle {360/\n*(\s - 1)}; \coordinate (u) at ([shift=({\angle}:\rad)]v); \draw (v) -- (u); } \filldraw (v) circle (1pt); \end{tikzpicture} } \fi)}(\hbar_{\text{ren}})$&$1$&$ \frac{1}{2}$&$- \frac{1}{2}$&$- \frac{1}{4}$&$-2$&$- \frac{29}{2}$\\
\hline
$z^{( \ifmmode \usebox{\fgsimplethreevtx} \else \newsavebox{\fgsimplethreevtx} \savebox{\fgsimplethreevtx}{ \begin{tikzpicture}[x=1ex,y=1ex,baseline={([yshift=-.5ex]current bounding box.center)}] \coordinate (v) ; \def \n {3}; \def \rad {.8}; \filldraw[white] (v) circle (\rad); \foreach \s in {1,...,5} { \def \angle {180+360/\n*(\s - 1)}; \coordinate (u) at ([shift=({\angle}:\rad)]v); \draw (v) -- (u); } \filldraw (v) circle (1pt); \end{tikzpicture} } \fi)}(\hbar_{\text{ren}})$&$1$&$-1$&$- \frac{1}{2}$&$-4$&$-29$&$- \frac{545}{2}$\\
\hline
\end{tabular}
\subcaption{Table of the first coefficients of the renormalization quantities in $\varphi^3$-theory.}
\label{tab:phi3ren}
\end{subtable}
\begin{subtable}[c]{\textwidth}
\centering
\tiny
\def\arraystretch{1.5}
\begin{tabular}{|c||c||c|c|c|c|c|c|}
\hline
&prefactor&$\hbar_{\text{ren}}^{0}$&$\hbar_{\text{ren}}^{1}$&$\hbar_{\text{ren}}^{2}$&$\hbar_{\text{ren}}^{3}$&$\hbar_{\text{ren}}^{4}$&$\hbar_{\text{ren}}^{5}$\\
\hline\hline
$\left(\asyOpV{\frac23}{0}{\hbar_{\text{ren}}} \hbar \right)(\hbar_{\text{ren}})$&$e^{- \frac{10}{3}} \frac{\hbar^{-1}}{2 \pi}$&$-16$&$ \frac{412}{3}$&$- \frac{3200}{9}$&$ \frac{113894}{81}$&$ \frac{765853}{243}$&$ \frac{948622613}{14580}$\\
\hline
$\left(\asyOpV{\frac23}{0}{\hbar_{\text{ren}}} z^{( \ifmmode \usebox{\fgsimpletwovtx} \else \newsavebox{\fgsimpletwovtx} \savebox{\fgsimpletwovtx}{ \begin{tikzpicture}[x=1ex,y=1ex,baseline={([yshift=-.5ex]current bounding box.center)}] \coordinate (v) ; \def \n {2}; \def \rad {.8}; \filldraw[white] (v) circle (\rad); \foreach \s in {1,...,\n} { \def \angle {360/\n*(\s - 1)}; \coordinate (u) at ([shift=({\angle}:\rad)]v); \draw (v) -- (u); } \filldraw (v) circle (1pt); \end{tikzpicture} } \fi)} \right)(\hbar_{\text{ren}})$&$e^{- \frac{10}{3}} \frac{\hbar^{-1}}{2 \pi}$&$-4$&$ \frac{64}{3}$&$ \frac{76}{9}$&$ \frac{13376}{81}$&$ \frac{397486}{243}$&$ \frac{284898947}{14580}$\\
\hline
$\left(\asyOpV{\frac23}{0}{\hbar_{\text{ren}}} z^{( \ifmmode \usebox{\fgsimplethreevtx} \else \newsavebox{\fgsimplethreevtx} \savebox{\fgsimplethreevtx}{ \begin{tikzpicture}[x=1ex,y=1ex,baseline={([yshift=-.5ex]current bounding box.center)}] \coordinate (v) ; \def \n {3}; \def \rad {.8}; \filldraw[white] (v) circle (\rad); \foreach \s in {1,...,5} { \def \angle {180+360/\n*(\s - 1)}; \coordinate (u) at ([shift=({\angle}:\rad)]v); \draw (v) -- (u); } \filldraw (v) circle (1pt); \end{tikzpicture} } \fi)} \right)(\hbar_{\text{ren}})$&$e^{- \frac{10}{3}} \frac{\hbar^{-2}}{2 \pi}$&$-8$&$ \frac{128}{3}$&$ \frac{152}{9}$&$ \frac{26752}{81}$&$ \frac{794972}{243}$&$ \frac{569918179}{14580}$\\
\hline
\end{tabular}
\subcaption{Table of the first coefficients of the asymptotics of the renormalization quantities in $\varphi^3$-theory.}
\label{tab:phi3renasymp}
\end{subtable}
\caption{Renormalization constants in $\varphi^3$-theory.}
\end{table}%
By an application of the $\asyOpV{}{}{}$-derivative and the product and chain rules from Section \ref{sec:ring2}, the asymptotics of $z^{( \ifmmode \usebox{\fgsimplethreevtx} \else \newsavebox{\fgsimplethreevtx} \savebox{\fgsimplethreevtx}{ \begin{tikzpicture}[x=1ex,y=1ex,baseline={([yshift=-.5ex]current bounding box.center)}] \coordinate (v) ; \def \n {3}; \def \rad {.8}; \filldraw[white] (v) circle (\rad); \foreach \s in {1,...,5} { \def \angle {180+360/\n*(\s - 1)}; \coordinate (u) at ([shift=({\angle}:\rad)]v); \draw (v) -- (u); } \filldraw (v) circle (1pt); \end{tikzpicture} } \fi)}$ are:
\begin{align} \begin{split} \label{eqn:asymptoticsZ} \asyOpV{\frac23}{0}{\hbar_\text{ren}} z^{( \ifmmode \usebox{\fgsimplethreevtx} \else \newsavebox{\fgsimplethreevtx} \savebox{\fgsimplethreevtx}{ \begin{tikzpicture}[x=1ex,y=1ex,baseline={([yshift=-.5ex]current bounding box.center)}] \coordinate (v) ; \def \n {3}; \def \rad {.8}; \filldraw[white] (v) circle (\rad); \foreach \s in {1,...,5} { \def \angle {180+360/\n*(\s - 1)}; \coordinate (u) at ([shift=({\angle}:\rad)]v); \draw (v) -- (u); } \filldraw (v) circle (1pt); \end{tikzpicture} } \fi)}(\hbar_\text{ren}) &= -\left[\left( \partial_{\varphi_c}^3 G^{\varphi^3} \big|_{\varphi_c = 0}(\hbar)\right)^{-2} e^{\frac23 \left(\frac{1}{\hbar_{\text{ren}}} - \frac{1}{\hbar}\right)} \left( \asyOpV{\frac23}{0}{\hbar} \partial_{\varphi_c}^3 G^{\varphi^3} \big|_{\varphi_c = 0}(\hbar)   \vphantom{ \left(\partial_\hbar \partial_{\varphi_c}^3 G^{\varphi^3} \big|_{\varphi_c = 0}(\hbar)\right) \frac{ \asyOpV{\frac23}{0}{\hbar} \left(\hbar \alpha(\hbar) \right) } { \partial_\hbar \left( \hbar \alpha(\hbar)\right) } }   \right. \right. \\ &- \left. \left.   \vphantom{ \left( \partial_{\varphi_c}^3 G^{\varphi^3} \big|_{\varphi_c = 0}(\hbar)\right)^{-2} e^{\frac23 \left(\frac{1}{\hbar_{\text{ren}}} - \frac{1}{\hbar}\right)} \left( \asyOpV{\frac23}{0}{\hbar} \partial_{\varphi_c}^3 G^{\varphi^3} \big|_{\varphi_c = 0}(\hbar) \right. }   \left(\partial_\hbar \partial_{\varphi_c}^3 G^{\varphi^3} \big|_{\varphi_c = 0}(\hbar)\right) \frac{ \asyOpV{\frac23}{0}{\hbar} \left(\hbar \alpha(\hbar) \right) } { \partial_\hbar \left( \hbar \alpha(\hbar)\right) } \right) \right]_{\hbar = \hbar(\hbar_\text{ren})}. \end{split} \end{align}
and for $z^{( \ifmmode \usebox{\fgsimpletwovtx} \else \newsavebox{\fgsimpletwovtx} \savebox{\fgsimpletwovtx}{ \begin{tikzpicture}[x=1ex,y=1ex,baseline={([yshift=-.5ex]current bounding box.center)}] \coordinate (v) ; \def \n {2}; \def \rad {.8}; \filldraw[white] (v) circle (\rad); \foreach \s in {1,...,\n} { \def \angle {360/\n*(\s - 1)}; \coordinate (u) at ([shift=({\angle}:\rad)]v); \draw (v) -- (u); } \filldraw (v) circle (1pt); \end{tikzpicture} } \fi)}$ analogously. Some coefficients of the renormalization constants and their asymptotics are given in Tables \ref{tab:phi3ren} and \ref{tab:phi3renasymp}. 

It was observed by Cvitanović et al.\ \cite{cvitanovic1978number} that $1-z^{( \ifmmode \usebox{\fgsimplethreevtx} \else \newsavebox{\fgsimplethreevtx} \savebox{\fgsimplethreevtx}{ \begin{tikzpicture}[x=1ex,y=1ex,baseline={([yshift=-.5ex]current bounding box.center)}] \coordinate (v) ; \def \n {3}; \def \rad {.8}; \filldraw[white] (v) circle (\rad); \foreach \s in {1,...,5} { \def \angle {180+360/\n*(\s - 1)}; \coordinate (u) at ([shift=({\angle}:\rad)]v); \draw (v) -- (u); } \filldraw (v) circle (1pt); \end{tikzpicture} } \fi)}(\hbar_\text{ren})$ is the generating function of \textit{skeleton} diagrams. Skeleton diagrams are 1PI diagrams without any superficially divergent subgraphs. 
This was proven in Chapter \ref{chap:hopf_algebra_of_fg} using the interpretation of subgraph structures as algebraic lattices. 
Applying Definition \ref{def:asymp}  and Corollary \ref{crll:threevalentprimitivecounting} directly gives a complete asymptotic expansion of the coefficients of $1-z^{( \ifmmode \usebox{\fgsimplethreevtx} \else \newsavebox{\fgsimplethreevtx} \savebox{\fgsimplethreevtx}{ \begin{tikzpicture}[x=1ex,y=1ex,baseline={([yshift=-.5ex]current bounding box.center)}] \coordinate (v) ; \def \n {3}; \def \rad {.8}; \filldraw[white] (v) circle (\rad); \foreach \s in {1,...,5} { \def \angle {180+360/\n*(\s - 1)}; \coordinate (u) at ([shift=({\angle}:\rad)]v); \draw (v) -- (u); } \filldraw (v) circle (1pt); \end{tikzpicture} } \fi)}(\hbar_\text{ren})$,
\begin{gather*} [\hbar_\text{ren}^n]( 1-z^{( \ifmmode \usebox{\fgsimplethreevtx} \else \newsavebox{\fgsimplethreevtx} \savebox{\fgsimplethreevtx}{ \begin{tikzpicture}[x=1ex,y=1ex,baseline={([yshift=-.5ex]current bounding box.center)}] \coordinate (v) ; \def \n {3}; \def \rad {.8}; \filldraw[white] (v) circle (\rad); \foreach \s in {1,...,5} { \def \angle {180+360/\n*(\s - 1)}; \coordinate (u) at ([shift=({\angle}:\rad)]v); \draw (v) -- (u); } \filldraw (v) circle (1pt); \end{tikzpicture} } \fi)}(\hbar_\text{ren})) = \sum_{k=0}^{R-1} c_{k} \left(\frac{2}{3}\right)^{-n-2+k} \Gamma(n+2-k) \\ + \bigO \left( \left(\frac{2}{3}\right)^{-n} \Gamma(n+2-R)\right) \intertext{for all $R \geq 0$, where $c_k = [\hbar_\text{ren}^k] \left(-\hbar_\text{ren}^2 \asyOpV{\frac23}{0}{\hbar_\text{ren}} z^{( \ifmmode \usebox{\fgsimplethreevtx} \else \newsavebox{\fgsimplethreevtx} \savebox{\fgsimplethreevtx}{ \begin{tikzpicture}[x=1ex,y=1ex,baseline={([yshift=-.5ex]current bounding box.center)}] \coordinate (v) ; \def \n {3}; \def \rad {.8}; \filldraw[white] (v) circle (\rad); \foreach \s in {1,...,5} { \def \angle {180+360/\n*(\s - 1)}; \coordinate (u) at ([shift=({\angle}:\rad)]v); \draw (v) -- (u); } \filldraw (v) circle (1pt); \end{tikzpicture} } \fi)}(\hbar_\text{ren}) \right)$. Or more explicit for large $n$,} [\hbar_\text{ren}^n]( 1-z^{( \ifmmode \usebox{\fgsimplethreevtx} \else \newsavebox{\fgsimplethreevtx} \savebox{\fgsimplethreevtx}{ \begin{tikzpicture}[x=1ex,y=1ex,baseline={([yshift=-.5ex]current bounding box.center)}] \coordinate (v) ; \def \n {3}; \def \rad {.8}; \filldraw[white] (v) circle (\rad); \foreach \s in {1,...,5} { \def \angle {180+360/\n*(\s - 1)}; \coordinate (u) at ([shift=({\angle}:\rad)]v); \draw (v) -- (u); } \filldraw (v) circle (1pt); \end{tikzpicture} } \fi)}(\hbar_\text{ren})) \underset{n\rightarrow \infty}{\sim} \frac{e^{-\frac{10}{3}}}{2\pi} \left(\frac{2}{3}\right)^{-n-2} \Gamma(n+2) \left( 8 - \frac{2}{3}\frac{128}{3} \frac{1}{n+1} \right. \\ \left. - \left(\frac{2}{3}\right)^2 \frac{152}{9}\frac{1}{n(n+1)} - \left(\frac{2}{3}\right)^3 \frac{26752}{81}\frac{1}{(n-1)n(n+1)} +\ldots \right). \end{gather*}
The constant coefficient of $\asyOpV{\frac23}{0}{\hbar_\text{ren}} z^{( \ifmmode \usebox{\fgsimplethreevtx} \else \newsavebox{\fgsimplethreevtx} \savebox{\fgsimplethreevtx}{ \begin{tikzpicture}[x=1ex,y=1ex,baseline={([yshift=-.5ex]current bounding box.center)}] \coordinate (v) ; \def \n {3}; \def \rad {.8}; \filldraw[white] (v) circle (\rad); \foreach \s in {1,...,5} { \def \angle {180+360/\n*(\s - 1)}; \coordinate (u) at ([shift=({\angle}:\rad)]v); \draw (v) -- (u); } \filldraw (v) circle (1pt); \end{tikzpicture} } \fi)}(\hbar_\text{ren})$ was also given in \cite{cvitanovic1978number}.

Using the first coefficients of $\asyOpV{\frac23}{0}{\hbar} \partial_{\varphi_c}^3 G^{\varphi^3}\big|_{\varphi_c=0}$ and $-\asyOpV{\frac23}{0}{\hbar_\text{ren}} z^{( \ifmmode \usebox{\fgsimplethreevtx} \else \newsavebox{\fgsimplethreevtx} \savebox{\fgsimplethreevtx}{ \begin{tikzpicture}[x=1ex,y=1ex,baseline={([yshift=-.5ex]current bounding box.center)}] \coordinate (v) ; \def \n {3}; \def \rad {.8}; \filldraw[white] (v) circle (\rad); \foreach \s in {1,...,5} { \def \angle {180+360/\n*(\s - 1)}; \coordinate (u) at ([shift=({\angle}:\rad)]v); \draw (v) -- (u); } \filldraw (v) circle (1pt); \end{tikzpicture} } \fi)}(\hbar_\text{ren})$, we may deduce that the proportion of skeleton diagrams in the set of all proper vertex diagrams is,
\begin{align*} \frac{ \frac{e^{-\frac{10}{3}}}{2\pi} \left(\frac{2}{3}\right)^{-n-2} \Gamma(n+2) \left( 8 - \frac23 \frac{128}{3} \frac{1}{n+1} + \ldots \right)}{ \frac{e^{-1}}{2\pi} \left(\frac{2}{3}\right)^{-n-2} \Gamma(n+2) \left( 8 - \frac23 \frac{100}{3} \frac{1}{n+1} + \ldots \right) }= e^{-\frac{7}{3}}\left(1-\frac{56}{9}\frac{1}{n} + \bigO\left(\frac{1}{n^2}\right)\right). \end{align*}
A random 1PI diagram in $\varphi^3$-theory is therefore a skeleton diagram with probability 
\begin{align*} e^{-\frac{7}{3}}\left(1-\frac{56}{9}\frac{1}{n} + \bigO\left(\frac{1}{n^2}\right)\right), \end{align*}
where $n$ is the loop number.

All results obtained in this section can be translated to the respective asymptotic results on cubic graphs. For instance, $\frac{1}{3!}(1-z^{( \ifmmode \usebox{\fgsimplethreevtx} \else \newsavebox{\fgsimplethreevtx} \savebox{\fgsimplethreevtx}{ \begin{tikzpicture}[x=1ex,y=1ex,baseline={([yshift=-.5ex]current bounding box.center)}] \coordinate (v) ; \def \n {3}; \def \rad {.8}; \filldraw[white] (v) circle (\rad); \foreach \s in {1,...,5} { \def \angle {180+360/\n*(\s - 1)}; \coordinate (u) at ([shift=({\angle}:\rad)]v); \draw (v) -- (u); } \filldraw (v) circle (1pt); \end{tikzpicture} } \fi)}(\hbar_\text{ren}))$ is the generating function of cyclically four-connected graphs with one distinguished vertex. In \cite{wormald1985enumeration}, the first coefficient of the asymptotic expansion of those graphs is given, which agrees with our expansion.
\subsection{\texorpdfstring{$\varphi^4$}{ϕ⁴}-theory}
\label{subsec:varphifourtheoryapplication}

In $\varphi^4$-theory the partition function is given by the formal integral,
\begin{align*} Z^{\varphi^4}(\hbar, j) &:= \int \frac{dx}{\sqrt{2 \pi \hbar}} e^{\frac{1}{\hbar} \left( -\frac{x^2}{2} + \frac{x^4}{4!} + x j \right) } \\ &= 1 + \frac{j^2}{2\hbar} + \frac{j^4}{24 \hbar} + \frac58 j^2 + \frac{1155}{128} j^4 + \frac{1}{8}\hbar + \ldots \end{align*}
In this case, it is not possible to completely absorb the $j$ dependents into the argument of $Z^{\varphi^4}_0$. We only can do so up to fourth order in $j$, which is still sufficient to obtain the generating functions which are necessary to calculate the renormalization constants:
\begin{align*} Z^{\varphi^4}(\hbar, j) &= \int \frac{dx}{\sqrt{2 \pi \hbar}} e^{\frac{1}{\hbar} \left( -\frac{x^2}{2} + \frac{x^4}{4!} + \hbar \log \cosh \frac{x j}{\hbar} \right) } \\ &= \int \frac{dx}{\sqrt{2 \pi \hbar}} e^{\frac{1}{\hbar} \left( -\frac{x^2}{2} + \frac{x^4}{4!} + \hbar \left( \frac{1}{2} \left( \frac{x j}{\hbar} \right)^2 - \frac{1}{12} \left(\frac{x j}{\hbar} \right)^4 + \bigO(j^6) \right) \right) } \\ &= \int \frac{dx}{\sqrt{2 \pi \hbar}} e^{\frac{1}{\hbar} \left( -\left(1 - \frac{j^2}{\hbar}\right) \frac{x^2}{2} + \left(1 - 2 \frac{j^4}{\hbar^3} \right)\frac{x^4}{4!} \right) } + \bigO(j^6) \\ &= \frac{1}{\sqrt{1 - \frac{j^2}{\hbar}}} Z^{\varphi^4}_0 \left(\hbar \frac{1 - 2 \frac{j^4}{\hbar^3}}{\left(1 - \frac{j^2}{\hbar}\right)^2}\right) + \bigO(j^6) \end{align*}
where $Z^{\varphi^4}_0(\hbar) := Z^{\varphi^4}(\hbar, 0) = \Fop\left[-\frac{x^2}{2} + \frac{x^4}{4!} \right](\hbar)$.

\begin{table}
\begin{subtable}[c]{\textwidth}
\centering
\tiny
\def\arraystretch{1.5}
\begin{tabular}{|c||c||c|c|c|c|c|c|}
\hline
&prefactor&$\hbar^{0}$&$\hbar^{1}$&$\hbar^{2}$&$\hbar^{3}$&$\hbar^{4}$&$\hbar^{5}$\\
\hline\hline
$\partial_j^{0} Z^{\varphi^4} \big|_{j=0}$&$\hbar^{0}$&$1$&$ \frac{1}{8}$&$ \frac{35}{384}$&$ \frac{385}{3072}$&$ \frac{25025}{98304}$&$ \frac{1616615}{2359296}$\\
\hline
$\partial_j^{2} Z^{\varphi^4} \big|_{j=0}$&$\hbar^{-1}$&$1$&$ \frac{5}{8}$&$ \frac{105}{128}$&$ \frac{5005}{3072}$&$ \frac{425425}{98304}$&$ \frac{11316305}{786432}$\\
\hline
$\partial_j^{4} Z^{\varphi^4} \big|_{j=0}$&$\hbar^{-2}$&$3$&$ \frac{35}{8}$&$ \frac{1155}{128}$&$ \frac{25025}{1024}$&$ \frac{8083075}{98304}$&$ \frac{260275015}{786432}$\\
\hline
\end{tabular}
\subcaption{The first coefficients of the bivariate generating function $Z^{\varphi^4}(\hbar, j)$.}
\label{tab:Zphi4}
\end{subtable}
\begin{subtable}[c]{\textwidth}
\centering
\tiny
\def\arraystretch{1.5}
\begin{tabular}{|c||c||c|c|c|c|c|c|}
\hline
&prefactor&$\hbar^{0}$&$\hbar^{1}$&$\hbar^{2}$&$\hbar^{3}$&$\hbar^{4}$&$\hbar^{5}$\\
\hline\hline
$\asyOpV{\frac32}{0}{\hbar} \partial_j^{0} Z^{\varphi^4} \big|_{j=0}$&$\frac{\hbar^{0}}{\sqrt{2}\pi}$&$1$&$- \frac{1}{8}$&$ \frac{35}{384}$&$- \frac{385}{3072}$&$ \frac{25025}{98304}$&$- \frac{1616615}{2359296}$\\
\hline
$\asyOpV{\frac32}{0}{\hbar} \partial_j^{2} Z^{\varphi^4} \big|_{j=0}$&$\frac{\hbar^{-2}}{\sqrt{2}\pi}$&$6$&$ \frac{1}{4}$&$- \frac{5}{64}$&$ \frac{35}{512}$&$- \frac{5005}{49152}$&$ \frac{85085}{393216}$\\
\hline
$\asyOpV{\frac32}{0}{\hbar} \partial_j^{4} Z^{\varphi^4} \big|_{j=0}$&$\frac{\hbar^{-4}}{\sqrt{2}\pi}$&$36$&$- \frac{9}{2}$&$ \frac{9}{32}$&$- \frac{35}{256}$&$ \frac{1155}{8192}$&$- \frac{15015}{65536}$\\
\hline
\end{tabular}
\subcaption{The first coefficients of the bivariate generating function $\asyOpV{\frac32}{0}{\hbar}Z^{\varphi^4}(\hbar, j)$.}
\label{tab:Zphi4asymp}
\end{subtable}
\caption{Partition function in $\varphi^4$-theory.}
\end{table}

The asymptotics of $Z^{\varphi^4}_0$ can be calculated directly by using Corollary \ref{crll:comb_int_asymp}: The action $\Sact(x) = -\frac{x^2}{2} + \frac{x^4}{4!}$ is real analytic and all critical points lie on the real axis. The non-trivial critical points of $\Sact(x) =-\frac{x^2}{2} + \frac{x^4}{4!}$ are $\tau_{\pm} = \pm \sqrt{3!}$. The value at the critical points is $\Sact(\tau_{\pm})= - \frac32$. These are the dominant singularities which both contribute. Therefore, $A= \frac{3}{2}$ and $\Sact(\tau_{\pm})-\Sact(x+\tau_{\pm})= -x^2 \pm \frac{x^3}{\sqrt{3!}} + \frac{x^4}{4!}$. 
\begin{align*} \asyOpV{\frac32}{0}{\hbar} Z^{\varphi^4}_0 (\hbar) &= \frac{1}{2 \pi} \left(\Fop\left[-x^2 + \frac{x^3}{\sqrt{3!}} + \frac{x^4}{4!}\right](-\hbar) + \Fop\left[-x^2 - \frac{x^3}{\sqrt{3!}} + \frac{x^4}{4!}\right](-\hbar) \right) \\ &= \frac{1}{\pi} \Fop\left[-x^2 + \frac{x^3}{\sqrt{3!}} + \frac{x^4}{4!}\right](-\hbar) \\ &= \frac{1}{\sqrt{2}\pi} \left( 1 - \frac{1}{8} \hbar + \frac{35}{384} \hbar^{2} - \frac{385}{3072} \hbar^{3} + \frac{25025}{98304} \hbar^{4}  + \ldots\right) \end{align*}
The combinatorial interpretation of this sequence is the following: Diagrams with three or four-valent vertices are weighted with a $\lambda_3=\sqrt{3!}$ for each three-valent vertex, $\lambda_4=1$ for each four-valent vertex, a factor $a=\frac12$ for each edge and a $(-1)$ for every loop in accordance to Proposition \ref{prop:diagraminterpretation}. The whole sequence is preceded by a factor of $\sqrt{a}=\frac{1}{\sqrt{2}}$ as required by the definition of $\Fop$.

The asymptotics for $Z^{\varphi^4}(\hbar, j)$ can again be obtained by utilizing the chain rule for $\asyOpV{}{}{}$:
\begin{align*} \asyOpV{\frac32}{0}{\hbar} Z^{\varphi^4}(\hbar, j) &= \frac{1}{\sqrt{1 - \frac{j^2}{\hbar}}} \left[ e^{ \frac32 \left( \frac{1}{\hbar} - \frac{1}{\widetilde \hbar} \right) } (\asyOpV{\frac32}{0}{\widetilde \hbar} Z^{\varphi^4}_0 )\left(\widetilde \hbar \right) \right]_{\widetilde \hbar = \hbar \frac{1 - 2 \frac{j^4}{\hbar^3}}{(1 - \frac{j^2}{\hbar})^2}} + \bigO(j^6) \end{align*}
The first coefficients of $Z^{\varphi^4}(\hbar, j)$ are given in Table \ref{tab:Zphi4} and the respective asymptotic expansions in Table \ref{tab:Zphi4asymp}.

The generating function of the connected graphs is given by,
\begin{align*} W^{\varphi^4}(\hbar, j) &:= \hbar \log Z^{\varphi^4}(\hbar, j) \\ &= \frac12 \hbar \log \frac{1}{1-\frac{j^2}{\hbar}} + \hbar \log Z^{\varphi^4}_0 \left(\hbar \frac{1 - 2 \frac{j^4}{\hbar^3}}{(1 - \frac{j^2}{\hbar})^2}\right) + \bigO(j^6) \end{align*}
\begin{table}
\begin{subtable}[c]{\textwidth}
\centering
\tiny
\def\arraystretch{1.5}
\begin{tabular}{|c||c|c|c|c|c|c|}
\hline
&$\hbar^{0}$&$\hbar^{1}$&$\hbar^{2}$&$\hbar^{3}$&$\hbar^{4}$&$\hbar^{5}$\\
\hline\hline
$\partial_j^{0} W^{\varphi^4} \big|_{j=0}$&$0$&$0$&$ \frac{1}{8}$&$ \frac{1}{12}$&$ \frac{11}{96}$&$ \frac{17}{72}$\\
\hline
$\partial_j^{2} W^{\varphi^4} \big|_{j=0}$&$1$&$ \frac{1}{2}$&$ \frac{2}{3}$&$ \frac{11}{8}$&$ \frac{34}{9}$&$ \frac{619}{48}$\\
\hline
$\partial_j^{4} W^{\varphi^4} \big|_{j=0}$&$1$&$ \frac{7}{2}$&$ \frac{149}{12}$&$ \frac{197}{4}$&$ \frac{15905}{72}$&$ \frac{107113}{96}$\\
\hline
\end{tabular}
\subcaption{The first coefficients of the bivariate generating function $W^{\varphi^4}(\hbar, j)$.}
\label{tab:Wphi4}
\end{subtable}
\begin{subtable}[c]{\textwidth}
\centering
\tiny
\def\arraystretch{1.5}
\begin{tabular}{|c||c||c|c|c|c|c|c|}
\hline
&prefactor&$\hbar^{0}$&$\hbar^{1}$&$\hbar^{2}$&$\hbar^{3}$&$\hbar^{4}$&$\hbar^{5}$\\
\hline\hline
$\asyOpV{\frac32}{0}{\hbar} \partial_j^{0} W^{\varphi^4} \big|_{j=0}$&$\frac{\hbar^{1}}{\sqrt{2} \pi}$&$1$&$- \frac{1}{4}$&$ \frac{1}{32}$&$- \frac{89}{384}$&$ \frac{353}{6144}$&$- \frac{10623}{8192}$\\
\hline
$\asyOpV{\frac32}{0}{\hbar} \partial_j^{2} W^{\varphi^4} \big|_{j=0}$&$\frac{\hbar^{-1}}{\sqrt{2} \pi}$&$6$&$- \frac{3}{2}$&$- \frac{13}{16}$&$- \frac{73}{64}$&$- \frac{2495}{1024}$&$- \frac{84311}{12288}$\\
\hline
$\asyOpV{\frac32}{0}{\hbar} \partial_j^{4} W^{\varphi^4} \big|_{j=0}$&$\frac{\hbar^{-3}}{\sqrt{2} \pi}$&$36$&$-45$&$- \frac{111}{8}$&$- \frac{687}{32}$&$- \frac{25005}{512}$&$- \frac{293891}{2048}$\\
\hline
\end{tabular}
\subcaption{The first coefficients of the bivariate generating function $\asyOpV{\frac32}{0}{\hbar}W^{\varphi^4}(\hbar, j)$.}
\label{tab:Wphi4asymp}
\end{subtable}
\caption{Free energy in $\varphi^4$-theory.}
\end{table}
and the asymptotics are,
\begin{align*} \asyOpV{\frac32}{0}{\hbar} W^{\varphi^4}(\hbar, j) &= \hbar \left[ e^{ \frac32 \left( \frac{1}{\hbar} - \frac{1}{\widetilde \hbar} \right) } \asyOpV{\frac32}{0}{\hbar} \log Z^{\varphi^4}_0\left(\widetilde \hbar \right) \right]_{\widetilde \hbar = \hbar \frac{1 - 2 \frac{j^4}{\hbar^3}}{(1 - \frac{j^2}{\hbar})^2}} + \bigO(j^6). \end{align*}
The first coefficients of the original generating function and the generating function for the asymptotics are given in Tables \ref{tab:Wphi4} and \ref{tab:Wphi4asymp}.

The effective action, which is the Legendre transform of $W^{\varphi^4}$,
\begin{align*} G^{\varphi^4}(\hbar, \varphi_c) &= W^{\varphi^4}(\hbar, j) - j \varphi_c \end{align*}
where $\varphi_c := \partial_j W^{\varphi^4}$,
is easy to handle in this case, as there are no graphs with exactly one external leg. Derivatives of $G^{\varphi^4}(\hbar, \varphi_c) $ with respect to $\varphi_c$ can be calculated by exploiting that $\varphi_c=0$ implies $j=0$.  
\begin{table}
\begin{subtable}[c]{\textwidth}
\centering
\tiny
\def\arraystretch{1.5}
\begin{tabular}{|c||c|c|c|c|c|c|}
\hline
&$\hbar^{0}$&$\hbar^{1}$&$\hbar^{2}$&$\hbar^{3}$&$\hbar^{4}$&$\hbar^{5}$\\
\hline\hline
$\partial_{\varphi_c}^{0} G^{\varphi^4} \big|_{\varphi_c=0}$&$0$&$0$&$ \frac{1}{8}$&$ \frac{1}{12}$&$ \frac{11}{96}$&$ \frac{17}{72}$\\
\hline
$\partial_{\varphi_c}^{2} G^{\varphi^4} \big|_{\varphi_c=0}$&$-1$&$ \frac{1}{2}$&$ \frac{5}{12}$&$ \frac{5}{6}$&$ \frac{115}{48}$&$ \frac{625}{72}$\\
\hline
$\partial_{\varphi_c}^{4} G^{\varphi^4} \big|_{\varphi_c=0}$&$1$&$ \frac{3}{2}$&$ \frac{21}{4}$&$ \frac{45}{2}$&$ \frac{1775}{16}$&$ \frac{4905}{8}$\\
\hline
\end{tabular}
\subcaption{The first coefficients of the bivariate generating function $G^{\varphi^4}(\hbar, j)$.}
\label{tab:Gphi4}
\end{subtable}
\begin{subtable}[c]{\textwidth}
\centering
\tiny
\def\arraystretch{1.5}
\begin{tabular}{|c||c||c|c|c|c|c|c|}
\hline
&prefactor&$\hbar^{0}$&$\hbar^{1}$&$\hbar^{2}$&$\hbar^{3}$&$\hbar^{4}$&$\hbar^{5}$\\
\hline\hline
$\asyOpV{\frac32}{0}{\hbar} \partial_{\varphi_c}^{0} G^{\varphi^4} \big|_{\varphi_c=0}$&$ \frac{\hbar^{1}}{\sqrt{2} \pi}$&$1$&$- \frac{1}{4}$&$ \frac{1}{32}$&$- \frac{89}{384}$&$ \frac{353}{6144}$&$- \frac{10623}{8192}$\\
\hline
$\asyOpV{\frac32}{0}{\hbar} \partial_{\varphi_c}^{2} G^{\varphi^4} \big|_{\varphi_c=0}$&$ \frac{\hbar^{-1}}{\sqrt{2} \pi}$&$6$&$- \frac{15}{2}$&$- \frac{45}{16}$&$- \frac{445}{64}$&$- \frac{22175}{1024}$&$- \frac{338705}{4096}$\\
\hline
$\asyOpV{\frac32}{0}{\hbar} \partial_{\varphi_c}^{4} G^{\varphi^4} \big|_{\varphi_c=0}$&$ \frac{\hbar^{-3}}{\sqrt{2} \pi}$&$36$&$-117$&$ \frac{369}{8}$&$- \frac{1671}{32}$&$- \frac{103725}{512}$&$- \frac{1890555}{2048}$\\
\hline
\end{tabular}
\subcaption{The first coefficients of the bivariate generating function $\asyOpV{\frac32}{0}{\hbar}G^{\varphi^4}(\hbar, j)$.}
\label{tab:Gphi4asymp}
\end{subtable}
\caption{Effective action in $\varphi^4$-theory.}
\end{table}
For instance,
\begin{align*} \left. G^{\varphi^4} \right|_{\varphi_c=0} &= W^{\varphi^4}(\hbar, 0) \\  \left. \partial_{\varphi_c}^2 G^{\varphi^4} \right|_{\varphi_c=0} &= \left. - \frac{\partial j}{\partial \varphi_c} \right|_{\varphi_c=0} = - \frac{1}{\left. \partial_j^2 W^{\varphi^4} \right|_{j=0}} \\          \left. \partial_{\varphi_c}^4 G^{\varphi^4} \right|_{\varphi_c=0} &= \frac{\left. \partial_j^4 W^{\varphi^4} \right|_{j=0}}{ \left( \left. \partial_j^2 W^{\varphi^4} \right|_{j=0} \right)^4}                     \end{align*}
The calculation of the asymptotic expansions can be performed by applying the $\asyOpV{}{}{}$-derivative on these expressions and using the product and chain rules to write them in terms of the asymptotics of $W^{\varphi^4}$.
Some coefficients of $G^{\varphi^4}(\hbar, j)$ are listed in Table \ref{tab:Gphi4} with the respective asymptotics in Table \ref{tab:Gphi4asymp}.

\begin{table}
\begin{subtable}[c]{\textwidth}
\centering
\tiny
\def\arraystretch{1.5}
\begin{tabular}{|c||c|c|c|c|c|c|}
\hline
&$\hbar_{\text{ren}}^{0}$&$\hbar_{\text{ren}}^{1}$&$\hbar_{\text{ren}}^{2}$&$\hbar_{\text{ren}}^{3}$&$\hbar_{\text{ren}}^{4}$&$\hbar_{\text{ren}}^{5}$\\
\hline\hline
$\hbar(\hbar_{\text{ren}})$&$0$&$1$&$- \frac{5}{2}$&$ \frac{25}{6}$&$- \frac{15}{2}$&$ \frac{25}{3}$\\
\hline
$z^{( \ifmmode \usebox{\fgsimpletwovtx} \else \newsavebox{\fgsimpletwovtx} \savebox{\fgsimpletwovtx}{ \begin{tikzpicture}[x=1ex,y=1ex,baseline={([yshift=-.5ex]current bounding box.center)}] \coordinate (v) ; \def \n {2}; \def \rad {.8}; \filldraw[white] (v) circle (\rad); \foreach \s in {1,...,\n} { \def \angle {360/\n*(\s - 1)}; \coordinate (u) at ([shift=({\angle}:\rad)]v); \draw (v) -- (u); } \filldraw (v) circle (1pt); \end{tikzpicture} } \fi)}(\hbar_{\text{ren}})$&$1$&$ \frac{1}{2}$&$- \frac{7}{12}$&$ \frac{1}{8}$&$- \frac{9}{16}$&$- \frac{157}{96}$\\
\hline
$z^{( \ifmmode \usebox{\fgsimplefourvtx} \else \newsavebox{\fgsimplefourvtx} \savebox{\fgsimplefourvtx}{ \begin{tikzpicture}[x=1ex,y=1ex,baseline={([yshift=-.5ex]current bounding box.center)}] \coordinate (v) ; \def \n {4}; \def \rad {.8}; \filldraw[white] (v) circle (\rad); \foreach \s in {1,...,5} { \def \angle {45+360/\n*(\s - 1)}; \coordinate (u) at ([shift=({\angle}:\rad)]v); \draw (v) -- (u); } \filldraw (v) circle (1pt); \end{tikzpicture} } \fi)}(\hbar_{\text{ren}})$&$1$&$- \frac{3}{2}$&$ \frac{3}{4}$&$- \frac{11}{8}$&$- \frac{45}{16}$&$- \frac{499}{32}$\\
\hline
\end{tabular}
\subcaption{Table of the first coefficients of the renormalization quantities in $\varphi^4$-theory.}
\label{tab:phi4ren}
\end{subtable}
\begin{subtable}[c]{\textwidth}
\centering
\tiny
\def\arraystretch{1.5}
\begin{tabular}{|c||c||c|c|c|c|c|c|}
\hline
&prefactor&$\hbar_{\text{ren}}^{0}$&$\hbar_{\text{ren}}^{1}$&$\hbar_{\text{ren}}^{2}$&$\hbar_{\text{ren}}^{3}$&$\hbar_{\text{ren}}^{4}$&$\hbar_{\text{ren}}^{5}$\\
\hline\hline
$\left(\asyOpV{\frac32}{0}{\hbar_{\text{ren}}} \hbar \right)(\hbar_{\text{ren}})$&$e^{- \frac{15}{4}} \frac{\hbar^{-2}}{\sqrt{2} \pi}$&$-36$&$ \frac{387}{2}$&$- \frac{13785}{32}$&$ \frac{276705}{256}$&$- \frac{8524035}{8192}$&$ \frac{486577005}{65536}$\\
\hline
$\left(\asyOpV{\frac32}{0}{\hbar_{\text{ren}}} z^{( \ifmmode \usebox{\fgsimpletwovtx} \else \newsavebox{\fgsimpletwovtx} \savebox{\fgsimpletwovtx}{ \begin{tikzpicture}[x=1ex,y=1ex,baseline={([yshift=-.5ex]current bounding box.center)}] \coordinate (v) ; \def \n {2}; \def \rad {.8}; \filldraw[white] (v) circle (\rad); \foreach \s in {1,...,\n} { \def \angle {360/\n*(\s - 1)}; \coordinate (u) at ([shift=({\angle}:\rad)]v); \draw (v) -- (u); } \filldraw (v) circle (1pt); \end{tikzpicture} } \fi)} \right)(\hbar_{\text{ren}})$&$e^{- \frac{15}{4}} \frac{\hbar^{-2}}{\sqrt{2} \pi}$&$-18$&$ \frac{219}{4}$&$ \frac{567}{64}$&$ \frac{49113}{512}$&$ \frac{8281053}{16384}$&$ \frac{397802997}{131072}$\\
\hline
$\left(\asyOpV{\frac32}{0}{\hbar_{\text{ren}}} z^{( \ifmmode \usebox{\fgsimplefourvtx} \else \newsavebox{\fgsimplefourvtx} \savebox{\fgsimplefourvtx}{ \begin{tikzpicture}[x=1ex,y=1ex,baseline={([yshift=-.5ex]current bounding box.center)}] \coordinate (v) ; \def \n {4}; \def \rad {.8}; \filldraw[white] (v) circle (\rad); \foreach \s in {1,...,5} { \def \angle {45+360/\n*(\s - 1)}; \coordinate (u) at ([shift=({\angle}:\rad)]v); \draw (v) -- (u); } \filldraw (v) circle (1pt); \end{tikzpicture} } \fi)} \right)(\hbar_{\text{ren}})$&$e^{- \frac{15}{4}} \frac{\hbar^{-3}}{\sqrt{2} \pi}$&$-36$&$ \frac{243}{2}$&$- \frac{729}{32}$&$ \frac{51057}{256}$&$ \frac{7736445}{8192}$&$ \frac{377172477}{65536}$\\
\hline
\end{tabular}
\subcaption{Table of the first coefficients of the asymptotics of the renormalization quantities in $\varphi^4$-theory.}
\label{tab:phi4renasymp}
\end{subtable}
\caption{Renormalization constants in $\varphi^4$-theory.}
\end{table}

Using the procedure established in Section \ref{sec:expl_hopfalgebra}, the renormalization constants can be calculated by defining the invariant charge as 
\begin{align*} \alpha(\hbar):= \left( \frac{ \left( \left. \partial_{\varphi_c}^4 G^{\varphi^4} \right|_{\varphi_c=0}(\hbar) \right)^{\frac12} }{ \left. \partial_{\varphi_c}^2 G^{\varphi^4} \right|_{\varphi_c=0} (\hbar) } \right)^2. \end{align*}
Having defined the invariant charge, the calculation of the renormalization constants is completely equivalent to the calculation for $\varphi^3$-theory. 
The results are given in Table \ref{tab:phi4ren} and \ref{tab:phi4renasymp}.

As already mentioned in the last chapter, Argyres, van Hameren, Kleiss and Papadopoulos remarked that $1-z^{( \ifmmode \usebox{\fgsimplefourvtx} \else \newsavebox{\fgsimplefourvtx} \savebox{\fgsimplefourvtx}{ \begin{tikzpicture}[x=1ex,y=1ex,baseline={([yshift=-.5ex]current bounding box.center)}] \coordinate (v) ; \def \n {4}; \def \rad {.8}; \filldraw[white] (v) circle (\rad); \foreach \s in {1,...,5} { \def \angle {45+360/\n*(\s - 1)}; \coordinate (u) at ([shift=({\angle}:\rad)]v); \draw (v) -- (u); } \filldraw (v) circle (1pt); \end{tikzpicture} } \fi)}(\hbar_\text{ren})$ does not count the number of skeleton diagrams in $\varphi^4$-theory as might be expected from analogy to $\varphi^3$-theory. The fact that this cannot by the case can be seen from the second term of $z^{( \ifmmode \usebox{\fgsimplefourvtx} \else \newsavebox{\fgsimplefourvtx} \savebox{\fgsimplefourvtx}{ \begin{tikzpicture}[x=1ex,y=1ex,baseline={([yshift=-.5ex]current bounding box.center)}] \coordinate (v) ; \def \n {4}; \def \rad {.8}; \filldraw[white] (v) circle (\rad); \foreach \s in {1,...,5} { \def \angle {45+360/\n*(\s - 1)}; \coordinate (u) at ([shift=({\angle}:\rad)]v); \draw (v) -- (u); } \filldraw (v) circle (1pt); \end{tikzpicture} } \fi)}(\hbar_\text{ren})$ which is positive (see Table \ref{tab:phi4ren}), destroying a counting function interpretation of $1-z^{( \ifmmode \usebox{\fgsimplefourvtx} \else \newsavebox{\fgsimplefourvtx} \savebox{\fgsimplefourvtx}{ \begin{tikzpicture}[x=1ex,y=1ex,baseline={([yshift=-.5ex]current bounding box.center)}] \coordinate (v) ; \def \n {4}; \def \rad {.8}; \filldraw[white] (v) circle (\rad); \foreach \s in {1,...,5} { \def \angle {45+360/\n*(\s - 1)}; \coordinate (u) at ([shift=({\angle}:\rad)]v); \draw (v) -- (u); } \filldraw (v) circle (1pt); \end{tikzpicture} } \fi)}(\hbar_\text{ren})$. In Example \ref{expl:phi4_lattices} it was shown that additionally to skeleton diagrams, also chains of one loop diagrams, $\begin{tikzpicture}[x=2ex,y=2ex,baseline={([yshift=-.5ex]current bounding box.center)}] \coordinate (v0); \coordinate [right=.5 of v0] (vm1); \coordinate [right=.5 of vm1] (v1); \node [right=.5 of v1] (v2) {$\ldots$}; \coordinate [right=.5 of v1] (v1m); \coordinate [right=.01 of v2] (v2m); \coordinate [right=.5 of v2m] (v3); \coordinate [right=.5 of v3] (vm2); \coordinate [right=.5 of vm2] (v4); \coordinate [above left=.5 of v0] (i0); \coordinate [below left=.5 of v0] (i1); \coordinate [above right=.5 of v4] (o0); \coordinate [below right=.5 of v4] (o1); \draw (vm1) circle(.5); \draw (vm2) circle(.5); \draw (i0) -- (v0); \draw (i1) -- (v0); \draw (o0) -- (v4); \draw (o1) -- (v4); \draw ([shift=(90:.5)]v1m) arc (90:270:.5); \draw ([shift=(-90:.5)]v2m) arc (-90:90:.5); \filldraw (v0) circle(1pt); \filldraw (v1) circle(1pt); \filldraw (v3) circle(1pt); \filldraw (v4) circle(1pt); \end{tikzpicture}$ contribute to the generating function $z^{( \ifmmode \usebox{\fgsimplefourvtx} \else \newsavebox{\fgsimplefourvtx} \savebox{\fgsimplefourvtx}{ \begin{tikzpicture}[x=1ex,y=1ex,baseline={([yshift=-.5ex]current bounding box.center)}] \coordinate (v) ; \def \n {4}; \def \rad {.8}; \filldraw[white] (v) circle (\rad); \foreach \s in {1,...,5} { \def \angle {45+360/\n*(\s - 1)}; \coordinate (u) at ([shift=({\angle}:\rad)]v); \draw (v) -- (u); } \filldraw (v) circle (1pt); \end{tikzpicture} } \fi)}(\hbar_\text{ren})$. The chains of one loop bubbles contribute with alternating sign.

Using the expression from Example \ref{expl:phi4_lattices}, the generating function of skeleton diagrams in $\varphi^4$ theory is given by,
\begin{align} 1 - z^{( \ifmmode \usebox{\fgsimplefourvtx} \else \newsavebox{\fgsimplefourvtx} \savebox{\fgsimplefourvtx}{ \begin{tikzpicture}[x=1ex,y=1ex,baseline={([yshift=-.5ex]current bounding box.center)}] \coordinate (v) ; \def \n {4}; \def \rad {.8}; \filldraw[white] (v) circle (\rad); \foreach \s in {1,...,5} { \def \angle {45+360/\n*(\s - 1)}; \coordinate (u) at ([shift=({\angle}:\rad)]v); \draw (v) -- (u); } \filldraw (v) circle (1pt); \end{tikzpicture} } \fi)}(\hbar_\text{ren}) + 3 \sum \limits_{n\geq 2} (-1)^n \left( \frac{\hbar_\text{ren} }{2} \right)^n, \end{align}
where we needed to include a factor of $4!$ to convert from Example \ref{expl:phi4_lattices} to the present notation of leg-fixed diagrams.
The first coefficients are,
\begin{gather*} 0, \frac{3}{2},0,1,3, \frac{31}{2}, \frac{529}{6}, \frac{2277}{4}, \frac{16281}{4}, \frac{254633}{8}, \frac{2157349}{8}, \frac{39327755}{16}, \frac{383531565}{16}, \ldots \end{gather*}
The asymptotic expansion of this sequence agrees with the one of $( 1-z^{( \ifmmode \usebox{\fgsimplefourvtx} \else \newsavebox{\fgsimplefourvtx} \savebox{\fgsimplefourvtx}{ \begin{tikzpicture}[x=1ex,y=1ex,baseline={([yshift=-.5ex]current bounding box.center)}] \coordinate (v) ; \def \n {4}; \def \rad {.8}; \filldraw[white] (v) circle (\rad); \foreach \s in {1,...,5} { \def \angle {45+360/\n*(\s - 1)}; \coordinate (u) at ([shift=({\angle}:\rad)]v); \draw (v) -- (u); } \filldraw (v) circle (1pt); \end{tikzpicture} } \fi)}(\hbar_\text{ren}))$,
\begin{align*} [\hbar_\text{ren}^n]( 1-z^{( \ifmmode \usebox{\fgsimplefourvtx} \else \newsavebox{\fgsimplefourvtx} \savebox{\fgsimplefourvtx}{ \begin{tikzpicture}[x=1ex,y=1ex,baseline={([yshift=-.5ex]current bounding box.center)}] \coordinate (v) ; \def \n {4}; \def \rad {.8}; \filldraw[white] (v) circle (\rad); \foreach \s in {1,...,5} { \def \angle {45+360/\n*(\s - 1)}; \coordinate (u) at ([shift=({\angle}:\rad)]v); \draw (v) -- (u); } \filldraw (v) circle (1pt); \end{tikzpicture} } \fi)}(\hbar_\text{ren})) \underset{n\rightarrow \infty}{\sim} \frac{e^{-\frac{15}{4}}}{\sqrt{2}\pi} \left(\frac{2}{3}\right)^{n+3} \Gamma(n+3) \left( 36 - \frac{3}{2}\frac{243}{2} \frac{1}{n+2} \right. \\ \left. + \left(\frac{3}{2}\right)^2 \frac{729}{32}\frac{1}{(n+1)(n+2)} - \left(\frac{3}{2}\right)^3 \frac{51057}{256}\frac{1}{n(n+1)(n+2)} +\ldots \right). \end{align*}
More coefficients are given in Table \ref{tab:phi4renasymp}.
\section{QED-type examples}
We will discuss more general theories with two types of `particles', which are of QED-type in the sense that we can interpret one particle as boson (in our case a photon $ \ifmmode \usebox{\fgsimplephotonprop} \else \newsavebox{\fgsimplephotonprop} \savebox{\fgsimplephotonprop}{ \begin{tikzpicture}[x=1ex,y=1ex,baseline={([yshift=-.5ex]current bounding box.center)}] \coordinate (v) ; \coordinate [right=1.2 of v] (u); \draw[photon] (v) -- (u); \end{tikzpicture} } \fi$ or a meson $ \ifmmode \usebox{\fgsimplemesonprop} \else \newsavebox{\fgsimplemesonprop} \savebox{\fgsimplemesonprop}{ \begin{tikzpicture}[x=1ex,y=1ex,baseline={([yshift=-.5ex]current bounding box.center)}] \coordinate (v) ; \coordinate [right=1.2 of v] (u); \draw[meson] (v) -- (u); \end{tikzpicture} } \fi$) and the other as fermion ($ \ifmmode \usebox{\fgsimplefermionprop} \else \newsavebox{\fgsimplefermionprop} \savebox{\fgsimplefermionprop}{ \begin{tikzpicture}[x=1ex,y=1ex,baseline={([yshift=-.5ex]current bounding box.center)}] \coordinate (v) ; \coordinate [right=1.2 of v] (u); \draw[fermion] (v) -- (u); \end{tikzpicture} } \fi$) with a fermion-fermion-boson vertex (either a fermion-fermion-photon $ \ifmmode \usebox{\fgsimpleqedvtx} \else \newsavebox{\fgsimpleqedvtx} \savebox{\fgsimpleqedvtx}{ \begin{tikzpicture}[x=1ex,y=1ex,baseline={([yshift=-.5ex]current bounding box.center)}] \coordinate (v) ; \def \rad {1}; \filldraw[white] (v) circle (\rad); \coordinate (u1) at ([shift=(180:\rad)]v); \coordinate (u2) at ([shift=(300:\rad)]v); \coordinate (u3) at ([shift=(60:\rad)]v); \draw[photon] (u1) -- (v); \draw[fermion] (u2) -- (v); \draw[fermion] (v) -- (u3); \filldraw (v) circle (1pt); \end{tikzpicture} } \fi$ or a fermion-fermion-meson vertex $ \ifmmode \usebox{\fgsimpleyukvtx} \else \newsavebox{\fgsimpleyukvtx} \savebox{\fgsimpleyukvtx}{ \begin{tikzpicture}[x=1ex,y=1ex,baseline={([yshift=-.5ex]current bounding box.center)}] \coordinate (v) ; \def \rad {1}; \filldraw[white] (v) circle (\rad); \coordinate (u1) at ([shift=(180:\rad)]v); \coordinate (u2) at ([shift=(300:\rad)]v); \coordinate (u3) at ([shift=(60:\rad)]v); \draw[meson] (u1) -- (v); \draw[fermion] (u2) -- (v); \draw[fermion] (v) -- (u3); \filldraw (v) circle (1pt); \end{tikzpicture} } \fi$). 

Consider the partition function
\begin{align*} Z(\hbar, j, \eta) &:= \int_\R \frac{dx}{\sqrt{2 \pi \hbar}} \int_\C\frac{dz d\bar z}{\pi \hbar} e^{\frac{1}{\hbar} \left( - \frac{x^2}{2} - |z|^2 + x |z|^2 + j x + \eta \bar z + \bar \eta z \right)}. \end{align*}
The Gaussian integration over $z$ and $\bar z$ can be performed immediately,
\begin{align} \begin{split} \label{eqn:qedprototype} Z(\hbar, j, \eta) &= \int_\R \frac{dx}{\sqrt{2 \pi \hbar}} \frac{dz d\bar z}{\pi \hbar} e^{\frac{1}{\hbar} \left( - \frac{x^2}{2} - (1-x)\left|z- \frac{\eta}{1-x}\right|^2 + j x + \frac{|\eta|^2}{1-x} \right)} \\ &= \int_\R \frac{dx}{\sqrt{2 \pi \hbar}} \frac{1}{1-x} e^{\frac{1}{\hbar} \left( - \frac{x^2}{2} + j x + \frac{|\eta|^2}{1-x} \right)} \\ &= \int_\R \frac{dx}{\sqrt{2 \pi \hbar}} e^{\frac{1}{\hbar} \left( - \frac{x^2}{2} + j x + \frac{|\eta|^2}{1-x} + \hbar \log \frac{1}{1-x} \right)} \end{split} \end{align}
Note that the transformation above has not been justified rigorously in the scope of formal integrals, but here it is sufficient to consider the last line in eq.\ \eqref{eqn:qedprototype} as input for our mathematical machinery and the previous as a physical motivation.
The combinatorial interpretation of this expression is the following: $\frac{|\eta|^2}{1-x}$ generates a fermion propagator line and $\hbar \log \frac{1}{1-x}$ generates a fermion loop, both with an arbitrary number of boson lines attached. 
The interpretation of the $j x$ and $-\frac{x^2}{2}$ terms are standard.

We will consider the following variations of this partition function:
\begin{labeling}{(Quenched QED)}
\item[(QED)]
In quantum electrodynamics (QED) all fermion loops have an even number of fermion edges, as Furry's theorem guarantees that diagrams with odd fermion loops vanish. The modification,
\begin{align*} \hbar \log \frac{1}{1-x} \rightarrow \hbar \frac12 \left( \log \frac{1}{1-x} + \log \frac{1}{1+x} \right) = \frac12\hbar \log \frac{1}{1-x^2}, \end{align*}
results in the required partition function \cite{cvitanovic1978number, itzykson2005quantum}.
\item[(Quenched QED)]
In the quenched approximation of QED, fermion loops are neglected altogether. This corresponds to the modification $\hbar \log \frac{1}{1-x} \rightarrow 0$.
\item[(Yukawa)]
We will also consider the integral without modification. Also odd fermion loops are allowed in this case. This can be seen as the zero-dimensional version of Yukawa theory. The bosons in Yukawa theory are usually mesons ($ \ifmmode \usebox{\fgsimplemesonprop} \else \newsavebox{\fgsimplemesonprop} \savebox{\fgsimplemesonprop}{ \begin{tikzpicture}[x=1ex,y=1ex,baseline={([yshift=-.5ex]current bounding box.center)}] \coordinate (v) ; \coordinate [right=1.2 of v] (u); \draw[meson] (v) -- (u); \end{tikzpicture} } \fi$) and not photons and we have a fermion-fermion-meson vertex ($ \ifmmode \usebox{\fgsimpleyukvtx} \else \newsavebox{\fgsimpleyukvtx} \savebox{\fgsimpleyukvtx}{ \begin{tikzpicture}[x=1ex,y=1ex,baseline={([yshift=-.5ex]current bounding box.center)}] \coordinate (v) ; \def \rad {1}; \filldraw[white] (v) circle (\rad); \coordinate (u1) at ([shift=(180:\rad)]v); \coordinate (u2) at ([shift=(300:\rad)]v); \coordinate (u3) at ([shift=(60:\rad)]v); \draw[meson] (u1) -- (v); \draw[fermion] (u2) -- (v); \draw[fermion] (v) -- (u3); \filldraw (v) circle (1pt); \end{tikzpicture} } \fi$). Mesons are depicted as dashed lines  as the example in Figure \ref{fig:graph_representations}.
\end{labeling}
\subsection{QED}
\label{sec:QED}
In QED the partition function in eq.\ \eqref{eqn:qedprototype} must be modified to
\begin{align*} Z^\text{QED}(\hbar, j, \eta):= \int_\R \frac{dx}{\sqrt{2 \pi \hbar}} e^{\frac{1}{\hbar} \left( - \frac{x^2}{2} + j x + \frac{|\eta|^2}{1-x} + \frac12 \hbar \log \frac{1}{1-x^2} \right)}. \end{align*}
As in $\varphi^4$-theory, we hide the dependence on the sources inside a composition:
\begin{align*} Z^\text{QED}(\hbar, j, \eta)&:= \left( 1 + \frac{j^2}{2\hbar} + \frac{|\eta|^2}{\hbar} \right)  Z_0^\text{QED} \left( \frac{\hbar \left( 1 + \frac{2 j |\eta|^2}{\hbar^2} \right)}{\left(1 - \frac{2 |\eta|^2}{\hbar}\right)\left( 1-\frac{j^2}{\hbar} \right)} \right) \\ &+ \bigO(j^4) + \bigO(j^2 |\eta|^2) + \bigO(|\eta|^4), \intertext{where} Z^\text{QED}_0(\hbar) &:= Z^\text{QED}(\hbar, 0, 0) = \int_\R \frac{dx}{\sqrt{2 \pi \hbar}} e^{\frac{1}{\hbar} \left( - \frac{x^2}{2} + \frac12 \hbar \log \frac{1}{1-x^2} \right)}. \end{align*}
Recall that this expression is meant to be expanded under the integral sign. Because $e^{\frac12 \log \frac{1}{1-x^2}} = \frac{1}{\sqrt{1-x^2}}$, we conclude, using the rules of Gaussian integration that
\begin{align*} Z^\text{QED}_0(\hbar)= \sum_{n=0}^\infty \hbar^{n} (2n-1)!! [x^{2n}] \frac{1}{\sqrt{1-x^2}}. \end{align*}
In Example \ref{expl:sinegordoncurve} it was shown using Proposition \ref{prop:formalchangeofvar} that this may be written as,
\begin{align*} Z^\text{QED}_0(\hbar)= \Fop\left[ -\frac{\sin^2(x)}{2}\right](\hbar). \end{align*}
The partition function of zero-dimensional QED without sources is therefore equal to the partition function of the zero-dimensional sine-Gordon model. 

Using Corollary \ref{crll:comb_int_asymp}, it is straightforward to calculate the all-order asymptotics. 
The saddle points of $-\frac{\sin^2(x)}{2}$ all lie on the real axis. The dominant saddles are at $\tau_\pm = \pm \frac{\pi}{2}$. We find that $A = -\frac{\sin^2(\tau_\pm)}{2} = \frac12$ and $\Sact(\tau_\pm) - \Sact(\tau_\pm+x) = -\frac{\sin^2(x)}{2}$. Therefore, $Z^\text{QED}_0 \in \fring{\hbar}{\frac12}{0}$ and
\begin{align*} \asyOpV{\frac12}{0}{\hbar} Z^\text{QED}_0(\hbar)= \asyOpV{\frac12}{0}{\hbar} \Fop\left[ -\frac{\sin^2(x)}{2}\right](\hbar) = \frac{2}{2\pi} \Fop\left[ -\frac{\sin^2(x)}{2}\right](-\hbar). \end{align*}
The calculation of the asymptotics of $Z^\text{QED}(\hbar, j, \eta)$ as well as setting up the free energy $W^\text{QED}(\hbar,j,\eta)$ and calculating its asymptotics are analogous to the preceding examples. The respective first coefficients are listed in Tables \ref{tab:Zqed} and \ref{tab:Wqed}.

\begin{table}
\begin{subtable}[c]{\textwidth}
\centering
\tiny
\def\arraystretch{1.5}
\begin{tabular}{|c||c||c|c|c|c|c|c|}
\hline
&prefactor&$\hbar^{0}$&$\hbar^{1}$&$\hbar^{2}$&$\hbar^{3}$&$\hbar^{4}$&$\hbar^{5}$\\
\hline\hline
$\partial_j^{0} (\partial_{\eta} \partial_{ \bar \eta} )^{0} Z^{\text{QED}} \big|_{\substack{j=0\\\eta=0}}$&$\hbar^{0}$&$1$&$ \frac{1}{2}$&$ \frac{9}{8}$&$ \frac{75}{16}$&$ \frac{3675}{128}$&$ \frac{59535}{256}$\\
\hline
$\partial_j^{2} (\partial_{\eta} \partial_{ \bar \eta} )^{0} Z^{\text{QED}} \big|_{\substack{j=0\\\eta=0}}$&$\hbar^{-1}$&$1$&$ \frac{3}{2}$&$ \frac{45}{8}$&$ \frac{525}{16}$&$ \frac{33075}{128}$&$ \frac{654885}{256}$\\
\hline
$\partial_j^{0} (\partial_{\eta} \partial_{ \bar \eta} )^{1} Z^{\text{QED}} \big|_{\substack{j=0\\\eta=0}}$&$\hbar^{-1}$&$1$&$ \frac{3}{2}$&$ \frac{45}{8}$&$ \frac{525}{16}$&$ \frac{33075}{128}$&$ \frac{654885}{256}$\\
\hline
$\partial_j^{1} (\partial_{\eta} \partial_{ \bar \eta} )^{1} Z^{\text{QED}} \big|_{\substack{j=0\\\eta=0}}$&$\hbar^{-1}$&$1$&$ \frac{9}{2}$&$ \frac{225}{8}$&$ \frac{3675}{16}$&$ \frac{297675}{128}$&$ \frac{7203735}{256}$\\
\hline
\end{tabular}
\subcaption{The first coefficients of the trivariate generating function $Z^{\text{QED}}(\hbar, j, \eta)$.}
\end{subtable}
\begin{subtable}[c]{\textwidth}
\centering
\tiny
\def\arraystretch{1.5}
\begin{tabular}{|c||c||c|c|c|c|c|c|}
\hline
&prefactor&$\hbar^{0}$&$\hbar^{1}$&$\hbar^{2}$&$\hbar^{3}$&$\hbar^{4}$&$\hbar^{5}$\\
\hline\hline
$\asyOpV{\frac12}{0}{\hbar} \partial_j^{0} (\partial_{\eta} \partial_{ \bar \eta} )^{0} Z^{\text{QED}} \big|_{\substack{j=0\\\eta=0}}$&$\frac{\hbar^{0}}{\pi}$&$1$&$- \frac{1}{2}$&$ \frac{9}{8}$&$- \frac{75}{16}$&$ \frac{3675}{128}$&$- \frac{59535}{256}$\\
\hline
$\asyOpV{\frac12}{0}{\hbar} \partial_j^{2} (\partial_{\eta} \partial_{ \bar \eta} )^{0} Z^{\text{QED}} \big|_{\substack{j=0\\\eta=0}}$&$\frac{\hbar^{-2}}{\pi}$&$1$&$ \frac{1}{2}$&$- \frac{3}{8}$&$ \frac{15}{16}$&$- \frac{525}{128}$&$ \frac{6615}{256}$\\
\hline
$\asyOpV{\frac12}{0}{\hbar} \partial_j^{0} (\partial_{\eta} \partial_{ \bar \eta} )^{1} Z^{\text{QED}} \big|_{\substack{j=0\\\eta=0}}$&$\frac{\hbar^{-2}}{\pi}$&$1$&$ \frac{1}{2}$&$- \frac{3}{8}$&$ \frac{15}{16}$&$- \frac{525}{128}$&$ \frac{6615}{256}$\\
\hline
$\asyOpV{\frac12}{0}{\hbar} \partial_j^{0} (\partial_{\eta} \partial_{ \bar \eta} )^{2} Z^{\text{QED}} \big|_{\substack{j=0\\\eta=0}}$&$\frac{\hbar^{-3}}{\pi}$&$1$&$- \frac{1}{2}$&$ \frac{1}{8}$&$- \frac{3}{16}$&$ \frac{75}{128}$&$- \frac{735}{256}$\\
\hline
\end{tabular}
\subcaption{The first coefficients of the trivariate generating function $\asyOpV{\frac12}{0}{\hbar}Z^{\text{QED}}(\hbar, j, \eta)$.}
\end{subtable}
\caption{Partition function in QED.}
\label{tab:Zqed}
\end{table}

\begin{table}
\begin{subtable}[c]{\textwidth}
\centering
\tiny
\def\arraystretch{1.5}
\begin{tabular}{|c||c|c|c|c|c|c|}
\hline
&$\hbar^{0}$&$\hbar^{1}$&$\hbar^{2}$&$\hbar^{3}$&$\hbar^{4}$&$\hbar^{5}$\\
\hline\hline
$\partial_j^{0} (\partial_{\eta} \partial_{ \bar \eta} )^{0} W^{\text{QED}} \big|_{\substack{j=0\\\eta=0}}$&$0$&$0$&$ \frac{1}{2}$&$1$&$ \frac{25}{6}$&$26$\\
\hline
$\partial_j^{2} (\partial_{\eta} \partial_{ \bar \eta} )^{0} W^{\text{QED}} \big|_{\substack{j=0\\\eta=0}}$&$1$&$1$&$4$&$25$&$208$&$2146$\\
\hline
$\partial_j^{0} (\partial_{\eta} \partial_{ \bar \eta} )^{1} W^{\text{QED}} \big|_{\substack{j=0\\\eta=0}}$&$1$&$1$&$4$&$25$&$208$&$2146$\\
\hline
$\partial_j^{1} (\partial_{\eta} \partial_{ \bar \eta} )^{1} W^{\text{QED}} \big|_{\substack{j=0\\\eta=0}}$&$1$&$4$&$25$&$208$&$2146$&$26368$\\
\hline
\end{tabular}
\subcaption{The first coefficients of the trivariate generating function $W^{\text{QED}}(\hbar, j, \eta)$.}
\end{subtable}
\begin{subtable}[c]{\textwidth}
\centering
\tiny
\def\arraystretch{1.5}
\begin{tabular}{|c||c||c|c|c|c|c|c|}
\hline
&prefactor&$\hbar^{0}$&$\hbar^{1}$&$\hbar^{2}$&$\hbar^{3}$&$\hbar^{4}$&$\hbar^{5}$\\
\hline\hline
$\asyOpV{\frac12}{0}{\hbar} \partial_j^{0} (\partial_{\eta} \partial_{ \bar \eta} )^{0} W^{\text{QED}} \big|_{\substack{j=0\\\eta=0}}$&$\frac{\hbar^{1}}{\pi}$&$1$&$-1$&$ \frac{1}{2}$&$- \frac{17}{2}$&$ \frac{67}{8}$&$- \frac{3467}{8}$\\
\hline
$\asyOpV{\frac12}{0}{\hbar} \partial_j^{2} (\partial_{\eta} \partial_{ \bar \eta} )^{0} W^{\text{QED}} \big|_{\substack{j=0\\\eta=0}}$&$\frac{\hbar^{-1}}{\pi}$&$1$&$-1$&$- \frac{3}{2}$&$- \frac{13}{2}$&$- \frac{341}{8}$&$- \frac{2931}{8}$\\
\hline
$\asyOpV{\frac12}{0}{\hbar} \partial_j^{0} (\partial_{\eta} \partial_{ \bar \eta} )^{1} W^{\text{QED}} \big|_{\substack{j=0\\\eta=0}}$&$\frac{\hbar^{-1}}{\pi}$&$1$&$-1$&$- \frac{3}{2}$&$- \frac{13}{2}$&$- \frac{341}{8}$&$- \frac{2931}{8}$\\
\hline
$\asyOpV{\frac12}{0}{\hbar} \partial_j^{1} (\partial_{\eta} \partial_{ \bar \eta} )^{1} W^{\text{QED}} \big|_{\substack{j=0\\\eta=0}}$&$\frac{\hbar^{-2}}{\pi}$&$1$&$-1$&$- \frac{3}{2}$&$- \frac{13}{2}$&$- \frac{341}{8}$&$- \frac{2931}{8}$\\
\hline
\end{tabular}
\subcaption{The first coefficients of the trivariate generating function $\asyOpV{\frac12}{0}{\hbar}W^{\text{QED}}(\hbar, j, \eta)$.}
\end{subtable}
\caption{Free energy in QED.}
\label{tab:Wqed}
\end{table}

The effective action is given by the two variable Legendre transformation of $W^{\text{QED}}$:
\begin{align*} G^{\text{QED}}(\hbar, \phi_c, \psi_c) &= W^{\text{QED}}(\hbar, j, \eta) - j \phi_c - \bar \eta \psi_c - \eta \bar \psi_c, \end{align*}
where $\phi_c = \partial_j W^{\text{QED}}$ and $\psi_c = \partial_{\bar\eta} W^{\text{QED}}$. 
The variable $\phi_c$ counts the number of photon legs $ \ifmmode \usebox{\fgsimplephotonprop} \else \newsavebox{\fgsimplephotonprop} \savebox{\fgsimplephotonprop}{ \begin{tikzpicture}[x=1ex,y=1ex,baseline={([yshift=-.5ex]current bounding box.center)}] \coordinate (v) ; \coordinate [right=1.2 of v] (u); \draw[photon] (v) -- (u); \end{tikzpicture} } \fi$ and the variables $\psi_c$ and $\bar \psi_c$ the numbers of in- and out-going fermion legs $ \ifmmode \usebox{\fgsimplefermionprop} \else \newsavebox{\fgsimplefermionprop} \savebox{\fgsimplefermionprop}{ \begin{tikzpicture}[x=1ex,y=1ex,baseline={([yshift=-.5ex]current bounding box.center)}] \coordinate (v) ; \coordinate [right=1.2 of v] (u); \draw[fermion] (v) -- (u); \end{tikzpicture} } \fi$ of the 1PI graphs.

Because there are no graphs with only one leg in QED, it follows that,
\begin{gather*} G^\text{QED} \big|_{\substack{\phi_c=0\\\psi_c=0}} = W^\text{QED}\big|_{\substack{j=0\\\eta=0}} \\ \partial_{\psi_c} \partial_{\bar \psi_c} G^\text{QED} \big|_{\substack{\phi_c=0\\\psi_c=0}} = - \frac{1}{\partial_\eta \partial_{\bar \eta} W^\text{QED}\big|_{\substack{j=0\\\eta=0}} } \\ \partial_{\phi_c}^2 G^\text{QED} \big|_{\substack{\phi_c=0\\\psi_c=0}} = - \frac{1}{\partial_j^2 W^\text{QED}\big|_{\substack{j=0\\\eta=0}} } \\ \partial_{\phi_c}\partial_{\psi_c} \partial_{\bar \psi_c} G^\text{QED} \big|_{\substack{\phi_c=0\\\psi_c=0}} = \frac{ \partial_j \partial_\eta \partial_{\bar \eta} W^\text{QED}\big|_{\substack{j=0\\\eta=0}} }{ \partial_j^2 W^\text{QED}\big|_{\substack{j=0\\\eta=0}} \left( \partial_\eta \partial_{\bar \eta} W^\text{QED}\big|_{\substack{j=0\\\eta=0}} \right)^2 }. \end{gather*}
The calculation of asymptotics is similar to the one for $\varphi^4$-theory. Coefficients for the effective action are listed in Table \ref{tab:GQED}.

\begin{table}
\begin{subtable}[c]{\textwidth}
\centering
\tiny
\def\arraystretch{1.5}
\begin{tabular}{|c||c|c|c|c|c|c|}
\hline
&$\hbar^{0}$&$\hbar^{1}$&$\hbar^{2}$&$\hbar^{3}$&$\hbar^{4}$&$\hbar^{5}$\\
\hline\hline
$\partial_{\phi_c}^{0} (\partial_{\psi_c} \partial_{ \bar \psi_c} )^{0} G^{\text{QED}} \big|_{\substack{\phi_c=0\\\psi_c=0}}$&$0$&$0$&$ \frac{1}{2}$&$1$&$ \frac{25}{6}$&$26$\\
\hline
$\partial_{\phi_c}^{2} (\partial_{\psi_c} \partial_{ \bar \psi_c} )^{0} G^{\text{QED}} \big|_{\substack{\phi_c=0\\\psi_c=0}}$&$-1$&$1$&$3$&$18$&$153$&$1638$\\
\hline
$\partial_{\phi_c}^{0} (\partial_{\psi_c} \partial_{ \bar \psi_c} )^{1} G^{\text{QED}} \big|_{\substack{\phi_c=0\\\psi_c=0}}$&$-1$&$1$&$3$&$18$&$153$&$1638$\\
\hline
$\partial_{\phi_c}^{1} (\partial_{\psi_c} \partial_{ \bar \psi_c} )^{1} G^{\text{QED}} \big|_{\substack{\phi_c=0\\\psi_c=0}}$&$1$&$1$&$7$&$72$&$891$&$12672$\\
\hline
\end{tabular}
\subcaption{The first coefficients of the trivariate generating function $G^{\text{QED}}(\hbar, \phi_c, \psi_c)$.}
\end{subtable}
\begin{subtable}[c]{\textwidth}
\centering
\tiny
\def\arraystretch{1.5}
\begin{tabular}{|c||c||c|c|c|c|c|c|}
\hline
&prefactor&$\hbar^{0}$&$\hbar^{1}$&$\hbar^{2}$&$\hbar^{3}$&$\hbar^{4}$&$\hbar^{5}$\\
\hline\hline
$\asyOpV{\frac12}{0}{\hbar} \partial_{\phi_c}^{0} (\partial_{\psi_c} \partial_{ \bar \psi_c} )^{0} G^{\text{QED}} \big|_{\substack{\phi_c=0\\\psi_c=0}}$&$\frac{\hbar^{1}}{\pi}$&$1$&$-1$&$ \frac{1}{2}$&$- \frac{17}{2}$&$ \frac{67}{8}$&$- \frac{3467}{8}$\\
\hline
$\asyOpV{\frac12}{0}{\hbar} \partial_{\phi_c}^{2} (\partial_{\psi_c} \partial_{ \bar \psi_c} )^{0} G^{\text{QED}} \big|_{\substack{\phi_c=0\\\psi_c=0}}$&$\frac{\hbar^{-1}}{\pi}$&$1$&$-3$&$- \frac{9}{2}$&$- \frac{57}{2}$&$- \frac{2025}{8}$&$- \frac{22437}{8}$\\
\hline
$\asyOpV{\frac12}{0}{\hbar} \partial_{\phi_c}^{0} (\partial_{\psi_c} \partial_{ \bar \psi_c} )^{1} G^{\text{QED}} \big|_{\substack{\phi_c=0\\\psi_c=0}}$&$\frac{\hbar^{-1}}{\pi}$&$1$&$-3$&$- \frac{9}{2}$&$- \frac{57}{2}$&$- \frac{2025}{8}$&$- \frac{22437}{8}$\\
\hline
$\asyOpV{\frac12}{0}{\hbar} \partial_{\phi_c}^{1} (\partial_{\psi_c} \partial_{ \bar \psi_c} )^{1} G^{\text{QED}} \big|_{\substack{\phi_c=0\\\psi_c=0}}$&$\frac{\hbar^{-2}}{\pi}$&$1$&$-7$&$- \frac{3}{2}$&$- \frac{75}{2}$&$- \frac{3309}{8}$&$- \frac{41373}{8}$\\
\hline
\end{tabular}
\subcaption{The first coefficients of the trivariate generating function $\asyOpV{\frac12}{0}{\hbar}G^{\text{QED}}(\hbar, \phi_c, \psi_c)$.}
\end{subtable}
\caption{Effective action in QED.}
\label{tab:GQED}
\end{table}

To calculate the renormalization constants we define the invariant charge\footnote{Note, that the $\partial_{\phi_c}^2$ corresponds to the photon propagator $ \ifmmode \usebox{\fgsimplephotonprop} \else \newsavebox{\fgsimplephotonprop} \savebox{\fgsimplephotonprop}{ \begin{tikzpicture}[x=1ex,y=1ex,baseline={([yshift=-.5ex]current bounding box.center)}] \coordinate (v) ; \coordinate [right=1.2 of v] (u); \draw[photon] (v) -- (u); \end{tikzpicture} } \fi$, the $\partial_{\psi_c} \partial_{\bar \psi_c}$ to the fermion propagator $ \ifmmode \usebox{\fgsimplefermionprop} \else \newsavebox{\fgsimplefermionprop} \savebox{\fgsimplefermionprop}{ \begin{tikzpicture}[x=1ex,y=1ex,baseline={([yshift=-.5ex]current bounding box.center)}] \coordinate (v) ; \coordinate [right=1.2 of v] (u); \draw[fermion] (v) -- (u); \end{tikzpicture} } \fi$ and the $\partial_{\phi_c}\partial_{\psi_c} \partial_{\bar \psi_c}$ to the fermion-fermion-photon vertex $ \ifmmode \usebox{\fgsimpleqedvtx} \else \newsavebox{\fgsimpleqedvtx} \savebox{\fgsimpleqedvtx}{ \begin{tikzpicture}[x=1ex,y=1ex,baseline={([yshift=-.5ex]current bounding box.center)}] \coordinate (v) ; \def \rad {1}; \filldraw[white] (v) circle (\rad); \coordinate (u1) at ([shift=(180:\rad)]v); \coordinate (u2) at ([shift=(300:\rad)]v); \coordinate (u3) at ([shift=(60:\rad)]v); \draw[photon] (u1) -- (v); \draw[fermion] (u2) -- (v); \draw[fermion] (v) -- (u3); \filldraw (v) circle (1pt); \end{tikzpicture} } \fi$.} as,
\begin{align*} \alpha(\hbar):= \left( \frac{ \partial_{\phi_c}\partial_{\psi_c} \partial_{\bar \psi_c} G^\text{QED} \big|_{\substack{\phi_c=0\\\psi_c=0}} } { \left( - \partial_{\phi_c}^2 G^\text{QED} \big|_{\substack{\phi_c=0\\\psi_c=0}} \right)^\frac12 \left( - \partial_{\psi_c} \partial_{\bar \psi_c} G^\text{QED}\big|_{\substack{\phi_c=0\\\psi_c=0}} \right) } \right)^2. \end{align*}
The first coefficients of the renormalization constants and their asymptotics are listed in Table \ref{tab:QEDren}.

\begin{table}
\begin{subtable}[c]{\textwidth}
\centering
\tiny
\def\arraystretch{1.5}
\begin{tabular}{|c||c|c|c|c|c|c|}
\hline
&$\hbar_{\text{ren}}^{0}$&$\hbar_{\text{ren}}^{1}$&$\hbar_{\text{ren}}^{2}$&$\hbar_{\text{ren}}^{3}$&$\hbar_{\text{ren}}^{4}$&$\hbar_{\text{ren}}^{5}$\\
\hline\hline
$\hbar(\hbar_{\text{ren}})$&$0$&$1$&$-5$&$14$&$-58$&$20$\\
\hline
$z^{\left( \ifmmode \usebox{\fgsimplephotonprop} \else \newsavebox{\fgsimplephotonprop} \savebox{\fgsimplephotonprop}{ \begin{tikzpicture}[x=1ex,y=1ex,baseline={([yshift=-.5ex]current bounding box.center)}] \coordinate (v) ; \coordinate [right=1.2 of v] (u); \draw[photon] (v) -- (u); \end{tikzpicture} } \fi\right)}(\hbar_{\text{ren}})$&$1$&$1$&$-1$&$-1$&$-13$&$-93$\\
\hline
$z^{\left( \ifmmode \usebox{\fgsimplefermionprop} \else \newsavebox{\fgsimplefermionprop} \savebox{\fgsimplefermionprop}{ \begin{tikzpicture}[x=1ex,y=1ex,baseline={([yshift=-.5ex]current bounding box.center)}] \coordinate (v) ; \coordinate [right=1.2 of v] (u); \draw[fermion] (v) -- (u); \end{tikzpicture} } \fi\right)}(\hbar_{\text{ren}})$&$1$&$1$&$-1$&$-1$&$-13$&$-93$\\
\hline
$z^{\left( \ifmmode \usebox{\fgsimpleqedvtx} \else \newsavebox{\fgsimpleqedvtx} \savebox{\fgsimpleqedvtx}{ \begin{tikzpicture}[x=1ex,y=1ex,baseline={([yshift=-.5ex]current bounding box.center)}] \coordinate (v) ; \def \rad {1}; \filldraw[white] (v) circle (\rad); \coordinate (u1) at ([shift=(180:\rad)]v); \coordinate (u2) at ([shift=(300:\rad)]v); \coordinate (u3) at ([shift=(60:\rad)]v); \draw[photon] (u1) -- (v); \draw[fermion] (u2) -- (v); \draw[fermion] (v) -- (u3); \filldraw (v) circle (1pt); \end{tikzpicture} } \fi\right)}(\hbar_{\text{ren}})$&$1$&$-1$&$-1$&$-13$&$-93$&$-1245$\\
\hline
\end{tabular}
\subcaption{Table of the first coefficients of the renormalization quantities in QED.}
\end{subtable}
\begin{subtable}[c]{\textwidth}
\centering
\tiny
\def\arraystretch{1.5}
\begin{tabular}{|c||c||c|c|c|c|c|c|}
\hline
&prefactor&$\hbar_{\text{ren}}^{0}$&$\hbar_{\text{ren}}^{1}$&$\hbar_{\text{ren}}^{2}$&$\hbar_{\text{ren}}^{3}$&$\hbar_{\text{ren}}^{4}$&$\hbar_{\text{ren}}^{5}$\\
\hline\hline
$\left(\asyOpV{\frac12}{0}{\hbar_{\text{ren}}} \hbar \right)(\hbar_{\text{ren}})$&$e^{- \frac{5}{2}} \frac{\hbar^{-1}}{\pi}$&$-2$&$24$&$- \frac{379}{4}$&$ \frac{6271}{12}$&$ \frac{38441}{64}$&$ \frac{17647589}{480}$\\
\hline
$\left(\asyOpV{\frac12}{0}{\hbar_{\text{ren}}} z^{\left( \ifmmode \usebox{\fgsimplephotonprop} \else \newsavebox{\fgsimplephotonprop} \savebox{\fgsimplephotonprop}{ \begin{tikzpicture}[x=1ex,y=1ex,baseline={([yshift=-.5ex]current bounding box.center)}] \coordinate (v) ; \coordinate [right=1.2 of v] (u); \draw[photon] (v) -- (u); \end{tikzpicture} } \fi\right)} \right)(\hbar_{\text{ren}})$&$e^{- \frac{5}{2}} \frac{\hbar^{-1}}{\pi}$&$-1$&$ \frac{13}{2}$&$ \frac{67}{8}$&$ \frac{5177}{48}$&$ \frac{513703}{384}$&$ \frac{83864101}{3840}$\\
\hline
$\left(\asyOpV{\frac12}{0}{\hbar_{\text{ren}}} z^{\left( \ifmmode \usebox{\fgsimplefermionprop} \else \newsavebox{\fgsimplefermionprop} \savebox{\fgsimplefermionprop}{ \begin{tikzpicture}[x=1ex,y=1ex,baseline={([yshift=-.5ex]current bounding box.center)}] \coordinate (v) ; \coordinate [right=1.2 of v] (u); \draw[fermion] (v) -- (u); \end{tikzpicture} } \fi\right)} \right)(\hbar_{\text{ren}})$&$e^{- \frac{5}{2}} \frac{\hbar^{-1}}{\pi}$&$-1$&$ \frac{13}{2}$&$ \frac{67}{8}$&$ \frac{5177}{48}$&$ \frac{513703}{384}$&$ \frac{83864101}{3840}$\\
\hline
$\left(\asyOpV{\frac12}{0}{\hbar_{\text{ren}}} z^{\left( \ifmmode \usebox{\fgsimpleqedvtx} \else \newsavebox{\fgsimpleqedvtx} \savebox{\fgsimpleqedvtx}{ \begin{tikzpicture}[x=1ex,y=1ex,baseline={([yshift=-.5ex]current bounding box.center)}] \coordinate (v) ; \def \rad {1}; \filldraw[white] (v) circle (\rad); \coordinate (u1) at ([shift=(180:\rad)]v); \coordinate (u2) at ([shift=(300:\rad)]v); \coordinate (u3) at ([shift=(60:\rad)]v); \draw[photon] (u1) -- (v); \draw[fermion] (u2) -- (v); \draw[fermion] (v) -- (u3); \filldraw (v) circle (1pt); \end{tikzpicture} } \fi\right)} \right)(\hbar_{\text{ren}})$&$e^{- \frac{5}{2}} \frac{\hbar^{-2}}{\pi}$&$-1$&$ \frac{13}{2}$&$ \frac{67}{8}$&$ \frac{5177}{48}$&$ \frac{513703}{384}$&$ \frac{83864101}{3840}$\\
\hline
\end{tabular}
\subcaption{Table of the first coefficients of the asymptotics of the renormalization quantities in QED.}
\end{subtable}
\caption{Renormalization constants in QED.}
\label{tab:QEDren}
\end{table}

As in the example of $\varphi^3$-theory, the $z$-factor for the vertex, $z^{\left( \ifmmode \usebox{\fgsimpleqedvtx} \else \newsavebox{\fgsimpleqedvtx} \savebox{\fgsimpleqedvtx}{ \begin{tikzpicture}[x=1ex,y=1ex,baseline={([yshift=-.5ex]current bounding box.center)}] \coordinate (v) ; \def \rad {1}; \filldraw[white] (v) circle (\rad); \coordinate (u1) at ([shift=(180:\rad)]v); \coordinate (u2) at ([shift=(300:\rad)]v); \coordinate (u3) at ([shift=(60:\rad)]v); \draw[photon] (u1) -- (v); \draw[fermion] (u2) -- (v); \draw[fermion] (v) -- (u3); \filldraw (v) circle (1pt); \end{tikzpicture} } \fi\right)}$ can be used to enumerate the number of skeleton diagrams, due to Theorem \ref{thm:threeQFTmoebvert}. Asymptotically, this number is given by,
\begin{gather*} [\hbar_\text{ren}^n]( 1-z^{\left( \ifmmode \usebox{\fgsimpleqedvtx} \else \newsavebox{\fgsimpleqedvtx} \savebox{\fgsimpleqedvtx}{ \begin{tikzpicture}[x=1ex,y=1ex,baseline={([yshift=-.5ex]current bounding box.center)}] \coordinate (v) ; \def \rad {1}; \filldraw[white] (v) circle (\rad); \coordinate (u1) at ([shift=(180:\rad)]v); \coordinate (u2) at ([shift=(300:\rad)]v); \coordinate (u3) at ([shift=(60:\rad)]v); \draw[photon] (u1) -- (v); \draw[fermion] (u2) -- (v); \draw[fermion] (v) -- (u3); \filldraw (v) circle (1pt); \end{tikzpicture} } \fi\right)}(\hbar_\text{ren})) \underset{n\rightarrow \infty}{\sim} \frac{e^{-\frac{5}{2}}}{\pi} \left(\frac{1}{2}\right)^{-n-2} \Gamma(n+2) \left( 1 - \frac{1}{2}\frac{13}{2} \frac{1}{n+1} \right. \\ \left. - \left(\frac{1}{2}\right)^2 \frac{67}{8}\frac{1}{n(n+1)} - \left(\frac{1}{2}\right)^3 \frac{5177}{48}\frac{1}{(n-1)n(n+1)} +\ldots \right), \end{gather*}
which can be read off Table \ref{tab:QEDren}. The first two coefficients of this expansion were also given in \cite{cvitanovic1978number} in a different notation.
\subsection{Quenched QED}
For the quenched approximation, we need to remove the $\log$-term in the partition function given in eq.\ \eqref{eqn:qedprototype}:
\begin{align*} Z^\text{QQED}(\hbar, j, \eta) := \int_\R \frac{dx}{\sqrt{2 \pi \hbar}} e^{\frac{1}{\hbar} \left( - \frac{x^2}{2} + j x + \frac{|\eta|^2}{1-x} \right)} \end{align*}
The partition function cannot be reduced to a generating function of diagrams without sources as the only diagram without sources is the empty diagram. 

To obtain the first order in $|\eta|^2$, the partition function can be rewritten as,
\begin{gather*} Z^\text{QQED}(\hbar, j, \eta) = \\ e^{\frac{j^2}{2\hbar}} \left( 1+ \frac{|\eta|^2}{\hbar(1-j)} \int_\R \frac{dx}{\sqrt{2 \pi \left( \frac{\hbar}{(1-j)^2} \right)}} \frac{1}{1-x} e^{-\frac{x^2}{2\left( \frac{\hbar}{(1-j)^2} \right)} } + \bigO(|\eta|^4) \right). \end{gather*}
The formal integral in this expression can be easily expanded:
\begin{align*} \int_\R \frac{dx}{\sqrt{2 \pi \hbar}} \frac{1}{1-x} e^{-\frac{x^2}{2\hbar} } &= \sum_{n=0}^\infty \hbar^n(2n-1)!! =: \chi(\hbar) \end{align*}
This is in fact the expression, we encountered in Example \ref{expl:counterexpl}, whose asymptotics cannot be calculated by Corollary \ref{crll:comb_int_asymp} or Theorem \ref{thm:comb_int_asymp}. But extracting the asymptotics `by hand' is trivial. Because $(2n-1)!! = \frac{2^{n+\frac12}}{\sqrt{2 \pi}}\Gamma\left(n+\frac12\right)$, we can write,
\begin{align*} \asyOpV{\frac12}{}{\hbar} \chi(\hbar) = \frac{1}{\sqrt{2 \pi \hbar}}, \end{align*}
in the language of the ring of factorially divergent power series. It follows that,
\begin{align*} Z^\text{QQED}(\hbar, j, \eta) &= e^{\frac{j^2}{2\hbar}} \left( 1+ \frac{|\eta|^2}{\hbar(1-j)} \chi\left(\frac{\hbar}{(1-j)^2}\right) + \bigO(|\eta|^4) \right) \\ \asyOpV{\frac12}{}{\hbar} Z^\text{QQED}(\hbar, j, \eta) &= \frac{|\eta|^2 e^{\frac{j^2}{2\hbar}}}{\hbar (1-j)} \left( \asyOpV{\frac12}{}{\hbar} \chi\left( \frac{\hbar}{(1-j)^2} \right) \right) (\hbar) + \bigO(|\eta|^4) \intertext{and by the chain rule for $\asyOpV{}{}{}$,} \asyOpV{\frac12}{}{\hbar} Z^\text{QQED}(\hbar, j, \eta) &= \frac{|\eta|^2 e^{\frac{j^2}{2\hbar}}}{\hbar (1-j)} \left[ e^{\frac12 \left( \frac{1}{\hbar} - \frac{1}{\widetilde \hbar} \right)} \asyOpV{\frac12}{}{\widetilde \hbar} \chi\left( \widetilde \hbar\right) \right]_{\widetilde \hbar = \frac{\hbar}{(1-j)^2}} + \bigO(|\eta|^4) \\ &= \frac{|\eta|^2 e^{\frac{j^2}{2\hbar}}}{\hbar (1-j)} \left[ e^{\frac12 \left( \frac{1}{\hbar} - \frac{1}{\widetilde \hbar} \right)} \frac{1}{\sqrt{2 \pi \widetilde \hbar}} \right]_{\widetilde \hbar = \frac{\hbar}{(1-j)^2}} + \bigO(|\eta|^4) \\ &= \frac{|\eta|^2 e^{\frac{j}{\hbar}}}{\sqrt{2\pi} \hbar^{\frac32}} + \bigO(|\eta|^4) \end{align*}
Obtaining the free energy, which is essentially equivalent to the partition function, is straightforward,
\begin{align*} W^\text{QQED}(\hbar, j, \eta) &= \hbar \log Z^\text{QQED}(\hbar, j, \eta) \\ &= \frac{j^2}{2}+ \frac{|\eta|^2}{1-j} \chi\left(\frac{\hbar}{(1-j)^2}\right) + \bigO(|\eta|^4) \\ \asyOpV{\frac12}{}{\hbar} W^\text{QQED}(\hbar, j, \eta) &= \frac{|\eta|^2e^{\frac{j-\frac{j^2}{2}}{\hbar}}}{\sqrt{2\pi \hbar}} + \bigO(|\eta|^4). \end{align*}
\begin{table}
\begin{subtable}[c]{\textwidth}
\centering
\tiny
\def\arraystretch{1.5}
\begin{tabular}{|c||c|c|c|c|c|c|}
\hline
&$\hbar^{0}$&$\hbar^{1}$&$\hbar^{2}$&$\hbar^{3}$&$\hbar^{4}$&$\hbar^{5}$\\
\hline\hline
$\partial_j^{0} (\partial_{\eta} \partial_{ \bar \eta} )^{1} W^{\text{QQED}} \big|_{\substack{j=0\\\eta=0}}$&$1$&$1$&$3$&$15$&$105$&$945$\\
\hline
$\partial_j^{1} (\partial_{\eta} \partial_{ \bar \eta} )^{1} W^{\text{QQED}} \big|_{\substack{j=0\\\eta=0}}$&$1$&$3$&$15$&$105$&$945$&$10395$\\
\hline
\end{tabular}
\subcaption{The first coefficients of the trivariate generating function $W^{\text{QQED}}(\hbar, j, \eta)$.}
\end{subtable}
\begin{subtable}[c]{\textwidth}
\centering
\tiny
\def\arraystretch{1.5}
\begin{tabular}{|c||c||c|c|c|c|c|c|}
\hline
&prefactor&$\hbar^{0}$&$\hbar^{1}$&$\hbar^{2}$&$\hbar^{3}$&$\hbar^{4}$&$\hbar^{5}$\\
\hline\hline
$\asyOpV{\frac12}{0}{\hbar} \partial_j^{0} (\partial_{\eta} \partial_{ \bar \eta} )^{1} W^{\text{QQED}} \big|_{\substack{j=0\\\eta=0}}$&$\frac{\hbar^{0}}{\sqrt{2 \pi \hbar}}$&$1$&$0$&$0$&$0$&$0$&$0$\\
\hline
$\asyOpV{\frac12}{0}{\hbar} \partial_j^{1} (\partial_{\eta} \partial_{ \bar \eta} )^{1} W^{\text{QQED}} \big|_{\substack{j=0\\\eta=0}}$&$\frac{\hbar^{-1}}{\sqrt{2 \pi \hbar}}$&$1$&$0$&$0$&$0$&$0$&$0$\\
\hline
\end{tabular}
\subcaption{The first coefficients of the trivariate generating function $\asyOpV{\frac12}{0}{\hbar}W^{\text{QQED}}(\hbar, j, \eta)$.}
\end{subtable}
\caption{Free energy in quenched QED.}
\label{tab:WQQED}
\end{table}%
The effective action obtained by the Legendre transformation of $W^{\text{QQED}}$ can also be expressed explicitly:
\begin{align*} G^\text{QQED}(\hbar, \phi_c, \psi_c) &= -\frac{\phi_c^2}{2}+ |\psi_c|^2\frac{(\phi_c-1)}{\chi\left(\frac{\hbar}{(1-\phi_c)^2}\right)} + \bigO(|\psi_c|^4) \\ \asyOpV{\frac12}{}{\hbar} G^\text{QQED}(\hbar, \phi_c, \psi_c) &= |\psi_c|^2\frac{e^{\frac{\phi_c - \frac{\phi_c^2}{2}}{\hbar}}}{\sqrt{2\pi \hbar}} \frac{(1-\phi_c)^2}{\chi\left(\frac{\hbar}{(1-\phi_c)^2}\right)^2} + \bigO(|\psi_c|^4). \end{align*}
The first coefficients of the free energy and effective action are listed in the Tables \ref{tab:WQQED} and \ref{tab:GQQED} together with the respective asymptotics.

\begin{table}
\begin{subtable}[c]{\textwidth}
\centering
\tiny
\def\arraystretch{1.5}
\begin{tabular}{|c||c|c|c|c|c|c|}
\hline
&$\hbar^{0}$&$\hbar^{1}$&$\hbar^{2}$&$\hbar^{3}$&$\hbar^{4}$&$\hbar^{5}$\\
\hline\hline
$\partial_{\phi_c}^{0} (\partial_{\psi_c} \partial_{ \bar \psi_c} )^{1} G^{\text{QQED}} \big|_{\substack{\phi_c=0\\\psi_c=0}}$&$-1$&$1$&$2$&$10$&$74$&$706$\\
\hline
$\partial_{\phi_c}^{1} (\partial_{\psi_c} \partial_{ \bar \psi_c} )^{1} G^{\text{QQED}} \big|_{\substack{\phi_c=0\\\psi_c=0}}$&$1$&$1$&$6$&$50$&$518$&$6354$\\
\hline
\end{tabular}
\subcaption{The first coefficients of the trivariate generating function $\Gamma^{\text{QQED}}(\hbar, \phi_c, \psi_c)$.}
\end{subtable}
\begin{subtable}[c]{\textwidth}
\centering
\tiny
\def\arraystretch{1.5}
\begin{tabular}{|c||c||c|c|c|c|c|c|}
\hline
&prefactor&$\hbar^{0}$&$\hbar^{1}$&$\hbar^{2}$&$\hbar^{3}$&$\hbar^{4}$&$\hbar^{5}$\\
\hline\hline
$\asyOpV{\frac12}{0}{\hbar} \partial_{\phi_c}^{0} (\partial_{\psi_c} \partial_{ \bar \psi_c} )^{1} G^{\text{QQED}} \big|_{\substack{\phi_c=0\\\psi_c=0}}$&$\frac{\hbar^{0}}{\sqrt{2\pi\hbar}}$&$1$&$-2$&$-3$&$-16$&$-124$&$-1224$\\
\hline
$\asyOpV{\frac12}{0}{\hbar} \partial_{\phi_c}^{1} (\partial_{\psi_c} \partial_{ \bar \psi_c} )^{1} G^{\text{QQED}} \big|_{\substack{\phi_c=0\\\psi_c=0}}$&$\frac{\hbar^{-1}}{\sqrt{2\pi\hbar}}$&$1$&$-4$&$-3$&$-22$&$-188$&$-1968$\\
\hline
\end{tabular}
\subcaption{The first coefficients of the trivariate generating function $\asyOpV{\frac12}{0}{\hbar}\Gamma^{\text{QQED}}(\hbar, \phi_c, \psi_c)$.}
\end{subtable}
\caption{Effective action in quenched QED.}
\label{tab:GQQED}
\end{table}

The invariant charge is defined as
\begin{align*} \alpha(\hbar):= \left( \frac{ \partial_{\phi_c}\partial_{\psi_c} \partial_{\bar \psi_c} G^\text{QED} \big|_{\substack{\phi_c=0\\\psi_c=0}} } { - \partial_{\psi_c} \partial_{\bar \psi_c} G^\text{QED}\big|_{\substack{\phi_c=0\\\psi_c=0}} } \right)^2, \end{align*}
and the calculation of the renormalization quantities works as before. Some coefficients are listed in Table \ref{tab:QQEDren}.
\begin{table}
\begin{subtable}[c]{\textwidth}
\centering
\tiny
\def\arraystretch{1.5}
\begin{tabular}{|c||c|c|c|c|c|c|}
\hline
&$\hbar_{\text{ren}}^{0}$&$\hbar_{\text{ren}}^{1}$&$\hbar_{\text{ren}}^{2}$&$\hbar_{\text{ren}}^{3}$&$\hbar_{\text{ren}}^{4}$&$\hbar_{\text{ren}}^{5}$\\
\hline\hline
$\hbar(\hbar_{\text{ren}})$&$0$&$1$&$-4$&$8$&$-28$&$-48$\\
\hline
$z^{( \ifmmode \usebox{\fgsimplephotonprop} \else \newsavebox{\fgsimplephotonprop} \savebox{\fgsimplephotonprop}{ \begin{tikzpicture}[x=1ex,y=1ex,baseline={([yshift=-.5ex]current bounding box.center)}] \coordinate (v) ; \coordinate [right=1.2 of v] (u); \draw[photon] (v) -- (u); \end{tikzpicture} } \fi)}(\hbar_{\text{ren}})$&$1$&$1$&$-1$&$-1$&$-7$&$-63$\\
\hline
$z^{\left( \ifmmode \usebox{\fgsimpleqedvtx} \else \newsavebox{\fgsimpleqedvtx} \savebox{\fgsimpleqedvtx}{ \begin{tikzpicture}[x=1ex,y=1ex,baseline={([yshift=-.5ex]current bounding box.center)}] \coordinate (v) ; \def \rad {1}; \filldraw[white] (v) circle (\rad); \coordinate (u1) at ([shift=(180:\rad)]v); \coordinate (u2) at ([shift=(300:\rad)]v); \coordinate (u3) at ([shift=(60:\rad)]v); \draw[photon] (u1) -- (v); \draw[fermion] (u2) -- (v); \draw[fermion] (v) -- (u3); \filldraw (v) circle (1pt); \end{tikzpicture} } \fi\right)}(\hbar_{\text{ren}})$&$1$&$-1$&$-1$&$-7$&$-63$&$-729$\\
\hline
\end{tabular}
\subcaption{Table of the first coefficients of the renormalization quantities in quenched QED.}
\end{subtable}
\begin{subtable}[c]{\textwidth}
\centering
\tiny
\def\arraystretch{1.5}
\begin{tabular}{|c||c||c|c|c|c|c|c|}
\hline
&prefactor&$\hbar_{\text{ren}}^{0}$&$\hbar_{\text{ren}}^{1}$&$\hbar_{\text{ren}}^{2}$&$\hbar_{\text{ren}}^{3}$&$\hbar_{\text{ren}}^{4}$&$\hbar_{\text{ren}}^{5}$\\
\hline\hline
$\left(\asyOpV{\frac12}{0}{\hbar_{\text{ren}}} \hbar \right)(\hbar_{\text{ren}})$&$e^{-2} \frac{\hbar^{0}}{\sqrt{2\pi\hbar}}$&$-2$&$20$&$-62$&$ \frac{928}{3}$&$ \frac{2540}{3}$&$ \frac{330296}{15}$\\
\hline
$\left(\asyOpV{\frac12}{0}{\hbar_{\text{ren}}} z^{( \ifmmode \usebox{\fgsimplephotonprop} \else \newsavebox{\fgsimplephotonprop} \savebox{\fgsimplephotonprop}{ \begin{tikzpicture}[x=1ex,y=1ex,baseline={([yshift=-.5ex]current bounding box.center)}] \coordinate (v) ; \coordinate [right=1.2 of v] (u); \draw[photon] (v) -- (u); \end{tikzpicture} } \fi)} \right)(\hbar_{\text{ren}})$&$e^{-2} \frac{\hbar^{0}}{\sqrt{2\pi\hbar}}$&$-1$&$6$&$4$&$ \frac{218}{3}$&$890$&$ \frac{196838}{15}$\\
\hline
$\left(\asyOpV{\frac12}{0}{\hbar_{\text{ren}}} z^{\left( \ifmmode \usebox{\fgsimpleqedvtx} \else \newsavebox{\fgsimpleqedvtx} \savebox{\fgsimpleqedvtx}{ \begin{tikzpicture}[x=1ex,y=1ex,baseline={([yshift=-.5ex]current bounding box.center)}] \coordinate (v) ; \def \rad {1}; \filldraw[white] (v) circle (\rad); \coordinate (u1) at ([shift=(180:\rad)]v); \coordinate (u2) at ([shift=(300:\rad)]v); \coordinate (u3) at ([shift=(60:\rad)]v); \draw[photon] (u1) -- (v); \draw[fermion] (u2) -- (v); \draw[fermion] (v) -- (u3); \filldraw (v) circle (1pt); \end{tikzpicture} } \fi\right)} \right)(\hbar_{\text{ren}})$&$e^{-2} \frac{\hbar^{-1}}{\sqrt{2\pi\hbar}}$&$-1$&$6$&$4$&$ \frac{218}{3}$&$890$&$ \frac{196838}{15}$\\
\hline
\end{tabular}
\subcaption{Table of the first coefficients of the asymptotics of the renormalization quantities in quenched QED.}
\end{subtable}
\caption{Renormalization constants in quenched QED.}
\label{tab:QQEDren}
\end{table}
The sequence generated by $1-z^{\left( \ifmmode \usebox{\fgsimpleqedvtx} \else \newsavebox{\fgsimpleqedvtx} \savebox{\fgsimpleqedvtx}{ \begin{tikzpicture}[x=1ex,y=1ex,baseline={([yshift=-.5ex]current bounding box.center)}] \coordinate (v) ; \def \rad {1}; \filldraw[white] (v) circle (\rad); \coordinate (u1) at ([shift=(180:\rad)]v); \coordinate (u2) at ([shift=(300:\rad)]v); \coordinate (u3) at ([shift=(60:\rad)]v); \draw[photon] (u1) -- (v); \draw[fermion] (u2) -- (v); \draw[fermion] (v) -- (u3); \filldraw (v) circle (1pt); \end{tikzpicture} } \fi\right)}(\hbar_\text{ren})$, which enumerates the number of skeleton quenched QED vertex diagrams (Theorem \ref{thm:threeQFTmoebvert}), was also given in \cite{broadhurst1999four}. It is entry \texttt{A049464} in the OEIS \cite{oeis}.
The asymptotics, read off from Table \ref{tab:QQEDren}, of this sequence are,
\begin{gather*} [\hbar_\text{ren}^n]( 1-z^{\left( \ifmmode \usebox{\fgsimpleqedvtx} \else \newsavebox{\fgsimpleqedvtx} \savebox{\fgsimpleqedvtx}{ \begin{tikzpicture}[x=1ex,y=1ex,baseline={([yshift=-.5ex]current bounding box.center)}] \coordinate (v) ; \def \rad {1}; \filldraw[white] (v) circle (\rad); \coordinate (u1) at ([shift=(180:\rad)]v); \coordinate (u2) at ([shift=(300:\rad)]v); \coordinate (u3) at ([shift=(60:\rad)]v); \draw[photon] (u1) -- (v); \draw[fermion] (u2) -- (v); \draw[fermion] (v) -- (u3); \filldraw (v) circle (1pt); \end{tikzpicture} } \fi\right)}(\hbar_\text{ren})) \underset{n\rightarrow \infty}{\sim} e^{-2} (2n+1)!! \left( 1 - \frac{6}{2n+1} \right. \\ \left. - \frac{4}{(2n-1)(2n+1)} - \frac{218}{3}\frac{1}{(2n-3)(2n-1)(2n+1)} +\ldots \right), \end{gather*}
where we used $(2n-1)!! =\frac{2^{n+\frac12}}{\sqrt{2\pi}}\Gamma(n+\frac12)$. 
The first five coefficients of this expansion have been conjectured by Broadhurst \cite{davidexpansion} based on numerical calculations.
\subsection{Yukawa theory}

Note that on its own, Yukawa theory is not renormalizable. A $\varphi^4$ coupling must be included to to absorb the primitive divergencies from the four-meson function in pure Yukawa theory, beginning with the four-meson box: $\begin{tikzpicture}[x=2ex,y=2ex,baseline={([yshift=-.5ex]current bounding box.center)}] \coordinate (vt1); \coordinate [below=1 of vt1] (vb1); \coordinate [right=1 of vt1] (vt2); \coordinate [right=1 of vb1] (vb2); \coordinate [above left=.5 of vt1] (i0); \coordinate [below left=.5 of vb1] (i1); \coordinate [above right=.5 of vt2] (o0); \coordinate [below right=.5 of vb2] (o1); \draw[fermion] (vt1) -- (vt2); \draw[fermion] (vt2) -- (vb2); \draw[fermion] (vb2) -- (vb1); \draw[fermion] (vb1) -- (vt1); \draw[meson] (i0) -- (vt1); \draw[meson] (i1) -- (vb1); \draw[meson] (o0) -- (vt2); \draw[meson] (o1) -- (vb2); \filldraw (vt1) circle(1pt); \filldraw (vt2) circle(1pt); \filldraw (vb1) circle(1pt); \filldraw (vb2) circle(1pt); \end{tikzpicture}$. In this work, we will limit ourselves to the divergences of the two and three-point functions that can be renormalized by modifying the Yukawa coupling alone. Then, the combinatorics are similar to the case of QED without Furry's theorem. Note that gauge invariance protects QED from a primitive divergence of the four-photon amplitude.\footnote{I wish to thank David Broadhurst for noting this important point of non-renormalizability of Yukawa theory.}

The partition function of Yukawa theory in zero-dimensions is given by, 
\begin{align*} Z^{\text{Yuk}}(\hbar, j, \eta) &:= \int \frac{dx}{\sqrt{2 \pi \hbar}} e^{\frac{1}{\hbar} \left( - \frac{x^2}{2} + j x + \frac{|\eta|^2}{1-x} + \hbar \log \frac{1}{1-x} \right)} \end{align*}
Similarly, to the case of quenched QED, we can rewrite this with $\chi(\hbar)= \sum_{n=0}^\infty (2n-1)!! \hbar^n$ as
\begin{align*} Z^{\text{Yuk}}(\hbar, j, \eta) &= \frac{e^{\frac{j^2}{2\hbar}}}{1-j-\frac{|\eta|^2}{\hbar}} \chi\left(\frac{\hbar}{\left(1-j-\frac{|\eta|^2}{\hbar}\right)^2} \right) + \bigO(|\eta|^4), \end{align*}
where we expanded up to first order in $|\eta|^2$.

\begin{table}
\begin{subtable}[c]{\textwidth}
\centering
\tiny
\def\arraystretch{1.5}
\begin{tabular}{|c||c||c|c|c|c|c|c|}
\hline
&prefactor&$\hbar^{0}$&$\hbar^{1}$&$\hbar^{2}$&$\hbar^{3}$&$\hbar^{4}$&$\hbar^{5}$\\
\hline\hline
$\partial_j^{0} (\partial_{\eta} \partial_{ \bar \eta} )^{0} Z^{\text{Yuk}} \big|_{\substack{j=0\\\eta=0}}$&$\hbar^{0}$&$1$&$1$&$3$&$15$&$105$&$945$\\
\hline
$\partial_j^{1} (\partial_{\eta} \partial_{ \bar \eta} )^{0} Z^{\text{Yuk}} \big|_{\substack{j=0\\\eta=0}}$&$\hbar^{0}$&$1$&$3$&$15$&$105$&$945$&$10395$\\
\hline
$\partial_j^{2} (\partial_{\eta} \partial_{ \bar \eta} )^{0} Z^{\text{Yuk}} \big|_{\substack{j=0\\\eta=0}}$&$\hbar^{-1}$&$1$&$3$&$15$&$105$&$945$&$10395$\\
\hline
$\partial_j^{0} (\partial_{\eta} \partial_{ \bar \eta} )^{1} Z^{\text{Yuk}} \big|_{\substack{j=0\\\eta=0}}$&$\hbar^{-1}$&$1$&$3$&$15$&$105$&$945$&$10395$\\
\hline
$\partial_j^{1} (\partial_{\eta} \partial_{ \bar \eta} )^{1} Z^{\text{Yuk}} \big|_{\substack{j=0\\\eta=0}}$&$\hbar^{-1}$&$2$&$12$&$90$&$840$&$9450$&$124740$\\
\hline
\end{tabular}
\subcaption{The first coefficients of the trivariate generating function $Z^{\text{Yuk}}(\hbar, j, \eta)$.}
\end{subtable}
\begin{subtable}[c]{\textwidth}
\centering
\tiny
\def\arraystretch{1.5}
\begin{tabular}{|c||c||c|c|c|c|c|c|}
\hline
&prefactor&$\hbar^{0}$&$\hbar^{1}$&$\hbar^{2}$&$\hbar^{3}$&$\hbar^{4}$&$\hbar^{5}$\\
\hline\hline
$\asyOpV{\frac12}{0}{\hbar} \partial_j^{0} (\partial_{\eta} \partial_{ \bar \eta} )^{0} Z^{\text{Yuk}} \big|_{\substack{j=0\\\eta=0}}$&$\frac{\hbar^{0}}{\sqrt{2\pi\hbar}}$&$1$&$0$&$0$&$0$&$0$&$0$\\
\hline
$\asyOpV{\frac12}{0}{\hbar} \partial_j^{1} (\partial_{\eta} \partial_{ \bar \eta} )^{0} Z^{\text{Yuk}} \big|_{\substack{j=0\\\eta=0}}$&$\frac{\hbar^{-1}}{\sqrt{2\pi\hbar}}$&$1$&$0$&$0$&$0$&$0$&$0$\\
\hline
$\asyOpV{\frac12}{0}{\hbar} \partial_j^{2} (\partial_{\eta} \partial_{ \bar \eta} )^{0} Z^{\text{Yuk}} \big|_{\substack{j=0\\\eta=0}}$&$\frac{\hbar^{-2}}{\sqrt{2\pi\hbar}}$&$1$&$0$&$0$&$0$&$0$&$0$\\
\hline
$\asyOpV{\frac12}{0}{\hbar} \partial_j^{0} (\partial_{\eta} \partial_{ \bar \eta} )^{1} Z^{\text{Yuk}} \big|_{\substack{j=0\\\eta=0}}$&$\frac{\hbar^{-2}}{\sqrt{2\pi\hbar}}$&$1$&$0$&$0$&$0$&$0$&$0$\\
\hline
$\asyOpV{\frac12}{0}{\hbar} \partial_j^{1} (\partial_{\eta} \partial_{ \bar \eta} )^{1} Z^{\text{Yuk}} \big|_{\substack{j=0\\\eta=0}}$&$\frac{\hbar^{-3}}{\sqrt{2\pi\hbar}}$&$1$&$-1$&$0$&$0$&$0$&$0$\\
\hline
\end{tabular}
\subcaption{The first coefficients of the trivariate generating function $\asyOpV{\frac12}{0}{\hbar}Z^{\text{Yuk}}(\hbar, j, \eta)$.}
\end{subtable}
\caption{Partition function in Yukawa theory.}
\label{tab:ZYuk}
\end{table}

It follows from $\asyOpV{\frac12}{0}{\hbar} \chi(\hbar) = \frac{1}{\sqrt{2\pi \hbar}}$ and the chain rule that,
\begin{align*} \asyOpV{\frac12}{0}{\hbar} Z^{\text{Yuk}}(\hbar, j, \eta) &= \frac{1}{\sqrt{2\pi \hbar}} e^{\frac{j}{\hbar}} \left( 1 + |\eta|^2 \frac{1-j}{\hbar^2} \right) + \bigO(|\eta|^4) \end{align*}
As in the case of quenched QED, the asymptotic expansions for each order in $j$ and $|\eta|$ up to $|\eta|^2$ of the disconnected diagrams are finite and therefore exact. Some coefficients are given in Table \ref{tab:ZYuk}.
\begin{table}
\begin{subtable}[c]{\textwidth}
\centering
\tiny
\def\arraystretch{1.5}
\begin{tabular}{|c||c|c|c|c|c|c|}
\hline
&$\hbar^{0}$&$\hbar^{1}$&$\hbar^{2}$&$\hbar^{3}$&$\hbar^{4}$&$\hbar^{5}$\\
\hline\hline
$\partial_j^{0} (\partial_{\eta} \partial_{ \bar \eta} )^{0} W^{\text{Yuk}} \big|_{\substack{j=0\\\eta=0}}$&$0$&$0$&$1$&$ \frac{5}{2}$&$ \frac{37}{3}$&$ \frac{353}{4}$\\
\hline
$\partial_j^{1} (\partial_{\eta} \partial_{ \bar \eta} )^{0} W^{\text{Yuk}} \big|_{\substack{j=0\\\eta=0}}$&$0$&$1$&$2$&$10$&$74$&$706$\\
\hline
$\partial_j^{2} (\partial_{\eta} \partial_{ \bar \eta} )^{0} W^{\text{Yuk}} \big|_{\substack{j=0\\\eta=0}}$&$1$&$1$&$6$&$50$&$518$&$6354$\\
\hline
$\partial_j^{0} (\partial_{\eta} \partial_{ \bar \eta} )^{1} W^{\text{Yuk}} \big|_{\substack{j=0\\\eta=0}}$&$1$&$2$&$10$&$74$&$706$&$8162$\\
\hline
$\partial_j^{1} (\partial_{\eta} \partial_{ \bar \eta} )^{1} W^{\text{Yuk}} \big|_{\substack{j=0\\\eta=0}}$&$1$&$6$&$50$&$518$&$6354$&$89782$\\
\hline
\end{tabular}
\subcaption{The first coefficients of the trivariate generating function $W^{\text{Yuk}}(\hbar, j, \eta)$.}
\end{subtable}
\begin{subtable}[c]{\textwidth}
\centering
\tiny
\def\arraystretch{1.5}
\begin{tabular}{|c||c||c|c|c|c|c|c|}
\hline
&prefactor&$\hbar^{0}$&$\hbar^{1}$&$\hbar^{2}$&$\hbar^{3}$&$\hbar^{4}$&$\hbar^{5}$\\
\hline\hline
$\asyOpV{\frac12}{0}{\hbar} \partial_j^{0} (\partial_{\eta} \partial_{ \bar \eta} )^{0} W^{\text{Yuk}} \big|_{\substack{j=0\\\eta=0}}$&$\frac{\hbar^{1}}{\sqrt{2\pi\hbar}}$&$1$&$-1$&$-2$&$-10$&$-74$&$-706$\\
\hline
$\asyOpV{\frac12}{0}{\hbar} \partial_j^{1} (\partial_{\eta} \partial_{ \bar \eta} )^{0} W^{\text{Yuk}} \big|_{\substack{j=0\\\eta=0}}$&$\frac{\hbar^{0}}{\sqrt{2\pi\hbar}}$&$1$&$-2$&$-3$&$-16$&$-124$&$-1224$\\
\hline
$\asyOpV{\frac12}{0}{\hbar} \partial_j^{2} (\partial_{\eta} \partial_{ \bar \eta} )^{0} W^{\text{Yuk}} \big|_{\substack{j=0\\\eta=0}}$&$\frac{\hbar^{-1}}{\sqrt{2\pi\hbar}}$&$1$&$-4$&$-3$&$-22$&$-188$&$-1968$\\
\hline
$\asyOpV{\frac12}{0}{\hbar} \partial_j^{0} (\partial_{\eta} \partial_{ \bar \eta} )^{1} W^{\text{Yuk}} \big|_{\substack{j=0\\\eta=0}}$&$\frac{\hbar^{-1}}{\sqrt{2\pi\hbar}}$&$1$&$-2$&$-3$&$-16$&$-124$&$-1224$\\
\hline
$\asyOpV{\frac12}{0}{\hbar} \partial_j^{1} (\partial_{\eta} \partial_{ \bar \eta} )^{1} W^{\text{Yuk}} \big|_{\substack{j=0\\\eta=0}}$&$\frac{\hbar^{-2}}{\sqrt{2\pi\hbar}}$&$1$&$-4$&$-3$&$-22$&$-188$&$-1968$\\
\hline
\end{tabular}
\subcaption{The first coefficients of the trivariate generating function $\asyOpV{\frac12}{0}{\hbar}W^{\text{Yuk}}(\hbar, j, \eta)$.}
\end{subtable}
\caption{Free energy in Yukawa theory.}
\label{tab:WYuk}
\end{table}%
The free energy is defined as usual,
\begin{gather*} W^{\text{Yuk}}(\hbar, j, \eta) = \hbar \log Z^{\text{Yuk}}(\hbar, j, \eta) = \\ \frac{j^2}{2} + \hbar \log\frac{1}{1-j-\frac{|\eta|^2}{\hbar}} + \hbar \log \chi \left(\frac{\hbar}{\left( 1 - j - \frac{|\eta|^2}{\hbar} \right)^2 }\right) + \bigO(|\eta|^4), \end{gather*}
Its asymptotics are given by,
\begin{align*} \asyOpV{\frac12}{0}{\hbar} W^{\text{Yuk}}(\hbar, j, \eta) &=       \frac{\hbar}{\sqrt{2\pi \hbar}} e^{\frac{j - \frac{j^2}{2} }{\hbar}} \frac{1-j- \frac{|\eta|^2}{\hbar} \left(1-\frac{(1-j)^2}{\hbar} \right)}{\chi\left( \frac{\hbar}{(1-j - \frac{|\eta|^2}{\hbar})^2} \right)} + \bigO(|\eta|^4) \end{align*}
Some coefficients are given in Table \ref{tab:WYuk}.
The 1PI effective action is given by the Legendre transformation of $W^{\text{Yuk}}(\hbar, j, \eta)$. 
\begin{align*} G^{\text{Yuk}}(\hbar, \phi_c, \psi_c) &= W^{\text{Yuk}}(\hbar, j, \eta) - j \phi_c - \bar \eta \psi_c - \eta \bar \psi_c, \end{align*}
where $j, \eta, \phi_c$ and $\psi_c$ are related by the equations,
$\phi_c = \partial_j W^{\text{Yuk}}$ and $\psi_c = \partial_{\bar\eta} W^{\text{Yuk}}$.
The $\phi_c$ variable counts the number of meson legs $ \ifmmode \usebox{\fgsimplemesonprop} \else \newsavebox{\fgsimplemesonprop} \savebox{\fgsimplemesonprop}{ \begin{tikzpicture}[x=1ex,y=1ex,baseline={([yshift=-.5ex]current bounding box.center)}] \coordinate (v) ; \coordinate [right=1.2 of v] (u); \draw[meson] (v) -- (u); \end{tikzpicture} } \fi$ and the variables $\psi_c$ and $\bar \psi_c$ the numbers of fermion legs as before. 

Performing this Legendre transform is non-trivial in contrast to the preceding three examples, because we can have graphs with one leg as in the case of $\varphi^3$-theory.

As for $\varphi^3$-theory, we define 
\begin{align*} \gamma^{\text{Yuk}}_0(\hbar) &:= \frac{G^{\text{Yuk}}\big|_{\substack{\phi_c=0\\ \psi_c=0}}}{\hbar} = \frac{W^{\text{Yuk}}\big|_{\substack{j=j_0\\ \eta = 0}}}{\hbar} \\ &= \frac{j_0(\hbar)^2}{2 \hbar} + \log\frac{1}{1-j_0(\hbar)} + \log \chi \left(\frac{\hbar}{\left( 1 - j_0(\hbar) \right)^2 }\right), \end{align*}
where $j_0(\hbar)$ is the power series solution of 
$0 = \partial_j W^{\text{Yuk}}\big|_{j=j_0(\hbar)}$.
This gives 
\begin{gather*} G^{\text{Yuk}}(\hbar, \phi_c, \psi_c) = \\ - \frac{\phi_c^2}{2} +\hbar \log \frac{1}{1-\phi_c} + \hbar \gamma^{\text{Yuk}}_0\left( \frac{\hbar}{(1-\phi_c)^2} \right) + \hbar \frac{|\psi_c|^2(1-\phi_c)}{j_0\left(\frac{\hbar}{(1-\phi_c)^2} \right)} + \bigO(|\psi_c|^4). \end{gather*}
This equation also has a simple combinatorial interpretation: Every fermion line of a vacuum diagram can be dressed with an arbitrary number of mesons legs associated to a $\frac{1}{1-\phi_c}$ factor. Every additional loop gives two additional fermion propagators. The first two terms compensate for the fact that there are no vacuum diagrams with zero or one loop.

The asymptotics result from an application of the $\asyOpV{}{}{}$-derivative. Some coefficients are listed in Table \ref{tab:GYuk}. These sequences were also studied in \cite{kuchinskii1998combinatorial}. They obtained the constant, $e^{-1}$, and the linear coefficients $-\frac92$ and $-\frac{5}{2}$ for the $1$ and $2$-point functions using a combination of numerical and analytic techniques. 
\begin{table}
\begin{subtable}[c]{\textwidth}
\centering
\tiny
\def\arraystretch{1.5}
\begin{tabular}{|c||c|c|c|c|c|c|}
\hline
&$\hbar^{0}$&$\hbar^{1}$&$\hbar^{2}$&$\hbar^{3}$&$\hbar^{4}$&$\hbar^{5}$\\
\hline\hline
$\partial_{\phi_c}^{0} (\partial_{\psi_c} \partial_{ \bar \psi_c} )^{0} G^{\text{Yuk}} \big|_{\substack{\phi_c=0\\\psi_c=0}}$&$0$&$0$&$ \frac{1}{2}$&$1$&$ \frac{9}{2}$&$31$\\
\hline
$\partial_{\phi_c}^{1} (\partial_{\psi_c} \partial_{ \bar \psi_c} )^{0} G^{\text{Yuk}} \big|_{\substack{\phi_c=0\\\psi_c=0}}$&$0$&$1$&$1$&$4$&$27$&$248$\\
\hline
$\partial_{\phi_c}^{2} (\partial_{\psi_c} \partial_{ \bar \psi_c} )^{0} G^{\text{Yuk}} \big|_{\substack{\phi_c=0\\\psi_c=0}}$&$-1$&$1$&$3$&$20$&$189$&$2232$\\
\hline
$\partial_{\phi_c}^{0} (\partial_{\psi_c} \partial_{ \bar \psi_c} )^{1} G^{\text{Yuk}} \big|_{\substack{\phi_c=0\\\psi_c=0}}$&$-1$&$1$&$3$&$20$&$189$&$2232$\\
\hline
$\partial_{\phi_c}^{1} (\partial_{\psi_c} \partial_{ \bar \psi_c} )^{1} G^{\text{Yuk}} \big|_{\substack{\phi_c=0\\\psi_c=0}}$&$1$&$1$&$9$&$100$&$1323$&$20088$\\
\hline
\end{tabular}
\subcaption{The first coefficients of the trivariate generating function $G^{\text{Yuk}}(\hbar, \phi_c, \psi_c)$.}
\end{subtable}
\begin{subtable}[c]{\textwidth}
\centering
\tiny
\def\arraystretch{1.5}
\begin{tabular}{|c||c||c|c|c|c|c|c|}
\hline
&prefactor&$\hbar^{0}$&$\hbar^{1}$&$\hbar^{2}$&$\hbar^{3}$&$\hbar^{4}$&$\hbar^{5}$\\
\hline\hline
$\asyOpV{\frac12}{0}{\hbar} \partial_{\phi_c}^{0} (\partial_{\psi_c} \partial_{ \bar \psi_c} )^{0} G^{\text{Yuk}} \big|_{\substack{\phi_c=0\\\psi_c=0}}$&$e^{-1} \frac{\hbar^{1}}{\sqrt{2\pi\hbar}}$&$1$&$- \frac{3}{2}$&$- \frac{31}{8}$&$- \frac{393}{16}$&$- \frac{28757}{128}$&$- \frac{3313201}{1280}$\\
\hline
$\asyOpV{\frac12}{0}{\hbar} \partial_{\phi_c}^{1} (\partial_{\psi_c} \partial_{ \bar \psi_c} )^{0} G^{\text{Yuk}} \big|_{\substack{\phi_c=0\\\psi_c=0}}$&$e^{-1} \frac{\hbar^{0}}{\sqrt{2\pi\hbar}}$&$1$&$- \frac{5}{2}$&$- \frac{43}{8}$&$- \frac{579}{16}$&$- \frac{44477}{128}$&$- \frac{5326191}{1280}$\\
\hline
$\asyOpV{\frac12}{0}{\hbar} \partial_{\phi_c}^{2} (\partial_{\psi_c} \partial_{ \bar \psi_c} )^{0} G^{\text{Yuk}} \big|_{\substack{\phi_c=0\\\psi_c=0}}$&$e^{-1} \frac{\hbar^{-1}}{\sqrt{2\pi\hbar}}$&$1$&$- \frac{9}{2}$&$- \frac{43}{8}$&$- \frac{751}{16}$&$- \frac{63005}{128}$&$- \frac{7994811}{1280}$\\
\hline
$\asyOpV{\frac12}{0}{\hbar} \partial_{\phi_c}^{0} (\partial_{\psi_c} \partial_{ \bar \psi_c} )^{1} G^{\text{Yuk}} \big|_{\substack{\phi_c=0\\\psi_c=0}}$&$e^{-1} \frac{\hbar^{-1}}{\sqrt{2\pi\hbar}}$&$1$&$- \frac{9}{2}$&$- \frac{43}{8}$&$- \frac{751}{16}$&$- \frac{63005}{128}$&$- \frac{7994811}{1280}$\\
\hline
$\asyOpV{\frac12}{0}{\hbar} \partial_{\phi_c}^{1} (\partial_{\psi_c} \partial_{ \bar \psi_c} )^{1} G^{\text{Yuk}} \big|_{\substack{\phi_c=0\\\psi_c=0}}$&$e^{-1} \frac{\hbar^{-2}}{\sqrt{2\pi\hbar}}$&$1$&$- \frac{17}{2}$&$ \frac{29}{8}$&$- \frac{751}{16}$&$- \frac{75021}{128}$&$- \frac{10515011}{1280}$\\
\hline
\end{tabular}
\subcaption{The first coefficients of the trivariate generating function $\asyOpV{\frac12}{0}{\hbar}G^{\text{Yuk}}(\hbar, \phi_c, \psi_c)$.}
\end{subtable}
\caption{Effective action in Yukawa theory.}
\label{tab:GYuk}
\end{table}

The calculation of the renormalization constants proceeds as in the other cases with the invariant charge defined as for QED. The first coefficients are listed in Table \ref{tab:Yukren}.

\begin{table}
\begin{subtable}[c]{\textwidth}
\centering
\tiny
\def\arraystretch{1.5}
\begin{tabular}{|c||c|c|c|c|c|c|}
\hline
&$\hbar_{\text{ren}}^{0}$&$\hbar_{\text{ren}}^{1}$&$\hbar_{\text{ren}}^{2}$&$\hbar_{\text{ren}}^{3}$&$\hbar_{\text{ren}}^{4}$&$\hbar_{\text{ren}}^{5}$\\
\hline\hline
$\hbar(\hbar_{\text{ren}})$&$0$&$1$&$-5$&$10$&$-36$&$-164$\\
\hline
$z^{\left( \ifmmode \usebox{\fgsimplemesonprop} \else \newsavebox{\fgsimplemesonprop} \savebox{\fgsimplemesonprop}{ \begin{tikzpicture}[x=1ex,y=1ex,baseline={([yshift=-.5ex]current bounding box.center)}] \coordinate (v) ; \coordinate [right=1.2 of v] (u); \draw[meson] (v) -- (u); \end{tikzpicture} } \fi\right)}(\hbar_{\text{ren}})$&$1$&$1$&$-1$&$-3$&$-13$&$-147$\\
\hline
$z^{\left( \ifmmode \usebox{\fgsimplefermionprop} \else \newsavebox{\fgsimplefermionprop} \savebox{\fgsimplefermionprop}{ \begin{tikzpicture}[x=1ex,y=1ex,baseline={([yshift=-.5ex]current bounding box.center)}] \coordinate (v) ; \coordinate [right=1.2 of v] (u); \draw[fermion] (v) -- (u); \end{tikzpicture} } \fi\right)}(\hbar_{\text{ren}})$&$1$&$1$&$-1$&$-3$&$-13$&$-147$\\
\hline
$z^{\left( \ifmmode \usebox{\fgsimpleyukvtx} \else \newsavebox{\fgsimpleyukvtx} \savebox{\fgsimpleyukvtx}{ \begin{tikzpicture}[x=1ex,y=1ex,baseline={([yshift=-.5ex]current bounding box.center)}] \coordinate (v) ; \def \rad {1}; \filldraw[white] (v) circle (\rad); \coordinate (u1) at ([shift=(180:\rad)]v); \coordinate (u2) at ([shift=(300:\rad)]v); \coordinate (u3) at ([shift=(60:\rad)]v); \draw[meson] (u1) -- (v); \draw[fermion] (u2) -- (v); \draw[fermion] (v) -- (u3); \filldraw (v) circle (1pt); \end{tikzpicture} } \fi\right)}(\hbar_{\text{ren}})$&$1$&$-1$&$-3$&$-13$&$-147$&$-1965$\\
\hline
\end{tabular}
\subcaption{Table of the first coefficients of the renormalization quantities in Yukawa theory.}
\end{subtable}
\begin{subtable}[c]{\textwidth}
\centering
\tiny
\def\arraystretch{1.5}
\begin{tabular}{|c||c||c|c|c|c|c|c|}
\hline
&prefactor&$\hbar_{\text{ren}}^{0}$&$\hbar_{\text{ren}}^{1}$&$\hbar_{\text{ren}}^{2}$&$\hbar_{\text{ren}}^{3}$&$\hbar_{\text{ren}}^{4}$&$\hbar_{\text{ren}}^{5}$\\
\hline\hline
$\left(\asyOpV{\frac12}{0}{\hbar_{\text{ren}}} \hbar \right)(\hbar_{\text{ren}})$&$e^{- \frac{7}{2}} \frac{\hbar^{-1}}{\sqrt{2 \pi\hbar}}$&$-2$&$26$&$- \frac{377}{4}$&$ \frac{963}{2}$&$ \frac{140401}{64}$&$ \frac{16250613}{320}$\\
\hline
$\left(\asyOpV{\frac12}{0}{\hbar_{\text{ren}}} z^{\left( \ifmmode \usebox{\fgsimplemesonprop} \else \newsavebox{\fgsimplemesonprop} \savebox{\fgsimplemesonprop}{ \begin{tikzpicture}[x=1ex,y=1ex,baseline={([yshift=-.5ex]current bounding box.center)}] \coordinate (v) ; \coordinate [right=1.2 of v] (u); \draw[meson] (v) -- (u); \end{tikzpicture} } \fi\right)} \right)(\hbar_{\text{ren}})$&$e^{- \frac{7}{2}} \frac{\hbar^{-1}}{\sqrt{2 \pi\hbar}}$&$-1$&$ \frac{15}{2}$&$ \frac{97}{8}$&$ \frac{1935}{16}$&$ \frac{249093}{128}$&$ \frac{42509261}{1280}$\\
\hline
$\left(\asyOpV{\frac12}{0}{\hbar_{\text{ren}}} z^{\left( \ifmmode \usebox{\fgsimplefermionprop} \else \newsavebox{\fgsimplefermionprop} \savebox{\fgsimplefermionprop}{ \begin{tikzpicture}[x=1ex,y=1ex,baseline={([yshift=-.5ex]current bounding box.center)}] \coordinate (v) ; \coordinate [right=1.2 of v] (u); \draw[fermion] (v) -- (u); \end{tikzpicture} } \fi\right)} \right)(\hbar_{\text{ren}})$&$e^{- \frac{7}{2}} \frac{\hbar^{-1}}{\sqrt{2 \pi\hbar}}$&$-1$&$ \frac{15}{2}$&$ \frac{97}{8}$&$ \frac{1935}{16}$&$ \frac{249093}{128}$&$ \frac{42509261}{1280}$\\
\hline
$\left(\asyOpV{\frac12}{0}{\hbar_{\text{ren}}} z^{\left( \ifmmode \usebox{\fgsimpleyukvtx} \else \newsavebox{\fgsimpleyukvtx} \savebox{\fgsimpleyukvtx}{ \begin{tikzpicture}[x=1ex,y=1ex,baseline={([yshift=-.5ex]current bounding box.center)}] \coordinate (v) ; \def \rad {1}; \filldraw[white] (v) circle (\rad); \coordinate (u1) at ([shift=(180:\rad)]v); \coordinate (u2) at ([shift=(300:\rad)]v); \coordinate (u3) at ([shift=(60:\rad)]v); \draw[meson] (u1) -- (v); \draw[fermion] (u2) -- (v); \draw[fermion] (v) -- (u3); \filldraw (v) circle (1pt); \end{tikzpicture} } \fi\right)} \right)(\hbar_{\text{ren}})$&$e^{- \frac{7}{2}} \frac{\hbar^{-2}}{\sqrt{2 \pi\hbar}}$&$-1$&$ \frac{15}{2}$&$ \frac{97}{8}$&$ \frac{1935}{16}$&$ \frac{249093}{128}$&$ \frac{42509261}{1280}$\\
\hline
\end{tabular}
\subcaption{Table of the first coefficients of the asymptotics of the renormalization quantities in Yukawa theory.}
\end{subtable}
\caption{Renormalization constants in Yukawa theory.}
\label{tab:Yukren}
\end{table}

In \cite{molinari2005hedin,molinari2006enumeration} various low-order coefficients, which were obtained in this section, were enumerated using Hedin's equations \cite{hedin1965new}. The numerical results for the asymptotics given in \cite{molinari2006enumeration} agree with the analytic results obtained here. The $\Gamma(x)$ expansion of  \cite{molinari2006enumeration} corresponds to the generating function $\partial_{\phi_c}\partial_{\psi_c} \partial_{\bar \psi_c} G^\text{Yuk} \big|_{\substack{\phi_c=0\\\psi_c=0}}(\hbar)$ and the $\Gamma(u)$ expansion to the generating function $2-z^{\left( \ifmmode \usebox{\fgsimpleyukvtx} \else \newsavebox{\fgsimpleyukvtx} \savebox{\fgsimpleyukvtx}{ \begin{tikzpicture}[x=1ex,y=1ex,baseline={([yshift=-.5ex]current bounding box.center)}] \coordinate (v) ; \def \rad {1}; \filldraw[white] (v) circle (\rad); \coordinate (u1) at ([shift=(180:\rad)]v); \coordinate (u2) at ([shift=(300:\rad)]v); \coordinate (u3) at ([shift=(60:\rad)]v); \draw[meson] (u1) -- (v); \draw[fermion] (u2) -- (v); \draw[fermion] (v) -- (u3); \filldraw (v) circle (1pt); \end{tikzpicture} } \fi\right)}(\hbar)$. The later is the generating function of all skeleton diagrams in Yukawa theory (Theorem \ref{thm:threeQFTmoebvert}). Written traditionally the asymptotics are, 
\begin{gather*} [\hbar_\text{ren}^n]( 1-z^{\left( \ifmmode \usebox{\fgsimpleyukvtx} \else \newsavebox{\fgsimpleyukvtx} \savebox{\fgsimpleyukvtx}{ \begin{tikzpicture}[x=1ex,y=1ex,baseline={([yshift=-.5ex]current bounding box.center)}] \coordinate (v) ; \def \rad {1}; \filldraw[white] (v) circle (\rad); \coordinate (u1) at ([shift=(180:\rad)]v); \coordinate (u2) at ([shift=(300:\rad)]v); \coordinate (u3) at ([shift=(60:\rad)]v); \draw[meson] (u1) -- (v); \draw[fermion] (u2) -- (v); \draw[fermion] (v) -- (u3); \filldraw (v) circle (1pt); \end{tikzpicture} } \fi\right)}(\hbar_\text{ren})) \underset{n\rightarrow \infty}{\sim} e^{-\frac{7}{2}} (2n+3)!! \left( 1 - \frac{15}{2} \frac{1}{2n+3} \right. \\ \left. -\frac{97}{8}\frac{1}{(2n+1)(2n+3)} - \frac{1935}{16}\frac{1}{(2n-1)(2n+1)(2n+3)} +\ldots \right). \end{gather*}

\printbibliography

\cleardoublepage
\markboth{\nomname}{\nomname}
\printnomenclature

\end{document}